\documentclass[%
 reprint,
 amsmath,amssymb,
 aps,
 superscriptaddress,pra
 ]{revtex4-2}

\usepackage{mypackages}
\usepackage[resetlabels]{multibib}
\newcites{SI}{Supplementary References}
\makeatletter\AtBeginDocument{\let\thebibliography\rtx@thebibliography\let\endthebibliography\endrtx@thebibliography}\makeatother

\graphicspath{{figures/}}

\usepackage{float}

\begin{document}

\title{Low-Overhead Quantum Error Correction with Boundary-Connected Planar Modules}

\author{Oscar Higgott}
\email{oscarhiggott@google.com}
\affiliation{Google Quantum AI}
\author{Hasan Sayginel}
\email{hasansayginel@google.com}
\affiliation{Google Quantum AI}
\author{Francisco J.H. Heras}
\affiliation{Google DeepMind}
\author{Zhiyang He}
\affiliation{Google Quantum AI}
\affiliation{Department of Mathematics, Massachusetts Institute of Technology}
\author{Tomas~Jochym-O'Connor}
\affiliation{Google Quantum AI}
\author{Andrew W. Senior}
\affiliation{Google DeepMind}
\author{Lei M. Zhang}
\affiliation{Google DeepMind}
\author{Thomas Edlich}
\affiliation{Google DeepMind}
\author{James S. Spencer}
\affiliation{Google DeepMind}
\author{Matt~McEwen}
\affiliation{Google Quantum AI}
\author{Craig Gidney}
\affiliation{Google Quantum AI}
\author{Johannes Bausch}
\affiliation{Google DeepMind}
\author{Pushmeet Kohli}
\affiliation{Google DeepMind}
\author{Hartmut Neven}
\affiliation{Google Quantum AI}

\date{\today}

\begin{abstract}
Realizing practical quantum computers requires quantum error correction, but the most widely implemented approach, the planar surface code, demands a substantial physical qubit overhead. 
Here, we demonstrate that partitioning quantum processors into manageable,
flat modules with non-local inter-module connections provides a natural
solution.
By wiring sparse, static boundary connections between Euclidean planar modules, we construct modular hyperbolic surface and color codes, based on new families of semi-hyperbolic codes. 
Using modular syndrome extraction circuits and efficient neural network and matching decoders, our circuit-level simulations show that this modular memory design can reduce the physical qubit overhead by tenfold or more compared to the surface code, even under elevated inter-module seam error rates. 
Projecting to larger system sizes, we find modular codes with encoding rate $k/n=1/16$ and distance $d\geq 22$, exceeding both the rate and distance of the $[[288,12,18]]$ two-gross bivariate bicycle code while using mostly nearest-neighbor gates within planar modules, and implying an over $30\times$ overhead reduction relative to surface codes.
We develop fault-tolerant walking circuits for implementing logic via code automorphisms, which we co-design with low-weight, highly symmetrical logical bases and a modular extractor system that is $\approx 4.5\times$ smaller than the code itself.
Combining these methods, our modular memory can be used as dense storage in a universal fault-tolerant architecture, unlocking the efficiency of high-rate codes without sacrificing the fabrication advantages of planar modules.
\end{abstract}
           
\maketitle

Quantum computers can solve certain computational problems exponentially faster than any known classical algorithm \cite{deutsch_jozsa,shor}, with anticipated applications in quantum simulation for material science, drug design and other areas \cite{lloyd,quantum_chemistry_review}, while also presenting significant challenges to current cryptographic infrastructure \cite{shor}. However, realizing these applications requires quantum error correction (QEC), in which many physical qubits are encoded into logical qubits to protect information from inevitable noise and imperfections in the hardware. 

\begin{figure*}
    \centering
    \includegraphics[width=\linewidth]{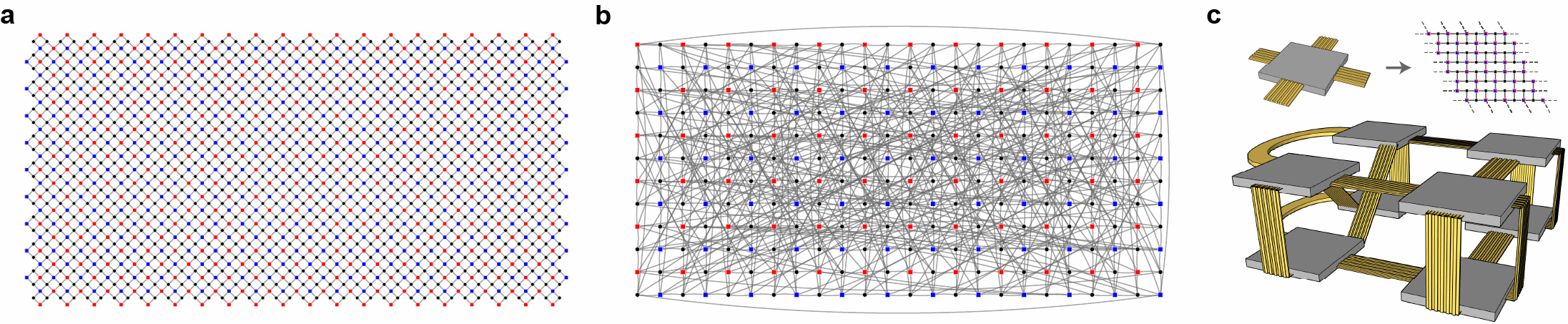}
    \caption{Hardware models. In all panels, black circles represent data qubits and red (blue) squares represent measure qubits for $X$ ($Z$) stabilizers. \textbf{a}, A monolithic grid architecture, in which qubits are placed on a plane with 2D nearest-neighbor grid connectivity. This geometry supports the Euclidean surface code, shown here. \textbf{b}, A general QLDPC architecture, in which each qubit is coupled to a constant number of other qubits, without additional constraints on geometric locality. Couplers may cross and may be long-range, as shown here for the Gross code~\cite{bivariate_bicycle_codes}. Each edge represents a coupler between qubits. \textbf{c},~A boundary-connected modular architecture. Planar modules are connected by flexible inter-module couplers routed without geometric constraints. The wiring pattern realizes a closed hyperbolic surface with genus $g\geq2$. The inset shows the planar connectivity within a single module.}
    \label{fig:hardware_model}
\end{figure*}

The surface code has led experimental efforts in this direction, owing to its compatibility with a square-grid layout on a planar chip, and recently enabled the first experimental demonstrations of exponential suppression of the logical error rate at physical error rates below threshold~\cite{surface_code_below_threshold}. 
However, the surface code requires a substantial resource overhead to achieve logical error rates relevant to practical applications, with utility-scale fault-tolerant quantum computers expected to require around 100k to 1 million qubits~\cite{gidney2025factor,low2026denserplanarsurfacecode}.
Furthermore, fabricating a monolithic chip at this scale becomes impractical, due to fundamental constraints on fabrication, yield and coupler density (see \Cref{fig:hardware_model}a), motivating modular architectures in which smaller units are networked together \cite{nickerson2013topological, ramette2024fault, strikis_2023}.

An alternative path to fault-tolerant quantum computation with lower overhead lies in the use of more general, high-rate quantum low-density parity-check (QLDPC) codes~\cite{breuckmanqldpc}.
In principle these codes can substantially reduce resource overheads relative to the surface code~\cite{bivariate_bicycle_codes}, however their implementation in solid state systems requires a high density of long-range crossing couplers (see \Cref{fig:hardware_model}b).

Here we demonstrate that the practical necessity of partitioning a quantum computer into modules can be leveraged to dramatically reduce the resource requirements of QEC.
We show that high encoding rates can be achieved using flat Euclidean modules of around 100 qubits, with sparse and fixed non-local couplers connecting qubits at module boundaries (see \Cref{fig:hardware_model}c).
This connectivity is motivated by state-of-the-art superconducting quantum computers, for which local planar gates now operate below the surface code threshold~\cite{surface_code_below_threshold}, while link resonators connecting chips through cables have been demonstrated to achieve error rates approaching 1\% \cite{Zhong_2021, Niu_2023, Qiu_2025}.

Our constructions are derived from color codes and surface codes on high-genus hyperbolic surfaces~\cite{freedman2002z2,breuckmann_2016_constructions,delfosse_tradeoffs,Breuckmann_2017_semihyperbolic,higgott_2024_floquet}.
Each module forms a locally flat Euclidean tile of the overall QLDPC code.
These tiles are connected using inter-module couplers to form the global connectivity of the hyperbolic tiling~\cite{higgott_2024_floquet,sutcliffe_2025_distributed}.
We design modular syndrome extraction circuits for these codes that are tailored to our distributed hardware model and develop efficient and accurate machine learning \cite{bausch2024learning, senior2025scalablerealtimeneuraldecoder} and matching decoders \cite{Higgott2025sparseblossom}.
Our circuit-level simulations demonstrate a $10\times$ reduction in physical qubit
overhead compared to the surface code at a logical error rate of $10^{-10}$,
assuming physical error rates expected for near-term experiments (SI1000 noise model with $p=0.1\%$, see Methods)
and inter-module gates that are ten times noisier ($p_{seam}=1\%$).
This is achieved with a modular hyperbolic surface code of 120 modules of 116 qubits each
(see \Cref{fig:frontier_codes_plots}), using multi-rate syndrome extraction circuits
that show substantial tolerance to slow and noisy inter-module couplers.
Projecting to larger system sizes, we find these savings scale to an over $30\times$
overhead reduction, requiring only 36 physical qubits per logical qubit,
including measure qubits, to achieve a distance of 20--24.
This is both lower overhead and higher distance than the [[288,12,18]] two-gross bivariate bicycle code (48 physical qubits per logical qubit, including check qubits)~\cite{bivariate_bicycle_codes,yoder2025tour}, despite requiring much simpler connectivity that is mostly local in 2D.
By comparison, a Euclidean surface code at the same distance requires over 1000
physical qubits per logical qubit.
Previous work using 2D nearest-neighbor gates~\cite{yoked_surface_code,low2026denserplanarsurfacecode,geher2025directional,gu2026nearest,nixon2026vine} or hyperbolic codes~\cite{higgott_2021_subsystem,higgott_2024_floquet,flag_proxy_hyperbolic_color} has demonstrated qubit overhead reductions of up to around $5\times$ at the circuit level.

Furthermore, we show how our modular QLDPC codes can be augmented to an architecture capable of implementing universal fault-tolerant computation.
We design fault-tolerant circuits for implementing logical gates via code automorphisms, corresponding to symmetries of the hyperbolic tiling that permute the modules, using only nearest-neighbor gates within each module and one additional seam of inter-module couplers on each side of a module.
We further design low-weight, highly symmetrical logical bases, as well as modular extractor systems that can measure logical Pauli operators via code surgery.
These techniques are combined to enable a single extractor system to access hundreds of distinct logical Pauli operators, allowing logical qubits to be teleported into and out of the code, and enabling universal computation with low space overhead.
All of these methods can be implemented using hardware components that have already been demonstrated with fidelity very close to those assumed here \cite{surface_code_below_threshold,Zhong_2021, Niu_2023, Qiu_2025}.

\section{Modular hyperbolic codes}

\begin{figure*}
    \centering
    \includegraphics[width=\linewidth]{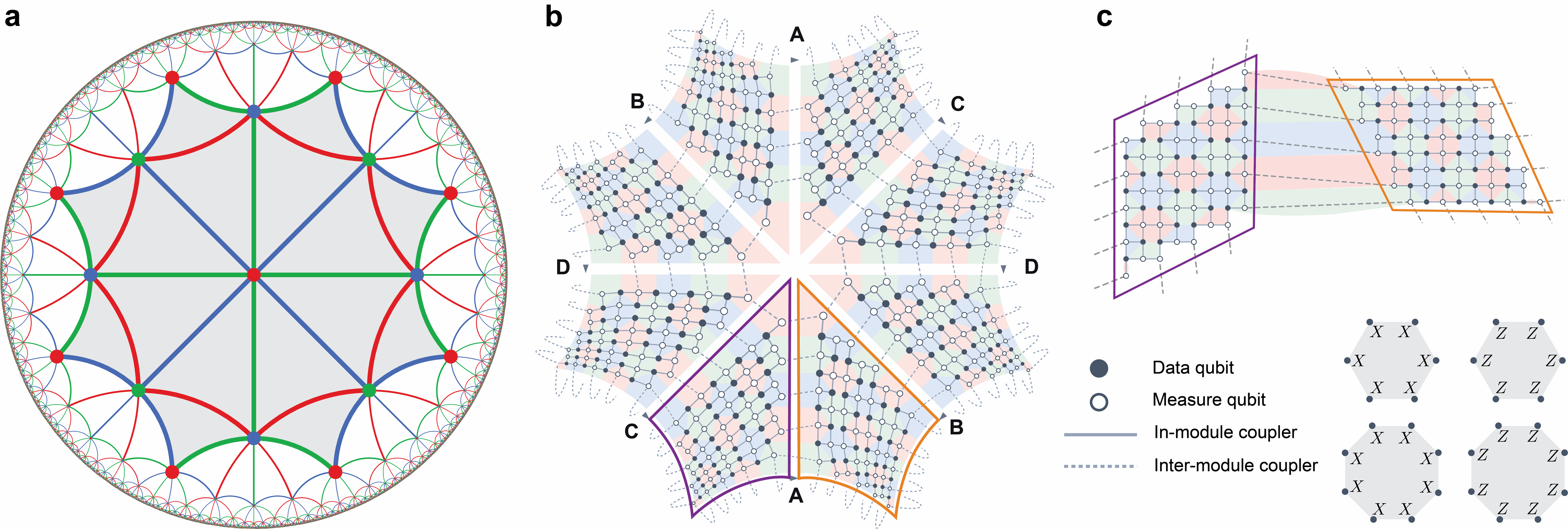}
    \caption{Construction of a modular hyperbolic color code, using the $[[256,8,12]]$ Bolza color code as an example. \textbf{a}, The base infinite hyperbolic tiling, shown in the Poincar\'e disk model, has triangular faces with 8 meeting at each vertex (face configuration V8.8.8). Its vertices are three-colorable (red, green, blue), with one vertex of each color per triangle. Each edge is assigned the third color, absent from its two vertices. The fundamental polygon of the genus-2 Bolza surface is shaded, and contains 16 triangles. \textbf{b}, The semi-hyperbolic color code obtained by replacing each triangular face with a Euclidean hexagonal tiling of side length $L=4$, giving a tiling of the surface by hexagons and octagons. A data qubit is associated with each vertex, and both an $X$ and a $Z$ stabilizer with each face. Sides of the fundamental polygon that are identified carry the same label. The code is partitioned into 8 modules, each derived from a pair of triangles of the base tiling sharing an edge of a chosen color (here red edges). \textbf{c}, The two modules outlined in \textbf{b} shown in detail. Two measure qubits are placed in each hexagonal face within a module, with additional measure qubits on the boundary. Qubits within a module are coupled by nearest-neighbor in-module couplers on a 2D grid (solid lines), while stabilizers spanning multiple modules are measured using inter-module couplers at the boundary (dashed lines).}
    \label{fig:bolza_color_code_overview}
\end{figure*}

Our codes are defined on tilings of closed hyperbolic surfaces, i.e.~closed surfaces of constant negative curvature, which have genus $g\geq2$.
Codes defined on such tilings encode $k$ logical qubits into $n$ physical qubits with a constant encoding rate $k/n$~\cite{freedman2002z2,delfosse_tradeoffs,breuckmann_2016_constructions}.
While other QLDPC code families are known to have constant rate asymptotically~\cite{tillich2014hgp,breuckmann2021product,panteleev2022good}, the rate for hyperbolic codes is unusually high in practice, approaching $1/2$ for color codes on 12.12.12 tilings~\cite{delfosse_tradeoffs,breuckmann_2016_constructions}.
For example, our search finds $[[n,k,d]]=[[648,328,8]]$ and $[[672,172,10]]$ hyperbolic color codes (see \Cref{tab:candidates_L1_color}), where $d$ is the code distance, which have rates comparable to recently proposed ultra-high-rate QLDPC codes~\cite{Kasai2026breaking,zhao2026ultrahighratequantumerrorcorrection}.
Although their distance grows logarithmically with $n$, for practical system sizes it is only a factor of two to three below the distances of around 20--24 needed for utility-scale quantum computation.

We close this gap by fine-graining: we replace each face of the base hyperbolic tiling with a patch of a Euclidean lattice of side length $L$ (with $L=1$ recovering the base tiling) to obtain a \emph{semi-hyperbolic} tiling, from which we construct a new family of semi-hyperbolic surface codes, distinct from those of Ref.~\cite{Breuckmann_2017_semihyperbolic}, and, for the first time, semi-hyperbolic color codes.
Fine-graining leaves $k$ unchanged while increasing $d$ by a factor of roughly $L$ and $n$ by a factor of roughly $L^2$, trading encoding rate for distance while approximately preserving the figure of merit $kd^2/n$.
Since the base hyperbolic codes have such a high rate, the resulting codes retain high encoding rates (as high as $k/n=1/16$ among the codes we find) at large distances ($d\geq22$).

Fine-graining also flattens the tiling locally, allowing the semi-hyperbolic code to be partitioned into Euclidean planar modules.
Semi-hyperbolic Floquet codes on the same class of tilings, including their modular implementation, have previously been studied~\cite{higgott_2024_floquet,sutcliffe_2025_distributed} (the latter using a generic graph-partitioning heuristic), but color codes have not previously been constructed on them.
We refer to the resulting modular semi-hyperbolic codes simply as \emph{modular hyperbolic} codes, and find many new examples by searching over suitable hyperbolic tilings, as described below.

A modular hyperbolic \textit{color} code is derived from a tiling of a hyperbolic surface with triangular faces.
An example of such a tiling of the infinite hyperbolic plane with 8 triangles meeting at each vertex is shown in \Cref{fig:bolza_color_code_overview}a. The vertices of the tiling are three-colorable, with one vertex of each color per triangle, and we assign each edge the third color absent from its two vertices.
A tiling of a finite, closed surface can be obtained by cutting a polygon (called the \emph{fundamental polygon}) out of this infinite plane and identifying (i.e.~gluing) pairs of its boundary edges together to fold the flat polygon into a closed manifold.
The fundamental polygon of the genus-2 Bolza surface is shaded in \Cref{fig:bolza_color_code_overview}a, with the identified pairs of edges labeled in \Cref{fig:bolza_color_code_overview}b.
Each triangular face is replaced with a Euclidean hexagonal tiling of side length $L$ to form a semi-hyperbolic tiling of hexagons and octagons~\cite{higgott_2024_floquet}.
An $X$ stabilizer and a $Z$ stabilizer generator are then defined on each face of the resulting three-face-colorable tiling.

We partition a semi-hyperbolic color code into modules by deriving each module from a contiguous set of triangular faces in the base hyperbolic tiling.
In \Cref{fig:bolza_color_code_overview}b-c we show an example of a \textit{quadrilateral} module partition, for which each module is defined by a pair of triangles in the base tiling that share an edge of a chosen color (e.g.~red edges).
By combining the eight triangles together that share a vertex of a given color (e.g.~red vertices), we can also obtain a larger \textit{octagon} module partition. 

We also construct modular hyperbolic \textit{surface} codes, derived instead from a hyperbolic tiling of quadrilateral faces.
Each quadrilateral face is replaced with a square-grid tiling of side length $L$.
Rather than placing data qubits on the edges of this fine-grained tiling~\cite{Breuckmann_2017_semihyperbolic}, we take the dual tiling and bicolor its faces red and blue.
We then associate a data qubit with each vertex of this dual tiling, an $X$ stabilizer with each red face and a $Z$ stabilizer with each blue face (see \Cref{fig:frontier_codes_plots}b), analogous to the rotated surface code.
Compared with placing data qubits on edges, this has two advantages. First, for odd $L$ it distributes the high-weight corner stabilizers between $X$ and $Z$ types (which are related by a symmetry when the tiling is regular, see \Cref{sec:asymmetry_X_Z}), which can help balance the $X$ and $Z$ distances and improve the code parameters. Second, no data qubits lie on the edges of the base quadrilaterals, which allows uniform module partitions.

Logical operators are supported on non-trivial cycles around a handle of the surface, with a genus-$g$ tiling supporting $4g$ logical qubits for the hyperbolic color codes and $2g$ logical qubits for the surface codes \cite{bombin_color,bombin_delgado_toric_surface_color}.
By the Gauss-Bonnet theorem, the genus of a closed hyperbolic surface grows linearly with its area.
Consequently, for a given code family, the number of logical qubits increases linearly with the number of modules, alongside an increasing distance (which grows with the tiling's combinatorial systole).
The negative curvature responsible for this scaling is absent within any module,
each of which is a flat Euclidean patch, and arises entirely from the
inter-module couplers connecting them.

\section{Modular Syndrome Extraction}
\begin{figure*}
    \centering
    \includegraphics[width=\linewidth]{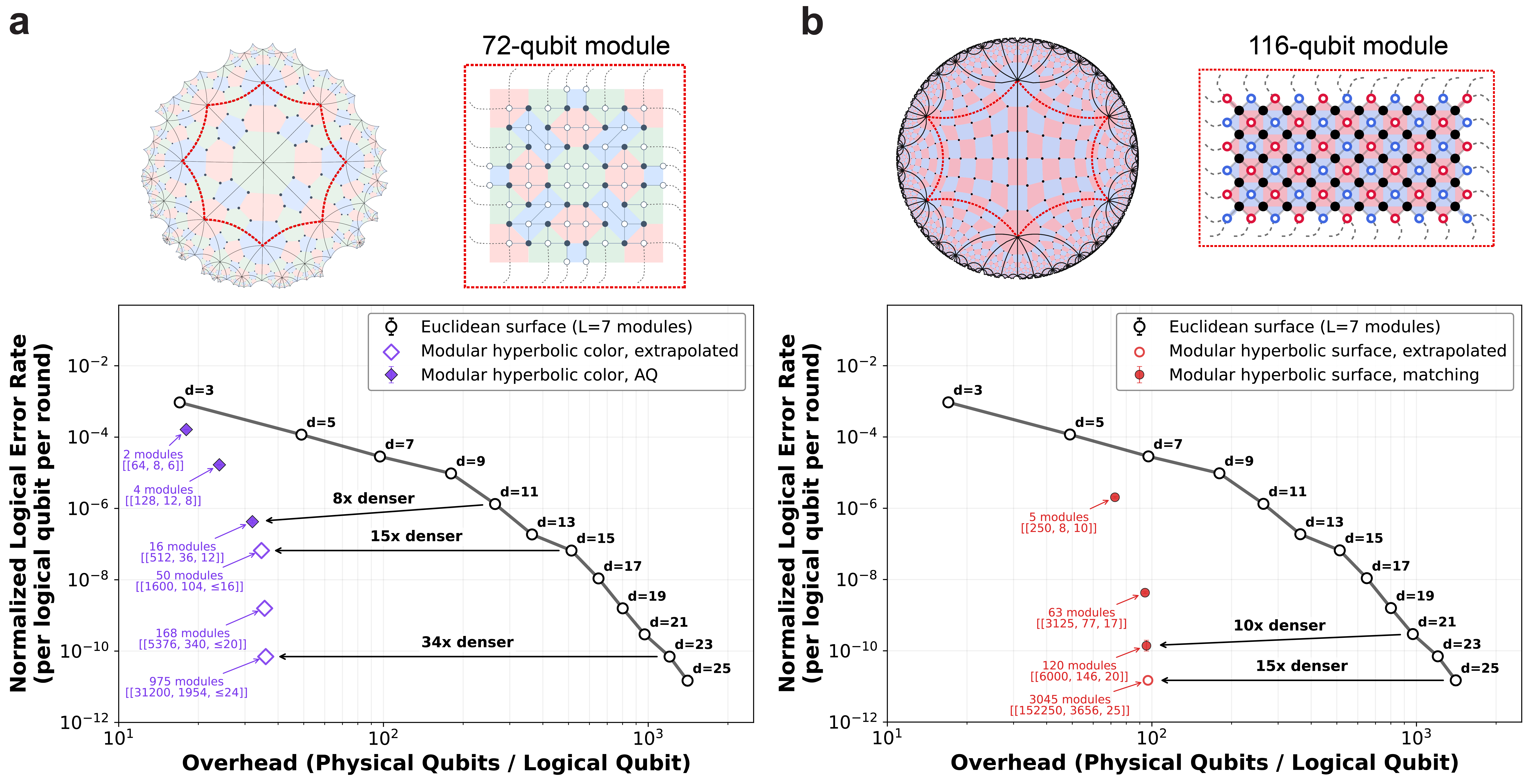}
    \caption{Simulation data for modular hyperbolic codes using an SI1000 noise model with $p=0.1\%$ and inter-module CZ gates adjusted to have a noise strength of $p_{\mathrm{seam}}=1\%$ (see Methods for details). \textbf{a}, Modular hyperbolic color codes derived from V8.8.8 hyperbolic tilings using $L=2$ and an octagon module partition. \textbf{b}, Modular hyperbolic surface codes derived from $\{4,10\}$ regular hyperbolic tilings with $L=5$ and a rectangle module partition. In both panels we compare with Euclidean surface codes partitioned into modules of a comparable size (with additional measure qubits on the boundaries).}
    \label{fig:frontier_codes_plots}
\end{figure*}
We design fault-tolerant syndrome extraction circuits that respect the connectivity constraints of the boundary-connected modular architecture, and benchmark them under a circuit-level noise model with elevated seam noise.
For the modular hyperbolic color codes, we place a pair of measure qubits in each hexagonal plaquette in the bulk and use them to prepare a Bell state.
The Bell state is coupled to the data qubits and used to extract the plaquette's $X$ stabilizer, then measured to obtain a stabilizer measurement along with a flag measurement \cite{chao_reichardt_flag,flag_QEC,takada_keisuke24}.
The process is then repeated in the opposite basis for the $Z$ stabilizers.
Stabilizers that span modules are measured using cat states (Greenberger--Horne--Zeilinger states) supported on $m$ measure qubits, which are prepared using inter-module couplers.
Once prepared, the $m$-qubit cat state is then coupled to the stabilizer's data qubits using in-module gates and then measured to obtain the stabilizer measurement and $m-1$ flag measurements.

The circuits for the modular hyperbolic surface codes are similar, with some important differences.
A single measure qubit (instead of a Bell state) is used to measure each square plaquette stabilizer in the bulk.
The $X$ and $Z$ stabilizers are also interleaved, leading to commutativity constraints.
We find valid low-depth extraction schedules by solving the relevant constraint satisfaction problem on the base tiling and lifting to translationally-invariant schedules within the modules.
Additionally, we design multi-rate circuits that measure the high-weight inter-module stabilizers less frequently (once every three QEC rounds), which we find leads to improved logical error rates while allowing for slower inter-module gates.

To decode efficiently we develop and train a variant of the AlphaQubit decoder (AQ) for the modular hyperbolic color codes~\cite{senior2025scalablerealtimeneuraldecoder}.
For the modular hyperbolic surface codes we decode using a correlated minimum-weight perfect matching (MWPM) decoder~\cite{Higgott2025sparseblossom}.

Our circuit-level simulations are presented in \Cref{fig:frontier_codes_plots}.
For the modular hyperbolic color codes (decoded with AlphaQubit) our simulations show an $8\times$ density improvement when compared to the Euclidean surface code partitioned into modules of a similar size.
By projecting to larger system sizes (details in Methods), we find that the savings increase to an over $30\times$ overhead reduction, with a modular hyperbolic color code requiring only 36 physical qubits per logical to achieve a distance of 20--24 using planar 72-qubit modules (see \Cref{tab:candidates_888}).
For modular hyperbolic surface codes, our simulations demonstrate a $10\times$ reduction in the number of physical qubits per logical qubit needed to reach a logical error rate of $10^{-10}$, increasing to a $15\times$ saving when extrapolating to larger system sizes.
While the modular hyperbolic surface codes are not as dense as their color code counterparts, they have the potential practical advantage of supporting fast and accurate MWPM decoding.
We find that our circuits for both hyperbolic color and surface codes are highly tolerant to noisy inter-module gates, with even $10\times$ higher seam errors than bulk errors leading to only a modest increase in logical error rate (see \Cref{fig:ler_vs_seam_noise}).

\section{Logical Control and Computation}
\begin{figure*}[t]
    \centering
    \includegraphics[width=\linewidth]{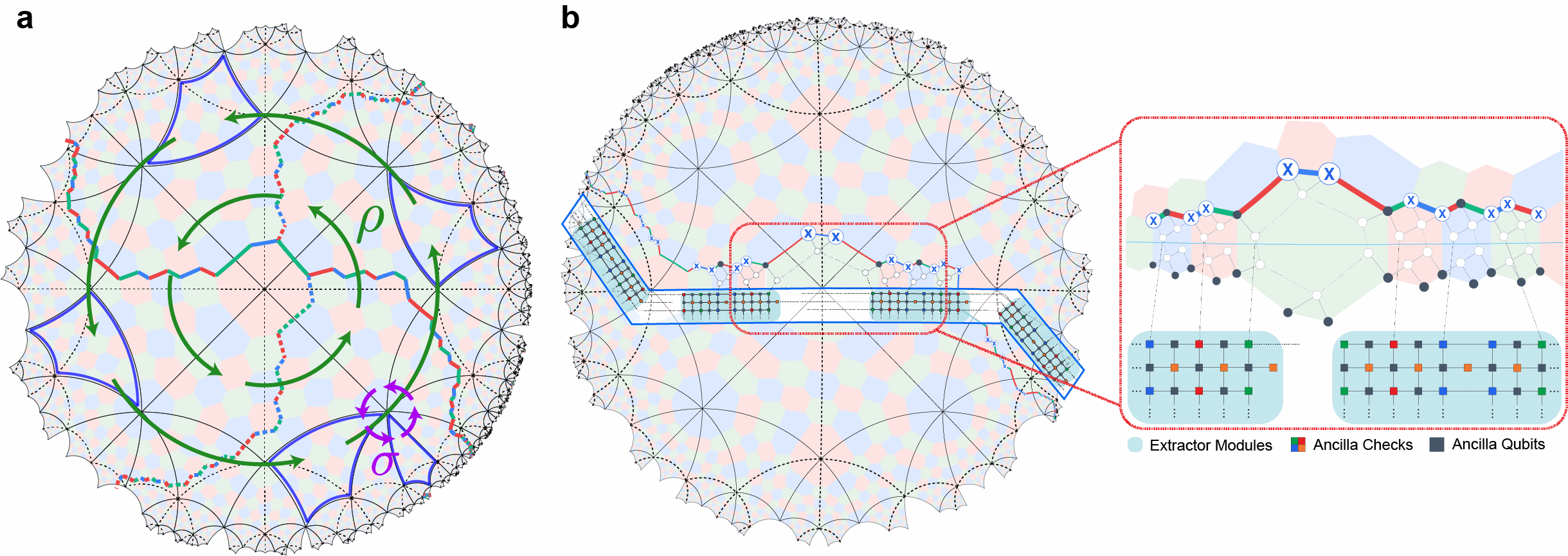}
    \caption{Logical control architecture: \emph{code automorphisms} and a \emph{ribbon extractor}. \textbf{a}, A symplectic pair of homologically non-trivial cycles on the semi-hyperbolic tiling of the $[[6400,104,30]]$ color code, drawn as solid and dashed lines. The two cycles are related by $\rho$, a global rigid rotation of the tiling by $\pi/2$ (green arrows). Together with a second rotation $\sigma$ (purple arrows), these generate all rotations of the tiling, each of which permutes the modules while preserving the codespace (a code automorphism). Logical Pauli operators are supported on the monochromatic substrings of a cycle, obtained by restricting it to edges of a single color. Two of the three colors (e.g. blue and red) give four independent logical operators per cycle. Operators on the same cycle commute, and anticommuting pairs are supported across the symplectic pair, so each pair defines four logical qubits. For the $[[6400,104,30]]$ code, the orbit of a single seed cycle under these rotations generates a symplectic basis for 100 of the 104 logical qubits. A single extractor then suffices for 100 qubits: it measures the operators on the seed cycle, and every other cycle in the orbit is rotated onto it. \textbf{b}, A ribbon extractor attached to a non-trivial cycle of the memory code, shown in detail in the inset. The extractor is itself boundary-connected and modular, containing ancilla qubits and ancilla checks. It couples to the memory code modules only along the cycle, at the boundaries. It can fault-tolerantly measure all $2^4-1=15$ non-trivial products of the four logical Pauli operators on the cycle, using roughly $22\%$ additional physical qubits relative to the memory.}
    \label{fig:logic}
\end{figure*}

To operate on the logical information stored in the modular hyperbolic memory we combine two ingredients: code surgery and code automorphisms. In code surgery~\cite{cohen2022low-overhead,cross2024improved,Williamson2026,ide2024fault,swaroop2026universal,he2025extractorsqldpcarchitecturesefficient}, a generalization of lattice surgery~\cite{horsman2012surface, landahl2014quantum}, an ancilla system of data and measure qubits is coupled to the memory code in order to measure its logical operators. 
These measurements implement fault-tolerant logic and allow logical information to be moved into and out of the memory~\cite{Breuckmann_2017_semihyperbolic,yoder2025tour}. Code automorphisms are symmetries that permute the physical qubits while preserving the codespace. Here they are global rotations and translations of the hyperbolic tiling, and they allow a single ancilla system to measure many different logical operators.

Combining these two ingredients with low overhead requires co-designing the logical basis of the code with both the surgery ancilla system and the circuits that implement automorphisms.
To teleport an individual logical qubit into and out of the memory, we must be able to measure single-qubit logical $\overline{X}_i$ and $\overline{Z}_i$ operators (see \Cref{sec:logical_measurement_methods}).
These single-qubit logical operators of the code are derived from a \textit{canonical homology basis} of the underlying hyperbolic surface.
On an orientable genus-$g$ surface, a full canonical homology basis consists of $g$ \emph{symplectic pairs} of homologically non-trivial cycles that cross each other an odd number of times (see \Cref{fig:logic}a for one such pair), while cycles from different pairs cross an even number of times.
In a hyperbolic color code, restricting a cycle to edges of one color defines a logical operator, and two colors give four independent, commuting $X$ and $Z$ logical operators per cycle. With anticommuting partners on the other cycle, each symplectic pair encodes four logical qubits~\cite{bombin_color} (two for a hyperbolic surface code), whose operators can be deformed onto module boundaries.
While it is known that a shortest homology basis can be found in polynomial time~\cite{Erickson_Whittlesey}, such a basis does not ensure that cycles form symplectic pairs, nor that they can be mapped onto each other via automorphisms.
To satisfy these additional constraints, we enumerate low-weight homologically non-trivial \emph{seed} cycles and generate the orbit of each seed cycle under the action of the automorphisms we can implement.
We then search for $g^\prime\leq g$ symplectic pairs within the orbit of each single seed loop (Methods).
For the $[[6400,104,30]]$ code ($g=26$) and the $[[3600,104,22]]$ code derived from the same tiling with $L=3$, the orbit of the seed cycle in \Cref{fig:logic}a contains $g^\prime=25$ symplectic pairs, covering 100 of the 104 logical qubits (\Cref{tab:candidates_888}).

We access this seed cycle by attaching a modular \emph{ribbon extractor}, an extractor system~\cite{he2025extractorsqldpcarchitecturesefficient} that we design for boundary-connected modules. The name comes from the fact that this system, which extracts logical information from the code, is a closed loop of non-trivial width attached to a cycle that wraps around the base code. 
Here we focus on ribbon extractors for modular hyperbolic color codes with the quadrilateral module partition; we expect the same methods to apply to other module partitions and to modular hyperbolic surface codes.
Like the memory itself, the extractor is built from planar modules joined at their boundaries, and it attaches to the memory along the cycle that supports the logical operators, using inter-module couplers at module boundaries as in the rest of the architecture. A single extractor can fault-tolerantly measure all $2^4-1=15$ non-trivial products of the four logical Pauli operators on the cycle it is attached to.
We prove (Supplementary Information) that this construction preserves the full distance $d$.
For both the $[[3600,104,22]]$ and $[[6400,104,30]]$ codes, the extractor is about $4.5\times$ smaller than the memory itself.

To measure the other logical operators with the ribbon extractor we implement code automorphisms using fault-tolerant \textit{walking} circuits.
Two rotations $\rho$ and $\sigma$ about neighboring vertices generate all orientation-preserving rotations and translations of the tiling (\Cref{fig:logic}a).
For quadrilateral modules, $\rho$ and $\sigma$ each map every module onto another module with the same internal layout and orientation.
Since the modules are in general not mapped to their neighbors in the code by these global rotations, we add an additional inter-module seam of couplers from the north boundary of each module to the south boundary of its image in $\rho$.
An $O(L)$-step walking circuit of SWAP gates implements the move while all stabilizers originally internal to a module continue to be measured at every step, and simulations of a simplified model of the resulting stabilizer dislocation suggest only a modest increase in logical error rate (Methods).
We implement the $\sigma$ rotation in a similar manner, using the east and west boundaries instead.
For the larger octagon modules, which would be internally rotated by $\rho$ and $\sigma$, we find that most tilings we consider admit an alternative, smaller group of symmetries that still moves every module onto every other while preserving internal module orientation, typically needing one additional seam on each side of a module (Supplementary Information). This smaller group requires two to four seed cycles rather than one to cover the same logical basis, which can still be measured by a single ribbon extractor by adding additional seams to its boundaries.
Bringing any basis cycle of the $[[3600,104,22]]$ or $[[6400,104,30]]$ code under the ribbon extractor, starting from any other, takes a mean of 4.4 and at most six $\rho$, $\rho^{-1}$, $\sigma$ or $\sigma^{-1}$ moves (\Cref{tab:candidates_888}), with each move transporting four logical operators simultaneously.

Because every logical qubit in this canonical basis has both its $\overline{X}_i$ and $\overline{Z}_i$ cycles in the orbit of the seed cycle, any individual logical qubit can be selectively teleported into and out of the hyperbolic memory by coupling the ribbon extractor to a small compute region, such as a Euclidean surface or color code, enabling universal computation with low space overhead.

As a benchmark, we consider a widely used arithmetic subroutine in quantum algorithms and compile a 50-bit ripple-carry adder~\cite{cuccaro2004new} onto the $[[3600,104,22]]$ ($L=3$) and $[[6400,104,30]]$ ($L=4$) memories derived from the same tiling, each using a single ribbon extractor to teleport operands into and out of a small compute region (Methods).
Since the symmetry group is non-Abelian, the number of moves between two cycles depends on where the extractor is attached (Methods).
By choosing the extractor location so that the two cycles of every symplectic pair are only a single automorphism move apart, and storing operands that are accessed in close succession in the same symplectic pair, we bound the time for the ribbon measurements and automorphisms by $21{,}402$ and $28{,}930$ QEC cycles respectively (428 and 579 per bit).

\section{Discussion and Outlook}

While partitioning quantum processors into manageable modules in solid state systems is known to be important for alleviating constraints on fabrication, yield and coupler density, our architecture demonstrates this modularity can also be exploited to deliver substantial efficiency gains in quantum error correction.
Realizing these savings in a superconducting quantum computer will require a fully integrated system that simultaneously operates high-performance local grids of qubits~\cite{surface_code_below_threshold} and high-fidelity inter-module couplers~\cite{Zhong_2021, Niu_2023, Qiu_2025}.
These systems must also be resilient to fabrication defects, such as broken qubits or couplers.
However, we expect that existing methods for handling fabrication defects in planar color codes~\cite{wolanski2025automated} and surface codes~\cite{auger2017fabrication,strikis2023fabrication,Debroy2025luciinsurfacecode,leroux2024snakes} should be naturally adaptable to modular hyperbolic codes, since the local code structure is similar.

While our implementation of neural network and correlated matching decoders demonstrates that efficient and accurate decoding is possible, real-time decoding at scale will require parallelization or specialized hardware implementations.
Since the full distance of a modular hyperbolic code is distributed across many modules, sub-distance parallelization strategies will likely be necessary, which generally requires parallelizing the decoding algorithm itself~\cite{wu_fusion_blossom,surface_code_below_threshold} rather than relying on parallel windowing~\cite{Skoric2023parallel,tan23scalable}.

Utility-scale quantum computation requires the ability to implement universal logic fault-tolerantly.
We have demonstrated that this is possible by providing the ingredients for a low-overhead quantum memory, using the ribbon extractor for logical measurement along with circuits for implementing automorphisms.
This architecture is universal and a natural fit for serial quantum algorithms and subroutines, such as addition circuits. However, we see several opportunities to improve on the current design. 
First, our ribbon extractor does not fully exploit the non-local connectivity available at module boundaries. This leaves open the possibility of building more space-efficient extractors that use this connectivity \cite{he2025extractorsqldpcarchitecturesefficient}.
Second, it would be desirable to design extractor systems capable of measuring more complex high-weight logical Pauli operators, either in serial or in parallel, in order to implement certain subroutines such as QROM lookup with lower time overhead.
Finally, these logical primitives should be co-designed with compilations of full quantum algorithms, building on our adder estimate (Methods), to provide a complete resource estimation of boundary-connected module architectures.

In summary, boundary-connected modules turn a constraint of scalable hardware into a resource for quantum error correction.
Our architecture builds hyperbolic codes from flat Euclidean modules, requiring connectivity similar to that of current superconducting processors designed for the surface code, with the hyperbolic geometry realized by the inter-module couplers at their boundaries.
Together with the modular syndrome extraction circuits, decoders and logical measurement methods we develop for this setting, this unlocks the efficiency of high-rate QLDPC codes on superconducting devices.
Therefore, boundary-connected modular architectures substantially lower the cost of reaching utility-scale fault-tolerant quantum computation using near-term hardware.

\clearpage

\setcounter{figure}{0}
\renewcommand{\thefigure}{M\arabic{figure}}
\setcounter{table}{0}
\renewcommand{\thetable}{M\arabic{table}}
\setcounter{equation}{0}
\renewcommand{\theequation}{M\arabic{equation}}
\renewcommand{\theHfigure}{M\arabic{figure}}
\renewcommand{\theHtable}{M\arabic{table}}
\renewcommand{\theHequation}{M\arabic{equation}}

\section*{Methods}
\setcounter{subsection}{0}
\renewcommand{\thesection}{M}
\renewcommand{\theHsubsection}{M.\Alph{subsection}}

\subsection{Symmetry groups of hyperbolic tilings}\label{sec:symmetry_group_methods}
\begin{figure}[b]
    \centering
    \includegraphics[width=\linewidth]{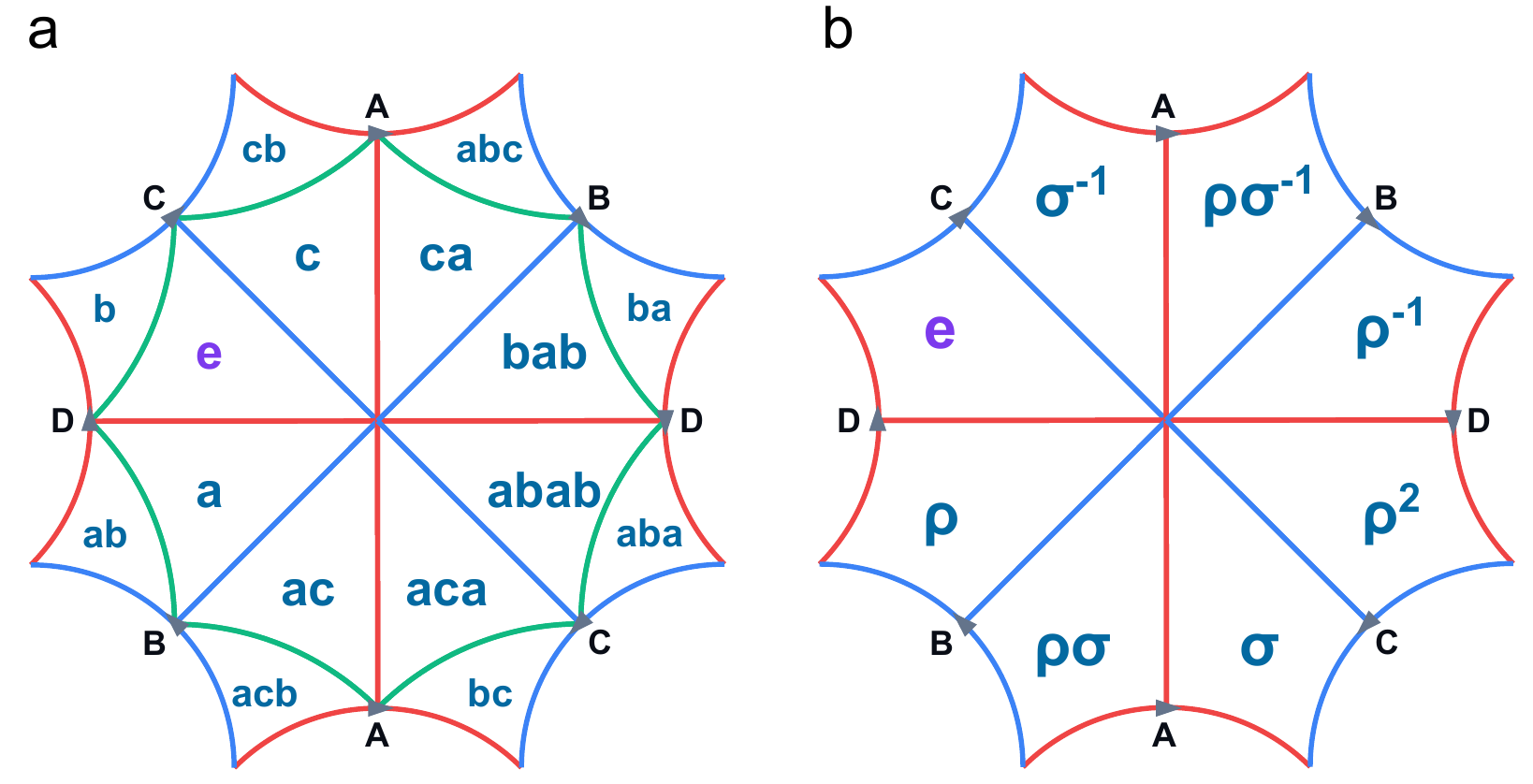}
    \caption{Symmetry groups of the $V8.8.8$ tiling of the genus-two Bolza surface. \textbf{a}, Reflecting locally across a red, green or blue edge corresponds to right-multiplication by $a$, $b$ or $c$ respectively. Taking one triangle labeled as the identity $e\in G(4,4,4)$ as the fundamental domain, each triangle is uniquely labeled with an element $g\in G(4,4,4)$. Opposite boundary edges (labeled with matching uppercase letters) are identified. \textbf{b}, We can instead take a pair of triangles related by a $b$ reflection as the fundamental domain of the orientation-preserving subgroup $G^+$ generated by $\rho\coloneqq ab$ and $\sigma\coloneqq bc$, and label the resultant quadrilaterals with elements of $G^+$.}
    \label{fig:bolza_symmetry_groups}
\end{figure}

Symmetry groups of hyperbolic tilings are used throughout this work for constructing the tilings, partitioning codes, finding logical bases and implementing automorphisms.
In this section, we introduce the relevant terminology and explain how the tilings are constructed.
For a more comprehensive overview, see the Supplementary Information as well as Refs.~\cite{Hatcher2002, breuckmann_2016_constructions, quantum_pin_codes, higgott_2024_floquet}.

A hyperbolic triangle group $\triangle(r,s,t)$ (with $r, s, t \in \mathbb{Z}_{\ge 2}$ satisfying $\frac{1}{r} + \frac{1}{s} + \frac{1}{t} < 1$) is generated by reflections across the sides of a hyperbolic triangle with internal angles $\pi/r$, $\pi/s$ and $\pi/t$.
This group has presentation
\begin{equation}\label{eq:triangle_group_methods}
\begin{split}
\triangle(r,s,t) = \bigl\langle\, a, b, c \;\big|\; & a^2 = b^2 = c^2 = e, \\
& (ab)^r = (bc)^s = (ca)^t = e \,\bigr\rangle,
\end{split}
\end{equation}
where the generators $a$, $b$ and $c$ represent reflections across the three sides of the triangle and $e$ is the identity element.
Reflecting the triangle across its bounding mirrors produces a triangular tiling of the hyperbolic plane $\mathbb{H}^2$ where each triangular tile has vertices with degrees $2r$, $2s$ and $2t$ (face configuration $V2r.2s.2t$).
An example of a $V8.8.8$ triangular tiling is shown in \Cref{fig:bolza_color_code_overview}\textbf{a}, where red, green and blue edges are the images of $a$, $b$ and $c$ mirrors, respectively.

We can construct a finite tiling $\mathcal{T}^\triangle$ of a closed hyperbolic surface by finding a torsion-free normal subgroup $\Gamma \triangleleft \triangle(r,s,t)$ of finite index. The quotient group
\begin{equation}
G(r,s,t) \coloneqq \triangle(r,s,t)/\Gamma
\end{equation}
is then the symmetry group of the finite $V2r.2s.2t$ triangular tiling on the closed hyperbolic surface $\mathbb{H}^2/\Gamma$.
See \Cref{fig:bolza_symmetry_groups}\textbf{a} for an example of a small finite $V8.8.8$ tiling (in this case of the Bolza surface).
Notice, for example, that $cbabca=e$ is a relation in $G(r,s,t)$ (passing through the identified boundary $A$), meaning $cbabca\in\Gamma$.
The orientation-preserving subgroup $G^+(r,s,t)\leq G(r,s,t)$ consists of all even-length words in $G$, and is generated by $\rho\coloneqq ab$ and $\sigma\coloneqq bc$ (see \Cref{fig:bolza_symmetry_groups}\textbf{b}).
For this work we enumerate normal subgroups $\Gamma \triangleleft \triangle(r,s,t)$ using the methods of Refs.~\cite{firth2005algorithm, lins} up to a finite index, and construct larger tilings by searching for surjective homomorphisms from $\triangle(r,s,t)$ onto candidate finite groups (such as $\mathrm{PSL}(2,q)$).

For the modular hyperbolic surface codes we use quadrilateral tilings, for which each face is instead a quadrilateral.
We construct many of these using \textit{quadrilateral} groups $\square(r,s,t,u)$, which are instead generated by reflections across the sides of a hyperbolic quadrilateral, see \Cref{sec:hyperbolic_tilings_intro} for details.
These include non-regular quadrilateral tilings with face configuration $V2r.2s.2t.2u$, substantially enlarging the space of tilings beyond the regular $\{p,q\}$ tilings used in previous hyperbolic surface code constructions~\cite{breuckmann_2016_constructions,Breuckmann_2017_semihyperbolic}.
In particular, the big square module partition described below requires each tile to have a corner at which four quadrilaterals meet, which is impossible for a regular hyperbolic $\{4,q\}$ tiling (see \Cref{tab:candidates_L1_surface,tab:candidates_L5_surface_ms2,tab:candidates_L3_surface_ms4_quad4} for examples of codes constructed from these tilings).

\subsection{Modular hyperbolic color codes}

\begin{table*}[ht]
\centering
\begin{tabular}{c@{\quad
\quad}c@{\quad}c@{\quad}c@{\quad}|@{\quad}c@{\quad}c@{\quad}c}
  & \multicolumn{3}{c}{\textbf{Hyperbolic Color Code Modules}} & \multicolumn{3}{c}{\textbf{Hyperbolic Surface Code Modules}} \\
   & \includegraphics[width=1.5cm]{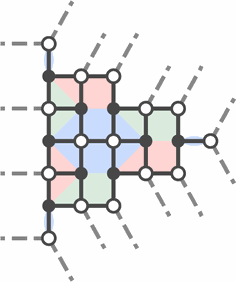}
   & \includegraphics[width=1.5cm]{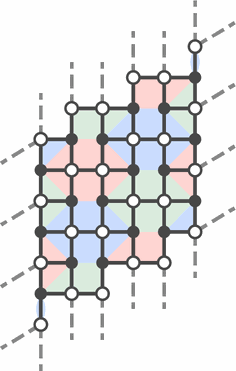}
   & \includegraphics[width=3cm]{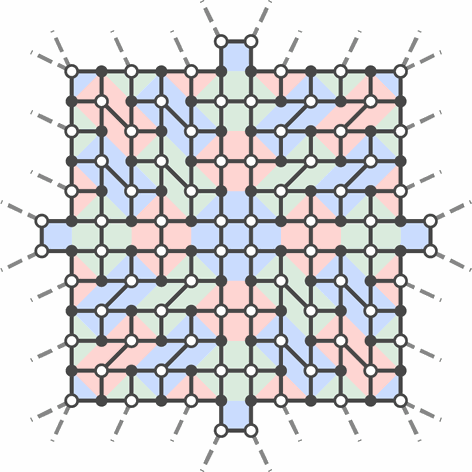}
   & \includegraphics[width=1.5cm]{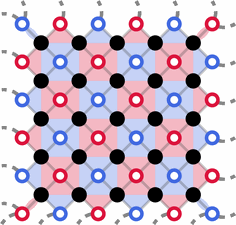}
   & \includegraphics[width=1.58cm]{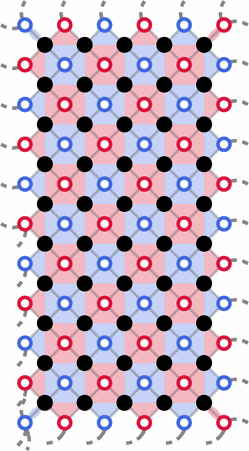}
   & \includegraphics[width=2.8cm]{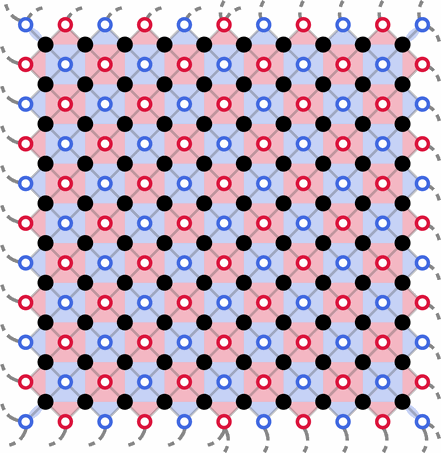} \\
\hline
 & \textbf{Triangle} & \textbf{Quadrilateral} & \textbf{Octagon*} & \textbf{Square} & \textbf{Rectangle} & \textbf{Big Square**} \\
$L$ & $2L^2+2L$ & $4L^2+2L$ & $16L^2+8$ & $2L^2+2L+1$ & $4L^2+3L+1$ & $8L^2+4L+1$ \\
\hline
2 & 12 & 20  & 72  & 13 & 23  & 41 \\
3 & 24 & 42  & 152 & 25 & 46  & 85 \\
4 & 40 & 72  & 264 & 41 & 77  & 145 \\
5 & 60 & 110 & 408 & 61 & 116 & 221 \\
\end{tabular}
\caption{Hyperbolic color and surface code module options. Number of qubits per module for different module shapes and values of the fine-graining parameter $L$. *The Octagon module is available for base triangular tilings with face configuration $V2r.2s.2t$ where at least one vertex has degree $8$ (e.g., $2r=8$). **The Big Square module is available for base quadrilateral tilings with face configuration $Va.b.c.d$ where at least one vertex degree is $4$ (e.g., $a=4$).}
\label{tab:module_sizes}
\end{table*}

Our modular hyperbolic color codes are constructed by fine-graining a triangular hyperbolic tiling $\mathcal{T}^\triangle$ (\Cref{sec:symmetry_group_methods}) and replacing each triangular face with a Euclidean hexagonal lattice of side length $L$.
This family of tilings was used to construct Floquet codes in \cite{higgott_2024_floquet}, but has not previously been studied for constructing color codes.
We associate a data qubit with each vertex and define an $X$ and $Z$ stabilizer on each face of the tiling, see Supplementary Information for more details on the code construction.

\Cref{tab:module_sizes} summarizes the different module layouts for modular hyperbolic codes, giving the total number of qubits per module as a function of the fine-graining parameter (side-length) $L$ for each module type. 
The color code modules partition the triangular tiles of the base triangular tiling ($\mathcal{T}^\triangle$) into triangle ($K=1$), quadrilateral ($K=2$) or octagon ($K=8$) modules.
Triangular modules are formed from a single triangle of the base tiling $\mathcal{T}^\triangle$, in bijection with the elements of $G(r,s,t)$.
Each quadrilateral module is a pair of triangles sharing an edge of a chosen color (e.g.~green, the images of $b$ mirrors; the other two choices are analogous), each labeled by a left coset $g\langle b\rangle$ (where $g\in G(r,s,t)$) or equivalently by an element of $G^+(r,s,t)$.
Each octagon module is formed from 8 triangles at a vertex type (e.g. meeting at vertex $b\cap c$, labeled by a left coset $g\langle b,c\rangle$ for $g\in G(r,s,t)$).

\subsection{Modular Hyperbolic Surface Codes}

\begin{figure}[b]
    \centering
    \includegraphics[width=\linewidth]{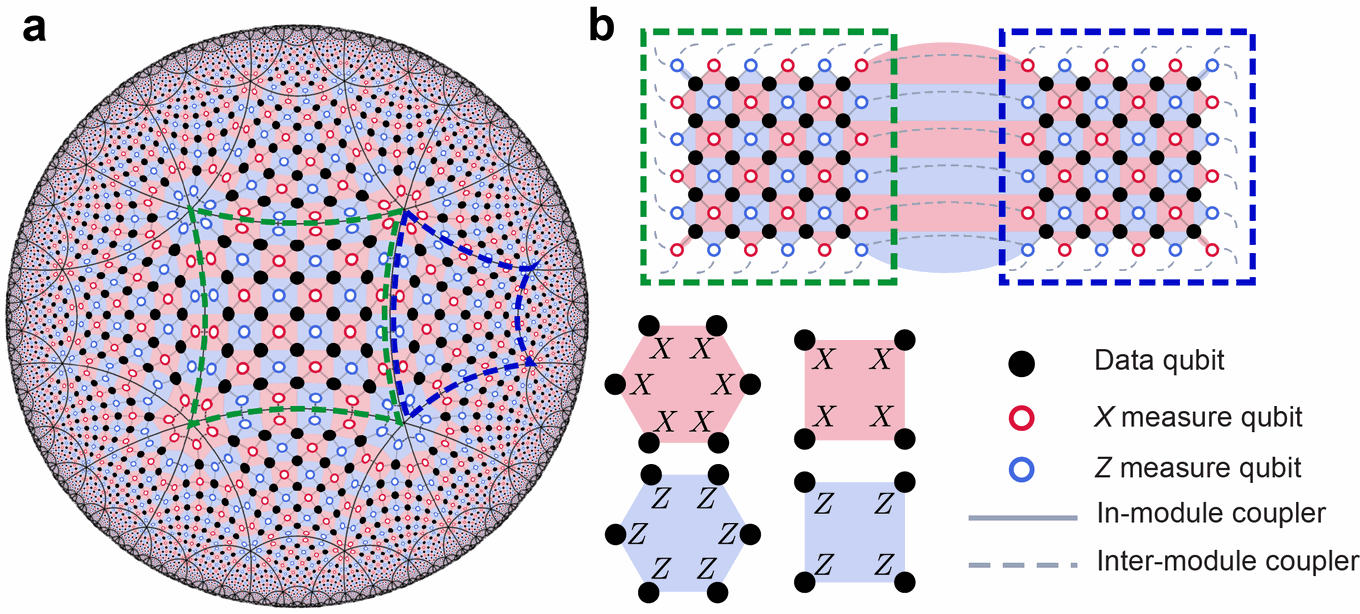}
    \caption{Construction of modular hyperbolic surface codes. \textbf{a}, A modular hyperbolic surface code derived from a quadrilateral tiling, in which six squares meet at each vertex (a regular $\{4,6\}$ tiling), shown in the Poincar\'e disk model. The code is partitioned into modules, each derived from a single square of the base quadrilateral tiling; two neighboring modules are outlined with dashed green and blue lines. 
    \textbf{b}, The two outlined modules of \textbf{a} shown in detail. 
    }
    \label{fig:hyperbolic_surface_code_overview}
\end{figure}

Our modular hyperbolic surface codes are derived from tilings of closed hyperbolic surfaces by quadrilateral faces (\Cref{fig:hyperbolic_surface_code_overview}a). We fine-grain each face of the quadrilateral tiling $\mathcal{T}^\square$ with an $L \times L$ Euclidean square grid to form a semi-hyperbolic tiling.
Rather than placing qubits on the edges of the tiling, as done in~\cite{Breuckmann_2017_semihyperbolic}, we instead take its dual tiling, bicolor the faces red and blue and associate a data qubit with each vertex, an $X$ stabilizer with each red face and a $Z$ stabilizer with each blue face. 
This corresponds to replacing each quadrilateral face of $\mathcal{T}^\square$ with a patch of square stabilizers resembling a rotated surface code in the bulk, so our construction could be considered a ``rotated'' analog of \cite{Breuckmann_2017_semihyperbolic}.
Stabilizers in the bulk of a tile then have weight four, locally identical to a planar surface code, while those at the corners of the tiles inherit the higher weight set by the vertex degree of $\mathcal{T}^\square$, which is six for the $\{4,6\}$ tiling shown in \Cref{fig:hyperbolic_surface_code_overview}.
Compared with Ref.~\cite{Breuckmann_2017_semihyperbolic}, where all high-weight corner stabilizers have the same Pauli type, for odd $L$ our construction distributes them between $X$ and $Z$ types (see \Cref{sec:asymmetry_X_Z}), and since no data qubits lie on the edges of the base quadrilaterals, the tiles can be grouped into uniform modules.

 We partition the code into modules by grouping contiguous quadrilateral tiles: a single tile gives a \emph{square} module ($K=1$), each labeled by an element of $G(r,s,t,u)\coloneqq \square(r,s,t,u)/\Gamma$.
 A \emph{rectangle} module ($K=2$) is formed from a pair of quadrilateral tiles sharing an edge (labeled by an element of the orientation-preserving subgroup $G^+(r,s,t,u)\leq G(r,s,t,u)$).
 If the quadrilateral tiling has face configuration $Va.b.c.d$ with at least one of $a,b,c,d$ equal to $4$, i.e.\ each quadrilateral tile has at least one corner at which four tiles meet, then we can form a \emph{big square} module ($K=4$) by choosing one such corner type and combining the four quadrilaterals that meet at each vertex of that type in $\mathcal{T}^\square$ (labeled by a left coset of the corresponding vertex stabilizer in $G(r,s,t,u)$).
 See \Cref{tab:module_sizes} for diagrams of the modules, as well as the sizes of the modules. 
 Qubits within a module are coupled by nearest-neighbor gates on a 2D grid, and only the stabilizers straddling tile boundaries require inter-module couplers (\Cref{fig:hyperbolic_surface_code_overview}b).

\subsection{Code Parameters and Tables} \label{sec:methods_code_candidates}

\begin{table*}[t]
\centering
\footnotesize
\setlength{\tabcolsep}{3.5pt}
\resizebox{\textwidth}{!}{%
\begin{tabular}{lrrccccccccccccc}
\toprule
 &  &  & \multicolumn{3}{c}{$L=2$, octagon ($K=8$)} & \multicolumn{3}{c}{$L=3$, quadrilateral ($K=2$)} & \multicolumn{3}{c}{$L=4$, quadrilateral ($K=2$)} & \multicolumn{4}{c}{Automorphism basis} \\
\cmidrule(lr){4-6}\cmidrule(lr){7-9}\cmidrule(lr){10-12}\cmidrule(lr){13-16}
Symmetry Group & $|G|$ & $k$ & $d$ & $N_{M8}/k$ & $kd^2/n$ & $d$ & $N_{M2}/k$ & $kd^2/n$ & $d$ & $N_{M2}/k$ & $kd^2/n$ & $k_{\mathrm{found}}/g$ & $w\,(w_1)$ & Pair & Cayley \\
\midrule
$\triangle(4,4,4)_{32}$ & $16$ & $8$ & $6$ & 18.0 & 4.5 & $8$ & 42.0 & 3.6 & $12$ & 72.0 & 4.5 & $2/2$ & $2{+}2\,(2{+}4)$ & 0.5 & 1.2/2 \\
$\triangle(4,4,4)_{40}$ & $32$ & $12$ & $8$ & 24.0 & 6.0 & $12$ & 56.0 & 6.0 & $16$ & 96.0 & 6.0 & $3/3$ & $3\,(4)$ & 1.3 & 2.4/4 \\
$\triangle(4,4,4)_{58}$ & $128$ & $36$ & $12$ & 32.0 & 10.1 & $16$ & 74.7 & 8.0 & $20\text{--}24$ & 128.0 & 7.0--10.1 & $8/9$ & $5\,(6)$ & 1 & 3.3/6 \\
$\triangle(4,4,4)_{96}$ & $336$ & $88$ & $14$ & 34.4 & 12.8 & $20$ & 80.2 & 11.6 & $26\text{--}28$ & 137.5 & 11.1--12.8 & $17/22$ & $6{+}6{+}6\,(6{+}6{+}8)$ & 2.3 & 4.1/6 \\
$\triangle(4,4,4)_{109}$ & $400$ & $104$ & $14\text{--}16$ & 34.6 & 12.7--16.6 & $22$ & 80.8 & 14.0 & $30$ & 138.5 & 14.6 & $25/26$ & $8\,(10)$ & 1 & 4.4/6 \\
$\triangle(4,4,4)_{120}$ & $512$ & $132$ & $16$ & 34.9 & 16.5 & $24$ & 81.5 & 16.5 & $32$ & 139.6 & 16.5 & $21/33$ & $6{+}7\,(8{+}8)$ & 2.4 & 4.9/8 \\
$\triangle(4,4,4)_{271}$ & $1344$ & $340$ & $18\text{--}20$ & 35.6 & 20.5--25.3 & $28$ & 83.0 & 22.0 & $36\text{--}38$ & 142.3 & 20.5--22.8 & $35/85$ & $8{+}8\,(10{+}10)$ & 4.1 & 6.0/10 \\
$\triangle(4,4,4)_{786}$ & $4096$ & $1028$ & $20$ & 35.9 & 25.1 & $28$ & 83.7 & 21.9 & $40$ & 143.4 & 25.1 & --- & --- & --- & --- \\
$\triangle(4,4,4)|\mathrm{PSL}(2,23)$ & $6072$ & $1522$ & $22$ & 35.9 & 30.3 & $32$ & 83.8 & 28.5 & $42\text{--}44$ & 143.6 & 27.6--30.3 & \multicolumn{4}{c}{Non-orientable} \\
$\triangle(4,4,4)|\mathrm{PSL}(2,25)$ & $7800$ & $1954$ & $22\text{--}24$ & 35.9 & 30.3--36.1 & $32$ & 83.8 & 28.5 & $42\text{--}46$ & 143.7 & 27.6--33.1 & \multicolumn{4}{c}{Non-orientable} \\
$\triangle(4,4,4)|\mathrm{PGL}(2,23)$ & $12144$ & $3040$ & $24$ & 36.0 & 36.0 & $36$ & 83.9 & 36.0 & $48$ & 143.8 & 36.0 & --- & --- & --- & --- \\
\bottomrule
\end{tabular}
}
\caption{Modular hyperbolic color codes derived from $8.8.8$ tilings, for side lengths $L=2$ (octagon modules, $K=8$), $L=3$ and $L=4$ (quadrilateral modules, $K=2$). Each row is a tiling on the Pareto frontier for at least one of these values of $L$, with symmetry group $G$ of order $|G|$, and $k$ is independent of $L$. For each $L$ we give $d$, the physical-to-logical qubit overhead $N_{MK}/k$ and $kd^2/n$, where $n=|G|L^2$ is the number of data qubits and $N_{MK}=(|G|/K)\,q_K(L)$ is the total number of physical qubits (data and ancilla) in the modular layout, with $q_K(L)$ the number of qubits per module given in \Cref{tab:module_sizes}. Distances quoted as a range are bounded below and above by the methods of \Cref{sec:distance_finding}, and $kd^2/n$ is then given for both bounds. The automorphism basis columns are independent of $L$ and describe the symmetric homology basis found using $G^+$ (showing the basis with the lowest total seed weight $w$ when multiple Pareto-optimal bases exist). 
$k_{\mathrm{found}}/g$ is the number of symplectic pairs found, out of the genus $g$, $w$ are the seed weights on $\mathcal{T}^\triangle$ and $w_1$ are the seed weights once restricted to a single edge color in $\mathcal{T}^\triangle$ (for compatibility with certain module layouts). Pair is the mean number of automorphism moves (right Cayley graph distance in the generators $\rho^{\pm1}$, $\sigma^{\pm1}$) between measuring the two cycles of a symplectic pair with a fixed extractor, and Cayley the mean and maximum of the same over all pairs of basis cycles. Separate tables for each $L$, as well as tables across all three-colorable tilings, are given in \Cref{sec:promising_candidate_tables}.}
\label{tab:candidates_888}
\end{table*}

\begin{table*}[pt]
\centering
\footnotesize
\setlength{\tabcolsep}{3.5pt}
\resizebox{\textwidth}{!}{%
\begin{tabular}{cccrrrlrccr}
\toprule
Tiling & Symmetry Group & $[[n,k,d]]$ & $M_{2}$ & $N_{M2}$ & $N_{M2}/k$ & Automorphism Bases & Cov. & Pair & Cayley & $kd^2/n$ \\
\midrule
\{4,10\} & $\triangle(2,4,10)_{24}$ & $[[250, 8, 10]]$ & $5$ & 580 & 72.5 & $4/4\,(w{=}2{+}2)$ & 100\% & --- & 0.7/1 & 3.20 \\
\{4,10\} & $\triangle(2,4,10)_{44}$ & $[[3125, 77, 17]]$ & $62\,({+}1)$ & 7253 & 94.2 & \text{Non-orientable} & --- & --- & --- & 7.12 \\
\{4,10\} & $\triangle(2,4,10)_{60}$ & $[[6000, 146, 20]]$ & $120$ & 13920 & 95.3 & $36/73\,(w{=}28{+}28)$ & 49\% & 3.3 & 3.7/6 & 9.73 \\
\{4,10\} & $\triangle(2,4,10)|\mathrm{PGL}(2,29) \times \mathrm{Z}_{2}|1$ & $[[152250, 3656, 25]]$ & $3045$ & 353220 & 96.6 & \text{Non-orientable} & --- & --- & --- & 15.01 \\
\{4,10\} & $\triangle(2,4,10)|\mathrm{PGL}(2,31) \times \mathrm{Z}_{2}$ & $[[186000, 4466, 30]]$ & $3720$ & 431520 & 96.6 & \text{Non-orientable} & --- & --- & --- & 21.61 \\
\bottomrule
\end{tabular}%
}
\caption{Modular hyperbolic surface codes for side length $L=5$ with rectangle modules ($K=2$ base quadrilaterals per module), restricted to $\{4,10\}$ regular tilings. $M_2$: number of rectangle modules (residual single square modules in parentheses if unmatched). $N_{M2}$: total physical qubits in the modular layout, including both data and ancilla qubits. $N_{M2}/k$: physical-to-logical qubit overhead. Automorphism Bases: logical basis operator coverage $k_{\mathrm{found}}/g$ and generator seed weights $w$, pruned to bases with $\ge 85\%$ coverage or the single highest-coverage basis otherwise. Cov.: maximum basis coverage percentage. Pair: mean number of automorphism moves between measuring the two cycles of a symplectic pair with a fixed extractor. Cayley: mean and maximum of the same over all pairs of basis loops. Both are right Cayley graph distances among basis automorphisms, with the extractor placed to minimise Pair (then the maximum, then the mean), for the listed basis with the highest coverage. $kd^2/n$: code figure of merit. A table across all quadrilateral tilings is given in \Cref{tab:candidates_L5_surface_ms2}.}
\label{tab:candidates_L5_surface_schlafli10}
\end{table*}

\Cref{tab:candidates_888,tab:candidates_L5_surface_schlafli10} list codes on the Pareto frontier for the two modular hyperbolic code families whose circuit-level simulations are shown in \Cref{fig:frontier_codes_plots}.
\Cref{tab:candidates_888} gives semi-hyperbolic color codes derived from $8.8.8$ tilings with side length $L=2$ and an octagon module partition, for which each module contains $16L^2+8=72$ physical qubits, as well as with side lengths $L=3$ and $L=4$ and a quadrilateral module partition (with $4L^2+2L=42$ and $72$ physical qubits per module, respectively), together with the logical bases found using $G^+$ (\Cref{sec:logicals}).
\Cref{tab:candidates_L5_surface_schlafli10} gives semi-hyperbolic surface codes derived from $\{4,10\}$ regular tilings with $L=5$ and a rectangle module partition, for which each module contains $4L^2+3L+1=116$ physical qubits.
Both tables are restricted to the Pareto frontier of the corresponding tiling family; broader tables spanning all tilings, module partitions and side lengths we considered are given in \Cref{sec:promising_candidate_tables}.

\subsection{Distance finding}

Distances of the hyperbolic color codes and surface codes are computed by finding the minimum length homologically non-trivial cycle (combinatorial systole) of certain tilings associated with the codes.
For the color codes, our string-like distance upper bound $d_{\mathrm{string}}\geq d$ is computed as twice the length of the smallest systole among the three shrunk lattices~\cite{bombin_color}.
In \Cref{sec:distance_finding} we prove that a lower bound on the distance is given by $d_{\mathrm{res}} \coloneqq \operatorname{median}\bigl(s_R^{\mathrm{res}},\, s_G^{\mathrm{res}},\, s_B^{\mathrm{res}}\bigr)\leq d$, where $s_C^{\mathrm{res}}$ is the systole of the restricted lattice~\cite{delfosse2014decoding,Kubica2023efficientcolorcode} of color $C\in\{R,G,B\}$.
Finally, the $X$ and $Z$ distances of the hyperbolic surface codes are computed exactly as the systole of a cell complex associated with the tiling.
All systoles are computed using a tree-cotree decomposition~\cite{tree_cotree_decomp}, accelerated by restricting the source nodes to a single fundamental domain (tile) of the tiling, which retains full accuracy due to the symmetries of the tiling.
In \Cref{sec:distance_finding} we explain our methods in more detail and show that our distance finding methods empirically scale linearly with code size and are orders of magnitude faster than alternative available methods for codes of size $n>10^3$--$10^4$.
Furthermore, from comparing to the exact connected-cluster method~\cite{dumer2017distance,dist_m4ri}, we observe that the upper bound $d_{\mathrm{string}}$ is usually tight ($d_{\mathrm{string}}=d$) with almost all discrepancies occurring for tilings with distance $d=2$ at $L=1$.

\subsection{Modular Syndrome Extraction Circuits}\label{sec:modular_syndrome_extraction_methods}
\begin{table}[b]
    \centering
    \small
    \setlength{\tabcolsep}{5pt}
    \begin{tabular}{@{}ll@{}}
    \toprule
    \textbf{Ideal operation} & \textbf{Noisy implementation} \\
    \midrule
    CZ, in-module              & $\mathrm{DEP}_2(p)\cdot\mathrm{CZ}$ \\
    CZ, inter-module           & $\mathrm{DEP}_2(p_{\mathrm{seam}})\cdot\mathrm{CZ}$ \\
    Single-qubit gate or idle  & $\mathrm{DEP}_1(p/10)\cdot U$ \\
    Reset, $R$               & $\mathrm{XERR}(2p)\cdot R$ \\
    Measurement, $M$         & $\mathrm{DEP}_1(p)\cdot\mathrm{MERR}(5p)$ \\
    Idle during $M$, $R$   & $\mathrm{DEP}_1(2p)$ \\
    \bottomrule
\end{tabular}
\caption{The superconducting-inspired modular SI1000 noise
model, where inter-module CZ gates carry an
independent error rate $p_{\mathrm{seam}}$; setting $p_{\mathrm{seam}}=p$ recovers
standard SI1000. Here $\mathrm{DEP}_1(q)$ and $\mathrm{DEP}_2(q)$ are one- and
two-qubit depolarizing channels of strength $q$, $\mathrm{XERR}(q)$ applies $X$ with
probability $q$, and $\mathrm{MERR}(q)$ flips the reported measurement outcome with
probability $q$. The single parameter $p$ sets the two-qubit gate error rate, with all
other rates defined relative to it.}
\label{tab:si1000}
\end{table}

\begin{figure}[t]
    \centering
    \includegraphics[width=\linewidth]{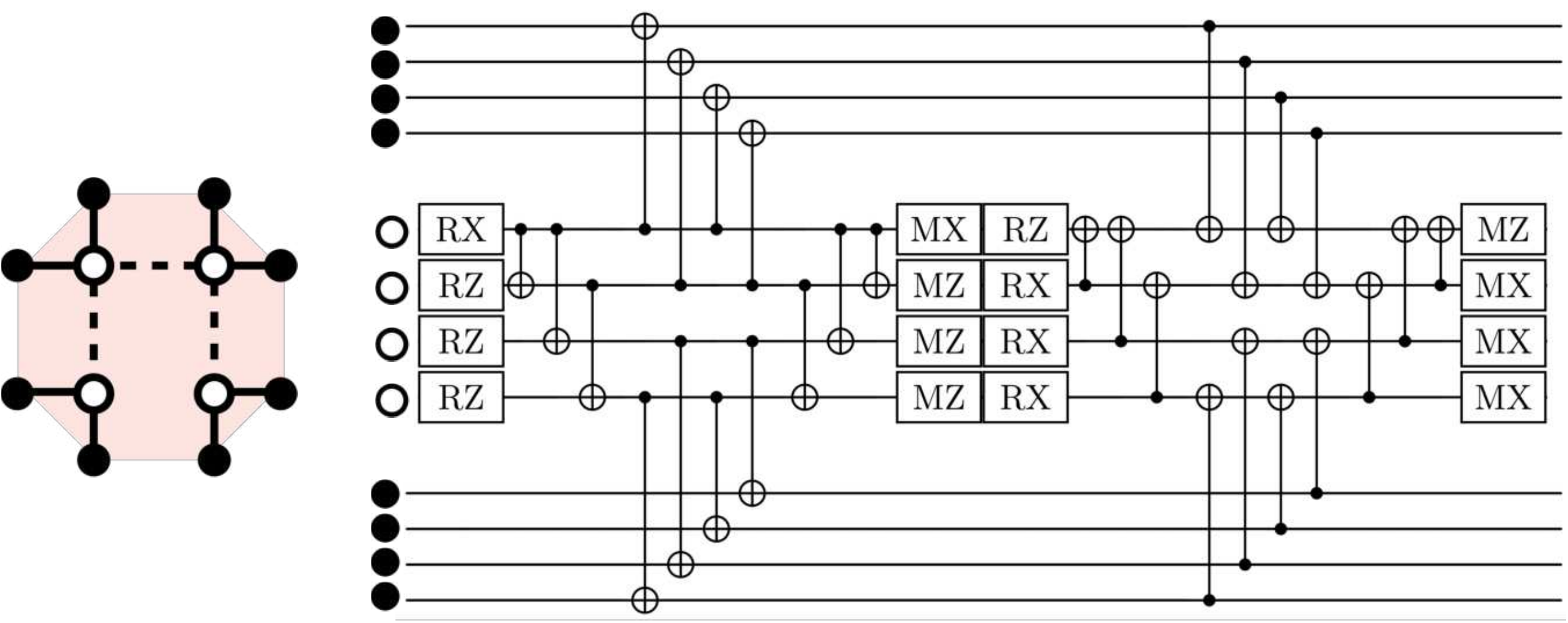}
    \caption{Cat-flagged syndrome extraction for a weight-$8$ corner stabilizer of a modular hyperbolic color code, such as those derived from $V8.8.8$ tilings.
    }
    \label{fig:cat_flagged_w8_a4}
\end{figure}

We design syndrome extraction circuits compatible with the planar module layout. In the
bulk of each module, stabilizers have weight four for surface codes and weight six for
color codes. We add one measure qubit per bulk face for surface codes and two measure
qubits per bulk face for color codes, with both cases giving a planar layout in which every
qubit couples to at most four nearest neighbors on a 2D Euclidean grid. At the edges of
modules, stabilizers span two modules; at the corners, a weight-$w$ stabilizer spans $m$
modules with one measure qubit per module, forming a linear chain of $m$ measure qubits.
We place all measure qubits using a graph flooding algorithm that preserves this layout.

To measure the higher-weight corner stabilizers while respecting the planar module layout
and existing inter-module connectivity, we develop cat-flagged circuits. We prepare a cat
state (a Greenberger--Horne--Zeilinger state) across the linear chain using the inter-module links, couple each measure qubit to
its local data qubits to extract the stabilizer, and measure the cat state. The linear
chain gives a cat-state preparation depth of $O(m)$. 

For a particular stabilizer on a face, one measure qubit reads out the syndrome and the
remaining $m-1$ act as flag qubits, which can detect hook errors. \Cref{fig:cat_flagged_w8_a4} shows a weight-eight
corner stabilizer spanning four modules with four measure qubits. For surface codes, we
interleave $X$ and $Z$ stabilizer extraction, ensuring that their interactions with
shared data qubits are consistently ordered. We find such a schedule by solving a constraint satisfaction problem on the base tiling, explained in Supplementary
Information. For color codes,
$X$ and $Z$ stabilizers are extracted sequentially. We schedule each extraction stage
using a translationally invariant plaquette schedule, assigning the same up-middle-down
ordering to every bulk plaquette within a module.

All stabilizers can be measured every cycle in the uniform-rate (measurement-aligned)
variant of our circuits. Inter-module stabilizers require additional time for cat or Bell
state preparation and measurement. In the multi-rate variant, we spread their
preparation, extraction and measurement across three consecutive cycles, interleaving
them with bulk stabilizer extraction so that the noisy inter-module links are used less
frequently.
We find that the multi-rate circuits have improved logical error rates relative to the uniform-rate circuits, see \Cref{sec:si_modular_circuits} for more details.

We benchmark these circuits under the modular SI1000 noise model of \Cref{tab:si1000}.
Circuits are first transpiled into a gate set of CZ gates, single-qubit rotations,
$Z$-basis resets and $Z$-basis measurements, and noise is applied to the transpiled
circuit. Every CZ gate acting between qubits in different modules is assigned the
elevated error rate $p_{\mathrm{seam}}$, while all other operations take their bulk rates
regardless of their position relative to a module boundary.

\begin{figure*}[ht]
    \centering
    \includegraphics[width=\linewidth]{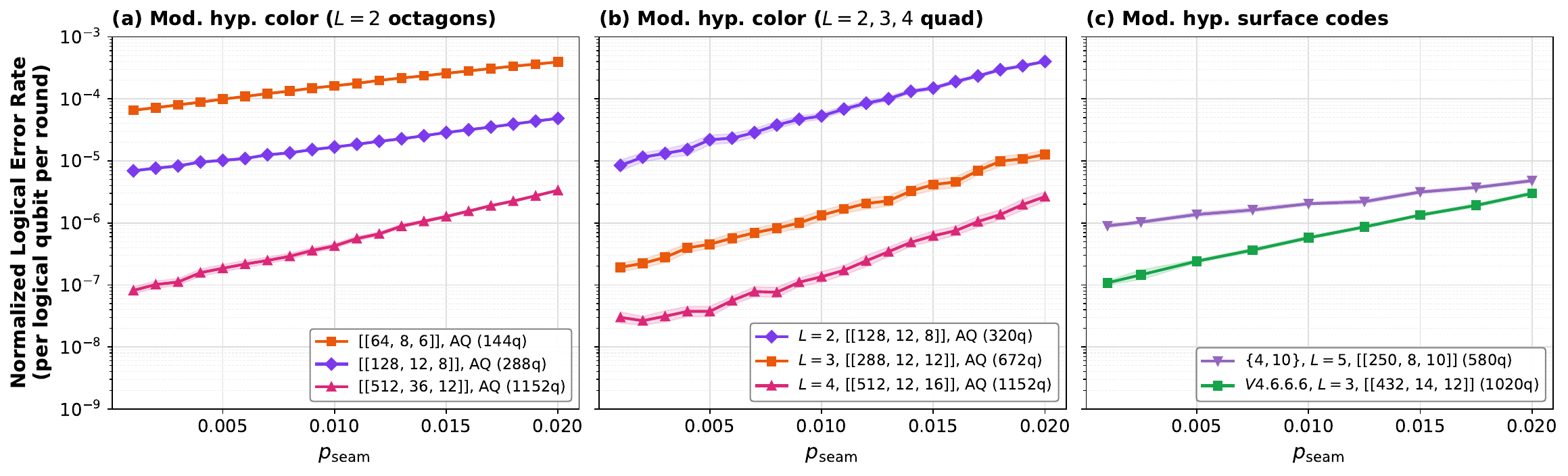}
    \caption{Logical error rate as a function of the inter-module error rate
    $p_{\mathrm{seam}}$, swept from $0.1\%$ to $2\%$ with the bulk error rate fixed at
    $p=0.1\%$. \textbf{a}, Modular hyperbolic color codes with $L=2$ octagon modules.
    \textbf{b}, Modular hyperbolic color codes with quadrilateral modules at $L=2,3,4$.
    \textbf{c}, Modular hyperbolic surface codes derived from two different tilings.}
    \label{fig:ler_vs_seam_noise}
\end{figure*}

\Cref{fig:ler_vs_seam_noise} shows the logical error rate as $p_{\mathrm{seam}}$ is swept from $0.1\%$ to $2\%$ with the bulk error rate fixed at $p=0.1\%$. For modular hyperbolic codes across all module partitions and all module sizes, the logical error rate is only weakly sensitive to $p_{\mathrm{seam}}$. This tolerance follows from boundary connectivity, since only the small fraction of gates that cross a module boundary is affected by the elevated error rates.

\subsection{Neural Network Decoding for Hyperbolic Color Codes}\label{sec:alphaqubit}

AlphaQubit (AQ) is a family of recurrent, attention-based decoders, mapping the error correction syndrome into a predicted probability that each logical observable is flipped~\cite{bausch2024learning,senior2025scalablerealtimeneuraldecoder}. We use here a variant that differs from what was previously published \cite{senior2025scalablerealtimeneuraldecoder} in several regards. 

The first difference is that the network is trained to produce several binary classification predictions, one per logical measurement in the circuit. We train each of them using a binary cross-entropy loss to independently predict each observable.

The second difference is that, rather than performing attention across all the stabilizer measurements, we can perform it across all triggered stabilizers (events) in a time window. Because the circuits have a few rounds only (16 rounds), here the time window corresponds to the whole experiment. 

The third major difference is the auxiliary targets we use during training. Instead of simulating stopping the experiment at every point in time~\cite{senior2025scalablerealtimeneuraldecoder}, we adapt an approach similar to~\cite{zhang2026learningdecodeparallelselfcoordinating}. 
In the detector error model, the prediction for each of the $k$ observables results from XOR’ing together the predictions for several error mechanisms. 
By assigning a time to each error mechanism (the time of the first event that it triggers), we can partition the error mechanisms that can flip each observable. By XOR’ing the errors in each partition, we produce $T\times k$ auxiliary binary targets for $T$ rounds. Predicting the auxiliary targets during training via auxiliary heads both improves training time and performance, even though those labels are not fault-tolerant observables in isolation. 

Another difference is our use of graph-derived spacetime coordinates (combining the round index with scaled spatial coordinates) for both input features and rotary position embeddings~\cite{reid2025rotary}. Unlike Poincar\'e projections, which introduce metric distortions and artificial boundary cuts, our spatial coordinates are obtained from a spectral embedding of the single-round detector error model hypergraph (using the leading non-trivial eigenvectors of a normalized diffusion operator), so that detectors that frequently co-occur in the noise model are placed close together in Euclidean space.

The resulting AQ network has around 55 million parameters. 
We train a different instance per circuit, using up to 3 billion simulated examples, from a mixture of SI1000 noise levels. After training, AQ can decode each 16-round example (wall time for the forward pass, on a single v6e Trillium TPU) from 140~$\mu$s to 6000~$\mu$s, depending on the size of the circuit.

\subsection{Correlated Matching Decoding for Hyperbolic Surface Codes}\label{sec:correlated_matching}

For the hyperbolic surface codes, we use a variant of the correlated sparse blossom decoder in pymatching~\cite{Higgott2025sparseblossom,fowler2013optimal}.
We use three passes, finding that this slightly improves accuracy relative to two-pass matching.

We use a custom decomposition of the detector error model by classifying the detectors in the circuit into four classes: $X$ flags, $X$ stabilizers, $Z$ flags and $Z$ stabilizers.
The $X$ ($Z$) stabilizer detectors are the ones formed from consecutive measurements of $X$ ($Z$) stabilizers in the circuit.
The flag detectors are formed from single measurements of qubits in the cat (or Bell) states that do not correspond to measurements of the code's stabilizers.
For example, when measuring an $X$ stabilizer using an $n$-qubit $X$ cat state (stabilized by $\langle X^{\otimes n},Z_0Z_1,Z_1Z_2,\ldots,Z_{n-2}Z_{n-1}\rangle$), the cat state is unencoded into a product state followed by single-qubit measurements, of which one is an $X$ stabilizer measurement and the remaining $n-1$ are $Z$ flag measurements.

Given these four detector classes, we decompose the detector error model (DEM) so that each component of a decomposed error triggers only detectors from a single class. Concretely, for each error mechanism $e$ in the DEM that triggers detectors belonging to more than one class, we decompose $e$ into components $e = e_1 \oplus e_2 \oplus \cdots$, where each $e_i$ triggers detectors from only one of the four classes $\{X\text{-stab},\, Z\text{-stab},\, X\text{-flag},\, Z\text{-flag}\}$. Each component $e_i$ is required to also appear as an independent error elsewhere in the DEM, which determines its detector and observable signature. The observable assignments are chosen so that the symmetric difference of the observables across all components matches the observables of the original undecomposed error.

This decomposition is finer than the standard two-class basis decomposition used for planar surface codes (which separates detectors only into $X$-type and $Z$-type). The finer four-class decomposition is necessary to ensure that each component of a decomposed error is matchable, i.e.\ that it triggers at most two detectors of its class. 
In each basis, a DEM error (or error component) is guaranteed to trigger two stabilizer detectors (since locally the code resembles a standard Euclidean surface code) and either one or two flag detectors (as shown by at most two flag detecting regions being supported on each qubit in \Cref{fig:cat_flagged_detectors}).
Without separating flag detectors from stabilizer detectors, a single component could trigger, for example, two $X$ stabilizer detectors together with one or two $X$ flag detectors, giving a component with three or four detectors that cannot be represented as an edge in the matching graph. 

The decomposed DEM is decoded using the correlated minimum-weight perfect matching decoder in PyMatching~\cite{Higgott2025sparseblossom}, which builds a single matching graph from the full decomposed DEM and accounts for correlations between the components of decomposed errors when computing the minimum-weight matching. We run the decoder for three passes, following the iterative re-weighting approach of Ref.~\cite{fowler2013optimal}: after each pass, the edge weights in the matching graph are updated based on the errors identified in the previous pass, and the matching is recomputed. We find that three passes slightly improves accuracy relative to two passes, with diminishing returns beyond three.

\subsection{Simulation methodology}\label{sec:simulation_methodology}

All logical error rates are estimated by Monte Carlo sampling of separate $X$ and $Z$ basis memory experiments under the modular SI1000 noise model (\Cref{tab:si1000}) with $p=0.1\%$ and, unless stated otherwise, $p_{\mathrm{seam}}=1\%$, using Stim~\cite{Gidney2021stimfaststabilizer}. 
From the shot failure probability $P_L$ we report the logical error rate per logical qubit per round, $\epsilon = 1-(1-P_L)^{1/(kr)}$, which is $\approx P_L/(kr)$ in the low-error regime $P_L\ll 1$ relevant here%
, and report 95\% Clopper--Pearson confidence intervals. 
Surface codes (hyperbolic and Euclidean) use $r=30$ and the correlated matching decoder (\Cref{sec:correlated_matching}). 
Hyperbolic color codes use $r=16$ and AlphaQubit (\Cref{sec:alphaqubit}). 
While the number of rounds differs between code families, we do not expect this to impact our comparisons, since we normalize the logical error rate by the number of rounds. 
For the hyperbolic surface codes in \Cref{fig:frontier_codes_plots} we use multi-rate circuits, alternate the extraction order forward and backward between rounds, and use a ZZ schedule within each module (see \Cref{sec:modular_syndrome_extraction_methods} and \Cref{sec:hyperbolic_surface_code_circuits_si}).

We compare to rotated surface codes of odd distance $3\le d\le 25$ using a standard NZ schedule. 
Each code is partitioned into modules of $7\times 7$ data qubits (98 qubits per module), every inter-module CZ gate is assigned $p_{\mathrm{seam}}$, and decoding uses the same correlated matching decoder. 
For simplicity, the simulated Euclidean circuits measure each stabilizer cut by a module boundary with a single measure qubit, coupled to the data qubits in neighboring modules by inter-module CZ gates, rather than using the cat-state and flag construction of our hyperbolic circuits. %
A modular partition of the code, however, needs a measure qubit in every module the stabilizer touches, for example to prepare the Bell pairs that mediate the inter-module gates, so to compare qubit counts fairly we count one extra measure qubit for each additional module touched by a cut stabilizer. This gives $2d^2-1+2wd+w^2$ qubits, with $L=7$ and $w=\lfloor (d-1)/L\rfloor$ cuts in each direction: each cut is crossed by $d$ stabilizers, and the stabilizer at each of the $w^2$ cut intersections touches four modules. The hyperbolic circuits include the corresponding qubits explicitly.

The overhead of a code is its total number of physical qubits (including ancilla qubits) divided by $k$.
For a simulated hyperbolic code, we compute the density improvement against the largest-distance surface code whose $\epsilon$ is no lower than that of the hyperbolic code (the improvement is that surface code's overhead divided by the hyperbolic code's overhead).
Hollow symbols mark codes that are too large to simulate or decode.
We plot these at the measured $\epsilon$ of the Euclidean surface code with the largest odd distance not exceeding $d$.
This extrapolation is supported by the fact that every simulated hyperbolic code except the smallest, $[[64,8,6]]$, reaches a lower $\epsilon$ than the odd-distance Euclidean surface code at distance $d$ or $d-1$.
For the color codes, $d$ is the stringlike upper bound of \Cref{sec:distance_finding}, which we expect to be tight.

\subsection{Fault-Tolerant Automorphism Gates} \label{sec:auts}

\begin{figure}[t]
    \centering
    \includegraphics[width=1.0\linewidth]{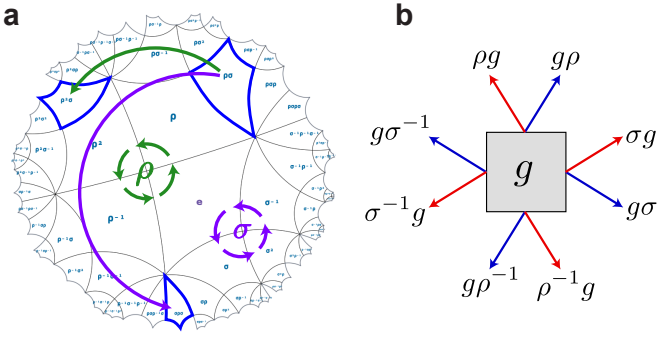}
    \caption{Modules and inter-module connectivity as group elements and action. \textbf{a}, The quadrilateral modules are in bijection with the elements of the group generated by the two rotations $\rho$ (green) and $\sigma$ (purple), shown here in the Poincar\'e disk with each module labeled by its group element. Taking one module as the identity $e$, every other module is labeled by a sequence of rotations mapping $e$ onto it. \textbf{b}, The two kinds of inter-module connection required at a module $g$. Right multiplication (blue) gives the four modules adjacent to $g$ in the tiling, whose shared boundaries carry the couplers used to measure stabilizers spanning two modules. Left multiplication (red) gives the modules that $g$ is mapped onto by the rotations, whose shared boundaries carry the couplers used by the walking circuits that implement the automorphisms.}
    \label{fig:module_connectivity}
\end{figure}

A code automorphism is a permutation of the physical qubits that preserves the stabilizer group, while implementing a (potentially nontrivial) logical operation~\cite{calderbank1998qec,moussa2016transversal,Breuckmann2024foldtransversal,Quintavalle2023partitioningqubits,autqec}.
For hyperbolic codes, the code automorphisms are derived from the symmetries of the hyperbolic tiling.
In this section we explain how certain code automorphisms can be implemented using \emph{walking circuits}.

Let's consider modular hyperbolic color codes with a quadrilateral module partition as an example, for which the modules are in bijection with the elements of $G^+$.
Left multiplication by an element $h\in G^+$ defines a rigid motion (isometry) of the tiling, mapping module $g$ to $hg$ (see \Cref{fig:module_connectivity}).
We define a consistent internal coordinate system of each module, such that $(g,x,y)$ denotes a qubit at coordinate $(x,y)$ in module $g$.
Notice that any automorphism $h\in G^+$ permutes the modules while preserving their internal coordinate system, mapping $(g,x,y)\rightarrow(hg,x,y)$.
We can implement such an automorphism by using inter-module couplers to connect the north boundary of module $g$ to the south boundary of module $hg$, using a walking circuit (consisting of SWAP gates) to map the quantum information in module $g$ to module $hg$.

\begin{figure}[ht]
    \centering
    \includegraphics[width=0.5\linewidth]{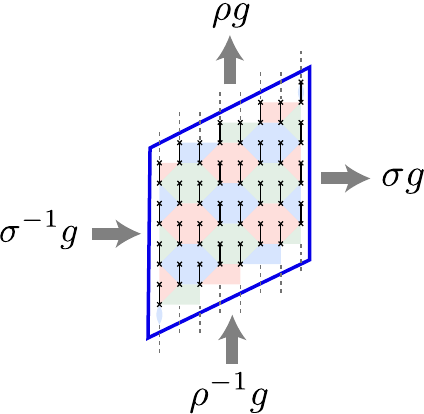}
    \caption{A permutation of the modules
    is implemented using walking circuits and additional seams on opposing boundaries,
    shown here for a single module $g$. Under the global rotation $\rho$, module $g$ is
    mapped onto $\rho g$: the north seam of $g$ connects to the south seam of $\rho g$.
    The east and west seams implement $\sigma$ in the same way, and the two rotations
    together generate the full orientation-preserving symmetry group. Grey arrows mark
    the four modules reached by left multiplication.}
    \label{fig:module_aut_connectivity}
\end{figure}

We can similarly implement an arbitrary automorphism in $G^+$ by adding a second seam to the east and west boundaries of each module.
Since $G^+$ is generated by the two elements $\rho\coloneqq ab$ and $\sigma\coloneqq bc$, we add additional couplers at the north and south boundaries to implement the automorphism $g\rightarrow \rho g$ and use the east and west boundaries to implement $g\rightarrow \sigma g$, see \Cref{fig:module_aut_connectivity}.
With these additional couplers we can use walking circuits to implement \textit{any} automorphism in $G^+$ using a sequence of automorphisms in $\{\sigma,\sigma^{-1},\rho,\rho^{-1}\}$.
Since the existing seams required for code connectivity (stabilizer measurement) correspond to right multiplication by $\rho$ and $\sigma$, the overall inter-module connectivity corresponds to a left-right Cayley graph of $G^+$, see \Cref{fig:module_connectivity}.

Implementing a single automorphism $\rho$ or $\sigma$ (an automorphism \textit{move}) with a walking circuit takes $O(L)$ steps, during which only the stabilizers originally fully internal to a module can be repeatedly measured.
The stabilizers originally on the boundary of the module become dislocated until the move is completed, after which they can be measured again.
We describe the automorphism walking circuits in more detail in \Cref{sec:automorphism_circuits}. In \Cref{sec:stabilizer_dislocation} we also simulate a simplified model in which the stabilizers on module boundaries are not measured for several rounds, providing evidence that this stabilizer dislocation causes only a modest increase in logical error rates.

For the larger octagon color code modules, each containing eight triangles, this construction requires modification. The octagon modules are in bijection with the left cosets $G/D = \{ gD \mid g \in G \}$, where $D\coloneqq \langle b,c\rangle\leq G$ is a dihedral subgroup of order eight, $|D|=8$.
The automorphism $\sigma=bc\in G^+$ would rotate the identity module $\langle b,c\rangle$ rather than leaving the internal coordinates fixed, which is not compatible with our walking circuits using fixed inter-module couplers.
We resolve this by finding a \textit{complement} $\Lambda\leq G$ of $D\leq G$, which by definition satisfies
\begin{equation}
    \Lambda D=G,\quad \Lambda\cap D=\{e\}.
\end{equation}
Each element $g\in G$ can then be factorized as $g=\lambda d$, where $\lambda\in\Lambda$ labels the octagon module that triangle $g$ belongs to and $d\in D$ is the \textit{slot} of $g$ in module $\lambda\in\Lambda$.
We refer to the identity slot $e\in D$ in each module as its \textit{frame}, which is used to define a fixed, consistent internal coordinate system $(x,y)$ for each module $\lambda\in\Lambda$.
Left multiplication by any element $\lambda^\prime\in\Lambda$ then maps $(\lambda d,x,y)\rightarrow (\lambda^\prime\lambda d,x,y)$, corresponding to an automorphism that permutes modules while leaving the internal coordinate system fixed, as desired.
Using additional seams this can be implemented using walking circuits in a similar way to how we described for quadrilateral modules.

It is not \textit{a priori} obvious whether a complement $\Lambda\leq G$ of $D\leq G$ exists for a given tiling $\mathcal{T}^\triangle$.
Additionally, it is desirable for $\Lambda$ to be two-generated, such that only one additional seam of inter-module couplers is required on each of the four boundaries of the module.
In \Cref{sec:module_complements} we provide an efficient method to enumerate the complements $\Lambda$ of $D$ and find that complements do exist for many of the tilings in \Cref{tab:candidates_888}, including those with symmetry group $\triangle(4,4,4)_i$ for $i\in \{32,40,58,109\}$, but not for $\triangle(4,4,4)_{96}$ (\Cref{tab:module_complements}).
Moreover, for each of these four tilings, we find a complement requiring at most two generators.

\subsection{Low-weight Symmetrical Logical Basis}\label{sec:logicals}

Logical operators are associated with homologically non-trivial cycles of the tiling, and identifying a \emph{canonical logical basis} reduces to finding a \emph{canonical homology basis} for the underlying hyperbolic tiling. Two non-trivial cycles form a symplectic pair if they cross each other an odd number of times. Such a pair supports two logical qubits for surface codes and four for color codes. A complete canonical homology basis consists of $g$ symplectic pairs, where cycles within each pair cross an odd number of times and cycles belonging to distinct pairs cross an even number of times. This gives $k = 2g$ logical qubits for surface codes and $k = 4g$ for color codes, where each cycle additionally supports operators of two independent colors. Shorter cycles give lower-weight logical operators, enabling more efficient Pauli measurements. Finding a shortest homology basis is tractable in polynomial time~\cite{Erickson_Whittlesey}, and any homology basis can be brought into canonical form by Gaussian elimination over $\mathbb{F}_2$ also in polynomial time. However, Gaussian elimination combines cycles, which may increase their length and generally yields cycles that are no longer related to one another by code automorphisms. 

We develop algorithms that address this by exploiting the symmetry structure of the tiling. We enumerate short homologically non-trivial cycles from every vertex by breadth-first search, retaining the shortest representative per homology class. One or two of the shortest cycles are selected as seed cycles. Their full orbit under $G^+$ is generated, producing a candidate set of cycles related by the tiling symmetries. We then identify mutually compatible symplectic pairs by a backtracking search on the candidate set. For most hyperbolic tilings we consider, one or two seed cycles suffice to generate a complete or near-complete basis yielding logical operators at or near minimum weight (see \Cref{tab:candidates_888}). The resulting basis is lifted from the hyperbolic tiling to the semi-hyperbolic tiling to obtain low-weight logical operators compatible with the module partition. Full details are in the Supplementary Information.

Since logical operators in the basis are related by $\rho$ and $\sigma$, a single extractor system \cite{he2025extractorsqldpcarchitecturesefficient} suffices per seed cycle. Applying walking circuits rotates between all logical operators in the orbit of that seed. Since the length of a homologically non-trivial cycle on a hyperbolic surface grows logarithmically with its area, each extractor system couples to only a logarithmic fraction of the total modules, provided the cycles in the basis are close to minimum length. With one or two seed cycles, a constant number of extractor systems, each coupled to $O(\log M)$ of the $M$ modules, is therefore sufficient to access all logical operators. If the extractor is attached to the cycle $h\gamma$, where $\gamma$ is the seed cycle and $h\in G^+$, then the cycle $x\gamma$ is brought under it by the automorphism $hx^{-1}$, so switching from measuring $x\gamma$ to measuring $y\gamma$ takes $|h\,x^{-1}y\,h^{-1}|$ moves, where $|\cdot|$ is the word length in the generators $\{\rho^{\pm1},\sigma^{\pm1}\}$. For $h=e$ this is the distance between $x$ and $y$ in the right Cayley graph of $G^+$, rather than the left Cayley distance $|yx^{-1}|$ (the length of the automorphism that maps $x\gamma$ onto $y\gamma$). Since $G^+$ is in general non-Abelian, the choice of $h$ affects the number of moves and can be optimized (\Cref{sec:ripple_carry_supplement}). For the $[[6400,104,30]]$ code, bringing any basis cycle under the extractor, starting from any other, takes a mean of 4.4 and at most six moves (\Cref{tab:candidates_888}).

\subsection{Logical Measurement}\label{sec:logical_measurement_methods}

Given a choice of seed cycle, we can build an ancillary extractor system to measure out the set of non-trivial logical Pauli operators associated with the given seed cycle. One important fact is that for any logical operator belonging to a given seed cycle in a semi-hyperbolic surface or color code tiling, we can always choose a representative that is only supported on the data qubits at the boundary of individual modules forming the hyperbolic code, see \Cref{fig:logic}b for an example.

We will construct an extractor system that will be itself formed of a collection of modules that will be connected through their respective boundary links. The extractor will then couple to the base code along the boundaries of the modules that support the logical operators corresponding to the chosen seed cycle. This coupling to the base code will use its own set of boundary links, and locally will have many of the features that are common to lattice surgery, see \Cref{fig:logic}b and~\Cref{sec:ribbon_extractor} for further details. In particular, there will be a new set of stabilizers that will have to be measured across qubits belonging to the measurement ancilla and base code, in a similar fashion to those measured across two different tiles in the base semi-hyperbolic code. 

Overall, the simplest such extractor will be topologically equivalent to a long ribbon, whose base connects to the base code in the aforementioned way. The ribbon extractor itself will be equivalent to a long strip of a closed surface code, built from a linear chain of connected modules, such that the product of all of the stabilizers of one type will determine the value of the desired logical measurement~\cite{cohen2022low-overhead}. Such an extractor would be able to measure any logical operator that is supported on the seed cycle, which for the color code tilings would be any of the 15 non-trivial logical Pauli representatives. 

To quantify the physical overhead of the ribbon extractor, we detail an explicit construction for a tiling with side length $L=4$ for the $[[6400,104,30]]$ hyperbolic color code derived from a $V8.8.8$ tiling, as given explicitly in \Cref{fig:logic}b and detailed in~\Cref{sec:ribbon_extractor} why such a ribbon extractor is distance-preserving. Each extractor module is built in layers, where the first layer will have 4~ancilla (data) qubits and 6~ancilla check qubits and the second layer will have 6 qubits of each. 
In order for the extractor to be provably distance preserving, we use $d/2$~odd layers and $d/2-1$~even layers (this is a well-studied approach called thickening~\cite{cohen2022low-overhead,Williamson2026,he2025extractorsqldpcarchitecturesefficient}), thus totaling $10d/2+12(d/2-1) = 11d-12$ qubits per extractor module. There will be one such extractor module for each tile boundary that is a part of the seed loop, which in the case of the given lattice is 10 such modules. This gives an overhead of 3180~qubits (both ancilla data and check qubits) for the ribbon extractor, which results in about a 22\%~overhead as the number of physical (data and measurement) qubits in the base code is 14400, see~\Cref{tab:candidates_L4_color_ms2_888}. 
Choosing $L=3$ instead, the same seed cycle gives the $[[3600,104,22]]$ code with 8400 physical qubits (\Cref{tab:candidates_L3_color_ms2_888}), for which the ribbon extractor requires 1880 qubits, about 22\% of the memory. 
Using the $L=2$ octagon partition of this tiling ($[[1600,104,14\text{--}16]]$, 3600 physical qubits), the ribbon extractor would require 1040 qubits (about 29\% of the memory). Here the smaller group of implementable automorphisms splits the same basis into four seed cycles rather than one, but since each extractor module uses only one seam of inter-module couplers to attach to the memory, a single extractor can serve all four by adding additional seams on the sides of the ribbon, one to each seed cycle (\Cref{sec:module_complements}). Since the complement is smaller than $G^+$ (50 rather than 200 elements), bringing any basis cycle under the extractor then takes a mean of 2.9 and at most four moves (\Cref{tab:module_complements}).
It should be noted that this is a rather overly-cautious bound on the physical overhead, as the number of layers of thickening is taken to be $d/2$~which is in all likelihood superfluous for the purpose of a rigorous distance proof. Namely, if we have some guarantee of the overlap of other logical operators with the target logical operator being measured, then this overhead could be significantly reduced. 

Finally, in addition to measuring the aforementioned logical qubits, the ribbon extractor can be used to facilitate the joint measurement of a logical qubit from the base code and another code patch~\cite{cross2024improved}. This is particularly useful if one were to treat the hyperbolic code as a form of logical memory and have some other code, say a surface code, that would be used as a compute region. The ribbon extractor would be used to perform joint logical measurements of the form~$\overline{X}_m\overline{X}_c$ and~$\overline{Z}_m\overline{Z}_c$, where $\overline{P}_m$ and $\overline{P}_c$ are logical Pauli operators for the memory and compute codes, respectively. The ribbon extractor would achieve this by coupling to the base hyperbolic code as discussed while additionally coupling its upper boundary to the compute code, for example a surface code, thereby jointly measuring the logical operators across both codes. In order to complete both of these measurements, one would perform an automorphism gate between measurements to rotate the appropriate conjugate basis element from the hyperbolic code to facilitate this joint measurement between code blocks. Together, these joint logical Pauli measurements enable fault-tolerant logical state and gate teleportation between the memory and compute blocks~\cite{gottesman1999demonstrating}. Universal quantum circuits can then be compiled and executed via Pauli-based computation~\cite{bravyi2016trading, litinski2019game}, consuming logical magic states supplied to the compute region via magic state distillation~\cite{bravyi2005universal, litinski2019magic} or cultivation~\cite{gidney2024magic}.

\subsection{Compiling an adder onto the hyperbolic memory}\label{sec:adder_methods}

We compile the $n$-bit CDKM ripple-carry adder~\cite{cuccaro2004new} onto the $[[3600,104,22]]$ ($L=3$) and $[[6400,104,30]]$ ($L=4$) modular hyperbolic color codes derived from the $\triangle(4,4,4)_{109}$ 8.8.8 tiling, storing the $a$ and $b$ registers in the memory and applying all gates in a compute region with $C$ operand slots.
Qubits are teleported between the two using the ribbon extractor (\Cref{sec:logical_measurement_methods}): a load uses the measurements $M_{\overline{X}_q\overline{X}_c}$ and $M_{\overline{Z}_q}$, and a store uses $M_{\overline{X}_q\overline{X}_c}$.
Retaining operands in the compute region between blocks, the adder requires $N_R=3(4n-C)$ ribbon measurements, and automorphism moves (\Cref{sec:auts}) bring each logical operator under the extractor.
With $\alpha d$ cycles per ribbon measurement and $\kappa L$ per move, this takes $T=\alpha d N_R+\kappa L N_A$ cycles, where $N_A$ is the number of moves.
Every symplectic pair in our basis is related by the same automorphism, so with an appropriate extractor location, switching between the two cycles of a pair takes a single move.
Performing the two measurements of each load consecutively, and using at most $D=6$ moves (the diameter of the Cayley graph of $G^+$ with generators $\{\rho^{\pm1},\sigma^{\pm1}\}$) for every other measurement, gives $N_A\leq(2D+1)(4n-C)$.
We further store the four operands $a_{2i},b_{2i},a_{2i+1},b_{2i+1}$ in the same symplectic pair $(\alpha_i,\beta_i)$, with each $\overline{X}_q$ on $\alpha_i$ and each $\overline{Z}_q$ on $\beta_i$.
With $C=3$, the loads and stores before each block then act on at most two pairs, each costing at most one jump of $\leq D$ moves and one switch, giving $N_A\leq 4(D+1)n$ with $N_R=3(4n-3)$.
For a $50$-bit adder with $\alpha=1$ and $\kappa=2$, this gives $T\leq 21{,}402$ cycles (428 per bit) at $L=3$ and $T\leq 28{,}930$ cycles (579 per bit) at $L=4$, compared with $28{,}080$ and $37{,}830$ cycles without grouping ($C=5$), and $33{,}930$ and $45{,}630$ cycles for the coarse bound $N_A\leq DN_R$.
The compute-region logic can largely run in parallel with the extractor measurements and automorphisms.
See \Cref{sec:ripple_carry_supplement} for details.

\section*{Acknowledgments}
We thank Earl Campbell, Sergio Boixo and Alex Gaunt for valuable feedback on the manuscript. We also thank Hussain Masoom for support, and Cody Jones, Mike Newman, Guang Hao Low and Leonid Pryadko for helpful discussions.

\section*{AI Usage}

In the later stages of the project, AI tools assisted with tasks including implementing parts of the research software, helping with technical details of the octagon module subgroup complements and graph-derived detector coordinates, and proposing the argument underlying the color code distance lower bound. The authors developed the key ideas and constructions, prepared the manuscript, verified all AI-assisted contributions and take full responsibility for the content.

\section*{Author Contributions}

O.H. conceived and guided the project, which was motivated by a boundary-connected hardware model that M.M. proposed.
H.S. and O.H. developed the modular hyperbolic color code constructions, cat-flagged circuits, and symmetric canonical logical bases. H.S. implemented the first versions of the hyperbolic color code circuits and the Euclidean surface code circuits used for comparison. O.H. carried out the code search, developed the qubit layouts, the hyperbolic surface code circuits and the distance-finding methods, adapted correlated matching to the hyperbolic surface codes, designed the automorphism walking circuits, performed the final numerical simulations, and compiled the ripple-carry adder. 
Z.H. and T.J.-O., with support from O.H., designed the ribbon extractor and proved that it preserves the code distance.
F.H., A.S., L.Z., T.E. and J.S. developed and extended the AlphaQubit decoder and ran decoding experiments, with guidance from J.B. and input from O.H. on the graph-derived detector coordinates. 
M.M. advised on hardware assumptions. C.G. assisted with the three-dimensional visualization in Fig. 1 and the adder compilation. 
O.H. and H.S. prepared most of the manuscript and figures, with Z.H. and T.J.-O. contributing text and figures on logical measurement and the Google DeepMind authors contributing the Methods on AlphaQubit. 
P.K. and H.N. provided leadership and support. All authors discussed the results and commented on the manuscript.

\section*{Additional Information}
\noindent\textbf{Supplementary Information} is available for this paper. \textbf{Correspondence and requests for materials} should be addressed to
O.H. (oscarhiggott@google.com) or H.S. (hasansayginel@google.com).

\bibliography{bibliography}

\providecommand{\noopsort}[1]{}\providecommand{\singleletter}[1]{#1}%
\begin{thebibliography}{48}%
\makeatletter
\providecommand \@ifxundefined [1]{%
 \@ifx{#1\undefined}
}%
\providecommand \@ifnum [1]{%
 \ifnum #1\expandafter \@firstoftwo
 \else \expandafter \@secondoftwo
 \fi
}%
\providecommand \@ifx [1]{%
 \ifx #1\expandafter \@firstoftwo
 \else \expandafter \@secondoftwo
 \fi
}%
\providecommand \natexlab [1]{#1}%
\providecommand \enquote  [1]{``#1''}%
\providecommand \bibnamefont  [1]{#1}%
\providecommand \bibfnamefont [1]{#1}%
\providecommand \citenamefont [1]{#1}%
\providecommand \href@noop [0]{\@secondoftwo}%
\providecommand \href [0]{\begingroup \@sanitize@url \@href}%
\providecommand \@href[1]{\@@startlink{#1}\@@href}%
\providecommand \@@href[1]{\endgroup#1\@@endlink}%
\providecommand \@sanitize@url [0]{\catcode `\\12\catcode `\$12\catcode
  `\&12\catcode `\#12\catcode `\^12\catcode `\_12\catcode `\%12\relax}%
\providecommand \@@startlink[1]{}%
\providecommand \@@endlink[0]{}%
\providecommand \url  [0]{\begingroup\@sanitize@url \@url }%
\providecommand \@url [1]{\endgroup\@href {#1}{\urlprefix }}%
\providecommand \urlprefix  [0]{URL }%
\providecommand \Eprint [0]{\href }%
\providecommand \doibase [0]{https://doi.org/}%
\providecommand \selectlanguage [0]{\@gobble}%
\providecommand \bibinfo  [0]{\@secondoftwo}%
\providecommand \bibfield  [0]{\@secondoftwo}%
\providecommand \translation [1]{[#1]}%
\providecommand \BibitemOpen [0]{}%
\providecommand \bibitemStop [0]{}%
\providecommand \bibitemNoStop [0]{.\EOS\space}%
\providecommand \EOS [0]{\spacefactor3000\relax}%
\providecommand \BibitemShut  [1]{\csname bibitem#1\endcsname}%
\let\auto@bib@innerbib\@empty
\bibitem [{\citenamefont {Bombin}\ and\ \citenamefont
  {Martin-Delgado}(2006)}]{bombin_color}%
  \BibitemOpen
  \bibfield  {author} {\bibinfo {author} {\bibfnamefont {H.}~\bibnamefont
  {Bombin}}\ and\ \bibinfo {author} {\bibfnamefont {M.~A.}\ \bibnamefont
  {Martin-Delgado}},\ }\bibfield  {title} {\bibinfo {title} {Topological
  quantum distillation},\ }\href
  {https://doi.org/10.1103/PhysRevLett.97.180501} {\bibfield  {journal}
  {\bibinfo  {journal} {Phys. Rev. Lett.}\ }\textbf {\bibinfo {volume} {97}},\
  \bibinfo {pages} {180501} (\bibinfo {year} {2006})}\BibitemShut {NoStop}%
\bibitem [{\citenamefont {Vuillot}\ and\ \citenamefont
  {Breuckmann}(2022)}]{quantum_pin_codes}%
  \BibitemOpen
  \bibfield  {author} {\bibinfo {author} {\bibfnamefont {C.}~\bibnamefont
  {Vuillot}}\ and\ \bibinfo {author} {\bibfnamefont {N.~P.}\ \bibnamefont
  {Breuckmann}},\ }\bibfield  {title} {\bibinfo {title} {Quantum pin codes},\
  }\href {https://doi.org/10.1109/TIT.2022.3170846} {\bibfield  {journal}
  {\bibinfo  {journal} {IEEE Transactions on Information Theory}\ }\textbf
  {\bibinfo {volume} {68}},\ \bibinfo {pages} {5955} (\bibinfo {year}
  {2022})}\BibitemShut {NoStop}%
\bibitem [{\citenamefont {Higgott}\ and\ \citenamefont
  {Breuckmann}(2024)}]{higgott_2024_floquet}%
  \BibitemOpen
  \bibfield  {author} {\bibinfo {author} {\bibfnamefont {O.}~\bibnamefont
  {Higgott}}\ and\ \bibinfo {author} {\bibfnamefont {N.~P.}\ \bibnamefont
  {Breuckmann}},\ }\bibfield  {title} {\bibinfo {title} {Constructions and
  performance of hyperbolic and semi-hyperbolic {Floquet} codes},\ }\href
  {https://doi.org/10.1103/PRXQuantum.5.040327} {\bibfield  {journal} {\bibinfo
   {journal} {PRX Quantum}\ }\textbf {\bibinfo {volume} {5}},\ \bibinfo {pages}
  {040327} (\bibinfo {year} {2024})}\BibitemShut {NoStop}%
\bibitem [{\citenamefont {Breuckmann}\ and\ \citenamefont
  {Terhal}(2016)}]{breuckmann_2016_constructions}%
  \BibitemOpen
  \bibfield  {author} {\bibinfo {author} {\bibfnamefont {N.~P.}\ \bibnamefont
  {Breuckmann}}\ and\ \bibinfo {author} {\bibfnamefont {B.~M.}\ \bibnamefont
  {Terhal}},\ }\bibfield  {title} {\bibinfo {title} {Constructions and noise
  threshold of hyperbolic surface codes},\ }\href
  {https://doi.org/10.1109/TIT.2016.2555700} {\bibfield  {journal} {\bibinfo
  {journal} {IEEE Transactions on Information Theory}\ }\textbf {\bibinfo
  {volume} {62}},\ \bibinfo {pages} {3731} (\bibinfo {year}
  {2016})}\BibitemShut {NoStop}%
\bibitem [{\citenamefont {Rober}(2024)}]{lins}%
  \BibitemOpen
  \bibfield  {author} {\bibinfo {author} {\bibfnamefont {F.}~\bibnamefont
  {Rober}},\ }\href@noop {} {\bibinfo {title} {{LINS}, provides an algorithm
  for computing the normal subgroups of a finitely presented group up to some
  given index bound., {V}ersion 0.9}},\ \bibinfo {howpublished} {\href
  {https://gap-packages.github.io/LINS/}
  {\texttt{https://gap-packages.github.io/}\discretionary
  {}{}{}\texttt{LINS/}}} (\bibinfo {year} {2024}),\ \bibinfo {note} {{GAP}
  package}\BibitemShut {NoStop}%
\bibitem [{GAP()}]{GAP4}%
  \BibitemOpen
  GAP,\ \href {https://www.gap-system.org} {\emph {\bibinfo {title} {{GAP} ---
  {Groups}, {Algorithms}, and {Programming}, {Version} 4.13.1}}},\ \bibinfo
  {organization} {The GAP~Group} (\bibinfo {year} {2024})\BibitemShut {NoStop}%
\bibitem [{\citenamefont {Bombin}\ and\ \citenamefont
  {Martin-Delgado}(2007)}]{bombin_delgado_toric_surface_color}%
  \BibitemOpen
  \bibfield  {author} {\bibinfo {author} {\bibfnamefont {H.}~\bibnamefont
  {Bombin}}\ and\ \bibinfo {author} {\bibfnamefont {M.~A.}\ \bibnamefont
  {Martin-Delgado}},\ }\bibfield  {title} {\bibinfo {title} {Optimal resources
  for topological two-dimensional stabilizer codes: Comparative study},\ }\href
  {https://doi.org/10.1103/PhysRevA.76.012305} {\bibfield  {journal} {\bibinfo
  {journal} {Phys. Rev. A}\ }\textbf {\bibinfo {volume} {76}},\ \bibinfo
  {pages} {012305} (\bibinfo {year} {2007})}\BibitemShut {NoStop}%
\bibitem [{\citenamefont {Delfosse}(2013)}]{delfosse_tradeoffs}%
  \BibitemOpen
  \bibfield  {author} {\bibinfo {author} {\bibfnamefont {N.}~\bibnamefont
  {Delfosse}},\ }\bibfield  {title} {\bibinfo {title} {Tradeoffs for reliable
  quantum information storage in surface codes and color codes},\ }in\ \href
  {https://doi.org/10.1109/ISIT.2013.6620360} {\emph {\bibinfo {booktitle}
  {2013 IEEE International Symposium on Information Theory}}}\ (\bibinfo {year}
  {2013})\ pp.\ \bibinfo {pages} {917--921}\BibitemShut {NoStop}%
\bibitem [{\citenamefont {Breuckmann}\ \emph {et~al.}(2017)\citenamefont
  {Breuckmann}, \citenamefont {Vuillot}, \citenamefont {Campbell},
  \citenamefont {Krishna},\ and\ \citenamefont
  {Terhal}}]{Breuckmann_2017_semihyperbolic}%
  \BibitemOpen
  \bibfield  {author} {\bibinfo {author} {\bibfnamefont {N.~P.}\ \bibnamefont
  {Breuckmann}}, \bibinfo {author} {\bibfnamefont {C.}~\bibnamefont {Vuillot}},
  \bibinfo {author} {\bibfnamefont {E.}~\bibnamefont {Campbell}}, \bibinfo
  {author} {\bibfnamefont {A.}~\bibnamefont {Krishna}},\ and\ \bibinfo {author}
  {\bibfnamefont {B.~M.}\ \bibnamefont {Terhal}},\ }\bibfield  {title}
  {\bibinfo {title} {Hyperbolic and semi-hyperbolic surface codes for quantum
  storage},\ }\href {https://doi.org/10.1088/2058-9565/aa7d3b} {\bibfield
  {journal} {\bibinfo  {journal} {Quantum Science and Technology}\ }\textbf
  {\bibinfo {volume} {2}},\ \bibinfo {pages} {035007} (\bibinfo {year}
  {2017})}\BibitemShut {NoStop}%
\bibitem [{\citenamefont {Sutcliffe}\ \emph {et~al.}(2025)\citenamefont
  {Sutcliffe}, \citenamefont {Jonnadula}, \citenamefont {Le~Gall},
  \citenamefont {Moylett},\ and\ \citenamefont
  {Westoby}}]{sutcliffe_2025_distributed}%
  \BibitemOpen
  \bibfield  {author} {\bibinfo {author} {\bibfnamefont {E.}~\bibnamefont
  {Sutcliffe}}, \bibinfo {author} {\bibfnamefont {B.}~\bibnamefont
  {Jonnadula}}, \bibinfo {author} {\bibfnamefont {C.}~\bibnamefont {Le~Gall}},
  \bibinfo {author} {\bibfnamefont {A.~E.}\ \bibnamefont {Moylett}},\ and\
  \bibinfo {author} {\bibfnamefont {C.~M.}\ \bibnamefont {Westoby}},\
  }\bibfield  {title} {\bibinfo {title} {Distributed quantum error correction
  based on hyperbolic {Floquet} codes},\ }in\ \href
  {https://doi.org/10.1109/QCE65121.2025.00076} {\emph {\bibinfo {booktitle}
  {2025 IEEE International Conference on Quantum Computing and Engineering
  (QCE)}}},\ Vol.~\bibinfo {volume} {01}\ (\bibinfo {year} {2025})\ pp.\
  \bibinfo {pages} {649--657}\BibitemShut {NoStop}%
\bibitem [{\citenamefont {Baireuther}\ \emph {et~al.}(2019)\citenamefont
  {Baireuther}, \citenamefont {Caio}, \citenamefont {Criger}, \citenamefont
  {Beenakker},\ and\ \citenamefont {O'Brien}}]{bell_flagged_circuit}%
  \BibitemOpen
  \bibfield  {author} {\bibinfo {author} {\bibfnamefont {P.}~\bibnamefont
  {Baireuther}}, \bibinfo {author} {\bibfnamefont {M.~D.}\ \bibnamefont
  {Caio}}, \bibinfo {author} {\bibfnamefont {B.}~\bibnamefont {Criger}},
  \bibinfo {author} {\bibfnamefont {C.~W.~J.}\ \bibnamefont {Beenakker}},\ and\
  \bibinfo {author} {\bibfnamefont {T.~E.}\ \bibnamefont {O'Brien}},\
  }\bibfield  {title} {\bibinfo {title} {Neural network decoder for topological
  color codes with circuit level noise},\ }\href
  {https://doi.org/10.1088/1367-2630/aaf29e} {\bibfield  {journal} {\bibinfo
  {journal} {New Journal of Physics}\ }\textbf {\bibinfo {volume} {21}},\
  \bibinfo {pages} {013003} (\bibinfo {year} {2019})}\BibitemShut {NoStop}%
\bibitem [{\citenamefont {Vittal}\ \emph {et~al.}(2024)\citenamefont {Vittal},
  \citenamefont {Javadi-Abhari}, \citenamefont {Cross}, \citenamefont
  {Bishop},\ and\ \citenamefont {Qureshi}}]{flag_proxy_hyperbolic_color}%
  \BibitemOpen
  \bibfield  {author} {\bibinfo {author} {\bibfnamefont {S.}~\bibnamefont
  {Vittal}}, \bibinfo {author} {\bibfnamefont {A.}~\bibnamefont
  {Javadi-Abhari}}, \bibinfo {author} {\bibfnamefont {A.~W.}\ \bibnamefont
  {Cross}}, \bibinfo {author} {\bibfnamefont {L.~S.}\ \bibnamefont {Bishop}},\
  and\ \bibinfo {author} {\bibfnamefont {M.}~\bibnamefont {Qureshi}},\
  }\bibfield  {title} {\bibinfo {title} {Flag-proxy networks: Overcoming the
  architectural, scheduling and decoding obstacles of quantum {LDPC} codes},\
  }in\ \href {https://doi.org/10.1109/MICRO61859.2024.00059} {\emph {\bibinfo
  {booktitle} {Proceedings of the 2024 57th IEEE/ACM International Symposium on
  Microarchitecture}}},\ \bibinfo {series and number} {MICRO '24}\ (\bibinfo
  {publisher} {IEEE Press},\ \bibinfo {year} {2024})\ p.\ \bibinfo {pages}
  {718–734}\BibitemShut {NoStop}%
\bibitem [{\citenamefont {Tremblay}\ \emph {et~al.}(2022)\citenamefont
  {Tremblay}, \citenamefont {Delfosse},\ and\ \citenamefont
  {Beverland}}]{tremblay2022constant}%
  \BibitemOpen
  \bibfield  {author} {\bibinfo {author} {\bibfnamefont {M.~A.}\ \bibnamefont
  {Tremblay}}, \bibinfo {author} {\bibfnamefont {N.}~\bibnamefont {Delfosse}},\
  and\ \bibinfo {author} {\bibfnamefont {M.~E.}\ \bibnamefont {Beverland}},\
  }\bibfield  {title} {\bibinfo {title} {Constant-overhead quantum error
  correction with thin planar connectivity},\ }\href
  {https://doi.org/10.1103/PhysRevLett.129.050504} {\bibfield  {journal}
  {\bibinfo  {journal} {Phys. Rev. Lett.}\ }\textbf {\bibinfo {volume} {129}},\
  \bibinfo {pages} {050504} (\bibinfo {year} {2022})}\BibitemShut {NoStop}%
\bibitem [{\citenamefont {Gidney}(2021)}]{Gidney2021stimfaststabilizer}%
  \BibitemOpen
  \bibfield  {author} {\bibinfo {author} {\bibfnamefont {C.}~\bibnamefont
  {Gidney}},\ }\bibfield  {title} {\bibinfo {title} {Stim: a fast stabilizer
  circuit simulator},\ }\href {https://doi.org/10.22331/q-2021-07-06-497}
  {\bibfield  {journal} {\bibinfo  {journal} {{Quantum}}\ }\textbf {\bibinfo
  {volume} {5}},\ \bibinfo {pages} {497} (\bibinfo {year} {2021})}\BibitemShut
  {NoStop}%
\bibitem [{\citenamefont {Dennis}\ \emph {et~al.}(2002)\citenamefont {Dennis},
  \citenamefont {Kitaev}, \citenamefont {Landahl},\ and\ \citenamefont
  {Preskill}}]{dennis2002topological}%
  \BibitemOpen
  \bibfield  {author} {\bibinfo {author} {\bibfnamefont {E.}~\bibnamefont
  {Dennis}}, \bibinfo {author} {\bibfnamefont {A.}~\bibnamefont {Kitaev}},
  \bibinfo {author} {\bibfnamefont {A.}~\bibnamefont {Landahl}},\ and\ \bibinfo
  {author} {\bibfnamefont {J.}~\bibnamefont {Preskill}},\ }\bibfield  {title}
  {\bibinfo {title} {Topological quantum memory},\ }\href
  {https://doi.org/10.1063/1.1499754} {\bibfield  {journal} {\bibinfo
  {journal} {Journal of Mathematical Physics}\ }\textbf {\bibinfo {volume}
  {43}},\ \bibinfo {pages} {4452} (\bibinfo {year} {2002})}\BibitemShut
  {NoStop}%
\bibitem [{\citenamefont {Tomita}\ and\ \citenamefont
  {Svore}(2014)}]{tomita2014lowdistance}%
  \BibitemOpen
  \bibfield  {author} {\bibinfo {author} {\bibfnamefont {Y.}~\bibnamefont
  {Tomita}}\ and\ \bibinfo {author} {\bibfnamefont {K.~M.}\ \bibnamefont
  {Svore}},\ }\bibfield  {title} {\bibinfo {title} {Low-distance surface codes
  under realistic quantum noise},\ }\href
  {https://doi.org/10.1103/PhysRevA.90.062320} {\bibfield  {journal} {\bibinfo
  {journal} {Phys. Rev. A}\ }\textbf {\bibinfo {volume} {90}},\ \bibinfo
  {pages} {062320} (\bibinfo {year} {2014})}\BibitemShut {NoStop}%
\bibitem [{\citenamefont {Gidney}(2025)}]{gidney2025alternating}%
  \BibitemOpen
  \bibfield  {author} {\bibinfo {author} {\bibfnamefont {C.}~\bibnamefont
  {Gidney}},\ }\href {https://quantumcomputing.stackexchange.com/a/41264}
  {\bibinfo {title} {Answer to ``preserving distance during stabilizer
  measurements by alternating interaction order from round to round?''}},\
  \bibinfo {howpublished} {Quantum Computing Stack Exchange} (\bibinfo {year}
  {2025}),\ \bibinfo {note} {answer 41264, posted 22 February 2025}\BibitemShut
  {NoStop}%
\bibitem [{\citenamefont {Berthusen}\ and\ \citenamefont
  {Gottesman}(2024)}]{Berthusen2024partialsyndrome}%
  \BibitemOpen
  \bibfield  {author} {\bibinfo {author} {\bibfnamefont {N.}~\bibnamefont
  {Berthusen}}\ and\ \bibinfo {author} {\bibfnamefont {D.}~\bibnamefont
  {Gottesman}},\ }\bibfield  {title} {\bibinfo {title} {Partial {S}yndrome
  {M}easurement for {H}ypergraph {P}roduct {C}odes},\ }\href
  {https://doi.org/10.22331/q-2024-05-14-1345} {\bibfield  {journal} {\bibinfo
  {journal} {{Quantum}}\ }\textbf {\bibinfo {volume} {8}},\ \bibinfo {pages}
  {1345} (\bibinfo {year} {2024})}\BibitemShut {NoStop}%
\bibitem [{\citenamefont {Gupta}\ \emph {et~al.}(2026)\citenamefont {Gupta},
  \citenamefont {Bhambay}, \citenamefont {Alavisamani}, \citenamefont
  {Walton},\ and\ \citenamefont {Vasantam}}]{gupta2026boundary}%
  \BibitemOpen
  \bibfield  {author} {\bibinfo {author} {\bibfnamefont {S.}~\bibnamefont
  {Gupta}}, \bibinfo {author} {\bibfnamefont {S.}~\bibnamefont {Bhambay}},
  \bibinfo {author} {\bibfnamefont {N.}~\bibnamefont {Alavisamani}}, \bibinfo
  {author} {\bibfnamefont {N.}~\bibnamefont {Walton}},\ and\ \bibinfo {author}
  {\bibfnamefont {T.}~\bibnamefont {Vasantam}},\ }\href
  {https://arxiv.org/abs/2604.22471} {\bibinfo {title} {Boundary-aware
  stabilizer scheduling for distributed quantum error correction}} (\bibinfo
  {year} {2026}),\ \Eprint {https://arxiv.org/abs/2604.22471}
  {arXiv:2604.22471} \BibitemShut {NoStop}%
\bibitem [{\citenamefont {Horsman}\ \emph {et~al.}(2012)\citenamefont
  {Horsman}, \citenamefont {Fowler}, \citenamefont {Devitt},\ and\
  \citenamefont {Meter}}]{horsman2012surface}%
  \BibitemOpen
  \bibfield  {author} {\bibinfo {author} {\bibfnamefont {D.}~\bibnamefont
  {Horsman}}, \bibinfo {author} {\bibfnamefont {A.~G.}\ \bibnamefont {Fowler}},
  \bibinfo {author} {\bibfnamefont {S.}~\bibnamefont {Devitt}},\ and\ \bibinfo
  {author} {\bibfnamefont {R.~V.}\ \bibnamefont {Meter}},\ }\bibfield  {title}
  {\bibinfo {title} {{Surface code quantum computing by lattice surgery}},\
  }\href {https://doi.org/10.1088/1367-2630/14/12/123011} {\bibfield  {journal}
  {\bibinfo  {journal} {New Journal of Physics}\ }\textbf {\bibinfo {volume}
  {14}},\ \bibinfo {pages} {123011} (\bibinfo {year} {2012})}\BibitemShut
  {NoStop}%
\bibitem [{\citenamefont {Cohen}\ \emph {et~al.}(2022)\citenamefont {Cohen},
  \citenamefont {Kim}, \citenamefont {Bartlett},\ and\ \citenamefont
  {Brown}}]{cohen2022low-overhead}%
  \BibitemOpen
  \bibfield  {author} {\bibinfo {author} {\bibfnamefont {L.~Z.}\ \bibnamefont
  {Cohen}}, \bibinfo {author} {\bibfnamefont {I.~H.}\ \bibnamefont {Kim}},
  \bibinfo {author} {\bibfnamefont {S.~D.}\ \bibnamefont {Bartlett}},\ and\
  \bibinfo {author} {\bibfnamefont {B.~J.}\ \bibnamefont {Brown}},\ }\bibfield
  {title} {\bibinfo {title} {{Low-overhead fault-tolerant quantum computing
  using long-range connectivity}},\ }\href
  {https://doi.org/10.1126/sciadv.abn1717} {\bibfield  {journal} {\bibinfo
  {journal} {Science Advances}\ }\textbf {\bibinfo {volume} {8}},\ \bibinfo
  {pages} {eabn1717} (\bibinfo {year} {2022})}\BibitemShut {NoStop}%
\bibitem [{\citenamefont {Cross}\ \emph {et~al.}(2024)\citenamefont {Cross},
  \citenamefont {He}, \citenamefont {Rall},\ and\ \citenamefont
  {Yoder}}]{cross2024improved}%
  \BibitemOpen
  \bibfield  {author} {\bibinfo {author} {\bibfnamefont {A.}~\bibnamefont
  {Cross}}, \bibinfo {author} {\bibfnamefont {Z.}~\bibnamefont {He}}, \bibinfo
  {author} {\bibfnamefont {P.}~\bibnamefont {Rall}},\ and\ \bibinfo {author}
  {\bibfnamefont {T.}~\bibnamefont {Yoder}},\ }\href
  {https://arxiv.org/abs/2407.18393} {\bibinfo {title} {{Improved {QLDPC}
  Surgery: Logical Measurements and Bridging Codes}}} (\bibinfo {year}
  {2024}),\ \Eprint {https://arxiv.org/abs/2407.18393} {arXiv:2407.18393
  [quant-ph]} \BibitemShut {NoStop}%
\bibitem [{\citenamefont {Williamson}\ and\ \citenamefont
  {Yoder}(2026)}]{Williamson2026}%
  \BibitemOpen
  \bibfield  {author} {\bibinfo {author} {\bibfnamefont {D.~J.}\ \bibnamefont
  {Williamson}}\ and\ \bibinfo {author} {\bibfnamefont {T.~J.}\ \bibnamefont
  {Yoder}},\ }\bibfield  {title} {\bibinfo {title} {Low-overhead fault-tolerant
  quantum computation by gauging logical operators},\ }\href
  {https://doi.org/10.1038/s41567-026-03220-8} {\bibfield  {journal} {\bibinfo
  {journal} {Nature Physics}\ }\textbf {\bibinfo {volume} {22}},\ \bibinfo
  {pages} {598–603} (\bibinfo {year} {2026})}\BibitemShut {NoStop}%
\bibitem [{\citenamefont {Ide}\ \emph {et~al.}(2025)\citenamefont {Ide},
  \citenamefont {Gowda}, \citenamefont {Nadkarni},\ and\ \citenamefont
  {Dauphinais}}]{ide2024fault}%
  \BibitemOpen
  \bibfield  {author} {\bibinfo {author} {\bibfnamefont {B.}~\bibnamefont
  {Ide}}, \bibinfo {author} {\bibfnamefont {M.~G.}\ \bibnamefont {Gowda}},
  \bibinfo {author} {\bibfnamefont {P.~J.}\ \bibnamefont {Nadkarni}},\ and\
  \bibinfo {author} {\bibfnamefont {G.}~\bibnamefont {Dauphinais}},\ }\bibfield
   {title} {\bibinfo {title} {{Fault-Tolerant Logical Measurements via
  Homological Measurement}},\ }\href
  {https://doi.org/10.1103/PhysRevX.15.021088} {\bibfield  {journal} {\bibinfo
  {journal} {Phys. Rev. X}\ }\textbf {\bibinfo {volume} {15}},\ \bibinfo
  {pages} {021088} (\bibinfo {year} {2025})}\BibitemShut {NoStop}%
\bibitem [{\citenamefont {Swaroop}\ \emph {et~al.}(2026)\citenamefont
  {Swaroop}, \citenamefont {Jochym-O'Connor},\ and\ \citenamefont
  {Yoder}}]{swaroop2026universal}%
  \BibitemOpen
  \bibfield  {author} {\bibinfo {author} {\bibfnamefont {E.}~\bibnamefont
  {Swaroop}}, \bibinfo {author} {\bibfnamefont {T.}~\bibnamefont
  {Jochym-O'Connor}},\ and\ \bibinfo {author} {\bibfnamefont {T.~J.}\
  \bibnamefont {Yoder}},\ }\bibfield  {title} {\bibinfo {title} {Universal
  adapters between quantum low-density parity check codes},\ }\href@noop {}
  {\bibfield  {journal} {\bibinfo  {journal} {PRX Quantum}\ }\textbf {\bibinfo
  {volume} {7}},\ \bibinfo {pages} {010324} (\bibinfo {year}
  {2026})}\BibitemShut {NoStop}%
\bibitem [{\citenamefont {He}\ \emph {et~al.}(2025)\citenamefont {He},
  \citenamefont {Cowtan}, \citenamefont {Williamson},\ and\ \citenamefont
  {Yoder}}]{he2025extractorsqldpcarchitecturesefficient}%
  \BibitemOpen
  \bibfield  {author} {\bibinfo {author} {\bibfnamefont {Z.}~\bibnamefont
  {He}}, \bibinfo {author} {\bibfnamefont {A.}~\bibnamefont {Cowtan}}, \bibinfo
  {author} {\bibfnamefont {D.~J.}\ \bibnamefont {Williamson}},\ and\ \bibinfo
  {author} {\bibfnamefont {T.~J.}\ \bibnamefont {Yoder}},\ }\href
  {https://arxiv.org/abs/2503.10390} {\bibinfo {title} {Extractors: {QLDPC}
  architectures for efficient {Pauli}-based computation}} (\bibinfo {year}
  {2025}),\ \Eprint {https://arxiv.org/abs/2503.10390} {arXiv:2503.10390
  [quant-ph]} \BibitemShut {NoStop}%
\bibitem [{\citenamefont {Yuan}\ \emph {et~al.}(2026)\citenamefont {Yuan},
  \citenamefont {Cowtan}, \citenamefont {He}, \citenamefont {Lin},\ and\
  \citenamefont {Williamson}}]{yuan2026parsimonious}%
  \BibitemOpen
  \bibfield  {author} {\bibinfo {author} {\bibfnamefont {A.~C.}\ \bibnamefont
  {Yuan}}, \bibinfo {author} {\bibfnamefont {A.}~\bibnamefont {Cowtan}},
  \bibinfo {author} {\bibfnamefont {Z.}~\bibnamefont {He}}, \bibinfo {author}
  {\bibfnamefont {T.-C.}\ \bibnamefont {Lin}},\ and\ \bibinfo {author}
  {\bibfnamefont {D.~J.}\ \bibnamefont {Williamson}},\ }\href
  {https://arxiv.org/abs/2603.05082} {\bibinfo {title} {Parsimonious quantum
  low-density parity-check code surgery}} (\bibinfo {year} {2026}),\ \Eprint
  {https://arxiv.org/abs/2603.05082} {arXiv:2603.05082 [quant-ph]} \BibitemShut
  {NoStop}%
\bibitem [{Note1()}]{Note1}%
  \BibitemOpen
  \bibinfo {note} {For readers familiar with modern code surgery methods, the
  construction we described is very similar to \protect \textit {thickening},
  which is a common technique used in various prior works on surgery~\cite
  {he2025extractorsqldpcarchitecturesefficient,cohen2022low-overhead,Williamson2026}.
  A technical difference is that some surgery checks in the first layer are
  wide (implemented with two ancilla check qubits), which act on two ancilla
  data qubits in the second layer. This design choice helps ensure the deformed
  code has sufficient distance in the surgery protocol.}\BibitemShut {Stop}%
\bibitem [{Note2()}]{Note2}%
  \BibitemOpen
  \bibinfo {note} {In the modern graph surgery language, this $Z$~operator on
  the extractor is a `cycle check'. Normally we need to measure these cycle
  checks, but in our case of the ribbon extractor this cycle check is a product
  of deformed stabilizers.~\cite
  {cross2024improved,Williamson2026,he2025extractorsqldpcarchitecturesefficient}}\BibitemShut
  {NoStop}%
\bibitem [{\citenamefont {Bravyi}\ \emph {et~al.}(2024)\citenamefont {Bravyi},
  \citenamefont {Cross}, \citenamefont {Gambetta}, \citenamefont {Maslov},
  \citenamefont {Rall},\ and\ \citenamefont {Yoder}}]{bivariate_bicycle_codes}%
  \BibitemOpen
  \bibfield  {author} {\bibinfo {author} {\bibfnamefont {S.}~\bibnamefont
  {Bravyi}}, \bibinfo {author} {\bibfnamefont {A.~W.}\ \bibnamefont {Cross}},
  \bibinfo {author} {\bibfnamefont {J.~M.}\ \bibnamefont {Gambetta}}, \bibinfo
  {author} {\bibfnamefont {D.}~\bibnamefont {Maslov}}, \bibinfo {author}
  {\bibfnamefont {P.}~\bibnamefont {Rall}},\ and\ \bibinfo {author}
  {\bibfnamefont {T.~J.}\ \bibnamefont {Yoder}},\ }\bibfield  {title} {\bibinfo
  {title} {High-threshold and low-overhead fault-tolerant quantum memory},\
  }\href {https://doi.org/10.1038/s41586-024-07107-7} {\bibfield  {journal}
  {\bibinfo  {journal} {Nature}\ }\textbf {\bibinfo {volume} {627}},\ \bibinfo
  {pages} {778} (\bibinfo {year} {2024})}\BibitemShut {NoStop}%
\bibitem [{\citenamefont {McEwen}\ \emph {et~al.}(2023)\citenamefont {McEwen},
  \citenamefont {Bacon},\ and\ \citenamefont
  {Gidney}}]{McEwen2023relaxinghardware}%
  \BibitemOpen
  \bibfield  {author} {\bibinfo {author} {\bibfnamefont {M.}~\bibnamefont
  {McEwen}}, \bibinfo {author} {\bibfnamefont {D.}~\bibnamefont {Bacon}},\ and\
  \bibinfo {author} {\bibfnamefont {C.}~\bibnamefont {Gidney}},\ }\bibfield
  {title} {\bibinfo {title} {Relaxing {H}ardware {R}equirements for {S}urface
  {C}ode {C}ircuits using {T}ime-dynamics},\ }\href
  {https://doi.org/10.22331/q-2023-11-07-1172} {\bibfield  {journal} {\bibinfo
  {journal} {{Quantum}}\ }\textbf {\bibinfo {volume} {7}},\ \bibinfo {pages}
  {1172} (\bibinfo {year} {2023})}\BibitemShut {NoStop}%
\bibitem [{\citenamefont {Gidney}\ and\ \citenamefont
  {Jones}(2023)}]{gidney2023new}%
  \BibitemOpen
  \bibfield  {author} {\bibinfo {author} {\bibfnamefont {C.}~\bibnamefont
  {Gidney}}\ and\ \bibinfo {author} {\bibfnamefont {C.}~\bibnamefont {Jones}},\
  }\href {https://arxiv.org/abs/2312.08813} {\bibinfo {title} {New circuits and
  an open source decoder for the color code}} (\bibinfo {year} {2023}),\
  \Eprint {https://arxiv.org/abs/2312.08813} {arXiv:2312.08813} \BibitemShut
  {NoStop}%
\bibitem [{\citenamefont {Yoshida}\ \emph {et~al.}(2025)\citenamefont
  {Yoshida}, \citenamefont {Gidney}, \citenamefont {McEwen},\ and\
  \citenamefont {Zalcman}}]{yoshida2025low}%
  \BibitemOpen
  \bibfield  {author} {\bibinfo {author} {\bibfnamefont {S.}~\bibnamefont
  {Yoshida}}, \bibinfo {author} {\bibfnamefont {C.}~\bibnamefont {Gidney}},
  \bibinfo {author} {\bibfnamefont {M.}~\bibnamefont {McEwen}},\ and\ \bibinfo
  {author} {\bibfnamefont {A.}~\bibnamefont {Zalcman}},\ }\href
  {https://arxiv.org/abs/2510.00370} {\bibinfo {title} {Low depth color code
  circuits with {CXSWAP} gate}} (\bibinfo {year} {2025}),\ \Eprint
  {https://arxiv.org/abs/2510.00370} {arXiv:2510.00370} \BibitemShut {NoStop}%
\bibitem [{\citenamefont {Shaw}\ and\ \citenamefont
  {Terhal}(2025)}]{shaw2025lowering}%
  \BibitemOpen
  \bibfield  {author} {\bibinfo {author} {\bibfnamefont {M.~H.}\ \bibnamefont
  {Shaw}}\ and\ \bibinfo {author} {\bibfnamefont {B.~M.}\ \bibnamefont
  {Terhal}},\ }\bibfield  {title} {\bibinfo {title} {Lowering connectivity
  requirements for bivariate bicycle codes using morphing circuits},\ }\href
  {https://doi.org/10.1103/PhysRevLett.134.090602} {\bibfield  {journal}
  {\bibinfo  {journal} {Phys. Rev. Lett.}\ }\textbf {\bibinfo {volume} {134}},\
  \bibinfo {pages} {090602} (\bibinfo {year} {2025})}\BibitemShut {NoStop}%
\bibitem [{\citenamefont {Cuccaro}\ \emph {et~al.}(2004)\citenamefont
  {Cuccaro}, \citenamefont {Draper}, \citenamefont {Kutin},\ and\ \citenamefont
  {Moulton}}]{cuccaro2004new}%
  \BibitemOpen
  \bibfield  {author} {\bibinfo {author} {\bibfnamefont {S.~A.}\ \bibnamefont
  {Cuccaro}}, \bibinfo {author} {\bibfnamefont {T.~G.}\ \bibnamefont {Draper}},
  \bibinfo {author} {\bibfnamefont {S.~A.}\ \bibnamefont {Kutin}},\ and\
  \bibinfo {author} {\bibfnamefont {D.~P.}\ \bibnamefont {Moulton}},\ }\href
  {https://arxiv.org/abs/quant-ph/0410184} {\bibinfo {title} {A new quantum
  ripple-carry addition circuit}} (\bibinfo {year} {2004}),\ \Eprint
  {https://arxiv.org/abs/quant-ph/0410184} {arXiv:quant-ph/0410184}
  \BibitemShut {NoStop}%
\bibitem [{\citenamefont {Belady}(1966)}]{belady1966replacement}%
  \BibitemOpen
  \bibfield  {author} {\bibinfo {author} {\bibfnamefont {L.~A.}\ \bibnamefont
  {Belady}},\ }\bibfield  {title} {\bibinfo {title} {A study of replacement
  algorithms for a virtual-storage computer},\ }\href
  {https://doi.org/10.1147/sj.52.0078} {\bibfield  {journal} {\bibinfo
  {journal} {IBM Systems Journal}\ }\textbf {\bibinfo {volume} {5}},\ \bibinfo
  {pages} {78} (\bibinfo {year} {1966})}\BibitemShut {NoStop}%
\bibitem [{\citenamefont {Gidney}\ and\ \citenamefont
  {Fowler}(2019)}]{gidney2019flexible}%
  \BibitemOpen
  \bibfield  {author} {\bibinfo {author} {\bibfnamefont {C.}~\bibnamefont
  {Gidney}}\ and\ \bibinfo {author} {\bibfnamefont {A.~G.}\ \bibnamefont
  {Fowler}},\ }\href {https://arxiv.org/abs/1905.08916} {\bibinfo {title}
  {Flexible layout of surface code computations using {AutoCCZ} states}}
  (\bibinfo {year} {2019}),\ \Eprint {https://arxiv.org/abs/1905.08916}
  {arXiv:1905.08916} \BibitemShut {NoStop}%
\bibitem [{\citenamefont {Eppstein}(2003)}]{tree_cotree_decomp}%
  \BibitemOpen
  \bibfield  {author} {\bibinfo {author} {\bibfnamefont {D.}~\bibnamefont
  {Eppstein}},\ }\bibfield  {title} {\bibinfo {title} {Dynamic generators of
  topologically embedded graphs},\ }in\ \href@noop {} {\emph {\bibinfo
  {booktitle} {Proceedings of the Fourteenth Annual ACM-SIAM Symposium on
  Discrete Algorithms}}},\ \bibinfo {series and number} {SODA '03}\ (\bibinfo
  {publisher} {Society for Industrial and Applied Mathematics},\ \bibinfo
  {address} {USA},\ \bibinfo {year} {2003})\ p.\ \bibinfo {pages}
  {599–608}\BibitemShut {NoStop}%
\bibitem [{\citenamefont {Erickson}\ and\ \citenamefont
  {Whittlesey}(2005)}]{Erickson_Whittlesey}%
  \BibitemOpen
  \bibfield  {author} {\bibinfo {author} {\bibfnamefont {J.}~\bibnamefont
  {Erickson}}\ and\ \bibinfo {author} {\bibfnamefont {K.}~\bibnamefont
  {Whittlesey}},\ }\bibfield  {title} {\bibinfo {title} {Greedy optimal
  homotopy and homology generators},\ }in\ \href@noop {} {\emph {\bibinfo
  {booktitle} {Proceedings of the Sixteenth Annual ACM-SIAM Symposium on
  Discrete Algorithms}}},\ \bibinfo {series and number} {SODA '05}\ (\bibinfo
  {publisher} {Society for Industrial and Applied Mathematics},\ \bibinfo
  {address} {USA},\ \bibinfo {year} {2005})\ p.\ \bibinfo {pages}
  {1038–1046}\BibitemShut {NoStop}%
\bibitem [{\citenamefont {Erickson}\ and\ \citenamefont
  {Har-Peled}(2004)}]{Erickson_HarPeled}%
  \BibitemOpen
  \bibfield  {author} {\bibinfo {author} {\bibfnamefont {J.}~\bibnamefont
  {Erickson}}\ and\ \bibinfo {author} {\bibfnamefont {S.}~\bibnamefont
  {Har-Peled}},\ }\bibfield  {title} {\bibinfo {title} {Optimally cutting a
  surface into a disk},\ }\href {https://doi.org/10.1007/s00454-003-2948-z}
  {\bibfield  {journal} {\bibinfo  {journal} {Discrete \& Computational
  Geometry}\ }\textbf {\bibinfo {volume} {31}},\ \bibinfo {pages} {37}
  (\bibinfo {year} {2004})}\BibitemShut {NoStop}%
\bibitem [{\citenamefont {Chambers}\ \emph {et~al.}(2008)\citenamefont
  {Chambers}, \citenamefont {{Colin de Verdi{\`e}re}}, \citenamefont
  {Erickson}, \citenamefont {Lazarus},\ and\ \citenamefont
  {Whittlesey}}]{CHAMBERS2008}%
  \BibitemOpen
  \bibfield  {author} {\bibinfo {author} {\bibfnamefont {E.~W.}\ \bibnamefont
  {Chambers}}, \bibinfo {author} {\bibfnamefont {{\'E}.}~\bibnamefont {{Colin
  de Verdi{\`e}re}}}, \bibinfo {author} {\bibfnamefont {J.}~\bibnamefont
  {Erickson}}, \bibinfo {author} {\bibfnamefont {F.}~\bibnamefont {Lazarus}},\
  and\ \bibinfo {author} {\bibfnamefont {K.}~\bibnamefont {Whittlesey}},\
  }\bibfield  {title} {\bibinfo {title} {Splitting (complicated) surfaces is
  hard},\ }\href {https://doi.org/10.1016/j.comgeo.2007.10.010} {\bibfield
  {journal} {\bibinfo  {journal} {Computational Geometry}\ }\textbf {\bibinfo
  {volume} {41}},\ \bibinfo {pages} {94} (\bibinfo {year} {2008})},\ \bibinfo
  {note} {special Issue on the 22nd European Workshop on Computational Geometry
  (EuroCG)}\BibitemShut {NoStop}%
\bibitem [{\citenamefont {Cameron}(1989)}]{induced_matchings}%
  \BibitemOpen
  \bibfield  {author} {\bibinfo {author} {\bibfnamefont {K.}~\bibnamefont
  {Cameron}},\ }\bibfield  {title} {\bibinfo {title} {Induced matchings},\
  }\href {https://doi.org/10.1016/0166-218X(92)90275-F} {\bibfield  {journal}
  {\bibinfo  {journal} {Discrete Applied Mathematics}\ }\textbf {\bibinfo
  {volume} {24}},\ \bibinfo {pages} {97} (\bibinfo {year} {1989})}\BibitemShut
  {NoStop}%
\bibitem [{\citenamefont {Eppstein}\ \emph {et~al.}(2010)\citenamefont
  {Eppstein}, \citenamefont {L{\"o}ffler},\ and\ \citenamefont
  {Strash}}]{maximal_clique}%
  \BibitemOpen
  \bibfield  {author} {\bibinfo {author} {\bibfnamefont {D.}~\bibnamefont
  {Eppstein}}, \bibinfo {author} {\bibfnamefont {M.}~\bibnamefont
  {L{\"o}ffler}},\ and\ \bibinfo {author} {\bibfnamefont {D.}~\bibnamefont
  {Strash}},\ }\bibfield  {title} {\bibinfo {title} {Listing all maximal
  cliques in sparse graphs in near-optimal time},\ }in\ \href@noop {} {\emph
  {\bibinfo {booktitle} {Algorithms and Computation}}},\ \bibinfo {editor}
  {edited by\ \bibinfo {editor} {\bibfnamefont {O.}~\bibnamefont {Cheong}},
  \bibinfo {editor} {\bibfnamefont {K.-Y.}\ \bibnamefont {Chwa}},\ and\
  \bibinfo {editor} {\bibfnamefont {K.}~\bibnamefont {Park}}}\ (\bibinfo
  {publisher} {Springer Berlin Heidelberg},\ \bibinfo {address} {Berlin,
  Heidelberg},\ \bibinfo {year} {2010})\ pp.\ \bibinfo {pages}
  {403--414}\BibitemShut {NoStop}%
\bibitem [{\citenamefont {Webster}\ \emph {et~al.}(2026)\citenamefont
  {Webster}, \citenamefont {Jacob},\ and\ \citenamefont
  {Higgott}}]{webster2026distance}%
  \BibitemOpen
  \bibfield  {author} {\bibinfo {author} {\bibfnamefont {M.}~\bibnamefont
  {Webster}}, \bibinfo {author} {\bibfnamefont {A.}~\bibnamefont {Jacob}},\
  and\ \bibinfo {author} {\bibfnamefont {O.}~\bibnamefont {Higgott}},\ }\href
  {https://arxiv.org/abs/2603.22532} {\bibinfo {title} {Distance-finding
  algorithms for quantum codes and circuits}} (\bibinfo {year} {2026}),\
  \Eprint {https://arxiv.org/abs/2603.22532} {arXiv:2603.22532} \BibitemShut
  {NoStop}%
\bibitem [{\citenamefont {Delfosse}(2014)}]{delfosse2014decoding}%
  \BibitemOpen
  \bibfield  {author} {\bibinfo {author} {\bibfnamefont {N.}~\bibnamefont
  {Delfosse}},\ }\bibfield  {title} {\bibinfo {title} {Decoding color codes by
  projection onto surface codes},\ }\href
  {https://doi.org/10.1103/PhysRevA.89.012317} {\bibfield  {journal} {\bibinfo
  {journal} {Phys. Rev. A}\ }\textbf {\bibinfo {volume} {89}},\ \bibinfo
  {pages} {012317} (\bibinfo {year} {2014})}\BibitemShut {NoStop}%
\bibitem [{\citenamefont {Kubica}\ and\ \citenamefont
  {Delfosse}(2023)}]{Kubica2023efficientcolorcode}%
  \BibitemOpen
  \bibfield  {author} {\bibinfo {author} {\bibfnamefont {A.}~\bibnamefont
  {Kubica}}\ and\ \bibinfo {author} {\bibfnamefont {N.}~\bibnamefont
  {Delfosse}},\ }\bibfield  {title} {\bibinfo {title} {Efficient color code
  decoders in {$d\geq 2$} dimensions from toric code decoders},\ }\href
  {https://doi.org/10.22331/q-2023-02-21-929} {\bibfield  {journal} {\bibinfo
  {journal} {{Quantum}}\ }\textbf {\bibinfo {volume} {7}},\ \bibinfo {pages}
  {929} (\bibinfo {year} {2023})}\BibitemShut {NoStop}%
\bibitem [{\citenamefont {Dumer}\ \emph {et~al.}(2017)\citenamefont {Dumer},
  \citenamefont {Kovalev},\ and\ \citenamefont {Pryadko}}]{dumer2017distance}%
  \BibitemOpen
  \bibfield  {author} {\bibinfo {author} {\bibfnamefont {I.}~\bibnamefont
  {Dumer}}, \bibinfo {author} {\bibfnamefont {A.~A.}\ \bibnamefont {Kovalev}},\
  and\ \bibinfo {author} {\bibfnamefont {L.~P.}\ \bibnamefont {Pryadko}},\
  }\bibfield  {title} {\bibinfo {title} {Distance verification for classical
  and quantum {LDPC} codes},\ }\href {https://doi.org/10.1109/TIT.2017.2690381}
  {\bibfield  {journal} {\bibinfo  {journal} {IEEE Transactions on Information
  Theory}\ }\textbf {\bibinfo {volume} {63}},\ \bibinfo {pages} {4675}
  (\bibinfo {year} {2017})}\BibitemShut {NoStop}%
\bibitem [{\citenamefont {Pryadko}(2026)}]{dist_m4ri}%
  \BibitemOpen
  \bibfield  {author} {\bibinfo {author} {\bibfnamefont {L.~P.}\ \bibnamefont
  {Pryadko}},\ }\href {https://github.com/QEC-pages/dist-m4ri} {\bibinfo
  {title} {dist-m4ri: Distance of a classical or quantum {CSS} code}},\
  \bibinfo {howpublished} {GitHub repository} (\bibinfo {year}
  {2026})\BibitemShut {NoStop}%
\end{thebibliography}%


\providecommand{\noopsort}[1]{}\providecommand{\singleletter}[1]{#1}%
\begin{thebibliography}{83}%
\makeatletter
\providecommand \@ifxundefined [1]{%
 \@ifx{#1\undefined}
}%
\providecommand \@ifnum [1]{%
 \ifnum #1\expandafter \@firstoftwo
 \else \expandafter \@secondoftwo
 \fi
}%
\providecommand \@ifx [1]{%
 \ifx #1\expandafter \@firstoftwo
 \else \expandafter \@secondoftwo
 \fi
}%
\providecommand \natexlab [1]{#1}%
\providecommand \enquote  [1]{``#1''}%
\providecommand \bibnamefont  [1]{#1}%
\providecommand \bibfnamefont [1]{#1}%
\providecommand \citenamefont [1]{#1}%
\providecommand \href@noop [0]{\@secondoftwo}%
\providecommand \href [0]{\begingroup \@sanitize@url \@href}%
\providecommand \@href[1]{\@@startlink{#1}\@@href}%
\providecommand \@@href[1]{\endgroup#1\@@endlink}%
\providecommand \@sanitize@url [0]{\catcode `\\12\catcode `\$12\catcode
  `\&12\catcode `\#12\catcode `\^12\catcode `\_12\catcode `\%12\relax}%
\providecommand \@@startlink[1]{}%
\providecommand \@@endlink[0]{}%
\providecommand \url  [0]{\begingroup\@sanitize@url \@url }%
\providecommand \@url [1]{\endgroup\@href {#1}{\urlprefix }}%
\providecommand \urlprefix  [0]{URL }%
\providecommand \Eprint [0]{\href }%
\providecommand \doibase [0]{https://doi.org/}%
\providecommand \selectlanguage [0]{\@gobble}%
\providecommand \bibinfo  [0]{\@secondoftwo}%
\providecommand \bibfield  [0]{\@secondoftwo}%
\providecommand \translation [1]{[#1]}%
\providecommand \BibitemOpen [0]{}%
\providecommand \bibitemStop [0]{}%
\providecommand \bibitemNoStop [0]{.\EOS\space}%
\providecommand \EOS [0]{\spacefactor3000\relax}%
\providecommand \BibitemShut  [1]{\csname bibitem#1\endcsname}%
\let\auto@bib@innerbib\@empty
\bibitem [{\citenamefont {Deutsch}\ and\ \citenamefont
  {Jozsa}(1992)}]{deutsch_jozsa}%
  \BibitemOpen
  \bibfield  {author} {\bibinfo {author} {\bibfnamefont {D.}~\bibnamefont
  {Deutsch}}\ and\ \bibinfo {author} {\bibfnamefont {R.}~\bibnamefont
  {Jozsa}},\ }\bibfield  {title} {\bibinfo {title} {Rapid solution of problems
  by quantum computation},\ }\href {https://doi.org/10.1098/rspa.1992.0167}
  {\bibfield  {journal} {\bibinfo  {journal} {Proceedings of the Royal Society
  of London. Series A: Mathematical and Physical Sciences}\ }\textbf {\bibinfo
  {volume} {439}},\ \bibinfo {pages} {553} (\bibinfo {year}
  {1992})}\BibitemShut {NoStop}%
\bibitem [{\citenamefont {Shor}(1997)}]{shor}%
  \BibitemOpen
  \bibfield  {author} {\bibinfo {author} {\bibfnamefont {P.~W.}\ \bibnamefont
  {Shor}},\ }\bibfield  {title} {\bibinfo {title} {Polynomial-time algorithms
  for prime factorization and discrete logarithms on a quantum computer},\
  }\href {https://doi.org/10.1137/S0097539795293172} {\bibfield  {journal}
  {\bibinfo  {journal} {SIAM J. Comput.}\ }\textbf {\bibinfo {volume} {26}},\
  \bibinfo {pages} {1484–1509} (\bibinfo {year} {1997})}\BibitemShut
  {NoStop}%
\bibitem [{\citenamefont {Lloyd}(1996)}]{lloyd}%
  \BibitemOpen
  \bibfield  {author} {\bibinfo {author} {\bibfnamefont {S.}~\bibnamefont
  {Lloyd}},\ }\bibfield  {title} {\bibinfo {title} {Universal quantum
  simulators},\ }\href {https://doi.org/10.1126/science.273.5278.1073}
  {\bibfield  {journal} {\bibinfo  {journal} {Science}\ }\textbf {\bibinfo
  {volume} {273}},\ \bibinfo {pages} {1073} (\bibinfo {year}
  {1996})}\BibitemShut {NoStop}%
\bibitem [{\citenamefont {Cao}\ \emph {et~al.}(2019)\citenamefont {Cao},
  \citenamefont {Romero}, \citenamefont {Olson}, \citenamefont {Degroote},
  \citenamefont {Johnson}, \citenamefont {Kieferov{\'a}}, \citenamefont
  {Kivlichan}, \citenamefont {Menke}, \citenamefont {Peropadre}, \citenamefont
  {Sawaya}, \citenamefont {Sim}, \citenamefont {Veis},\ and\ \citenamefont
  {Aspuru-Guzik}}]{quantum_chemistry_review}%
  \BibitemOpen
  \bibfield  {author} {\bibinfo {author} {\bibfnamefont {Y.}~\bibnamefont
  {Cao}}, \bibinfo {author} {\bibfnamefont {J.}~\bibnamefont {Romero}},
  \bibinfo {author} {\bibfnamefont {J.~P.}\ \bibnamefont {Olson}}, \bibinfo
  {author} {\bibfnamefont {M.}~\bibnamefont {Degroote}}, \bibinfo {author}
  {\bibfnamefont {P.~D.}\ \bibnamefont {Johnson}}, \bibinfo {author}
  {\bibfnamefont {M.}~\bibnamefont {Kieferov{\'a}}}, \bibinfo {author}
  {\bibfnamefont {I.~D.}\ \bibnamefont {Kivlichan}}, \bibinfo {author}
  {\bibfnamefont {T.}~\bibnamefont {Menke}}, \bibinfo {author} {\bibfnamefont
  {B.}~\bibnamefont {Peropadre}}, \bibinfo {author} {\bibfnamefont {N.~P.~D.}\
  \bibnamefont {Sawaya}}, \bibinfo {author} {\bibfnamefont {S.}~\bibnamefont
  {Sim}}, \bibinfo {author} {\bibfnamefont {L.}~\bibnamefont {Veis}},\ and\
  \bibinfo {author} {\bibfnamefont {A.}~\bibnamefont {Aspuru-Guzik}},\
  }\bibfield  {title} {\bibinfo {title} {Quantum chemistry in the age of
  quantum computing},\ }\href {https://doi.org/10.1021/acs.chemrev.8b00803}
  {\bibfield  {journal} {\bibinfo  {journal} {Chemical Reviews}\ }\textbf
  {\bibinfo {volume} {119}},\ \bibinfo {pages} {10856} (\bibinfo {year}
  {2019})}\BibitemShut {NoStop}%
\bibitem [{\citenamefont {{Google Quantum AI and
  Collaborators}}(2025)}]{surface_code_below_threshold}%
  \BibitemOpen
  \bibfield  {author} {\bibinfo {author} {\bibnamefont {{Google Quantum AI and
  Collaborators}}},\ }\bibfield  {title} {\bibinfo {title} {Quantum error
  correction below the surface code threshold},\ }\href
  {https://doi.org/10.1038/s41586-024-08449-y} {\bibfield  {journal} {\bibinfo
  {journal} {Nature}\ }\textbf {\bibinfo {volume} {638}},\ \bibinfo {pages}
  {920} (\bibinfo {year} {2025})}\BibitemShut {NoStop}%
\bibitem [{\citenamefont {Gidney}(2025)}]{gidney2025factor}%
  \BibitemOpen
  \bibfield  {author} {\bibinfo {author} {\bibfnamefont {C.}~\bibnamefont
  {Gidney}},\ }\href {https://arxiv.org/abs/2505.15917} {\bibinfo {title} {How
  to factor 2048 bit {RSA} integers with less than a million noisy qubits}}
  (\bibinfo {year} {2025}),\ \Eprint {https://arxiv.org/abs/2505.15917}
  {arXiv:2505.15917} \BibitemShut {NoStop}%
\bibitem [{\citenamefont {Low}\ \emph {et~al.}(2026)\citenamefont {Low},
  \citenamefont {Huggins}, \citenamefont {Berry}, \citenamefont {Khattar},
  \citenamefont {White}, \citenamefont {Rubin},\ and\ \citenamefont
  {Babbush}}]{low2026denserplanarsurfacecode}%
  \BibitemOpen
  \bibfield  {author} {\bibinfo {author} {\bibfnamefont {G.~H.}\ \bibnamefont
  {Low}}, \bibinfo {author} {\bibfnamefont {W.~J.}\ \bibnamefont {Huggins}},
  \bibinfo {author} {\bibfnamefont {D.~W.}\ \bibnamefont {Berry}}, \bibinfo
  {author} {\bibfnamefont {T.}~\bibnamefont {Khattar}}, \bibinfo {author}
  {\bibfnamefont {A.~F.}\ \bibnamefont {White}}, \bibinfo {author}
  {\bibfnamefont {N.~C.}\ \bibnamefont {Rubin}},\ and\ \bibinfo {author}
  {\bibfnamefont {R.}~\bibnamefont {Babbush}},\ }\href
  {https://arxiv.org/abs/2605.30455} {\bibinfo {title} {A denser planar surface
  code}} (\bibinfo {year} {2026}),\ \Eprint {https://arxiv.org/abs/2605.30455}
  {arXiv:2605.30455 [quant-ph]} \BibitemShut {NoStop}%
\bibitem [{\citenamefont {Nickerson}\ \emph {et~al.}(2013)\citenamefont
  {Nickerson}, \citenamefont {Li},\ and\ \citenamefont
  {Benjamin}}]{nickerson2013topological}%
  \BibitemOpen
  \bibfield  {author} {\bibinfo {author} {\bibfnamefont {N.~H.}\ \bibnamefont
  {Nickerson}}, \bibinfo {author} {\bibfnamefont {Y.}~\bibnamefont {Li}},\ and\
  \bibinfo {author} {\bibfnamefont {S.~C.}\ \bibnamefont {Benjamin}},\
  }\bibfield  {title} {\bibinfo {title} {Topological quantum computing with a
  very noisy network and local error rates approaching one percent},\ }\href
  {https://doi.org/10.1038/ncomms2773} {\bibfield  {journal} {\bibinfo
  {journal} {Nature Communications}\ }\textbf {\bibinfo {volume} {4}},\
  \bibinfo {pages} {1756} (\bibinfo {year} {2013})}\BibitemShut {NoStop}%
\bibitem [{\citenamefont {Ramette}\ \emph {et~al.}(2024)\citenamefont
  {Ramette}, \citenamefont {Sinclair}, \citenamefont {Breuckmann},\ and\
  \citenamefont {Vuleti{\'c}}}]{ramette2024fault}%
  \BibitemOpen
  \bibfield  {author} {\bibinfo {author} {\bibfnamefont {J.}~\bibnamefont
  {Ramette}}, \bibinfo {author} {\bibfnamefont {J.}~\bibnamefont {Sinclair}},
  \bibinfo {author} {\bibfnamefont {N.~P.}\ \bibnamefont {Breuckmann}},\ and\
  \bibinfo {author} {\bibfnamefont {V.}~\bibnamefont {Vuleti{\'c}}},\
  }\bibfield  {title} {\bibinfo {title} {Fault-tolerant connection of
  error-corrected qubits with noisy links},\ }\href
  {https://doi.org/10.1038/s41534-024-00855-4} {\bibfield  {journal} {\bibinfo
  {journal} {npj Quantum Information}\ }\textbf {\bibinfo {volume} {10}},\
  \bibinfo {pages} {58} (\bibinfo {year} {2024})}\BibitemShut {NoStop}%
\bibitem [{\citenamefont {Strikis}\ and\ \citenamefont
  {Berent}(2023)}]{strikis_2023}%
  \BibitemOpen
  \bibfield  {author} {\bibinfo {author} {\bibfnamefont {A.}~\bibnamefont
  {Strikis}}\ and\ \bibinfo {author} {\bibfnamefont {L.}~\bibnamefont
  {Berent}},\ }\bibfield  {title} {\bibinfo {title} {Quantum low-density
  parity-check codes for modular architectures},\ }\href
  {https://doi.org/10.1103/PRXQuantum.4.020321} {\bibfield  {journal} {\bibinfo
   {journal} {PRX Quantum}\ }\textbf {\bibinfo {volume} {4}},\ \bibinfo {pages}
  {020321} (\bibinfo {year} {2023})}\BibitemShut {NoStop}%
\bibitem [{\citenamefont {Bravyi}\ \emph {et~al.}(2024)\citenamefont {Bravyi},
  \citenamefont {Cross}, \citenamefont {Gambetta}, \citenamefont {Maslov},
  \citenamefont {Rall},\ and\ \citenamefont {Yoder}}]{bivariate_bicycle_codes}%
  \BibitemOpen
  \bibfield  {author} {\bibinfo {author} {\bibfnamefont {S.}~\bibnamefont
  {Bravyi}}, \bibinfo {author} {\bibfnamefont {A.~W.}\ \bibnamefont {Cross}},
  \bibinfo {author} {\bibfnamefont {J.~M.}\ \bibnamefont {Gambetta}}, \bibinfo
  {author} {\bibfnamefont {D.}~\bibnamefont {Maslov}}, \bibinfo {author}
  {\bibfnamefont {P.}~\bibnamefont {Rall}},\ and\ \bibinfo {author}
  {\bibfnamefont {T.~J.}\ \bibnamefont {Yoder}},\ }\bibfield  {title} {\bibinfo
  {title} {High-threshold and low-overhead fault-tolerant quantum memory},\
  }\href {https://doi.org/10.1038/s41586-024-07107-7} {\bibfield  {journal}
  {\bibinfo  {journal} {Nature}\ }\textbf {\bibinfo {volume} {627}},\ \bibinfo
  {pages} {778} (\bibinfo {year} {2024})}\BibitemShut {NoStop}%
\bibitem [{\citenamefont {Breuckmann}\ and\ \citenamefont
  {Eberhardt}(2021{\natexlab{a}})}]{breuckmanqldpc}%
  \BibitemOpen
  \bibfield  {author} {\bibinfo {author} {\bibfnamefont {N.~P.}\ \bibnamefont
  {Breuckmann}}\ and\ \bibinfo {author} {\bibfnamefont {J.~N.}\ \bibnamefont
  {Eberhardt}},\ }\bibfield  {title} {\bibinfo {title} {Quantum low-density
  parity-check codes},\ }\href {https://doi.org/10.1103/PRXQuantum.2.040101}
  {\bibfield  {journal} {\bibinfo  {journal} {PRX Quantum}\ }\textbf {\bibinfo
  {volume} {2}},\ \bibinfo {pages} {040101} (\bibinfo {year}
  {2021}{\natexlab{a}})}\BibitemShut {NoStop}%
\bibitem [{\citenamefont {Zhong}\ \emph {et~al.}(2021)\citenamefont {Zhong},
  \citenamefont {Chang}, \citenamefont {Bienfait} \emph {et~al.}}]{Zhong_2021}%
  \BibitemOpen
  \bibfield  {author} {\bibinfo {author} {\bibfnamefont {Y.}~\bibnamefont
  {Zhong}}, \bibinfo {author} {\bibfnamefont {H.-S.}\ \bibnamefont {Chang}},
  \bibinfo {author} {\bibfnamefont {A.}~\bibnamefont {Bienfait}}, \emph
  {et~al.},\ }\bibfield  {title} {\bibinfo {title} {Deterministic multi-qubit
  entanglement in a quantum network},\ }\href
  {https://doi.org/10.1038/s41586-021-03288-7} {\bibfield  {journal} {\bibinfo
  {journal} {Nature}\ }\textbf {\bibinfo {volume} {590}},\ \bibinfo {pages}
  {571–575} (\bibinfo {year} {2021})}\BibitemShut {NoStop}%
\bibitem [{\citenamefont {Niu}\ \emph {et~al.}(2023)\citenamefont {Niu},
  \citenamefont {Zhang}, \citenamefont {Liu} \emph {et~al.}}]{Niu_2023}%
  \BibitemOpen
  \bibfield  {author} {\bibinfo {author} {\bibfnamefont {J.}~\bibnamefont
  {Niu}}, \bibinfo {author} {\bibfnamefont {L.}~\bibnamefont {Zhang}}, \bibinfo
  {author} {\bibfnamefont {Y.}~\bibnamefont {Liu}}, \emph {et~al.},\ }\bibfield
   {title} {\bibinfo {title} {Low-loss interconnects for modular
  superconducting quantum processors},\ }\href
  {https://doi.org/10.1038/s41928-023-00925-z} {\bibfield  {journal} {\bibinfo
  {journal} {Nature Electronics}\ }\textbf {\bibinfo {volume} {6}},\ \bibinfo
  {pages} {235–241} (\bibinfo {year} {2023})}\BibitemShut {NoStop}%
\bibitem [{\citenamefont {Qiu}\ \emph {et~al.}(2025)\citenamefont {Qiu},
  \citenamefont {Liu}, \citenamefont {Hu} \emph {et~al.}}]{Qiu_2025}%
  \BibitemOpen
  \bibfield  {author} {\bibinfo {author} {\bibfnamefont {J.}~\bibnamefont
  {Qiu}}, \bibinfo {author} {\bibfnamefont {Y.}~\bibnamefont {Liu}}, \bibinfo
  {author} {\bibfnamefont {L.}~\bibnamefont {Hu}}, \emph {et~al.},\ }\bibfield
  {title} {\bibinfo {title} {Deterministic quantum state and gate teleportation
  between distant superconducting chips},\ }\href
  {https://doi.org/10.1016/j.scib.2024.11.047} {\bibfield  {journal} {\bibinfo
  {journal} {Science Bulletin}\ }\textbf {\bibinfo {volume} {70}},\ \bibinfo
  {pages} {351–358} (\bibinfo {year} {2025})}\BibitemShut {NoStop}%
\bibitem [{\citenamefont {Freedman}\ \emph {et~al.}(2002)\citenamefont
  {Freedman}, \citenamefont {Meyer},\ and\ \citenamefont
  {Luo}}]{freedman2002z2}%
  \BibitemOpen
  \bibfield  {author} {\bibinfo {author} {\bibfnamefont {M.~H.}\ \bibnamefont
  {Freedman}}, \bibinfo {author} {\bibfnamefont {D.~A.}\ \bibnamefont
  {Meyer}},\ and\ \bibinfo {author} {\bibfnamefont {F.}~\bibnamefont {Luo}},\
  }\bibfield  {title} {\bibinfo {title} {{$\mathbb{Z}_2$}-systolic freedom and
  quantum codes},\ }in\ \href@noop {} {\emph {\bibinfo {booktitle} {Mathematics
  of quantum computation}}}\ (\bibinfo  {publisher} {Chapman and Hall/CRC},\
  \bibinfo {year} {2002})\ pp.\ \bibinfo {pages} {303--338}\BibitemShut
  {NoStop}%
\bibitem [{\citenamefont {Breuckmann}\ and\ \citenamefont
  {Terhal}(2016)}]{breuckmann_2016_constructions}%
  \BibitemOpen
  \bibfield  {author} {\bibinfo {author} {\bibfnamefont {N.~P.}\ \bibnamefont
  {Breuckmann}}\ and\ \bibinfo {author} {\bibfnamefont {B.~M.}\ \bibnamefont
  {Terhal}},\ }\bibfield  {title} {\bibinfo {title} {Constructions and noise
  threshold of hyperbolic surface codes},\ }\href
  {https://doi.org/10.1109/TIT.2016.2555700} {\bibfield  {journal} {\bibinfo
  {journal} {IEEE Transactions on Information Theory}\ }\textbf {\bibinfo
  {volume} {62}},\ \bibinfo {pages} {3731} (\bibinfo {year}
  {2016})}\BibitemShut {NoStop}%
\bibitem [{\citenamefont {Delfosse}(2013)}]{delfosse_tradeoffs}%
  \BibitemOpen
  \bibfield  {author} {\bibinfo {author} {\bibfnamefont {N.}~\bibnamefont
  {Delfosse}},\ }\bibfield  {title} {\bibinfo {title} {Tradeoffs for reliable
  quantum information storage in surface codes and color codes},\ }in\ \href
  {https://doi.org/10.1109/ISIT.2013.6620360} {\emph {\bibinfo {booktitle}
  {2013 IEEE International Symposium on Information Theory}}}\ (\bibinfo {year}
  {2013})\ pp.\ \bibinfo {pages} {917--921}\BibitemShut {NoStop}%
\bibitem [{\citenamefont {Breuckmann}\ \emph {et~al.}(2017)\citenamefont
  {Breuckmann}, \citenamefont {Vuillot}, \citenamefont {Campbell},
  \citenamefont {Krishna},\ and\ \citenamefont
  {Terhal}}]{Breuckmann_2017_semihyperbolic}%
  \BibitemOpen
  \bibfield  {author} {\bibinfo {author} {\bibfnamefont {N.~P.}\ \bibnamefont
  {Breuckmann}}, \bibinfo {author} {\bibfnamefont {C.}~\bibnamefont {Vuillot}},
  \bibinfo {author} {\bibfnamefont {E.}~\bibnamefont {Campbell}}, \bibinfo
  {author} {\bibfnamefont {A.}~\bibnamefont {Krishna}},\ and\ \bibinfo {author}
  {\bibfnamefont {B.~M.}\ \bibnamefont {Terhal}},\ }\bibfield  {title}
  {\bibinfo {title} {Hyperbolic and semi-hyperbolic surface codes for quantum
  storage},\ }\href {https://doi.org/10.1088/2058-9565/aa7d3b} {\bibfield
  {journal} {\bibinfo  {journal} {Quantum Science and Technology}\ }\textbf
  {\bibinfo {volume} {2}},\ \bibinfo {pages} {035007} (\bibinfo {year}
  {2017})}\BibitemShut {NoStop}%
\bibitem [{\citenamefont {Higgott}\ and\ \citenamefont
  {Breuckmann}(2024)}]{higgott_2024_floquet}%
  \BibitemOpen
  \bibfield  {author} {\bibinfo {author} {\bibfnamefont {O.}~\bibnamefont
  {Higgott}}\ and\ \bibinfo {author} {\bibfnamefont {N.~P.}\ \bibnamefont
  {Breuckmann}},\ }\bibfield  {title} {\bibinfo {title} {Constructions and
  performance of hyperbolic and semi-hyperbolic {Floquet} codes},\ }\href
  {https://doi.org/10.1103/PRXQuantum.5.040327} {\bibfield  {journal} {\bibinfo
   {journal} {PRX Quantum}\ }\textbf {\bibinfo {volume} {5}},\ \bibinfo {pages}
  {040327} (\bibinfo {year} {2024})}\BibitemShut {NoStop}%
\bibitem [{\citenamefont {Sutcliffe}\ \emph {et~al.}(2025)\citenamefont
  {Sutcliffe}, \citenamefont {Jonnadula}, \citenamefont {Le~Gall},
  \citenamefont {Moylett},\ and\ \citenamefont
  {Westoby}}]{sutcliffe_2025_distributed}%
  \BibitemOpen
  \bibfield  {author} {\bibinfo {author} {\bibfnamefont {E.}~\bibnamefont
  {Sutcliffe}}, \bibinfo {author} {\bibfnamefont {B.}~\bibnamefont
  {Jonnadula}}, \bibinfo {author} {\bibfnamefont {C.}~\bibnamefont {Le~Gall}},
  \bibinfo {author} {\bibfnamefont {A.~E.}\ \bibnamefont {Moylett}},\ and\
  \bibinfo {author} {\bibfnamefont {C.~M.}\ \bibnamefont {Westoby}},\
  }\bibfield  {title} {\bibinfo {title} {Distributed quantum error correction
  based on hyperbolic {Floquet} codes},\ }in\ \href
  {https://doi.org/10.1109/QCE65121.2025.00076} {\emph {\bibinfo {booktitle}
  {2025 IEEE International Conference on Quantum Computing and Engineering
  (QCE)}}},\ Vol.~\bibinfo {volume} {01}\ (\bibinfo {year} {2025})\ pp.\
  \bibinfo {pages} {649--657}\BibitemShut {NoStop}%
\bibitem [{\citenamefont {Bausch}\ \emph {et~al.}(2024)\citenamefont {Bausch},
  \citenamefont {Senior}, \citenamefont {Heras}, \citenamefont {Edlich},
  \citenamefont {Davies}, \citenamefont {Newman}, \citenamefont {Jones},
  \citenamefont {Satzinger}, \citenamefont {Niu}, \citenamefont {Blackwell}
  \emph {et~al.}}]{bausch2024learning}%
  \BibitemOpen
  \bibfield  {author} {\bibinfo {author} {\bibfnamefont {J.}~\bibnamefont
  {Bausch}}, \bibinfo {author} {\bibfnamefont {A.~W.}\ \bibnamefont {Senior}},
  \bibinfo {author} {\bibfnamefont {F.~J.}\ \bibnamefont {Heras}}, \bibinfo
  {author} {\bibfnamefont {T.}~\bibnamefont {Edlich}}, \bibinfo {author}
  {\bibfnamefont {A.}~\bibnamefont {Davies}}, \bibinfo {author} {\bibfnamefont
  {M.}~\bibnamefont {Newman}}, \bibinfo {author} {\bibfnamefont
  {C.}~\bibnamefont {Jones}}, \bibinfo {author} {\bibfnamefont
  {K.}~\bibnamefont {Satzinger}}, \bibinfo {author} {\bibfnamefont {M.~Y.}\
  \bibnamefont {Niu}}, \bibinfo {author} {\bibfnamefont {S.}~\bibnamefont
  {Blackwell}}, \emph {et~al.},\ }\bibfield  {title} {\bibinfo {title}
  {Learning high-accuracy error decoding for quantum processors},\ }\href
  {https://doi.org/10.1038/s41586-024-08148-8} {\bibfield  {journal} {\bibinfo
  {journal} {Nature}\ }\textbf {\bibinfo {volume} {635}},\ \bibinfo {pages}
  {834} (\bibinfo {year} {2024})}\BibitemShut {NoStop}%
\bibitem [{\citenamefont {Senior}\ \emph {et~al.}(2025)\citenamefont {Senior},
  \citenamefont {Edlich}, \citenamefont {Heras} \emph
  {et~al.}}]{senior2025scalablerealtimeneuraldecoder}%
  \BibitemOpen
  \bibfield  {author} {\bibinfo {author} {\bibfnamefont {A.~W.}\ \bibnamefont
  {Senior}}, \bibinfo {author} {\bibfnamefont {T.}~\bibnamefont {Edlich}},
  \bibinfo {author} {\bibfnamefont {F.~J.~H.}\ \bibnamefont {Heras}}, \emph
  {et~al.},\ }\href {https://arxiv.org/abs/2512.07737} {\bibinfo {title} {A
  scalable and real-time neural decoder for topological quantum codes}}
  (\bibinfo {year} {2025}),\ \Eprint {https://arxiv.org/abs/2512.07737}
  {arXiv:2512.07737 [quant-ph]} \BibitemShut {NoStop}%
\bibitem [{\citenamefont {Higgott}\ and\ \citenamefont
  {Gidney}(2025)}]{Higgott2025sparseblossom}%
  \BibitemOpen
  \bibfield  {author} {\bibinfo {author} {\bibfnamefont {O.}~\bibnamefont
  {Higgott}}\ and\ \bibinfo {author} {\bibfnamefont {C.}~\bibnamefont
  {Gidney}},\ }\bibfield  {title} {\bibinfo {title} {Sparse {B}lossom:
  correcting a million errors per core second with minimum-weight matching},\
  }\href {https://doi.org/10.22331/q-2025-01-20-1600} {\bibfield  {journal}
  {\bibinfo  {journal} {{Quantum}}\ }\textbf {\bibinfo {volume} {9}},\ \bibinfo
  {pages} {1600} (\bibinfo {year} {2025})}\BibitemShut {NoStop}%
\bibitem [{\citenamefont {Yoder}\ \emph {et~al.}(2025)\citenamefont {Yoder},
  \citenamefont {Schoute}, \citenamefont {Rall}, \citenamefont {Pritchett},
  \citenamefont {Gambetta}, \citenamefont {Cross}, \citenamefont {Carroll},\
  and\ \citenamefont {Beverland}}]{yoder2025tour}%
  \BibitemOpen
  \bibfield  {author} {\bibinfo {author} {\bibfnamefont {T.~J.}\ \bibnamefont
  {Yoder}}, \bibinfo {author} {\bibfnamefont {E.}~\bibnamefont {Schoute}},
  \bibinfo {author} {\bibfnamefont {P.}~\bibnamefont {Rall}}, \bibinfo {author}
  {\bibfnamefont {E.}~\bibnamefont {Pritchett}}, \bibinfo {author}
  {\bibfnamefont {J.~M.}\ \bibnamefont {Gambetta}}, \bibinfo {author}
  {\bibfnamefont {A.~W.}\ \bibnamefont {Cross}}, \bibinfo {author}
  {\bibfnamefont {M.}~\bibnamefont {Carroll}},\ and\ \bibinfo {author}
  {\bibfnamefont {M.~E.}\ \bibnamefont {Beverland}},\ }\href
  {https://arxiv.org/abs/2506.03094} {\bibinfo {title} {Tour de gross: A
  modular quantum computer based on bivariate bicycle codes}} (\bibinfo {year}
  {2025}),\ \Eprint {https://arxiv.org/abs/2506.03094} {arXiv:2506.03094}
  \BibitemShut {NoStop}%
\bibitem [{\citenamefont {Gidney}\ \emph {et~al.}(2025)\citenamefont {Gidney},
  \citenamefont {Newman}, \citenamefont {Brooks},\ and\ \citenamefont
  {Jones}}]{yoked_surface_code}%
  \BibitemOpen
  \bibfield  {author} {\bibinfo {author} {\bibfnamefont {C.}~\bibnamefont
  {Gidney}}, \bibinfo {author} {\bibfnamefont {M.}~\bibnamefont {Newman}},
  \bibinfo {author} {\bibfnamefont {P.}~\bibnamefont {Brooks}},\ and\ \bibinfo
  {author} {\bibfnamefont {C.}~\bibnamefont {Jones}},\ }\bibfield  {title}
  {\bibinfo {title} {Yoked surface codes},\ }\href
  {https://doi.org/10.1038/s41467-025-59714-1} {\bibfield  {journal} {\bibinfo
  {journal} {Nature Communications}\ }\textbf {\bibinfo {volume} {16}},\
  \bibinfo {pages} {4498} (\bibinfo {year} {2025})}\BibitemShut {NoStop}%
\bibitem [{\citenamefont {Gehér}\ \emph {et~al.}(2026)\citenamefont {Gehér},
  \citenamefont {Byfield},\ and\ \citenamefont {Ruban}}]{geher2025directional}%
  \BibitemOpen
  \bibfield  {author} {\bibinfo {author} {\bibfnamefont {G.~P.}\ \bibnamefont
  {Gehér}}, \bibinfo {author} {\bibfnamefont {D.}~\bibnamefont {Byfield}},\
  and\ \bibinfo {author} {\bibfnamefont {A.}~\bibnamefont {Ruban}},\ }\href
  {https://arxiv.org/abs/2507.19430} {\bibinfo {title} {Directional codes: a
  new family of quantum {LDPC} codes on hexagonal- and square-grid connectivity
  hardware}} (\bibinfo {year} {2026}),\ \Eprint
  {https://arxiv.org/abs/2507.19430} {arXiv:2507.19430 [quant-ph]} \BibitemShut
  {NoStop}%
\bibitem [{\citenamefont {Gu}\ \emph {et~al.}(2026)\citenamefont {Gu},
  \citenamefont {Noszko}, \citenamefont {Steffan}, \citenamefont {Eberhardt},
  \citenamefont {Roffe}, \citenamefont {Eisert},\ and\ \citenamefont
  {Koutsioumpas}}]{gu2026nearest}%
  \BibitemOpen
  \bibfield  {author} {\bibinfo {author} {\bibfnamefont {B.}~\bibnamefont
  {Gu}}, \bibinfo {author} {\bibfnamefont {T.}~\bibnamefont {Noszko}}, \bibinfo
  {author} {\bibfnamefont {V.}~\bibnamefont {Steffan}}, \bibinfo {author}
  {\bibfnamefont {J.~N.}\ \bibnamefont {Eberhardt}}, \bibinfo {author}
  {\bibfnamefont {J.}~\bibnamefont {Roffe}}, \bibinfo {author} {\bibfnamefont
  {J.}~\bibnamefont {Eisert}},\ and\ \bibinfo {author} {\bibfnamefont
  {S.}~\bibnamefont {Koutsioumpas}},\ }\href {https://arxiv.org/abs/2606.19482}
  {\bibinfo {title} {Nearest-neighbour gates are all you need: High-rate
  quantum low-density parity-check codes on a planar grid}} (\bibinfo {year}
  {2026}),\ \Eprint {https://arxiv.org/abs/2606.19482} {arXiv:2606.19482
  [quant-ph]} \BibitemShut {NoStop}%
\bibitem [{\citenamefont {Nixon}\ \emph {et~al.}(2026)\citenamefont {Nixon},
  \citenamefont {McLauchlan},\ and\ \citenamefont {van Rest}}]{nixon2026vine}%
  \BibitemOpen
  \bibfield  {author} {\bibinfo {author} {\bibfnamefont {G.~M.}\ \bibnamefont
  {Nixon}}, \bibinfo {author} {\bibfnamefont {C.~K.}\ \bibnamefont
  {McLauchlan}},\ and\ \bibinfo {author} {\bibfnamefont {C.~C.~L.}\
  \bibnamefont {van Rest}},\ }\href {https://arxiv.org/abs/2606.20263}
  {\bibinfo {title} {Vine codes: Low-overhead quantum {LDPC} codes on a planar
  square grid}} (\bibinfo {year} {2026}),\ \Eprint
  {https://arxiv.org/abs/2606.20263} {arXiv:2606.20263 [quant-ph]} \BibitemShut
  {NoStop}%
\bibitem [{\citenamefont {Higgott}\ and\ \citenamefont
  {Breuckmann}(2021)}]{higgott_2021_subsystem}%
  \BibitemOpen
  \bibfield  {author} {\bibinfo {author} {\bibfnamefont {O.}~\bibnamefont
  {Higgott}}\ and\ \bibinfo {author} {\bibfnamefont {N.~P.}\ \bibnamefont
  {Breuckmann}},\ }\bibfield  {title} {\bibinfo {title} {Subsystem codes with
  high thresholds by gauge fixing and reduced qubit overhead},\ }\href
  {https://doi.org/10.1103/PhysRevX.11.031039} {\bibfield  {journal} {\bibinfo
  {journal} {Phys. Rev. X}\ }\textbf {\bibinfo {volume} {11}},\ \bibinfo
  {pages} {031039} (\bibinfo {year} {2021})}\BibitemShut {NoStop}%
\bibitem [{\citenamefont {Vittal}\ \emph {et~al.}(2024)\citenamefont {Vittal},
  \citenamefont {Javadi-Abhari}, \citenamefont {Cross}, \citenamefont
  {Bishop},\ and\ \citenamefont {Qureshi}}]{flag_proxy_hyperbolic_color}%
  \BibitemOpen
  \bibfield  {author} {\bibinfo {author} {\bibfnamefont {S.}~\bibnamefont
  {Vittal}}, \bibinfo {author} {\bibfnamefont {A.}~\bibnamefont
  {Javadi-Abhari}}, \bibinfo {author} {\bibfnamefont {A.~W.}\ \bibnamefont
  {Cross}}, \bibinfo {author} {\bibfnamefont {L.~S.}\ \bibnamefont {Bishop}},\
  and\ \bibinfo {author} {\bibfnamefont {M.}~\bibnamefont {Qureshi}},\
  }\bibfield  {title} {\bibinfo {title} {Flag-proxy networks: Overcoming the
  architectural, scheduling and decoding obstacles of quantum {LDPC} codes},\
  }in\ \href {https://doi.org/10.1109/MICRO61859.2024.00059} {\emph {\bibinfo
  {booktitle} {Proceedings of the 2024 57th IEEE/ACM International Symposium on
  Microarchitecture}}},\ \bibinfo {series and number} {MICRO '24}\ (\bibinfo
  {publisher} {IEEE Press},\ \bibinfo {year} {2024})\ p.\ \bibinfo {pages}
  {718–734}\BibitemShut {NoStop}%
\bibitem [{\citenamefont {Tillich}\ and\ \citenamefont
  {Zémor}(2014)}]{tillich2014hgp}%
  \BibitemOpen
  \bibfield  {author} {\bibinfo {author} {\bibfnamefont {J.-P.}\ \bibnamefont
  {Tillich}}\ and\ \bibinfo {author} {\bibfnamefont {G.}~\bibnamefont
  {Zémor}},\ }\bibfield  {title} {\bibinfo {title} {Quantum {LDPC} codes with
  positive rate and minimum distance proportional to the square root of the
  blocklength},\ }\href {https://doi.org/10.1109/TIT.2013.2292061} {\bibfield
  {journal} {\bibinfo  {journal} {IEEE Transactions on Information Theory}\
  }\textbf {\bibinfo {volume} {60}},\ \bibinfo {pages} {1193} (\bibinfo {year}
  {2014})}\BibitemShut {NoStop}%
\bibitem [{\citenamefont {Breuckmann}\ and\ \citenamefont
  {Eberhardt}(2021{\natexlab{b}})}]{breuckmann2021product}%
  \BibitemOpen
  \bibfield  {author} {\bibinfo {author} {\bibfnamefont {N.~P.}\ \bibnamefont
  {Breuckmann}}\ and\ \bibinfo {author} {\bibfnamefont {J.~N.}\ \bibnamefont
  {Eberhardt}},\ }\bibfield  {title} {\bibinfo {title} {Balanced product
  quantum codes},\ }\href {https://doi.org/10.1109/TIT.2021.3097347} {\bibfield
   {journal} {\bibinfo  {journal} {IEEE Transactions on Information Theory}\
  }\textbf {\bibinfo {volume} {67}},\ \bibinfo {pages} {6653} (\bibinfo {year}
  {2021}{\natexlab{b}})}\BibitemShut {NoStop}%
\bibitem [{\citenamefont {Panteleev}\ and\ \citenamefont
  {Kalachev}(2022)}]{panteleev2022good}%
  \BibitemOpen
  \bibfield  {author} {\bibinfo {author} {\bibfnamefont {P.}~\bibnamefont
  {Panteleev}}\ and\ \bibinfo {author} {\bibfnamefont {G.}~\bibnamefont
  {Kalachev}},\ }\bibfield  {title} {\bibinfo {title} {Asymptotically good
  quantum and locally testable classical {LDPC} codes},\ }in\ \href
  {https://doi.org/10.1145/3519935.3520017} {\emph {\bibinfo {booktitle}
  {Proceedings of the 54th Annual ACM SIGACT Symposium on Theory of
  Computing}}},\ \bibinfo {series and number} {STOC 2022}\ (\bibinfo
  {publisher} {Association for Computing Machinery},\ \bibinfo {address} {New
  York, NY, USA},\ \bibinfo {year} {2022})\ p.\ \bibinfo {pages}
  {375–388}\BibitemShut {NoStop}%
\bibitem [{\citenamefont {Kasai}(2026)}]{Kasai2026breaking}%
  \BibitemOpen
  \bibfield  {author} {\bibinfo {author} {\bibfnamefont {K.}~\bibnamefont
  {Kasai}},\ }\bibfield  {title} {\bibinfo {title} {Breaking the
  {O}rthogonality {B}arrier in {Q}uantum {LDPC} {C}odes},\ }\href
  {https://doi.org/10.22331/q-2026-09-09-2205} {\bibfield  {journal} {\bibinfo
  {journal} {{Quantum}}\ }\textbf {\bibinfo {volume} {10}},\ \bibinfo {pages}
  {2205} (\bibinfo {year} {2026})}\BibitemShut {NoStop}%
\bibitem [{\citenamefont {Zhao}\ \emph {et~al.}(2026)\citenamefont {Zhao},
  \citenamefont {Duckering}, \citenamefont {Gu}, \citenamefont {Maskara},\ and\
  \citenamefont {Zhou}}]{zhao2026ultrahighratequantumerrorcorrection}%
  \BibitemOpen
  \bibfield  {author} {\bibinfo {author} {\bibfnamefont {C.}~\bibnamefont
  {Zhao}}, \bibinfo {author} {\bibfnamefont {C.}~\bibnamefont {Duckering}},
  \bibinfo {author} {\bibfnamefont {A.}~\bibnamefont {Gu}}, \bibinfo {author}
  {\bibfnamefont {N.}~\bibnamefont {Maskara}},\ and\ \bibinfo {author}
  {\bibfnamefont {H.}~\bibnamefont {Zhou}},\ }\href
  {https://arxiv.org/abs/2604.16209} {\bibinfo {title} {Towards ultra-high-rate
  quantum error correction with reconfigurable atom arrays}} (\bibinfo {year}
  {2026}),\ \Eprint {https://arxiv.org/abs/2604.16209} {arXiv:2604.16209
  [quant-ph]} \BibitemShut {NoStop}%
\bibitem [{\citenamefont {Bombin}\ and\ \citenamefont
  {Martin-Delgado}(2006)}]{bombin_color}%
  \BibitemOpen
  \bibfield  {author} {\bibinfo {author} {\bibfnamefont {H.}~\bibnamefont
  {Bombin}}\ and\ \bibinfo {author} {\bibfnamefont {M.~A.}\ \bibnamefont
  {Martin-Delgado}},\ }\bibfield  {title} {\bibinfo {title} {Topological
  quantum distillation},\ }\href
  {https://doi.org/10.1103/PhysRevLett.97.180501} {\bibfield  {journal}
  {\bibinfo  {journal} {Phys. Rev. Lett.}\ }\textbf {\bibinfo {volume} {97}},\
  \bibinfo {pages} {180501} (\bibinfo {year} {2006})}\BibitemShut {NoStop}%
\bibitem [{\citenamefont {Bombin}\ and\ \citenamefont
  {Martin-Delgado}(2007)}]{bombin_delgado_toric_surface_color}%
  \BibitemOpen
  \bibfield  {author} {\bibinfo {author} {\bibfnamefont {H.}~\bibnamefont
  {Bombin}}\ and\ \bibinfo {author} {\bibfnamefont {M.~A.}\ \bibnamefont
  {Martin-Delgado}},\ }\bibfield  {title} {\bibinfo {title} {Optimal resources
  for topological two-dimensional stabilizer codes: Comparative study},\ }\href
  {https://doi.org/10.1103/PhysRevA.76.012305} {\bibfield  {journal} {\bibinfo
  {journal} {Phys. Rev. A}\ }\textbf {\bibinfo {volume} {76}},\ \bibinfo
  {pages} {012305} (\bibinfo {year} {2007})}\BibitemShut {NoStop}%
\bibitem [{\citenamefont {Chao}\ and\ \citenamefont
  {Reichardt}(2018)}]{chao_reichardt_flag}%
  \BibitemOpen
  \bibfield  {author} {\bibinfo {author} {\bibfnamefont {R.}~\bibnamefont
  {Chao}}\ and\ \bibinfo {author} {\bibfnamefont {B.~W.}\ \bibnamefont
  {Reichardt}},\ }\bibfield  {title} {\bibinfo {title} {Quantum error
  correction with only two extra qubits},\ }\href
  {https://doi.org/10.1103/PhysRevLett.121.050502} {\bibfield  {journal}
  {\bibinfo  {journal} {Phys. Rev. Lett.}\ }\textbf {\bibinfo {volume} {121}},\
  \bibinfo {pages} {050502} (\bibinfo {year} {2018})}\BibitemShut {NoStop}%
\bibitem [{\citenamefont {Chamberland}\ and\ \citenamefont
  {Beverland}(2018)}]{flag_QEC}%
  \BibitemOpen
  \bibfield  {author} {\bibinfo {author} {\bibfnamefont {C.}~\bibnamefont
  {Chamberland}}\ and\ \bibinfo {author} {\bibfnamefont {M.~E.}\ \bibnamefont
  {Beverland}},\ }\bibfield  {title} {\bibinfo {title} {Flag fault-tolerant
  error correction with arbitrary distance codes},\ }\href
  {https://doi.org/10.22331/q-2018-02-08-53} {\bibfield  {journal} {\bibinfo
  {journal} {{Quantum}}\ }\textbf {\bibinfo {volume} {2}},\ \bibinfo {pages}
  {53} (\bibinfo {year} {2018})}\BibitemShut {NoStop}%
\bibitem [{\citenamefont {Takada}\ and\ \citenamefont
  {Fujii}(2024)}]{takada_keisuke24}%
  \BibitemOpen
  \bibfield  {author} {\bibinfo {author} {\bibfnamefont {Y.}~\bibnamefont
  {Takada}}\ and\ \bibinfo {author} {\bibfnamefont {K.}~\bibnamefont {Fujii}},\
  }\bibfield  {title} {\bibinfo {title} {Improving threshold for fault-tolerant
  color-code quantum computing by flagged weight optimization},\ }\href
  {https://doi.org/10.1103/PRXQuantum.5.030352} {\bibfield  {journal} {\bibinfo
   {journal} {PRX Quantum}\ }\textbf {\bibinfo {volume} {5}},\ \bibinfo {pages}
  {030352} (\bibinfo {year} {2024})}\BibitemShut {NoStop}%
\bibitem [{\citenamefont {Cohen}\ \emph {et~al.}(2022)\citenamefont {Cohen},
  \citenamefont {Kim}, \citenamefont {Bartlett},\ and\ \citenamefont
  {Brown}}]{cohen2022low-overhead}%
  \BibitemOpen
  \bibfield  {author} {\bibinfo {author} {\bibfnamefont {L.~Z.}\ \bibnamefont
  {Cohen}}, \bibinfo {author} {\bibfnamefont {I.~H.}\ \bibnamefont {Kim}},
  \bibinfo {author} {\bibfnamefont {S.~D.}\ \bibnamefont {Bartlett}},\ and\
  \bibinfo {author} {\bibfnamefont {B.~J.}\ \bibnamefont {Brown}},\ }\bibfield
  {title} {\bibinfo {title} {{Low-overhead fault-tolerant quantum computing
  using long-range connectivity}},\ }\href
  {https://doi.org/10.1126/sciadv.abn1717} {\bibfield  {journal} {\bibinfo
  {journal} {Science Advances}\ }\textbf {\bibinfo {volume} {8}},\ \bibinfo
  {pages} {eabn1717} (\bibinfo {year} {2022})}\BibitemShut {NoStop}%
\bibitem [{\citenamefont {Cross}\ \emph {et~al.}(2024)\citenamefont {Cross},
  \citenamefont {He}, \citenamefont {Rall},\ and\ \citenamefont
  {Yoder}}]{cross2024improved}%
  \BibitemOpen
  \bibfield  {author} {\bibinfo {author} {\bibfnamefont {A.}~\bibnamefont
  {Cross}}, \bibinfo {author} {\bibfnamefont {Z.}~\bibnamefont {He}}, \bibinfo
  {author} {\bibfnamefont {P.}~\bibnamefont {Rall}},\ and\ \bibinfo {author}
  {\bibfnamefont {T.}~\bibnamefont {Yoder}},\ }\href
  {https://arxiv.org/abs/2407.18393} {\bibinfo {title} {{Improved {QLDPC}
  Surgery: Logical Measurements and Bridging Codes}}} (\bibinfo {year}
  {2024}),\ \Eprint {https://arxiv.org/abs/2407.18393} {arXiv:2407.18393
  [quant-ph]} \BibitemShut {NoStop}%
\bibitem [{\citenamefont {Williamson}\ and\ \citenamefont
  {Yoder}(2026)}]{Williamson2026}%
  \BibitemOpen
  \bibfield  {author} {\bibinfo {author} {\bibfnamefont {D.~J.}\ \bibnamefont
  {Williamson}}\ and\ \bibinfo {author} {\bibfnamefont {T.~J.}\ \bibnamefont
  {Yoder}},\ }\bibfield  {title} {\bibinfo {title} {Low-overhead fault-tolerant
  quantum computation by gauging logical operators},\ }\href
  {https://doi.org/10.1038/s41567-026-03220-8} {\bibfield  {journal} {\bibinfo
  {journal} {Nature Physics}\ }\textbf {\bibinfo {volume} {22}},\ \bibinfo
  {pages} {598–603} (\bibinfo {year} {2026})}\BibitemShut {NoStop}%
\bibitem [{\citenamefont {Ide}\ \emph {et~al.}(2025)\citenamefont {Ide},
  \citenamefont {Gowda}, \citenamefont {Nadkarni},\ and\ \citenamefont
  {Dauphinais}}]{ide2024fault}%
  \BibitemOpen
  \bibfield  {author} {\bibinfo {author} {\bibfnamefont {B.}~\bibnamefont
  {Ide}}, \bibinfo {author} {\bibfnamefont {M.~G.}\ \bibnamefont {Gowda}},
  \bibinfo {author} {\bibfnamefont {P.~J.}\ \bibnamefont {Nadkarni}},\ and\
  \bibinfo {author} {\bibfnamefont {G.}~\bibnamefont {Dauphinais}},\ }\bibfield
   {title} {\bibinfo {title} {{Fault-Tolerant Logical Measurements via
  Homological Measurement}},\ }\href
  {https://doi.org/10.1103/PhysRevX.15.021088} {\bibfield  {journal} {\bibinfo
  {journal} {Phys. Rev. X}\ }\textbf {\bibinfo {volume} {15}},\ \bibinfo
  {pages} {021088} (\bibinfo {year} {2025})}\BibitemShut {NoStop}%
\bibitem [{\citenamefont {Swaroop}\ \emph {et~al.}(2026)\citenamefont
  {Swaroop}, \citenamefont {Jochym-O'Connor},\ and\ \citenamefont
  {Yoder}}]{swaroop2026universal}%
  \BibitemOpen
  \bibfield  {author} {\bibinfo {author} {\bibfnamefont {E.}~\bibnamefont
  {Swaroop}}, \bibinfo {author} {\bibfnamefont {T.}~\bibnamefont
  {Jochym-O'Connor}},\ and\ \bibinfo {author} {\bibfnamefont {T.~J.}\
  \bibnamefont {Yoder}},\ }\bibfield  {title} {\bibinfo {title} {Universal
  adapters between quantum low-density parity check codes},\ }\href@noop {}
  {\bibfield  {journal} {\bibinfo  {journal} {PRX Quantum}\ }\textbf {\bibinfo
  {volume} {7}},\ \bibinfo {pages} {010324} (\bibinfo {year}
  {2026})}\BibitemShut {NoStop}%
\bibitem [{\citenamefont {He}\ \emph {et~al.}(2025)\citenamefont {He},
  \citenamefont {Cowtan}, \citenamefont {Williamson},\ and\ \citenamefont
  {Yoder}}]{he2025extractorsqldpcarchitecturesefficient}%
  \BibitemOpen
  \bibfield  {author} {\bibinfo {author} {\bibfnamefont {Z.}~\bibnamefont
  {He}}, \bibinfo {author} {\bibfnamefont {A.}~\bibnamefont {Cowtan}}, \bibinfo
  {author} {\bibfnamefont {D.~J.}\ \bibnamefont {Williamson}},\ and\ \bibinfo
  {author} {\bibfnamefont {T.~J.}\ \bibnamefont {Yoder}},\ }\href
  {https://arxiv.org/abs/2503.10390} {\bibinfo {title} {Extractors: {QLDPC}
  architectures for efficient {Pauli}-based computation}} (\bibinfo {year}
  {2025}),\ \Eprint {https://arxiv.org/abs/2503.10390} {arXiv:2503.10390
  [quant-ph]} \BibitemShut {NoStop}%
\bibitem [{\citenamefont {Horsman}\ \emph {et~al.}(2012)\citenamefont
  {Horsman}, \citenamefont {Fowler}, \citenamefont {Devitt},\ and\
  \citenamefont {Meter}}]{horsman2012surface}%
  \BibitemOpen
  \bibfield  {author} {\bibinfo {author} {\bibfnamefont {D.}~\bibnamefont
  {Horsman}}, \bibinfo {author} {\bibfnamefont {A.~G.}\ \bibnamefont {Fowler}},
  \bibinfo {author} {\bibfnamefont {S.}~\bibnamefont {Devitt}},\ and\ \bibinfo
  {author} {\bibfnamefont {R.~V.}\ \bibnamefont {Meter}},\ }\bibfield  {title}
  {\bibinfo {title} {{Surface code quantum computing by lattice surgery}},\
  }\href {https://doi.org/10.1088/1367-2630/14/12/123011} {\bibfield  {journal}
  {\bibinfo  {journal} {New Journal of Physics}\ }\textbf {\bibinfo {volume}
  {14}},\ \bibinfo {pages} {123011} (\bibinfo {year} {2012})}\BibitemShut
  {NoStop}%
\bibitem [{\citenamefont {Landahl}\ and\ \citenamefont
  {Ryan-Anderson}(2014)}]{landahl2014quantum}%
  \BibitemOpen
  \bibfield  {author} {\bibinfo {author} {\bibfnamefont {A.~J.}\ \bibnamefont
  {Landahl}}\ and\ \bibinfo {author} {\bibfnamefont {C.}~\bibnamefont
  {Ryan-Anderson}},\ }\href {https://arxiv.org/abs/1407.5103} {\bibinfo {title}
  {Quantum computing by color-code lattice surgery}} (\bibinfo {year} {2014}),\
  \Eprint {https://arxiv.org/abs/1407.5103} {arXiv:1407.5103} \BibitemShut
  {NoStop}%
\bibitem [{\citenamefont {Erickson}\ and\ \citenamefont
  {Whittlesey}(2005)}]{Erickson_Whittlesey}%
  \BibitemOpen
  \bibfield  {author} {\bibinfo {author} {\bibfnamefont {J.}~\bibnamefont
  {Erickson}}\ and\ \bibinfo {author} {\bibfnamefont {K.}~\bibnamefont
  {Whittlesey}},\ }\bibfield  {title} {\bibinfo {title} {Greedy optimal
  homotopy and homology generators},\ }in\ \href@noop {} {\emph {\bibinfo
  {booktitle} {Proceedings of the Sixteenth Annual ACM-SIAM Symposium on
  Discrete Algorithms}}},\ \bibinfo {series and number} {SODA '05}\ (\bibinfo
  {publisher} {Society for Industrial and Applied Mathematics},\ \bibinfo
  {address} {USA},\ \bibinfo {year} {2005})\ p.\ \bibinfo {pages}
  {1038–1046}\BibitemShut {NoStop}%
\bibitem [{\citenamefont {Cuccaro}\ \emph {et~al.}(2004)\citenamefont
  {Cuccaro}, \citenamefont {Draper}, \citenamefont {Kutin},\ and\ \citenamefont
  {Moulton}}]{cuccaro2004new}%
  \BibitemOpen
  \bibfield  {author} {\bibinfo {author} {\bibfnamefont {S.~A.}\ \bibnamefont
  {Cuccaro}}, \bibinfo {author} {\bibfnamefont {T.~G.}\ \bibnamefont {Draper}},
  \bibinfo {author} {\bibfnamefont {S.~A.}\ \bibnamefont {Kutin}},\ and\
  \bibinfo {author} {\bibfnamefont {D.~P.}\ \bibnamefont {Moulton}},\ }\href
  {https://arxiv.org/abs/quant-ph/0410184} {\bibinfo {title} {A new quantum
  ripple-carry addition circuit}} (\bibinfo {year} {2004}),\ \Eprint
  {https://arxiv.org/abs/quant-ph/0410184} {arXiv:quant-ph/0410184}
  \BibitemShut {NoStop}%
\bibitem [{\citenamefont {Wolanski}(2025)}]{wolanski2025automated}%
  \BibitemOpen
  \bibfield  {author} {\bibinfo {author} {\bibfnamefont {S.}~\bibnamefont
  {Wolanski}},\ }\href {https://arxiv.org/abs/2512.01943} {\bibinfo {title}
  {Automated compilation including dropouts: Tolerating defective components in
  stabiliser codes}} (\bibinfo {year} {2025}),\ \Eprint
  {https://arxiv.org/abs/2512.01943} {arXiv:2512.01943} \BibitemShut {NoStop}%
\bibitem [{\citenamefont {Auger}\ \emph {et~al.}(2017)\citenamefont {Auger},
  \citenamefont {Anwar}, \citenamefont {Gimeno-Segovia}, \citenamefont
  {Stace},\ and\ \citenamefont {Browne}}]{auger2017fabrication}%
  \BibitemOpen
  \bibfield  {author} {\bibinfo {author} {\bibfnamefont {J.~M.}\ \bibnamefont
  {Auger}}, \bibinfo {author} {\bibfnamefont {H.}~\bibnamefont {Anwar}},
  \bibinfo {author} {\bibfnamefont {M.}~\bibnamefont {Gimeno-Segovia}},
  \bibinfo {author} {\bibfnamefont {T.~M.}\ \bibnamefont {Stace}},\ and\
  \bibinfo {author} {\bibfnamefont {D.~E.}\ \bibnamefont {Browne}},\ }\bibfield
   {title} {\bibinfo {title} {Fault-tolerance thresholds for the surface code
  with fabrication errors},\ }\href
  {https://doi.org/10.1103/PhysRevA.96.042316} {\bibfield  {journal} {\bibinfo
  {journal} {Phys. Rev. A}\ }\textbf {\bibinfo {volume} {96}},\ \bibinfo
  {pages} {042316} (\bibinfo {year} {2017})}\BibitemShut {NoStop}%
\bibitem [{\citenamefont {Strikis}\ \emph {et~al.}(2023)\citenamefont
  {Strikis}, \citenamefont {Benjamin},\ and\ \citenamefont
  {Brown}}]{strikis2023fabrication}%
  \BibitemOpen
  \bibfield  {author} {\bibinfo {author} {\bibfnamefont {A.}~\bibnamefont
  {Strikis}}, \bibinfo {author} {\bibfnamefont {S.~C.}\ \bibnamefont
  {Benjamin}},\ and\ \bibinfo {author} {\bibfnamefont {B.~J.}\ \bibnamefont
  {Brown}},\ }\bibfield  {title} {\bibinfo {title} {Quantum computing is
  scalable on a planar array of qubits with fabrication defects},\ }\href
  {https://doi.org/10.1103/PhysRevApplied.19.064081} {\bibfield  {journal}
  {\bibinfo  {journal} {Phys. Rev. Appl.}\ }\textbf {\bibinfo {volume} {19}},\
  \bibinfo {pages} {064081} (\bibinfo {year} {2023})}\BibitemShut {NoStop}%
\bibitem [{\citenamefont {Debroy}\ \emph {et~al.}(2025)\citenamefont {Debroy},
  \citenamefont {McEwen}, \citenamefont {Gidney}, \citenamefont {Shutty},\ and\
  \citenamefont {Zalcman}}]{Debroy2025luciinsurfacecode}%
  \BibitemOpen
  \bibfield  {author} {\bibinfo {author} {\bibfnamefont {D.~M.}\ \bibnamefont
  {Debroy}}, \bibinfo {author} {\bibfnamefont {M.}~\bibnamefont {McEwen}},
  \bibinfo {author} {\bibfnamefont {C.}~\bibnamefont {Gidney}}, \bibinfo
  {author} {\bibfnamefont {N.}~\bibnamefont {Shutty}},\ and\ \bibinfo {author}
  {\bibfnamefont {A.}~\bibnamefont {Zalcman}},\ }\bibfield  {title} {\bibinfo
  {title} {{LUCI} in the {S}urface {C}ode with {D}ropouts},\ }\href
  {https://doi.org/10.22331/q-2025-12-11-1936} {\bibfield  {journal} {\bibinfo
  {journal} {{Quantum}}\ }\textbf {\bibinfo {volume} {9}},\ \bibinfo {pages}
  {1936} (\bibinfo {year} {2025})}\BibitemShut {NoStop}%
\bibitem [{\citenamefont {Leroux}\ \emph {et~al.}(2024)\citenamefont {Leroux},
  \citenamefont {Lin}, \citenamefont {Bienias}, \citenamefont {Sankar},
  \citenamefont {Benhemou}, \citenamefont {Kubica},\ and\ \citenamefont
  {Iverson}}]{leroux2024snakes}%
  \BibitemOpen
  \bibfield  {author} {\bibinfo {author} {\bibfnamefont {C.}~\bibnamefont
  {Leroux}}, \bibinfo {author} {\bibfnamefont {S.~F.}\ \bibnamefont {Lin}},
  \bibinfo {author} {\bibfnamefont {P.}~\bibnamefont {Bienias}}, \bibinfo
  {author} {\bibfnamefont {K.~R.}\ \bibnamefont {Sankar}}, \bibinfo {author}
  {\bibfnamefont {A.}~\bibnamefont {Benhemou}}, \bibinfo {author}
  {\bibfnamefont {A.}~\bibnamefont {Kubica}},\ and\ \bibinfo {author}
  {\bibfnamefont {J.~K.}\ \bibnamefont {Iverson}},\ }\href
  {https://arxiv.org/abs/2412.11504} {\bibinfo {title} {Snakes and ladders:
  Adapting the surface code to defects}} (\bibinfo {year} {2024}),\ \Eprint
  {https://arxiv.org/abs/2412.11504} {arXiv:2412.11504} \BibitemShut {NoStop}%
\bibitem [{\citenamefont {Wu}\ and\ \citenamefont
  {Zhong}(2023)}]{wu_fusion_blossom}%
  \BibitemOpen
  \bibfield  {author} {\bibinfo {author} {\bibfnamefont {Y.}~\bibnamefont
  {Wu}}\ and\ \bibinfo {author} {\bibfnamefont {L.}~\bibnamefont {Zhong}},\
  }\bibfield  {title} {\bibinfo {title} {Fusion blossom: Fast {MWPM} decoders
  for {QEC}},\ }in\ \href {https://doi.org/10.1109/QCE57702.2023.00107} {\emph
  {\bibinfo {booktitle} {2023 IEEE International Conference on Quantum
  Computing and Engineering (QCE)}}},\ Vol.~\bibinfo {volume} {01}\ (\bibinfo
  {year} {2023})\ pp.\ \bibinfo {pages} {928--938}\BibitemShut {NoStop}%
\bibitem [{\citenamefont {Skoric}\ \emph {et~al.}(2023)\citenamefont {Skoric},
  \citenamefont {Browne}, \citenamefont {Barnes}, \citenamefont {Gillespie},\
  and\ \citenamefont {Campbell}}]{Skoric2023parallel}%
  \BibitemOpen
  \bibfield  {author} {\bibinfo {author} {\bibfnamefont {L.}~\bibnamefont
  {Skoric}}, \bibinfo {author} {\bibfnamefont {D.~E.}\ \bibnamefont {Browne}},
  \bibinfo {author} {\bibfnamefont {K.~M.}\ \bibnamefont {Barnes}}, \bibinfo
  {author} {\bibfnamefont {N.~I.}\ \bibnamefont {Gillespie}},\ and\ \bibinfo
  {author} {\bibfnamefont {E.~T.}\ \bibnamefont {Campbell}},\ }\bibfield
  {title} {\bibinfo {title} {Parallel window decoding enables scalable fault
  tolerant quantum computation},\ }\href
  {https://doi.org/10.1038/s41467-023-42482-1} {\bibfield  {journal} {\bibinfo
  {journal} {Nature Communications}\ }\textbf {\bibinfo {volume} {14}},\
  \bibinfo {pages} {7040} (\bibinfo {year} {2023})}\BibitemShut {NoStop}%
\bibitem [{\citenamefont {Tan}\ \emph {et~al.}(2023)\citenamefont {Tan},
  \citenamefont {Zhang}, \citenamefont {Chao}, \citenamefont {Shi},\ and\
  \citenamefont {Chen}}]{tan23scalable}%
  \BibitemOpen
  \bibfield  {author} {\bibinfo {author} {\bibfnamefont {X.}~\bibnamefont
  {Tan}}, \bibinfo {author} {\bibfnamefont {F.}~\bibnamefont {Zhang}}, \bibinfo
  {author} {\bibfnamefont {R.}~\bibnamefont {Chao}}, \bibinfo {author}
  {\bibfnamefont {Y.}~\bibnamefont {Shi}},\ and\ \bibinfo {author}
  {\bibfnamefont {J.}~\bibnamefont {Chen}},\ }\bibfield  {title} {\bibinfo
  {title} {Scalable surface-code decoders with parallelization in time},\
  }\href {https://doi.org/10.1103/PRXQuantum.4.040344} {\bibfield  {journal}
  {\bibinfo  {journal} {PRX Quantum}\ }\textbf {\bibinfo {volume} {4}},\
  \bibinfo {pages} {040344} (\bibinfo {year} {2023})}\BibitemShut {NoStop}%
\bibitem [{\citenamefont {Hatcher}(2002)}]{Hatcher2002}%
  \BibitemOpen
  \bibfield  {author} {\bibinfo {author} {\bibfnamefont {A.}~\bibnamefont
  {Hatcher}},\ }\href {https://pi.math.cornell.edu/~hatcher/AT/AT.pdf} {\emph
  {\bibinfo {title} {Algebraic Topology}}}\ (\bibinfo  {publisher} {Cambridge
  University Press},\ \bibinfo {address} {Cambridge},\ \bibinfo {year}
  {2002})\BibitemShut {NoStop}%
\bibitem [{\citenamefont {Vuillot}\ and\ \citenamefont
  {Breuckmann}(2022)}]{quantum_pin_codes}%
  \BibitemOpen
  \bibfield  {author} {\bibinfo {author} {\bibfnamefont {C.}~\bibnamefont
  {Vuillot}}\ and\ \bibinfo {author} {\bibfnamefont {N.~P.}\ \bibnamefont
  {Breuckmann}},\ }\bibfield  {title} {\bibinfo {title} {Quantum pin codes},\
  }\href {https://doi.org/10.1109/TIT.2022.3170846} {\bibfield  {journal}
  {\bibinfo  {journal} {IEEE Transactions on Information Theory}\ }\textbf
  {\bibinfo {volume} {68}},\ \bibinfo {pages} {5955} (\bibinfo {year}
  {2022})}\BibitemShut {NoStop}%
\bibitem [{\citenamefont {Firth}(2005)}]{firth2005algorithm}%
  \BibitemOpen
  \bibfield  {author} {\bibinfo {author} {\bibfnamefont {D.}~\bibnamefont
  {Firth}},\ }\emph {\bibinfo {title} {An Algorithm to Find Normal Subgroups of
  a Finitely Presented Group, up to a Given Finite Index}},\ \href@noop {}
  {Ph.D. thesis},\ \bibinfo  {school} {University of Warwick}, \bibinfo
  {address} {Coventry, UK} (\bibinfo {year} {2005})\BibitemShut {NoStop}%
\bibitem [{\citenamefont {Rober}(2024)}]{lins}%
  \BibitemOpen
  \bibfield  {author} {\bibinfo {author} {\bibfnamefont {F.}~\bibnamefont
  {Rober}},\ }\href@noop {} {\bibinfo {title} {{LINS}, provides an algorithm
  for computing the normal subgroups of a finitely presented group up to some
  given index bound., {V}ersion 0.9}},\ \bibinfo {howpublished} {\href
  {https://gap-packages.github.io/LINS/}
  {\texttt{https://gap-packages.github.io/}\discretionary
  {}{}{}\texttt{LINS/}}} (\bibinfo {year} {2024}),\ \bibinfo {note} {{GAP}
  package}\BibitemShut {NoStop}%
\bibitem [{\citenamefont {Delfosse}(2014)}]{delfosse2014decoding}%
  \BibitemOpen
  \bibfield  {author} {\bibinfo {author} {\bibfnamefont {N.}~\bibnamefont
  {Delfosse}},\ }\bibfield  {title} {\bibinfo {title} {Decoding color codes by
  projection onto surface codes},\ }\href
  {https://doi.org/10.1103/PhysRevA.89.012317} {\bibfield  {journal} {\bibinfo
  {journal} {Phys. Rev. A}\ }\textbf {\bibinfo {volume} {89}},\ \bibinfo
  {pages} {012317} (\bibinfo {year} {2014})}\BibitemShut {NoStop}%
\bibitem [{\citenamefont {Kubica}\ and\ \citenamefont
  {Delfosse}(2023)}]{Kubica2023efficientcolorcode}%
  \BibitemOpen
  \bibfield  {author} {\bibinfo {author} {\bibfnamefont {A.}~\bibnamefont
  {Kubica}}\ and\ \bibinfo {author} {\bibfnamefont {N.}~\bibnamefont
  {Delfosse}},\ }\bibfield  {title} {\bibinfo {title} {Efficient color code
  decoders in {$d\geq 2$} dimensions from toric code decoders},\ }\href
  {https://doi.org/10.22331/q-2023-02-21-929} {\bibfield  {journal} {\bibinfo
  {journal} {{Quantum}}\ }\textbf {\bibinfo {volume} {7}},\ \bibinfo {pages}
  {929} (\bibinfo {year} {2023})}\BibitemShut {NoStop}%
\bibitem [{\citenamefont {Eppstein}(2003)}]{tree_cotree_decomp}%
  \BibitemOpen
  \bibfield  {author} {\bibinfo {author} {\bibfnamefont {D.}~\bibnamefont
  {Eppstein}},\ }\bibfield  {title} {\bibinfo {title} {Dynamic generators of
  topologically embedded graphs},\ }in\ \href@noop {} {\emph {\bibinfo
  {booktitle} {Proceedings of the Fourteenth Annual ACM-SIAM Symposium on
  Discrete Algorithms}}},\ \bibinfo {series and number} {SODA '03}\ (\bibinfo
  {publisher} {Society for Industrial and Applied Mathematics},\ \bibinfo
  {address} {USA},\ \bibinfo {year} {2003})\ p.\ \bibinfo {pages}
  {599–608}\BibitemShut {NoStop}%
\bibitem [{\citenamefont {Dumer}\ \emph {et~al.}(2017)\citenamefont {Dumer},
  \citenamefont {Kovalev},\ and\ \citenamefont {Pryadko}}]{dumer2017distance}%
  \BibitemOpen
  \bibfield  {author} {\bibinfo {author} {\bibfnamefont {I.}~\bibnamefont
  {Dumer}}, \bibinfo {author} {\bibfnamefont {A.~A.}\ \bibnamefont {Kovalev}},\
  and\ \bibinfo {author} {\bibfnamefont {L.~P.}\ \bibnamefont {Pryadko}},\
  }\bibfield  {title} {\bibinfo {title} {Distance verification for classical
  and quantum {LDPC} codes},\ }\href {https://doi.org/10.1109/TIT.2017.2690381}
  {\bibfield  {journal} {\bibinfo  {journal} {IEEE Transactions on Information
  Theory}\ }\textbf {\bibinfo {volume} {63}},\ \bibinfo {pages} {4675}
  (\bibinfo {year} {2017})}\BibitemShut {NoStop}%
\bibitem [{\citenamefont {Pryadko}(2026)}]{dist_m4ri}%
  \BibitemOpen
  \bibfield  {author} {\bibinfo {author} {\bibfnamefont {L.~P.}\ \bibnamefont
  {Pryadko}},\ }\href {https://github.com/QEC-pages/dist-m4ri} {\bibinfo
  {title} {dist-m4ri: Distance of a classical or quantum {CSS} code}},\
  \bibinfo {howpublished} {GitHub repository} (\bibinfo {year}
  {2026})\BibitemShut {NoStop}%
\bibitem [{\citenamefont {Zhang}\ \emph {et~al.}(2026)\citenamefont {Zhang},
  \citenamefont {Yi}, \citenamefont {Guo}, \citenamefont {Kong}, \citenamefont
  {Wang}, \citenamefont {Zhan}, \citenamefont {He}, \citenamefont {Lin},
  \citenamefont {Jiang}, \citenamefont {Gao} \emph
  {et~al.}}]{zhang2026learningdecodeparallelselfcoordinating}%
  \BibitemOpen
  \bibfield  {author} {\bibinfo {author} {\bibfnamefont {K.}~\bibnamefont
  {Zhang}}, \bibinfo {author} {\bibfnamefont {Z.}~\bibnamefont {Yi}}, \bibinfo
  {author} {\bibfnamefont {S.}~\bibnamefont {Guo}}, \bibinfo {author}
  {\bibfnamefont {L.}~\bibnamefont {Kong}}, \bibinfo {author} {\bibfnamefont
  {S.}~\bibnamefont {Wang}}, \bibinfo {author} {\bibfnamefont {X.}~\bibnamefont
  {Zhan}}, \bibinfo {author} {\bibfnamefont {T.}~\bibnamefont {He}}, \bibinfo
  {author} {\bibfnamefont {W.}~\bibnamefont {Lin}}, \bibinfo {author}
  {\bibfnamefont {T.}~\bibnamefont {Jiang}}, \bibinfo {author} {\bibfnamefont
  {D.}~\bibnamefont {Gao}}, \emph {et~al.},\ }\href
  {https://arxiv.org/abs/2601.09921} {\bibinfo {title} {Learning to decode in
  parallel: Self-coordinating neural network for real-time quantum error
  correction}} (\bibinfo {year} {2026}),\ \Eprint
  {https://arxiv.org/abs/2601.09921} {arXiv:2601.09921} \BibitemShut {NoStop}%
\bibitem [{\citenamefont {Reid}\ \emph {et~al.}(2025)\citenamefont {Reid},
  \citenamefont {Sehanobish}, \citenamefont {H{\"o}fs}, \citenamefont
  {Mlodozeniec}, \citenamefont {Vulpius}, \citenamefont {Barbero},
  \citenamefont {Weller}, \citenamefont {Choromanski}, \citenamefont {Turner},\
  and\ \citenamefont {Veli{\v{c}}kovi{\'c}}}]{reid2025rotary}%
  \BibitemOpen
  \bibfield  {author} {\bibinfo {author} {\bibfnamefont {I.}~\bibnamefont
  {Reid}}, \bibinfo {author} {\bibfnamefont {A.}~\bibnamefont {Sehanobish}},
  \bibinfo {author} {\bibfnamefont {C.}~\bibnamefont {H{\"o}fs}}, \bibinfo
  {author} {\bibfnamefont {B.}~\bibnamefont {Mlodozeniec}}, \bibinfo {author}
  {\bibfnamefont {L.}~\bibnamefont {Vulpius}}, \bibinfo {author} {\bibfnamefont
  {F.}~\bibnamefont {Barbero}}, \bibinfo {author} {\bibfnamefont
  {A.}~\bibnamefont {Weller}}, \bibinfo {author} {\bibfnamefont
  {K.}~\bibnamefont {Choromanski}}, \bibinfo {author} {\bibfnamefont {R.~E.}\
  \bibnamefont {Turner}},\ and\ \bibinfo {author} {\bibfnamefont
  {P.}~\bibnamefont {Veli{\v{c}}kovi{\'c}}},\ }\href
  {https://arxiv.org/abs/2509.22259} {\bibinfo {title} {Rotary position
  encodings for graphs}} (\bibinfo {year} {2025}),\ \Eprint
  {https://arxiv.org/abs/2509.22259} {arXiv:2509.22259} \BibitemShut {NoStop}%
\bibitem [{\citenamefont {Fowler}(2013)}]{fowler2013optimal}%
  \BibitemOpen
  \bibfield  {author} {\bibinfo {author} {\bibfnamefont {A.~G.}\ \bibnamefont
  {Fowler}},\ }\href {https://arxiv.org/abs/1310.0863} {\bibinfo {title}
  {Optimal complexity correction of correlated errors in the surface code}}
  (\bibinfo {year} {2013}),\ \Eprint {https://arxiv.org/abs/1310.0863}
  {arXiv:1310.0863} \BibitemShut {NoStop}%
\bibitem [{\citenamefont {Gidney}(2021)}]{Gidney2021stimfaststabilizer}%
  \BibitemOpen
  \bibfield  {author} {\bibinfo {author} {\bibfnamefont {C.}~\bibnamefont
  {Gidney}},\ }\bibfield  {title} {\bibinfo {title} {Stim: a fast stabilizer
  circuit simulator},\ }\href {https://doi.org/10.22331/q-2021-07-06-497}
  {\bibfield  {journal} {\bibinfo  {journal} {{Quantum}}\ }\textbf {\bibinfo
  {volume} {5}},\ \bibinfo {pages} {497} (\bibinfo {year} {2021})}\BibitemShut
  {NoStop}%
\bibitem [{\citenamefont {Calderbank}\ \emph {et~al.}(1998)\citenamefont
  {Calderbank}, \citenamefont {Rains}, \citenamefont {Shor},\ and\
  \citenamefont {Sloane}}]{calderbank1998qec}%
  \BibitemOpen
  \bibfield  {author} {\bibinfo {author} {\bibfnamefont {A.}~\bibnamefont
  {Calderbank}}, \bibinfo {author} {\bibfnamefont {E.}~\bibnamefont {Rains}},
  \bibinfo {author} {\bibfnamefont {P.}~\bibnamefont {Shor}},\ and\ \bibinfo
  {author} {\bibfnamefont {N.}~\bibnamefont {Sloane}},\ }\bibfield  {title}
  {\bibinfo {title} {Quantum error correction via codes over {GF(4)}},\ }\href
  {https://doi.org/10.1109/18.681315} {\bibfield  {journal} {\bibinfo
  {journal} {IEEE Transactions on Information Theory}\ }\textbf {\bibinfo
  {volume} {44}},\ \bibinfo {pages} {1369} (\bibinfo {year}
  {1998})}\BibitemShut {NoStop}%
\bibitem [{\citenamefont {Moussa}(2016)}]{moussa2016transversal}%
  \BibitemOpen
  \bibfield  {author} {\bibinfo {author} {\bibfnamefont {J.~E.}\ \bibnamefont
  {Moussa}},\ }\bibfield  {title} {\bibinfo {title} {Transversal {Clifford}
  gates on folded surface codes},\ }\href
  {https://doi.org/10.1103/PhysRevA.94.042316} {\bibfield  {journal} {\bibinfo
  {journal} {Phys. Rev. A}\ }\textbf {\bibinfo {volume} {94}},\ \bibinfo
  {pages} {042316} (\bibinfo {year} {2016})}\BibitemShut {NoStop}%
\bibitem [{\citenamefont {Breuckmann}\ and\ \citenamefont
  {Burton}(2024)}]{Breuckmann2024foldtransversal}%
  \BibitemOpen
  \bibfield  {author} {\bibinfo {author} {\bibfnamefont {N.~P.}\ \bibnamefont
  {Breuckmann}}\ and\ \bibinfo {author} {\bibfnamefont {S.}~\bibnamefont
  {Burton}},\ }\bibfield  {title} {\bibinfo {title} {Fold-{T}ransversal
  {C}lifford {G}ates for {Q}uantum {C}odes},\ }\href
  {https://doi.org/10.22331/q-2024-06-13-1372} {\bibfield  {journal} {\bibinfo
  {journal} {{Quantum}}\ }\textbf {\bibinfo {volume} {8}},\ \bibinfo {pages}
  {1372} (\bibinfo {year} {2024})}\BibitemShut {NoStop}%
\bibitem [{\citenamefont {Quintavalle}\ \emph {et~al.}(2023)\citenamefont
  {Quintavalle}, \citenamefont {Webster},\ and\ \citenamefont
  {Vasmer}}]{Quintavalle2023partitioningqubits}%
  \BibitemOpen
  \bibfield  {author} {\bibinfo {author} {\bibfnamefont {A.~O.}\ \bibnamefont
  {Quintavalle}}, \bibinfo {author} {\bibfnamefont {P.}~\bibnamefont
  {Webster}},\ and\ \bibinfo {author} {\bibfnamefont {M.}~\bibnamefont
  {Vasmer}},\ }\bibfield  {title} {\bibinfo {title} {Partitioning qubits in
  hypergraph product codes to implement logical gates},\ }\href
  {https://doi.org/10.22331/q-2023-10-24-1153} {\bibfield  {journal} {\bibinfo
  {journal} {{Quantum}}\ }\textbf {\bibinfo {volume} {7}},\ \bibinfo {pages}
  {1153} (\bibinfo {year} {2023})}\BibitemShut {NoStop}%
\bibitem [{\citenamefont {Sayginel}\ \emph {et~al.}(2025)\citenamefont
  {Sayginel}, \citenamefont {Koutsioumpas}, \citenamefont {Webster},
  \citenamefont {Rajput},\ and\ \citenamefont {Browne}}]{autqec}%
  \BibitemOpen
  \bibfield  {author} {\bibinfo {author} {\bibfnamefont {H.}~\bibnamefont
  {Sayginel}}, \bibinfo {author} {\bibfnamefont {S.}~\bibnamefont
  {Koutsioumpas}}, \bibinfo {author} {\bibfnamefont {M.}~\bibnamefont
  {Webster}}, \bibinfo {author} {\bibfnamefont {A.}~\bibnamefont {Rajput}},\
  and\ \bibinfo {author} {\bibfnamefont {D.~E.}\ \bibnamefont {Browne}},\
  }\bibfield  {title} {\bibinfo {title} {Fault-tolerant logical {Clifford}
  gates from code automorphisms},\ }\href {https://doi.org/10.1103/vf7v-cpq9}
  {\bibfield  {journal} {\bibinfo  {journal} {PRX Quantum}\ }\textbf {\bibinfo
  {volume} {6}},\ \bibinfo {pages} {030343} (\bibinfo {year}
  {2025})}\BibitemShut {NoStop}%
\bibitem [{\citenamefont {Gottesman}\ and\ \citenamefont
  {Chuang}(1999)}]{gottesman1999demonstrating}%
  \BibitemOpen
  \bibfield  {author} {\bibinfo {author} {\bibfnamefont {D.}~\bibnamefont
  {Gottesman}}\ and\ \bibinfo {author} {\bibfnamefont {I.~L.}\ \bibnamefont
  {Chuang}},\ }\bibfield  {title} {\bibinfo {title} {Demonstrating the
  viability of universal quantum computation using teleportation and
  single-qubit operations},\ }\href {https://doi.org/10.1038/46503} {\bibfield
  {journal} {\bibinfo  {journal} {Nature}\ }\textbf {\bibinfo {volume} {402}},\
  \bibinfo {pages} {390} (\bibinfo {year} {1999})},\ \Eprint
  {https://arxiv.org/abs/quant-ph/9908010} {arXiv:quant-ph/9908010}
  \BibitemShut {NoStop}%
\bibitem [{\citenamefont {Bravyi}\ \emph {et~al.}(2016)\citenamefont {Bravyi},
  \citenamefont {Smith},\ and\ \citenamefont {Smolin}}]{bravyi2016trading}%
  \BibitemOpen
  \bibfield  {author} {\bibinfo {author} {\bibfnamefont {S.}~\bibnamefont
  {Bravyi}}, \bibinfo {author} {\bibfnamefont {G.}~\bibnamefont {Smith}},\ and\
  \bibinfo {author} {\bibfnamefont {J.~A.}\ \bibnamefont {Smolin}},\ }\bibfield
   {title} {\bibinfo {title} {Trading classical and quantum computational
  resources},\ }\href {https://doi.org/10.1103/PhysRevX.6.021043} {\bibfield
  {journal} {\bibinfo  {journal} {Physical Review X}\ }\textbf {\bibinfo
  {volume} {6}},\ \bibinfo {pages} {021043} (\bibinfo {year} {2016})},\ \Eprint
  {https://arxiv.org/abs/1506.01396} {arXiv:1506.01396} \BibitemShut {NoStop}%
\bibitem [{\citenamefont {Litinski}(2019{\natexlab{a}})}]{litinski2019game}%
  \BibitemOpen
  \bibfield  {author} {\bibinfo {author} {\bibfnamefont {D.}~\bibnamefont
  {Litinski}},\ }\bibfield  {title} {\bibinfo {title} {A {G}ame of {S}urface
  {C}odes: {L}arge-{S}cale {Q}uantum {C}omputing with {L}attice {S}urgery},\
  }\href {https://doi.org/10.22331/q-2019-03-05-128} {\bibfield  {journal}
  {\bibinfo  {journal} {Quantum}\ }\textbf {\bibinfo {volume} {3}},\ \bibinfo
  {pages} {128} (\bibinfo {year} {2019}{\natexlab{a}})},\ \Eprint
  {https://arxiv.org/abs/1808.02892} {arXiv:1808.02892} \BibitemShut {NoStop}%
\bibitem [{\citenamefont {Bravyi}\ and\ \citenamefont
  {Kitaev}(2005)}]{bravyi2005universal}%
  \BibitemOpen
  \bibfield  {author} {\bibinfo {author} {\bibfnamefont {S.}~\bibnamefont
  {Bravyi}}\ and\ \bibinfo {author} {\bibfnamefont {A.}~\bibnamefont
  {Kitaev}},\ }\bibfield  {title} {\bibinfo {title} {Universal quantum
  computation with ideal {C}lifford gates and noisy ancillas},\ }\href
  {https://doi.org/10.1103/PhysRevA.71.022316} {\bibfield  {journal} {\bibinfo
  {journal} {Physical Review A}\ }\textbf {\bibinfo {volume} {71}},\ \bibinfo
  {pages} {022316} (\bibinfo {year} {2005})},\ \Eprint
  {https://arxiv.org/abs/quant-ph/0403025} {arXiv:quant-ph/0403025}
  \BibitemShut {NoStop}%
\bibitem [{\citenamefont {Litinski}(2019{\natexlab{b}})}]{litinski2019magic}%
  \BibitemOpen
  \bibfield  {author} {\bibinfo {author} {\bibfnamefont {D.}~\bibnamefont
  {Litinski}},\ }\bibfield  {title} {\bibinfo {title} {Magic {S}tate
  {D}istillation: {N}ot as {C}ostly as {Y}ou {T}hink},\ }\href
  {https://doi.org/10.22331/q-2019-12-02-205} {\bibfield  {journal} {\bibinfo
  {journal} {Quantum}\ }\textbf {\bibinfo {volume} {3}},\ \bibinfo {pages}
  {205} (\bibinfo {year} {2019}{\natexlab{b}})},\ \Eprint
  {https://arxiv.org/abs/1905.06903} {arXiv:1905.06903} \BibitemShut {NoStop}%
\bibitem [{\citenamefont {Gidney}\ \emph {et~al.}(2024)\citenamefont {Gidney},
  \citenamefont {Shutty},\ and\ \citenamefont {Jones}}]{gidney2024magic}%
  \BibitemOpen
  \bibfield  {author} {\bibinfo {author} {\bibfnamefont {C.}~\bibnamefont
  {Gidney}}, \bibinfo {author} {\bibfnamefont {N.}~\bibnamefont {Shutty}},\
  and\ \bibinfo {author} {\bibfnamefont {C.}~\bibnamefont {Jones}},\ }\href
  {https://arxiv.org/abs/2409.17595} {\bibinfo {title} {Magic state
  cultivation: growing {T} states as cheap as {CNOT} gates}} (\bibinfo {year}
  {2024}),\ \Eprint {https://arxiv.org/abs/2409.17595} {arXiv:2409.17595}
  \BibitemShut {NoStop}%
\end{thebibliography}%

\clearpage
\onecolumngrid
\makeatletter
\def\@citex[#1][#2]#3{\global\let\@save@auxout\@auxout\let\@auxout\@auxoutSI\std@@citex[#1][#2]{#3}\let\@auxout\@save@auxout}
\def\@extra@b@citeb{@SI}
\renewcommand\NAT@wrout[5]{\if@filesw{\let\protect\noexpand\let~\relax\immediate\write\@auxout{\string\bibcite{#5\@extra@b@citeb}{{#1}{#2}{{#3}}{{#4}}}}}\fi\ignorespaces}
\makeatother
\setcounter{page}{1}
\thispagestyle{empty}
\renewcommand{\thepage}{\arabic{page}}
\setcounter{figure}{0}
\renewcommand{\thefigure}{S\arabic{figure}}
\setcounter{table}{0}
\renewcommand{\thetable}{S\arabic{table}}
\setcounter{equation}{0}
\renewcommand{\theequation}{S\arabic{equation}}
\renewcommand{\theHfigure}{S\arabic{figure}}
\renewcommand{\theHtable}{S\arabic{table}}
\renewcommand{\theHequation}{S\arabic{equation}}
\setcounter{section}{0}
\renewcommand{\thesection}{S\arabic{section}}
\renewcommand{\thesubsection}{\Alph{subsection}}
\setcounter{secnumdepth}{3}
\renewcommand{\thesubsubsection}{\roman{subsubsection}}

\begin{center}
\textbf{\large Supplementary Information: Low-Overhead Quantum Error Correction \\ with Boundary-Connected Planar Modules\\}
\vspace{12pt}
{Oscar Higgott$^{1,*}$, Hasan Sayginel$^{1,\dagger}$, Francisco J.H. Heras$^{2}$,
Zhiyang He$^{1,3}$, Tomas~Jochym-O'Connor$^{1}$,\\ Andrew W. Senior$^{2}$, Lei M. Zhang$^{2}$,
Thomas Edlich$^{2}$, James S. Spencer$^{2}$, Matt McEwen$^{1}$, Craig Gidney$^{1}$, \\
Johannes Bausch$^{2}$, Pushmeet Kohli$^{2}$, and Hartmut Neven$^{1}$}\\[6pt]
{\small $^{1}$\textit{Google Quantum AI} \\ $^{2}$\textit{Google DeepMind}\\
$^{3}$\textit{Department of Mathematics, Massachusetts Institute of Technology}}
\end{center}

\titlecontents{section}[0pt]{\addvspace{2pt}\small}{\thecontentslabel\quad}{}{\titlerule*[6pt]{.}\contentspage}
\titlecontents{subsection}[1.5em]{\footnotesize}{\thecontentslabel\quad}{}{\titlerule*[6pt]{.}\contentspage}
\titlecontents{subsubsection}[1.5em]{\footnotesize}{\thecontentslabel\quad}{}{\titlerule*[6pt]{.}\contentspage}

\startcontents[supplementary]
\printcontents[supplementary]{}{1}[2]
{%
    \noindent\textbf{CONTENTS}\par%
}
\bigskip

\vfill
\noindent\rule{0.4\columnwidth}{0.4pt}\\
{\footnotesize $^{*}$oscarhiggott@google.com \\ $^{\dagger}$hasansayginel@google.com}

\pagebreak
\section{Hyperbolic Tilings and Symmetry Groups}\label{sec:hyperbolic_tilings_intro}
We detail the construction of hyperbolic and semi-hyperbolic codes derived from closed hyperbolic tilings. Throughout this work, the symmetry groups of these tilings play a central role. Beyond code construction, they guide module partitioning, inter-module connectivity, and the implementation of fault-tolerant logical gates. 

A tiling of the hyperbolic plane $\mathbb{H}^2$ is a covering of the space by closed polygonal faces that intersect only along their edges. Due to the plane's constant negative curvature, the interior angles of a hyperbolic polygon are smaller than those of its Euclidean counterpart, permitting vertex configurations that cannot be realized on flat surfaces. Hyperbolic tilings can be generated from subgroups of the isometry group of $\mathbb{H}^2$. Such a group acts on the plane with a \emph{fundamental domain}, a closed region whose images under the group cover the plane and overlap only along their boundaries. The tiling is the orbit of this domain under the group, with each tile a distinct image of it, so that the tiles are in bijection with the group elements and the combinatorial structure of the tiling is determined by the structure of the group. 

The tilings we consider arise from \emph{Coxeter groups} acting on a space of constant negative curvature. Their generators are order-two reflections across the edges of a \emph{Coxeter polygon}. This polygon serves as a fundamental domain whose interior angles are of the form $\pi/n_i$ for integers $n_i\geq 2$. The geometry completely determines the group's relations, since composing reflections across two mirrors intersecting at angle $\pi/n_i$ yields a rotation of order $n_i$. In this work, we focus on two of the simplest cases of these structures: \emph{triangle groups} and \emph{quadrilateral groups}.

\subsection{Triangle Groups}\label{sec:triangle_groups}

A hyperbolic triangle group, denoted $\triangle(r,s,t)$, is generated by reflections $a,b,c$ in the three sides of a hyperbolic triangle with interior angles $\pi/r, \pi/s, \pi/t$ satisfying $1/r + 1/s + 1/t < 1$. Taking this triangle as the fundamental domain and repeatedly reflecting it across its edges produces an infinite tiling (see \Cref{fig:triangle_group}). In the resulting tiling, $2r$, $2s$ and $2t$ triangles meet at the respective vertices, and each triangle copy corresponds uniquely to a group element of $\triangle(r,s,t)$. The group presentation is given by: 
\begin{equation}\label{eq:triangle_group}
    \triangle(r,s,t) = \left\langle\, a, b, c \;\middle|\;
    a^2 = b^2 = c^2 = (ab)^r = (bc)^s = (ca)^t = e
    \,\right\rangle,
\end{equation}
where $a$, $b$, $c$ are reflections in the three edges of the triangle and $e$ is the identity element. Composing two adjacent reflections yields a rotation about their shared vertex. This defines the corner rotations $\rho = ab$, $\sigma = bc$, $\tau = ca$ by angles $2\pi/r$, $2\pi/s$, $2\pi/t$ respectively. The reflections $a$, $b$, $c$ are orientation-reversing and the rotations $\rho$, $\sigma$, $\tau$ are orientation-preserving, therefore $\triangle(r,s,t)$ contains both orientation-preserving and orientation-reversing isometries.  The orientation-preserving subgroup, denoted $\triangle^+(r,s,t)$, is an index-2 subgroup generated by the corner rotations. Since $\rho\sigma\tau = (ab)(bc)(ca) = e$, the third rotation is redundant, $\tau = (\rho\sigma)^{-1}$, and two generators suffice:
\begin{equation}
    \triangle^+(r,s,t) = \left\langle\, \rho, \sigma \;\middle|\;
    \rho^r = \sigma^s = (\rho\sigma)^t= e
    \,\right\rangle.
\end{equation}
Geometrically, $\triangle^+(r,s,t)$ acts on a fundamental domain consisting of two reflected copies of the original triangle glued along a shared edge, so that each of its elements is a word of even length in $a, b, c$ (see \Cref{fig:orientation_preserving_poincare_cayley_graph_it58}).

\subsubsection{Wythoff Construction}

Uniform tilings of $\mathbb{H}^2$ can be generated from a triangle group by the \emph{Wythoff kaleidoscopic construction}. A uniform tiling can be described by its vertex configuration $p_1.p_2\dots p_k$, which specifies the sequence of $p_i$-sided polygons meeting in cyclic order around every vertex. Alternatively, for regular tilings composed of a single polygon type with $q$ regular $p$-gons meeting at every vertex, we can denote them more compactly by the Schl\"{a}fli symbol $\{p,q\}$.

\begin{figure}[h]
    \centering
    \includegraphics[width=0.70\linewidth]{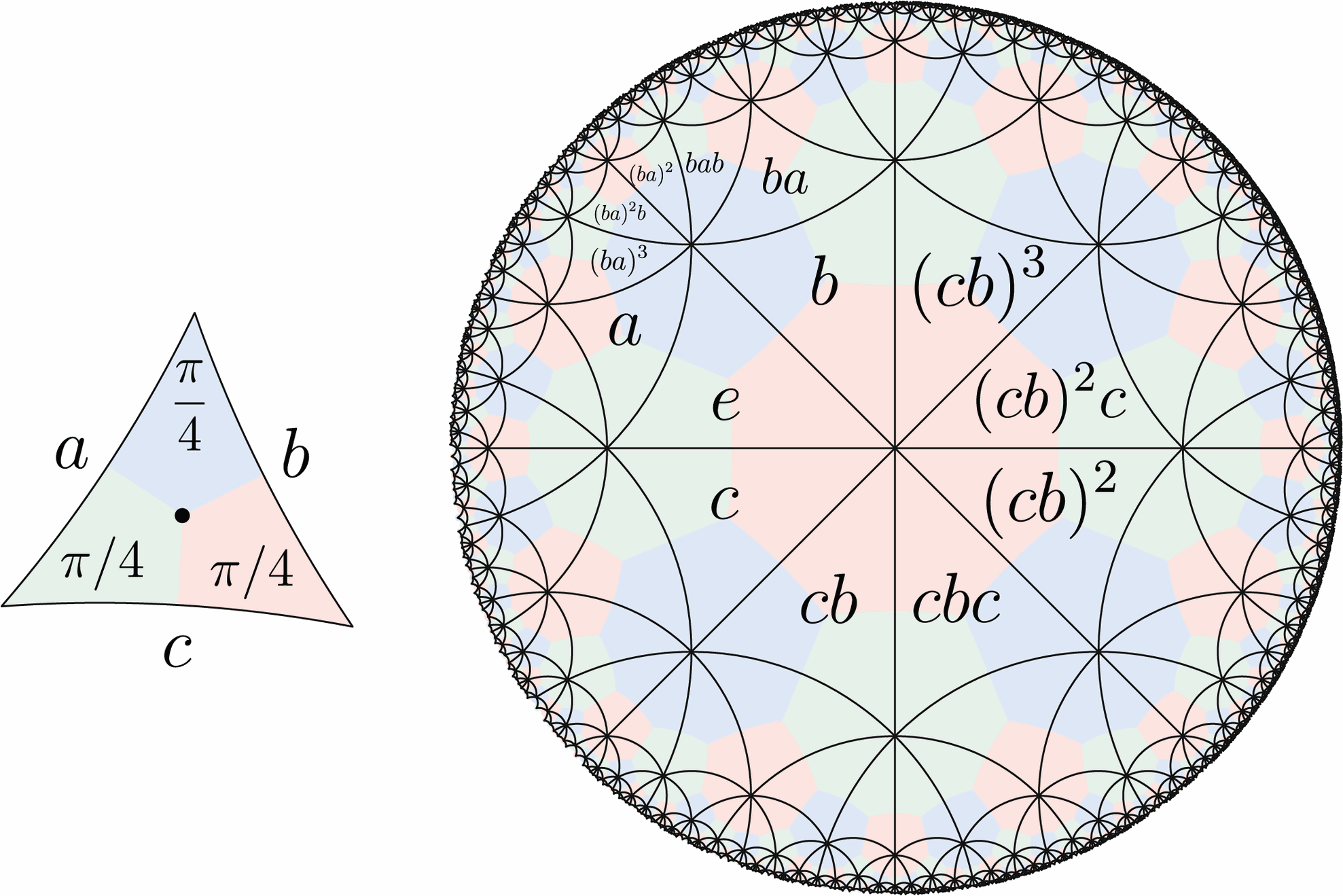}
    \caption{The $\triangle(4,4,4)$ triangle group and its associated three-colorable infinite $8.8.8$ (or $\{8,3\}$) hyperbolic tiling. \textbf{(Left)} The fundamental triangle with internal angles $\pi/4$, $\pi/4$, $\pi/4$ and generator point in the interior. \textbf{(Right)} The tiling generated by the orbit of the fundamental triangle under $\triangle(4,4,4)$, shown in the Poincar\'e disk. Each triangle corresponds to a unique group element, labeled by a word in the generators $a$, $b$, $c$, with the identity element $e$ marked.}
    \label{fig:triangle_group}
\end{figure}

The Wythoff construction proceeds by placing a vertex inside the fundamental triangle, called the \emph{generator point}, and drawing a perpendicular line from it to each triangle edge. The orbit of the generator point and its associated perpendicular lines under the group action of $\triangle(r,s,t)$ produces the vertex and edge sets of the tiling $\mathcal{T}=(V,E,F)$. The specific Wythoff tiling produced depends on the placement of the generator point relative to the triangle's edges, as summarized in \Cref{tab:wythoff}. When the point lies in the interior of the fundamental triangle, it rests on no mirrors. Consequently, every corner contributes its full ring of reflected copies, yielding faces with $2r, 2s,$ and $2t$ sides. This generates the omnitruncated tiling with a vertex configuration of $2r.2s.2t$, meaning polygons with $2r, 2s,$ and $2t$ sides meet at every vertex in cyclic order. \Cref{fig:triangle_group} illustrates this interior placement for the $\triangle(4,4,4)$ group, where $r=s=t=4$ yields the regular 8.8.8 (or $\{8,3\}$) tiling in which three octagons meet at every vertex. 

\begin{figure}[H]
    \centering
    \includegraphics[width=1.0\linewidth]{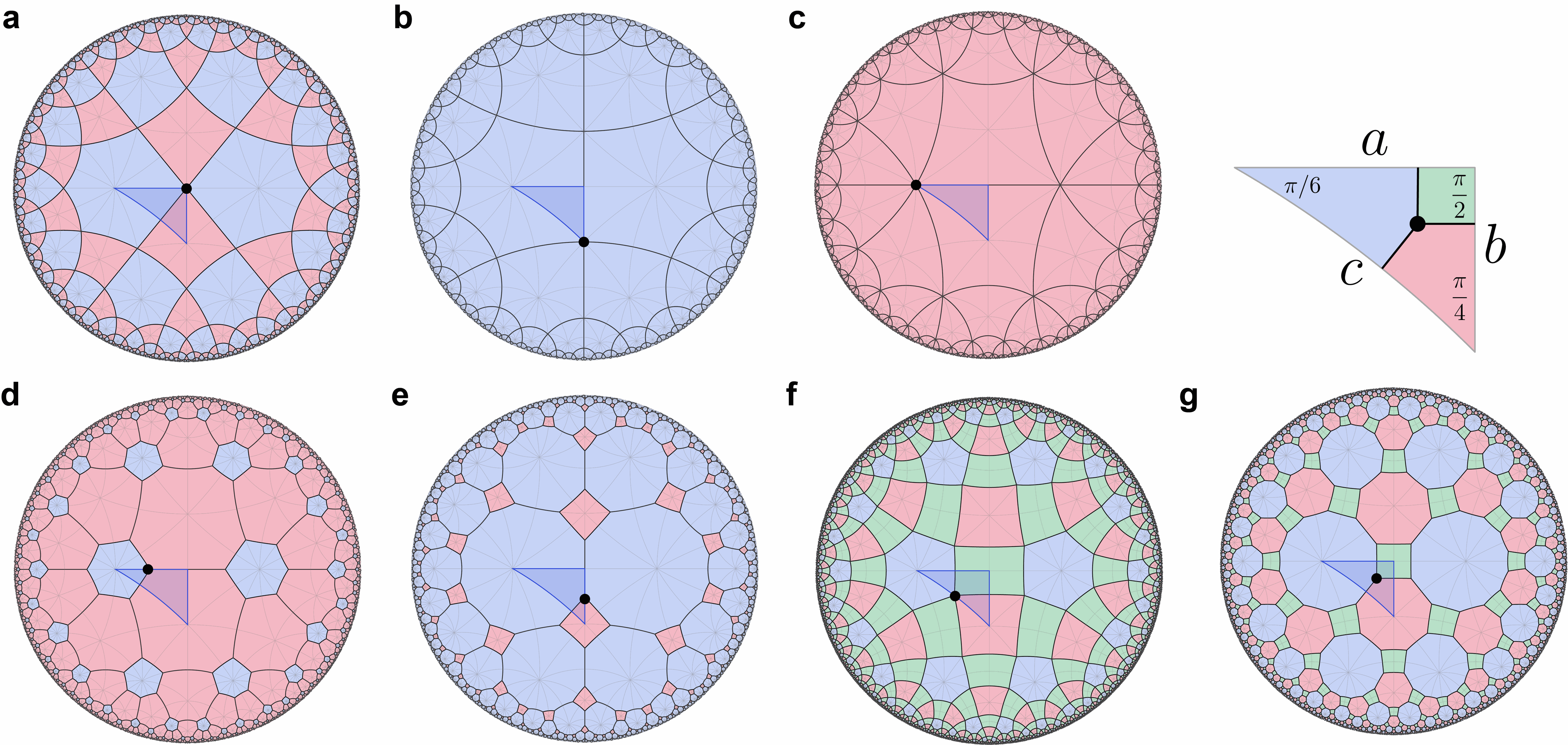}
    \caption{
    The seven uniform tilings generated from the triangle group $\triangle(2,4,6)$ by the Wythoff construction, one for each position of the generator point (black dot) in the fundamental triangle (blue highlight), following \Cref{tab:wythoff} with $(r,s,t) = (2,4,6)$. The three corner positions give the quasi-regular tiling \textbf{a} $4.6.4.6$ (corner $a \cap b$) and the regular tilings \textbf{b} $6.6.6.6$ (or $\{6,4\}$, corner $b \cap c$) and \textbf{c} $4.4.4.4.4.4$ (or $\{4,6\}$, corner $c \cap a$). The three mirror positions give the truncated tilings \textbf{d} $6.8.8$ (mirror $a$) and \textbf{e} $4.12.12$ (mirror $b$) and the cantellated tiling \textbf{f} $4.4.4.6$ (mirror $c$). The interior position gives the omnitruncated tiling \textbf{g} $4.8.12$. Since $r=2$ (i.e. the mirrors $a$ and $b$ intersect at a right angle $\pi/2$), reflections across these orthogonal mirrors generate digons which collapse into single edges resulting in the regular tilings in \textbf{b,c} and truncated tilings in \textbf{d,e}.}
    \label{fig:wythoff_246}
\end{figure}

Moving the generator point onto a mirror (triangle edge) or into a corner (triangle vertex) produces the other uniform tilings of the group. When the point lies on a single mirror, it remains fixed under the corresponding reflection. As a result, the polygonal rings around the two corners on that mirror are halved, while the ring at the opposite corner remains full. If the point sits precisely at a corner, it lies on two mirrors simultaneously: the rings at the other two corners are halved, and the occupied corner itself no longer generates a face. These seven possible generator positions and their corresponding tilings are detailed in \Cref{tab:wythoff}. As an illustrative example, \Cref{fig:wythoff_246} depicts the complete Wythoff family for the $\triangle(2,4,6)$ group.

\begin{table}[h]
\centering
\setlength{\tabcolsep}{15pt}
\renewcommand{\arraystretch}{1}
\begin{tabular}{lll}
\toprule
\textbf{Generator point} & \textbf{Vertex configuration} & \textbf{Type} \\
\midrule
corner $a \cap b$      & $(s.t)^{r}$   & \emph{regular} if $s=2$, $t=2$, or $s=t$; otherwise \emph{quasi-regular} \\
corner $b \cap c$      & $(t.r)^{s}$   & \emph{regular} if $t=2$, $r=2$, or $t=r$; otherwise \emph{quasi-regular} \\
corner $c \cap a$      & $(r.s)^{t}$   & \emph{regular} if $r=2$, $s=2$, or $r=s$; otherwise \emph{quasi-regular} \\
\midrule
mirror $a$             & $r.2s.t.2s$   & \emph{truncated} if $r=2$ or $t=2$; otherwise \emph{cantellated} \\
mirror $b$             & $r.2t.s.2t$   & \emph{truncated} if $r=2$ or $s=2$; otherwise \emph{cantellated} \\
mirror $c$             & $s.2r.t.2r$   & \emph{truncated} if $s=2$ or $t=2$; otherwise \emph{cantellated} \\
\midrule
interior               & $2r.2s.2t$    & \emph{omnitruncated} \\
\bottomrule
\end{tabular}
\caption{The seven positions of the generator point in the fundamental triangle of $\triangle(r,s,t)$ and the vertex configuration of the resulting uniform Wythoff tiling for a hyperbolic triangle group ($1/r + 1/s + 1/t < 1$). A configuration entry equal to $2$ denotes a degenerate digon (a two-sided face) that contributes an edge rather than a polygon, so that its two adjacent faces merge. A corner position gives a regular tiling when the two remaining orders are equal or at least one of them is $2$, and a quasi-regular tiling otherwise; a mirror position gives a cantellated tiling in general, or a truncated tiling when one of the orders on that mirror is $2$.}
\label{tab:wythoff}
\end{table}

\subsection{Quadrilateral Groups}\label{sec:quadrilateral_groups}
\begin{figure}[H]
    \centering
    \includegraphics[width=0.6\linewidth]{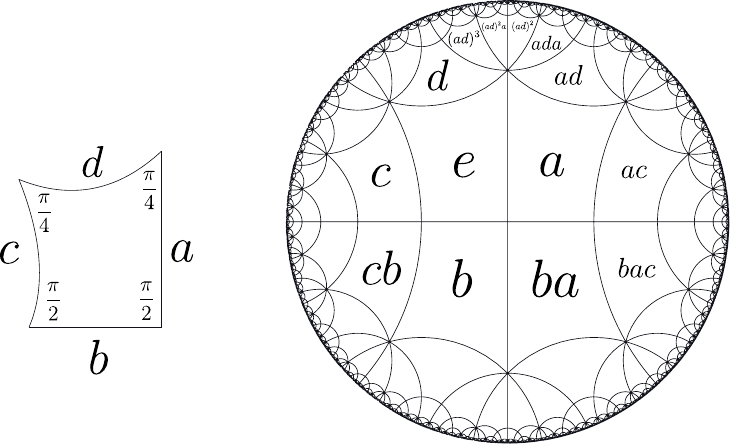}
    \caption{The $\square(2,2,4,4)$ quadrilateral group and its associated quadrilateral tiling of the hyperbolic plane. \textbf{(Left)} The fundamental quadrilateral with interior angles $\pi/2$, $\pi/2$, $\pi/4$, $\pi/4$ and reflection generators $a$, $b$, $c$, $d$ labeling its four sides. \textbf{(Right)} The tiling generated by the orbit of the fundamental quadrilateral under $\square(2,2,4,4)$, shown in the Poincar\'e disk. Each quadrilateral corresponds to a group element, labeled by a word in the generators $a$, $b$, $c$, $d$, with the identity element $e$ marked. The resulting tiling has face configuration $V4.4.8.8$, with vertex degrees $4$, $4$, $8$, $8$ at the four corners of each tile.}
    \label{fig:quadrilateral_group}
\end{figure}

A quadrilateral group is generated by reflections $a, b, c, d$ across the four sides of a quadrilateral with interior angles $\pi/r$, $\pi/s$, $\pi/t$, $\pi/u$. The group presentation is
\begin{equation}\label{eq:quad_group}
    \square(r,s,t,u) = \left\langle\, a, b, c, d \;\middle|\;
    a^2 = b^2 = c^2 = d^2 = (ab)^r = (bc)^s = (cd)^t = (da)^u = e
    \,\right\rangle,
\end{equation}
where $r, s, t, u$ are the orders of the corner rotations $\rho = ab$, $\sigma = bc$, $\tau = cd$, $\upsilon = da$ respectively, with $\rho\sigma\tau\upsilon = e$. The quadrilateral is hyperbolic when $1/r + 1/s + 1/t + 1/u < 2$. The orientation-preserving subgroup $\square^+(r,s,t,u)$ is an index-2 subgroup, with presentation
\begin{equation}\label{eq:orientable_quad_group}
    \square^+(r,s,t,u) = \left\langle\, \rho, \sigma, \tau \;\middle|\;
    \rho^r = \sigma^s = \tau^t = (\rho\sigma\tau)^u = e
    \,\right\rangle,
\end{equation}
where we have used $\upsilon = (\rho\sigma\tau)^{-1}$. Reflecting the fundamental quadrilateral in its four sides produces a tiling with face configuration $V2r.2s.2t.2u$, where every vertex degree is even (i.e.\ the vertex degrees of a quadrilateral tile are $2r$, $2s$, $2t$, $2u$). Each tile corresponds to a unique group element of $\square(r,s,t,u)$, labeled by a word in the generators $a, b, c, d$ (see \Cref{fig:quadrilateral_group}).

We note that quadrilateral tilings of the hyperbolic plane can also arise from triangle groups, as we discuss in \Cref{sec:quadrilateral_tilings}. The two families of groups are complementary: quadrilateral groups produce tilings whose vertex degrees are always even, including non-regular tilings such as $V4.4.8.8$ from $\square(2,2,4,4)$ (\Cref{fig:quadrilateral_group}), whereas triangle groups can produce only regular or quasi-regular (dual) quadrilateral tilings but can accommodate both even and odd vertex degrees. We also note that a Wythoff construction on the quadrilateral group is also possible, giving the other uniform tilings of the group, but we do not use it in this work.

\subsection{Dual Tilings}
\begin{figure}[H]
    \centering
    \includegraphics[width=0.95\linewidth]{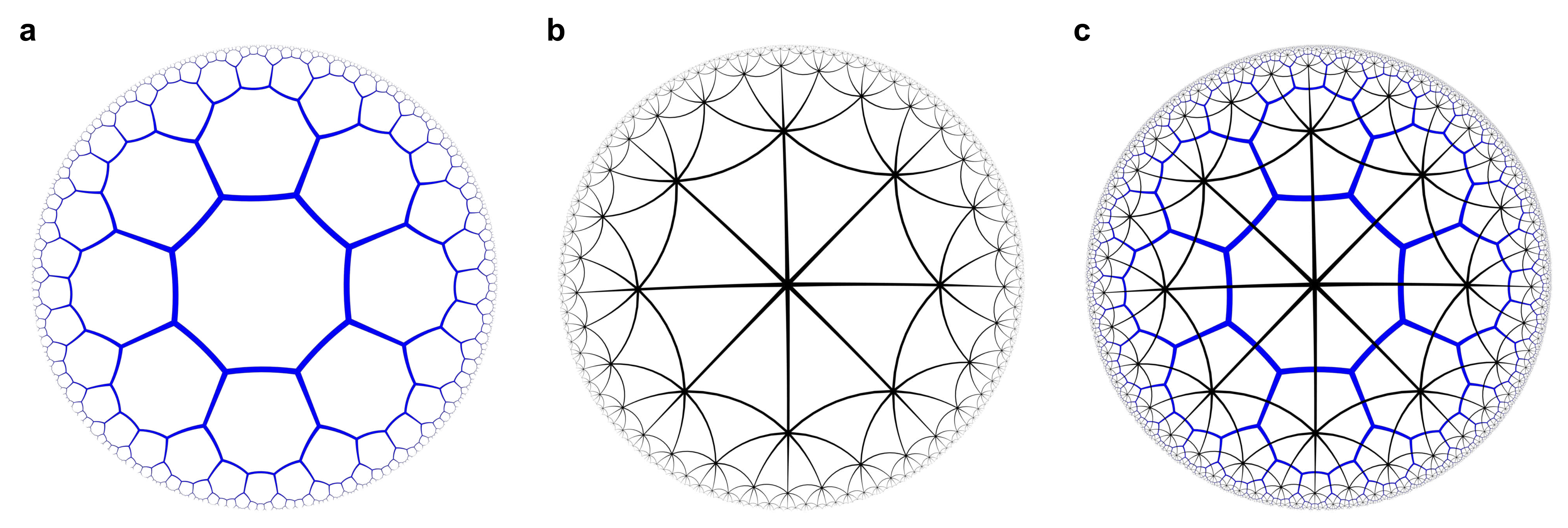}
    \caption{The $8.8.8$ (or $\{8,3\}$) tiling and its dual $V8.8.8$ (or $\{3,8\}$) tiling, shown in the Poincar\'e disk. The notation $8.8.8$ denotes the vertex configuration in which three octagons meet at each vertex; the prefix $V$ denotes the dual, whose vertices correspond to the faces of $8.8.8$. \textbf{a} The primal tiling $\mathcal{T} = 8.8.8$. \textbf{b} The dual tiling $\mathcal{T}^* = V8.8.8$. \textbf{c} $\mathcal{T}$ and $\mathcal{T}^*$ superimposed.}
    \label{fig:octagonal_tiling_and_dual}
\end{figure}

Every tiling $\mathcal{T} = (V, E, F)$ has a dual tiling $\mathcal{T}^*$, obtained by placing a vertex at the center of each face of $\mathcal{T}$ and connecting two such vertices whenever the corresponding faces share an edge. This exchanges the roles of vertices and faces: the faces of $\mathcal{T}$ become the vertices of $\mathcal{T}^*$, the vertices of $\mathcal{T}$ become the faces of $\mathcal{T}^*$, and each edge of $\mathcal{T}$ is crossed by exactly one dual edge of $\mathcal{T}^*$. Taking the dual twice recovers the original tiling, $(\mathcal{T}^*)^* = \mathcal{T}$. We sometimes use the face-configuration notation to name dual tilings: if $\mathcal{T}$ has vertex configuration $r.s.t$, we write $Vr.s.t$ for its dual $\mathcal{T}^*$. \Cref{fig:octagonal_tiling_and_dual} shows the primal $8.8.8$ (or $\{8,3\}$) tiling, in which three octagons meet at each vertex, together with its dual $V8.8.8$ (or $\{3,8\}$), in which eight triangles meet at each vertex.

\pagebreak
\subsection{Color Code Tilings}\label{sec:color_code_tilings}

To define a valid color code on a two-dimensional tiled surface, the tiling must satisfy two geometric conditions. The tiling must be \emph{trivalent} (three edges meeting at each vertex) and its faces must be 
\emph{three-colorable} such that no two faces that share an edge have the same color \cite{bombin_color}.  We refer to tilings that satisfy these specific geometric properties as \emph{color code tilings}.

To construct hyperbolic color codes, we systematically obtain valid color code tilings for hyperbolic surfaces using the Wythoff construction explained in \Cref{sec:triangle_groups}  \cite{quantum_pin_codes,higgott_2024_floquet}. As shown in \Cref{tab:wythoff}, the interior generator point yields a tiling with the vertex configuration $2r.2s.2t$. Since exactly three faces meet at every vertex, the resulting graph is trivalent. Moreover, the faces of the omnitruncated tiling correspond directly to the three distinct corners of the fundamental triangle. This natural tripartite division guarantees that the faces are three-colorable, making them color code tilings. \Cref{fig:triangle_group} illustrates an example of a color code tiling with octagon faces. The dual of the color tiling is the underlying triangular tiling which we denote $\mathcal{T^\triangle}$. The triangular tiles of $\mathcal{T^\triangle}$ are associated with group elements of $\triangle{(r,s,t)}$. Equivalently, the vertices and edges of the color code tiling $\mathcal{T}$ form the Cayley graph of $\triangle(r,s,t)$: the vertices are the group elements and they are joined by an edge whenever they are related by a generator $a$, $b$ or $c$.

\subsection{Quadrilateral Tilings} \label{sec:quadrilateral_tilings}
A \emph{quadrilateral tiling} $\mathcal{T}^\square$ is a tiling of quadrilateral faces. In this work, we use quadrilateral tilings to construct hyperbolic surface codes. Specifically, we define a hyperbolic surface code on the dual $\mathcal{T} = (\mathcal{T}^\square)^*$ of a quadrilateral tiling $\mathcal{T}^\square$ by associating stabilizer generators with the faces of $\mathcal{T}$. For $\mathcal{T}$ to support a valid surface code, its faces must be \emph{two-colorable}: each face is assigned one of two colors (e.g. red and blue) such that no two faces sharing an edge receive the same color. This ensures that $X$- and $Z$-type stabilizer generators can be assigned to faces of different color, so that stabilizers of opposite type with overlapping support always share an even number of data qubits and therefore commute. The faces of $\mathcal{T}$ are two-colorable whenever the vertices of $\mathcal{T}^\square$ are two-colorable, i.e.\ its edges form a bipartite graph. On the infinite plane this is always the case: every face of $\mathcal{T}^\square$ has even length, so every cycle has even length and the graph is bipartite. On the closed surfaces of \Cref{sec:closed_surfaces} this can fail since the tiling on the surface now has homologically non-trivial cycles, which may sometimes have odd length. Therefore, we check each closed quadrilateral tiling and keep only those that remain bipartite.

Quadrilateral tilings $\mathcal{T}^\square$ of the hyperbolic plane can arise from triangle groups, quadrilateral groups, or in many cases from both, and we construct codes using tilings from both symmetry groups.

The first class of tilings we consider are those that have quadrilateral reflection symmetry. These are generated by a quadrilateral group $\square(r,s,t,u)$ and correspond to tilings where every vertex degree is even, giving the face configuration $V2r.2s.2t.2u$. In this case, a single quadrilateral face is a valid Coxeter polygon and can be taken as the fundamental domain. The tiles are in bijection with the group elements of $\square(r,s,t,u)$, and adjacent tiles are related by right multiplication by a generator (see \Cref{fig:quadrilateral_group}).

The second class of tilings are regular quadrilateral tilings $\{4,q\}$ (or $Vq.q.q.q$), where $q$ quadrilaterals meet at every vertex and $q$ may be odd. When $q$ is odd (such as $\{4,5\}$), the quadrilateral cannot act as the fundamental domain for a quadrilateral reflection group. Nevertheless, $\{4,q\}$ tilings can be systematically constructed from triangle groups, such as from $\triangle(2,4,q)$ groups via the Wythoff construction by placing the generator point at the corner $c \cap a$ (see \Cref{fig:wythoff_246}c for the construction of $\{4,6\}$ from $\triangle(2,4,6)$). Algebraically, the quadrilateral tiles in this case correspond to the left cosets $\triangle(2,4,q)/D$ of the subgroup $D = \langle b,c \rangle \cong D_4$ of order $8$, with each quadrilateral formed by the union of the $8$ images of the fundamental triangle belonging to a coset $gD$ for group elements $g \in \triangle(2,4,q)$.

\begin{figure}[H]
\centering
\includegraphics[width=0.60\linewidth]{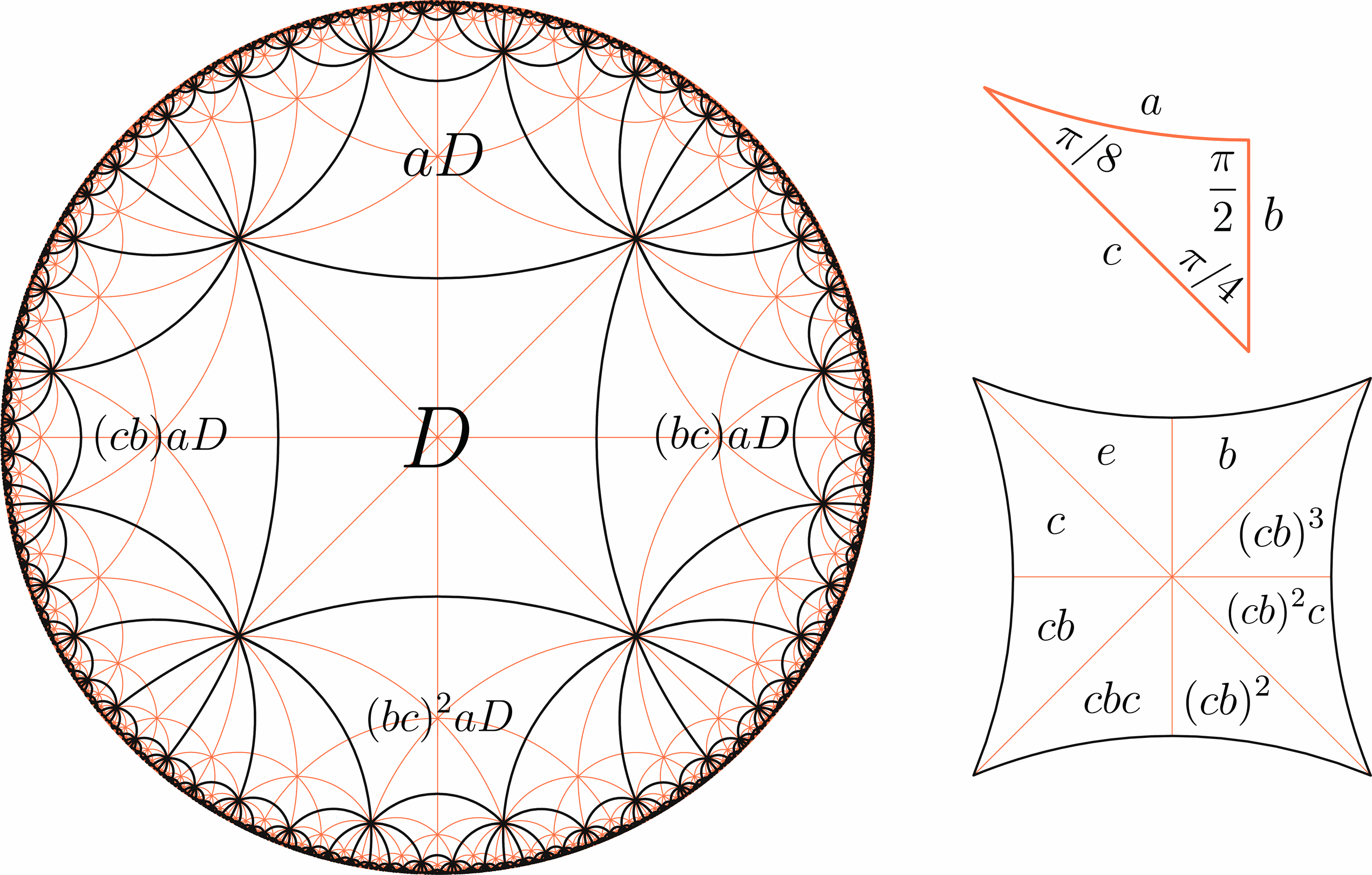}
\caption{The $\{4,8\}$ quadrilateral tiling of the hyperbolic plane and its coset decomposition under the triangle group $\triangle(2,4,8)$. \textbf{(Left)} The tiling generated on the Poincaré disk. Each quadrilateral tile (black lines) is formed by the union of $8$ triangles with interior angles $\pi/2, \pi/4, \pi/8$ (thin orange lines) and corresponds to a left coset $gD \in \triangle(2,4,8)/D$ of the dihedral subgroup $D = \langle b, c \rangle \cong D_4$. This tiling can equivalently be generated as the orbit of the fundamental domain of the quadrilateral group $\square(4,4,4,4)$, which is an index-$8$ subgroup of $\triangle(2,4,8)$. \textbf{(Right)} The fundamental triangle of $\triangle(2,4,8)$ and a quadrilateral tile (coset $eD = D$) with internal corner angles $\pi/4$. The center of the quadrilateral tile corresponds to the corner $b \cap c$ of the fundamental triangle (corner with angle $\pi/4$), the four outer boundaries correspond to mirror $a$, and the four vertices correspond to the corner $c \cap a$ with interior angle $\pi/8$.}
\label{fig:triangle_quad_48_tiling}
\end{figure}

Regular tilings with even vertex degrees support both algebraic descriptions simultaneously. For example, the $\{4,8\}$ tiling (\Cref{fig:triangle_quad_48_tiling}) can be generated either from the quadrilateral group $\square(4,4,4,4)$ or from the triangle group $\triangle(2,4,8)$. $\square(4,4,4,4)$ is an index-$8$ subgroup of $\triangle(2,4,8)$. \Cref{fig:triangle_quad_48_tiling} shows how the tiles are labeled by cosets. The fundamental triangle is the identity $e$, and every other triangle is an image of it under some $g \in \triangle(2,4,8)$. The $8$ images forming one quadrilateral are exactly those differing by an element of $D$, so they share a coset $gD$ and the tile is labeled by a chosen representative $g$, e.g. the central tile is $eD = D$. Each quadrilateral is thus both an image of the fundamental domain of $\square(4,4,4,4)$ and a coset of $D$ in $\triangle(2,4,8)$.

\subsection{Closed Hyperbolic Surfaces from Quotient Groups}\label{sec:closed_surfaces} 
The triangle and quadrilateral groups both tile the infinite hyperbolic plane, and closed hyperbolic surfaces can be obtained from their quotient groups. We describe the construction for a triangle group $\triangle(r,s,t)$ here. The quadrilateral case is identical, with $\square(r,s,t,u)$ in place of $\triangle(r,s,t)$ throughout.

To obtain a finite tiling on a closed surface we identify a finite-index normal subgroup $\Gamma \trianglelefteq \triangle(r,s,t)$ and form the quotient $\Sigma = \mathbb{H}^2/\Gamma$, which identifies points of $\mathbb{H}^2$ that lie in the same $\Gamma$-orbit. For $\Sigma$ to be a smooth closed surface, $\Gamma$ must act freely, that is without fixed points. Reflections fix their mirror lines and rotations fix their centers, so a free $\Gamma$ can contain no reflections or rotations. $\Gamma$ can consist only of the fixed-point-free isometries of $\mathbb{H}^2$ which are hyperbolic translations and glide reflections. The identifications imposed by such a $\Gamma$ fold the infinite tiling into a finite tiling on the compact surface $\Sigma$.

The quotient group $G(r,s,t) = \triangle(r,s,t)/\Gamma$ acts as the symmetry group of the tiled closed hyperbolic surface, with presentation
\begin{equation}
    G(r,s,t) = \left\langle\, a, b, c \;\middle|\;
    a^2 = b^2 = c^2 = (ab)^r = (bc)^s = (ca)^t = e,\;
    \Gamma\text{-relations}
    \,\right\rangle,
\end{equation}
where the $\Gamma$-relations are the additional identifications that render $G$ finite and correspond to generators of $\Gamma$. Each element of $G$ corresponds to a distinct copy of the fundamental triangle and the total number of triangles in the finite tiling equals the index $|G| = [\triangle(r,s,t):\Gamma]$. Adjacent triangles are related by right multiplication by a generator (see \Cref{fig:orientation_preserving_poincare_cayley_graph_it58}), so the adjacency structure of the tiling is the Cayley graph of $G$ with respect to $\{a, b, c\}$.

We use two complementary computational methods to enumerate quotient groups: 
\begin{enumerate}
    \item \textbf{Low-Index Normal Subgroups (LINS)}: The first approach is an exhaustive search based on Todd--Coxeter coset enumeration \cite{breuckmann_2016_constructions,higgott_2024_floquet}. We use the \texttt{LINS} package \cite{lins} in \texttt{GAP} \cite{GAP4} to systematically find all normal subgroups $\Gamma \trianglelefteq \triangle(r,s,t)$ up to a given index. This guarantees that we enumerate every finite tiling up to that size, but it must enumerate all normal subgroups to do so, and there are many of them, so reaching large index (large $|G|$) becomes prohibitively expensive.
    \item \textbf{Epimorphism Search}: To reach larger indices than the exhaustive search can, we instead search for epimorphisms onto a chosen target group. We fix a candidate finite group $H$ (such as a projective special linear group $\mathrm{PSL}(2,q)$) and search for a surjective homomorphism, or \emph{epimorphism}, $\phi : \triangle(r,s,t) \to H$. Its kernel $\Gamma = \ker(\phi)$ is a normal subgroup with quotient $G(r,s,t) = \triangle(r,s,t)/\Gamma \cong H$. By targeting a specific $H$ we obtain a large quotient directly, without enumerating subgroups up to its index.
\end{enumerate}

\subsubsection{Topology and Orientability of the Surface}

The topology of the closed surface $\Sigma = \mathbb{H}^2/\Gamma$ follows from the Gauss--Bonnet theorem. In the constant negative curvature of $\mathbb{H}^2$ each fundamental triangle has area $\pi(1 - 1/r - 1/s - 1/t)$, and $\Sigma$ is tiled by $|G|$ images of it, so
\begin{equation}\label{eq:gauss_bonnet}
    \mathrm{Area}(\Sigma) = |G|\,\pi\!\left(1 - \frac{1}{r} -
    \frac{1}{s} - \frac{1}{t}\right) = -2\pi\chi(\Sigma),
\end{equation}
where $\chi(\Sigma)$ is the Euler characteristic. Since $1/r + 1/s + 1/t < 1$ the area factor is positive, and $|\chi|$ grows linearly with $|G|$.

The surface $\Sigma$ is \textit{orientable} when $\Gamma$ preserves orientation, and non-orientable when it does not. Every isometry of $\mathbb{H}^2$ is either orientation-preserving (translations, rotations) or orientation-reversing (reflections, glide reflections), the latter being the odd-length words in $a,b,c$. Thus $\Sigma$ is orientable if and only if $\Gamma$ contains only even-length words, that is $\Gamma \le \triangle^+(r,s,t)$, the index-2 orientation-preserving subgroup. A free $\Gamma$ that meets this condition contains neither rotations (excluded by freeness) nor glide reflections (excluded by orientation), leaving only hyperbolic translations. Our enumeration produces both orientable and non-orientable quotients, and we construct hyperbolic color and surface codes on both. We report their parameters in \Cref{sec:promising_candidate_tables}.

\begin{figure}[h]
    \centering
    \includegraphics[width=0.90\linewidth]{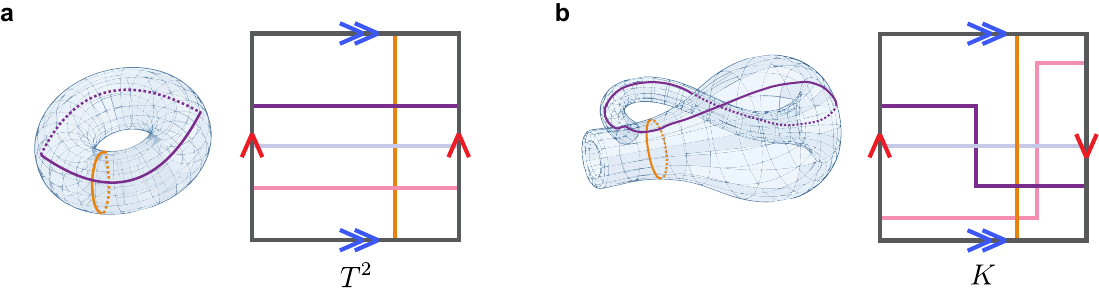}
    \caption{The torus $T^2$ and the Klein bottle $K$ as the simplest orientable and non-orientable Euclidean closed surfaces. Each is shown as a surface and as a fundamental polygon whose opposite edges are identified according to the arrows. \textbf{a}~The torus is orientable: both pairs of opposite edges are glued without a flip (all arrows on identified edges point in the same direction). The purple and orange cycles are homologically non-trivial and cross exactly once, forming a symplectic pair. On the fundamental polygon, the purple, lilac and pink lines are three distinct representatives of the same homology class. \textbf{b}~The Klein bottle is non-orientable: one pair of edges is glued with a flip (the red arrows point in opposite directions). On the fundamental polygon, the purple, lilac and pink lines are again representatives of the same homology class, but the orientation-reversing identification forces them to cross each other an odd number of times. Therefore, the non-trivial cycles do not arrange into symplectic pairs as they do on the torus.}
    \label{fig:klein_vs_torus}
\end{figure}

Orientable and non-orientable surfaces constructed from the same tiling look identical locally and differ only in how they close up. This affects the structure of the homologically non-trivial cycles of the surface, which, as we explain in \Cref{sec:logical_basis}, are associated with the logical operators of the code, so the difference carries over to the logical structure. The number of independent homologically non-trivial cycles is
\begin{equation}
    \dim H_1(\Sigma;\mathbb{F}_2) = 2 - \chi(\Sigma),
\end{equation}
which gives $2g$ cycles for an orientable surface of genus $g$ (where $\chi = 2-2g$) and $c$ cycles for a non-orientable surface with cross-cap number $c$ (where $\chi = 2-c$). \Cref{fig:klein_vs_torus} shows a torus and a Klein bottle as the smallest \emph{Euclidean} examples of an orientable and a non-orientable surface, each with its fundamental polygon. The distinction lies in how their cycles cross: on the orientable torus the non-trivial cycles form a \emph{symplectic pair}, crossing each other an odd number of times and any other representative of their own class an even number of times, whereas on the non-orientable Klein bottle a cycle can cross another representative of its own class an odd number of times, so no such pairing exists. The hyperbolic surfaces we use are the higher-genus and higher-cross-cap analogues of these two.

In this work, we mostly focus on the orientable surfaces, for which we build a canonical logical basis by finding symplectic pairs of homologically non-trivial cycles (\Cref{sec:logical_basis}). In the basis we design, non-trivial cycles are also related to one another by the orientation-preserving symmetries of the tiling, so that the corresponding logical operators can be mapped onto each other by the fault-tolerant walking circuits of \Cref{sec:automorphism_circuits} for logical measurement (\Cref{sec:ribbon_extractor}). The symplectic pairing of cycles and their symmetries are what makes this basis well-defined. 

The non-orientable surfaces also yield codes with favorable parameters, in some cases outperforming those of orientable surfaces of comparable size (see \Cref{sec:promising_candidate_tables}). We expect most of the methods developed in this work to apply to them directly, however we leave the finer details of a modular QEC architecture based on non-orientable surfaces to future work.

\subsubsection{Orientable Surfaces}
For an orientable $\Sigma$, the quotient group $G$ corresponds to a compact surface $\Sigma_g$ of genus $g \geq 2$ with $\chi = 2 - 2g$, and its orientation-preserving symmetries form the index-2 subgroup
\begin{equation}
    G^+(r,s,t) = \triangle^+(r,s,t)/\Gamma
    = \left\langle\, \rho, \sigma \;\middle|\;
    \rho^r = \sigma^s = (\rho\sigma)^t = e,\;
    \Gamma\text{-relations}
    \,\right\rangle,
\end{equation}
with $|G| = 2\,|G^+|$. Each element of $G^+$ corresponds to an image of the fundamental domain of $\triangle^+(r,s,t)$, that is, a pair of adjacent triangles glued along a shared edge forming a quadrilateral, and the adjacency of these quadrilaterals is given by the Cayley graph of $G^+$ with respect to $\{\rho, \sigma\}$ (see \Cref{fig:orientation_preserving_poincare_cayley_graph_it58}). Substituting $\chi = 2 - 2g$ and $|G| = 2|G^+|$ into \Cref{eq:gauss_bonnet} gives the genus,
\begin{equation}\label{eq:genus}
    g = 1 + \frac{|G^+|}{2}\left(1 - \frac{1}{r} - \frac{1}{s} -
    \frac{1}{t}\right).
\end{equation}

\subsubsection{Double Covers}\label{sec:double_cover}

Double covers provide an additional method to construct closed surfaces, enlarging the pool of tilings found by the epimorphism search. A double cover of $\Sigma$ is a surface $\widetilde{\Sigma}$ with a two-to-one map $\widetilde{\Sigma} \to \Sigma$, with two points of $\widetilde{\Sigma}$ lying above each point of $\Sigma$. We use double covers to obtain orientable surfaces from non-orientable ones and to enforce the vertex two-colorability (bipartiteness) required for valid surface codes on closed quadrilateral tilings (see \Cref{sec:quadrilateral_tilings}), thereby enlarging the pool of candidate tilings.

\paragraph{Orientable double cover.}
Algebraically, the orientable double cover discards the orientation-reversing elements of $\Gamma$ and replaces $\Gamma$ with
\begin{equation}
    \tilde{\Gamma}\;\coloneqq\; \Gamma \cap \triangle^+(r,s,t),
\end{equation}
which retains only the orientation-preserving elements. Since $\tilde{\Gamma} \le \triangle^+(r,s,t)$ by construction, the quotient $\widetilde{\Sigma} = \mathbb{H}^2/\tilde{\Gamma}$ is orientable. When $\Sigma$ is non-orientable, $\tilde{\Gamma}$ has index $2$ in $\Gamma$ (since $[\triangle : \triangle^+] = 2$), so it imposes half as many identifications and yields an orientable surface with twice as many tiles, $|\tilde{G}| = 2|G|$. 
When $\Sigma$ is already orientable, $\Gamma \le \triangle^+$ and the construction returns $\tilde{\Gamma} = \Gamma$ unchanged.

Computationally, we obtain $\tilde{\Gamma}$ by constructing the group $\tilde{G} \le G \times \mathbb{Z}_2$ generated by $\tilde{a} = (a, 1)$, $\tilde{b} = (b, 1)$, and $\tilde{c} = (c, 1)$, where $1$ is the non-identity element of $\mathbb{Z}_2$. The $\mathbb{Z}_2$ factor tracks word-length parity, ensuring that an element of $\triangle(r,s,t)$ maps to the identity in $\tilde{G}$ if and only if it belongs to $\Gamma$ and has even word length. Consequently, the kernel of the epimorphism $\triangle(r,s,t) \to \tilde{G}$ is precisely $\Gamma \cap \triangle^+(r,s,t)$.

\paragraph{Bipartite double cover.}
For regular quadrilateral tilings $\{4,q\}$ constructed from a triangle group $\triangle(2,4,q)$, the generators $a$ and $c$ both fix the vertices of the quadrilateral tiling, and only $b$ moves between vertices (see \Cref{fig:triangle_quad_48_tiling}). The vertex two-colorability required for a valid surface code (\Cref{sec:quadrilateral_tilings}) therefore depends on the parity of $b$ in each element of $\Gamma$: the closed tiling is bipartite if and only if every element of $\Gamma$ contains an even number of $b$ generators. When this condition fails, we can restore bipartiteness by taking a double cover that twists $b$ by the non-trivial element of $\mathbb{Z}_2$. Specifically, we construct $\tilde{G} \le G \times \mathbb{Z}_2$ generated by $\tilde{a} = (a, 0)$, $\tilde{b} = (b, 1)$, and $\tilde{c} = (c, 0)$, so that the $\mathbb{Z}_2$ factor tracks only the parity of $b$. We note that even when the base quadrilateral tiling is not bipartite, the fine-grained semi-hyperbolic tilings that we introduce later in \Cref{sec:semi_hyperbolic_surface_codes} can still be bipartite if $L$ is even, in which case valid semi-hyperbolic surface codes can be defined without taking a double cover.

We apply these double cover constructions to the quotients found by our searches, keeping both the original tilings and the covers as code tiling candidates.

\subsection{Left and Right Multiplication in Hyperbolic Symmetry Groups}\label{sec:left_right_mult}

\begin{figure}[h]
    \centering
    \includegraphics[width=0.95\linewidth]{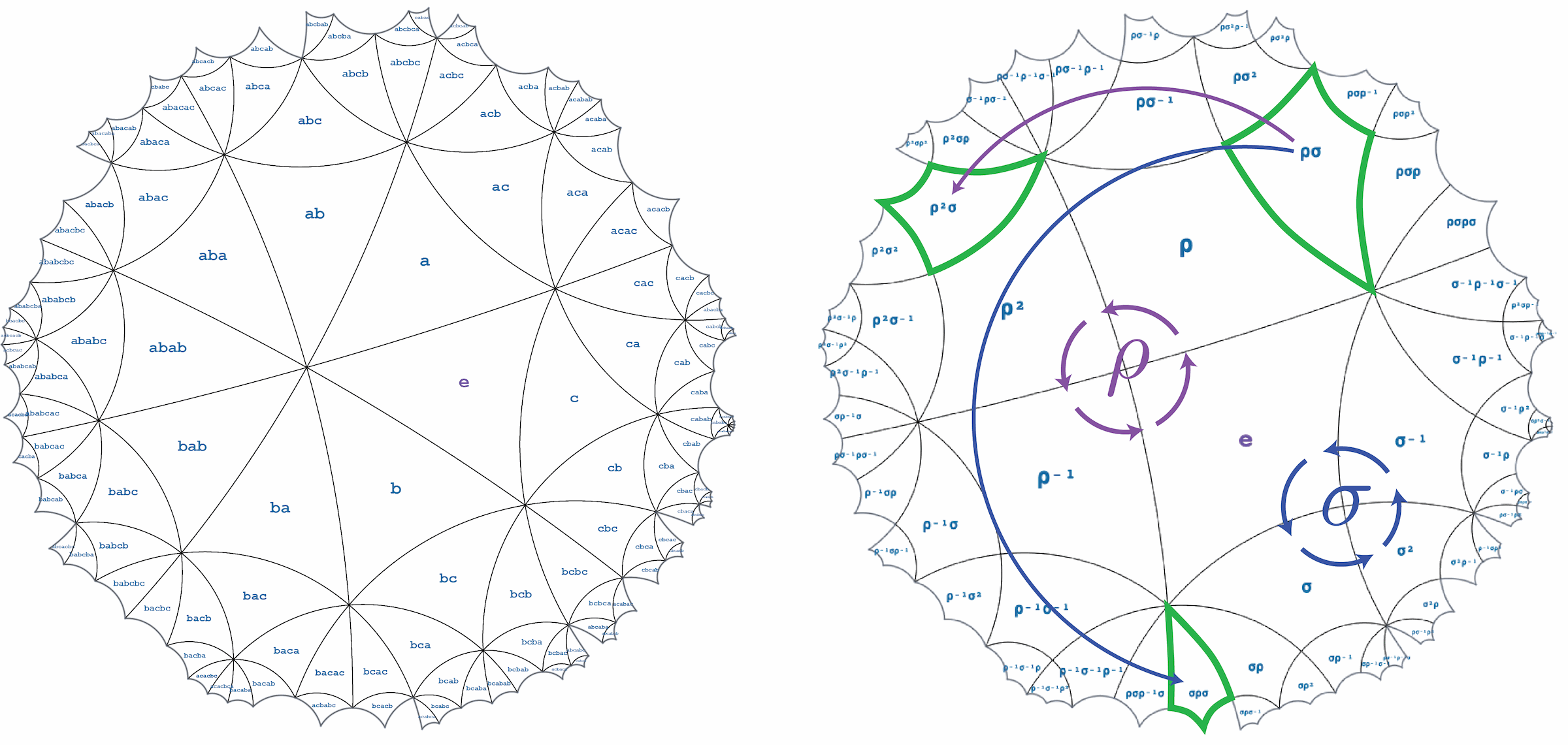}
    \caption{Example of a quotient triangle group $\triangle(4,4,4)/\Gamma$ and its orientation-preserving subgroup. The full quotient triangle group has order 128 and defines a hyperbolic color code with parameters [[128,36,8]]. \textbf{(Left)} The full quotient group defines a V8.8.8 tiling of a closed hyperbolic surface. Each group element is associated with a distinct triangle. Right multiplication by a generator ($a$, $b$ or $c$) corresponds to a reflection across a local mirror to an adjacent triangle, defining the local connectivity. Left multiplication corresponds to a global isometry (reflection, rotation, or translation) of the entire tiled surface. \textbf{(Right)} The orientation-preserving subgroup $\triangle^+(4,4,4)/\Gamma$. Its fundamental domains are quadrilaterals, each formed by joining a pair of adjacent triangles (related by a $b$ reflection). The subgroup generators $\rho\coloneqq ab$ and $\sigma \coloneqq bc$ are corner rotations. Left multiplication by each of these subgroup generators acts as a global isometry (as illustrated by the arrows connecting quadrilaterals), transforming the entire tiled surface while preserving its structure. Right multiplication instead defines the local adjacency between quadrilaterals.}
 \label{fig:orientation_preserving_poincare_cayley_graph_it58}
\end{figure}

As explained in \Cref{sec:triangle_groups}, when the hyperbolic surface is tiled by a fundamental domain of a  triangle group, we obtain a tiling with face configuration $V2r.2s.2t$ (referred to as the dual tiling in \Cref{fig:octagonal_tiling_and_dual}b).
As illustrated in \Cref{fig:orientation_preserving_poincare_cayley_graph_it58}, the fundamental triangle is labeled by the identity element $e$.
Any other triangle in the tiling is reached by applying a specific sequence of reflections, corresponding to a unique group element $g\in G$.

The algebraic structure of the quotient group $G=\triangle(r,s,t)/\Gamma$ and its orientation-preserving subgroup $G^+$ defines both the local geometric structure and global symmetries of the corresponding hyperbolic tilings.
This dual role can be understood through the distinction between left and right group multiplication.

Left multiplication defines global symmetries of the tiling.
As per convention, the group acts on the tiled surface via left multiplication.
Consider the operation that maps every tile $g$ in the tiling to $hg$, where $h\in G$ is a fixed group element.
Geometrically, this left multiplication applies a global isometry (a translation, rotation, or reflection) to the entire tiled surface, mapping the tiling onto itself.

Right multiplication defines local adjacency, corresponding to a local move to an adjacent tile.
Considering a specific triangular tile $g$ as an example, its immediate neighbors are the triangles that share its edges.
The reflections across the three edges of the fundamental triangle $e$ are the generators $a$, $b$ and $c$.
The reflections across the corresponding edges of the displaced triangle $g$ are given instead by the conjugate operators $gag^{-1}$, $gbg^{-1}$ and $gcg^{-1}$.
To see why, consider the action of reflecting across an edge of $g$: one first maps $g$ back to the identity $e$ (via $g^{-1}$), performs the reflection (e.g., $a$), and then maps back to $g$ (via $g$). 
Applying one of these local reflections to the tile $g$ gives its neighbor $(gag^{-1})g=ga$.
Therefore, right multiplication by a generator corresponds to crossing a local mirror (tile boundary) to an adjacent tile.

This geometric interpretation of left and right multiplication generalizes in the expected way to the orientation-preserving subgroup $G^{+}=\triangle^+(r,s,t)/\Gamma$, shown in the right panel of \Cref{fig:orientation_preserving_poincare_cayley_graph_it58}.
Here, the fundamental domain is no longer a single triangle but a quadrilateral formed by joining a pair of adjacent triangles (specifically the identity triangle $e$ and its neighbor $b$, related by a reflection).
Each quadrilateral in this new tiling is identified with an element $p\in G^{+}$, representing a pair of triangles $p$ and $pb$.
Left multiplication by elements of $G^{+}$ defines the global, orientation-preserving isometries of the closed surface.
For the two rotation generators $\rho\coloneqq ab$ and $\sigma\coloneqq bc$, multiplying on the left corresponds to a global rotation (with $\rho$ and $\sigma$ each centered on different vertices), while right multiplication by the same generators corresponds to a local move to an adjacent tile.

In the context of modular hyperbolic codes, as discussed later in this supplement, we will see that the global symmetries define automorphisms of the code used for implementing logic (and the connectivity required for this) whereas right multiplication defines the inter-module connectivity required to implement syndrome extraction circuits.

\section{Hyperbolic and Semi-Hyperbolic Code Constructions}\label{sec:code_constructions}
In this section we explain how we construct hyperbolic codes from tilings of closed hyperbolic surfaces, and how we obtain semi-hyperbolic codes by fine-graining the base hyperbolic tiling with a Euclidean tiling.

\subsection{Hyperbolic and Semi-Hyperbolic Color Codes}\label{sec:semi_hyperbolic_color_codes}

Color codes are defined on color code tilings, with data qubits associated with vertices and both an $X$- and a $Z$-type stabilizer generator supported on each face~\cite{bombin_color}. We start with a color code tiling $\mathcal{T} = (V, E, F)$ on a closed hyperbolic surface $\Sigma$, whose dual is the base triangular tiling $\mathcal{T}^\triangle = (V^\triangle,E^\triangle,F^\triangle)$ associated with the symmetry group $G = \triangle(r,s,t)/\Gamma$. We write $|F^\triangle|$ for the number of base triangular tiles. Each triangle face in $\mathcal{T}^\triangle$ corresponds to a group element $g
\in G$, so the number of triangular tiles is equal to the order of the quotient group $|F^\triangle| = |G|$. The tiling $\mathcal{T}$ has vertex configuration $2r.2s.2t$ and defines a hyperbolic color code, which we label $2r.2s.2t$-$|F^\triangle|$. Since $\mathcal{T} = (\mathcal{T}^\triangle)^*$ is the dual of the base triangular tiling, its vertices, edges and faces are those of $\mathcal{T}^\triangle$ with the roles of vertices and faces exchanged, $|V| = |F^\triangle|$, $|E| = |E^\triangle|$ and $|F| = |V^\triangle|$. Using $|F^\triangle| = |G|$, that $\mathcal{T}^\triangle$ is a triangulation ($|E^\triangle| = \tfrac32 |F^\triangle|$), and that each vertex of $\mathcal{T}^\triangle$ has degree $2r$, $2s$ or $2t$,
\begin{equation}
    |V| = |F^\triangle|, \qquad
    |E| = \frac{3|F^\triangle|}{2}, \qquad
    |F| = \frac{|F^\triangle|}{2}\left(\frac1r + \frac1s + \frac1t\right).
\end{equation}
The dimension of the first homology group then follows from the Euler characteristic $\chi = |V| - |E| + |F|$,
\begin{equation}\label{eq:color_dimH1}
    \dim H_1(\Sigma;\mathbb{F}_2) = 2 - \chi
    = 2 + \frac{|F^\triangle|}{2}\left(1 - \frac1r - \frac1s - \frac1t\right).
\end{equation}
The hyperbolic color code has $n = |V| = |F^\triangle|$ data qubits, one on each vertex of $\mathcal{T}$. Since each homologically non-trivial cycle supports independent logical operators for two of the three colors~\cite{bombin_color,bombin_delgado_toric_surface_color} (see \Cref{sec:logical_basis}), the number of logical qubits is
\begin{equation}\label{eq:color_k}
    k = 2\dim H_1(\Sigma;\mathbb{F}_2)
    = 4 + |F^\triangle|\left(1 - \frac1r - \frac1s - \frac1t\right).
\end{equation}
On an orientable surface of genus $g$ 
\Cref{eq:color_k} reduces to $k = 4g$. The code distance of hyperbolic color codes grows logarithmically with the number of data qubits, $d = O(\log n)$~\cite{delfosse_tradeoffs}.

\begin{figure}[H]
\centering
\includegraphics[width=0.85\textwidth]{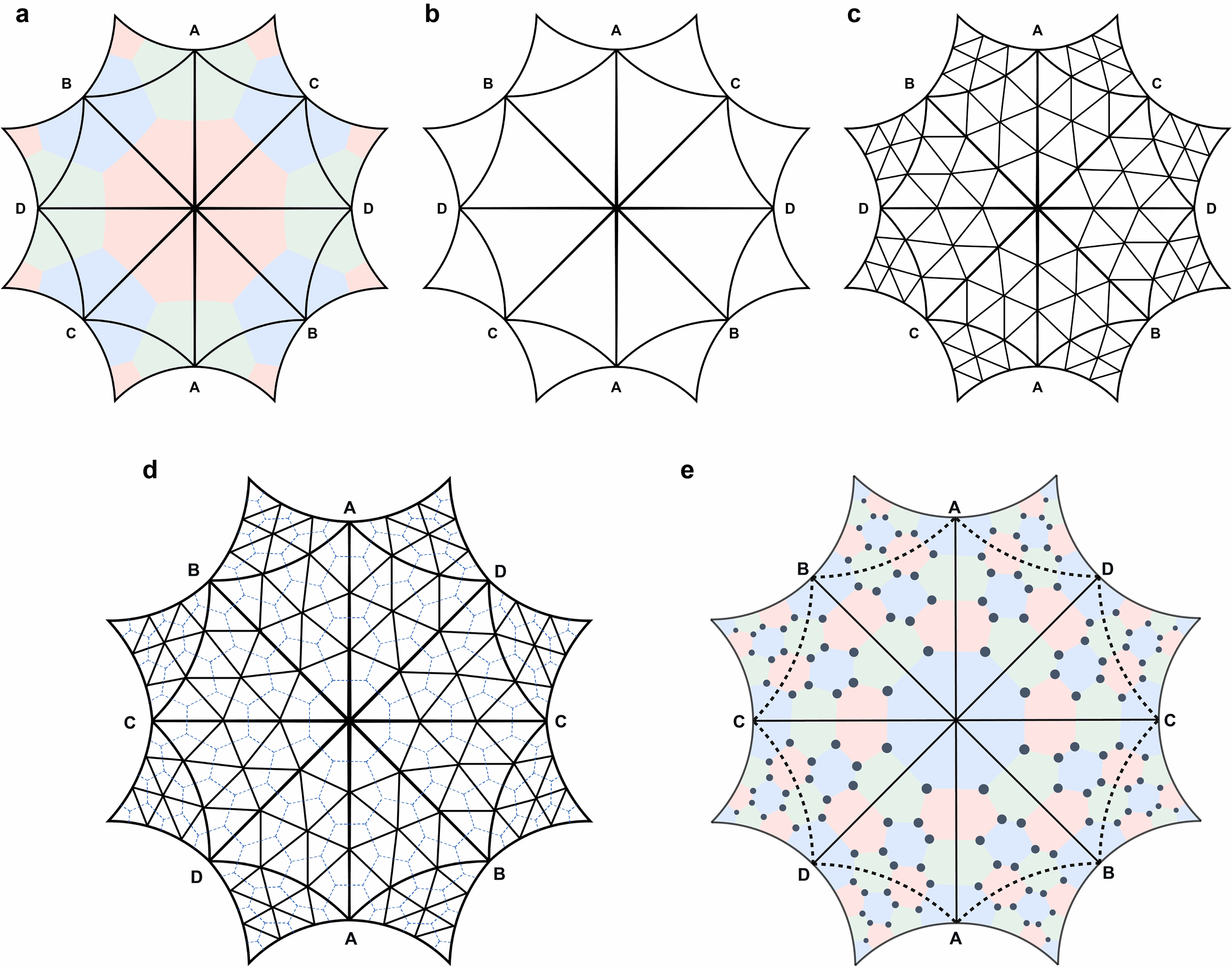}
\caption{Semi-hyperbolic color code construction in stages, shown for the Bolza surface, the closed genus-$2$ surface obtained from $\triangle(4,4,4)$ at index $|G| = 16$ by identifying opposite boundary edges. \textbf{a} The three-colorable tiling $\mathcal{T}$ on a closed hyperbolic surface, on which we define the hyperbolic color code denoted $8.8.8$-$16$ (equivalently $8.8.8$-$16$-$L1$). \textbf{b} The base triangular tiling $\mathcal{T}^\triangle$. \textbf{c} The fine-grained base triangular tiling $\mathcal{T}^\triangle_L$, shown for fine-graining parameter $L=3$. \textbf{d} The dual of the fine-grained tiling, $\mathcal{T}_L = (\mathcal{T}^\triangle_L)^*$. \textbf{e} The semi-hyperbolic $8.8.8$-$16$-$L3$ color code with parameters $[[144,8,8]]$ associated with $\mathcal{T}_L$.}
\label{fig:finegraining_color}
\end{figure}

We construct semi-hyperbolic color codes by fine-graining the hyperbolic tiling~\cite{higgott_2024_floquet}, illustrated in stages in \Cref{fig:finegraining_color}. We note that these semi-hyperbolic tilings were used previously to construct semi-hyperbolic Floquet codes~\cite{higgott_2024_floquet}, but have not previously been used to construct color codes. Following \cite{higgott_2024_floquet}, we start from the base triangular tiling $\mathcal{T}^\triangle$ (\Cref{fig:finegraining_color}b), the dual of the color code tiling $\mathcal{T}$ (\Cref{fig:finegraining_color}a). We subdivide every edge into $L \geq 1$ segments, where $L$ is the fine-graining parameter (\emph{side-length}), and fill each triangular tile with a Euclidean $\{3,6\}$ tiling, giving the fine-grained tiling $\mathcal{T}^\triangle_L$ (\Cref{fig:finegraining_color}c). Its dual gives the semi-hyperbolic color code tiling $\mathcal{T}_L = (\mathcal{T}^\triangle_L)^*$ (\Cref{fig:finegraining_color}d), on which we place the color code (\Cref{fig:finegraining_color}e). At $L = 1$ no fine-graining takes place and we recover the hyperbolic color code, while $L \geq 2$ gives a semi-hyperbolic tiling and the corresponding semi-hyperbolic color code. We label the resulting codes as $2r.2s.2t$-$|F^\triangle|$-$L$.

The faces of the semi-hyperbolic color code tiling $\mathcal{T}_L$ are of two types. Bulk faces lie in the interior of a fine-grained region and are hexagons, supporting weight-six stabilizers locally identical to a Euclidean planar $6.6.6$ color code. Corner faces, on the other hand, are inherited from the vertex configuration of the hyperbolic tiling $\mathcal{T}$ supporting stabilizers of weight $2r$, $2s$ or $2t$. The number of high-weight corner stabilizers is fixed by the underlying hyperbolic surface and does not grow with $L$.

Fine-graining does not change the topology of $\Sigma$, so $\dim H_1(\Sigma;\mathbb{F}_2)$ and hence $k$ are unchanged, while the number of physical data qubits grows quadratically in $L$. On the other hand, code distance scales roughly linearly with $L$. The numbers of vertices, edges and faces of the semi-hyperbolic color code tiling $\mathcal{T}_L$ are
\begin{align}
    |V_L| &= L^2 |F^\triangle|, \\
    |E_L| &= \frac{3L^2 |F^\triangle|}{2}, \\
    |F_L| &= \frac{|F^\triangle|}{2}\left(L^2 - 1 + \frac{1}{r} + \frac{1}{s} +
             \frac{1}{t}\right),
\end{align}
so the semi-hyperbolic color code has $n = |V_L| = L^2 |F^\triangle|$ data qubits~\cite{higgott_2024_floquet}. Therefore, the encoding rate of a $2r.2s.2t$-$|F^\triangle|$-$L$ code is given as
\begin{align}\label{eq:color_rate}
    \frac{k}{n} &= \frac{1}{L^2}\left(1 - \frac1r - \frac1s - \frac1t\right)
    + \frac{4}{L^2 |F^\triangle|} \\
    &\;\xrightarrow[|F^\triangle|\to\infty]{}\;
    \frac{1}{L^2}\left(1 - \frac1r - \frac1s - \frac1t\right). \nonumber
\end{align}
The bracket is always positive for a hyperbolic tiling with $1/r + 1/s + 1/t < 1$. Code distances scale as $d = O(L \log |F^\triangle|)$, so $L$ tunes a distance--rate trade-off: increasing $L$ raises the distance while $k$ is fixed by the topology and $n$ grows as $L^2$. Although fine-graining reduces the encoding rate quadratically, it boosts the distance, therefore these codes can retain a constant-factor density advantage in $kd^2/n$ over the surface code. We report this figure of merit for the resulting code families in \Cref{sec:promising_candidate_tables}. We also note that we developed efficient and accurate methods to compute code distances as explained in \Cref{sec:distance_finding}.

As an example, the $8.8.8$-$16$-$L3$ code defined on the Bolza surface shown in \Cref{fig:finegraining_color}e, corresponds to a semi-hyperbolic color code with code parameters $[[144,8,8]]$.

\subsection{Hyperbolic and Semi-Hyperbolic Surface Codes}\label{sec:semi_hyperbolic_surface_codes}

Surface codes can be defined on two-colorable tilings, with data qubits associated with vertices, $X$-type stabilizer generators supported on each face of one color (e.g.\ red faces) and $Z$-type stabilizer generators on each face of another color (e.g. blue faces). We start with a base quadrilateral tiling $\mathcal{T}^\square = (V^\square, E^\square, F^\square)$ on a closed hyperbolic surface $\Sigma$ and define the hyperbolic surface code on its dual $\mathcal{T} = (\mathcal{T}^\square)^*$. We write $|F^\square|$ for the number of base quadrilateral tiles. As explained in \Cref{sec:quadrilateral_tilings}, we use quadrilateral tilings with face configuration $V2r.2s.2t.2u$ that arise from quadrilateral groups  or regular $\{4,q\}$ quadrilateral tilings from a triangle group. The underlying symmetry group determines how the number of quadrilateral tiles $|F^\square|$ relates to the quotient order $|G|$. In either case $|F^\square|$ is the quantity associated with the number of physical data qubits. For generality, let us use $Va.b.c.d$ for the face configuration of an arbitrary quadrilateral tiling $\mathcal{T}^\square$ in this section, where $a,b,c,d$ are the vertex degrees of a tile.  We label the hyperbolic surface code $Va.b.c.d$-$|F^\square|$. For the regular tilings $\{4,q\}$, we also use the shorthand $\{4,q\}$-$|F^\square|$.

Since $\mathcal{T} = (\mathcal{T}^\square)^*$ is the dual of the base quadrilateral tiling, its vertices, edges and faces are those of $\mathcal{T}^\square$ with the roles of vertices and faces exchanged, $|V| = |F^\square|$, $|E| = |E^\square|$ and $|F| = |V^\square|$. Using that $\mathcal{T}^\square$ has quadrilateral faces ($|E^\square| = 2|F^\square|$) and vertices of degree $a$, $b$, $c$ or $d$,
\begin{equation}
    |V| = |F^\square|, \qquad
    |E| = 2|F^\square|, \qquad
    |F| = |F^\square|\left(\frac1a + \frac1b + \frac1c + \frac1d\right).
\end{equation}
The dimension of the first homology group then follows from the Euler characteristic $\chi = |V| - |E| + |F|$,
\begin{equation}\label{eq:surface_dimH1}
    \dim H_1(\Sigma;\mathbb{F}_2) = 2 - \chi
    = 2 + |F^\square|\left(1 - \frac1a - \frac1b - \frac1c - \frac1d\right).
\end{equation}
The hyperbolic surface code has $n = |V| = |F^\square|$ data qubits, one on each vertex of $\mathcal{T}$, and encodes one logical qubit per independent homology class (see \Cref{sec:logical_basis}), so
\begin{equation}\label{eq:surface_k_general}
    k = \dim H_1(\Sigma;\mathbb{F}_2)
    = 2 + |F^\square|\left(1 - \frac1a - \frac1b - \frac1c - \frac1d\right).
\end{equation}
On an orientable surface of genus $g$, \Cref{eq:surface_k_general} reduces to $k = 2g$. As for the color codes, the distance grows logarithmically in the number of data qubits, $d = O(\log n)$ \cite{delfosse_tradeoffs}.

\begin{figure}[H]
    \centering
    \includegraphics[width=0.85\textwidth]{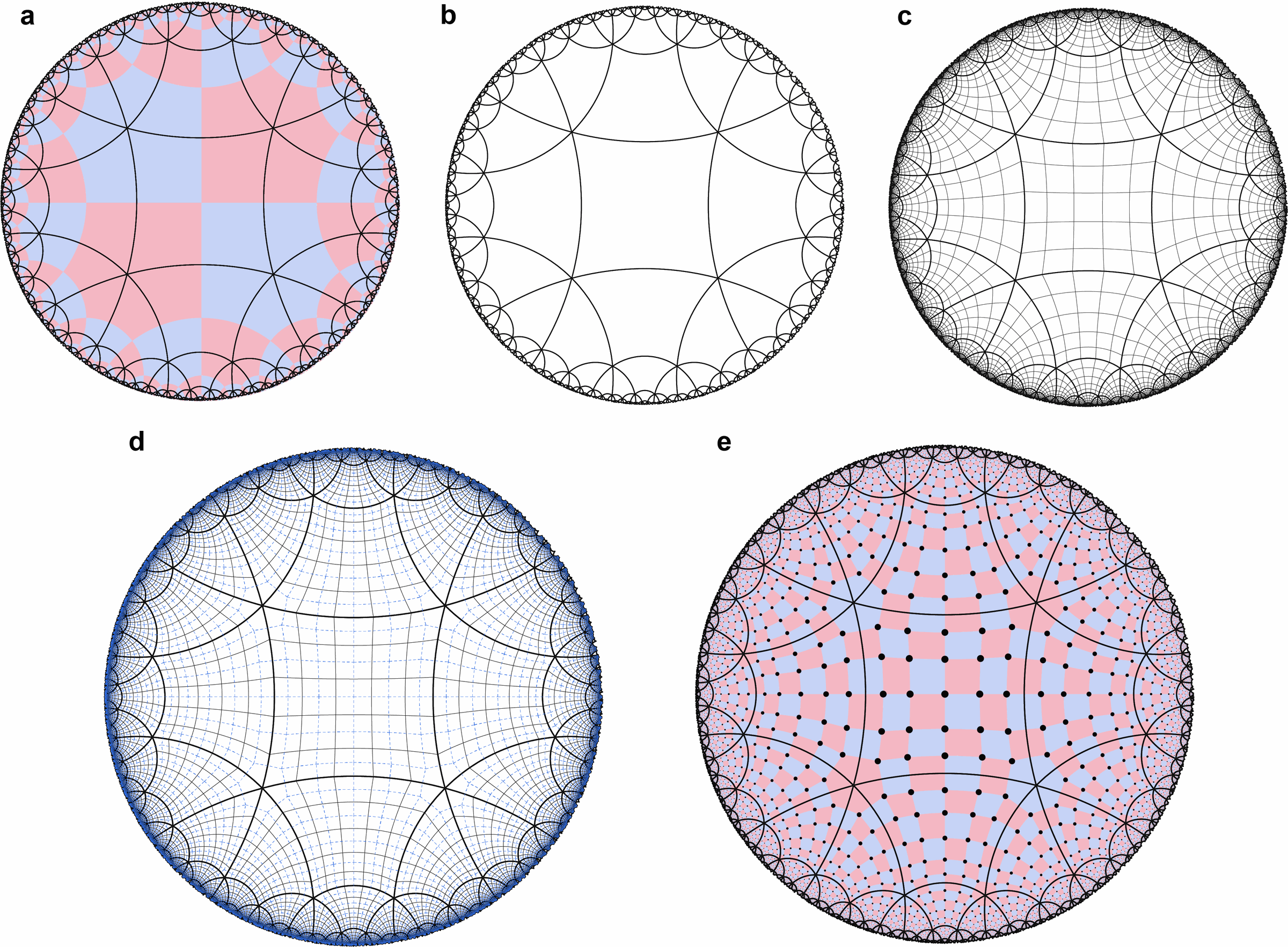}
    \caption{Semi-hyperbolic surface code construction in stages, shown for a base quadrilateral tiling $\mathcal{T}^\square = V6.6.6.6$ on a closed hyperbolic surface of genus $g$ with $|F^\square|$ quadrilateral tiles. \textbf{a} The two-colorable dual tiling $\mathcal{T} = (\mathcal{T}^\square)^*$ (the $6.6.6.6$ tiling) on the closed surface, on which we define the hyperbolic $V6.6.6.6$-$|F^\square|$-$L1$ surface code. \textbf{b} The base quadrilateral tiling $\mathcal{T}^\square$, whose dual is $\mathcal{T}$. \textbf{c} The fine-grained tiling $\mathcal{T}^\square_L$, each quadrilateral face replaced by an $L \times L$ Euclidean square tiling, shown for fine-graining parameter $L=5$. \textbf{d} The dual of the fine-grained tiling, $\mathcal{T}_L = (\mathcal{T}^\square_L)^*$. \textbf{e} The semi-hyperbolic $V6.6.6.6$-$|F^\square|$-$L5$ surface code (also denoted $\{4,6\}$-$|F^\square|$-$L5$), with a data qubit on each vertex, an $X$-stabilizer generator on each red face and a $Z$-stabilizer generator on each blue face.}
    \label{fig:finegraining_surface}
\end{figure}

We construct semi-hyperbolic surface codes by fine-graining the hyperbolic tiling as for the color codes, but with square tilings instead of triangular ones, illustrated in stages in \Cref{fig:finegraining_surface}. Our fine-graining procedure is similar to that of Ref.~\cite{Breuckmann_2017_semihyperbolic}, but the resulting codes differ: we place data qubits on the vertices of $(\mathcal{T}^\square_L)^*$, with the two stabilizer types on faces of different color, whereas Ref.~\cite{Breuckmann_2017_semihyperbolic} places data qubits on the edges of $\mathcal{T}^\square_L$, analogous to the distinction between the rotated and unrotated planar surface codes. With data qubits on the edges of $\mathcal{T}^\square_L$, every stabilizer of one type is a weight-four face, while all high-weight corner stabilizers (on vertices of degree greater than four) are of the other type, and the resulting $X$ and $Z$ distances generally differ (see Table 1 of Ref.~\cite{Breuckmann_2017_semihyperbolic}). In contrast, for odd $L$ our construction distributes the high-weight corner stabilizers between both $X$ and $Z$ types, and for regular tilings the $X$ and $Z$ stabilizers are related by a symmetry of the tiling (see \Cref{sec:asymmetry_X_Z}). Furthermore, with data qubits on the edges of $\mathcal{T}^\square_L$, those on the edges of the base quadrilaterals would lie on module boundaries, with no natural choice of module to assign them to, leading to non-uniform modules, whereas in our construction every data qubit lies strictly inside a base quadrilateral. We start from the base quadrilateral tiling $\mathcal{T}^\square$ (\Cref{fig:finegraining_surface}b), the dual of the surface code tiling $\mathcal{T}$ (\Cref{fig:finegraining_surface}a). We subdivide every edge into $L \geq 1$ segments, where $L$ is the fine-graining parameter, and fill each quadrilateral tile with a Euclidean square tiling, giving the fine-grained tiling $\mathcal{T}^\square_L$ (\Cref{fig:finegraining_surface}c). Its dual gives the semi-hyperbolic surface code tiling $\mathcal{T}_L = (\mathcal{T}^\square_L)^*$ (\Cref{fig:finegraining_surface}d), on which we place the surface code (\Cref{fig:finegraining_surface}e). At $L=1$ no fine-graining takes place and we recover the hyperbolic surface code, while $L \geq 2$ gives a semi-hyperbolic tiling and the corresponding semi-hyperbolic surface code. We label the resulting codes as $Va.b.c.d$-$|F^\square|$-$L$ (or $\{4,q\}$-$|F^\square|$-$L$ for regular tilings).

As in the color-code case, the faces of $(\mathcal{T}^\square_L)^*$ are of two types. Bulk faces are weight-four squares, locally identical to a Euclidean planar surface code, while the corner faces inherited from the vertices of $\mathcal{T}^\square$ carry higher-weight stabilizers of weight $a$, $b$, $c$ or $d$. The number of high-weight corner stabilizers is fixed by the underlying hyperbolic surface and does not grow with $L$.

Fine-graining does not change the topology of $\Sigma$, so $\dim H_1(\Sigma;\mathbb{F}_2)$ and hence $k$ are unchanged, while the number of data qubits grows quadratically in $L$. Since $(\mathcal{T}^\square_L)^*$ is four-valent, its counts are
\begin{align}
    |V_L| &= L^2 |F^\square|, \\
    |E_L| &= 2L^2 |F^\square|, \\
    |F_L| &= |F^\square|\left(L^2 - 1 + \frac1a + \frac1b + \frac1c + \frac1d\right),
\end{align}
so the semi-hyperbolic surface code has $n = |V_L| = L^2 |F^\square|$ data qubits. Dividing $k$ from \Cref{eq:surface_k_general} by $n$ gives the encoding rate
\begin{align}\label{eq:surface_rate_general}
    \frac{k}{n} &= \frac{1}{L^2}\left(1 - \frac1a - \frac1b - \frac1c - \frac1d\right)
    + \frac{2}{L^2 |F^\square|} \\
    &\;\xrightarrow[|F^\square|\to\infty]{}\;
    \frac{1}{L^2}\left(1 - \frac1a - \frac1b - \frac1c - \frac1d\right). \nonumber
\end{align}
The bracket is always positive for a hyperbolic tiling with $1/a + 1/b + 1/c + 1/d < 1$. The regular tilings $\{4,q\}$ are the special case with $a=b=c=d=q$, for which
\begin{equation}\label{eq:surface_rate_regular}
    k = 2 + |F^\square|\left(1 - \frac4q\right),
    \qquad
    \frac{k}{n} = \frac{1}{L^2}\left(1 - \frac4q\right) + \frac{2}{L^2 |F^\square|},
\end{equation}
and hyperbolic when $q > 4$. We explain how accurate code distances for hyperbolic surface codes can be computed efficiently in \Cref{sec:distance_finding}, and report the parameters of the resulting code families in \Cref{sec:promising_candidate_tables}.

A caveat concerns which fine-grained tilings define valid surface codes. The surface code requires the tiling $(\mathcal{T}^\square_L)^*$ to be two-colorable, which holds when $\mathcal{T}^\square_L$ is bipartite. A non-bipartite base tiling $\mathcal{T}^\square$ can become bipartite after fine-graining at certain values of $L$, so we check each fine-grained tiling directly and keep only those that are two-colorable. When the base tiling itself is not bipartite, it can also be made so by taking a bipartite double cover as described in \Cref{sec:double_cover}.

\section{Module Partitions and Ancilla Placement}\label{sec:module_partitions}
The hyperbolic code constructions of \Cref{sec:code_constructions} are defined on closed hyperbolic surfaces. In this section we describe how to map these codes onto a modular QEC architecture of boundary-connected planar chips. We present the available module partitions for color codes and surface codes. The module partitions and the inter-module connectivity follow from the algebraic structure of the symmetry groups of the tiling as we will see below. Modular implementations of semi-hyperbolic Floquet codes were discussed in Ref.~\cite{higgott_2024_floquet}, using modules similar to the triangle partition below, and simulated in Ref.~\cite{sutcliffe_2025_distributed}, using recursive spectral bisection of the code's connectivity graph, which does not in general produce identical modules or a layout with nearest-neighbor gates on a 2D grid. In contrast, each of our modules is a small grouping of base tiles that can be laid out on a 2D grid with nearest-neighbor couplers.

\subsection{Semi-Hyperbolic Color Code Modules}\label{sec:color_code_modules}

\begin{figure}[h]
    \centering
    \includegraphics[width=0.8\linewidth]{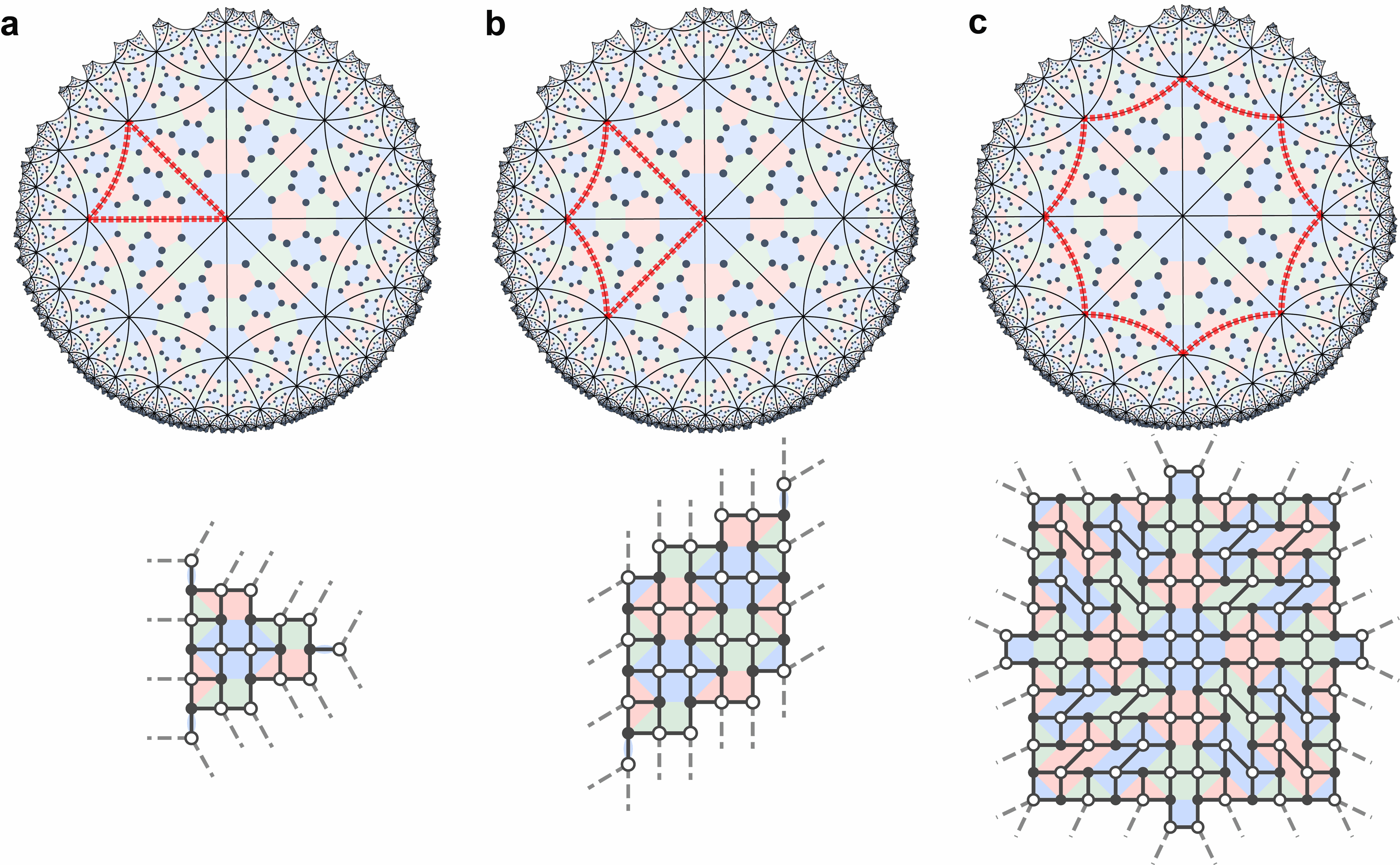}
    \caption{Module partition options for semi-hyperbolic color codes, illustrated on the $V8.8.8$ base triangular tiling from $G=\triangle(4,4,4)/\Gamma$ with $|G|=128$ (the same tiling shown with group element labels in \Cref{fig:orientation_preserving_poincare_cayley_graph_it58}). \textbf{Top row:} The base tiling shown in the Poincar\'e disk, with the red dashed lines tracing the boundaries of a single module. \textbf{Bottom row:} The corresponding fine-grained planar module (shown here for $L=3$), with data qubits (filled circles) and ancilla qubits (open circles). Solid black lines indicate in-module couplers whereas dashed gray lines indicate inter-module couplers. \textbf{a}~\textit{Triangle module} ($K{=}1$): each triangular tile forms its own module and is an image $gF$ of a chosen fundamental domain $F$ of~$G$, so the modules are in bijection with the elements of $G$. \textbf{b}~\textit{Quadrilateral module} ($K{=}2$): pairs of adjacent triangles are merged across an edge of the chosen edge color (e.g. here it is a blue edge), each an image of a fundamental domain of the orientation-preserving subgroup~$G^+$. \textbf{c}~\textit{Octagon module} ($K{=}8$): eight triangles meeting at a vertex are grouped together, each module corresponding to a coset.}
    \label{fig:color_code_modules}
\end{figure}

The semi-hyperbolic color codes are constructed from a base triangular tiling $\mathcal{T}^\triangle$, whose $|F^\triangle|=|G|$ tiles are in bijection with elements of the quotient triangle group $G = \triangle(r,s,t)/\Gamma$. A module partition groups the triangular tiles, and their associated data qubits and ancilla qubits into planar modules. \Cref{fig:color_code_modules} illustrates three color code module types and how they are derived from the base triangular tiling $\mathcal{T}^\triangle$. The top row shows the module partition on the base tiling in the Poincar\'e disk and the bottom row shows the corresponding fine-grained planar modules. We describe the different partitions with parameter $K$, which corresponds to the number of triangular tiles grouped into a single module. \Cref{fig:color_code_modules}\textbf{a}-\textbf{c} correspond to triangle ($K=1$), quadrilateral ($K=2$) and octagon ($K=8$) partitions respectively.

\subsubsection{Triangle Modules (\texorpdfstring{$K=1$}{K=1})}\label{sec:triangle_modules}
The simplest partition assigns each tile of $\mathcal{T}^\triangle$ to its own module, giving $|G|$ triangular modules (\Cref{fig:color_code_modules}{a}). Each tile is an image $gF$ of a chosen fundamental domain $F$ of the quotient group~$G$, so the modules are in bijection with the elements of $G$. After fine-graining with parameter~$L$, each triangular module contains $L^2$ data qubits and $N = 2L^2 + 2L$ total qubits including ancillas. This is the finest possible partition: it minimizes module size but maximizes the number of inter-module seams.

\subsubsection{Quadrilateral Modules (\texorpdfstring{$K=2$}{K=2})}\label{sec:quadrilateral_modules}
The quadrilateral partition we use is associated with the orientation-preserving subgroup $G^+=\triangle^+(r,s,t)/\Gamma$ and pairs adjacent triangles into quadrilateral modules (\Cref{fig:color_code_modules}b). The construction exploits the three-edge-coloring of the base triangular tiling $\mathcal{T}^\triangle$, where the three edge colors correspond to the three reflection generators $a$, $b$, $c$ of $\triangle(r,s,t)$, so that two adjacent tiles sharing a color-$m$ edge ($m\in \{a,b,c\}$) are related by right multiplication by the reflection $m$.

Choosing any one of the three colors, or equivalently one of the three reflections $m\in\{a,b,c\}$, and pairing each tile $g$ with its neighbor $gm$ across the corresponding edge, partitions the $|G|$ tiles into $|G|/2$ left cosets $gD = \{g, gm\}$ of the order-two subgroup $D = \langle m \rangle \le G$. 

\begin{figure}[h]
    \centering
    \includegraphics[width=0.6\linewidth]{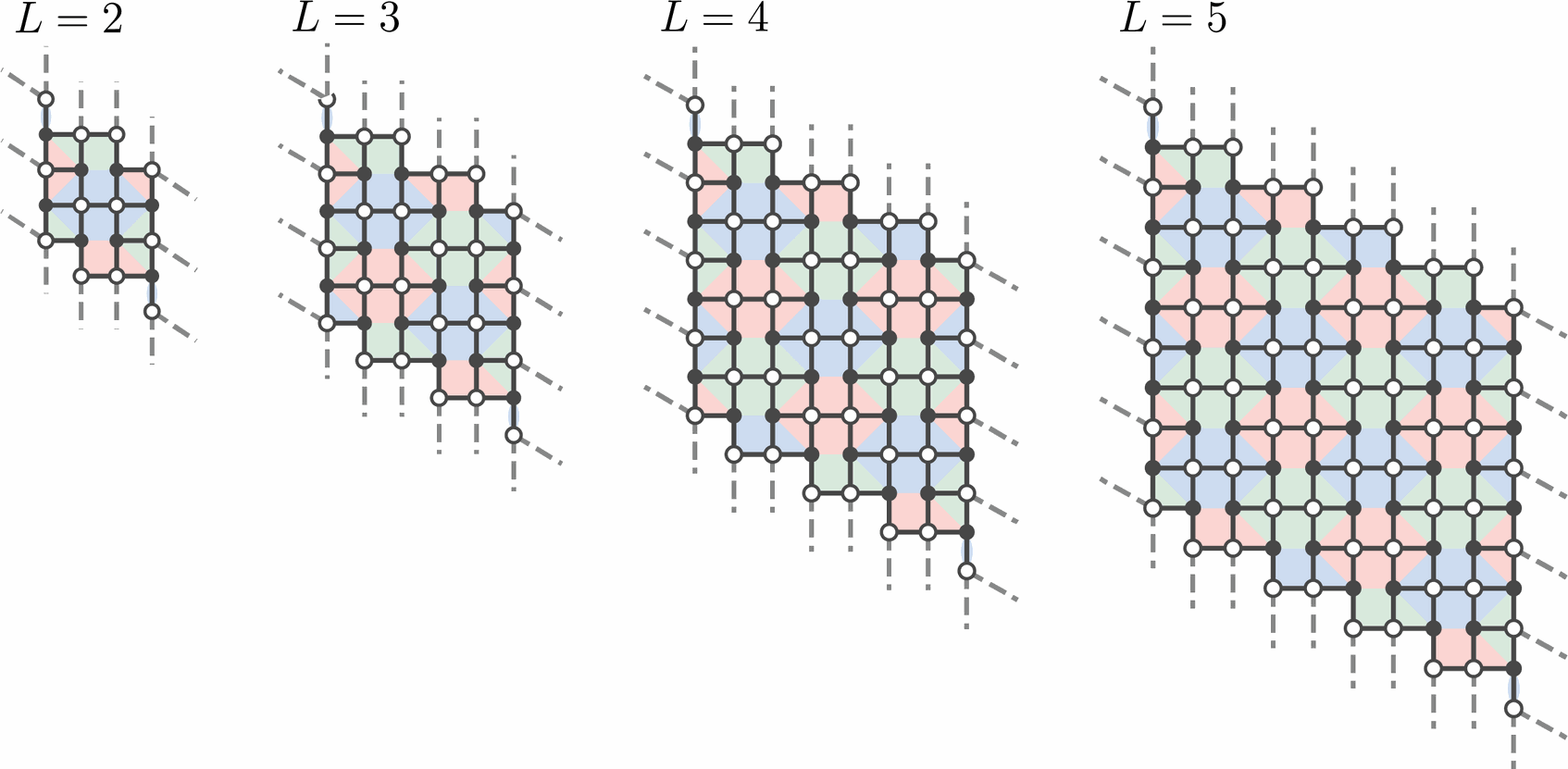}
    \caption{Quadrilateral (honeycomb) hyperbolic color code modules with increasing fine-graining parameter $L$.}
    \label{fig:color_code_modules_finegraining}
\end{figure}

Because $m$ reverses orientation, each coset $\{g, gm\}$ contains exactly one element $g \in G^+ = \langle \rho, \sigma \rangle$, so each module is an image of a fundamental domain of $G^+$ and can be labeled by its orientation-preserving element $g \in G^+$ (see \Cref{fig:orientation_preserving_poincare_cayley_graph_it58}). The resulting quadrilateral modules each contain $2L^2$ data qubits and $N=4L^2+2L$ total qubits. \Cref{fig:color_code_modules_finegraining} illustrates how a quadrilateral module grows with the fine-graining parameter $L$.

This partition is the one we use most, because the bijection between modules and elements of $G^+$ allows us to design fault-tolerant circuits with minimal additional inter-module connectivity in order to implement orientation-preserving symmetries of the tiling. As explained in \Cref{sec:left_right_mult}, neighboring modules  are reached by right multiplication by the generators $\rho$ and $\sigma$, which fixes the inter-module connectivity, while a global symmetry of the surface acts by left multiplication and permutes the modules as whole units. A global automorphism therefore maps each module onto another module rather than cutting across module boundaries, and this allows it to be realized as an automorphism circuit (\Cref{sec:automorphism_circuits}). We use these automorphism circuits to map between the logical operators of the code by choosing a suitable symmetrical canonical logical basis (\Cref{sec:logical_basis,sec:color_code_logical_basis}) and design a ribbon extractor for logical readout (\Cref{sec:ribbon_extractor}). We explain these constructions in the later sections. We can also choose to leave some of the paired triangles split, keeping them as two separate triangle modules rather than merging them into a quadrilateral. This introduces additional module boundaries where more inter-module seams can be attached, which is useful when we need extra connectivity for particular modules, for instance for logical readout using an extractor.

\begin{figure}[h]
    \centering
    \includegraphics[width=0.50\linewidth]{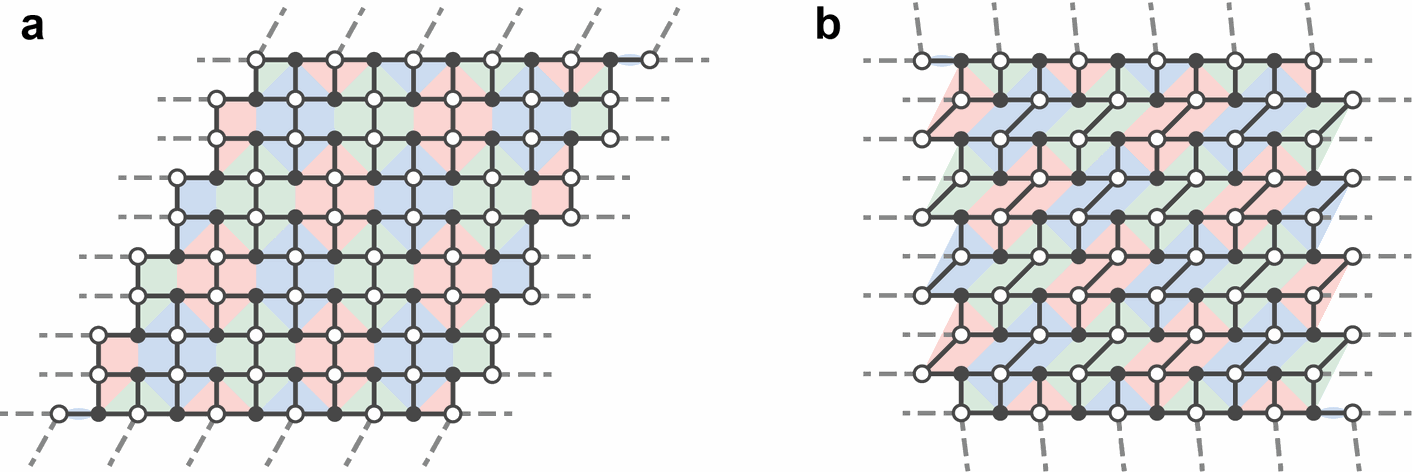}
    \caption{Two plaquette template options for color code modules shown on a quadrilateral module at $L=5$. \textbf{a} \textit{Honeycomb layout}. \textbf{b} \textit{Diagonal rectangle layout}.}
    \label{fig:diag_rect}
\end{figure}

Each quadrilateral module can be laid out on the square grid in two ways as shown in \Cref{fig:diag_rect}. The honeycomb layout uses honeycomb plaquettes and results in a rhombus-shaped module, while the diagonal rectangle layout uses diagonal couplers to pack the qubits into a rectangular footprint. Both layouts lead to modules with an equal number of qubits, and the choice affects only the physical arrangement of qubits and couplers.

\subsubsection{Octagon Modules (\texorpdfstring{$K=8$}{K=8})}\label{sec:octagon_modules}
A coarser partition groups eight triangles into a single octagonal module (\Cref{fig:color_code_modules}\textbf{c}). These eight tiles are the ones meeting at a vertex of $\mathcal{T}^\triangle$ of a chosen color, or equivalently the eight triangles making up one octagonal face of the color code tiling $\mathcal{T}$ of a chosen color, for instance a red face. 
Associating red, green and blue edges with $a$, $b$ and $c$ mirrors, respectively, a red face corresponds to eight triangular tiles of $\mathcal{T}^\triangle$ related by $b$ and $c$ reflections (right multiplication by $b$ and $c$).
Each such group of tiles is a left coset $gD = g\langle b, c \rangle$ of the order-eight dihedral subgroup $D = \langle b, c \rangle \cong D_4 \le G$, so in general the octagonal modules are in bijection with the $|G|/8$ left cosets of $D$ rather than with single group elements of $G^+$. After fine-graining, each octagon module contains $8L^2$ data qubits and $N = 16L^2 + 8$ total qubits.

This module is available only when the color code tiling actually has octagonal faces, which happens for a $2r.2s.2t$ tiling when one of $r$, $s$, $t$ equals $4$, so that the corresponding face has $8$ sides. For the $8.8.8$ family all three face colors are octagons, and any of them may be used. Grouping tiles into an octagon module reduces the number of inter-module seams at the cost of a larger planar chip.
Octagon modules require a modified set of automorphisms that permute modules without rotating them internally. In \Cref{sec:auts,sec:module_complements} we explain how we find these, together with symmetric logical bases, for the tilings we consider.

\subsection{Semi-Hyperbolic Surface Code Modules}\label{sec:surface_code_modules}
\begin{figure}[h]
    \centering
    \includegraphics[width=0.8\linewidth]{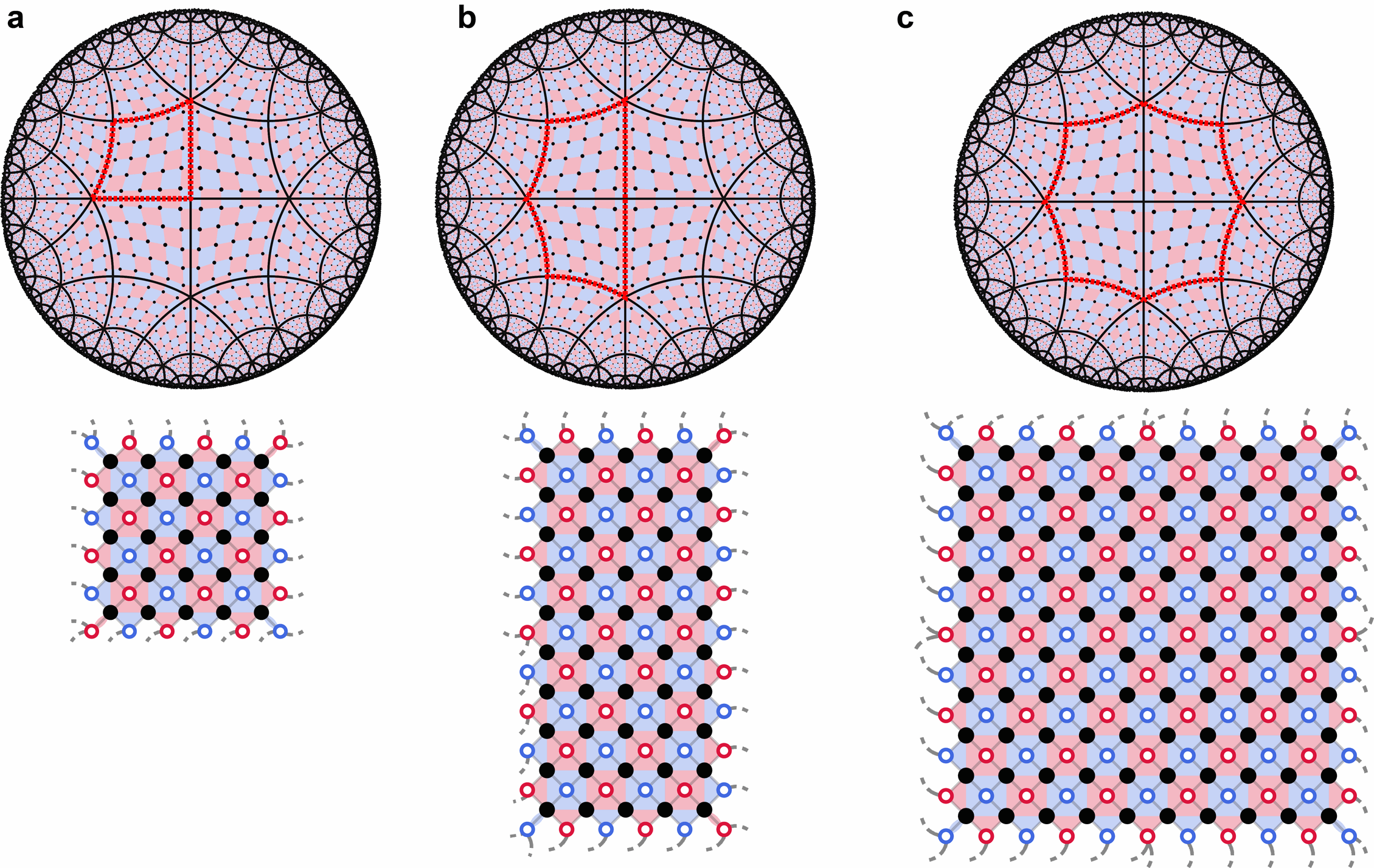}
    \caption{Module partition options for semi-hyperbolic surface codes, illustrated on a $V4.6.4.6$ base quadrilateral tiling. \textbf{Top row:} The base tiling shown in the Poincar\'e disk, with the red dashed lines tracing the boundaries of a single module. \textbf{Bottom row:} The corresponding planar module (shown here for $L=5$), with data qubits (filled circles) and ancilla qubits (open circles). Solid black lines indicate in-module couplers whereas dashed gray lines indicate inter-module couplers. \textbf{a}~\textit{Square module} ($K{=}1$): each quadrilateral tile forms its own module. \textbf{b}~\textit{Rectangle module} ($K{=}2$): pairs of adjacent quadrilateral tiles are merged across a shared edge. \textbf{c}~\textit{Big square module} ($K{=}4$): this requires a base tiling $Va.b.c.d$ with at least one of $a,b,c,d$ equal to $4$; choosing one such corner type, the four tiles meeting at each vertex of that type are grouped together.}
    \label{fig:surface_code_modules}
\end{figure}

The semi-hyperbolic surface codes are constructed from a base quadrilateral tiling $\mathcal{T}^\square$. A module partition groups the quadrilateral tiles, together with their data and ancilla qubits, into planar modules, and we again label the partitions by the number $K$ of base tiles grouped into each module. \Cref{fig:surface_code_modules} shows the three surface code module types, with $K=1$, $K=2$ and $K=4$ giving the square, rectangle and big square modules respectively. As in the color code case, the module structure follows from the symmetry group of the tiling, though here the details depend on whether the quadrilateral tiling comes from a quadrilateral group or from a triangle group, as discussed in \Cref{sec:quadrilateral_tilings}.

\subsubsection{Square Modules (\texorpdfstring{$K=1$}{K=1})}\label{sec:square_modules}
The simplest partition assigns each quadrilateral tile of $\mathcal{T}^\square$ to its own module, giving one square module per tile (\Cref{fig:surface_code_modules}a). When the tiling comes from a quadrilateral group, each tile is an image $gF$ of a chosen fundamental domain $F$ of $G = \square(r,s,t,u)/\Gamma$, so the modules are in bijection with the group elements. When the quadrilateral tiling instead arises from a triangle group, each tile is associated with a coset rather than a single group element (see \Cref{sec:quadrilateral_tilings}). After fine-graining, each square module contains $L^2$ data qubits and $N = 2L^2 + 2L + 1$ total qubits. Like the triangle module for the color code, this is the finest partition we use, minimizing module size.

\subsubsection{Rectangle Modules (\texorpdfstring{$K=2$}{K=2})}\label{sec:rectangle_modules}
The rectangle partition pairs two adjacent quadrilateral tiles into a single module (\Cref{fig:surface_code_modules}b). When the tiling has quadrilateral reflection symmetry, each pair is an image of a fundamental domain of the orientation-preserving subgroup $G^+$, and the modules are in bijection with its elements. In this case the same picture as for the color code applies, with adjacent modules reached by right multiplication and automorphisms acting by left multiplication, so we can again implement automorphism circuits, described with a connectivity given by the left Cayley graph of $G^+$. After fine-graining, each rectangle module contains $2L^2$ data qubits and $N = 4L^2 + 3L + 1$ total qubits.

Pairing the square tiles into rectangles is not always possible when the quadrilateral tiling does not have quadrilateral reflection symmetry. We use a maximum matching on the Cayley graph of the tiles in those cases and any tile left unmatched remains a lone square ($K=1$) module. 

\subsubsection{Big Square Modules (\texorpdfstring{$K=4$}{K=4})}\label{sec:big_square_modules}
The coarsest surface code partition collects four quadrilateral tiles into a single big square module (\Cref{fig:surface_code_modules}c). After fine-graining, each big square module contains $4L^2$ data qubits and $N = 8L^2 + 4L + 1$ total qubits. This module requires the four tiles to meet in a way that forms a larger square, which is available only for base quadrilateral tilings $Va.b.c.d$ in which one of $a$, $b$, $c$, $d$ equals $4$.

\section{Modular Cat-Flagged Syndrome Extraction Circuits}\label{sec:si_modular_circuits}
We now explain the modular fault-tolerant syndrome extraction circuits we use to benchmark the performance of the semi-hyperbolic ($L \geq 2$) color and surface codes as a quantum memory. In all the circuits we design, every two-qubit gate is either local to a module and acts between neighboring qubits, or it is a long-range gate between modules acting via an inter-module coupler.

\subsection{Hyperbolic Color Code Circuits}
Modular hyperbolic color codes are associated with a $2r.2s.2t$ hyperbolic tiling as explained in \Cref{sec:code_constructions}. We focus on the quadrilateral module partition ($K{=}2$) in the explanation here, though the same circuit techniques apply to the other partitions. We can classify the stabilizers into two categories: \textit{bulk} and \textit{boundary} stabilizers. Bulk stabilizers are supported entirely within a single module and are all weight-$6$ in a quadrilateral module. Weight-$6$ bulk stabilizers are measured using standard Bell-flagged color code circuits introduced for Euclidean planar color codes \cite{bell_flagged_circuit}. Boundary stabilizers have support on data qubits in more than one module. Most boundary stabilizers are also weight-$6$, but they also include the \emph{corner stabilizers} of weight $2r$, $2s$ or $2t$, inherited from the faces of the underlying $2r.2s.2t$ hyperbolic tiling. We measure corner stabilizers with a generalization of Bell-flagged circuits that we call \emph{modular cat-flagged circuits}. Flag-based circuits for (non-fine-grained, non-modular) hyperbolic color codes were also studied in Ref.~\cite{flag_proxy_hyperbolic_color}.

The idea for modular cat-flagged syndrome extraction is that for a stabilizer allocated $m$ ancillas across the modules it touches, we unitarily prepare an $m$-qubit cat state on the ancillas using the inter-module couplers. Each ancilla then couples to the data qubits associated with the stabilizer within its own module  using nearest-neighbor gates on the square grid. Finally we apply the inverse of the cat-state preparation circuit followed by single-qubit measurements, which measures the stabilizer on a single ancilla qubit. The remaining $m-1$ ancillas give the flag measurements that detect hook errors. One syndrome extraction round consists of an $X$ subround followed by a $Z$ subround. Each subround carries out a cat-state preparation, a data--ancilla coupling stage, and cat-state measurement.

\begin{figure}[H]
    \centering
    \includegraphics[width=0.9\linewidth]{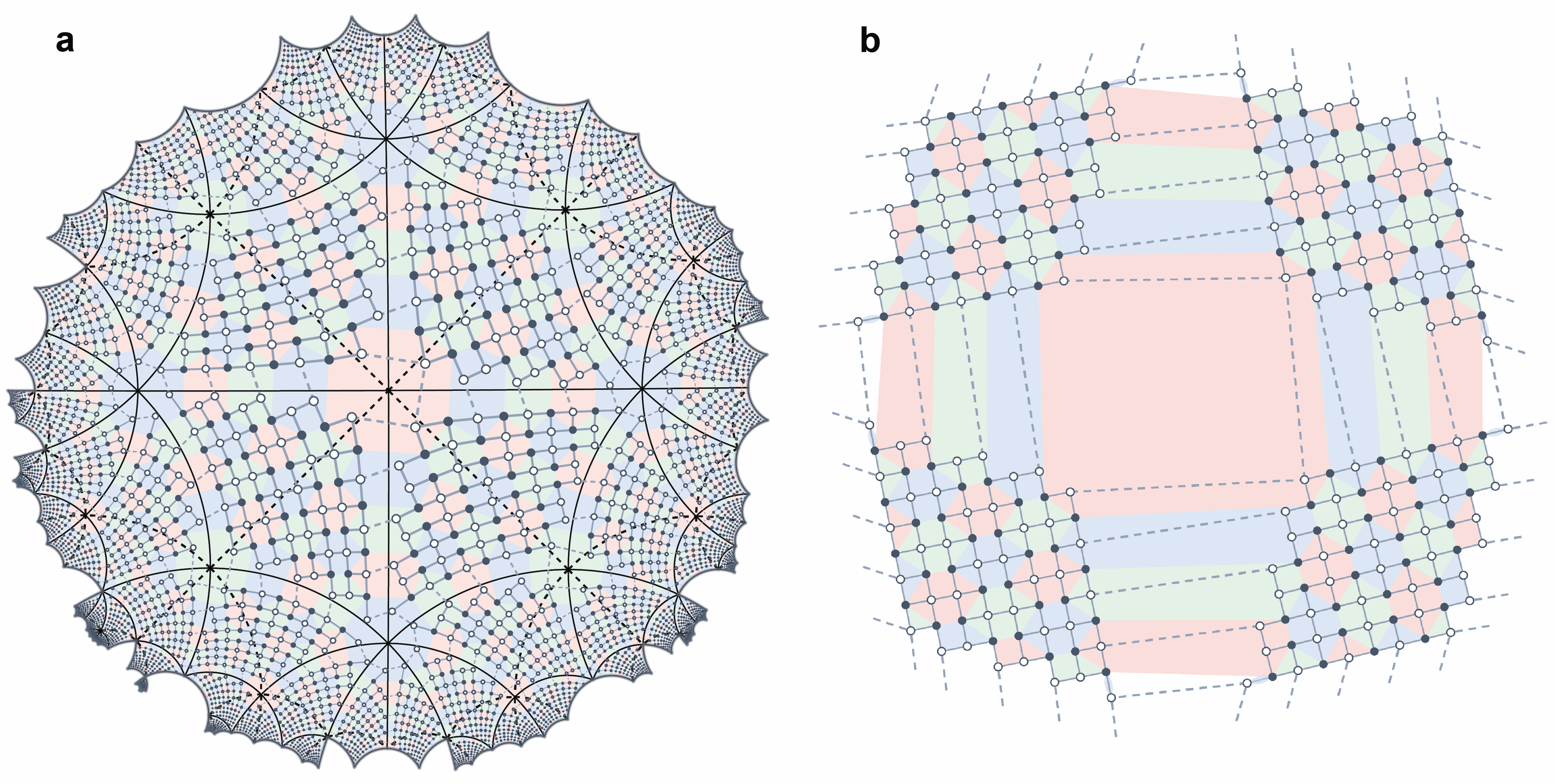}
    \caption{Long-range stabilizers of four boundary-connected planar modules of an
    $8.8.8$ modular hyperbolic color code. The four modules are identical and related to their neighbors by a $90^\circ$ rotation. Data qubits (black) and ancilla qubits (white) sit on the square grid within each module. Dashed lines show the inter-module links. Stabilizers spanning across modules may get skewed relative to the square-grid layout, and since ancillas are placed on the square grid the number available varies between stabilizers, even among those of equal weight. The cat-flagged circuits accommodate this. A stabilizer allocated $m$ ancillas is extracted using an $m$-qubit cat state.}
    \label{fig:four_modules}
\end{figure}

\begin{figure}[H]
    \centering
    \includegraphics[width=0.3\linewidth]{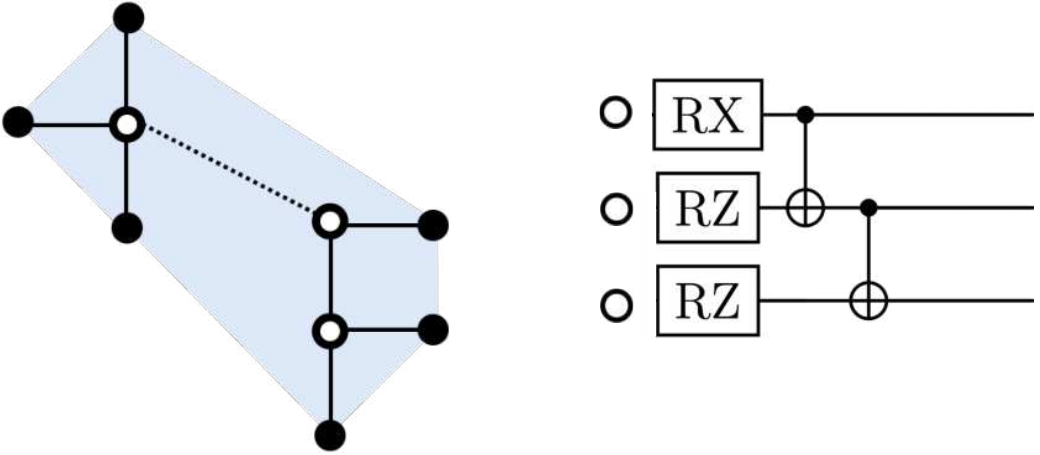}
    \caption{A skewed weight-$6$ boundary stabilizer and the $Z$-basis cat-state preparation
    circuit used to measure the $X$ stabilizer, using a $3$-qubit cat state prepared in
    CNOT-depth $2$.}
    \label{fig:cat_flagged_w6_a3}
\end{figure}

\begin{figure}[H]
    \centering
    \includegraphics[width=0.9\linewidth]{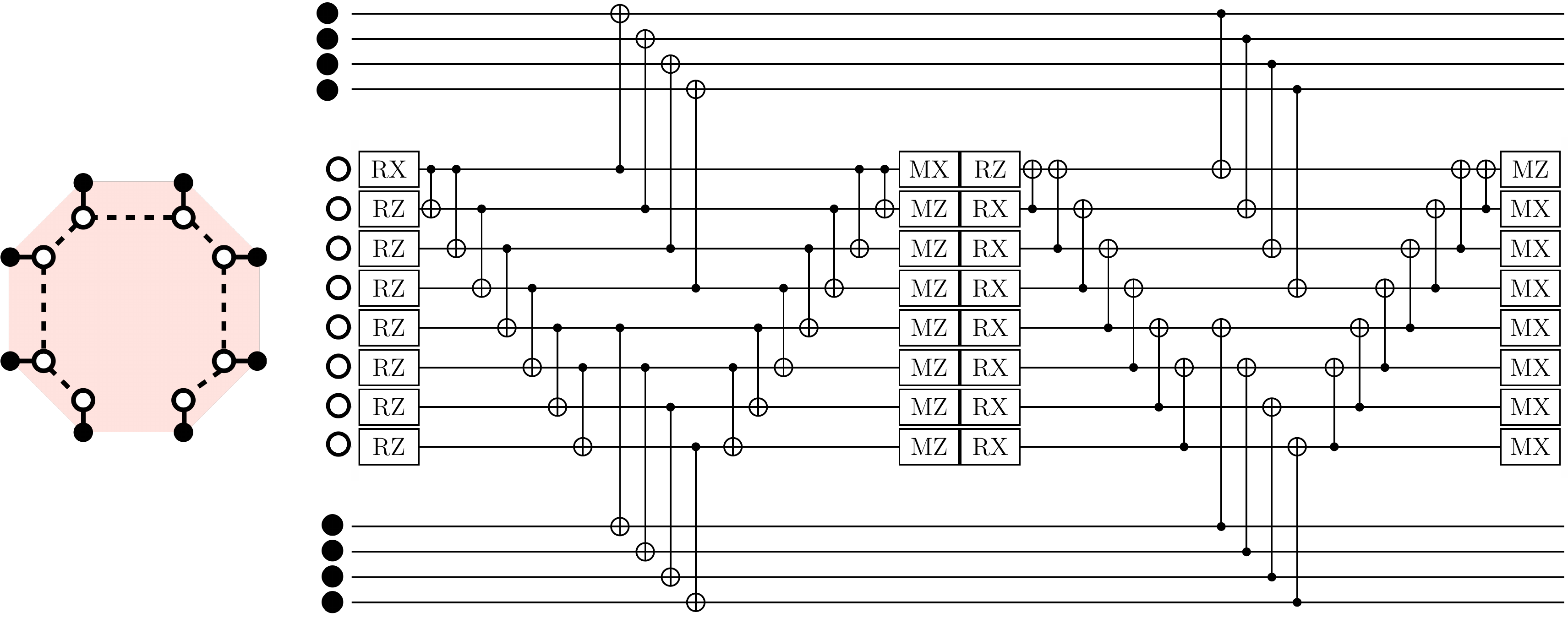}
    \caption{A weight-$8$ stabilizer spanning $8$ modules, with $1$ ancilla per module, and the cat-flagged circuits used to measure its $X$ and $Z$ stabilizers, using an $8$-qubit cat state. The cat state is prepared and measured along a linear chain in CNOT-depth $4$, and syndrome CNOTs in depth $3$.}
    \label{fig:cat_flagged_w8_a8_lin}
\end{figure}

The number of ancillas associated with a stabilizer $m$ is fixed by the boundary geometry rather than the stabilizer weight, and varies even among stabilizers of equal weight, as \Cref{fig:four_modules} shows. \Cref{fig:cat_flagged_w6_a3}, for example, shows a weight-$6$ boundary stabilizer on a skewed boundary present in \Cref{fig:four_modules}, allocated $m=3$ ancillas (instead of $2$) to enforce nearest-neighbor two-qubit gates. The corner stabilizers are the most demanding case in the syndrome extraction, since they require preparation of larger cat states. \Cref{fig:cat_flagged_w8_a4} shows a weight-$8$ corner stabilizer shared among $4$ modules, extracted with a $4$-qubit cat state. \Cref{fig:cat_flagged_w8_a8_lin} shows another weight-$8$ stabilizer that instead spans $8$ modules with a single ancilla in each. 

\begin{figure}[H]
    \centering
    \includegraphics[width=0.35\linewidth]{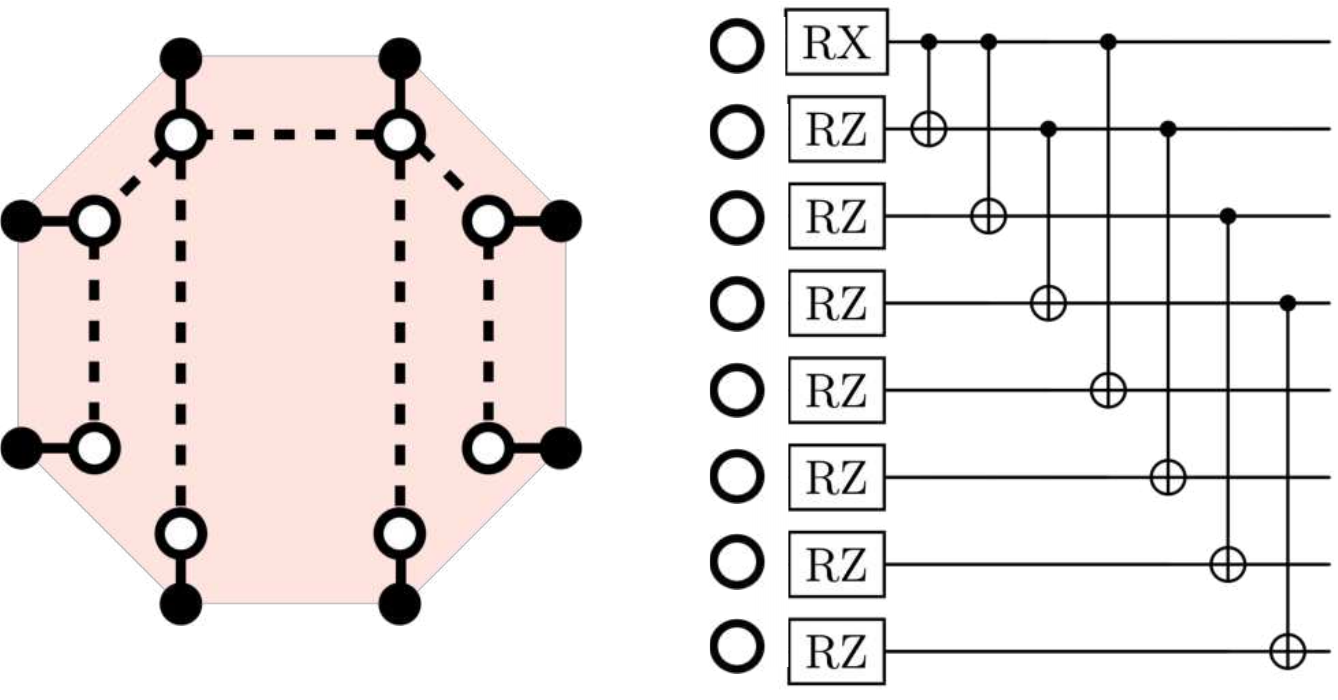}
    \caption{Log-depth preparation of the $8$-qubit $Z$-basis cat state. Using additional inter-module connectivity, the cat state is prepared in CNOT-depth $3$ rather than depth $4$. The measurement circuit is the inverse of the preparation circuit above.}
    \label{fig:cat_flagged_w8_a8_log}
\end{figure}

The cost of an $m$-qubit cat state depends on the inter-module connectivity available. Prepared along a linear chain of seams it takes depth $\lceil m/2\rceil$. Adding further inter-module connectivity as shown in \Cref{fig:cat_flagged_w8_a8_log} reduces this to log-depth $\lceil\log_2 m\rceil$, trading extra inter-module connectivity for a shallower circuit. 

\begin{figure}[H]
    \centering
    \includegraphics[width=0.65\linewidth]{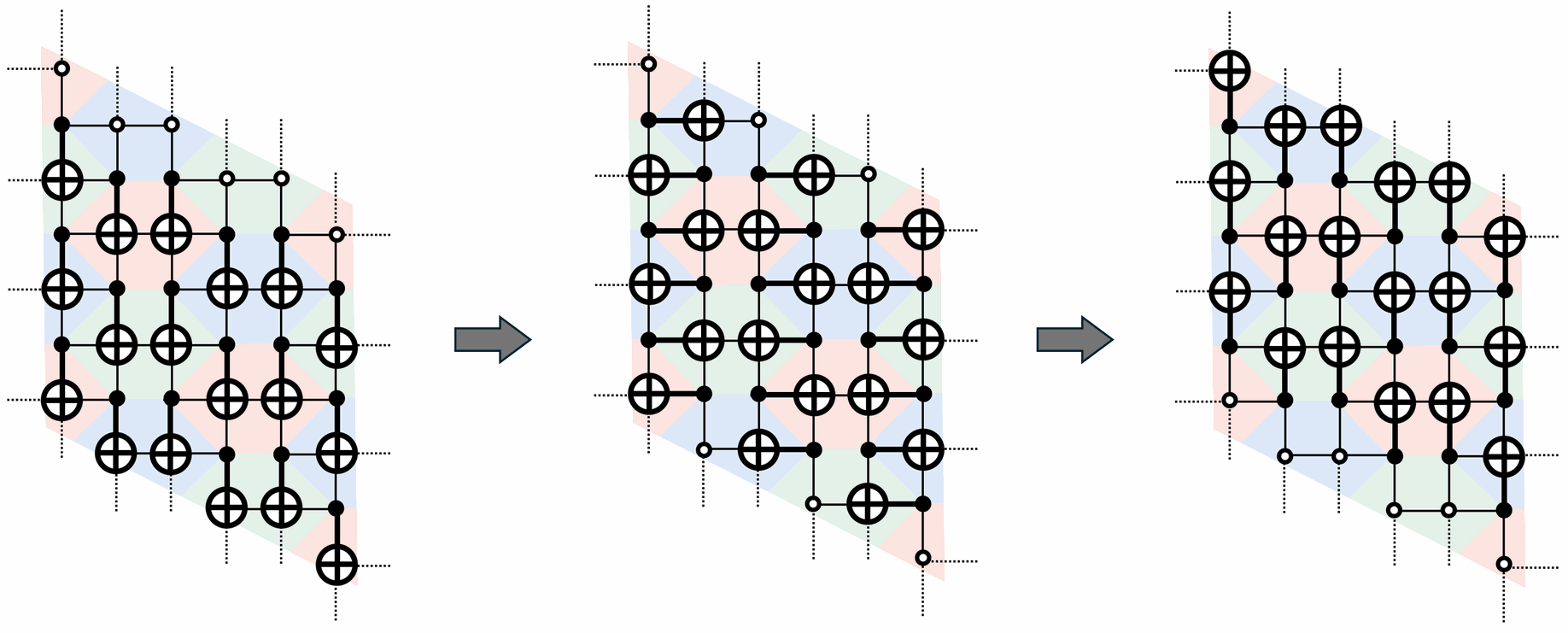}
    \caption{The up--middle--down plaquette schedule for the local data--ancilla coupling stage. On an orientable surface, the same up--middle--down ordering is applied identically to every plaquette. The three panels, left to right, show its successive time steps. The schedule shown extracts the $Z$ stabilizers; the $X$ stabilizers are extracted by the same schedule with the CNOT data--ancilla directions reversed.}
    \label{fig:hyperbolic_color_cnot_schedule}
\end{figure}

The data--ancilla coupling gates within a block of $X$- (or $Z$-) extraction all commute, and any edge coloring of the bipartite data--ancilla interaction graph yields a valid schedule, of depth at most the maximum vertex degree of that graph~\cite{tremblay2022constant}. Since each data qubit couples to three ancillas, three time steps suffice. Additionally, on an orientable surface we can define a globally consistent direction, and use the translationally-invariant up--middle--down plaquette schedule of the planar color code rather than an unstructured circuit found by edge coloring. The up-middle-down circuit shown in \Cref{fig:hyperbolic_color_cnot_schedule} is applied identically to every face and every module.

\begin{figure}[H]
    \centering
    \includegraphics[width=\linewidth]{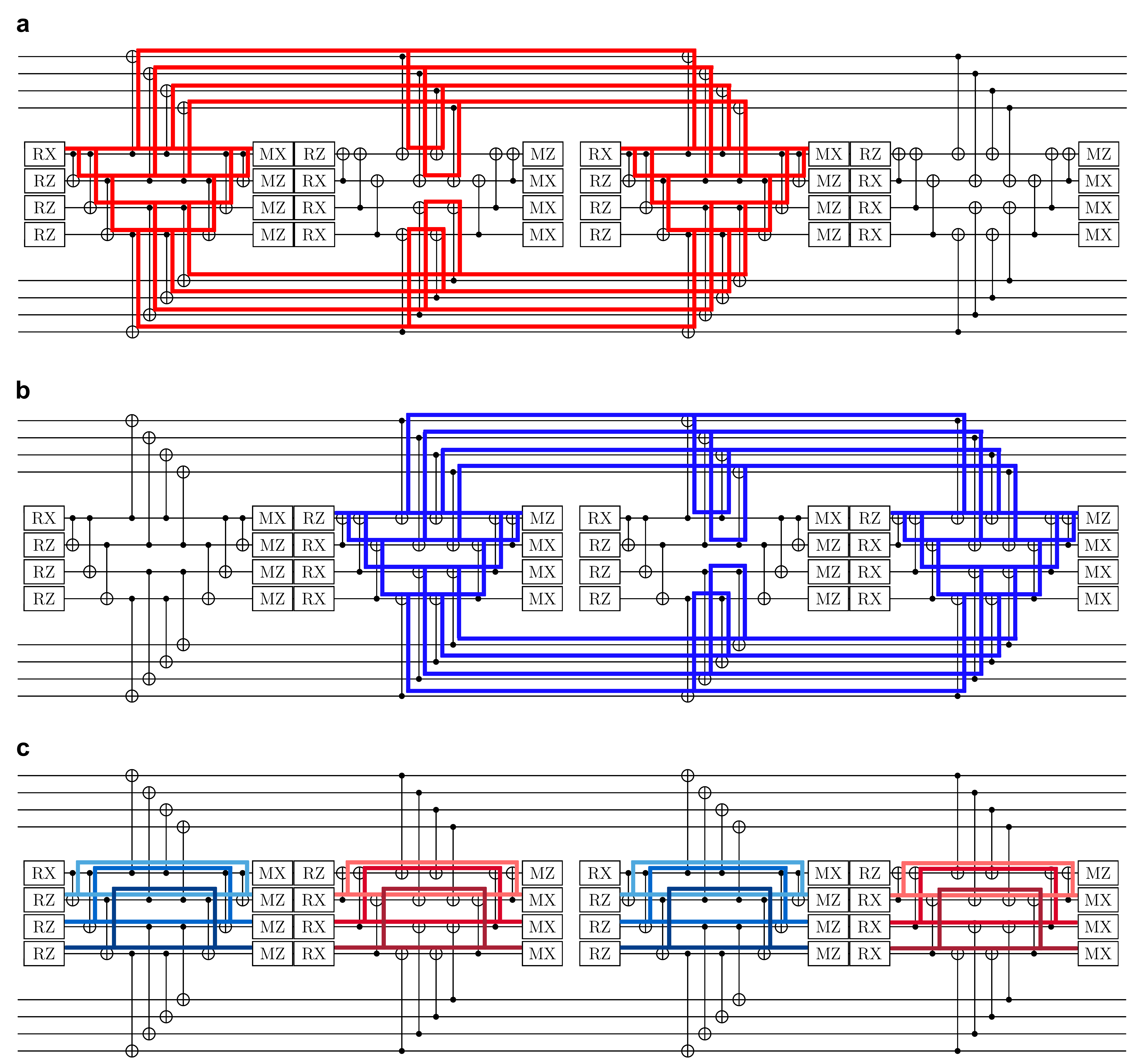}
    \caption{Detecting regions of the cat-flagged syndrome extraction circuit for a weight-$8$ stabilizer measured with $m=4$ ancillas, shown over two rounds each consisting of an $X$ and $Z$ subround. \textbf{a}~$X$ plaquette stabilizer detecting region (red), sensitive to $Z$ errors. Each detector compares the plaquette stabilizer ancilla outcomes of two consecutive rounds. \textbf{b}~$Z$ plaquette stabilizer detecting region (blue), sensitive to $X$ errors. \textbf{c}~Flag detecting regions, shown in different shades of blue and red to distinguish independent detectors. Each of the $m-1$ flag ancillas is measured in the opposite basis to the plaquette stabilizer ancilla, and yields a deterministic outcome every sub-round. The flag detectors detect hook errors that would otherwise propagate onto multiple data qubits.}
    \label{fig:cat_flagged_detectors}
\end{figure}

The detectors of cat-flagged circuits are of two types, as shown in \Cref{fig:cat_flagged_detectors}. A \emph{plaquette detector} compares a stabilizer's syndrome measurement in one round with its measurement in the previous round. In the absence of errors, this parity is deterministic, so the detector spans two consecutive rounds. In an $X$ subround these are the $X$-plaquette detecting regions (\Cref{fig:cat_flagged_detectors}a), sensitive to $Z$ errors, and in a $Z$ subround they are the $Z$-plaquette detecting regions (\Cref{fig:cat_flagged_detectors}b), sensitive to $X$ errors. A \emph{flag detector} arises from each of the $m-1$ ancillas that are not used to read out the plaquette stabilizer syndrome. The cat-state measurement collapses the syndrome onto a single ancilla, and the remaining $m-1$ ancillas are measured in the opposite basis. Each such measurement has a deterministic outcome in the absence of errors, yielding $m-1$ flag detectors per round (\Cref{fig:cat_flagged_detectors}c). These flag detectors fire on hook errors that would otherwise propagate from a single fault on one ancilla qubit onto multiple data qubits and reduce the effective distance.

\subsection{Hyperbolic Surface Code Circuits}\label{sec:hyperbolic_surface_code_circuits_si}

The modular hyperbolic surface codes use cat-flagged syndrome extraction that is conceptually similar to that of the modular hyperbolic color codes.
As was done for the modular hyperbolic color code circuits, we unitarily prepare an $m$-qubit cat state on the $m$ ancilla qubits of each stabilizer that is supported on module boundaries (e.g.~each stabilizer that spans more than one module, or that spans both sides of a module that wraps around a handle of the surface).
Once the cat states have been prepared, each ancilla couples to the data qubits of its stabilizer within its own module using nearest-neighbor gates on the square grid.
Finally we apply the inverse of the unitary cat state preparation circuit followed by single-qubit measurements to measure the stabilizer along with $m-1$ flag measurements.
\begin{figure}[h]
    \centering
    \includegraphics[width=0.5\linewidth]{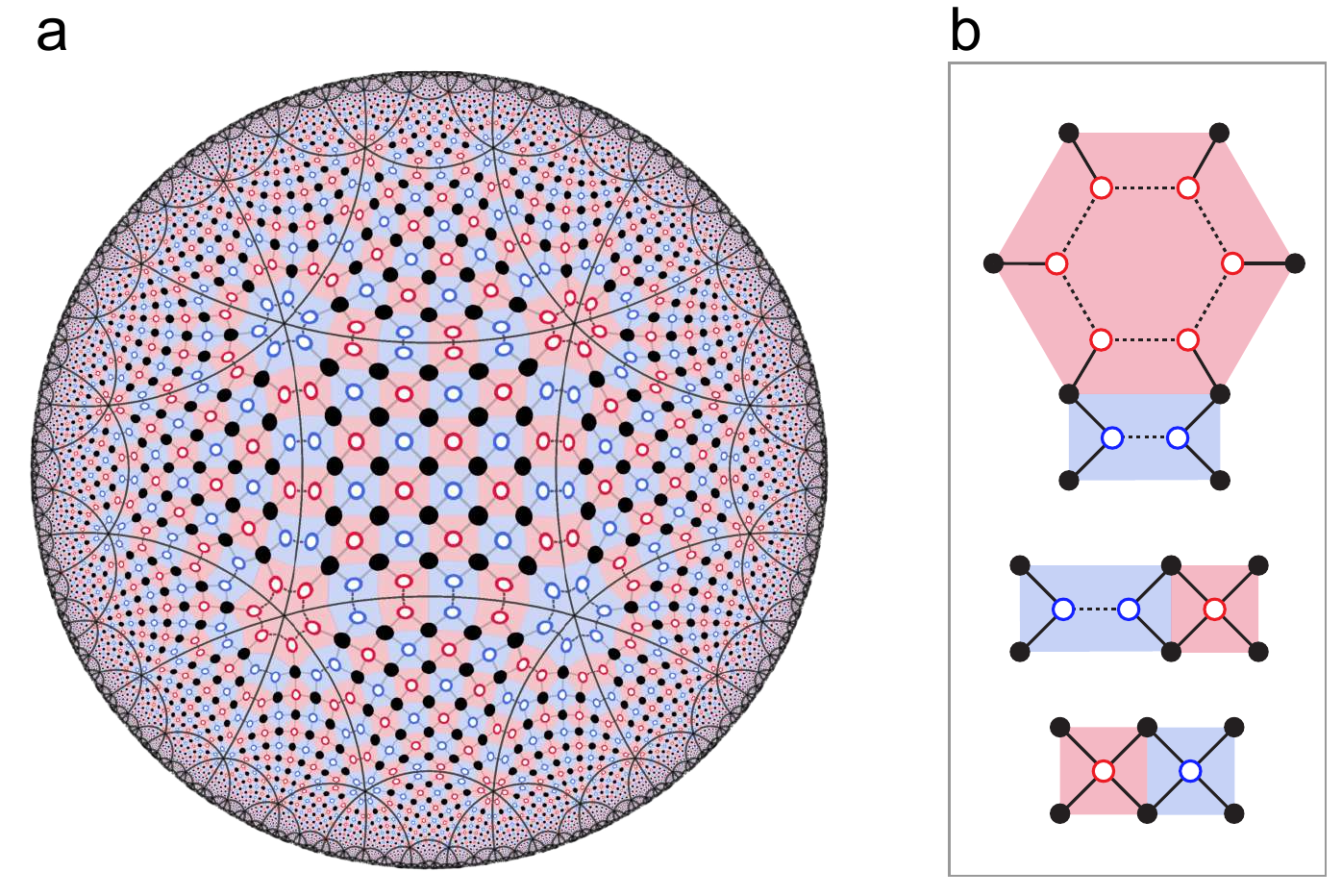}
    \caption{Adjacent $X$ and $Z$ stabilizers in modular hyperbolic surface codes. \textbf{a}, The stabilizer structure of a modular hyperbolic surface code derived from a $\{4,6\}$ regular tiling with $L=5$ fine-graining. Here we show the structure on the infinite hyperbolic plane (rather than a fundamental polygon corresponding to a specific closed surface). \textbf{b}, Examples of $X$ and $Z$ stabilizers that share support on data qubits in the code. If an $X$ stabilizer $i$ intersects with a $Z$ stabilizer $j$, the two stabilizers always have exactly two qubits in common. Let stabilizer $i$ be supported on an $n_i$-gon plaquette containing an $m_i$-qubit cat state. Similarly stabilizer $j$ is supported on an $n_j$-gon plaquette containing an $m_j$-qubit cat state. In general note that $n_i\neq n_j$ and $m_i\neq m_j$, and furthermore note that $m_k=1$ corresponds to the trivial case of a single-qubit $\ket{+}$ or $\ket{0}$ ancilla qubit for an $X$ or $Z$ stabilizer respectively.}
\label{fig:hyperbolic_surface_adjacent_plaquettes}
\end{figure}

\begin{figure}
    \centering
    \includegraphics[width=0.65\linewidth]{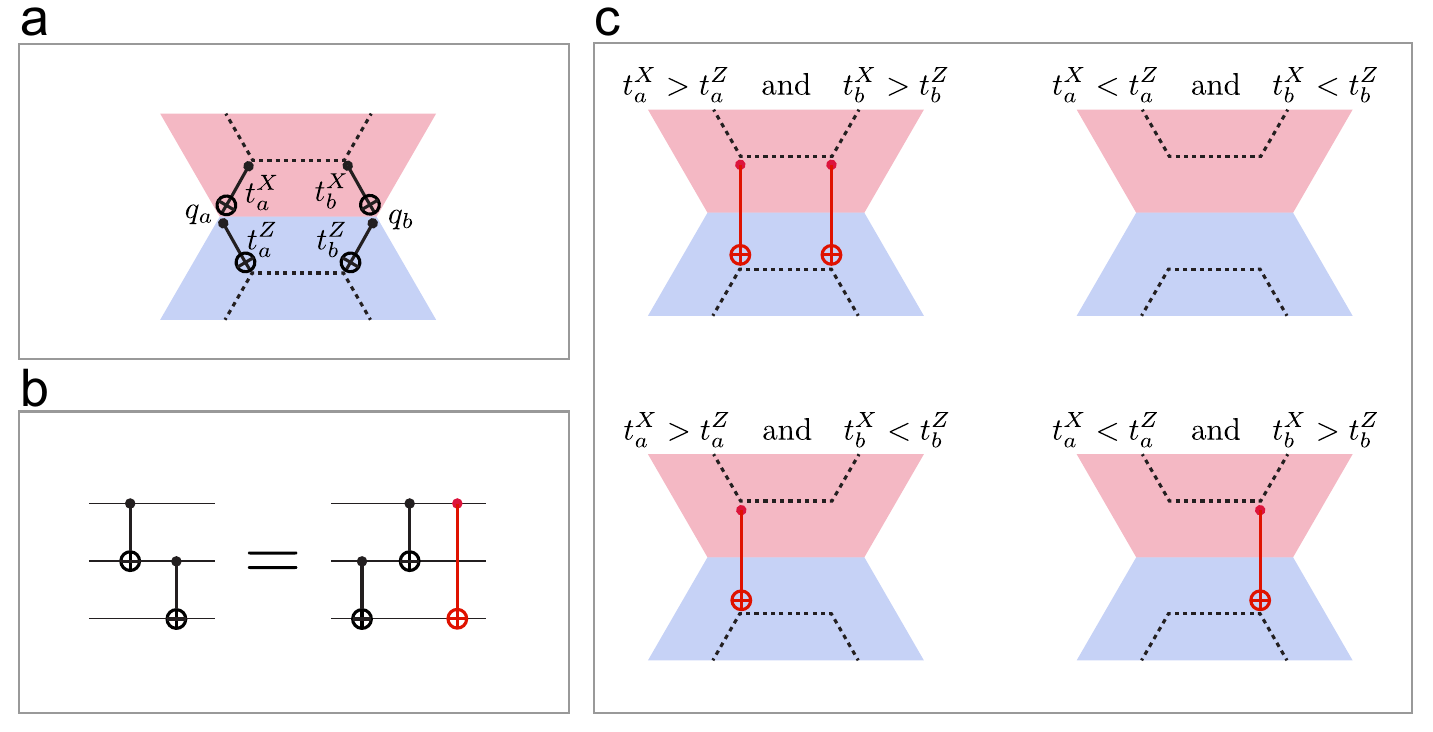}
    \caption{\textbf{a}, Any $X$ and $Z$ stabilizer that intersect in a modular hyperbolic surface code share exactly two data qubits and are each measured by coupling an $m_i$-qubit cat state to the stabilizer's data qubits using CNOT gates. Note that the size $m_i\geq 1$ of the cat state may differ for the two stabilizers. The CNOT gates are targeted on the data qubits for the $X$ stabilizer and controlled on them for the $Z$ stabilizer. We have labeled each CNOT gate with the time step it is applied in the extraction circuit. \textbf{b}, Since the CNOTs act as a target on the data qubits for the $X$ stabilizer and as a control for $Z$, they do not commute. Swapping the order of a CNOT gate $
    \mathrm{CNOT}(a_X,q)$ for the $X$ stabilizer with another gate $
    \mathrm{CNOT}(q,a_Z)$ for the $Z$ stabilizer introduces an effective CNOT interaction $
    \mathrm{CNOT}(a_X,a_Z)$ controlled on an ancilla (cat state) qubit $a_X$ of the $X$ stabilizer and targeted on an ancilla qubit $a_Z$ of the $Z$ stabilizer. \textbf{c}, We can verify the correctness of an extraction circuit by starting with an extraction circuit that is not interleaved and therefore trivially correct (extract $X$ stabilizers followed by $Z$ stabilizers), and then commuting CNOT gates past each other until the schedule's CNOT ordering is reached. We can then inspect the action of the effective CNOT gates (in red) to check if their action is trivial. In this subfigure we show the four possible cases. If the number of effective (red) CNOT gates coupling the ancillas in the two plaquettes is even (top row) then their action is trivial.}
    \label{fig:adjacent_cat_measurement_gate_orders}
\end{figure}

One important difference with the hyperbolic surface code circuits is that we interleave the $X$ and $Z$ syndrome extraction, unlike the hyperbolic color codes for which $X$ and $Z$ are measured sequentially.
In other words, we extract $X$ and $Z$ stabilizers simultaneously after unitarily preparing cat states for stabilizers of both bases.
This leads to additional commutativity constraints for the hyperbolic surface code circuits that did not arise for the hyperbolic color code circuits.
These commutativity constraints arise wherever an $X$ stabilizer shares support with a $Z$ stabilizer, since the CNOT gates used to extract adjacent $X$ and $Z$ stabilizers do not commute.

If an $X$ and $Z$ stabilizer have data qubits in common, they always share exactly two data qubits, as shown in \Cref{fig:hyperbolic_surface_adjacent_plaquettes}.
We can determine if a particular interleaved extraction schedule (CNOT ordering) is valid using the commutation of the CNOTs in neighboring $X$ and $Z$ plaquettes.
As illustrated in \Cref{fig:adjacent_cat_measurement_gate_orders}, swapping the order of two CNOT gates $\mathrm{CNOT}(a_X,q)$ and $\mathrm{CNOT}(q,a_Z)$ in adjacent $X$ and $Z$ stabilizers introduces an additional effective CNOT gate $\mathrm{CNOT}(a_X,a_Z)$ coupling the ancillas of the two plaquettes.
We can determine if a given schedule is correct by starting with a schedule that is not interleaved and which we therefore know to be correct (extract $X$ stabilizers followed by $Z$ stabilizers), and then commute CNOT gates past each other until the schedule's gate ordering is reached.
If the action of the effective CNOT gates introduced from the commutation in \Cref{fig:adjacent_cat_measurement_gate_orders}\textbf{b} is trivial, then we know that the schedule is valid~\cite{tremblay2022constant}.
As shown in \Cref{fig:adjacent_cat_measurement_gate_orders}\textbf{c}, if the schedule gate ordering in \Cref{fig:adjacent_cat_measurement_gate_orders}\textbf{a}, defined by the time steps $(t_a^X, t_b^X, t_a^Z, t_b^Z)$, satisfies
\begin{equation}\label{eq:timesteps_neighboring_X_and_Z_stab}
    (t_a^X < t_a^Z) \iff (t_b^X < t_b^Z)
\end{equation}
then either 0 or 2 CNOT gates coupling the $X$ and $Z$ ancillas are introduced from commuting the CNOT gates in the extraction circuit past one another, which have trivial action on the cat states.

To see why, consider the state of a cat state that is supported on the ancilla qubits $a_X^i\in Q_X^{\mathrm{anc}}$, with stabilizer group
\begin{equation}
\mathcal{S}_X^{\mathrm{anc}} = \left\langle X^{\otimes m_X}, \{Z_{a_X^i} Z_{a_X^{i+1}}\}_{i=0}^{m_X-2} \right\rangle,
\end{equation}
as well as the state of the opposite-basis cat state supported on the $Z$ stabilizer ancilla qubits $a_Z^j\in Q_Z^{\mathrm{anc}}$, with stabilizer group
\begin{equation}
\mathcal{S}_Z^{\mathrm{anc}} = \left\langle Z^{\otimes m_Z}, \{X_{a_Z^j} X_{a_Z^{j+1}}\}_{j=0}^{m_Z-2} \right\rangle.
\end{equation}
Note that a single CNOT gate $\mathrm{CNOT}(a_X^i,a_Z^j)$ between the two cat states maps the global transversal stabilizers as $X^{\otimes m_X}\rightarrow X^{\otimes m_X}X_{a_Z^j}$ and $Z^{\otimes m_Z}\rightarrow Z^{\otimes m_Z}Z_{a_X^i}$. 

If there are exactly two effective CNOT gates introduced by the commutation, for example $\mathrm{CNOT}(a_X^{i_1},a_Z^{j_1})$ and $\mathrm{CNOT}(a_X^{i_2},a_Z^{j_2})$, their combined action maps the global $X$ stabilizer of the $X$-basis cat state to:
\begin{equation}
    X^{\otimes m_X} \rightarrow X^{\otimes m_X} \otimes X_{a_Z^{j_1}} X_{a_Z^{j_2}}.
\end{equation}
However the operator $X_{a_Z^{j_1}} X_{a_Z^{j_2}}$ is an element of the stabilizer group $\mathcal{S}_Z^{\mathrm{anc}}$ (as it can be formed by a product of the adjacent $XX$ parity operators). Therefore, it acts trivially on the $Z$ plaquette's cat state. 

Similarly, the global $Z$ stabilizer of the $Z$-basis cat state transforms as:
\begin{equation}
    Z^{\otimes m_Z} \rightarrow Z_{a_X^{i_1}} Z_{a_X^{i_2}} \otimes Z^{\otimes m_Z},
\end{equation}
and because $Z_{a_X^{i_1}} Z_{a_X^{i_2}} \in \mathcal{S}_X^{\mathrm{anc}}$, this also acts trivially on the $X$ plaquette's cat state. 
Since all generators of the joint stabilizer group $\mathcal{S}_X^{\mathrm{anc}} \otimes \mathcal{S}_Z^{\mathrm{anc}}$ map to themselves up to multiplication by existing stabilizers, any even number of effective CNOT gates between the two cat states leaves their joint state invariant.

\subsubsection{Lifted schedules}

We now show how we find a valid CNOT gate ordering for a low-depth syndrome extraction circuit for the entire modular hyperbolic code (i.e.~such that the gate ordering for every pair of adjacent $X$ and $Z$ stabilizers satisfies \Cref{eq:timesteps_neighboring_X_and_Z_stab}).
We simplify the scheduling problem by choosing to use a translationally invariant schedule within each face of the base tiling, so that the schedule is fully determined by its values at the four corners of each face of the base tiling. 
Translational invariance is not a requirement, but it significantly simplifies the overall constraint satisfaction problem (by reducing its size by a factor of $L^2$) and leads to a more structured circuit.

Recall that the semi-hyperbolic surface code is constructed by subdividing each face of the base quadrilateral tiling $\mathcal{T}^\square$ into an $L \times L$ grid of fine squares, and then taking the dual. The fine tiling within each base face is therefore a regular square lattice, and we choose a schedule within each base face that is translationally invariant on this fine-grained lattice (every face in the fine-grained lattice is assigned the same time steps, regardless of its basis).

\begin{figure}
    \centering
    \includegraphics[width=0.5\linewidth]{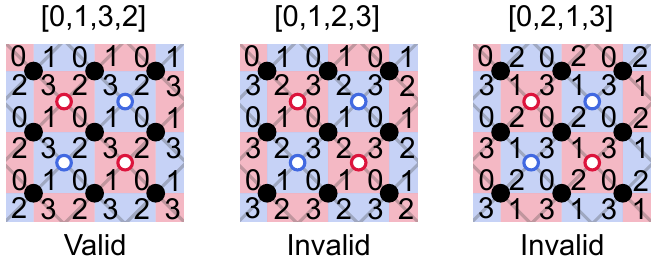}
    \caption{Valid translationally-invariant schedules for a square plaquette. Of the $24$ permutations of $\{0,1,2,3\}$, only the $8$ in the dihedral orbit of $[0,1,3,2]$ under $D_4$ yield valid translationally-invariant schedules on the fine-grained square lattice. The remaining schedules are in the dihedral orbit of $[0,1,2,3]$ or $[0,2,1,3]$, neither of which yield valid schedules.}
    \label{fig:translationally_invariant_D4_seeds}
\end{figure}

Not every permutation of $\{0,1,2,3\}$ assigned to the $4$ corners of a square yields a valid translationally-invariant schedule that satisfies \Cref{eq:timesteps_neighboring_X_and_Z_stab} for pairs of adjacent $X$ and $Z$ stabilizers. Only those permutations in the dihedral orbit of the base permutation $[0,1,3,2]$ under $D_4$ do (\Cref{fig:translationally_invariant_D4_seeds}). This gives $8$ valid permutations per face: the $4$ rotations and $4$ reflections of $[0,1,3,2]$.

Assigning one of these $8$ valid permutations to a face of the base tiling uniquely determines a translationally-invariant fine schedule within that face (\Cref{fig:lifted_schedule_centered_edge_of_4_6}\textbf{b}). 
We call this the \emph{lifted schedule}: the coarse assignment at the $4$ corners of each base face is ``lifted'' to the full $L \times L$ fine grid by translational invariance. The scheduling problem thus reduces to a constraint satisfaction problem (CSP) on the base tiling, with one variable per face and a domain of $8$ valid permutations.

Optionally, the domain can be extended by allowing a \emph{global shift}: adding a constant offset $\delta$ to all $4$ timesteps of a face, giving permutations of $\{\delta, \delta{+}1, \delta{+}3, \delta{+}2\}$. This enlarges the domain from $8$ to $8(1 + \delta_{\max})$ permutations per face, increasing the likelihood of finding a satisfying assignment at the cost of additional extraction layers.
For the circuits used in this work, we initially try $\delta=0$.
If that does not yield a solution we try $\delta=1$ and finally fall back to the boundary-first depth-6 schedule explained in \Cref{sec:boundary_first_fallback} if that also fails.

\subsubsection{Compatibility constraints between adjacent faces}

\begin{figure}
    \centering
    \includegraphics[width=\linewidth]{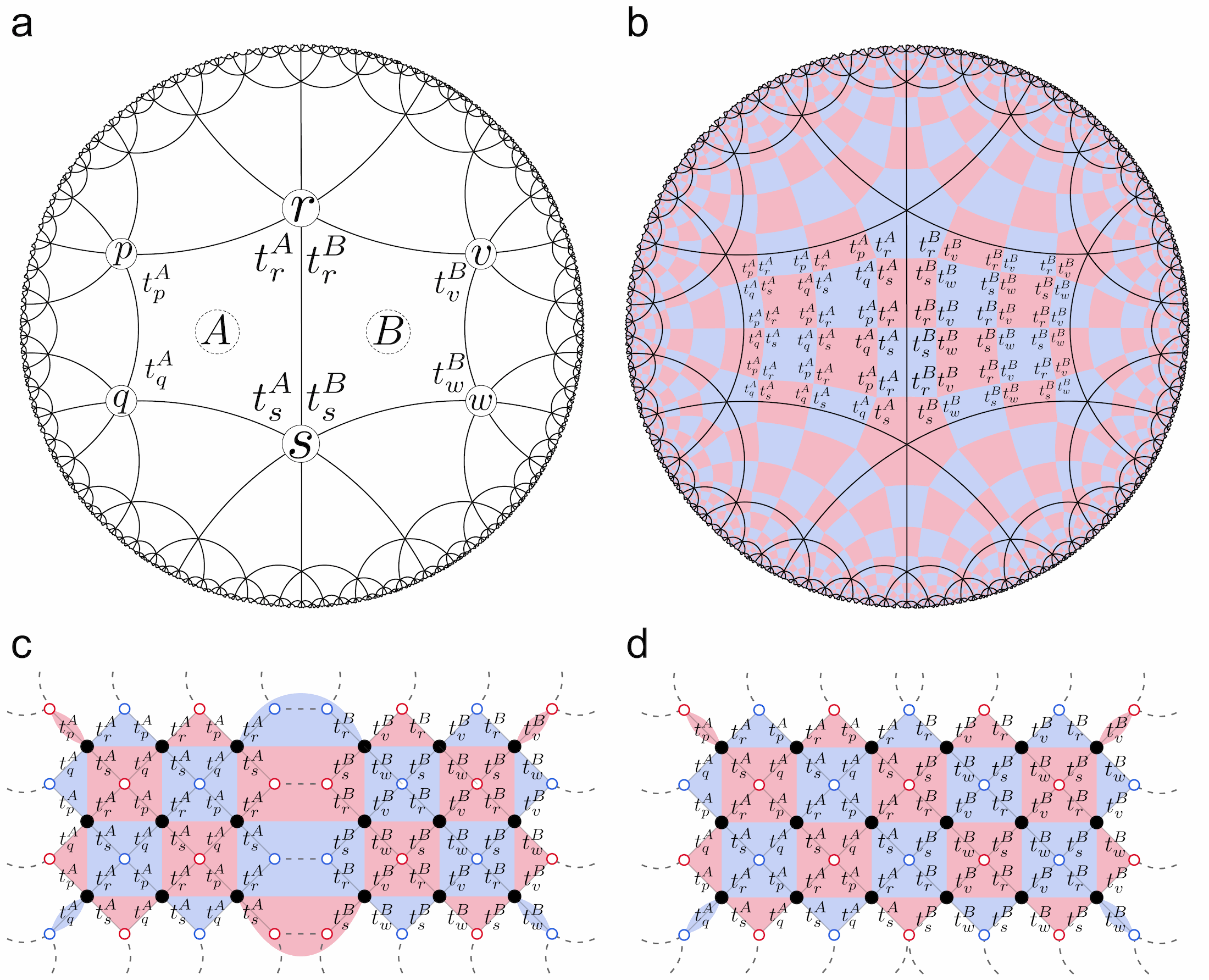}
    \caption{Scheduling constraints for syndrome extraction on modular hyperbolic surface codes. \textbf{a}, Two adjacent faces $A$ and $B$ of a $\{4,6\}$ hyperbolic tiling used to derive a modular hyperbolic surface code. The faces share two vertices $r$ and $s$ and each corner of each face is assigned a timestep. \textbf{b}, The timesteps in the corner of each face of the base tiling are lifted to a translationally-invariant schedule on the fine-grained tiling within the face. \textbf{c}, The translationally invariant schedules within adjacent faces must be compatible. If the two faces belong to different modules, they need only satisfy the commutativity constraint. \textbf{d}, If the two faces additionally belong to the same module, then we must additionally impose time step collision constraints on the stabilizers that span the two base tiling faces $A$ and $B$ (ensuring that the ancillas are acted on at most once in a time step).}
    \label{fig:lifted_schedule_centered_edge_of_4_6}
\end{figure}

When two base faces $A$ and $B$ share an edge with vertices $r$ and $s$, as in \Cref{fig:lifted_schedule_centered_edge_of_4_6}\textbf{a}, the lifted schedules within $A$ and $B$ must be compatible. Let $t_x^F$ denote the time step assigned to vertex $x$ in face $F$ (see \Cref{fig:lifted_schedule_centered_edge_of_4_6}\textbf{a}).
Two types of constraints arise:

\paragraph{Commutativity constraint (all adjacent faces).}
The relative ordering of the timesteps at the shared vertices must be consistent across the two faces (\Cref{fig:lifted_schedule_centered_edge_of_4_6}\textbf{c}). Concretely, for two base faces $A$ and $B$ sharing vertices $r$ and $s$ we require:
\begin{equation}\label{eq:edge_ordering}
    \bigl(t_r^A < t_s^A\bigr) \iff \bigl(t_r^B < t_s^B\bigr).
\end{equation}
This ensures that the lifted translationally-invariant schedule satisfies the commutativity condition \Cref{eq:timesteps_neighboring_X_and_Z_stab} for all pairs of adjacent $X$ and $Z$ stabilizers that lie on the border of the two coarse base faces $A$ and $B$.

\paragraph{Timestep collision constraint (same-module faces).}
When two base faces additionally belong to the same module, there exist fine stabilizers along the shared boundary whose support spans both faces' territories (\Cref{fig:lifted_schedule_centered_edge_of_4_6}\textbf{d}). Each such stabilizer has a single ancilla that is coupled to data qubits in both faces, and this ancilla can participate in at most one CNOT per timestep. This requires that the four timesteps at the shared vertices are all distinct:
\begin{equation}\label{eq:vertex_collision}
    \bigl|\{t_r^A,\, t_s^A,\, t_r^B,\, t_s^B\}\bigr| = 4.
\end{equation}

\subsubsection{Constraint satisfaction via backtracking}

We cast the scheduling problem as a binary CSP and solve it with a backtracking search enhanced by arc consistency. Each face of the base tiling is a variable with a domain of $8$ valid permutations (or more with shifts). Each shared edge imposes the commutativity constraint~\eqref{eq:edge_ordering}, and each shared edge between same-module faces additionally imposes the collision constraint~\eqref{eq:vertex_collision}. The solver uses the AC-3 algorithm to maintain arc consistency, pruning the domain of each variable after every assignment. Variables are selected using the minimum remaining values (MRV) heuristic, and values are ordered by the least constraining value (LCV) heuristic.

\subsubsection{Boundary-first fallback}\label{sec:boundary_first_fallback}

For some base tilings and module partitions, no assignment satisfying both the commutativity constraint~\eqref{eq:edge_ordering} and the collision constraint~\eqref{eq:vertex_collision} exists. In this case we relax the commutativity constraint on cross-module edges (edges shared by faces in different modules), while retaining it on intra-module edges. 
The resulting schedule may have inconsistent gate orderings across module boundaries, which we handle by using a \emph{boundary-first} extraction strategy: the boundary CNOTs used for extraction of inter-module $X$ stabilizers are implemented in earlier time steps, scheduled via a bipartite edge coloring~\cite{tremblay2022constant} to avoid time step collisions, before the remaining CNOTs proceed with the standard lifted schedule.
By measuring the inter-module $X$ stabilizers first, this ensures that the $X$ and $Z$ inter-module stabilizer measurements are not interleaved and hence satisfy the necessary commutativity constraints with their neighbors. 
Since the degree of the Tanner graph restricted to the inter-module $X$ stabilizers is two, these stabilizers can be scheduled in two time steps~\cite{tremblay2022constant}.
Combined with the remaining four time steps of the subsequent lifted schedule this leads to a total extraction circuit depth of six.

\subsubsection{Scheduler benchmarks}

\begin{figure}
    \centering
    \includegraphics[width=\linewidth]{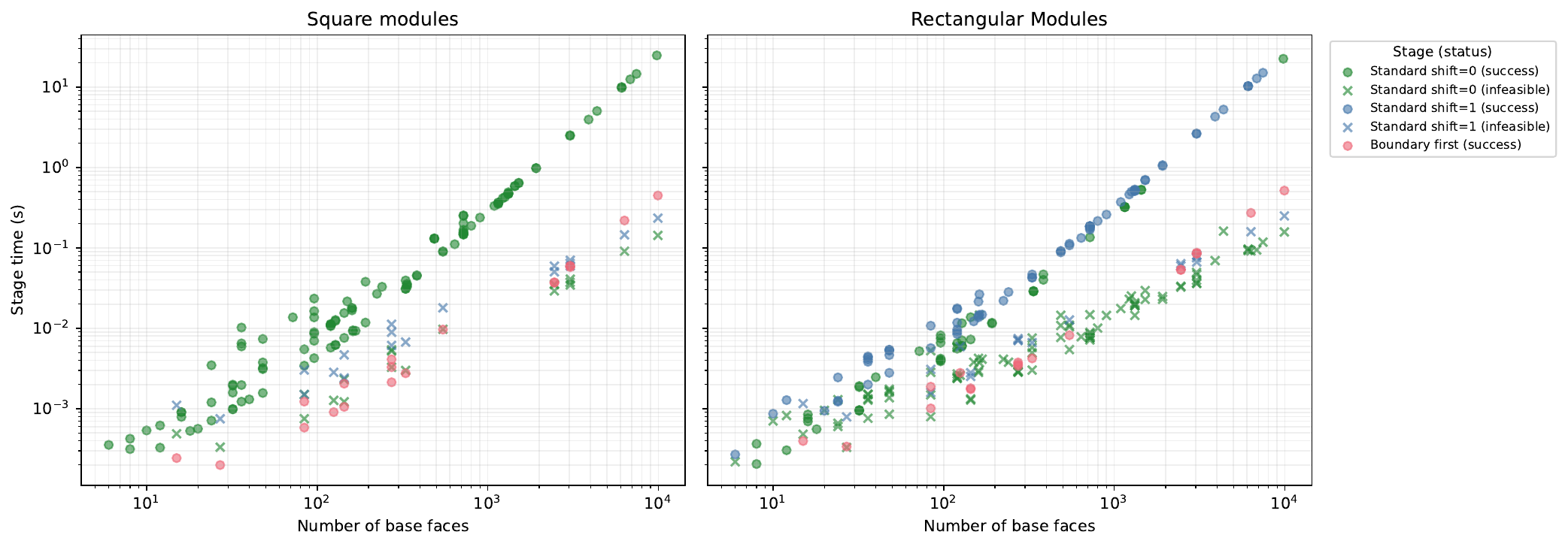}
    \caption{Runtime per stage of the hyperbolic surface code scheduler for modular hyperbolic surface code tilings on the Pareto frontier for $\{4,s\}$ regular tilings with $s\in\{5,6,\ldots,12\}$ and quadrilateral tilings $V4.x.y.z$ for $x,y,z\in \{4,5,\ldots, 16\}$. A tiling is on the Pareto frontier for a side length $L$ if there does not exist (among the tilings we have constructed) a hyperbolic surface code in the same tiling family with fewer faces that has a larger $kd^2/n$. We define the Pareto frontier here for $L=5$ although the scheduling does not use $L$ (it operates directly on the base tiling). The $y$ axis shows the time taken for a given stage of the scheduler per tiling. The $x$ axis is the number of faces in the base hyperbolic tiling. For each tiling, we first try to find a standard interleaved schedule with a shift $\delta=0$. If no schedule exists (``infeasible'') we instead try a shift of $\delta=1$ (resulting in an additional marker on the plot for the tiling, showing the duration of only this stage). If this is also infeasible or times out, we instead fall back to the boundary first fallback, which is guaranteed to succeed and uses 6 time steps for extraction. Again, the additional marker on the plot for the boundary first fallback shows the duration of this stage only. We carry out this analysis for both the square module partition (left figure) and rectangular module partition (right figure), using the same set of tilings for both.}
    \label{fig:scheduler_runtime_per_stage}
\end{figure}

\begin{figure}
    \centering
    \includegraphics[width=0.9\linewidth]{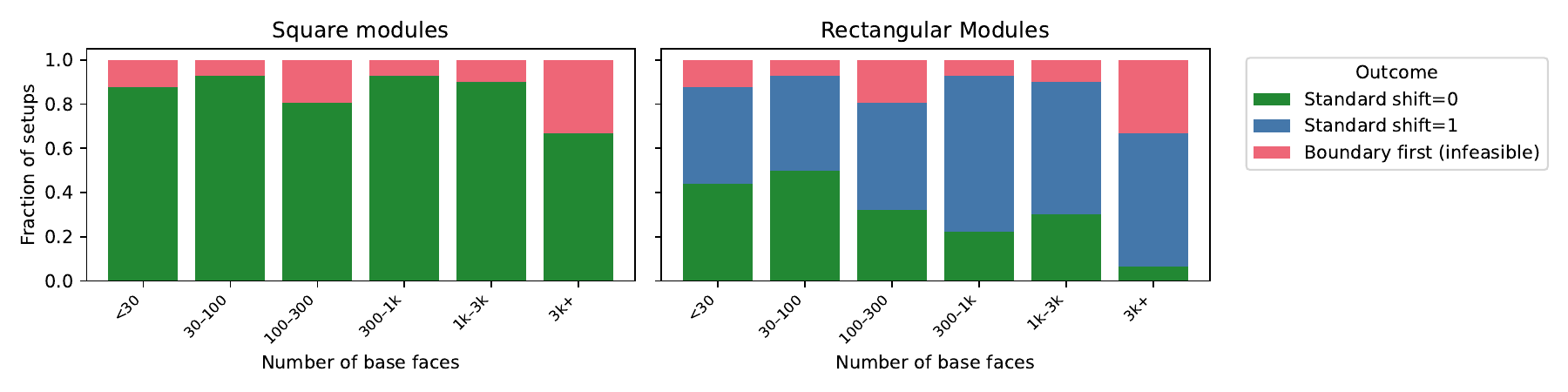}
    \caption{For the same tilings and benchmarks run in \Cref{fig:scheduler_runtime_per_stage}, we show the fraction of tilings that reach each outcome using the modular hyperbolic surface code scheduler. We bin the tilings by their size ($x$ axis), and the ``outcome'' (legend) is the stage of the scheduler that successfully scheduled the circuit. For example the outcome ``Standard shift=1'' means that a shift $\delta=1$ successfully scheduled the circuit (despite the first stage $\delta=0$ failing). The ``Boundary first'' outcome (schedule) is used if the attempted $\delta\in\{
0
,1\}$ standard schedules were infeasible.}
    \label{fig:scheduler_outcome_proportions}
\end{figure}

We benchmarked the scheduler for 137 unique tilings, corresponding to the tilings on a Pareto frontier for each family of regular tilings, as well as quadrilateral tilings.
Specifically, we defined a tiling to be on the Pareto frontier for a given family and side length $L$ if there does not exist in our database a tiling with fewer faces and higher $kd^2/n$ for a modular hyperbolic code with the chosen $L$.
For each tiling, we run 10 random seeds, each of which randomises the tie-breaking order in the MRV and LCV heuristics, and report the median runtime.
The families we choose are regular $\{4,s\}$ tilings for $s\in\{5,6,\ldots,12\}$ and quadrilateral tilings $V4.x.y.z$ for $x,y,z\in \{4,5,\ldots, 16\}$.
For each of these tilings we apply our scheduling method, first attempting to use a standard schedule with $\delta=0$, then trying a standard $\delta=1$ schedule (if $\delta=0$ failed), before finally using the boundary first fallback described in \Cref{sec:boundary_first_fallback} if $\delta=1$ was also infeasible with the standard schedule.
In \Cref{fig:scheduler_runtime_per_stage} we show the runtime of our scheduler (written in Python) for each tiling and each stage of the scheduling procedure, and in \Cref{fig:scheduler_outcome_proportions} we show the fraction of tilings scheduled at each stage.
For all base tilings we consider (with up to $10^4$ quadrilateral faces, i.e.~defining modular hyperbolic codes with up to $10^4$ square modules or $10^4/2$ rectangular modules), we find that the scheduler finds a schedule in less than a minute.
For base tilings with fewer than $10^3$ faces, the scheduler finds a schedule in less than a second.

\subsubsection{Hook errors, NZ and alternating schedules}\label{sec:forward_backward_schedules}

\begin{figure}
    \centering
    \includegraphics[width=0.2\linewidth]{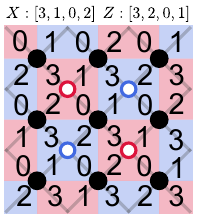}
    \caption{NZ schedule, following the standard schedule for the Euclidean rotated surface code.}
    \label{fig:nz_schedule}
\end{figure}

\begin{table*}[t]
\centering
\small
\setlength{\tabcolsep}{4pt}
\resizebox{0.8\textwidth}{!}{%
\begin{tabular}{lcccccccccccccc}
\toprule
\textbf{Tiling} & \textbf{Module Type} & \textbf{Mods} & $N$ & $k$ & \textbf{Basis} & $d_{\text{code}}$ & \multicolumn{2}{c}{\textbf{ZZ Schedule}} & \multicolumn{2}{c}{\textbf{NZ Schedule}} & \multicolumn{2}{c}{\textbf{Shift} $\delta$} & \multicolumn{2}{c}{\textbf{Boundary Fallback}} \\
\cmidrule(lr){8-9} \cmidrule(lr){10-11} \cmidrule(lr){12-13} \cmidrule(lr){14-15}
 & & & & & & & Std & Alt (fb) & Std & Alt (fb) & ZZ & NZ & ZZ & NZ \\
\midrule
$\{4, 6\}$ & Square & 6 & 366 & 4 & X & 10 & 6 & 6 & 6 & \textbf{8} (+2) & 0 & 0 & No & No \\
$\{4, 6\}$ & Square & 6 & 366 & 4 & Z & 10 & 6 & 6 & 6 & \textbf{8} (+2) & 0 & 0 & No & No \\
$\{4, 8\}$ & Square & 8 & 488 & 6 & X & 10 & 6 & 8 & 6 & 8 (=) & 0 & 0 & No & No \\
$\{4, 8\}$ & Square & 8 & 488 & 6 & Z & 10 & 6 & 8 & 6 & 8 (=) & 0 & 0 & No & No \\
$\{4, 10\}$ & Square & 10 & 610 & 8 & X & 10 & 6 & 6 & 6 & \textbf{8} (+2) & 0 & 0 & No & No \\
$\{4, 10\}$ & Square & 10 & 610 & 8 & Z & 10 & 6 & 6 & 6 & \textbf{8} (+2) & 0 & 0 & No & No \\
$\{4, 12\}$ & Square & 12 & 732 & 10 & X & 10 & 6 & 6 & 6 & \textbf{8} (+2) & 0 & 0 & No & No \\
$\{4, 12\}$ & Square & 12 & 732 & 10 & Z & 10 & 6 & 6 & 6 & \textbf{8} (+2) & 0 & 0 & No & No \\
\midrule
$\{4, 6\}$ & Rectangular & 3 & 348 & 4 & X & 10 & 5 & 6 & 6 & \textbf{7} (+1) & 1 & 0 & No & Yes \\
$\{4, 6\}$ & Rectangular & 3 & 348 & 4 & Z & 10 & 5 & 6 & 6 & \textbf{8} (+2) & 1 & 0 & No & Yes \\
$\{4, 8\}$ & Rectangular & 4 & 464 & 6 & X & 10 & 7 & 8 & 7 & 8 (=) & 0 & 0 & No & No \\
$\{4, 8\}$ & Rectangular & 4 & 464 & 6 & Z & 10 & 7 & 8 & 7 & 8 (=) & 0 & 0 & No & No \\
$\{4, 10\}$ & Rectangular & 5 & 580 & 8 & X & 10 & 5 & 6 & 6 & \textbf{7} (+1) & 1 & 0 & No & Yes \\
$\{4, 10\}$ & Rectangular & 5 & 580 & 8 & Z & 10 & 5 & 6 & 6 & \textbf{8} (+2) & 1 & 0 & No & Yes \\
$\{4, 12\}$ & Rectangular & 6 & 696 & 10 & X & 10 & 7 & 8 & 7 & 8 (=) & 0 & 0 & No & No \\
$\{4, 12\}$ & Rectangular & 6 & 696 & 10 & Z & 10 & 7 & 8 & 7 & 8 (=) & 0 & 0 & No & No \\
\bottomrule
\end{tabular}
}
\caption{Circuit-level distance comparison between standard ZZ and NZ scheduling strategies, comparing standard (non-alternating) rounds with alternating forward-backward (fb) rounds for multi-rate hyperbolic modular surface codes at side length $L=5$ ($r=5$ rounds). Distances are computed using a search for graphlike errors on the detector error model (using the stim method \texttt{stim.Circuit.shortest\_graphlike\_error}~\cite{Gidney2021stimfaststabilizer}). For Square modules, non-alternating rounds cap the circuit distance to $d_{\text{circ}}=6$ for all tilings; round alternation alone is insufficient to raise the distance of the ZZ schedule on $\{4, 6\}$, $\{4, 10\}$, and $\{4, 12\}$, whereas the NZ schedule combined with round alternation recovers $d_{\text{circ}}=8$ across all tilings. For Rectangular modules on $\{4, 6\}$ and $\{4, 10\}$ codes, zero-shift collision-free scheduling is infeasible: the standard ZZ schedule resolves collisions by introducing a shift offset ($\delta=1$), whereas the NZ schedule disables $\delta=1$ and instead uses the boundary-first fallback strategy (yielding an asymmetry where $Z$-basis distance outperforms $X$-basis distance).}
\label{tab:circuit_distances_L5}
\end{table*}

In stabilizer measurement circuits for the Euclidean surface code, an $X$ error on an $X$ stabilizer ancilla half-way through the cycle propagates onto two data qubits (and likewise for $Z$ errors in the middle of a $Z$ stabilizer measurement).
These types of errors are commonly known as hook errors~\cite{dennis2002topological}.
For the rotated surface code, it is known that some gate orderings can lead to hook errors that reduce the circuit distance below the code distance~\cite{tomita2014lowdistance}.

\begin{figure}
    \centering
    \includegraphics[width=\linewidth]{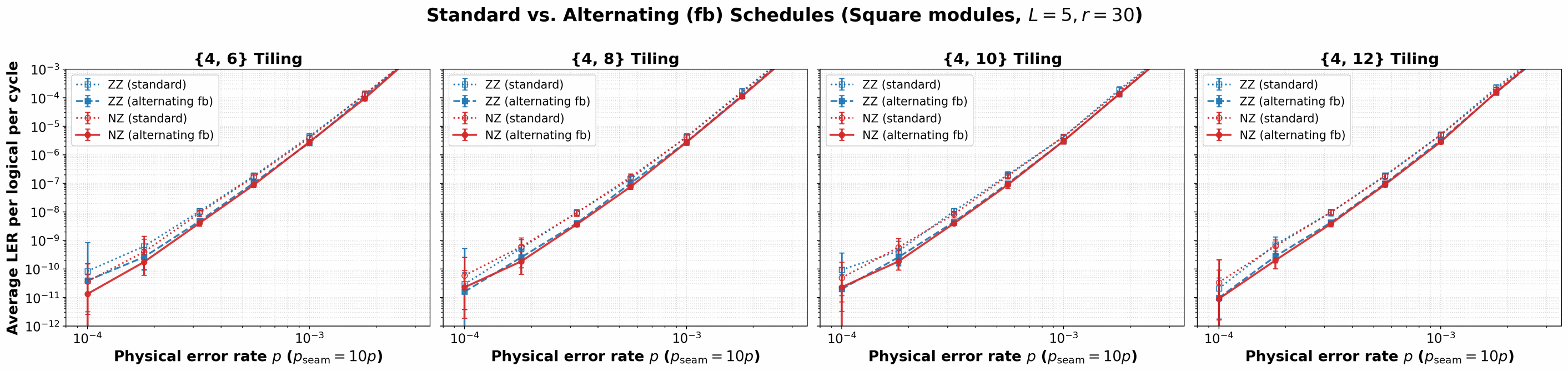}
    \caption{Comparison between standard and alternating forward-backward (fb) schedules for ZZ and NZ across all four tilings for square modules (side length $L=5$, $r=30$ syndrome measurement rounds, averaged over $X$ and $Z$ bases).}
    \label{fig:schedule_comparison_square_modules}
\end{figure}

\begin{figure}
    \centering
    \includegraphics[width=\linewidth]{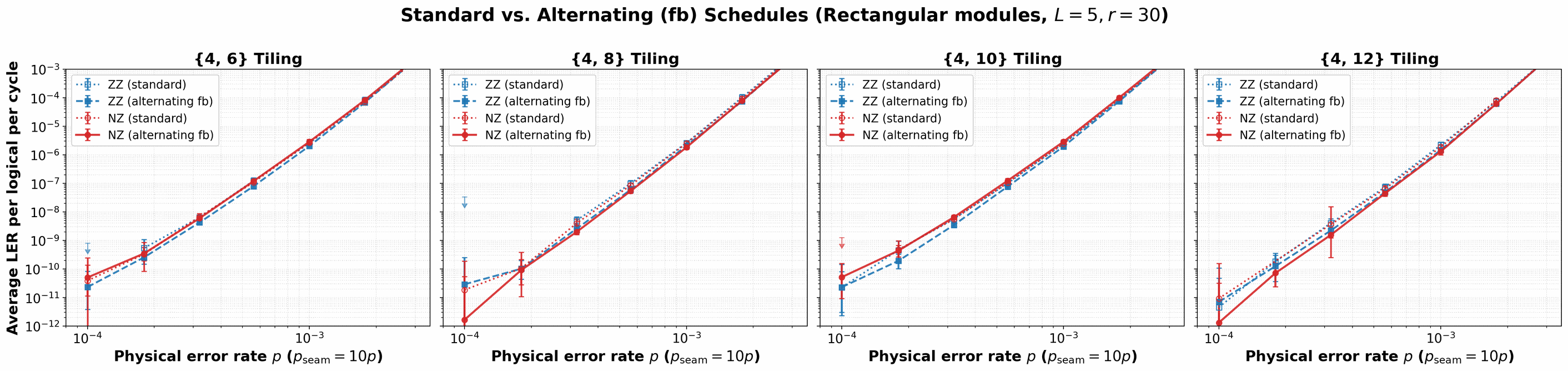}
    \caption{Comparison between standard (non-alternating) and alternating forward--backward (fb) schedules for ZZ and NZ across all four tilings for rectangular modules ($ms=2$, side length $L=5$, $r=30$ syndrome measurement rounds, averaged over $X$ and $Z$ bases).}
    \label{fig:schedule_comparison_rectangular_modules}
\end{figure}

\begin{figure}
    \centering
    \includegraphics[width=0.4\linewidth]{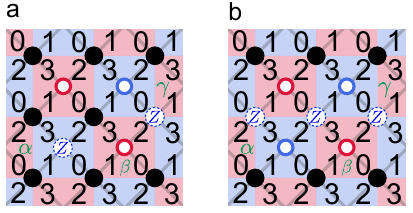}
    \caption{Hook errors in a ZZ schedule of a modular hyperbolic surface code. The two $Z$ errors in \textbf{a}, both occurring between time steps 1 and 2 of round $r$ of the syndrome extraction circuit, are equivalent to the three $Z$ errors in \textbf{b}, all occurring immediately after time step 3 (data qubit errors occurring between round $r$ and round $r+1$). Notice that the left $Z$ error in \textbf{a} flips stabilizer $\alpha$ in round $r+1$ and flips stabilizer $\beta$ in round $r$. This error is therefore diagonal in spacetime, however the right $Z$ error in \textbf{a} \textit{also} flips stabilizer $\beta$ in round $r$, canceling with the left $Z$ error's syndrome. Therefore the combined effect of the two $Z$ errors is a pair of stabilizers $\alpha$ and $\gamma$ occurring in the same round $r+1$, and equivalent to the three data qubit errors occurring between rounds $r$ and $r+1$ in \textbf{b}.}
    \label{fig:zz_hook_errors}
\end{figure}

\begin{figure}
    \centering
    \includegraphics[width=\linewidth]{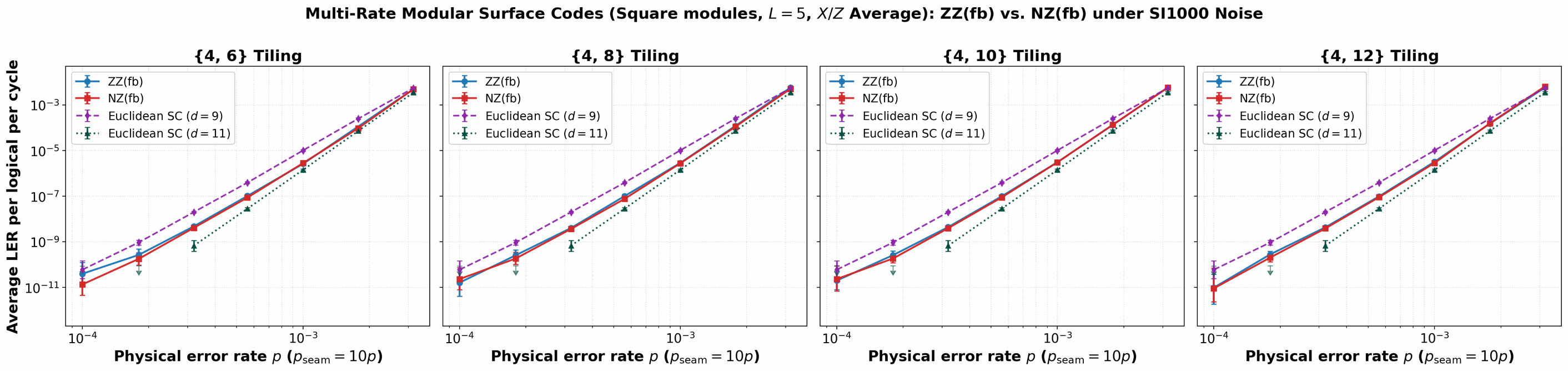}
    \caption{Comparison of alternating ZZ and NZ schedules, for $d=10$, $L=5$ modular hyperbolic surface codes with square module partitions, averaged over $X$ and $Z$ bases. We compare with $d=9$ and $d=11$ Euclidean rotated surface codes using a standard NZ schedule, with a single vertical and horizontal seam (CZ gates that cross the seam have a noise strength $p_{seam}$).}
    \label{fig:nz_ler_comparison_square_modules}
\end{figure}

\begin{figure}
    \centering
    \includegraphics[width=\linewidth]{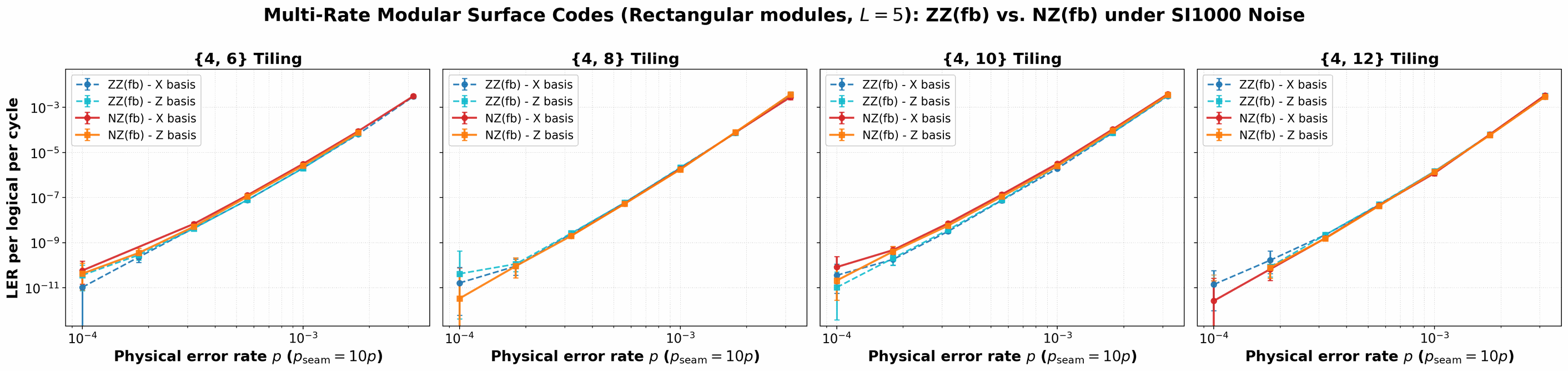}
    \caption{Comparison of alternating ZZ and NZ schedules, for $d=10$, $L=5$ modular hyperbolic surface codes with rectangular module partitions.}
    \label{fig:nz_ler_comparison_rectangular_modules}
\end{figure}

One can avoid distance-reducing hook errors in the Euclidean rotated surface code by choosing a schedule such that the hook errors propagate orthogonal to the relevant logical operator.
However, even a bad gate ordering, with distance-reducing hook errors, can mitigate against some hook errors using an \textit{alternating schedule}~\cite{gidney2025alternating}.
If one starts with a circuit with bad hook errors, such that the circuit distance is $\lceil d/2\rceil$ for a distance $d$ rotated surface code, the circuit distance can be improved to $d-1$ by alternating the order of CNOT gates in consecutive cycles (i.e.~use a reversed ordering of the CNOT gates in odd cycles).
This trick exploits the fact that problematic hook errors in rotated surface code circuits correspond to diagonal errors in the spacetime decoding graph (the problematic errors trigger a pair of detectors that are separated in time by one cycle)~\cite{gidney2025alternating}.
Furthermore, the shortest logical error in such a circuit with hook errors (and without an alternating schedule) consists of a path of these diagonal errors that spans $\lceil d/2\rceil$ cycles.

For our modular hyperbolic surface codes, we would expect any schedule for our qubit layout (single bare ancilla per square plaquette in the bulk) to have distance-reducing hook errors.
To understand why, we first note that any homologically non-trivial cycle on the hyperbolic code supports both an $X$ \textit{and} a $Z$ logical operator.
Suppose a low-weight logical $X$ operator crosses a module of a modular hyperbolic surface code horizontally.
Naively we might try to avoid hooks just by using a standard ``NZ'' rotated surface code schedule (with a different schedule for $X$ and $Z$ plaquettes, see \Cref{fig:nz_schedule}), in the hope of aligning problematic $X$ hook errors orthogonal to the minimum weight logical $X$ operator while keeping $Z$ hook errors parallel.
However, there is \textit{also} a low-weight logical $Z$ operator crossing the module horizontally, so this approach will not work.
Furthermore, for any tiling with a symmetry group that includes the internal symmetries of the square module (e.g.~those derived from $\{4,s\}$ regular tilings), a minimum weight logical operator that passes horizontally through the module can be mapped to another minimum weight logical operator passing vertically across the module via an automorphism (a rotation centered on the module).
Therefore we would not expect a standard ``NZ'' schedule to avoid distance-reducing hook errors in our modular hyperbolic codes, since it is impossible to align hook errors orthogonal to logical operators.

We confirmed the presence of hook errors numerically by finding the shortest graphlike error for some of the circuits of interest, see \Cref{tab:circuit_distances_L5}.
The shortest graphlike error gives an upper bound on the full circuit distance and is efficient to compute (we used the stim method \texttt{stim.Circuit.shortest\_graphlike\_error}~\cite{Gidney2021stimfaststabilizer}).
We denote by ``ZZ'' the standard translationally-invariant schedule for the modular hyperbolic surface codes we have introduced so far, where we give every plaquette within a module the same CNOT ordering regardless of its basis (each ``Z'' denotes the shape drawn by the CNOT order in each of the $X$ and $Z$ basis plaquettes).
As shown in \Cref{tab:circuit_distances_L5}, using the ZZ schedule for $d=10$, $L=5$ regular $\{4,s\}$ modular hyperbolic surface codes, the circuit distance drops to 5, 6 or 7, well below the full code distance of 10.

We mitigate against this distance reduction by making two further modifications to the circuit.
First, we use an alternating schedule by reversing the order of the CNOT gates that couple data qubits to ancilla qubits in the odd rounds.
As shown in \Cref{tab:circuit_distances_L5}, for the square module partition this boosts the distance from 6 to 8 for the $\{4,8\}$ code, however the distance remains 6 for the other three codes.
For the rectangular module partition, the alternating schedule increases the distance by one (up to 6 or 8, depending on the tiling).
The reason that the alternating schedule does not mitigate hook errors in the ZZ schedule as well as it does for the NZ schedule in the Euclidean surface code is shown in \Cref{fig:zz_hook_errors}.
Specifically, \Cref{fig:zz_hook_errors}\textbf{a} shows a pair of $Z$ errors that trigger only two detectors, both in the \textit{same} cycle, and which are equivalent to three $Z$ errors on data qubits lying parallel with a logical operator.
Since chains of hook errors of this form are not diagonal in the spacetime detector error model (the chain does not traverse time), the alternating schedule is unable to mitigate against it.
Nevertheless, the alternating schedule does still lead to a modest improvement in the logical error rate, see \Cref{fig:schedule_comparison_square_modules} and \Cref{fig:schedule_comparison_rectangular_modules}.

Despite the circuit distance dropping below the code distance even with the alternating schedule, the logical error rate performance is still excellent, matching what we would expect for using a Euclidean surface code at a distance equal to the hyperbolic code's full code distance.
This is shown in \Cref{fig:nz_ler_comparison_square_modules}, where we see that the ZZ alternating schedule circuits outperform $d=9$ Euclidean surface codes and almost match $d=11$ Euclidean surface code performance, despite the circuit distance being 6 for the $\{4,6\}$, $\{4,10\}$ and $\{4,12\}$ codes.
Remarkably, our simulations of the distance 20 $\{4,10\}$ modular hyperbolic surface code, shown in \Cref{fig:frontier_codes_plots} of the main text, show that its logical error rate outperforms a distance-21 Euclidean surface code, despite the inevitable presence of hook errors that reduce the circuit distance below the code distance of 20 (we used ZZ alternating multi-rate schedules for these simulations).

While the ZZ alternating schedule circuits already have good performance, we also implemented an ``NZ'' alternating schedule, in order to further mitigate against hook errors, and in the hope that this might additionally translate into improved logical error rates.
Our NZ schedule adapts the standard rotated surface code extraction pattern~\cite{tomita2014lowdistance} to modular hyperbolic surface codes by assigning distinct CNOT interaction orders to the $X$ and $Z$ plaquettes within one base face of the original hyperbolic tiling, see \Cref{fig:nz_schedule}.
In cyclic order around each fine-grained square plaquette, $X$ stabilizers follow an ``N''-shaped schedule (traversing vertically first, i.e.~time steps $P_X\coloneqq [3,1,0,2]$ for corners [NW,NE,SE,SW]), while $Z$ stabilizers follow a ``Z''-shaped schedule (traversing horizontally first, e.g.~$P_Z\coloneqq [3,2,0,1]$).
This NZ circuit does not have the problematic, purely ``spatial'', pair of hook errors shown in \Cref{fig:zz_hook_errors}, allowing the alternating schedule to be more effective at hook error mitigation in the bulk of a base face.

To lift this schedule to the full, fine-grained modular hyperbolic code, each face of the base tiling is assigned a joint $X$/$Z$ schedule pair $(g\cdot P_X, g\cdot P_Z)$, obtained by applying the same dihedral symmetry $g\in D_4$ (one of the 4 rotations or reflections) simultaneously to both seed schedules (each a permutation of $\{0,1,2,3\}$).
This preserves the relative orientation between $X$ and $Z$ schedules across the fine grid within each base face.
The global schedule is found using the same backtracking CSP solver described in \Cref{sec:hyperbolic_surface_code_circuits_si}: each base face represents a variable with a domain of the 8 joint $D_4$ permutations, with edge commutativity constraints adapted to the NZ schedule to ensure \Cref{eq:timesteps_neighboring_X_and_Z_stab} holds for $X$ and $Z$ stabilizers spanning adjacent base modules.
For adjacent base faces within a module, rather than using only the time step collision constraint (\Cref{eq:vertex_collision}), we instead enforce that different base faces within the same module have translationally invariant schedules (i.e.~all $X$ plaquettes within a module have the same translationally invariant schedule, and similarly for $Z$).
Since we wanted to enforce translational invariance within a module, we fixed the shift $\delta=0$ (rather than falling back to $\delta=1$ as we did for the ZZ schedule).
However when the $\delta=0$ NZ scheduling fails, we do still use the boundary first fallback.

The circuit distances of the NZ schedule, with and without using an alternating schedule, are shown in \Cref{tab:circuit_distances_L5}.
Across the codes benchmarked, the NZ alternating schedule has the best circuit distance among the different schedules, as expected, typically achieving a circuit distance of 8 (for a distance 10 code).
This corresponds to losing a distance of 1 across each of the two modules that wraps around a handle of the hyperbolic surface.
When the boundary fallback is used, the distance drops to 7.
This is expected, since the boundary fallback un-interleaves the module boundary stabilizer measurement CNOTs in one basis, so a hook error that would have been diagonal in spacetime becomes purely spacelike (and as such the alternating schedule cannot mitigate against it).
Regardless, as shown in \Cref{fig:nz_ler_comparison_square_modules} and \Cref{fig:nz_ler_comparison_rectangular_modules}, we find that the NZ alternating schedule has essentially the same logical error rate performance as the ZZ alternating schedule, despite the different circuit distances.
This further confirms that, while hook errors are present in these modular hyperbolic surface code circuits, they are so improbable that they have no noticeable impact on logical error rates at practical physical error rates.

\subsubsection{Multi-rate measurement schedules}\label{sec:multi_rate_circuits}\label{sec:pipelined_circuits}

\begin{figure}[t]
    \centering
    \includegraphics[width=\linewidth]{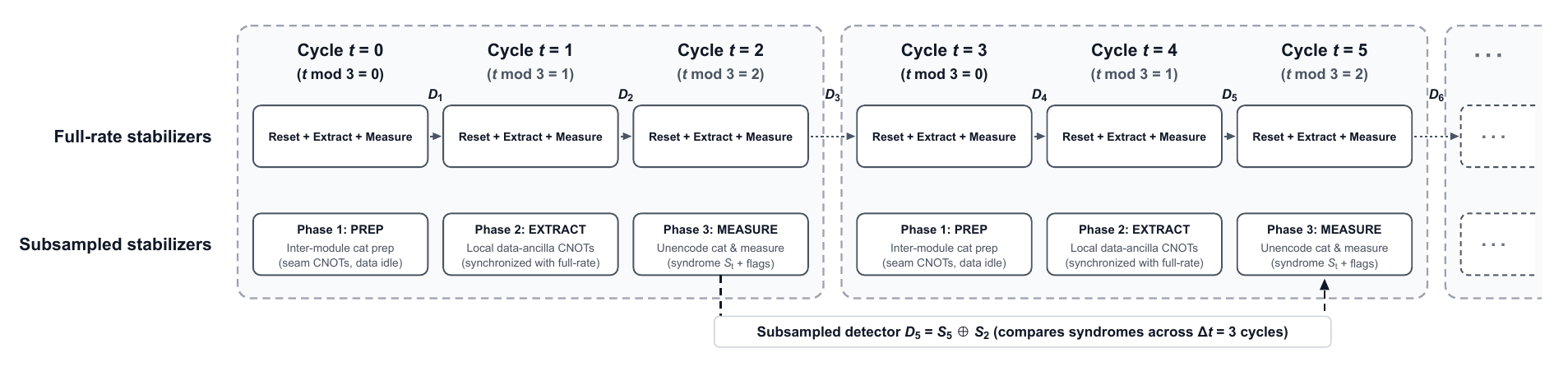}
    \caption{Timeline of a multi-rate syndrome extraction circuit for modular hyperbolic codes.
    Full-rate stabilizers (including intra-module bulk stabilizers and inter-module edge stabilizers extracted via two-qubit Bell pairs) are extracted and measured in every cycle ($t = 0, 1, 2, \dots$), with detectors $D_t$ comparing syndromes between consecutive cycles ($D_t = S_t \oplus S_{t-1}$).
    Subsampled corner stabilizers, which have higher weight and require multi-qubit cat states with preparation and measurement depth scaling with the cat size across multiple module seams, are executed across a 3-cycle sequence ($t \bmod 3 \in \{0, 1, 2\}$):
    Phase~1 ($t \bmod 3 = 0$) prepares the multi-qubit cat state across seams while data qubits remain idle;
    Phase~2 ($t \bmod 3 = 1$) applies local data-ancilla CNOT gates, synchronized with the full-rate schedules;
    Phase~3 ($t \bmod 3 = 2$) unencodes the cat state and measures the root ancilla (syndrome $S_t$) alongside leaf flag ancillas.
    Subsampled detectors compare syndromes across $\Delta t = 3$ cycles ($D_t = S_t \oplus S_{t-3}$, e.g.~$D_5 = S_5 \oplus S_2$).}
    \label{fig:multi_rate_circuit_timeline}
    \label{fig:pipelined_circuit_timeline}
\end{figure}

Rather than measuring all stabilizers in every cycle, we measure some higher-weight stabilizers less frequently.
Concretely, any stabilizer in the code associated with a vertex of the original base hyperbolic tiling is measured once every three cycles.
As an example, consider a square module partition of a $\{4,s\}$ regular hyperbolic surface code.
In the modular hyperbolic code, each degree $s$ vertex in the base tiling lifts to a weight-$s$ stabilizer in the semi-hyperbolic code, which is shared by the $s$ square modules at their corners (see \Cref{fig:hyperbolic_surface_adjacent_plaquettes}\textbf{a} for the stabilizer structure of a $\{4,s\}$, $L=5$ modular hyperbolic code, as an example).
Preparing the cat state with the linear encoding circuit we use requires $\lceil \frac{s}{2}\rceil$ layers of inter-module gates.
We will refer to these high-weight stabilizers as ``corner'' stabilizers.
Since the corner stabilizers require both high-weight cat states \textit{and} the use of inter-module couplers, which are expected to be noisier and slower, they are therefore much slower to measure than the weight-four stabilizers fully supported within a module.
If we were to measure all stabilizers every cycle, the data qubits would be idling while the corner stabilizers' cat states are being prepared.

Instead, we measure the higher-weight, non-local stabilizers less frequently~\cite{Berthusen2024partialsyndrome,gupta2026boundary}.
Concretely, in our \textit{multi-rate} circuit, the corner stabilizers are measured once every three cycles, while all remaining stabilizers are measured every cycle.
A timeline of a multi-rate syndrome extraction circuit is shown in \Cref{fig:multi_rate_circuit_timeline}.
The ``full-rate'' stabilizers (all stabilizers aside from corner stabilizers) are measured in full every cycle.
The ``subsampled'' stabilizers (corner stabilizers) are measured across three cycles.
In a cycle $t$ for which $t\mod 3 = 0$, the ancillas of the subsampled stabilizers are reset and inter-module gates are used to unitarily prepare a cat state on the shared ancilla qubits.
In cycles where $t\mod 3 = 1$, local CNOT (or CZ) gates within a module are used to couple data and ancilla qubits, synchronized for all stabilizers (subsampled and full-rate).
Finally, for cycles $t\mod 3 = 2$, the inverse of the unitary cat-state encoding circuit is applied to the ancilla qubits of the subsampled stabilizers, followed by measurement of all the ancillas.
One of these ancillas is used to measure the stabilizer, while the remaining $s-1$ ancillas are flag qubits.
We still use the alternating (forward-backward) schedule for the multi-rate circuits by reversing the order of the data-ancilla CNOT (or CZ) gates in cycles for which $t\mod 2 = 1$.

\begin{figure}
    \centering
    \includegraphics[width=\linewidth]{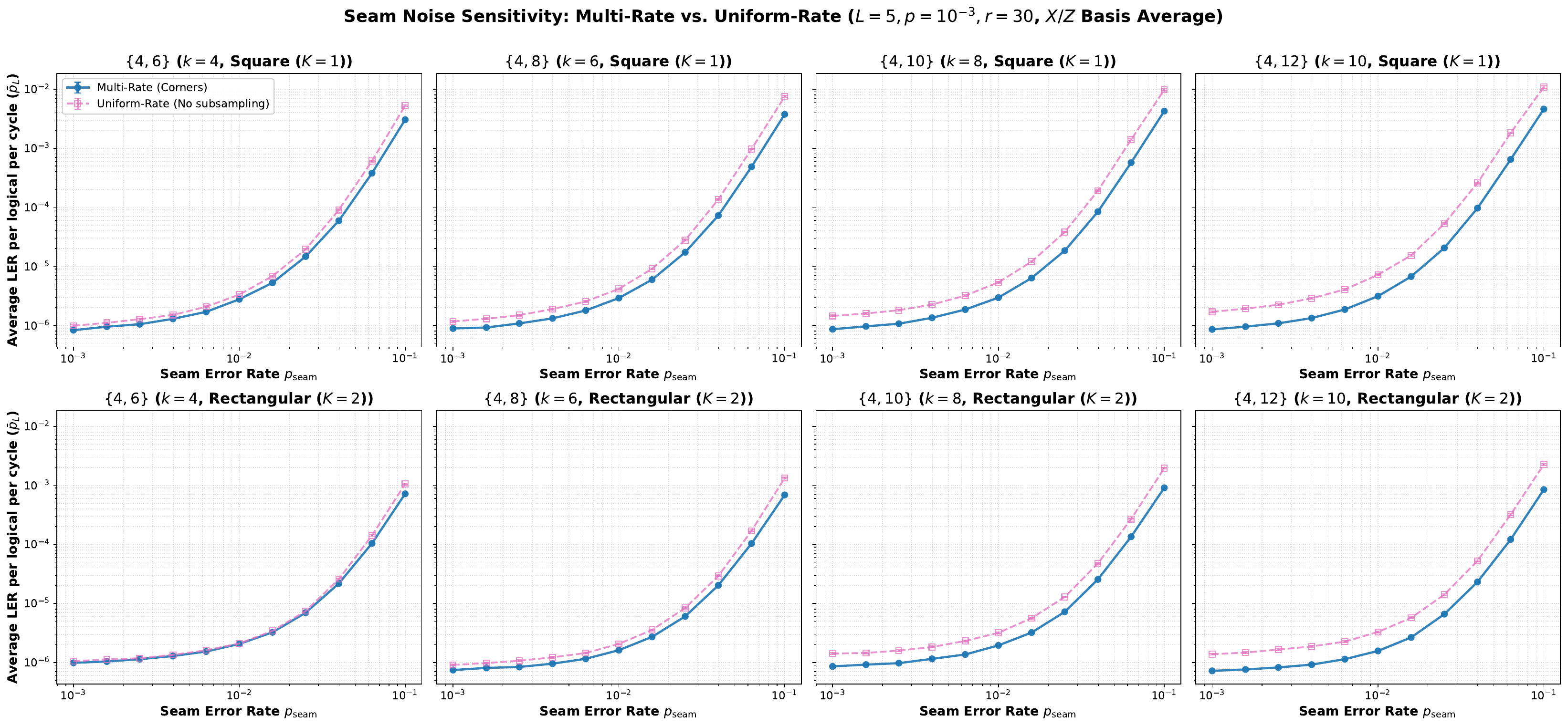}
    \caption{Simulations comparing uniform and multi-rate syndrome extraction circuits for $L=5$, $d=10$ regular $\{4,s\}$ modular hyperbolic surface codes. The top panel of four plots uses a square module partitioning, whereas the bottom four plots use a rectangular module partitioning. The multi-rate circuits measure the corner stabilizers once every three cycles, whereas the uniform-rate schedule measures all stabilizers every cycle. For all circuits we use an alternating (forward-backward) schedule for the data-ancilla CNOT/CZ ordering.}
    \label{fig:uniform_vs_multi_rate_ler_vs_p_seam}
\end{figure}

In \Cref{fig:uniform_vs_multi_rate_ler_vs_p_seam} we show simulations that compare the multi-rate and uniform rate syndrome extraction circuits for small regular $\{4,s\}$ modular hyperbolic surface codes with $L=5$ and for $s\in\{6,8,10,12\}$.
We find that the multi-rate circuits, for which corner stabilizers are measured every three cycles, have better (lower) logical error rates than the uniform circuits for which every stabilizer is measured each cycle.
The improvement increases with the stabilizer weight $s$.
This improvement can be explained by the fact that the uniform schedule requires the data qubits to idle while the corner stabilizers' cat states are encoded and unencoded.

\section{Symmetrical Canonical Logical Basis for Hyperbolic Codes}\label{sec:logical_basis}
The logical operators for hyperbolic color and surface codes are associated with the topology of the underlying closed hyperbolic tiling. In particular, they are related to the homologically non-trivial cycles of the surface. 

\begin{figure}[H]
    \centering
    \includegraphics[width=0.6\linewidth]{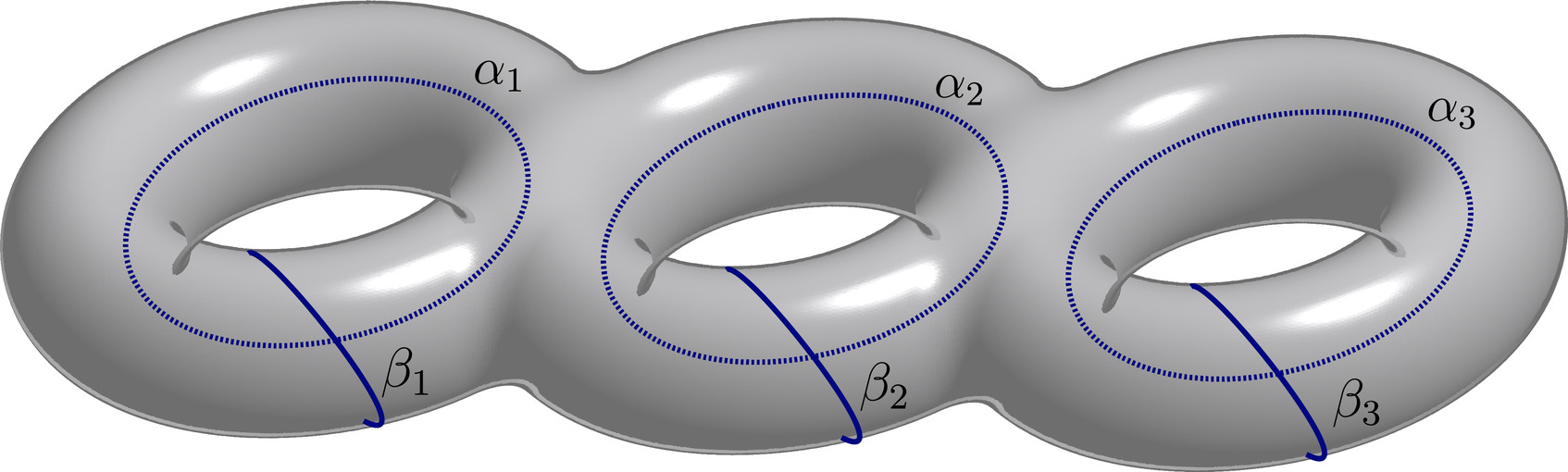}
    \caption{Canonical homology basis on an orientable genus-3 surface $\Sigma_3$. Each handle $i$ supports a symplectic pair of cycles $(\alpha_i,\beta_i)$. Partner cycles in each pair cross each other an odd number of times and cross cycles of other symplectic pairs an even number of times.}
    \label{fig:genus-3}
\end{figure}

A closed \emph{orientable} surface $\Sigma_g$ of genus $g$ has $2g$ independent homologically non-trivial cycles and these cycles can be grouped into $g$ pairs, called \emph{symplectic pairs}, in order to define a \emph{canonical homology basis} $\mathcal{B} = \{(\alpha_1, \beta_1), \dots, (\alpha_g, \beta_g)\}$. We define the \emph{intersection number} $\langle \alpha, \beta \rangle \in \mathbb{F}_2$ of two non-trivial cycles $\alpha$, $\beta$ as their number of geometric crossings modulo $2$ and we say that the two cycles \emph{cross} when this number is $1$ (when they geometrically cross each other an odd number of times). In a canonical homology basis, symplectic pairs cross an odd number of times only with their partner cycle and cross an even number of times with any cycle from another symplectic pair. \Cref{fig:genus-3} illustrates an example for an arbitrary genus-3 surface.

On a tiled surface, a cycle is a closed loop of edges. The cycle is homologically trivial if the edges of the cycle form the boundary of a set of tiling faces. Otherwise, it is homologically non-trivial (it wraps around a handle of the surface). Since stabilizers are defined on the faces of the underlying hyperbolic tiling for hyperbolic color and surface codes, homologically trivial cycles of the hyperbolic tiling correspond to the stabilizers of the code, while the non-trivial cycles are associated with the logical operators. However, the precise way a logical Pauli operator is constructed from a non-trivial cycle differs between color and surface codes, as we explain in the rest of this section.

\subsection{Color Code Logical Basis}\label{sec:color_code_logical_basis}
\begin{figure}[H]
    \centering
    \includegraphics[height=0.95\textheight]{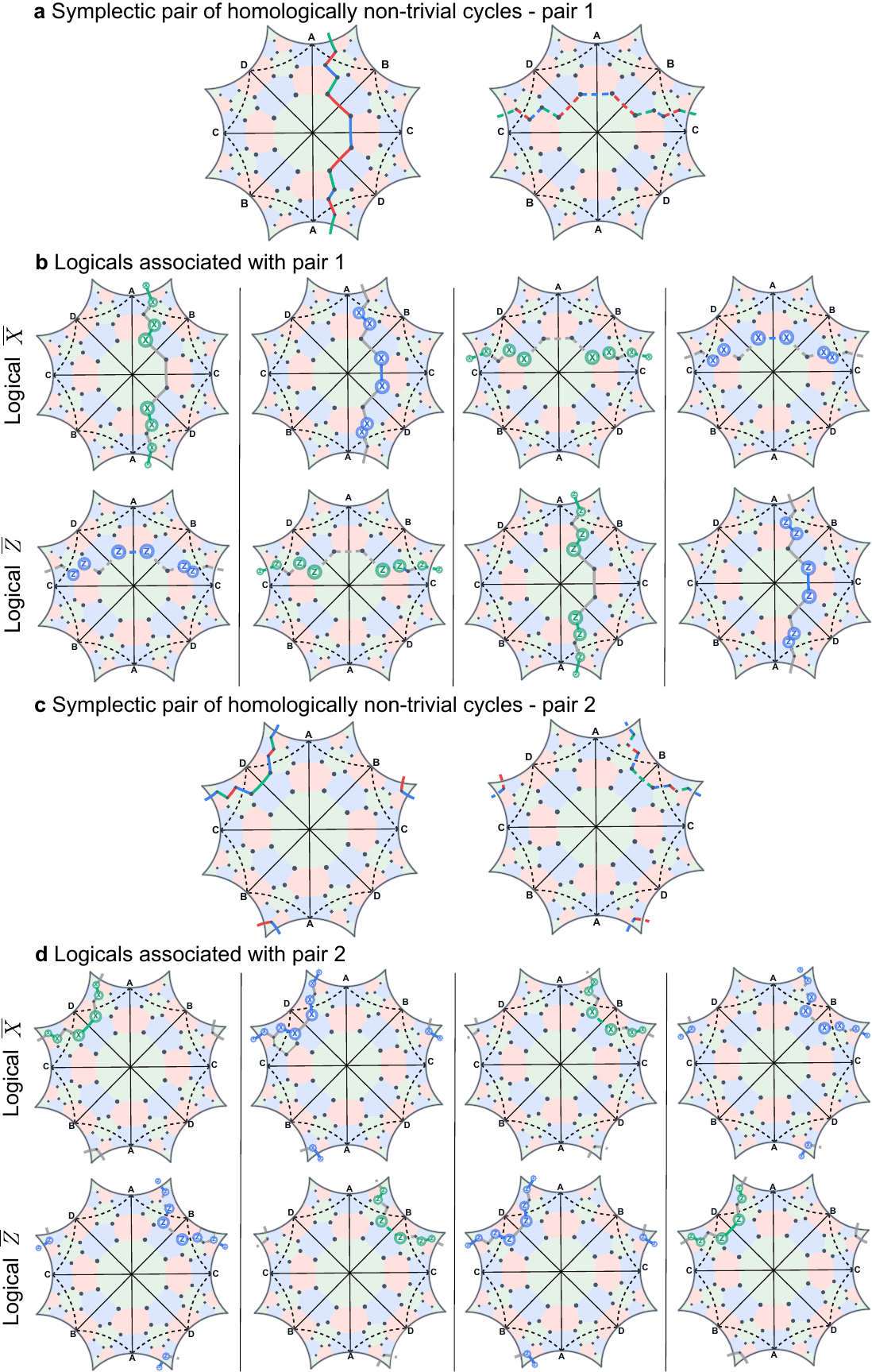}
    \caption{Canonical logical basis for the genus-$2$ semi-hyperbolic color code $8.8.8$-$16$-$L2$. \textbf{(a,c)} Two symplectic pairs of homologically non-trivial cycles on the tiling. \textbf{(b,d)} The $4$ logical $\overline{X}$ (top) and $4$ logical $\overline{Z}$ (bottom) operators derived from each pair by restricting the cycle support to green and blue edges, giving $8$ logical qubits in total.}
    \label{fig:bolza_l2_logical_basis}
\end{figure}

For hyperbolic color codes, we associate the logical operators with the homologically non-trivial cycles on the primal three-colorable tiling $\mathcal{T}$ (see \Cref{fig:bolza_l2_logical_basis}), using the colored structure of the edges. Every edge of the color code tiling separates two faces of the same color, and we assign it that color through the map $c : E \to \{R, G, B\}$, so that red faces are separated by red edges, and likewise for green and blue. A homologically non-trivial cycle $\alpha$ consists of edges of all three colors, whereas a cycle whose edges use only two colors bounds a set of faces and is homologically trivial~\cite{bombin_color,quantum_pin_codes}.

Following Ref.~\cite{bombin_color}, a logical Pauli operator of the color code is obtained from a non-trivial cycle $\alpha$ by restricting its support to a single color. For each color $C \in \{R, G, B\}$, the \emph{monochromatic substring}
\begin{equation}
    \alpha_C \coloneqq \{\, e \in \alpha \mid c(e) = C \,\}
\end{equation}
is the set of color-$C$ edges of $\alpha$, and we write $V(\alpha_C) \subseteq V$ for the set of vertices incident to an edge in $\alpha_C$. The substring $\alpha_C$ supports both an $\overline{X}$- and a $\overline{Z}$-type logical operator,
\begin{equation}
    \overline{X}_C^{\alpha} \coloneqq \bigotimes_{v \in V(\alpha_C)} X_v,
    \qquad
    \overline{Z}_C^{\alpha} \coloneqq \bigotimes_{v \in V(\alpha_C)} Z_v,
\end{equation}
obtained by placing an $X$ (respectively $Z$) on the two data qubits at the endpoints of each color-$C$ edge of $\alpha$.

This restriction to color-$C$ edges is equivalent to the shrunk-lattice construction of Ref.~\cite{bombin_color}. There, a color-$C$ string is defined on the shrunk lattice, obtained by contracting every color-$C$ face of $\mathcal{T}$ to a single vertex and keeping only the color-$C$ edges. We  define logical operators on monochromatic substrings throughout, and it underlies the weighted shortest-cycle search we use to compute color-code distances in \Cref{sec:color_distance_upper_bound}.

Monochromatic substring operators of $X$ or $Z$ type supported on the same non-trivial cycle are not all independent. The product of any two of the same Pauli type equals the third up to stabilizers, for example $\overline{X}^{\alpha}_{R} \simeq \overline{X}^{\alpha}_{B}\,\overline{X}^{\alpha}_{G}$. The cycle $\alpha$ therefore carries four independent logical operators, for example $\overline{X}^{\alpha}_{B}$, $\overline{X}^{\alpha}_{G}$, $\overline{Z}^{\alpha}_{B}$ and $\overline{Z}^{\alpha}_{G}$. The operators on $\alpha$ mutually commute, since any two substrings on the same cycle overlap on an even number of qubits. The anticommuting operators are instead supported across the symplectic pair $(\alpha,\beta)$, where a substring on $\alpha$ and a substring on $\beta$ of different color and different Pauli type overlap an odd number of times, e.g. $\overline{X}^{\alpha}_{B}$ anticommutes with $\overline{Z}^{\beta}_{G}$, $\overline{X}^{\alpha}_{G}$ with $\overline{Z}^{\beta}_{B}$, and similarly with the roles of $\alpha$ and $\beta$ exchanged (see \Cref{fig:bolza_l2_logical_basis}). This groups the eight operators obtained from $\alpha$ and $\beta$ into four anticommuting pairs, defining four logical qubits per symplectic pair,
\begin{equation}
\begin{aligned}
    \overline{X}_1 &:= \overline{X}^{\alpha}_{G}, & \overline{Z}_1 &:= \overline{Z}^{\beta}_{B}, \\
    \overline{X}_2 &:= \overline{X}^{\alpha}_{B}, & \overline{Z}_2 &:= \overline{Z}^{\beta}_{G}, \\
    \overline{X}_3 &:= \overline{X}^{\beta}_{G}, & \overline{Z}_3 &:= \overline{Z}^{\alpha}_{B}, \\
    \overline{X}_4 &:= \overline{X}^{\beta}_{B}, & \overline{Z}_4 &:= \overline{Z}^{\alpha}_{G}.
\end{aligned}
\end{equation}

\begin{figure}[H]
    \centering
    \includegraphics[width=\linewidth]{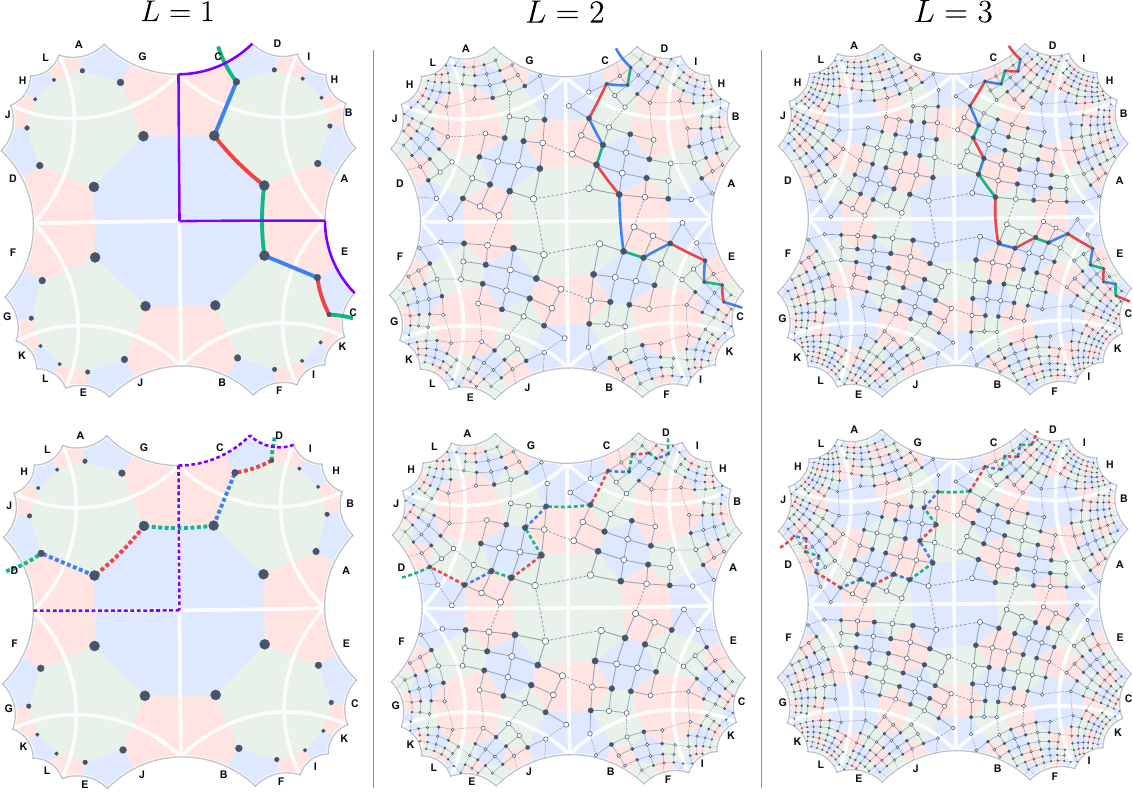}
    \caption{A homologically non-trivial seed cycle (solid RGB lines) and its symplectic partner (dashed RGB line) on the $8.8.8$-$32$ semi-hyperbolic color code, shown for fine-graining parameters $L=1$, $L=2$ and $L=3$ using a quadrilateral ($K{=}2$) module partition with $M=|G^+|=16$ modules. At $L=1$ (left), both the primal cycle (solid RGB lines) and the corresponding dual cycle (purple) are shown on the base hyperbolic tiling. At $L=2$ (middle) and $L=3$ (right), the primal cycle is lifted to the fine-grained tiling, where it follows the module boundaries along the path traced by the dual cycle. Together, the symplectic pair $(\alpha,\beta)$ defines four logical qubits via the monochromatic substrings of any two of the three edge colors.}
    \label{fig:iter40_finegrained_seedloop}
\end{figure}

Starting from a canonical homology basis $\mathcal{B} = \{(\alpha_1, \beta_1), \dots, (\alpha_g, \beta_g)\}$ of $g$ symplectic pairs and defining the Pauli operators on their monochromatic substrings using two colors yields a symplectic basis for the $k = 4g$ logical qubits of the hyperbolic color code. Furthermore, fine-graining a hyperbolic tiling with a Euclidean tiling does not change the topology of the surface, therefore, we can always search for a canonical homology basis on the primal hyperbolic tiling and then lift its non-trivial cycles to the fine-grained semi-hyperbolic tiling (we describe the lifting procedure in detail in \Cref{sec:homology_basis}). Fine-graining does, however, increase the length of the non-trivial cycles, and hence the number of qubits in the support of the logical operators, boosting the code distance. \Cref{fig:iter40_finegrained_seedloop} illustrates this for a symplectic pair $(\alpha,\beta)$ of the $8.8.8$-$32$ semi-hyperbolic color code: the same pair of non-trivial cycles is shown for fine-graining parameters $L=1$, $L=2$ and $L=3$. As $L$ increases, each cycle is lifted to a longer path on the fine-grained tiling, so that the logical operators in its support grow in weight while their homology class is preserved.

In our work we require an additional structure from the canonical logical basis. We construct it using the symmetries of the tiling, particularly the orientation-preserving automorphisms $\mathrm{Aut}^+(\mathcal{T}) = G^+ = \langle \rho, \sigma \rangle$ that correspond to global rotations and translations of the tiling. We restrict to the orientation-preserving symmetries since we can realize these symmetries as fault-tolerant automorphism circuits with minimal additional seam connectivity (described in \Cref{sec:automorphism_circuits}). The module partition we focus on is quadrilateral ($K=2$) modules for which the modules are in correspondence with group elements of $G^+$. Our aim is therefore a \emph{symmetric} basis, in which the homologically non-trivial cycles are related to one another by the action of $G^+$, so that the corresponding logical operators can be mapped onto one another by these automorphism circuits. We build the basis from \emph{seed cycles}. First we enumerate a pool of short seed-cycle candidates: we perform a graph search from every vertex of the tiling, find and classify each non-trivial cycle by its homology class, and keep the shortest representative of each class. We order the cycles in the candidate pool by their length, starting from the systole which is the shortest non-trivial cycle of the tiling. We then take the orbit of a seed cycle under $G^+$. Rotating the tiling maps the seed to other non-trivial cycles of equal length, sweeping it across many homology classes and giving a pool of equal-length candidates from which we search for $g$ cycles that pair up into a symplectic basis. We try the systole first, then proceed through successive seed cycles in order of increasing weight, until the orbits together contain a complete basis. We give a complete account of how the symmetric canonical homology basis is constructed in \Cref{sec:homology_basis}, explaining non-trivial cycle enumeration, the generation of seed-cycle orbits under the orientation-preserving group $G^+ = \langle \rho, \sigma \rangle$, and the symplectic-pair search.

\begin{figure}[H]
    \centering
    \includegraphics[width=0.75\linewidth]{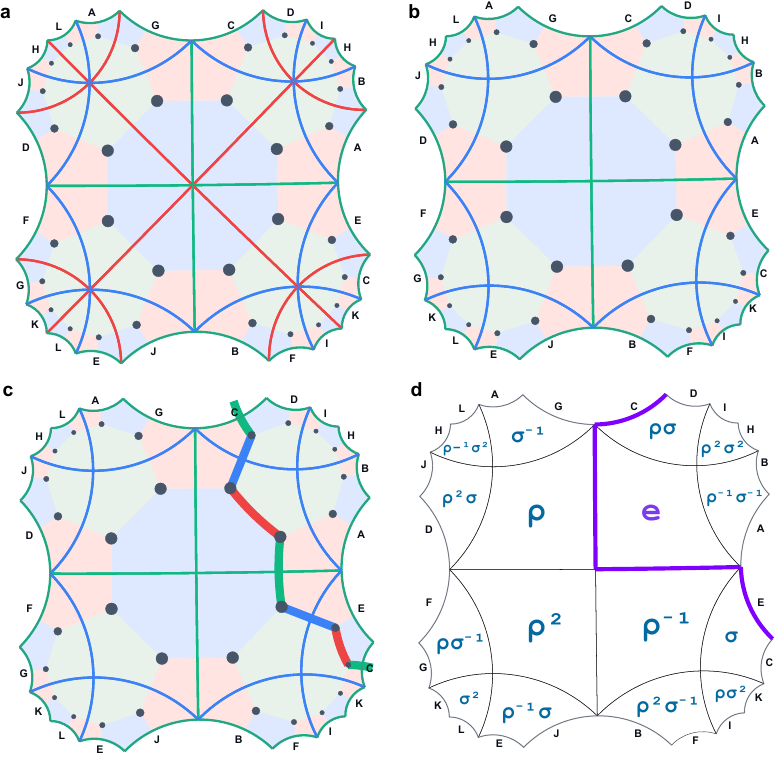}
    \caption{Construction of a dual cycle along module boundaries for the $8.8.8$-$32$ color code with a quadrilateral ($K{=}2$) module partition. \textbf{a}~The base triangular tiling $\mathcal{T}^\triangle$ with its three edge colors (overlaid on the three-colorable primal tiling $\mathcal{T}$). In the $K{=}2$ partition, each quadrilateral module is formed by pairing adjacent triangles across edges of a chosen color, here the red edges. \textbf{b}~Removing the red edges from the dual tiling leaves a reduced graph consisting of only the blue and green dual edges, on which dual cycles lying on module boundaries can be found. \textbf{c}~A homologically non-trivial primal cycle (RGB edges). \textbf{d}~The corresponding non-trivial dual cycle (purple) on the quadrilateral module partition, with each module labeled by its group element $g \in G^+= \langle \rho, \sigma \rangle$.}
    \label{fig:iter40_dual_loops_along_module_boundaries}
\end{figure}

\begin{figure}[H]
    \centering
    \includegraphics[width=0.8\linewidth]{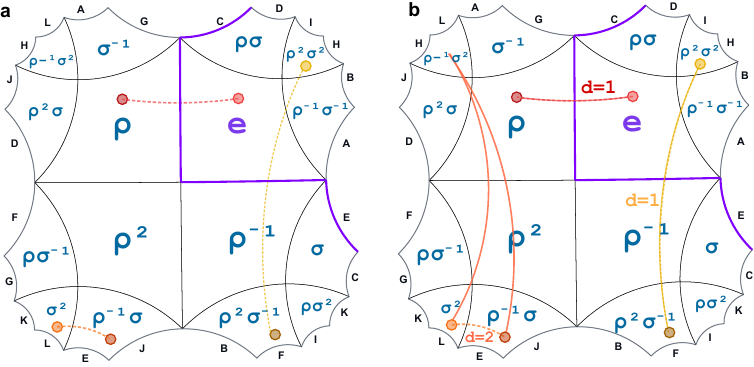}
    \caption{Symplectic pairs and automorphism paths for the $8.8.8$-$32$ color code (genus $g=3$) with a quadrilateral ($K{=}2$) module partition. Each quadrilateral module is labeled by an element of $G^+ = \langle \rho, \sigma \rangle$ with $|G^+|=16$. \textbf{a}~The dual seed cycle $\gamma$ (purple) is associated with the identity element $e$. Its orbit under $G^+$ contains six homologically non-trivial cycles of three symplectic pairs, where each cycle $g\gamma$ is labeled by the group element $g$ that maps the seed cycle to it via left multiplication (the automorphism). For example, the cycle labeled by $\rho$ is the image of $\gamma$ under a global $\rho$ rotation. Dashed lines connect the two cycles within each symplectic pair. \textbf{b}~Automorphism paths between symplectic partners on the Cayley graph of $G^+$. Each label $d$ for a pair $(g\gamma,h\gamma)$ denotes the distance between $g$ and $h$ on the right Cayley graph, corresponding to the minimum number of elementary automorphism moves $\rho, \rho^{-1}, \sigma, \sigma^{-1}$ needed to bring one cycle under an extractor that is measuring its symplectic partner. These distances determine the depth of the automorphism circuits required to rotate between cycles within a symplectic pair. In general $G^+$ is not Abelian, although the quotient $G^+ \cong \mathbb{Z}_4 \times \mathbb{Z}_4$ defining this particular tiling happens to be Abelian.}
    \label{fig:iter40_autpaths}
\end{figure}

Since the modules of the quadrilateral partition correspond to elements of $G^+$, the symmetric structure of the canonical logical basis has a natural description on the module partition, which we illustrate in \Cref{fig:iter40_dual_loops_along_module_boundaries} and \Cref{fig:iter40_autpaths} for the $8.8.8$-$32$ color code. Recall from \Cref{sec:quadrilateral_modules} that each quadrilateral module is formed by pairing adjacent triangles of the base triangular tiling across an edge of one chosen color (here the red edges). Removing the dual edges of that color leaves a reduced dual graph whose remaining (blue and green) dual edges lie along the module boundaries, so a non-trivial cycle can be represented by its dual cycle running along these boundaries. Lifting a primal cycle to the fine-grained tiling, it follows the module boundaries along the path traced by this dual cycle, and the corresponding logical operator can be supported on either side of the boundary. Furthermore, the non-trivial dual cycles can be labeled by group elements: associating a chosen seed cycle with the identity element $e$, its orbit under $G^+$ produces the remaining cycles in its orbit, each labeled by the group element $g\in G^+$ that maps the seed onto it. The group elements corresponding to the non-trivial cycles that form a canonical basis are shown as colored nodes in \Cref{fig:iter40_autpaths} which makes explicit how every cycle in the basis is obtained from the seed by a global rotation of the tiling. For the $8.8.8$-$32$ color code (genus $g=3$) the orbit of a \emph{single} seed cycle under $G^+$ produces six non-trivial cycles that form all three symplectic pairs, so this one seed suffices to construct the full logical basis. The dashed lines in \Cref{fig:iter40_autpaths}a show the partner cycles of each symplectic pair and \Cref{fig:iter40_autpaths}b illustrates the right Cayley distance between them. If the ribbon extractor is attached to the seed cycle $\gamma$, the cycle $x\gamma$ is brought under it by the automorphism $x^{-1}$, so switching from measuring $x\gamma$ to measuring $y\gamma$ requires the automorphism $y^{-1}x$. Its word length $|x^{-1}y|$ in the generators $\{\rho^{\pm1},\sigma^{\pm1}\}$ is the distance between $x$ and $y$ in the right Cayley graph of $G^+$. This is the minimum number of elementary moves $\rho, \rho^{-1}, \sigma, \sigma^{-1}$ needed to bring a cycle under the extractor when it is measuring the partner cycle, and determines the depth of the automorphism circuits required to switch between the cycles within a symplectic pair. Note that this in general differs from the left Cayley distance $|yx^{-1}|$, which is the length of the automorphism $yx^{-1}$ mapping $x\gamma$ onto $y\gamma$. We discuss the general case of an extractor attached to a cycle $h\gamma$ in \Cref{sec:ripple_carry_supplement} (see \Cref{eq:move_cost}).

\begin{figure}[H]
    \centering
    \includegraphics[width=1.0\linewidth]{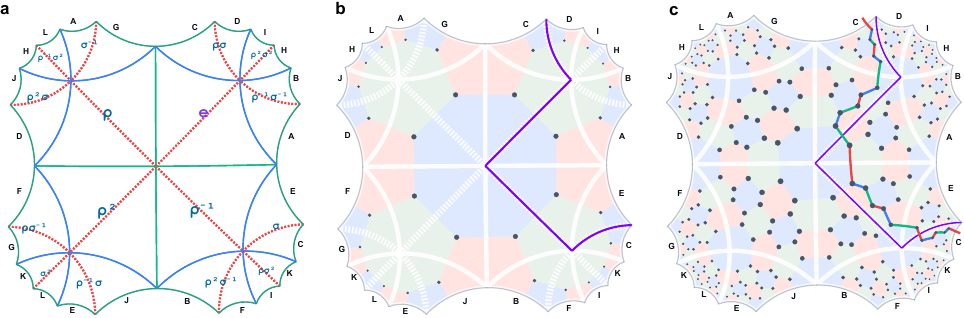}
    \caption{An alternative dual cycle that cuts through the middle of the modules, for the $8.8.8$-$32$ color code with a quadrilateral ($K{=}2$) module partition. This is the counterpart to \Cref{fig:iter40_dual_loops_along_module_boundaries}, where the dual cycle instead runs \emph{along} the module boundaries. \textbf{a}~The base triangular tiling $\mathcal{T}^\triangle$ with each quadrilateral module labeled by its group element $g \in G^+ = \langle \rho, \sigma \rangle$. Each module pairs two triangular tiles across a red edge. Rather than removing the red edges as in \Cref{fig:iter40_dual_loops_along_module_boundaries}, here we search for the non-trivial dual cycle \emph{on} the red edges, so that it passes through the interior of the modules. \textbf{b}~The resulting non-trivial dual cycle (purple) on the hyperbolic tiling ($L=1$), passing through the middle of the modules in its support. \textbf{c}~The dual cycle (purple) together with its primal seed cycle (RGB edges) lifted to the fine-grained $8.8.8$-$32$-$L3$ tiling. Only the modules in the support of the dual cycle are split into two triangular modules, exposing an internal boundary for logical measurement, while all other modules remain quadrilateral.}
    \label{fig:seed_loop_module_middle}
\end{figure}

The dual cycle described above runs \emph{along} the module boundaries, obtained by removing the red edges and searching the reduced blue--green dual graph. An alternative is to search on the red edges instead, giving a dual cycle that passes through the \emph{interior} of the modules in its support, as shown in \Cref{fig:seed_loop_module_middle}. The modules are still defined by pairing triangles across red edges, but now each module crossed by the cycle is split into its two constituent triangular modules along that red edge. This exposes an internal module boundary through which the logical operator can be accessed, which is convenient for logical measurement with the ribbon extractor of \Cref{sec:ribbon_extractor}, since it separates the seam used for logical readout from the seams used for memory and automorphisms. Only the modules in the support of the seed cycle are split in this way, while all other modules remain quadrilateral.

A single seed orbit does not always yield a complete basis. When it does not, we may accept a partial basis, in which fewer than $g$ symplectic pairs, and hence fewer than the full $k = 4g$ logical qubits, are covered by the orbit of a single cycle. This may sometimes be a favorable trade-off. The covered logicals are all related to one seed cycle by automorphisms, and can be measured with a single ribbon extractor (\Cref{sec:ribbon_extractor}), with the remaining logicals handled separately. For the $8.8.8$-$400$ code, for example, one seed cycle covers $100$ of the $104$ logical qubits, so a single extractor suffices for $100$ logical qubits. To search for a complete basis, we can enlarge the candidate pool by taking the combined orbits of pairs of seed cycles, then triplets, and so on. \Cref{tab:homology_automorphism_logicals} reports how many seed cycles are used for a range of $8.8.8$ color codes and how many logicals their orbits generate.

To summarize, the canonical basis of symplectic pairs is built from the orbits of seed cycles, and this structure significantly simplifies logical measurement. Each seed cycle and its orbit under $G^+$ is handled by a single ribbon extractor (\Cref{sec:ribbon_extractor}), which measures the logical operators on the seed cycle, while the automorphism circuits (\Cref{sec:automorphism_circuits}) map the images in the orbit back onto it. The logical operators are associated with non-trivial cycles, the smallest of which is the systole, the shortest non-trivial cycle of the tiling. Since the systole scales as $O(\log|G|)$, this upper bounds the weight of the shortest logical operators to $O(\log|G|)$ base tiles (see \Cref{sec:distance_finding}). After fine-graining, these tiles are grouped into modules, so the shortest logical operators are supported on only $O(\log M)$ of the $M$ modules, with weight $O(L\log M)$ at fine-graining parameter $L$. Each such operator can moreover be made to lie on the module boundaries (\Cref{fig:iter40_dual_loops_along_module_boundaries}) or cut through the middle of the module which is then split into two triangular modules (\Cref{fig:seed_loop_module_middle}). Logical measurement then requires only minimal additional seam connectivity, never addressing qubits in the bulk, and the modules themselves remain planar.

\begin{table}[H]
    \centering
    \setlength{\tabcolsep}{10pt}
    \renewcommand{\arraystretch}{1.2}
    \begin{tabular}{lccccc}
        \toprule
        \textbf{Code} & \textbf{$[[n,k,d]]$} & \textbf{Seeds} & \textbf{Logicals} & \textbf{Coverage} & \textbf{Weight} \\
        \midrule
        $8.8.8$-$16$  & $[[16, 8, 2]]$   & $2$ & $4 + 4$ & $8/8$     & $4$ \\
        $8.8.8$-$32$  & $[[32, 12, 4]]$  & $1$ & $12$    & $12/12$   & $4$ \\
        $8.8.8$-$128$ & $[[128, 36, 8]]$ & $1$ & $32$    & $32/36$   & $8$ \\
        $8.8.8$-$400$ & $[[400, 104, 8]]$& $1$ & $100$   & $100/104$ & $12$ \\
        \bottomrule
    \end{tabular}
    \caption{Symmetric canonical logical bases for selected $8.8.8$ hyperbolic color codes, constructed from the automorphism orbits of seed cycles. \textbf{Seeds} is the number of seed cycles used. \textbf{Logicals} is the number of logical operators generated by their orbits under the orientation-preserving automorphisms $\langle\rho,\sigma\rangle$, all of equal \textbf{Weight}. \textbf{Coverage} gives these against the total number of logical qubits of the code $k = 4g$.}
    \label{tab:homology_automorphism_logicals}
\end{table}

\subsection{Surface Code Logical Basis} \label{sec:surface_code_logical_basis}
\begin{figure}[h]
    \centering
    \includegraphics[width=0.85\linewidth]{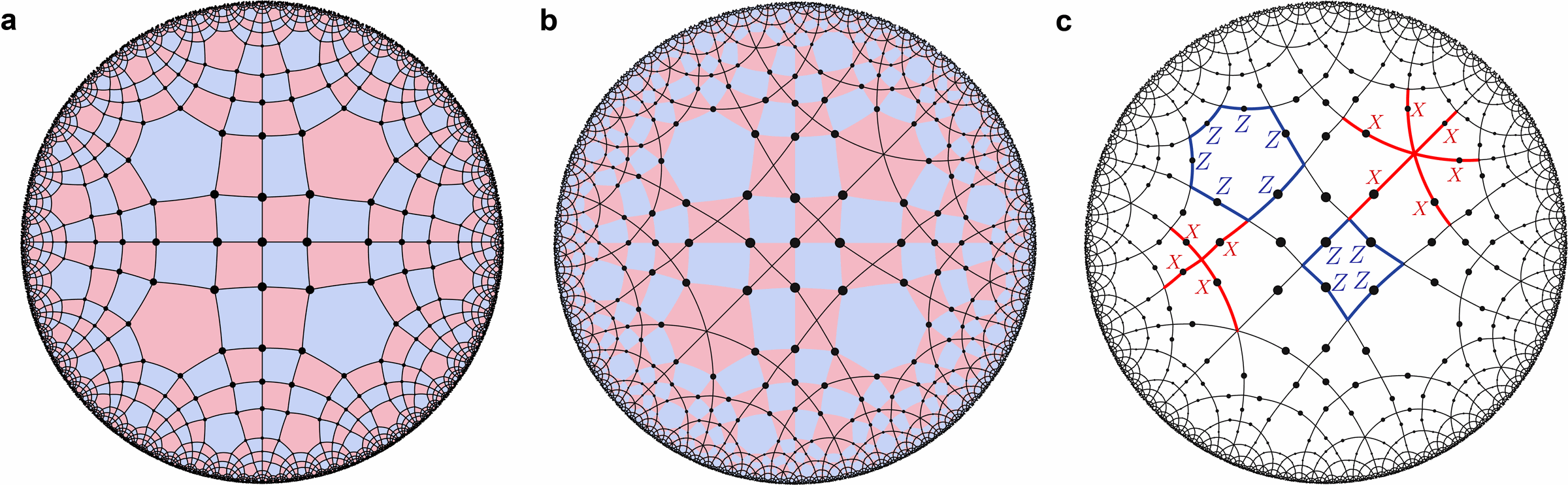}
    \caption{Cell complex description of a semi-hyperbolic surface code. \textbf{(a)} The two-colorable tiling  $(\mathcal{T}^\square_L)^*$ with red and blue faces. \textbf{(b)} Cell complex $\mathcal{C}$ overlaid on the two-colorable $(\mathcal{T}^\square_L)^*$ tiling. Red faces are shrunk to vertices and new edges connect the new vertices. Qubits are now associated with edges of the tiling. \textbf{(c)} Cell complex tiling $\mathcal{T}_{\textit{cell}}$: $Z$-stabilizers are associated with faces, data qubits with edges and $X$-stabilizers are associated with vertices. Examples of weight-$4$ and weight-$6$ stabilizer generators are shown. Homologically trivial cycles of $\mathcal{C}$ are products of $Z$-stabilizers and non-trivial cycles give logical $\overline{Z}$ operators, while on the dual cell complex $\mathcal{C}^*$ trivial cycles are products of $X$-stabilizers and non-trivial cycles give logical $\overline{X}$ operators.}
    \label{fig:surface_code_cell_complex_tiling}
\end{figure}

Unlike color codes, hyperbolic surface codes are not self-dual, and their $X$- and $Z$-type operators are supported on different structures. In the two-colorable representation, the stabilizers are supported on the faces of the tiling, with the two face colors carrying the two stabilizer types, e.g. red faces support $X$-stabilizers and blue faces support $Z$-stabilizers. To describe the logical operators it is convenient to pass to the equivalent cell complex representation, obtained by collapsing every red face to a vertex and associating data qubits with the edges connecting the new vertices, as illustrated in \Cref{fig:surface_code_cell_complex_tiling}. In the resulting cell complex $\mathcal{C}$, the $2$-cells (faces) support $Z$-stabilizers, the $1$-cells (edges) support data qubits, and the $0$-cells (vertices) support $X$-stabilizers. This is captured by the chain complex
\begin{equation}
    C_2 \xrightarrow{\;\partial_2\;} C_1 \xrightarrow{\;\partial_1\;} C_0,
\end{equation}
where $H_Z = \partial_2^{\mathsf{T}}$ and $H_X = \partial_1$.

On $\mathcal{C}$, homologically trivial cycles are products of $Z$-stabilizers while homologically non-trivial cycles give logical $\overline{Z}$ operators. On the dual cell complex $\mathcal{C}^*$, trivial cycles are products of $X$-stabilizers while non-trivial cycles give logical $\overline{X}$ operators. A logical $\overline{Z}$ on a cycle and a logical $\overline{X}$ on a cocycle anticommute precisely when the cycle and cocycle cross, i.e. when their intersection number is $1$. To build the canonical basis we take $g$ symplectic pairs of non-trivial cycles $(\alpha_1,\beta_1),\dots,(\alpha_g,\beta_g)$ on $\mathcal{C}$. Each $\alpha_i,\beta_i$ carries a logical $\overline{Z}$, giving $2g$ mutually commuting $\overline{Z}$ operators. The anticommuting $\overline{X}$ partners are the dual cocycles: $\alpha_i^*$ crosses $\beta_i$ and $\beta_i^*$ crosses $\alpha_i$, so $(\overline{Z}_{\beta_i},\overline{X}_{\alpha_i^*})$ and $(\overline{Z}_{\alpha_i},\overline{X}_{\beta_i^*})$ form the $2g$ anticommuting logical qubit pairs, giving $k = 2g$ logical qubits on a genus-$g$ surface.

\Cref{fig:surface_code_cell_complex_tiling} shows a semi-hyperbolic surface code together with its cell complex description. For the canonical basis, however, we work with the base quadrilateral tiling $\mathcal{T}^\square$: we search for a canonical homology basis there, related by automorphisms as for the color codes, and then interpret the $X$- and $Z$-type logical operators by relating the cycles to the cell complex tiling. The combinatorial problem we solve is therefore the same for hyperbolic color and surface codes, and is further explained in \Cref{sec:homology_basis}.

\section{Ribbon Extractor for Logical Measurement} \label{sec:ribbon_extractor}

We utilize a technique called code surgery, which is a generalization of surface code lattice surgery~\cite{horsman2012surface}, to access the logical information stored in our modular hyperbolic memory. 
This access is provided in the form of fault-tolerant measurements of logical Pauli operators, which can be utilized to implement different computational primitives such as teleportation.
The high-level idea can be summarized as follows: given a logical operator we intend to measure, we initialize a set of ancilla data qubits and define a new stabilizer code called the \textit{deformed code} (supported on the original data qubits and ancilla data qubits). 
In the deformed code, the operator we intend to measure (which is of Hamming weight at least the code distance $d$) will become a product of new stabilizer checks (which are often of constant Hamming weight). 
Therefore, by measuring the constant-weight checks of the deformed code, we also measure the high-weight logical operator of the original code.
Fault tolerance of this procedure is then guaranteed by guaranteeing the distance of the deformed code is not reduced and by repeating the measurement of the deformed code for sufficiently many rounds (typically $d$ rounds).

The space- and time-costs of surgery procedures are determined by the ancilla systems and the associated deformed codes.
Construction of these systems and codes has been the focus of a line of recent works~\cite{cohen2022low-overhead,cross2024improved,Williamson2026,ide2024fault,swaroop2026universal,he2025extractorsqldpcarchitecturesefficient,yuan2026parsimonious}, for which we refer readers to section~3 of Ref.~\cite{he2025extractorsqldpcarchitecturesefficient} as a brief review. 
For our semi-hyperbolic codes, we construct surgery gadgets called \textit{ribbon extractors} using techniques from~\cite{he2025extractorsqldpcarchitecturesefficient,cohen2022low-overhead,cross2024improved}. 
In this section, we detail the construction of these extractors, the deformed codes we use for measurements, and their architectural utilities.
Our presentation will be specific to ribbon extractors on hyperbolic color codes with $8.8.8$ tiling. 
The same methods can be applied to hyperbolic color codes with other tilings or hyperbolic surface codes.

\subsection{Construction of the Extractor Systems}
\begin{figure}[H]
    \centering
    \includegraphics[width=0.8\linewidth]{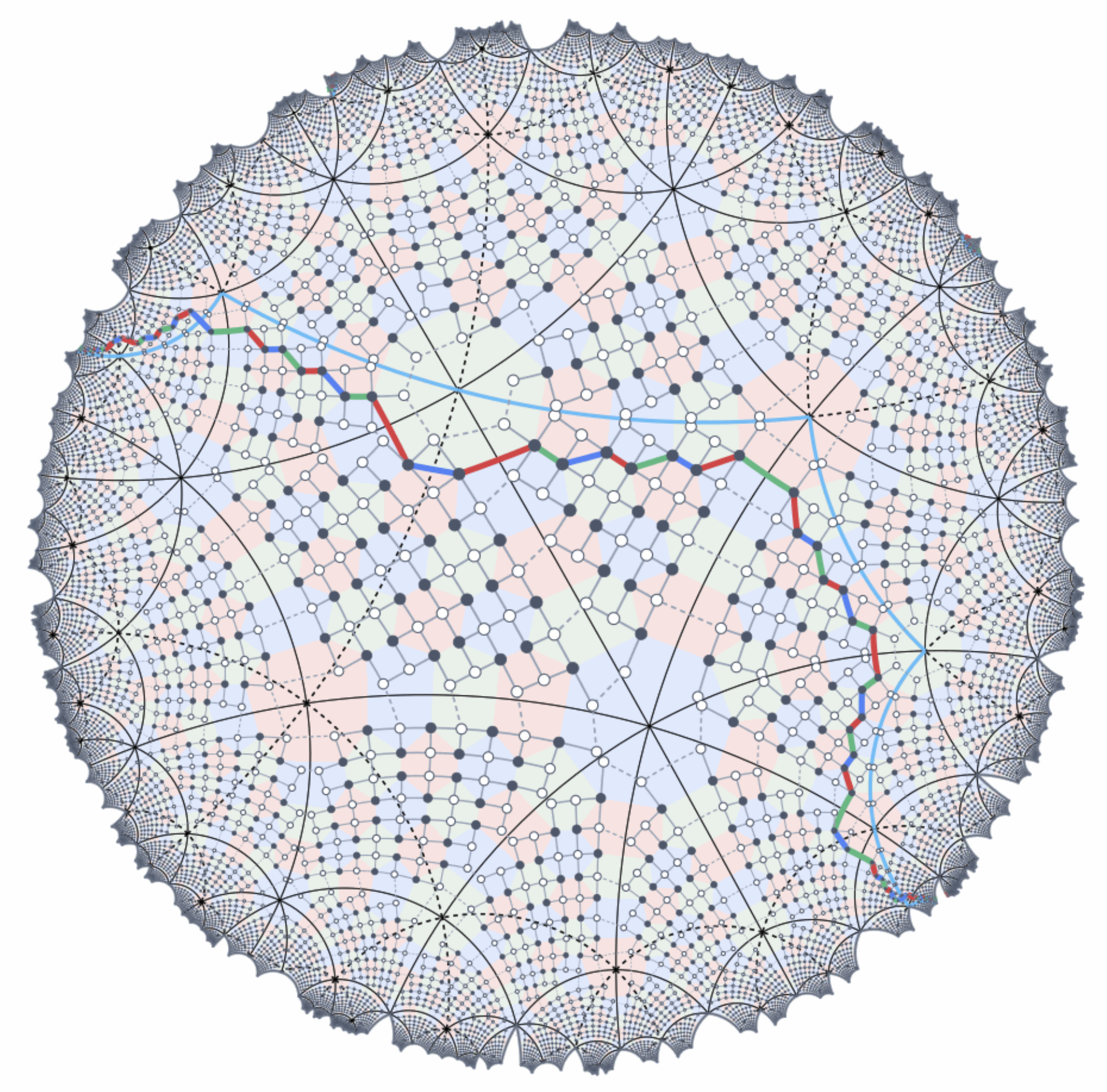}
    \caption{Choosing a seed loop that contains a logical representative in the semi-hyperbolic color code.
    }
    \label{fig:seed-loop}
\end{figure}

\begin{figure}[H]
    \centering
    \includegraphics[width=0.95\linewidth]{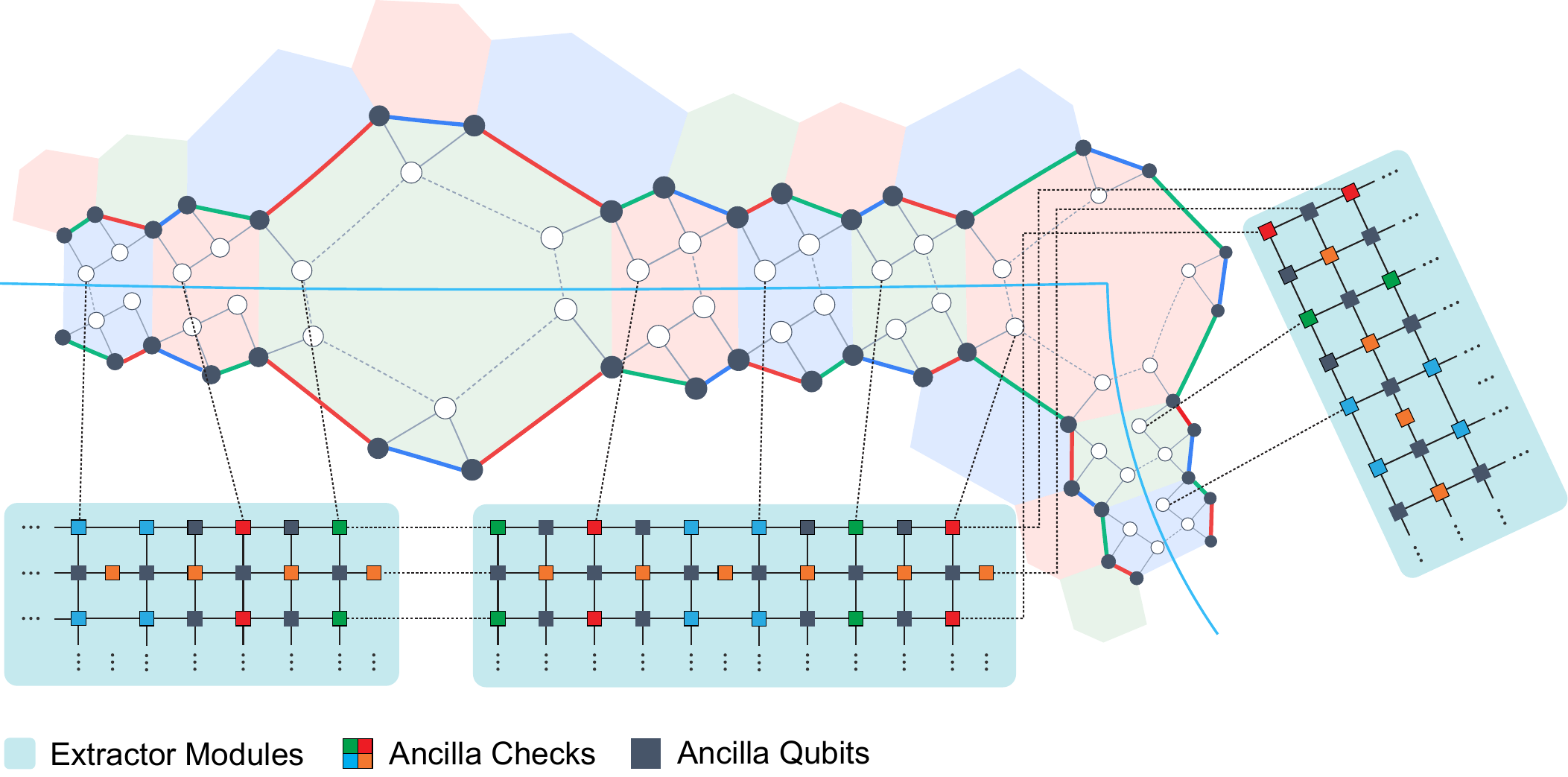}
    \caption{Extractor connectivity for measuring logical operators along a seed loop. Solid lines represent in-module connectivity while dashed lines represent inter-module connectivity. Surgery checks on the first row of surgery gadget can couple to the qubits of the corresponding color in the base code via inter-module connection to the associated base code ancilla measurement qubits. Each shaded extractor module box corresponds to an individual module which are connected in a 1D-chain manner to one another along with the base code, thus forming a ribbon extractor.}
    \label{fig:surgery-conn}
\end{figure}

We begin by finding a non-trivial cycle in the base code. This cycle and its associated logical operators will be supported on the boundary qubits of a collection of modules.
One such cycle of interest is given in \Cref{fig:seed-loop}; we will refer to such a cycle (given in cyan) as a seed~loop. The highlighted set of edges represent a set of qubits where a given logical operator is supported, such qubits all belong to stabilizer checks that are at the boundaries of modules forming the code. Note, there is a freedom in choosing a logical representative residing along a seed loop, namely one can choose equivalent logical representatives to be on either, or both sides, of the seed loop. Logical representatives of the same Pauli type and color (as will be described below) will be logically equivalent. See \Cref{fig:surgery-conn} for an example of a set of connected edges on either side of the seed loop that can harbor logical operators.

Along each seed loop there exist four different independent logical Pauli operators, two of both $X$ and $Z$ type, all of which commute. These logical operators can be chosen to  live along the edges of a given color, that is they form a continuous set of edges connecting faces of the same color. As a result, we can label the resulting logical operators as $\overline{X}_B, \overline{X}_R, \overline{Z}_B, \overline{Z}_R$ (where the $B$ and $R$ labels refer to the blue and red logicals, we forgo the superscript label from~\Cref{sec:color_code_logical_basis} as we are focusing on a single loop). Note, that while there is also a green logical, it would be equivalent to the product of the logicals of the other two colors $\overline{X}_G \simeq \overline{X}_B\overline{X}_R$.

We will denote the stabilizers that straddle both sides of the seed loop, and as such share data qubits between more than one module, as \emph{boundary stabilizers}. 
Note that all the weight-6 boundary stabilizers have half of their data qubit support on one module, and half on another module on either side of the boundary seed loop. For the weight-8 \emph{corner} stabilizers, the stabilizer can either have a 4-4 or 2-6 split on either side of the seed loop.

We will now describe the connectivity of the extractor gadget (see \Cref{fig:surgery-conn}) that will be used to implement logical measurements via code surgery. 
We will detail the exact stabilizers to be measured for a given desired logical in the subsequent section.

The ribbon extractor consists of layers of ancilla check qubits and ancilla data qubits. 
Note that a surgery check, which is a stabilizer we use to perform surgery measurement, can be implemented by one or more ancilla check qubits.
For readability, in the rest of this section we always differentiate between `checks' and `check qubits'. 
There will be a total of $2\ell+1$ layers, for an integer $\ell$ to be specified later. 
The first layer of the extractor is constructed as follows.
\begin{itemize}
    \item For every boundary stabilizer, we will associate a surgery check that can access a subset of qubits belonging to the corresponding stabilizer through the associated ancilla check qubits (white circles).
    In \Cref{fig:surgery-conn}, these surgery check qubits are colored according to the colors of the boundary stabilizers.
    For reasons which we will explain later, the surgery checks associated to the blue boundary stabilizers are implemented using two ancilla check qubits, as well as the surgery checks associated to the corner stabilizers. 
    We will refer to these surgery checks as `wide'. 
    The remaining `standard' surgery checks are each implemented with one ancilla check qubit.
    
    \item Each surgery check acts on two ancilla data qubits on the first layer. Surgery checks corresponding to neighboring boundary stabilizers share one such qubit.
\end{itemize}
In particular, the first layer of the ribbon extractor is a closed loop of alternating surgery checks (which are further connected to check qubits in the code) and ancilla data qubits.

The second layer of the extractor is constructed as follows. 
\begin{itemize}
    \item 
    For each standard surgery check from the first layer, we add one ancilla data qubit of support. 
    For each wide surgery check from the first layer, we add two ancilla data qubits of support.     
    
    \item For each ancilla data qubit from the first layer, we add one ancilla check qubit (colored orange in \Cref{fig:surgery-conn}). 
    This orange check will act on the data qubit from the first layer, as well as two adjacent ancilla qubits in the second layer.

    \item For each wide surgery check from the first layer, we add one orange check qubit acting on the two corresponding ancilla data qubits in the second layer.

\end{itemize}
Similar to the first layer, the second layer is also a closed loop of alternating ancilla checks and data qubits, connected to the first layer in an almost 1-to-1 fashion. 

Once we construct the first two layers, we repeat them for $\ell-1$ times such that the $(2i+1)$-th layer is connected to the $2i$-th layer in an almost 1-to-1 fashion. Finally, we add one final copy of the first layer as the $(2\ell+1)$-th layer, and connect it to the $2\ell$-th layer. 
This completes the construction of the ribbon extractor~\footnote{For readers familiar with modern code surgery methods, the construction we described is very similar to \textit{thickening}, which is a common technique used in various prior works on surgery~\cite{he2025extractorsqldpcarchitecturesefficient,cohen2022low-overhead,Williamson2026}. A technical difference is that some surgery checks in the first layer are wide (implemented with two ancilla check qubits), which act on two ancilla data qubits in the second layer. This design choice helps ensure the deformed code has sufficient distance in the surgery protocol.}.

We can calculate the space-cost of ribbon extractors (total number of ancilla data and check qubits) from the parameters of corresponding seed loops. 
Suppose the chosen seed loop traverses the boundaries of $t$ triangles. 
A boundary of a triangle has $L+1$ adjacent plaquettes, and a seed loop traverses $tL$ plaquettes, $t$ of which are corner plaquettes.
We will choose the seed loops such that the corner plaquettes are always green or red. 
As a result, out of the $tL$ plaquettes traversed $t$ of them are blue, and none of them are corners. 
On the first layer of a ribbon extractor, we therefore have $tL+2t$ ancilla check qubits (accounting for the blue and corner wide checks) and $tL$ ancilla data qubits.
On the second layer, we have $tL+2t$ ancilla data qubits and $tL+2t$ ancilla (orange) check qubits, because each wide check from the first layer needs an orange check qubit acting on its two adjacent data qubits in the second layer. 
In total, on the first two layers, we need
$t(2L+4)$ ancilla check qubits and $t(2L+2)$ ancilla data qubits.
We repeat these two layers for $\ell - 1$ times, and finish with another copy of the first layer. 
As a result, the total number of ancilla qubits in a ribbon extractor is
\begin{equation}\label{eq:ribbon-extractor-cost-1}
    \text{Space cost of ribbon extractor} = [t(2L+4) + t(2L+2)]\cdot \ell + t(2L+2) = t((4L+6)\ell + 2L + 2).
\end{equation}
As we will discuss in \Cref{sec:ribbon-extractor-distance-proof}, for a code of distance $d$ we will choose $\ell = \lceil d/2\rceil-1$ to ensure that the surgery procedure has phenomenological fault distance $d$. With this choice, the above equation simplifies to
\begin{equation}\label{eq:ribbon-extractor-cost-2}
    \text{Space cost of ribbon extractor} = \begin{cases}
        t((2L+3)(d-1) -1) & \text{for $d$ even,}\\ 
        t((2L+3)d -1) & \text{for $d$ odd}.
    \end{cases}
\end{equation}
We see that the leading factor is $2tLd$. 
The example code we used for~\Cref{fig:surgery-conn} has parameters $[[6400, 104, 30]]$, with $L = 4$, $t = 10$ and total qubit count $14400$. Putting these numbers in, we see that the qubit count of a ribbon extractor is up to $3180$, which is roughly $22.1\%$ of that of the code.  
Choosing $L = 3$ instead of $4$, we get a $[[3600, 104, 22]]$ code, with total qubit count $8400$ and the same seed loop where $t = 10$. The associated ribbon extractor uses up to $1880$ qubits, which is roughly $22.4\%$ of that of the code. It should be noted that the choice of wide checks is motivated for tilings of $L>3$ and as such may be suboptimal for $L=3$, therefore the associated overhead could potentially be reduced.  

\subsection{Logical Measurement via Surgery}

Given the ribbon extractor, to measure a logical operator $\overline{P}_C$, we will define a new stabilizer group supported on the code and extractor. 
We first describe this stabilizer group for measuring an $X$~logical operator of any color~$\overline{X}_C$, the case for $Z$~logical operators will follow by symmetry. 
We will return to measuring the product of an $X$ and $Z$~logical later. 

\begin{figure}[H]
    \centering
    \includegraphics[width=0.95\linewidth]{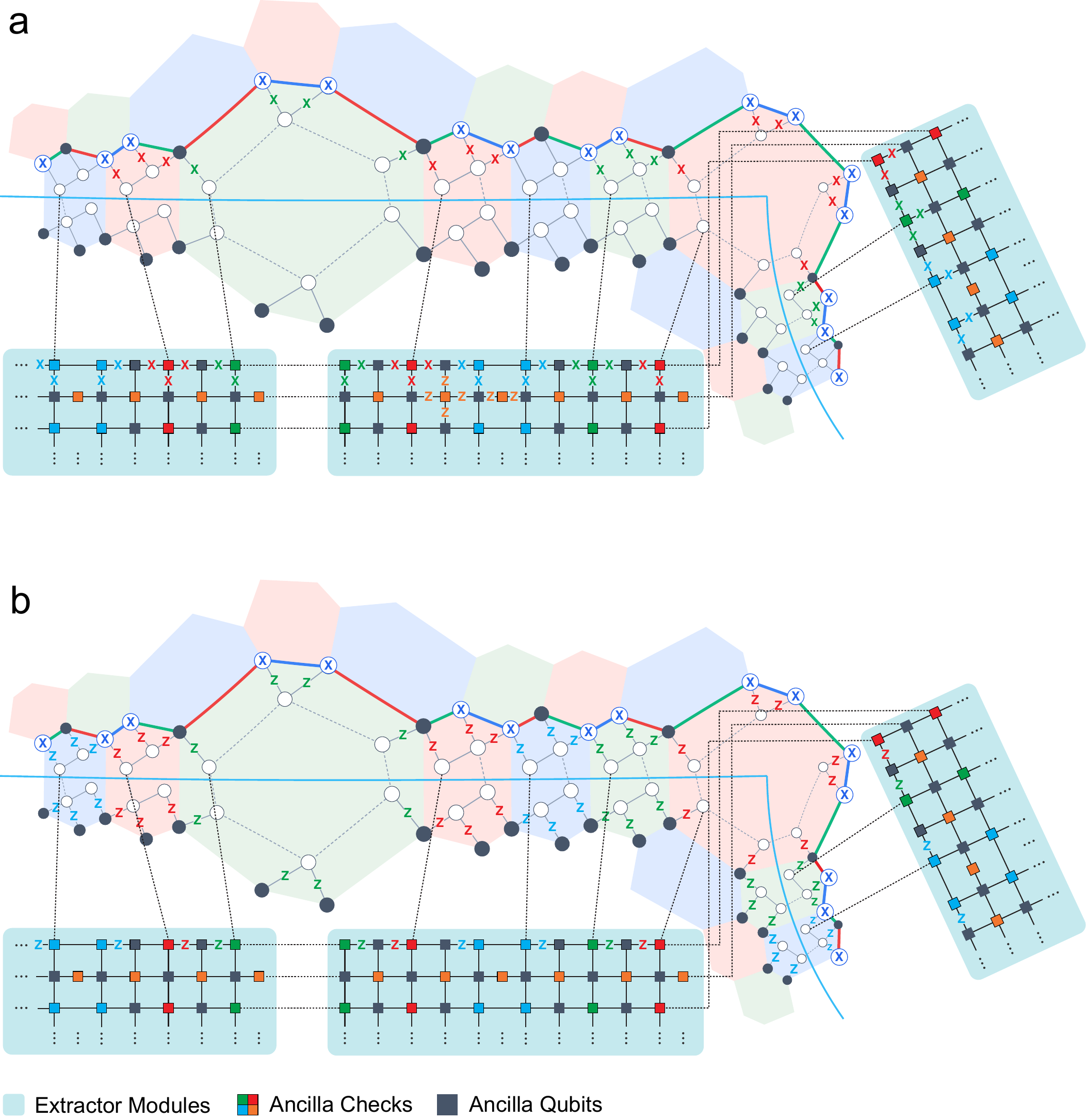}
    \caption{Measurement of a blue logical~$X$ operator, $\overline{X}_B$. (a)~Surgery checks that are red or green have support on the corresponding data qubits on one side of the seed loop. Blue surgery checks only retain support in the ancilla system. (b)~$Z$~boundary stabilizers are modified to gain support in the surgery ancilla qubits in order to commute with the new surgery checks.}
    \label{fig:surgery-Xb}
\end{figure}

Prepare all ancilla data qubits in the surgery gadget in~$\ket{0}$. To measure the $\overline{X}_{C_1}$ (color $C_1$): 

\begin{itemize}
    \item Choose one side of the seed loop. On the first layer of the ribbon extractor, for every surgery check of the other two colors $C_2, C_3$, the surgery check will have $X$~support on the qubits on the chosen side of the seed loop belonging to the associated boundary stabilizer. It will also have
    $X$~support on the (three to four) neighboring ancilla data qubits in the extractor modules. 
    For the surgery check of color $C_1$, only measure the $X$~support on the extractor module (no support on the data qubits of the original code), see \Cref{fig:surgery-Xb}(a) for an example of the support for measuring~$\overline{X}_B$.
    
    \item As the surgery checks on the first layer have $X$~support on the data qubits of the code, there will be $Z$~stabilizers, namely the boundary $Z$~stabilizers, whose original support will anti-commute with these measurements.
    To restore commutativity, they will have to gain support on the ancilla data qubits on the first layer, see \Cref{fig:surgery-Xb}(b). They will gain support on a given ancilla data qubit in the first layer if their corresponding neighboring $X$~surgery check had been extended into the data qubits of the code.

    \item For subsequent odd layers, the surgery checks (colored red, green or blue) act on adjacent data qubits by $X$. Note that the regular surgery checks have weight $4$, while the wide surgery checks have weight $6$. 

    \item On the even layers of the ribbon extractor, for each orange ancilla check qubit, we define a $Z$~stabilizer that acts on its adjacent data qubits. These $Z$ checks have weight $2$ or $4$, and they commute with the surgery checks defined above.    
\end{itemize}
We now prove the following crucial claim.

\begin{claim}
    The product of all surgery checks (from all odd layers) equals the $X$~logical operator of color $C_1$ on the base code ($\overline{X}_{C_1}$). 
\end{claim}

\begin{proof}
    We begin with a few definitions: we say an edge is adjacent to a given face if it is one of the boundary edges of the face, while an edge is incident to a face if it touches one and only one of the vertices of the face.
    We have chosen a path for the logical operators which is a continuous set of edges, alternating in color (where the color of an edge is defined in the standard way for color codes as an edge incident to two faces of that color). Each edge can be characterized in one of the following two ways: 
    \begin{enumerate}
        \item An edge is adjacent to a single boundary face and is not incident to any other boundary face.
        \item An edge is adjacent to a single boundary face and is incident to one or two other boundary faces.
    \end{enumerate}
    
    Without loss of generality, let us assume we want to measure a blue~($B$) logical $X$~operator. We want to argue that by multiplying the surgery checks, of which the red~($R$) and green~($G$) colored surgery checks on layer one act on the base code non-trivially, the resulting support will be the support of all the blue edges along the logical path. 
    
    Let us first show that any qubit belonging to a blue edge will have support. We will work this out case by case among the edges above.
    
    \begin{enumerate}
    \item If a blue edge is only adjacent to a single boundary face, that face must be of color $R$ or $G$ and as such the 
    corresponding surgery check on layer one
    will have support on this edge. Moreover, since it is not incident to any other boundary face, this support cannot be canceled out by another surgery check.

    \item Similarly to the case above, the support of the edge will have to be measured by the $R$ or $G$ surgery check to which it is adjacent. Any other face to which this edge is incident must be a blue face (since this is a blue edge). Since the blue surgery checks do not act on the base code, the support on these blue edges cannot be canceled out.
    
    \end{enumerate}

    Next, we must show that any qubit along the logical boundary path that does not belong to a $B$ edge, that is the endpoint of an $R$ and $G$ edge, must not have support in the overall measurement. There are two cases for such qubits: 
    \begin{enumerate}
        \item The only boundary stabilizer to which they belong is blue, as such the measurement has no support on such qubits.
        \item The qubit belongs to two boundary stabilizers. Given the qubit lies at the end point of red and green edges, the corresponding boundary stabilizers to which it belongs must be red and green, and the two corresponding surgery checks must cancel out its support.
    \end{enumerate}

    Therefore we have established that on the data qubits in the base code, the product of all the surgery checks is supported exactly on the blue $X$~logical operator.
    On the ancilla data qubits in the ribbon extractor, we note that every ancilla data qubit is acted on by exactly two surgery checks, whose Pauli action cancels on the qubit. 
    We conclude that the product of all the surgery checks equals the blue $X$~logical operator.
\end{proof}

\begin{corollary}
    Measurement of the $Z$~logical of color $C_1$ is identical in construction to that of the $X$~logical except that the support of the surgery checks on the code qubits is of $Z$-type (however, it remains of $X$-type on the ancilla qubits). As a result, only the $X$~boundary stabilizers are now deformed to gain $Z$~support on the ancilla qubits.
\end{corollary}

\begin{corollary}
    Measurement of the $Y$~logical of color $C_1$ is identical in construction to that of the $X$~logical except that the support of the surgery checks on the code qubits is of $Y$-type (however, it remains of $X$-type on the ancilla qubits). As a result, both the $X$ and $Z$~boundary stabilizers are now deformed to gain $Z$~support on the ancilla qubits.
\end{corollary}

\begin{figure}[H]
    \centering
    \includegraphics[width=0.95\linewidth]{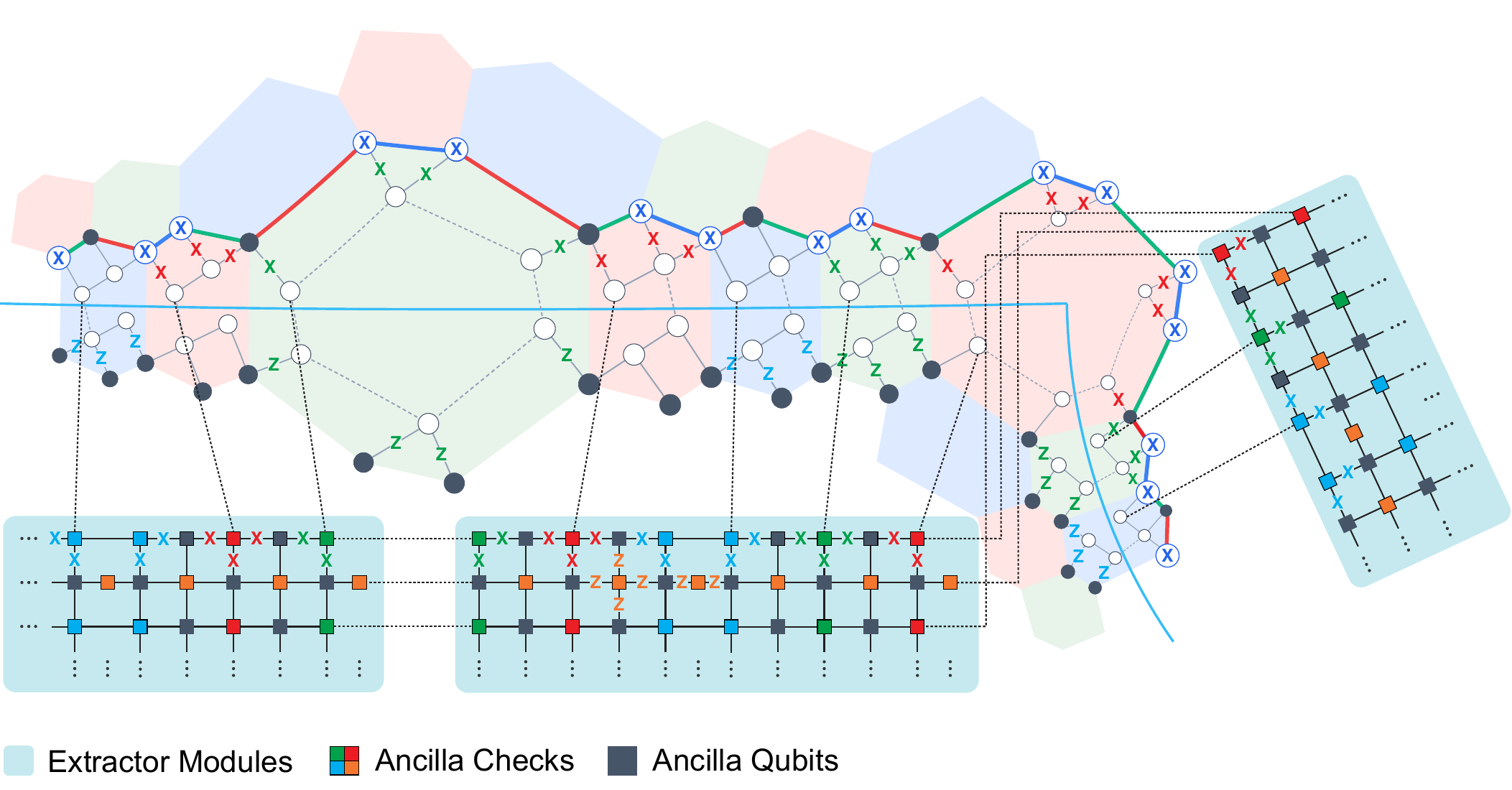}
    \caption{Measurement of $\overline{X}_B \overline{Z}_R$. Green surgery checks gain support of $X$ and $Z$ type on the data qubits on respective sides of the seed loop, while red (blue) surgery checks only gain $X(Z)$~support. All surgery checks only have $X$~support on the surgery ancilla qubits.}
    \label{fig:surgery-XbZr}
\end{figure}

In order to measure the product of two logical operators of different types, we combine the technique of measuring each logical~$X$ and~$Z$. Namely, the surgery checks will gain $X$~support on one side of the seed loop and $Z$~support on the other (depending on the color), see \Cref{fig:surgery-XbZr} for an example. The original boundary stabilizers would have to be deformed in a similar manner to guarantee commutativity.

\begin{corollary}
    \label{cor:ribbonXZmeasurement}
    In order to measure the product operator~$\overline{X}_{C_1} \overline{Z}_{C_2}$, that is the product of logical~$X$ of color~$C_1$ and logical~$Z$ of color~$C_2$, all surgery checks on layer one of colors~$C_2$ and~$C_3$ will have $X$~support on the data qubits on one side of the seed loop, while surgery checks on layer one of color~$C_1$ and~$C_3$ will have $Z$~support on the data qubits on the other side of the seed loop.
    All surgery checks will have their associated $X$~support on the ancilla data qubits in the extractor. 
    The original $X$ and $Z$~boundary stabilizers of the code will gain~$Z$ support on the ancilla qubits following the same procedure as the measurement of $\overline{X}_{C_1}$ and $\overline{Z}_{C_2}$ individually.
\end{corollary}

\subsection{Fault Tolerance}
\label{sec:ribbon-extractor-distance-proof}

Given the extractor system and the deformed codes, we now show that the surgery protocol is fault-tolerant. 
Specifically, we will show that the deformed codes have the same distance $d$ as the original code. 
Following standard arguments~\cite{cross2024improved,Williamson2026}, this implies that the surgery protocol has phenomenological fault distance $d$.

In order to prove that the deformed codes have sufficient distances, one must show that the original logical operators of the code cannot be weight-reduced by multiplying by the new stabilizers of the deformed code or any new gauge logical operators. We will begin by arguing that no new gauge logical qubits are introduced by the ribbon extractor before arguing that the new stabilizers of the code cannot reduce the weight of a logical operator to be below that of the distance of the original code.

\begin{claim}
    The ribbon extractor measures a target logical operator~$P$ in the base code without introducing any other new logical qubits.
\end{claim}

\begin{proof}

    From our previous calculations around \Cref{eq:ribbon-extractor-cost-1}, we can count the number of new data qubits and stabilizers introduced by the ribbon extractor. 
    There are $\ell+1$ odd layers, each with $tL$ new data qubits and $tL$ new stabilizers, and $\ell$ even layers, each with $t(L+2)$ new data qubits and $t(L+2)$ new stabilizers.
    It is straightforward to verify the following claim of independence: there is no product of deformed code stabilizers, with a non-trivial combination of extractor stabilizers, that results in identity. 
    Therefore, the number of logical degrees of freedom in the overall system cannot be more than $k$.

    However, this is not quite the end of the story. As when we take the product of all stabilizers whose support is of $X$~type in the ribbon extractor, the resulting support is the logical operator~$P$ from the base code to be measured, therefore we explicitly have eliminated one of the logical degrees of freedom of the original code! 
    Equivalently, the rank of stabilizers should have increased by one more than the number of new stabilizers. 
    We can observe this directly.
    In the original code, the $Z$~stabilizers are redundant in the sense that the product of all blue and red/green stabilizers give identity. 
    However, in the deformed code, the product of all blue and red/green (original and deformed) stabilizers equals the operator that is all~$Z$ on the data qubits of the first layer of the ribbon extractor~\footnote{In the modern graph surgery language, this $Z$~operator on the extractor is a `cycle check'. Normally we need to measure these cycle checks, but in our case of the ribbon extractor this cycle check is a product of deformed stabilizers.~\cite{cross2024improved,Williamson2026,he2025extractorsqldpcarchitecturesefficient}}.
    Therefore, this degree of redundancy is removed as we deform the code and we conclude that the deformed code has $k-1$ logical degrees of freedom.
\end{proof}

\begin{claim}
    Given a thickening of at least $d/2$ layers, the distance of the deformed code through surgery measurement remains distance~$d$.
\end{claim}

\begin{proof}
    Suppose we are given some logical operator $Q$ of weight~$w$ in the base code, we would like to show that the weight of this operator cannot be reduced to be less than the distance~$d$ by multiplying by the new surgery stabilizers of the code. 

    Let us first consider the case when the ancilla system is added to measure a blue~($B$) logical~$X$. Given none of the newly introduced $Z$~stabilizers in the ribbon extractor are supported on the base code, they cannot contribute to a distance reduction of the $Z$ logical operator. Moreover, multiplying by any deformed $Z$~stabilizer will only end up increasing the support of the logical~$Z$ operator, as such any logical~$Z$ cannot see their distance reduced by a ribbon extractor built for measuring a logical~$X$ operator.
    
    We will consider the set of neighboring $R_l,\ G_c, \ R_r$ surgery stabilizers where $G_c$ is the corner stabilizer, and $R_l,\ R_r$ are the left- and right-neighboring surgery stabilizers of $G_c$ (associated with the neighboring boundary stabilizers). Without loss of generality our arguments will also apply for $GRG$ faces. We will consider different scenarios of multiplying by the new surgery stabilizers and how support of the logical operator is deformed on the base code plus the first two layers of the ancilla graph.

    Consider multiplying by only the $G_c$~surgery stabilizers, that is the one associated with a corner stabilizer of the original code. The surgery stabilizer can have support of weight-2, 4, or 6 on the base code. Note the support of $G_c$ will be weight-4 on the ancilla code.

    \begin{itemize}
        \item Weight-2 support on base code: the weight of $QG_c$ must increase by at least 2 when compared to $Q$, as the support on the base code could be canceled but 4 qubits of support is gained in the ancilla system.
        \item Weight-4 support on base code: the weight of $QG_c$ must be at least as large as $Q$.
        \item Weight-6 support on base code: If the overlap between $G_c$ and $Q$ is at most 5, then the weight of $QG_c$ must be at least as large as $Q$. If the overlap is 6, then indeed the weight of $QG_c$ is decreased by 2, however in such a case since the overlap of $Q$ with $G_c$ is 6, it means that the overlap of $Q$ with the associated original $G$~$X$~stabilizer of the code is 6, as such by multiplying by that stabilizer we can find an equivalent logical operator of weight $w-4$, thus in such a scenario the weight~$w$ must be lower bounded by $d+4$. Therefore reducing the weight of $QG_c$ by 2 still bounds the weight of such an operator to be at least of weight~$d+2$.
    \end{itemize}

    Therefore, multiplying $Q$ by the surgery stabilizer $G_c$ associated with a corner stabilizer either cannot reduce the weight of $Q$ or bounds the product of $QG_c$ to be of weight at least $d+2$.

    Now we will consider the multiplication by a single $R$~surgery stabilizer, without loss of generality assume $R_l$. Since $R_l$ has weight-3 support on the base code and weight-3 in the ancilla code, if the overlap of $Q$ and $R_l$ is at most 2, then the weight of $QR_l$ will increase by 2. If their overlap is three, then the weight of $QR_l$ remains~$w$, however in such a scenario $Q$ must have at least weight-4 overlap with the associated $R$~boundary stabilizer of the base code, as such the weight of $Q$ would be at least $d+2$ in such a scenario (as multiplying by that boundary stabilizer could reduce the weight of the chosen representative $Q$ by 2). Therefore, $QR_l$ will be of weight at least $w$ or $d+2$.

    Now let us consider multiplying $Q$ by $R_lG_c$. Again there are three scenarios to consider based on the support of $G_c$ on the base code being either 2, 4, or 6. Note the support of $R_lG_c$ will be weight-5 on the ancilla code.

    \begin{itemize}
        \item $R_lG_c$ weight-3 support on base code: the weight of $QR_lG_c$ must increase by at least 2 when compared to $Q$, as the support on the base code could be canceled but 5 qubits of support is gained in the ancilla system.
        \item $R_lG_c$ weight-5 support on base code: the weight of $QR_lG_c$ must be at least as large as $Q$.
        \item $R_lG_c$ weight-7 support on base code: If the overlap between $R_lG_c$ and $Q$ is at most 6, then the weight of $QR_lG_c$ must be at least as large as $Q$. If the overlap is 7, then indeed the weight of $QR_lG_c$ is decreased by 2, however in such a case since the overlap of $Q$ with $G_c$ is at least 5, it means that the overlap of $Q$ with the associated original $G$~$X$~stabilizer of the code is 6, as such by multiplying by that stabilizer we can find an equivalent logical operator of weight $w-4$, thus in such a scenario the weight~$w$ must be lower bounded by $d+4$. Therefore reducing the weight of $QR_lG_c$ by 2 still bounds the weight of such an operator to be at least of weight~$d+2$.
    \end{itemize}

    Finally, let us consider multiplying $Q$ by $R_lG_cR_r$. Again there are three scenarios to consider based on the support of $G_c$ on the base code. Note the support of $R_lG_cR_r$ will be weight-6 on the ancilla code.

    \begin{itemize}
        \item $R_lG_cR_r$ weight-4 support on base code: the weight of $QR_lG_cR_r$ must increase by at least 2 when compared to $Q$.
        \item $R_lG_cR_r$ weight-6 support on base code: the weight of $QR_lG_cR_r$ must be at least as large as $Q$.
        \item $R_lG_cR_r$ weight-8 support on base code: If the overlap between $R_lG_cR_r$ and $Q$ is at most 7, then the weight of $QR_lG_cR_r$ must be at least as large as $Q$. If the overlap is 8, then indeed the weight of $QR_lG_cR_r$ is decreased by 2, however in such a case the weight-8 overlap is a result of 4 edges of the same color whose overall endpoint would be faces of the same color that could be connected via two other edges of the same color in the base code. As a result, in this case we can lower bound the weight~$w$ of $Q$ by~$d+4$. Therefore reducing the weight of $QR_lG_cR_r$ by 2 still bounds the weight of such an operator to be at least of weight~$d+2$.
    \end{itemize}

    \begin{figure}
    \centering
    \includegraphics[width=0.85\linewidth]{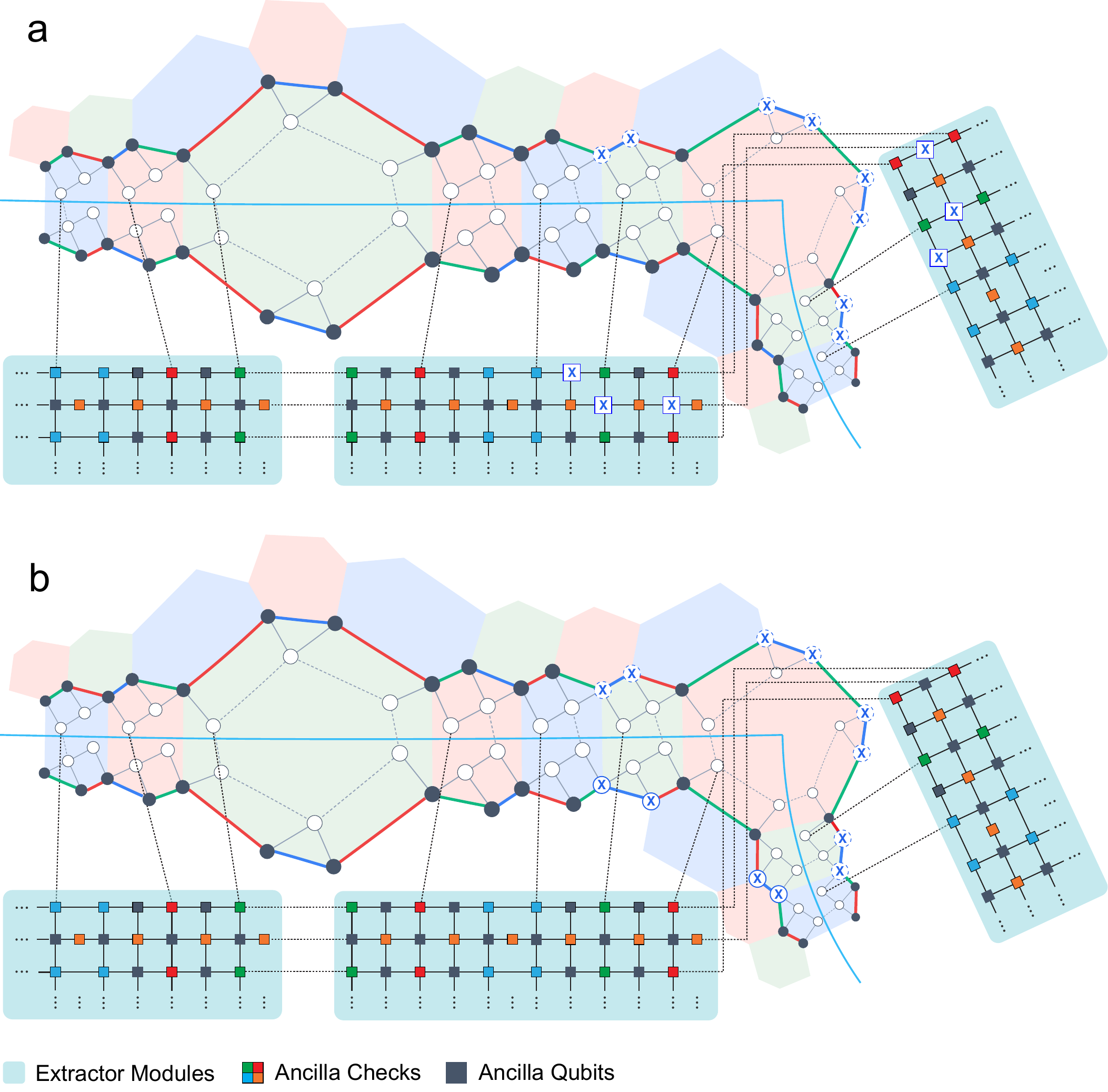}
    \caption{Reduction of support of a blue $X$~string on the base code. (a)~The weight-8 $X$~string (faded and dashed) on the base code can be reduced to a weight-6 operator on the ancilla qubits (square) of the surgery gadgets by multiplying by the ancilla checks. (b)~The original weight-8 $X$~string (faded and dashed) on the base code has an equivalent representation as a weight-4 $X$~string (solid circle) as their endpoints are the same. The equivalence can be seen under multiplication by the three corresponding boundary stabilizers.}
    \end{figure}

    Therefore we can conclude that multiplying $Q$ by any combination of $R_l, \ G_c,\ R_r$ is lower bounded by the weight~$w$ of $Q$ or $d+2$, and the case of $d+2$ is due to the existence of a local deformation of the representative~$Q$ that bounds the weight away from the distance~$d$ of the code. These bounds will apply for any sequence of $R_lG_cR_r$ or $G_lR_cG_r$ ancilla checks between blue ancilla checks which we have chosen to be weight-4 and are only supported on the ancilla code, therefore they cannot result in a reduction of weight. This concludes that the first layer of checks cannot reduce the weight of any other logical operator~$Q$ when measuring a blue logical~$X$. 

    We now consider the surgery checks associated with measuring a red~($R$) logical~$X$. Here, the $B$ and $G$~surgery checks will be extended into the base code following the procedure previously laid out. We would again like to consider another logical operator~$Q$ in the base code and show that by multiplying by the new surgery checks we cannot reduce the weight of~$Q$ to be below the distance of the code.

    Consider first neighboring $B_b$ and $G_b$~surgery checks that do not correspond to those associated with a corner stabilizer (we label them with the subscript~$b$ for bulk). Since $G_b$ has weight-3 support on the base code and weight-3 in the ancilla code, if the overlap of $Q$ and $G_b$ is at most 2, then the weight of $QG_b$ will increase by 2. If their overlap is three, then the weight of $QG_b$ remains~$w$. The exact same argument holds true for multiplying by only $B_b$, however $QB_b$ would have an additional qubit of support since it has weight-4 in the ancilla, thus $QB_b$ will be of weight at least $w+1$. 
    Finally, since $B_bG_b$ has weight-5 support in the ancilla and only weight-4 support on the base code, $QB_bG_b$ will always be at least of weight~$w+1$.

    Finally, as argued previously, multiplying by only a surgery check~$G_c$ associated with a corner stabilizer either cannot reduce the weight of $Q$ or bounds the product of $QG_c$ to be of weight at least $d+2$ (due to the possibility of local weight reduction from the associated corner stabilizer).

    Thus, considering the potential multiplication of different combinations of $B_b, \ G_b$ pairs and $G_c$ surgery checks, individually none of them can cause the distance to be reduced below the distance~$d$. The final piece is potentially multiplying by red surgery checks that only have support on the  ancilla system. Multiplying by red surgery checks can only reduce the weight if the red~surgery check is weight-3 and both neighboring blue and green surgery checks have also multiplied the logical operator~$Q$ (they have to be blue and green because the only weight-3 red surgery checks are those associated with non-corner stabilizers). However, as we have previously seen, if $Q$ is multiplied by any $B_b$~surgery check then the weight of such an operator is guaranteed to have already increased by 1, thus the potential decrease in weight from multiplying by red surgery check is harmless.

    As a final part to prove the distance preservation, we must argue that the weight obtained on the second layer of ancilla checks cannot be reduced without gaining some additional support in the ancilla. Indeed, any support of a representative on the second layer of ancilla qubits can be multiplied with the associated surgery check of $X$-type on the third layer and move its support up to the fourth layer. This process can be repeated up until the point of multiplying by the top layer of surgery $X$~stabilizers, however support will be also gained at the endpoints of chains of surgery checks, and that support will be as large as the number of $X$~stabilizer layers in the ancilla system. Therefore, to guarantee that the distance of a representative is lower bounded by~$d$, we need the height of the number of $X$~stabilizers in the ancilla system to be~$d/2$ since the support of~$d/2$ is gained at both endpoints.

    We have argued that the ribbon extractor for measuring a blue or red logical~$X$ operator cannot reduce the distance of the other logical operators in the base code. By symmetry between green and red colors, measuring a green logical~$X$ will also not cause a distance reduction. In addition, due to the symmetry of the construction, measurement of a logical~$Z$ operator using the ribbon extractor will also not result in a distance reduction. Therefore, the only remaining case to be considered would be the measurement of a product of logical~$X$ and~$Z$ operators of different color. However, as presented in Corollary~\ref{cor:ribbonXZmeasurement}, the construction of the ribbon extractor for the measurement of the product operator~$\overline{X}_{C_1}\overline{Z}_{C_2}$ will have surgery checks whose support on the base code will mirror that of the individual logical $X$ and $Z$ constructions, only jointly. This implies by the same arguments previously outlined, that the remaining logical~$X$ and $Z$ representatives of the base code cannot have their distance reduced since if such a reduction existed it would imply existence in the ribbon extractors for the individual measurements of logical~$X$ or~$Z$.
\end{proof}

\subsection{Teleporting Logical Information to an External Code}

In the previous subsections we have described measuring logical Pauli operators of the modular hyperbolic memory. These forms of measurement can be used in a model of measurement-based logical computation, however the computation would have to be facilitated with the introduction of resource states to complete universal computation. This could be achieved by preparing ancillary resource states in external codes, via techniques such as magic state cultivation or distillation. In addition, it may be advantageous to have an external compute region, which would be a set of logical qubits encoded in some other code that is specifically designed for the purpose of facilitating computation. Regardless of which of the above models are used to complete the universal set of operations required to run a quantum algorithm, we require a means to do joint logical Pauli measurements between logical qubits in the hyperbolic code and some external code. We can re-purpose the ribbon extractor for this task, using the top boundary of the ribbon extractor to couple to a logical qubit of the external code, thereby using the extractor to jointly measure the product of logical Paulis between the hyperbolic memory, $\overline{P}_m$, and the compute code~$\overline{P}_c$.

\begin{figure}[H]
    \centering
    \includegraphics[width=0.60\linewidth]{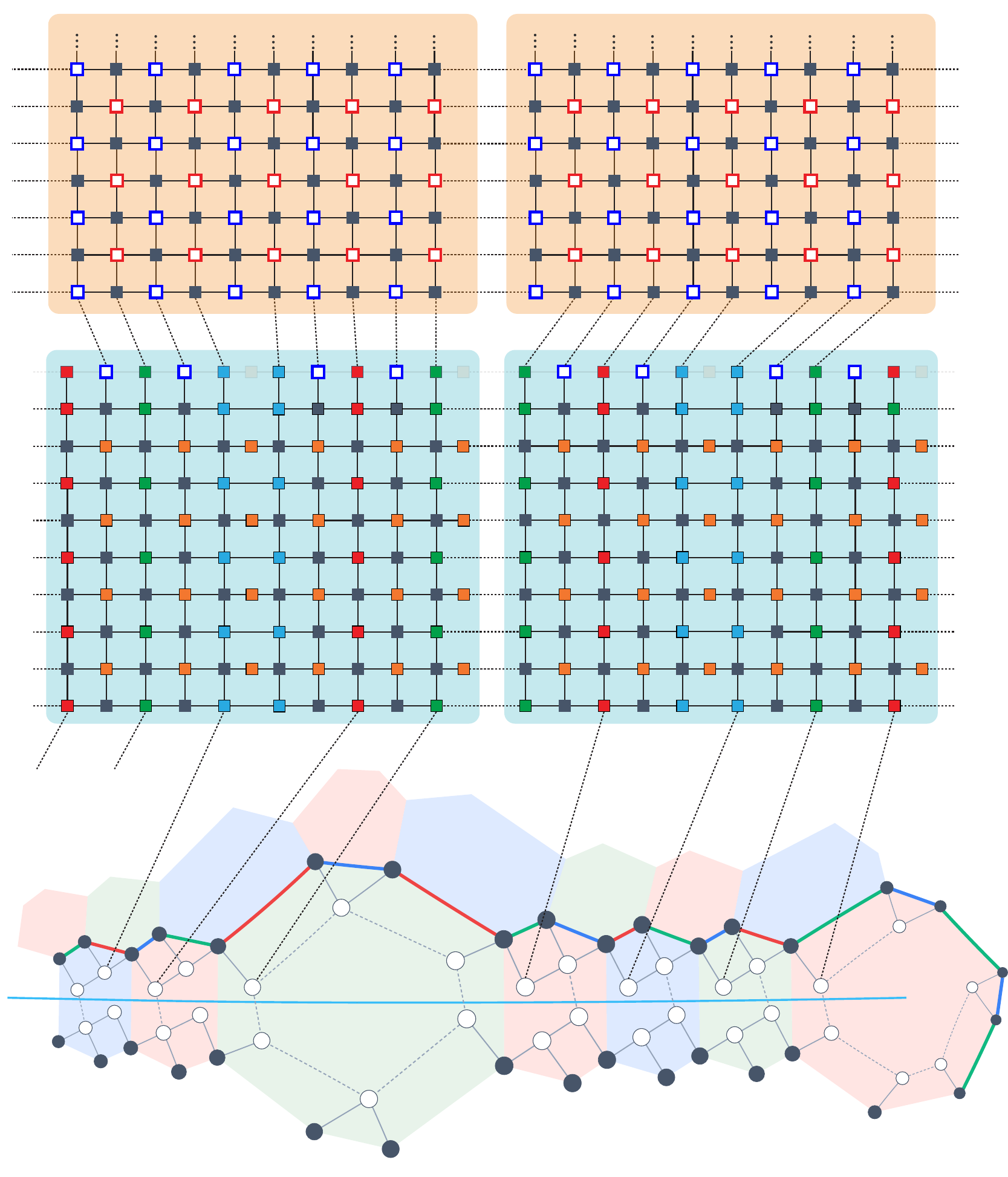}
    \caption{Ribbon extractor to measure the product $\overline{P}_m\overline{X}_c$ of a logical Pauli operator~$\overline{P}_m$ of the hyperbolic code and the logical~$\overline{X}_c$ of the surface code. The ribbon extractor is formed from connected modules~(cyan) whose bottom boundary is connected to the hyperbolic code, while the top boundary is connected to a surface code that itself is formed from a connected set of modules~(orange). The ribbon extractor has a total of $2\ell+1$ layers, with one extra layer at top boundary used to connect to the surface code.
    }
     \label{fig:Extractor-Surface}
\end{figure}

We provide an example of such a design in Fig.~\ref{fig:Extractor-Surface}, where the bottom boundary of the ribbon extractor is coupled to the hyperbolic memory as previously discussed. However, we now couple the top boundary to another code block, in the presented example it is a surface code. The connectivity at the top boundary would then be used to extend the surgery checks onto the support of a logical operator (in the example it would be~$\overline{X}_c$) of the surface code given by connected modules. Therefore, the product of all of the surgery checks would result in the measurement of~$\overline{P}_m\overline{X}_c$.

\section{Automorphism Circuits}\label{sec:automorphism_circuits}

Code automorphisms can be used to implement certain logical Clifford gates in modular hyperbolic codes.
A code automorphism is a permutation of the physical qubits that maps the stabilizer group onto itself.

For modular hyperbolic color codes, the global symmetries of the base tiling correspond to code automorphisms (since the stabilizer generators are defined from the faces of the tiling, and the codes are self-dual CSS codes).
The symmetry group of the base tiling maps the set of generators of the stabilizer group to itself (rather than just mapping the stabilizer group to itself) so is in general a subgroup of the full automorphism group.
However, for all but the smallest tilings, we find that the tiling's symmetry group is isomorphic to the group of code automorphisms.

The modular hyperbolic \textit{surface codes} are not self dual, and recall that the basis of a stabilizer ($X$ or $Z$) is determined by the color of its corresponding face in the tiling (red or blue).
A global isometry that is a symmetry of the base tiling either preserves the face coloring, in which case it directly corresponds to a code automorphism, or swaps the red and blue face colors, in which case it is a code automorphism up to a global application of a Hadamard gate on each physical qubit.
The specific subgroup of the automorphism group that we propose to implement does preserve face colors (and therefore does not require a layer of Hadamards).

\subsection{Automorphism Circuits for Color Codes}\label{sec:automorphism_circuits_color}

\begin{figure}
    \centering
    \includegraphics[width=\linewidth]{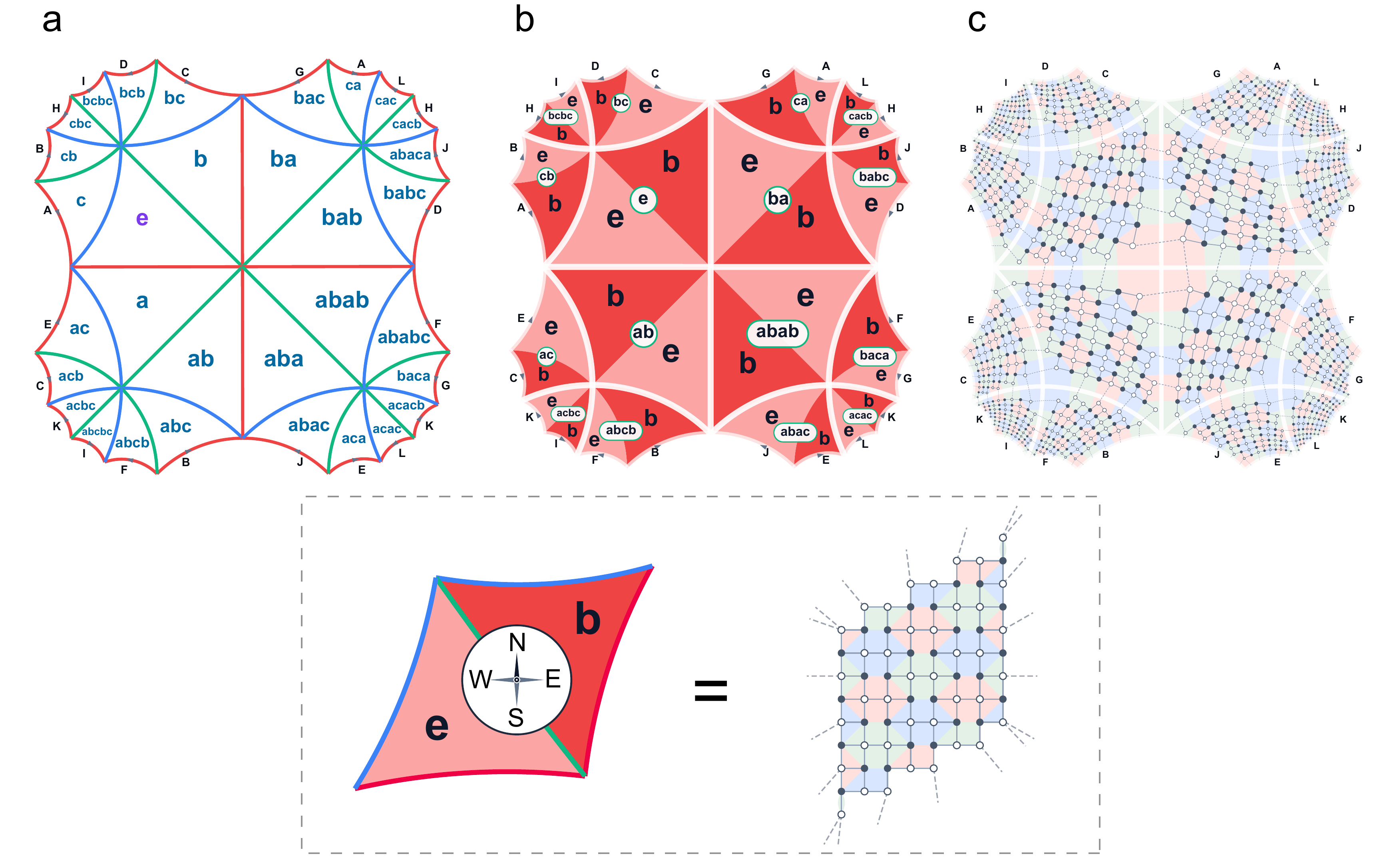}
    \caption{Quadrilateral ($K{=}2$) modules of the $8.8.8$-$32$ color code, labeled by the orientation-preserving subgroup $G^+ = \triangle^+(4,4,4)/\Gamma$. \textbf{a}~The base triangular tiling $\mathcal{T}^\triangle$, with each triangle labeled by a word in the reflections $a$, $b$, $c$ that reaches it from the identity triangle $e$ (purple), and each edge colored by its mirror ($a$ red, $b$ green, $c$ blue). Boundaries labeled with the same letter $A$--$L$ are identified. \textbf{b}~The $16$ quadrilateral modules, formed by pairing adjacent triangles across their green edge ($b$ mirror) into left cosets $g\langle b \rangle$. Because $b$ reverses orientation, each module contains exactly one even-length word $g \in G^+$ (circled in green, in the middle of each module), which labels the module. Left multiplication by any $h \in G^+$ maps $g \mapsto hg$ and $gb \mapsto (hg)b$, permuting the $16$ modules while preserving the internal coordinate system (compass, inset) of every module. \textbf{c}~The same partition on the fine-grained semi-hyperbolic code, with data qubits (filled circles), ancilla qubits (open circles), and inter-module seams (dashed lines).}
    \label{fig:quad_module_complements_it40}
\end{figure}

\begin{figure}
    \centering
    \includegraphics[width=0.8\linewidth]{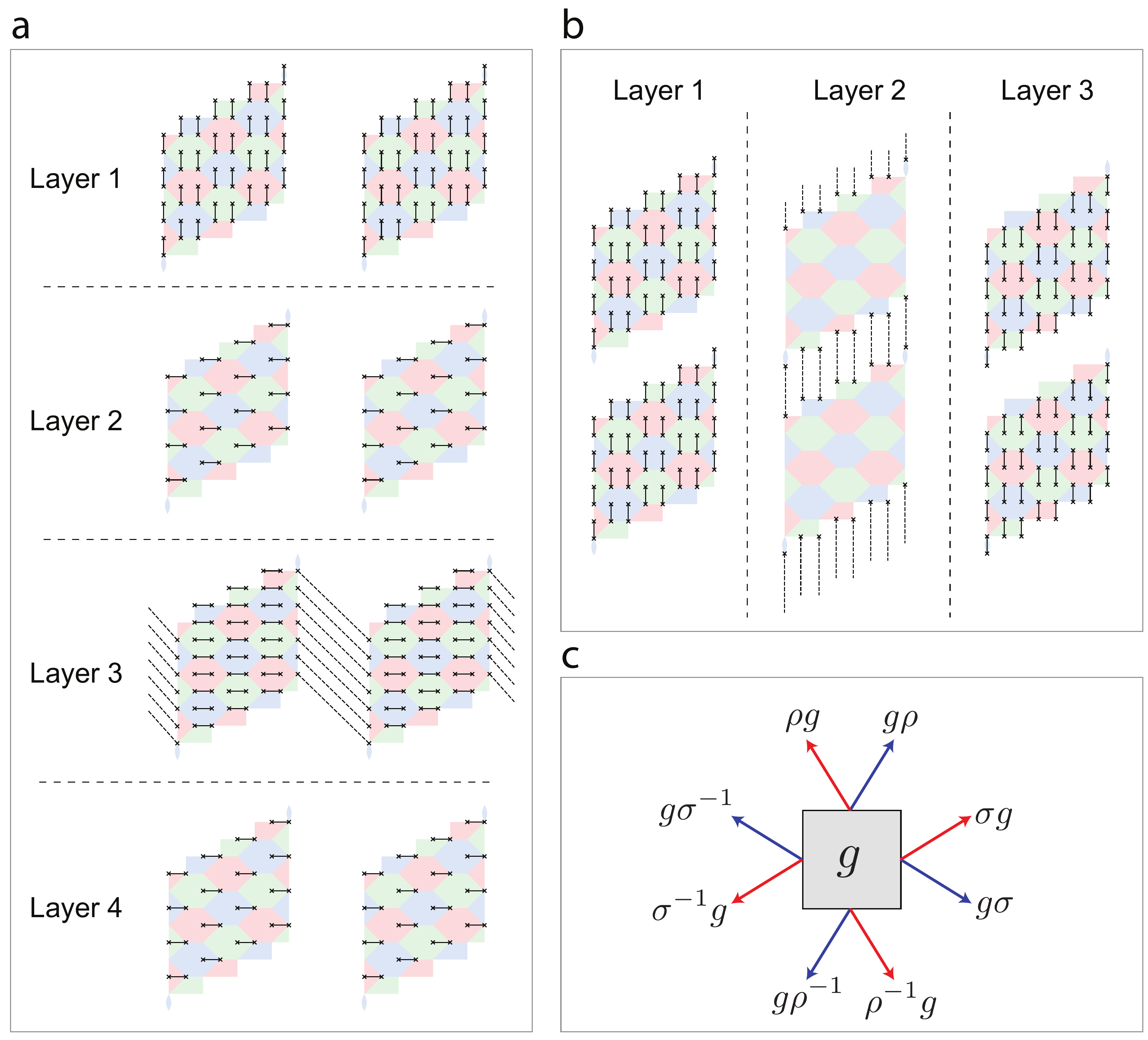}
    \caption{Examples of circuits that implement a permutation of the modules that preserves their internal rotation and orientation. We can use these circuits to implement automorphisms in modular hyperbolic codes. \textbf{a}, By adding additional seams (inter-module couplers) on the east and west boundary of each module we can use SWAP gates to implement one such permutation. The four layers of gates shown in the figure implement one step of the automorphism circuit (repeating this step $L$ times completes the automorphism, and in this example $L=4$). \textbf{b}, By adding additional seams on the north and south boundaries we can implement a different automorphism. Here the three layers of gates implement one step of the automorphism circuit, and again repeating it $L=4$ times completes the automorphism. \textbf{c}, Using each of these two circuits to implement one of the two generators of the automorphism group $\triangle^+(r,s,t)/\Gamma$, we can implement the full automorphism group using fixed connectivity. The connectivity required uses two seams on each side of the module. One seam is used for automorphisms (connectivity defined by left multiplication) and the other for syndrome extraction (connectivity defined by right multiplication). The overall connectivity of the modules then corresponds to the union of the edge sets in a Left Cayley graph and Right Cayley graph. One node (module) of this vertex-transitive graph is shown in \textbf{c}, where each arrow represents a seam of inter-module couplers leaving the edge and connecting the same pair of modules.}
    \label{fig:automorphism_circuits_horizontal_and_vertical_quad_only}
\end{figure}

We first explain our automorphism circuits using the modular hyperbolic color codes. The hyperbolic surface codes use a similar approach with quadrilateral groups instead of triangle groups.
We use a quadrilateral module partition of the modular hyperbolic color codes (\Cref{fig:quad_module_complements_it40} and right panel of \Cref{fig:orientation_preserving_poincare_cayley_graph_it58}), in which adjacent triangles of $\mathcal{T}^\triangle$ are paired across a chosen edge color (green, a $b$ mirror, in \Cref{fig:quad_module_complements_it40}\textbf{b}) into left cosets $g\langle b \rangle$.
Because every generator $a, b, c$ reverses orientation, each quadrilateral module $\{g, gb\}$ contains one triangle $g \in G^+ = \triangle^+(r,s,t)/\Gamma$ of even word length and one triangle $gb$ of odd word length.
Choosing the even-length triangle $g \in G^+$ as the reference triangle of each module labels the modules by the elements of $G^+$ and fixes a consistent internal 2D coordinate system $(g, x, y)$ across all modules (the compass in the inset of \Cref{fig:quad_module_complements_it40}).
As explained in \Cref{sec:left_right_mult}, an automorphism $\mathcal{A}_h$ defined by left multiplication by $h \in G^+$ maps the reference triangle $g$ of module $g$ to the reference triangle $hg \in G^+$ of module $hg$, and maps $gb$ to $(hg)b$.
Applying $\mathcal{A}_h$ therefore maps $(g,x,y)$ to $(hg,x,y)$, permuting modules while preserving the internal rotation and orientation of every module.

In \Cref{fig:automorphism_circuits_horizontal_and_vertical_quad_only} we show how such an automorphism can be implemented with physical circuits.
\Cref{fig:automorphism_circuits_horizontal_and_vertical_quad_only}\textbf{a} shows the local structure of a circuit on an $L=4$ modular hyperbolic color code that uses an additional seam of inter-module couplers on the east and west boundary of the module to implement one \textit{step} of one such permutation.
Repeating this circuit for $L=4$ steps completes the permutation.
\Cref{fig:automorphism_circuits_horizontal_and_vertical_quad_only}\textbf{b} shows a circuit that instead uses the north and south boundaries of the modules to implement a permutation of the modules.
Since each SWAP gate swaps a data qubit and an ancilla qubit, we can use the standard identity 
\begin{center}
    \includegraphics[width=4cm]{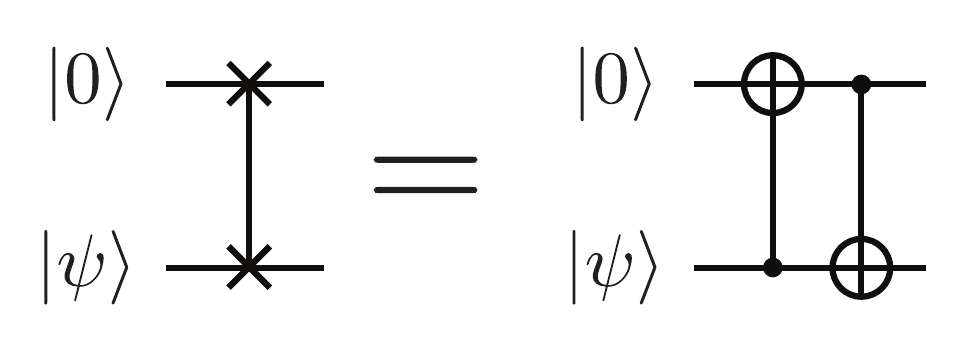}
\end{center}
to implement each SWAP gate with two CNOT gates by initializing the ancilla in $\ket{0}$.
Furthermore, we can ensure fault-tolerance by applying a $Z$ reset to each qubit in the $\ket{0}$ state between SWAPs, ensuring that a single fault does not propagate to more than one data qubit~\cite{bivariate_bicycle_codes}.

To implement the automorphisms corresponding to the full orientation-preserving subgroup $\triangle^+(r,s,t)/\Gamma$ we need only implement automorphisms $\mathcal{A}_\rho$ and $\mathcal{A}_\sigma$ corresponding to left multiplication by the generators $\rho$ and $\sigma$.
For a standard quadrilateral partition of a modular hyperbolic color code, we add a new seam from the north boundary of each module $g$ to the south boundary of module $\rho g$.
We also add a new seam from the east boundary of each module $g$ to the west boundary of module $\sigma g$.
This allows us to use walking circuits with the north/south seams to implement $\mathcal{A}_\rho$ and a walking circuit with the east/west seams to implement $\mathcal{A}_\sigma$.
Note that the existing inter-module couplers required for the code connectivity (syndrome extraction) instead use right multiplication (i.e.~each module $g$ is connected to $g\rho$, $g\rho^{-1}$, $g\sigma$ and $g\sigma^{-1}$).
Therefore, the overall connectivity of the modular architecture required for syndrome extraction and automorphisms corresponds to that of a superposed left Cayley graph and right Cayley graph, where each module is a node and left and right edges correspond to automorphism and memory seams, respectively (see \Cref{fig:automorphism_circuits_horizontal_and_vertical_quad_only}\textbf{c}).

\subsection{Automorphism Circuits for Surface Codes}\label{sec:automorphism_circuits_surface}

\begin{figure}
    \centering
    \includegraphics[width=0.9\linewidth]{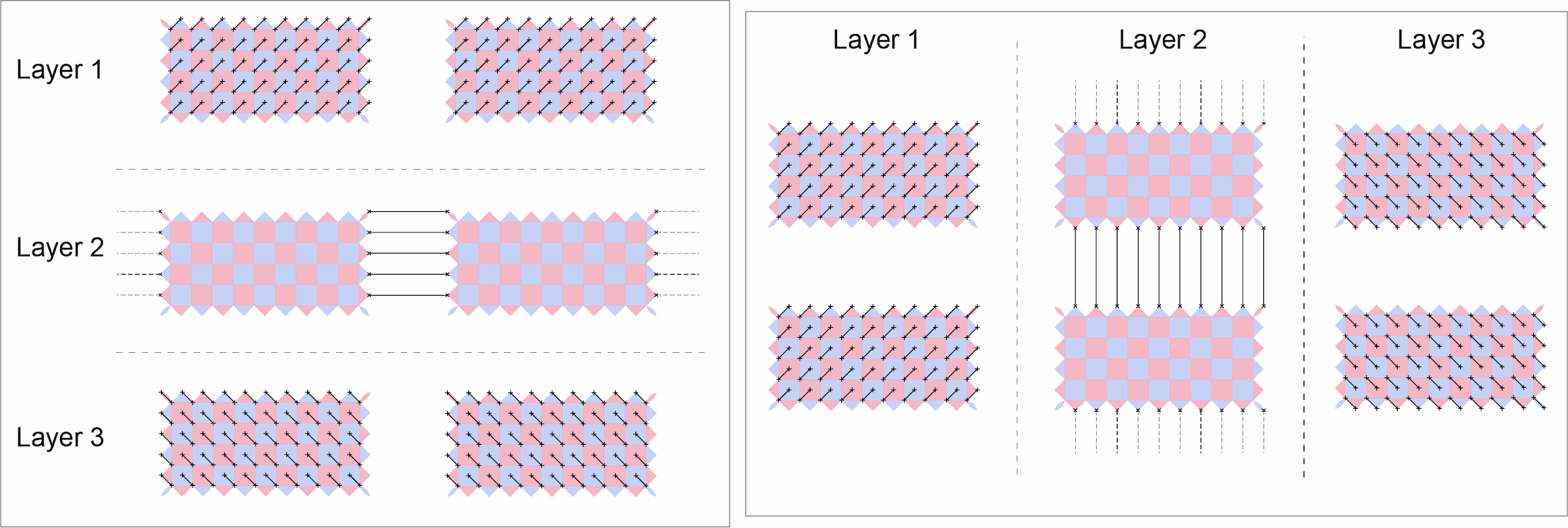}
    \caption{Circuits for implementing automorphisms in a modular hyperbolic surface code, using a rectangular partition of an $L=5$ modular hyperbolic surface code as an example. Left: One step of an ``east-west'' automorphism circuit that uses additional seams on the east and west boundaries of the module to implement a permutation of the modules. Each step of the circuit consists of three layers, and implementing the full automorphism requires repeating this step $2L=10$ times. Right: One step of a ``north-south'' automorphism circuit that uses additional seams on the north and south boundaries of the module to implement a permutation of the modules. Each step of the circuit requires three layers (shown) and repeating this step $L=5$ times completes the permutation.}
    \label{fig:automorphism_circuit_hyperbolic_surface_code}
\end{figure}

In \Cref{fig:automorphism_circuit_hyperbolic_surface_code} we show analogous circuits for modular hyperbolic \textit{surface} codes, capable of implementing automorphisms using either additional inter-module couplers on the east and west boundaries (left panel) or on the north and south boundary (right panel).
Let us assume that the quadrilateral tiling has the symmetries of an orientation-preserving quadrilateral group $\square^+(r,s,t,u)/\Gamma$ (see \Cref{sec:quadrilateral_groups}), i.e.~the dual of the quadrilateral tiling is the Cayley graph of $\square^+(r,s,t,u)/\Gamma$.
Let us also suppose that we use the rectangular module partition, for which we choose a module as the fundamental domain of $\square^+(r,s,t,u)/\Gamma$ and label each module with an element $g\in\square^+(r,s,t,u)/\Gamma$.
It is preferable to implement only two distinct automorphisms so that only one additional seam needs to be added to each module boundary.
For example, we can implement a generator $\mathcal{A}_\rho$ (where $\rho\in\square^+(r,s,t,u)/\Gamma$) using the east-west automorphism circuit and $\mathcal{A}_\sigma$ (where $\sigma\in\square^+(r,s,t,u)/\Gamma$) using the north-south automorphism circuit.
However note that, unlike the triangle groups, the quadrilateral groups are not necessarily two-generated (in general they require three generators $\rho$, $\sigma$ and $\tau$, see \Cref{sec:quadrilateral_groups}), and so $\rho$ and $\sigma$ alone in general generate a subgroup of $\square^+(r,s,t,u)/\Gamma$.
However, in practice we find that many quadrilateral groups $\square^+(r,s,t,u)/\Gamma$ are in fact two-generated, and so just the east-west and north-south circuits alone are enough to generate the full group.

\subsection{Stabilizer Dislocation and Logical Error Rates}\label{sec:stabilizer_dislocation}

\begin{figure}
    \centering
    \includegraphics[width=\linewidth]{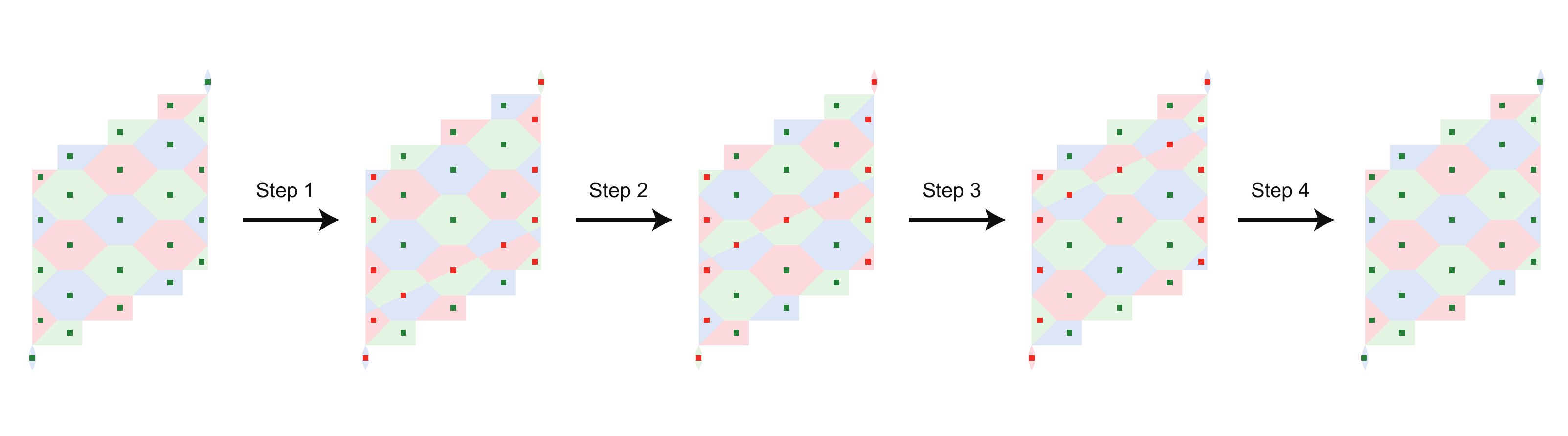}
    \caption{An illustration of the stabilizer structure within a module for each step of an automorphism circuit, for an $L=4$ modular hyperbolic color code with a quadrilateral module partition. Each step of the circuit (\Cref{fig:automorphism_circuits_horizontal_and_vertical_quad_only}\textbf{b}) moves the quantum information up by one plaquette within the module. In this example we show each of the $L=4$ steps of the automorphism circuit that uses seams on the north and south boundaries. Some of the stabilizers can still be measured between consecutive steps of the automorphism circuit, however some of the stabilizer generators become \textit{dislocated}. A stabilizer generator is dislocated if the data qubits in its support are no longer adjacent to the necessary measure qubits using the fixed connectivity of the device (and therefore cannot be measured). In each step of the automorphism circuit we have annotated the plaquettes supporting dislocated stabilizer generators with a red square, whereas the plaquettes supporting stabilizers that can be measured using device connectivity within the time step are marked with a green square. Each plaquette is given the color of the corresponding pair of $X$ and $Z$ stabilizers it supports (i.e.~the colors move with the quantum information) and plaquettes that support qubits spanning two dislocated stabilizers are bicolored (for each dislocated stabilizer generator within the plaquette, we color that portion of the plaquette with the stabilizer's color).  Data qubits are not explicitly drawn but are at the vertices of the hexagonal grid.}
    \label{fig:automorphism_circuits_stabilizer_dislocation}
\end{figure}

\begin{figure}[h]
    \centering
    \includegraphics[width=0.65\linewidth]{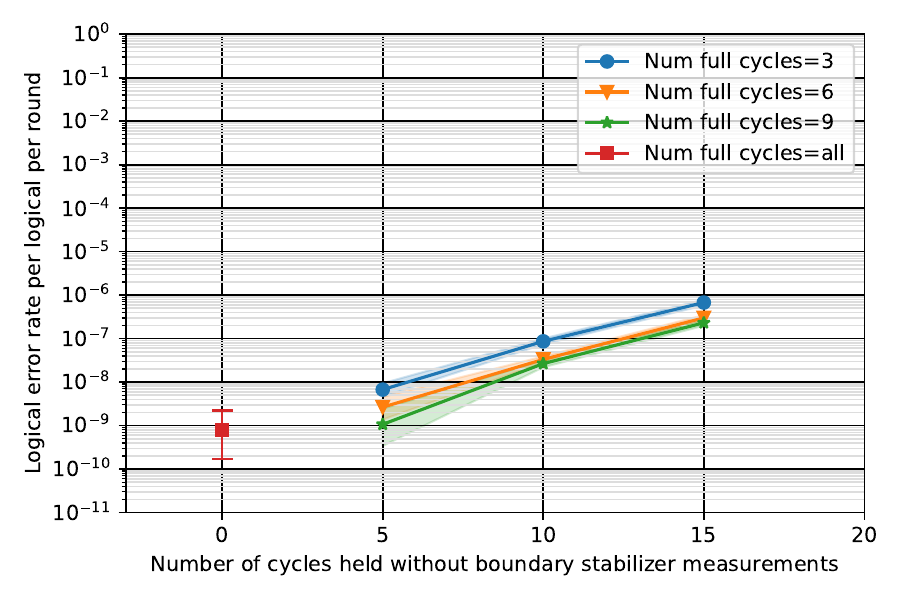}
    \caption{Simulations of a distance 20 modular hyperbolic surface code derived from a $\{4,6\}$ regular tiling with 48 faces, $L=5$ fine-graining and a square module partition. We use a modular SI1000 noise model with $p=0.1\%$ and $p_{\mathrm{seam}}=1\%$. Each circuit is 100 rounds and consists of alternating between $num_{\mathrm{gap}}$ cycles of syndrome extraction where stabilizers on the module boundary are not measured, and $num_{\mathrm{full}}$ cycles of full syndrome extraction. Here full syndrome extraction corresponds to measuring the weight-six stabilizers at the corners of the modules once every three cycles, and measuring all other stabilizers (including those on the sides of the modules) every cycle.}
    \label{fig:automorphism_emulation}
\end{figure}

While the automorphism circuits preserve the code distance (since errors cannot propagate between data qubits), we expect their logical error rates (LERs) to be higher than that of syndrome extraction alone (idling).
One reason that the LER can increase is that only a subset of the stabilizers can continue to be measured while the automorphism is being implemented.
Concretely, between two of the $L$ consecutive steps of the automorphism circuit, we can continue to measure the stabilizers that were contained fully within a module (before the automorphism circuit was implemented). 
Each of these stabilizer measurements can be compared to a prior measurement of the same stabilizer to form a detector.
However, some of the stabilizer generators become \textit{dislocated} during the automorphism circuit (marked with red squares in \Cref{fig:automorphism_circuits_stabilizer_dislocation}).
These dislocated stabilizer generators can only be measured again after all $L$ steps of the automorphism circuit have been completed.
While we have not yet simulated the full automorphism circuits, we have simulated modified memory experiments that turn off the stabilizers on the boundary of each module for a number of cycles, as a simple way to model the impact of stabilizer dislocation.
As is shown in \Cref{fig:automorphism_emulation}, turning off the boundary stabilizers for a few rounds only leads to a modest increase in logical error rate ($<10\times$ increase when turning off boundary stabilizers for 5 rounds).
We leave further optimization and full simulation of the automorphisms to future work.
In particular, we expect that the use of time-dynamic circuits~\cite{McEwen2023relaxinghardware,gidney2023new,yoshida2025low,shaw2025lowering} may lead to reduced circuit depth relative to the SWAP-gate-based circuits we have presented, as was demonstrated with the walking circuits for the Euclidean surface code in \cite{McEwen2023relaxinghardware}.

\subsection{Non-Uniform Module Partitions}\label{sec:automorphism_nonuniform_partitions}

\begin{figure}[H]
    \centering
    \includegraphics[width=0.7\linewidth]{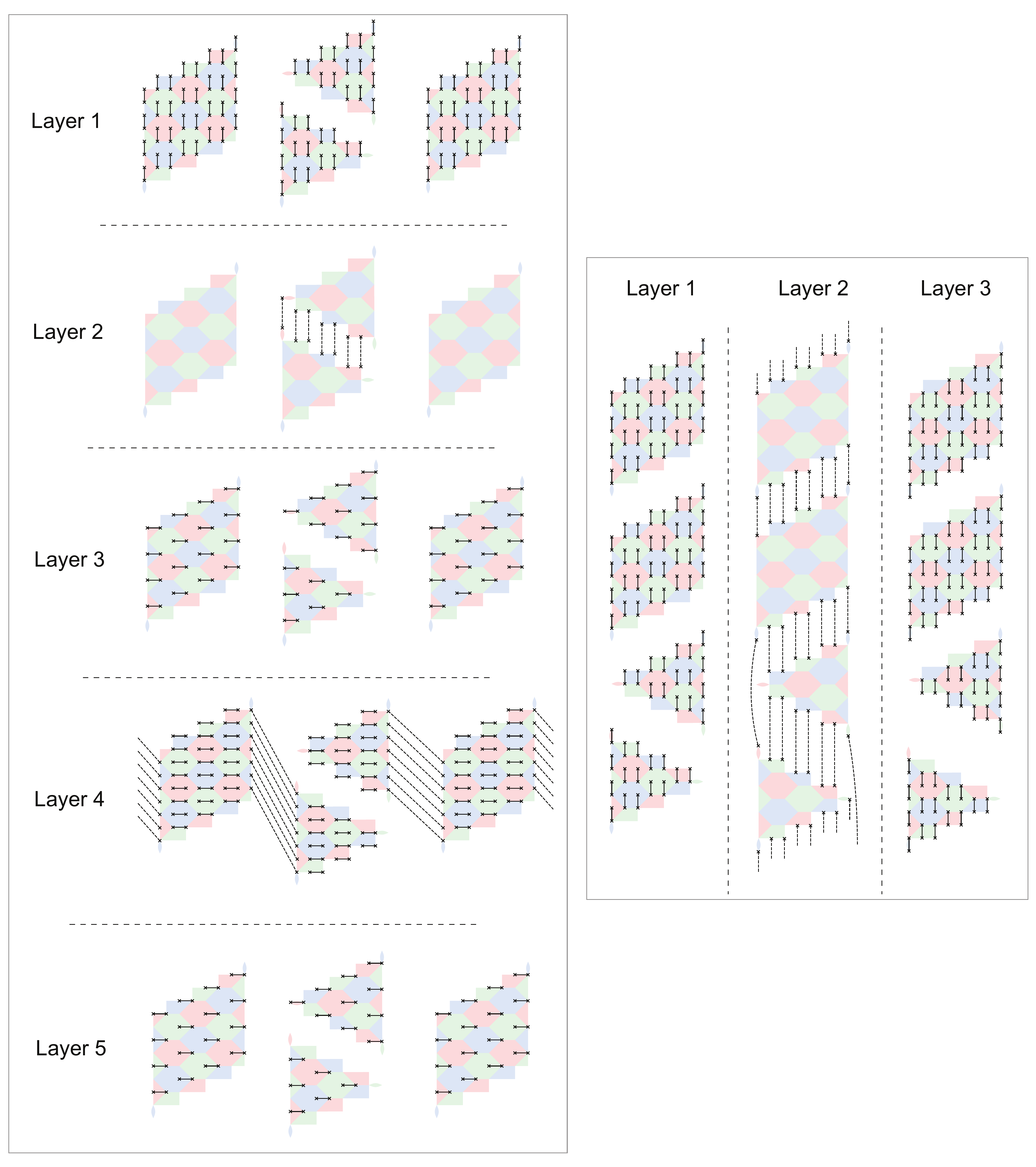}
    \caption{We can use a non-uniform module partition that uses a mixture of quadrilateral and triangular modules in a modular hyperbolic code. This can be useful for reducing the total qubit connectivity degree while implementing both automorphisms and surgery. For example, we can choose a representative of a logical operator that cuts through the middle of quadrilateral modules, and then we can split these modules into triangular modules. In this figure we show how the automorphism circuits of \Cref{fig:automorphism_circuits_horizontal_and_vertical_quad_only} can be generalized to handle such a non-uniform partition where some quadrilateral modules have been split into a pair of triangular modules.}
\label{fig:automorphism_circuits_horizontal_and_vertical_with_triangle_modules}
\end{figure}

In order to reduce the overall boundary connectivity of the architecture when a surgery ancilla system is added, we can split some of the quadrilateral modules each into two triangular modules.
This allows us to use the boundary of the triangular modules (corresponding to the interior of the quadrilateral module it was derived from) as an access point for the ribbon extractor, without increasing the overall boundary connectivity of the architecture.
The automorphism circuit can be adapted to this non-uniform partition, as shown in \Cref{fig:automorphism_circuits_horizontal_and_vertical_with_triangle_modules}.

\subsection{Module Frames and Complements of the Module Subgroup}\label{sec:module_complements}

\begin{figure}
    \centering
    \includegraphics[width=\linewidth]{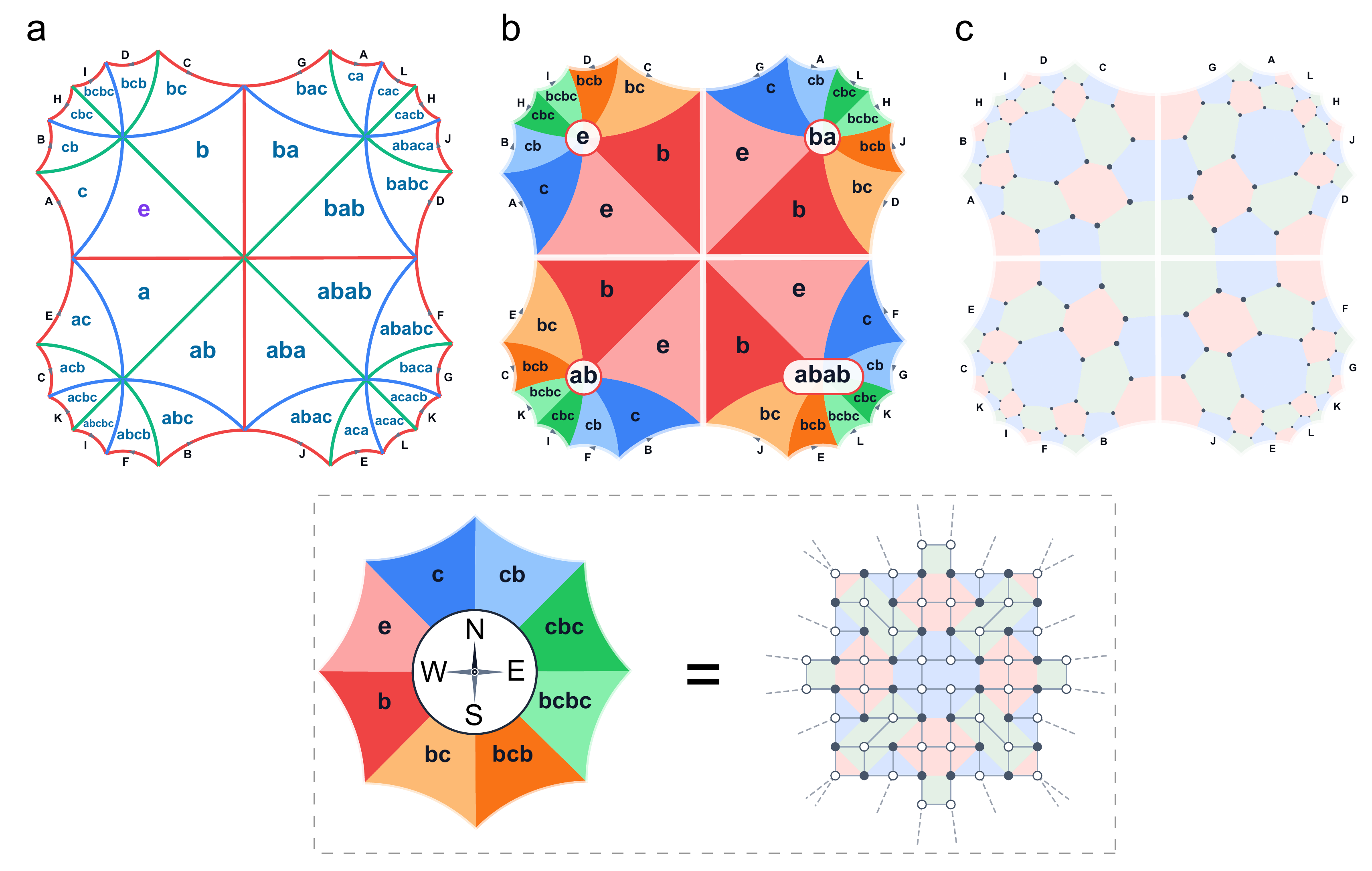}
    \caption{Octagon ($K{=}8$) modules of the $8.8.8$-$32$ color code and a complement of the module subgroup (\Cref{sec:module_complements}), with panels as in \Cref{fig:quad_module_complements_it40}. \textbf{a}~The base triangular tiling $\mathcal{T}^\triangle$, labeled as in \Cref{fig:quad_module_complements_it40}\textbf{a}. \textbf{b}~The four octagon modules, which are the left cosets $g\langle b,c \rangle$ of the order-eight dihedral subgroup $D = \langle b,c \rangle \cong D_4$: the eight triangles meeting at a corner $b \cap c$, or equivalently the eight triangles of one red octagonal face of $\mathcal{T}$ (\Cref{sec:octagon_modules}). Each triangle is labeled and colored by its slot $d \in D$, and the word at the center of each module (circled in red) is its element of the complement $\Lambda$. Of the $12$ complements of $D$ for this code we show $\Lambda = \langle \rho \rangle = \{e, ab, abab, ba\} \cong \mathbb{Z}_4$, the cyclic subgroup generated by the rotation $\rho = ab$, which is both normal ($\varphi(a)=b$) and orientation preserving. \textbf{c}~The corresponding partition of the fine-grained color code tiling $\mathcal{T}$ for $L=2$, in which each module is the set of qubits residing within one red octagonal face. \textbf{Inset}~A single octagon module with its eight slots and the internal coordinate system (compass) shared by all modules, together with the corresponding fine-grained planar module for $L=2$, including data qubits (filled circles), ancilla qubits (hollow circles) and couplers (lines).}
    \label{fig:octagon_module_complements_it40}
\end{figure}

The automorphism circuits described above rely on each module being labeled by an element of a subgroup of $G$, such that implementing an automorphism corresponds to left multiplication by an element of the same group.
This ensures that the automorphism corresponds to a permutation of the modules that leaves the internal coordinates $(x,y)$ of each module fixed, which is a requirement of the walking circuits.
This property is guaranteed to hold for the quadrilateral ($K=2$) module partition on orientable surfaces, where each module is labeled with an element of the orientation-preserving subgroup $G^+$.
In general, however, it is not guaranteed that a module partitioning scheme will admit a suitable labeling of each module by an element of a subgroup.
In particular, in this section we will focus our attention on the octagon ($K=8$) module partition, where each module is a \textit{left coset} of the order-eight dihedral subgroup $g\langle b,c\rangle \cong D_4\le G$.
If we were to attempt to implement automorphisms as left multiplication by elements of $G^+$, for example, this would require implementing the automorphism $g\rightarrow \sigma g$.
However, $\sigma\coloneqq bc\in G^+$ rotates the identity octagon module around its center, which does not preserve its internal coordinate system and cannot be implemented using our walking circuits.

In this section we will show that, even with octagon module partitions (or any partition into cosets), it can often still be possible to implement automorphisms using walking circuits.
As we will show, the requirement is that we can find a subgroup $\Lambda\le G$ such that each module contains exactly one element of $\Lambda$, and which is used to define its internal coordinate system.
The implementable automorphisms are then left multiplication by elements of $\Lambda$.
Equivalently, for a module partition defined by left cosets of a subgroup $D\le G$ (e.g.~$|D|=2$ for quadrilaterals, or $|D|=8$ for octagons) this corresponds to finding a \textit{subgroup complement} $\Lambda$ of $D$ in $G$, by definition satisfying:
\begin{equation}\label{eq:module_complement}
    \Lambda \cap D = \{e\}, \qquad \Lambda D = G .
\end{equation}
We will further show that we can find complements $\Lambda$ for octagon partitions of several of the most promising modular hyperbolic color codes considered in this work.

We will explain this condition using octagon module partitions as an example, following \Cref{fig:octagon_module_complements_it40}, but note that the same condition holds for any partition into left cosets.
Let us first assume that $\Lambda\subseteq G$ is just a subset of the elements of $G$, such that each module $gD$ ($D=\langle b,c\rangle\cong D_4$) contains exactly one triangle corresponding to an element of $\Lambda$.
We will refer to the triangle in a module corresponding to an element of $\Lambda$ as the module's \textit{frame}.
We further label each triangle in a module by its \textit{slot} $d\in D$, where by definition the module's \textit{frame} is assigned the slot $e$, and the remaining slots are determined by right multiplication by the generators $b,c$ of $D=\langle b,c\rangle$.
The frame alone (along with the generator labels $\{a,b,c\}$ of its three sides) is sufficient to define the module's internal coordinate system.
The inset of \Cref{fig:octagon_module_complements_it40} shows an octagon module with each of its triangles labeled by its slot $d\in D$, and with the compass signifying its internal coordinate system.
Note that we normalize $\Lambda$ such that it contains the identity element $e\in\Lambda$ corresponding to the frame of the identity module.

Notice that every triangle $g\in G$ factorizes as
\begin{equation}
    g = \lambda d \qquad (\lambda \in \Lambda, \; d \in D),
\end{equation}
where $\lambda\in \Lambda$ labels the module and $d\in D$ is the slot of $g$ in the module.
As an example, see the example of the octagon module partition of the $8.8.8$-$32$ code in \Cref{fig:octagon_module_complements_it40}, with triangles labeled by their group element $g\in G$ in \Cref{fig:octagon_module_complements_it40}\textbf{a} and factorized into their frame $\lambda\in\Lambda$ and slot $d\in D$ in \Cref{fig:octagon_module_complements_it40}\textbf{b}.

Now consider left multiplication by some group element $h\in G$, corresponding to an automorphism.
Our automorphism circuits require that this preserves the internal coordinate system of each module, which corresponds to mapping module frames to module frames.
Since $e\in\Lambda$, requiring that the identity frame maps to a frame, $e\rightarrow he \in \Lambda$, implies that the allowable automorphisms must be a subset of $\Lambda$.
Furthermore, if we want \textit{every} element $\lambda\in\Lambda$ to map frames to frames as an automorphism, this requires that every element $\lambda\in\Lambda$ maps a frame $\lambda^\prime\in\Lambda$ to a frame $\lambda\lambda^\prime\in\Lambda$, which is true if and only if $\Lambda$ is closed under multiplication.
Therefore a partition into left cosets of $D$ admits a full group of $|\Lambda|=|G|/|D|$ frame-preserving automorphisms $\mathcal{A}_\lambda\colon (\lambda^\prime, x, y)\rightarrow (\lambda \lambda^\prime, x, y)$ if and only if $\Lambda$ is a subgroup complement of $D$ in $G$ (\Cref{eq:module_complement}).

For quadrilateral modules ($D=\langle b\rangle$) partitioning an orientable surface, we can always use $\Lambda\coloneqq G^+$, which automatically satisfies \Cref{eq:module_complement}.
Furthermore, $G^+=\langle \rho,\sigma\rangle$ is two-generated, and therefore only requires one additional seam of inter-module couplers at each of the four boundaries of the module (one for each generator and its inverse, see \Cref{fig:automorphism_circuits_horizontal_and_vertical_quad_only}\textbf{c}).
For octagon modules ($D=\langle b,c\rangle$, $|D|=8$) there is no guarantee that a complement $\Lambda$ satisfying \Cref{eq:module_complement} will exist for any given tiling, nor that such a complement would be two-generated.
However we will now show that two-generated complements often do exist, which we find by a combinatorial search.

Given a group $G$ and a module partitioning into left cosets $gD$, a naive search for a complement $\Lambda$ might attempt to find the elements of $\Lambda$ by brute force, picking one element from each coset $gD$ and checking for group closure.
However, such a brute-force search would clearly be intractable; after fixing the identity element in the identity module, there remain $|D|^{|\Lambda|-1}$ sets to try, amounting to $8^{49}\approx10^{44}$ combinations for the $8.8.8$-$400$ code, for example.
Instead, we enumerate all complements efficiently by exploiting structure in the problem.
Rather than searching for elements of $\Lambda$ directly, we instead search for consistent \textit{slot} labelings using an efficient generalized graph coloring algorithm, and from that we extract $\Lambda$ as the set of triangles assigned slot $e$.

Our search makes use of the following observation.
Take any triangle $g$, which has a unique factorization $g=\lambda d$ (where $\lambda \in \Lambda, \; d \in D$) defining its slot label $d$.
If we multiply $g$ on the right by any element $x\in G$ we obtain $gx=\lambda d x=\lambda\lambda^\prime d^\prime$, where here $\lambda^\prime d^\prime$ is the unique factorization of $dx$.
Since $\lambda\lambda^\prime\in\Lambda$ (from group closure of $\Lambda$) we see that the slot $d^\prime$ of $gx$ depends \textit{only} on $x$ and the slot $d$ of $g$, and is completely independent of which module $g$ is in.
As a consequence, for any valid complement $\Lambda$, every element $x\in G$ defines a permutation of the slot labels $\pi_x$.
Any complement $\Lambda$ is therefore fully defined by how the three generators $a$, $b$ and $c$ of $G$ permute the slot labels.
Since each generator is self-inverse (a reflection), we know that $\pi_x:x\in\{a,b,c\}$ is an \textit{involution} of the slot labels $D$.
Additionally, for the octagon modules where $D=\langle b,c \rangle$, we already know how $b$ and $c$ permute the slot labels by considering their action on the identity module $\langle b,c\rangle$ (they are fixed).
Therefore enumerating complements of $D$ corresponds to enumerating possible involutions $\pi_a$ of the slot labels $D$ defined by generator $a$.
There are 764 possible involutions of the 8 elements of $D$ to check when enumerating octagon module complements, and this is fixed irrespective of the size of the tiling, unlike the brute-force approach.
We can check whether each of these 764 involutions defines a valid complement $\Lambda$ using a breadth-first search (BFS) on the tiling $\mathcal{T}^\triangle$.
For each candidate $\pi_a$, we start the BFS by labeling the identity triangle with the identity slot label $e$ and then propagate the slot labels across $\mathcal{T}^\triangle$ by applying the involution $\pi_x$ each time a mirror $x\in\{a,b,c\}$ is crossed, where $\pi_a$ is the candidate involution and $\pi_b$ and $\pi_c$ are fixed by the known action of $b$ and $c$ on slot labels.
Each time a mirror is crossed that closes a cycle (reaching a triangle that already has a slot label) we check for consistency of the slot label, and break if an inconsistent slot label is observed.

We do not require that $\Lambda$ be a \textit{normal} subgroup $\Lambda \trianglelefteq G$, but if $\Lambda$ \textit{is} normal then the slot labeling takes a particularly simple form.
To see why, take two elements $g_1,g_2\in G$ which factorize as $g_1=\lambda_1 d_1$ and $g_2=\lambda_2 d_2$ (where $\lambda_1,\lambda_2\in\Lambda$ and $d_1,d_2\in D$).
Taking their product, we have
\begin{equation}
    g_1g_2=(\lambda_1 d_1)\cdot (\lambda_2 d_2)=\lambda_1\cdot(d_1\lambda_2 d_1^{-1})\cdot (d_1d_2).
\end{equation}
Now if $\Lambda$ is normal then $d_1\lambda_2 d_1^{-1}\in \Lambda$, which implies that the slot of $g_1g_2$ is just the product $d_1d_2$ of the slots of the individual triangles $g_1$ and $g_2$.
Letting $\varphi \colon G \to D$ denote the map from each triangle to its slot, 
we have shown that the normality of $\Lambda$ implies that
\[
  \varphi(xy) = \varphi(x)\varphi(y) \quad \text{for all } x, y \in G.
\]
Therefore $\varphi$ is a group homomorphism, and the slot $\varphi(g)$ of any $g\in G$ is the product of the slots $\varphi(x)$ of the generators $x\in\{a,b,c\}$ in the word of $g$.
Since $\varphi(b)=b$ and $\varphi(c)=c$ are fixed for octagon modules, a normal complement $\Lambda$ is fully defined by the slot of $\varphi(a)$, which must be one of the six involutions $\{e,b,c,bcb,cbc,bcbc\}$ in $D=\langle b,c\rangle$.
An example of a normal complement is shown in \Cref{fig:octagon_module_complements_it40}\textbf{b}, which in this case is defined by $\varphi(a)=b$.
Note that normality of $\Lambda$ also implies that $G$ is the semidirect product $G=\Lambda \rtimes D$.

\begin{table*}[t]
    \centering
    \footnotesize
    \setlength{\tabcolsep}{3.5pt}
    \renewcommand{\arraystretch}{1.15}
    \resizebox{\textwidth}{!}{%
    \begin{tabular}{lllcccccccccccc}
    \toprule
    \textbf{Code} & \textbf{Symmetry group} & \textbf{Partition} & \textbf{Modules} & \textbf{Complements} & \textbf{Normal} & \textbf{Or.-preserving} & \textbf{Min.\ generators} & $k_{\mathrm{found}}/g$ & $s$ & $s^{\star}$ & $w$ & $w_1$ & \textbf{Pair} & \textbf{Cayley} \\
    \midrule
    $8.8.8$-$16$ & $\triangle(4,4,4)_{32}$ & Quadrilateral & $8$ & $4$ & $4$ & $1$ & $2$ & $2/2$ & $2$ & $2$ & $2$ & $2{+}4$ & $1$ & $0.7/1$ \\
     &  & Octagon & $2$ & $2$ & $0$ & $0$ & $1$ & $2/2$ & $3$ & $3$ & $2$ & $2{+}2{+}4$ & $0.5$ & $0.5/1$ \\
    \midrule
    $8.8.8$-$32$ & $\triangle(4,4,4)_{40}$ & Quadrilateral & $16$ & $4$ & $4$ & $1$ & $2$ & $3/3$ & $1$ & $1$ & $3$ & $4$ & $1.3$ & $2.4/4$ \\
     &  & Octagon & $4$ & $12$ & $4$ & $4$ & $1$ & $3/3$ & $3$ & $2$ & $3$ & $4$ & $0.3$ & $1.1/2$ \\
    \midrule
    $8.8.8$-$128$ & $\triangle(4,4,4)_{58}$ & Quadrilateral & $64$ & $4$ & $4$ & $1$ & $2$ & $8/9$ & $1$ & $1$ & $5$ & $6$ & $1.5$ & $3.2/6$ \\
     &  & Octagon & $16$ & $28$ & $4$ & $6$ & $2$ & $8/9$ & $3$ & $2$ & $5$ & $6$ & $0.8$ & $2.4/4$ \\
    \midrule
    $8.8.8$-$336$ & $\triangle(4,4,4)_{96}$ & Quadrilateral & $168$ & $1$ & $1$ & $1$ & $2$ & $17/22$ & $3$ & -- & $6$ & $6{+}6{+}8$ & $2.8$ & $4.1/6$ \\
     &  & Octagon & $42$ & $0$ & $0$ & $0$ & -- & -- & -- & -- & -- & -- & -- & -- \\
    \midrule
    $8.8.8$-$400$ & $\triangle(4,4,4)_{109}$ & Quadrilateral & $200$ & $4$ & $4$ & $1$ & $2$ & $25/26$ & $1$ & $1$ & $8$ & $10$ & $1$ & $4.4/6$ \\
     &  & Octagon & $50$ & $2$ & $0$ & $0$ & $2$ & $25/26$ & $4$ & -- & $8$ & $10$ & $0$ & $2.9/4$ \\
    \midrule
    $8.8.8$-$1344$ & $\triangle(4,4,4)_{271}$ & Quadrilateral & $672$ & $4$ & $4$ & $1$ & $2$ & -- & -- & -- & -- & -- & -- & -- \\
     &  & Octagon & $168$ & $8$ & $0$ & $0$ & $3$ & -- & -- & -- & -- & -- & -- & -- \\
    \bottomrule
\end{tabular}
    }
    \caption{Complements $\Lambda$ of the module subgroup $D$ for the quadrilateral ($|D|{=}2$) and octagon ($|D|{=}8$) partitions of selected $8.8.8$ hyperbolic color codes. The symmetry group is the Coxeter quotient $\triangle(4,4,4)_i$ that defines the code, written as in the extended candidate tables of \Cref{sec:promising_candidate_tables}; its order $|G|$ is the number in the code's name, and is also the number of data qubits at $L=1$. The \textit{Modules} column is $|G|/|D|$, which is also the order $|\Lambda|$ of every complement and hence the number of frame-preserving automorphisms it provides. The \textit{Complements} column gives the number of complements found by exhaustive enumeration. The \textit{Normal} column gives the number of complements $\Lambda$ that are normal subgroups of $G$, giving $G = \Lambda \rtimes D$. The \textit{Or.-preserving} column counts the number of complements contained in $G^+$. \textit{Min.\ generators} is the fewest generators any complement needs, which determines the number of additional automorphism seams needed at module boundaries. $k_{\mathrm{found}}/g$ is the number of symplectic pairs, out of the genus $g$ that a complete basis would reach, for the symmetric homology basis used in the subsequent columns. $s$ is the fewest seed cycles that cover such a basis when restricted to using the left action of one of the available complements (needing at most two generators) for the frame-preserving automorphism circuits, and $s^\star$ is the same count for a basis of the same $k_{\mathrm{found}}$ searched directly under $\Lambda$ instead of inherited from $G^+$. $w$ is the weight of a seed cycle, measured as the length of a homologous cycle in the dual tiling; there is one weight per seed cycle, and a single entry means all $s$ of them are that weight rather than that they sum to it. $w_1$ is the same weight for the shortest homologous dual cycle restricted to dual edges of a single colour, with one colour for all seeds chosen to minimise the total, listed in the same seed order as $w$. Pair gives the mean right Cayley graph distance, in the generators of the complement, between the two loops of each symplectic pair, i.e.\ the number of automorphism moves needed to switch a fixed extractor from measuring one cycle of the pair to measuring the other, with each seed's extractor placed on whichever translate of the seed minimises it. Cayley gives the mean and maximum of the same distance over all pairs of basis loops at that placement. For the quadrilateral partition, Pair and Cayley in $G^+$ with generators $\rho^{\pm1}$, $\sigma^{\pm1}$ are given in \Cref{tab:candidates_888}.}
    \label{tab:module_complements}
\end{table*}

\begin{figure}
    \centering
    \includegraphics[width=\linewidth]{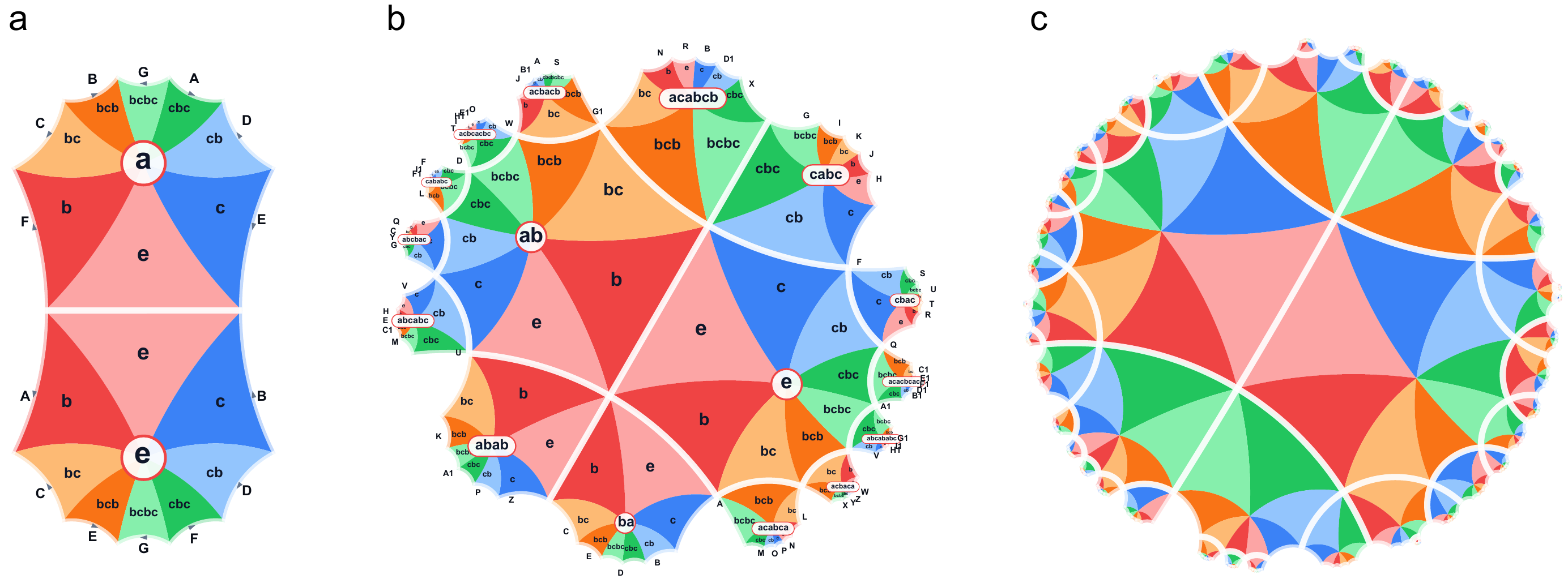}
    \caption{Examples of complements $\Lambda\leq G$ of the octagon module dihedral subgroup $D$ (where $|D|=8$) for three 8.8.8 hyperbolic tilings with Coxeter quotients $\triangle(4,4,4)_{i}$ for $i\in\{32,58,109\}$. As done in \Cref{fig:octagon_module_complements_it40} we label each octagon module with its element of $\Lambda$ (circled in red) and each triangle is labeled with its slot $d\in D$. For each tiling we provide one example of a complement, of the many counted in \Cref{tab:module_complements}. \textbf{a}, The $\triangle(4,4,4)_{32}$ tiling, for which there is no $\Lambda$ that is either orientation-preserving or normal. \textbf{b}, The $\triangle(4,4,4)_{58}$ tiling, with a normal $\Lambda$ shown satisfying $\varphi(a)=b$. \textbf{c}, The $\triangle(4,4,4)_{109}$ tiling, which has two octagon module complements, neither of which are orientation-preserving or normal. We show one such complement, with the triangles colored by their slot label.}
    \label{fig:octagon_complements_888}
\end{figure}

We enumerated the complements of the octagon module subgroup for the most promising 8.8.8 color code tilings at $L=2$, and present the results in \Cref{tab:module_complements}, including a comparison with quadrilateral module partitions.
We find that complements $\Lambda$ for $|D|=8$ do exist for most of these codes, including the smallest four codes in \Cref{fig:frontier_codes_plots} of the main text, corresponding to Coxeter quotient groups $\triangle(4,4,4)_{i}$ for $i\in\{32,40,58,109\}$.
Examples of complements for these tilings are visualized in \Cref{fig:octagon_module_complements_it40} and \Cref{fig:octagon_complements_888}.

For the code with group $\triangle(4,4,4)_{109}$, recall that we have an excellent symmetrical logical basis which allows a single seed cycle to be used to access 100 of the 104 logical qubits in the code using the quadrilateral module partition.
We find that this same tiling does admit two complements of its octagon module subgroup, listed in the $8.8.8$-$400$ row of \Cref{tab:module_complements}, which means that we can implement frame-preserving automorphism circuits.
Since the octagon complement $\Lambda$ is $4\times$ smaller than $G^+$ (here $|\Lambda|=50$ and $|G^+|=200$) we in general might expect to need more than one seed cycle to access all the logicals in the same basis using this smaller number of automorphisms.
We compute the number of seed cycles numerically by assigning two cycles in the basis to the same seed cycle class if they are related by left multiplication by an element $\lambda\in\Lambda$.
We find that we need 4 seed cycles, instead of one, to cover all the cycles in this same homology basis using the octagon modules.
One could simply construct 4 ribbon extractors, one to measure each of these seed cycles, however we note that each module in the ribbon extractor has only one outgoing seam of inter-module couplers at its boundary, and additional inter-module couplers could be used to couple the same ribbon extractor to all four seed cycles (using additional seams on both sides of the cylindrical ribbon).

\section{Compilation and Resource Estimate of a Ripple-Carry Adder}\label{sec:ripple_carry_supplement}

In order to estimate the time overhead of implementing quantum algorithms using our modular hyperbolic memory architecture, we compile a quantum arithmetic circuit to automorphism moves, ribbon extractor measurements and compute region operations.
We use the ripple-carry adder of Ref.~\cite{cuccaro2004new}, which we refer to as the CDKM adder.

\subsection{The CDKM Addition Circuit}

\begin{figure}[!htb]
\centering
\begin{minipage}{0.47\linewidth}\centering
\begin{quantikz}[row sep=0.25cm,column sep=0.3cm]
\lstick{$c$} & & \targ{} & \ctrl{2} & \\
\lstick{$b$} & \targ{} & & \ctrl{1} & \\
\lstick{$a$} & \ctrl{-1} & \ctrl{-2} & \targ{} &
\end{quantikz}\\[2pt]
{\small $\mathrm{MAJ}(c,b,a)$}
\end{minipage}\hfill
\begin{minipage}{0.47\linewidth}\centering
\begin{quantikz}[row sep=0.25cm,column sep=0.3cm]
\lstick{$c$} & \ctrl{2} & \targ{} & \ctrl{1} & \\
\lstick{$b$} & \ctrl{1} & & \targ{} & \\
\lstick{$a$} & \targ{} & \ctrl{-2} & &
\end{quantikz}\\[2pt]
{\small $\mathrm{UMA}(c,b,a)$}
\end{minipage}

\vspace{8pt}
\begin{quantikz}[row sep=0.3cm,column sep=0.3cm]
\lstick{$c_0$} & \gate[3]{\mathrm{MAJ}_0} & & & & & & \gate[3]{\mathrm{UMA}_0} & \\
\lstick{$b_0$} & & & & & & & & \\
\lstick{$a_0$} & & \gate[3]{\mathrm{MAJ}_1} & & & & \gate[3]{\mathrm{UMA}_1} & & \\
\lstick{$b_1$} & & & & & & & & \\
\lstick{$a_1$} & & & \gate[3]{\mathrm{MAJ}_2} & & \gate[3]{\mathrm{UMA}_2} & & & \\
\lstick{$b_2$} & & & & & & & & \\
\lstick{$a_2$} & & & & \ctrl{1} & & & & \\
\lstick{$z$} & & & & \targ{} & & & &
\end{quantikz}
\caption{Top: the MAJ and UMA blocks. Bottom: the $n=3$ adder. $c_0$ and $z$ never leave the compute region; the $2n$ qubits $a_i,b_i$ live in the memory.}
\label{fig:cuccaro}
\end{figure}
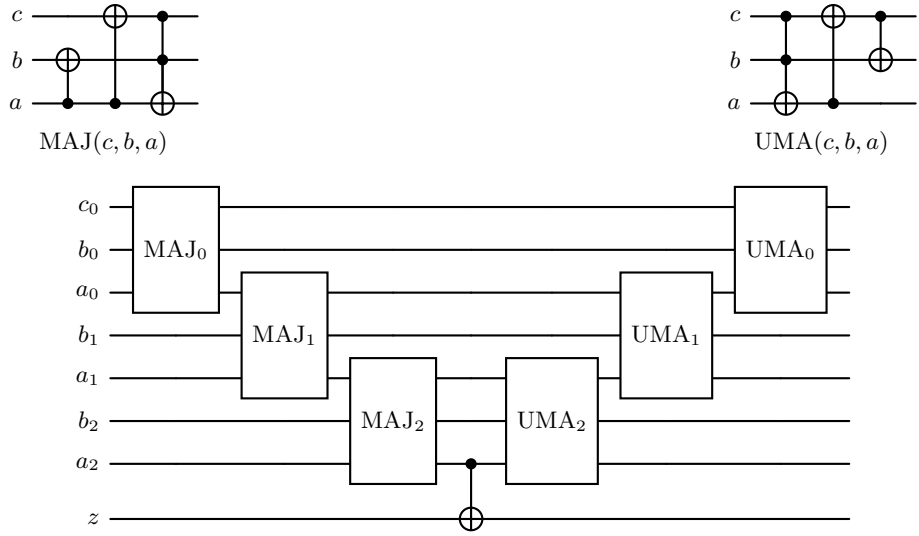

The CDKM adder is a ripple-carry quantum addition circuit introduced in Ref.~\cite{cuccaro2004new}.
It adds two $n$-bit numbers $a$ and $b$ in place: the low $n$ bits of the sum are written to the $b$ register, and the carry-out $c_n$ (the high bit of the sum) is XORed into an extra qubit $z$.
On computational basis states it acts as
\begin{equation}
    \lvert a\rangle\,\lvert b\rangle\,\lvert c_0{=}0\rangle\,\lvert z\rangle
  \;\mapsto\;
  \lvert a\rangle\,\lvert (a+b) \bmod 2^n\rangle\,\lvert 0\rangle\,\lvert z\oplus c_n\rangle,
  \qquad c_n = \bigl\lfloor (a+b)/2^n \bigr\rfloor ,
\end{equation}
where $c_0$ is the single ancilla (the carry-in), initialised in and returned to $\ket{0}$.
With $z$ initialised to $\ket{0}$, the $b$ register together with $z$ holds the full $(n{+}1)$-bit sum $a+b$.
An example of a CDKM adder, for $n=3$, is shown in \Cref{fig:cuccaro}.
The adder is composed primarily of two types of gates (or ``blocks''): the MAJ (majority) and UMA (unmajority-and-add) gates, each consisting of one Toffoli and two CNOTs (see \Cref{fig:cuccaro}), and acting on basis states as
\begin{align}
  \mathrm{MAJ}:&\quad \lvert c,\, b,\, a\rangle \;\mapsto\; \lvert c\oplus a,\; b\oplus a,\; \mathrm{maj}(a,b,c)\rangle, \\
  \mathrm{UMA}:&\quad \lvert c\oplus a,\; b\oplus a,\; \mathrm{maj}(a,b,c)\rangle \;\mapsto\; \lvert c,\; a\oplus b\oplus c,\; a\rangle,
\end{align}
where $\mathrm{maj}(a,b,c) = ab \oplus bc \oplus ca$ is the majority function.

\subsection{Compilation Overview}

\begin{figure}
\centering
\begin{minipage}{0.49\linewidth}\centering
\begin{quantikz}[row sep=0.35cm,column sep=0.25cm]
\lstick{$q$ (memory)} & \gate[2,style={fill=black!10}]{M_{\overline{X}_q \overline{X}_c}} & \gate[style={fill=black!10}]{M_{\overline{Z}_q}} & \rstick{$\ket{m}$} \\
\lstick{$\ket{0}_c$} & & \gate[style={dashed}]{P} & \rstick{$\ket{\psi}$}
\end{quantikz}\\[2pt]
{\small \textbf{a}\enspace Load}
\end{minipage}\hfill
\begin{minipage}{0.49\linewidth}\centering
\begin{quantikz}[row sep=0.35cm,column sep=0.25cm]
\lstick{$q$: $\ket{m}$} & \gate[2,style={fill=black!10}]{M_{\overline{X}_q \overline{X}_c}} & \gate[style={dashed}]{P} & \rstick{$\ket{\psi}$} \\
\lstick{$\ket{\psi}_c$} & & \gate{M_{\overline{Z}_c}} & \rstick{\scriptsize compute}
\end{quantikz}\\[2pt]
{\small \textbf{b}\enspace Store}
\end{minipage}
\caption{Standard circuits for teleporting a qubit into and out of memory using Pauli measurements. Measurements which touch the memory, and therefore use the ribbon extractor, are shaded in grey. The dashed boxes signify Pauli feedback that can be tracked by the classical control system in the accumulated Pauli frame. \textbf{a},~A \textit{load} operation, teleporting a qubit $q$ from memory into the compute region. A joint measurement $\overline{X}_q\overline{X}_c$ followed by a single measurement $\overline{Z}_q$. \textbf{b},~A \textit{store} operation, which teleports a qubit from the compute region back into the memory. We measure $\overline{X}_q\overline{X}_c$ followed by a single-qubit measurement $M_{\overline{Z}_c}$ in the compute region (e.g.~a transversal measurement of a surface code patch).}
\label{fig:si-teleport}
\end{figure}
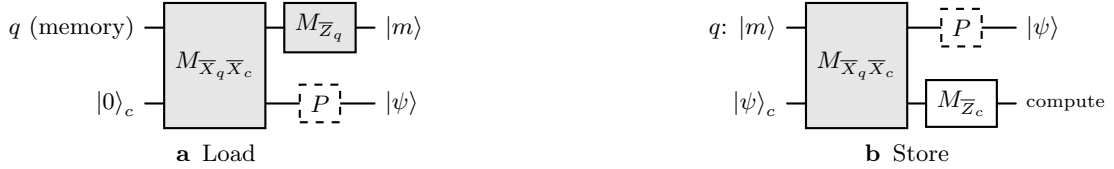

We implement the adder by combining the modular hyperbolic memory with a small compute region.
The ribbon extractor and automorphisms are used to teleport logical qubits between the memory and compute region, and all logical gates (CNOTs and Toffoli gates) are applied directly in the compute region.

To implement the teleportation we use the standard circuit identities shown in \Cref{fig:si-teleport}, for teleporting a memory qubit $q$ to or from a compute qubit $c$.
A \textit{load} operation requires two ribbon measurements, a joint $\overline{X}_q\overline{X}_c$ measurement followed by a single-qubit $\overline{Z}_q$ measurement.
The \textit{store} operation requires only a single $\overline{X}_q\overline{X}_c$ measurement followed by a single-qubit $\overline{Z}_c$ measurement in the compute region (which can just be a transversal $Z$ measurement of a surface code, for example).
The teleportation (load and store combined) therefore costs \textit{three} ribbon measurements.
Additionally, automorphisms are needed to move the $\overline{X}_q$ and $\overline{Z}_q$ logical operators into the extractor's slot, which corresponds to moving between the two cycles in a symplectic pair.
For simplicity we will leave the $c_0$ and $z$ qubits in the compute region, and only store the $a$ and $b$ registers in the hyperbolic memory.

We will assume that each ribbon extractor measurement takes $\alpha d$ cycles to achieve the desired logical fidelity, where $d$ is the distance of the hyperbolic memory and where we expect $\alpha\approx 1$.
Each basic automorphism move (implementing one of the generators $\{\rho,\rho^{-1},\sigma,\sigma^{-1}\}$) is a walking circuit described in \Cref{sec:automorphism_circuits}, which we assume takes $\kappa L$ cycles.
We will focus on the quadrilateral module partition of the hyperbolic color codes, for which we estimate that $\kappa\approx 2$ is achievable with some additional circuit optimizations inspired by Ref.~\cite{McEwen2023relaxinghardware}.
We will let $N_R$ denote the number of ribbon measurements used for the adder, and let $N_A$ denote the number of automorphisms.
The time attributed to ribbon measurements and automorphisms is then
\begin{equation}
    T=\alpha dN_R + \kappa L N_A.
\end{equation}

For each $\mathrm{MAJ}_i$ or $\mathrm{UMA}_i$ block in the adder, at least one of the three qubits it acts on (one of its operands) is already located in the compute region at the end of the prior operation (or initialized there, in the case of $c_0$).
Therefore we only need to load at most \textit{two} of the three operands before each block, and to store at most two operands to make room for the operands loaded for the next block.
Furthermore, we allocate $C\geq 3$ qubits in the compute region for memory qubits ($C$ \textit{operand slots}).
In the middle of the adder (between $\mathrm{MAJ}_{n-1}$ and $\mathrm{UMA}_{n-1}$) we can therefore retain $C$ qubits in the compute region, since the last $C$ operands used by $\mathrm{MAJ}_i$ blocks are the first $C$ needed by the $\mathrm{UMA}_i$ blocks.
This saves one store and one load for each of the $C$ operand slots, while the remaining $2n-C$ operand qubits each require the full two loads and two stores throughout the adder.
Since each combined load and store requires three ribbon measurements, this leads to
\begin{equation}
    N_R=3(4n-C)
\end{equation}
ribbon measurements needed for an $n$-bit adder.
We will choose $C=5$, and hence have $N_R=12n-15$.
Note that this corresponds to Belady's cache replacement policy~\cite{belady1966replacement} (when a slot is needed, storing the operand whose next use is furthest in the future), which minimizes the number of loads, and hence stores (each loaded qubit is stored exactly once), for a fixed order of blocks.

With $N_R$ determined (and independent of the hyperbolic memory code used) the rest of this section will be focused on estimating the number of automorphism moves $N_A$, which does depend on the tiling and logical basis.
We also make the simplifying assumption that the ribbon extractor can measure any of the $C$ operand slot $X$ logicals in the compute region for teleportation.
This could be achieved, for example, by using additional inter-module couplers, or instead by allocating additional routing space in the compute region (such that the ribbon extractor only needs to measure a single logical there to access all $C$ qubits).
We leave a detailed study of this aspect of the compilation to future work.

\subsection{Coarse Upper Bound on the Number of Automorphism Moves}

Before optimizing the number of automorphism moves, we will first compute a very coarse upper bound on $N_A$.
We will be focusing primarily on the 8.8.8 modular hyperbolic color code with symmetry group $\triangle(4,4,4)_{109}$ (the $8.8.8$-$400$ code in \Cref{tab:homology_automorphism_logicals}), using the logical basis generated by the orbit of a single seed cycle $\gamma$ under $G^+=\langle\rho,\sigma\rangle$.
This tiling is used to construct modular hyperbolic color codes with parameters $[[3600,104,22]]$ at $L=3$ and $[[6400,104,30]]$ at $L=4$.

For this code, the seed cycle $\gamma$ has trivial stabilizer in $G^+$, so each cycle in its orbit can be written uniquely as $C_x=x\gamma$ for some $x\in G^+$.
An automorphism move by a generator $s\in\{\rho^{\pm1},\sigma^{\pm1}\}$ maps every cycle $y\gamma$ to $sy\gamma$, so a sequence of moves $s_1,\dots,s_k$ applies the automorphism $s_k\cdots s_1$.
If we attach the ribbon extractor to the cycle $h\gamma$, then the cycle $C_x$ is under the extractor once we have applied a sequence of moves corresponding to the automorphism $hx^{-1}$.
Switching from measuring $C_{x_1}$ to measuring $C_{x_2}$ therefore requires a sequence of moves with product $(hx_2^{-1})(hx_1^{-1})^{-1}=h\,x_2^{-1}x_1\,h^{-1}$, i.e.
\begin{equation}\label{eq:move_cost}
  M(x_1,x_2) = \bigl|\,h\,x_1^{-1}x_2\,h^{-1}\bigr|
\end{equation}
automorphism moves, where $|\cdot|$ is the word length in the generators $\{\rho^{\pm1},\sigma^{\pm1}\}$ (a word and its inverse have the same length since this set is closed under inverses).
Equivalently, each move $s$ changes the label of the cycle under the extractor from $x$ to $x\,h^{-1}s^{-1}h$, i.e.~by right multiplication.
For $h=e$, $M(x_1,x_2)$ is therefore the \textit{right} Cayley distance $|x_1^{-1}x_2|$ between $x_1$ and $x_2$, rather than the left Cayley distance $|x_2x_1^{-1}|$ (the length of the automorphism mapping $C_{x_1}$ onto $C_{x_2}$).
For general $h$ it is the right Cayley distance with respect to the conjugated generators $h^{-1}sh$, and so choosing $h$ is equivalent to choosing the centres of rotation of the generators relative to the extractor.
Since conjugation by $h$ is an isomorphism of Cayley graphs, the maximum number of automorphism moves needed to move from one logical operator to another under the ribbon extractor is, for any $h$, the diameter of the Cayley graph,
\begin{equation}
  D = \operatorname{diam}\,\mathrm{Cay}\bigl(G^+, \{\rho^{\pm1}, \sigma^{\pm1}\}\bigr).
\end{equation}
Given that each ribbon extractor measurement requires at most $D$ automorphism moves, we obtain the coarse upper bound $N_A\leq D N_R$.
This gives a total time attributed to ribbon measurements and automorphisms of at most
\begin{equation}\label{eq:coarse_adder_cycles_ub}
    T\leq 3\left(\alpha d + \kappa D L \right)(4n-C)
\end{equation}
cycles.

For the $\triangle(4,4,4)_{109}$ tiling we have $D=6$.
For a $50$-bit adder, and assuming $\alpha=1$, $\kappa=2$ and $C=5$, \Cref{eq:coarse_adder_cycles_ub} gives $T\leq 33{,}930$ cycles (679 per bit) at $L=3$ and $T\leq 45{,}630$ cycles (913 per bit) at $L=4$.
Notice that around $62\%$ of the cycles are attributable to automorphisms at both $L=3$ and $L=4$.
However, we will now explain how a small number of important but straightforward optimizations can significantly reduce the number of automorphisms required.

\subsection{Minimizing Within-Pair Switches}

We will call an automorphism that moves the extractor (i.e.~changes which cycle is under the extractor) from a cycle in one symplectic pair to a cycle in a different symplectic pair a \textit{jump}.
A jump takes at most $D$ basic automorphism moves.
We will refer to an automorphism that moves the extractor from one cycle to its partner within a symplectic pair a \textit{switch}.

Each load operation requires measuring the $\overline{X}_q$ and $\overline{Z}_q$ logical operators of the same logical qubit $q$ stored in the hyperbolic memory, where one of these is measured jointly with a compute region logical (e.g.~$M_{\overline{X}_q\overline{X}_c}$ in \Cref{fig:si-teleport}).
Measuring these two logical operators consecutively requires a switch.
Given that this can be a common operation, it makes sense to optimize it.
As we will now explain, for this code (using our symmetric basis for it), an appropriate choice of extractor location can reduce the number of moves needed for a switch from 5 to 1.

In our basis, every symplectic pair has the same automorphism mapping one cycle to its partner: the $i$-th pair consists of the cycles $\alpha_i = y_i\gamma$ and $\beta_i = y_i w^{\pm1}\gamma$ for some $y_i\in G^+$, where $w=\sigma^2\rho\,\sigma^2$.
By \Cref{eq:move_cost}, a switch requires $M(y_i, y_iw^{\pm1}) = |h w^{\pm1} h^{-1}|$ moves, which is independent of $i$.
For $h=e$ this is $|w|=5$.
However, since $\sigma^4=e$ we have $w=\sigma^2\rho\,\sigma^{-2}$, and so choosing the extractor location $h=\sigma^2$ gives $hwh^{-1}=\rho$.
With this choice, every switch is a single $\rho$ or $\rho^{-1}$ move (as for the pair shown in \Cref{fig:logic}\textbf{a} of the main text).
Of the $200$ possible extractor locations $h\in G^+$, $8$ have this property.

We can choose an ordering of ribbon measurements where the two measurements in each load operation are always implemented consecutively (i.e.~$M_{\overline{Z}_q}$ immediately follows $M_{\overline{X}_q\overline{X}_c}$).
Moving the ribbon extractor from $M_{\overline{X}_q\overline{X}_c}$ to $M_{\overline{Z}_q}$ then always costs one switch (a single automorphism move).
All other automorphisms cost at most one jump.
This leads to at most $(4n-C)D$ moves for the stores, $(4n-C)D$ moves for the $M_{\overline{X}_q\overline{X}_c}$ in each load, and $(4n-C)$ for the $M_{\overline{Z}_q}$ in each load.
In total this gives $N_A\leq (2D+1)(4n-C)=\frac{2D+1}{3}N_R$ automorphism moves, or
\begin{equation}\label{eq:within_pair_adder_cycles_ub}
    T\leq 3\left(\alpha d + \frac{2D+1}{3} \kappa L \right)(4n-C)
\end{equation}
cycles.
Using the same assumptions as before, the number of cycles drops to 28,080 at $L=3$ (562 per bit) and 37,830 at $L=4$ (757 per bit).

\subsection{Grouping Qubits into Pairs}

We can reduce the number of moves further by storing operand qubits that are accessed in close succession in the same symplectic pair of homologically nontrivial cycles.
Each symplectic pair of cycles $(\alpha_i,\beta_i)$ stores four logical qubits, which we now use to store operand qubits $a_{2i}$, $b_{2i}$, $a_{2i+1}$ and $b_{2i+1}$, choosing $\overline{X}_q$ supported on $\alpha_i$ and $\overline{Z}_q$ on $\beta_i$ for all four.
This swaps the logical Pauli bases of two of the qubits relative to the convention in \Cref{sec:logical_basis}, purely for notational convenience; equivalently, we could keep that convention and interchange $X$ and $Z$ in the teleportation circuits of those two qubits, so that their joint measurement is $M_{\overline{Z}_q \overline{Z}_c}$ on $\alpha_i$ and their single measurement is $M_{\overline{X}_q}$ on $\beta_i$.

We refer to the loads and stores performed before a block (or at the end, when all qubits are returned to memory) as a \textit{step}.
With $C=3$, the qubits in each step have at most three consecutive bit indices (e.g.~before $\mathrm{MAJ}_i$ we store $a_{i-2}$ and $b_{i-1}$ and load $b_i$ and $a_i$), so each step involves at most two symplectic pairs.
Moreover, at most one of these pairs is acted on by both stores and loads, as can be checked for each type of step.
We can therefore handle the pairs one at a time while performing every store before every load, which frees space in the compute region: first a pair with only stores (if any), then a pair with both, then a pair with only loads.
On each pair, the stores and the joint measurements of the loaded qubits both measure $M_{\overline{X}_q\overline{X}_c}$, acting on $\alpha_i$, and are followed by the measurements $M_{\overline{Z}_q}$ on $\beta_i$.
Each pair therefore costs at most one jump to reach it and one switch from $\alpha_i$ to $\beta_i$ (a pair with only stores needs no switch), so each step requires at most $2D+2$ automorphism moves.
There are $2n$ non-empty steps, giving $N_A \leq 4(D+1)n \approx \frac{D+1}{3}N_R$.
Here we have used the memory schedule for $C=3$, for which $N_R=3(4n-3)$ (only six more ribbon measurements than for $C=5$), so that
\begin{equation}\label{eq:grouped_adder_cycles_ub}
    T \leq 3\alpha d\,(4n-3) + 4(D+1)\,\kappa L\, n
\end{equation}
cycles.
Using the same assumptions as before, this gives 21,402 cycles at $L=3$ (428 per bit) and 28,930 cycles at $L=4$ (579 per bit).

\subsection{Compute Region}

The logic in the compute region can be implemented in parallel to the ribbon extractor measurements and automorphisms, and has a negligible impact on the overall runtime of the adder.
After a load's joint measurement $M_{\overline{X}_q\overline{X}_c}$, the subsequent $M_{\overline{Z}_q}$ only determines a Pauli correction to the qubit in the compute region.
The next $\mathrm{MAJ}_i$ or $\mathrm{UMA}_i$ block can therefore begin immediately after the joint measurements, and the Pauli frame can be propagated through its CNOTs once it has been determined.
The Toffoli, implemented with a $\ket{CCZ}$ state, only needs the Pauli frame before the Clifford corrections.
Meanwhile, the extractor performs the $M_{\overline{Z}_q}$ measurements and the automorphism moves to the next step's stores (which act only on the memory), taking roughly $2\alpha d + (D+1)\kappa L$ cycles.
This is comparable to the duration of a block ($\approx 4d$ plus the decoder reaction time), so only a small fraction of the compute time remains exposed.
We also note that the duration of the block can be reduced to just the reaction time using AutoCCZ states and a sufficiently large compute region~\cite{gidney2019flexible}.

\section{Algorithms for Finding a Short Canonical Homology Basis}\label{sec:homology_basis}
As explained in \Cref{sec:logical_basis}, finding a canonical logical basis for a hyperbolic code reduces to finding a \emph{canonical homology basis} of the underlying tiling: for an orientable surface this corresponds to a set of $g$ symplectic pairs of homologically non-trivial cycles. For the color codes we search for these cycles directly on the three-colorable hyperbolic tiling $\mathcal{T}$, whereas for the surface codes we use the base quadrilateral tiling $\mathcal{T}^\square$. In either case the homology basis search is the same combinatorial problem on a hyperbolic tiling. We therefore drop the color/surface distinction for the remainder of this section and simply consider a finite, closed and orientable hyperbolic tiling denoted $\mathcal{T}$.

We fix the following notation. We write $\mathcal{T}$ for the hyperbolic tiling and $\mathcal{T}^*$ for its dual, and denote the primal graph by $\mathcal{G} = (V, E, F)$, with vertices $V$, edges $E$ and faces $F$. The dual graph $\mathcal{G}^*$ has one vertex per face of $\mathcal{T}$ and one dual edge $e^*\in E^*$ per primal edge $e\in E$, with a one-to-one correspondence between primal and dual edges $E^* \leftrightarrow E$. A \emph{cycle} is a closed loop of primal edges, and a closed loop of dual edges is a \emph{cocycle}. Two cycles combine by their symmetric difference $\gamma_i \oplus \gamma_j$ (the cycle of edges belonging to exactly one of them, with edges common to both removed). Under this operation the cycles form a group, and its quotient by the cycles that bound sets of faces (the trivial cycles) is the first homology group $H_1(\Sigma_g;\,\mathbb{F}_2)$. Taking the symmetric difference of a trivial cycle with a non-trivial $\gamma$ leaves the homology class of the non-trivial cycle unchanged, so many different cycles represent the same class. Two such cycles are \emph{homologous}, and there are $2g$ independent non-trivial classes. We write $[\gamma]$ for the class of $\gamma$.

We also recall the intersection number of \Cref{sec:logical_basis}. For two cycles $\gamma_i, \gamma_j$ we write $\langle \gamma_i, \gamma_j \rangle \in \mathbb{F}_2$, equal to $1$ if they cross an odd number of times and $0$ otherwise. This number depends only on the homology classes, so homologous cycles have the same intersection number with every other cycle. For a candidate set $\{\gamma_1, \ldots, \gamma_m\}$ we collect these numbers into the \emph{cycle intersection matrix} $M \in \{0,1\}^{m \times m}$, the symmetric zero-diagonal matrix with $M_{ij} = \langle \gamma_i, \gamma_j \rangle$. Finding a canonical homology basis then amounts to selecting $2g$ cycles and reordering the rows and columns of the corresponding $2g \times 2g$ submatrix of $M$ so that it equals the \emph{symplectic matrix}
\begin{equation}\label{eqn:symplectic_form_intro}
    \Omega \;=\; \bigoplus_{i=1}^{g} \begin{pmatrix} 0 & 1 \\ 1 & 0
    \end{pmatrix} \;\in\; \{0,1\}^{2g \times 2g}.
\end{equation}

A homology basis of $2g$ independent non-trivial cycles can be computed efficiently, for example via tree--cotree decomposition \cite{tree_cotree_decomp}, and brought into canonical form by Gaussian elimination over $\mathbb{F}_2$. This does not, however, yield geometrically short cycles. Each cycle $\gamma \subseteq E$ has a length $|\gamma|$ given by its number of edges, and Gaussian elimination corresponds to taking symmetric differences of cycles, which typically lengthens them. Conversely, Erickson and Whittlesey give a polynomial-time algorithm for the \emph{shortest homology basis}, but without the canonical constraint~\cite[Theorem~4.3]{Erickson_Whittlesey}, 
We are not aware of an efficient algorithm that returns a shortest basis in canonical form directly, and related problems are NP-hard \cite{Erickson_HarPeled,CHAMBERS2008}. In any case, we need more than a short canonical basis: we need one that is additionally \emph{symmetrical}, in the sense that its cycles lie in the orbit of a small number of seed cycles as described in \Cref{sec:logical_basis}. In this section, we describe algorithms to find a \emph{short canonical homology basis} of $\mathcal{T}$ by enumerating short non-trivial cycles from every vertex.

The symmetries we use are the orientation-preserving automorphisms $\mathrm{Aut}^+(\mathcal{T}) = G^+ = \langle \rho, \sigma \rangle$, that can be implemented as fault-tolerant automorphism circuits
(\Cref{sec:automorphism_circuits}). Relating the basis cycles by these rotations means that only a seed cycle needs its own ribbon extractor for logical measurement (\Cref{sec:ribbon_extractor}), while every other cycle in its orbit is mapped onto the seed by an automorphism circuit. The automorphisms map non-trivial cycles to non-trivial cycles of the same length and preserve intersection numbers, so the orbit of a cycle consists of cycles of equal length in potentially distinct homology classes.

At a high level, our algorithms enumerate non-trivial cycles by graph search from every vertex of $\mathcal{T}$, recording the length $|\gamma|$ and class $[\gamma]$ of each and keeping only the shortest representative per homology class encountered. We use these as seed cycles. Starting from the shortest one or two, we generate their full orbit under $\mathrm{Aut}^+(\mathcal{T}) = \langle \rho, \sigma \rangle$, producing a candidate set whose cycles are related by tiling symmetries. This ensures the resulting canonical pairs map to each other by global rotations. We then compute the intersection number for all candidate pairs and extract a subset forming $\leq g$ mutually compatible symplectic pairs by a clique or backtracking search.

\begin{figure}[H]
\centering
\begin{tikzpicture}[
    node distance = 5mm and 14mm,
    every node/.style = {font=\small},
    phasebox/.style = {
        rectangle, rounded corners=6pt,
        minimum width=74mm, minimum height=10mm,
        text centered, draw, thick
    },
    stepbox/.style = {
        rectangle, rounded corners=4pt,
        minimum width=66mm, minimum height=9mm,
        text centered, draw
    },
    p1box/.style  = {phasebox, fill=purple!8,  draw=purple!50},
    p1step/.style = {stepbox,  fill=purple!5,  draw=purple!30},
    p2box/.style  = {phasebox, fill=teal!8,    draw=teal!50},
    p2step/.style = {stepbox,  fill=teal!5,    draw=teal!30},
    p3box/.style  = {phasebox, fill=cyan!8,    draw=cyan!40},
    p3step/.style = {stepbox,  fill=cyan!5,    draw=cyan!30},
    p4box/.style  = {phasebox, fill=orange!8,  draw=orange!40},
    p4step/.style = {stepbox,  fill=orange!5,  draw=orange!30},
    algtag/.style  = {font=\scriptsize\itshape, text=gray},
    arr/.style     = {-{Stealth[length=4pt]}, thick, gray!60},
    arrthin/.style = {-{Stealth[length=3.5pt]}, gray!50},
]

\node[p1box] (p1)
    {\textbf{Phase 1:} \textbf{Homology Coordinate System}};

\node[p1step, below=4mm of p1] (tc)
    {Tree-Cotree Decomposition};
\node[algtag, right=2mm of tc]
    {Alg.~\ref{alg:tree-cotree}, \cite{tree_cotree_decomp}};

\node[p1step, below=3mm of tc] (hl)
    {Homology Label Assignment};
\node[algtag, right=2mm of hl]
    {Alg.~\ref{alg:signatures}, \cite{Erickson_Whittlesey}};

\node[algtag, below=1mm of hl, text=purple!60, font=\scriptsize]
    (p1out) {output: $\phi : E \to \{0,1\}^{2g}$};

\node[p2box, below=8mm of p1out] (p2)
    {\textbf{Phase 2:} \textbf{Non-Trivial Cycle Enumeration}};

\node[p2step, below=4mm of p2] (le)
    {Cycle Enumeration};
\node[algtag, right=2mm of le]
    {Alg.~\ref{alg:cycle-enum}};

\node[algtag, below=1mm of le, text=teal!60, font=\scriptsize]
    (p2out) {output: $\mathcal{L}$, shortest cycle per homology class encountered};

\node[p3box, below=8mm of p2out] (p3)
    {\textbf{Phase 3:} \textbf{Automorphism Orbit Generation}};

\node[p3step, below=4mm of p3] (aut)
    {Seed Cycle Orbit Generation};
\node[algtag, right=2mm of aut]
    {Alg.~\ref{alg:aut_basis}};

\node[algtag, below=1mm of aut, text=cyan!60, font=\scriptsize]
    (p3out) {output: $\mathcal{L}_{\mathrm{aut}}$, orbit of seed cycles};

\node[p4box, below=8mm of p3out] (p4)
    {\textbf{Phase 4:} \textbf{Canonical Homology Basis}};

\node[p4step, below=4mm of p4] (cm)
    {Cycle Intersection Matrix};
\node[algtag, right=2mm of cm]
    {Alg.~\ref{alg:crossing_matrix}};

\node[p4step, below=3mm of cm] (cs)
    {Symplectic Pair Search};
\node[algtag, right=2mm of cs]
    {Alg.~\ref{alg:max-clique}/\ref{alg:backtrack}};

\node[algtag, below=1mm of cs, text=orange!70, font=\scriptsize]
    (p4out) {output: $k_{\mathrm{found}} \leq g$ symplectic pairs
             $\mathcal{B} = \{(\alpha_1,\beta_1),\ldots,(\alpha_{k_{\mathrm{found}}},\beta_{k_{\mathrm{found}}})\}$};

\draw[arr]    (p1)    -- (tc);
\draw[arrthin](tc)    -- (hl);
\draw[arr]    (p1out) -- (p2);
\draw[arr]    (p2)    -- (le);
\draw[arr]    (p2out) -- (p3);
\draw[arr]    (p3)    -- (aut);
\draw[arr]    (p3out) -- (p4);
\draw[arr]    (p4)    -- (cm);
\draw[arrthin](cm)    -- (cs);

\end{tikzpicture}
\caption{Algorithm outline for constructing a canonical homology basis of a
hyperbolic tiling $\mathcal{T}$ of genus $g$
(Sections~\ref{sec:homology_coordinate_system}--\ref{sec:canonical_homology_basis}). \textbf{Phase~1} builds a homology coordinate system via a tree-cotree decomposition
and homology label assignment, producing the label map $\phi : E \to \{0,1\}^{2g}$
(\algoref{alg:tree-cotree}--\algoref{alg:signatures}).
\textbf{Phase~2} enumerates homologically non-trivial cycles by BFS from every vertex,
retaining the shortest representative per homology class encountered (\algoref{alg:cycle-enum}).
\textbf{Phase~3} generates the orbit of seed cycles under the orientation-preserving
automorphism group $\mathrm{Aut}^+(\mathcal{T}) = \langle \rho, \sigma \rangle$
(\algoref{alg:aut_basis}).
\textbf{Phase~4} constructs a canonical basis of $k_{\mathrm{found}} \leq g$ symplectic pairs from
the cycle intersection matrix via a clique or backtracking search
(\algoref{alg:max-clique}--\algoref{alg:backtrack}).}
\label{fig:hyperbolic_canonical_homology_basis}
\end{figure}
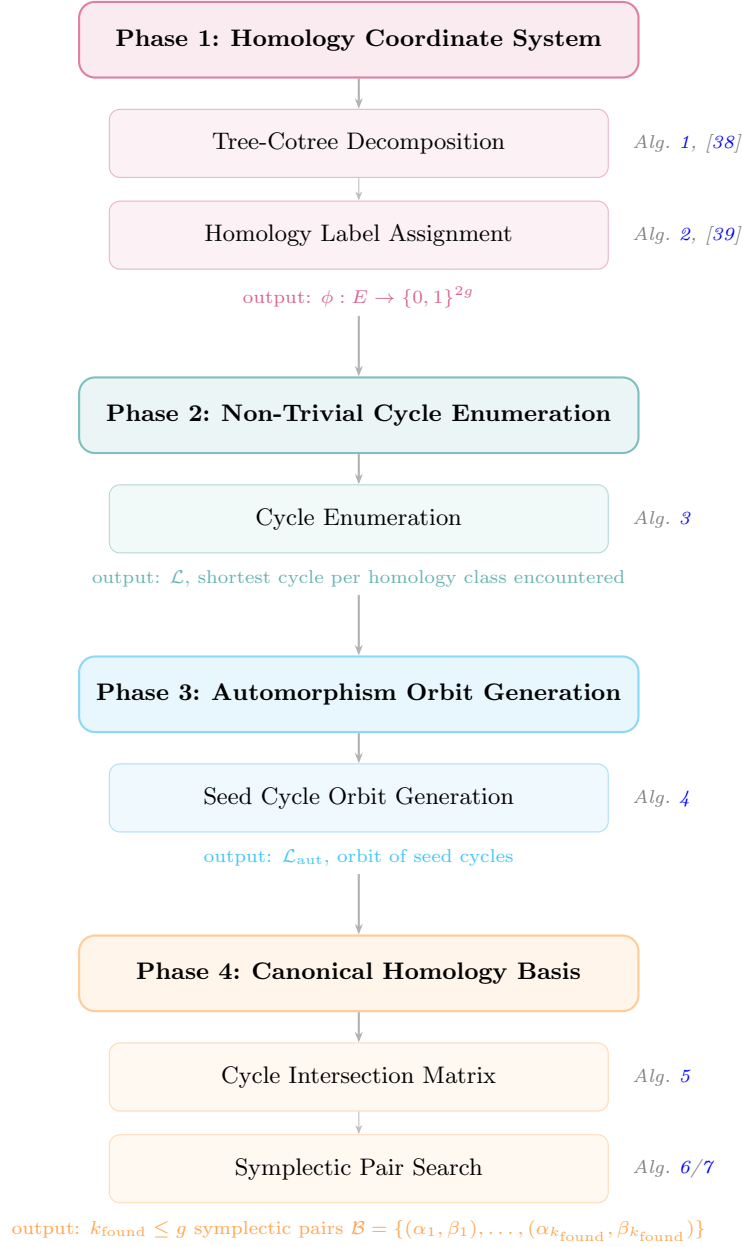

\Cref{fig:hyperbolic_canonical_homology_basis} outlines our algorithms for constructing a canonical homology basis of a hyperbolic tiling $\mathcal{T}$ of genus $g$. Phase~1 builds a homology coordinate system via a tree-cotree decomposition and homology label assignment for every edge of the primal graph (Algorithms~\ref{alg:tree-cotree}--\ref{alg:signatures}, explained in \Cref{sec:homology_coordinate_system}). Phase~2 enumerates non-trivial cycles from every vertex using breadth-first search (BFS) (Algorithm~\ref{alg:cycle-enum}, \Cref{sec:homology_cycle_enumeration}). Phase~3 generates the orbit of seed cycles under $\mathrm{Aut}^+(\mathcal{T}) = \langle \rho, \sigma \rangle$ (Algorithm~\ref{alg:aut_basis}, \Cref{sec:automorphism_basis}). Phase~4 constructs a canonical basis of $k_{\mathrm{found}} \leq g$ symplectic pairs from the cycle intersection matrix via a clique or backtracking search (Algorithms~\ref{alg:max-clique}--\ref{alg:backtrack}, \Cref{sec:canonical_homology_basis}).

\subsection{Homology Coordinate System}\label{sec:homology_coordinate_system}
We start with the primal graph $\mathcal{G} = (V, E, F)$ of the hyperbolic tiling 
$\mathcal{T}$. In order to classify the homology class $[\gamma]$ of any cycle we enumerate, we construct a map $\phi : E \to \{0,1\}^{2g}$ assigning to each edge a $2g$-bit vector. The homology class of any cycle $\gamma$ is then identified by the bit-vector label
\[
    h(\gamma) \;=\; \bigoplus_{e \in \gamma} \phi(e) \;\in\; \{0,1\}^{2g},
\]
obtained by taking the bit-wise XOR of the labels of its edges. The induced map $h$ on homology classes is an isomorphism $H_1(\Sigma_g;\,\mathbb{F}_2) \cong \mathbb{F}_2^{2g}$: the $2^{2g}-1$ non-trivial homology classes correspond to the $2^{2g}-1$ non-zero elements of $\{0,1\}^{2g}$, and the trivial class corresponds to the zero vector $\mathbf{0} \in \{0,1\}^{2g}$. Two cycles are homologous if and only if they share the same label. We describe how we construct $\phi$ using \algoref{alg:tree-cotree} and \algoref{alg:signatures}, closely following Erickson and Whittlesey~\cite{Erickson_Whittlesey}.

\textbf{Tree-cotree decomposition (\algoref{alg:tree-cotree})}. We begin by identifying a set of $2g$ edges of $\mathcal{G}$ whose associated cycles form a basis of $H_1(\Sigma_g;\,\mathbb{F}_2)$ via the tree-cotree decomposition~\cite{tree_cotree_decomp}. We first find a spanning tree $T$ of $\mathcal{G}$. Every non-tree edge $e \notin T$ closes exactly one cycle, formed by $e$ together with the unique path in $T$ connecting its endpoints. We then take a spanning cotree $C^*$ of the dual graph $\mathcal{G}^*$, restricted to the dual edges that do not cross any edge of $T$. The edges lying in neither $T$ nor in the cotree $C^*$ are the \emph{generator edges} $L = \{\ell_1, \ldots, \ell_{2g}\}$, and the cycle closed by each $\ell_i$ is a generator cycle.

There are exactly $2g$ generator edges, by a simple count. The spanning tree $T$ uses $|V|-1$ edges, and the cotree $C^*$, being a spanning tree of the dual graph on its $|F|$ vertices, uses $|F|-1$ edges. Every primal edge lies in exactly one of $T$, the cotree $C^*$ (as a dual edge), or the leftover set $L$, so
\[
    |L| = |E| - (|V|-1) - (|F|-1) = |E| - |V| - |F| + 2 = 2 - \chi = 2g,
\]
using the Euler characteristic $\chi = |V| - |E| + |F| = 2 - 2g$ of a closed orientable surface of genus $g$. This count matches $\dim H_1(\Sigma_g,\mathbb{F}_2)$ and since the generator cycles are independent, they form a basis ~\cite{tree_cotree_decomp}.  Consequently every cycle on $\Sigma_g$ is homologous to a symmetric difference of generator cycles, so its homology class is the sum of the corresponding generator classes. This makes the generators the natural basis for the homology label map $\phi$, where each bit of $\phi(e)$ records whether edge $e$ crosses the cocycle associated with one generator.

As described in \algoref{alg:tree-cotree}, $T$ is obtained by breadth-first search (BFS) from a root vertex $r \in V$. The dual graph $\mathcal{G}^*$ has one vertex per face of $\mathcal{T}$ and one edge $e^*$ per primal edge $e$. Removing the dual edges crossed by $T$ leaves a subgraph $\tilde{E}^* = E^* \setminus \{e^* : e \in T\}$, and any spanning tree of $\tilde{E}^*$ is a valid cotree $C^*$. The generator edges are then the edges that remain: $L = E \setminus (T \cup \{e : e^* \in C^*\})$.

\begin{algorithm}[h]
\caption{Tree-Cotree Decomposition~\cite{tree_cotree_decomp}}\label{alg:tree-cotree}
\KwIn{}
\begin{enumerate}
    \item A primal graph $\mathcal{G} = (V, E, F)$ embedded on a closed orientable surface
          $\Sigma_g$ of genus $g$;
    \item The corresponding dual graph $\mathcal{G}^* = (V^*, E^*, F^*)$.
\end{enumerate}
\KwOut{}
\begin{enumerate}
    \item A primal spanning tree $T \subseteq E$;
    \item A dual spanning cotree $C^* \subseteq E^*$;
    \item A set of generator edges $L \subseteq E$ with $|L| = 2g$, one per
          independent non-trivial cycle.
\end{enumerate}
\SetKwProg{Method}{Method}{:}{}
\Method{}{
\begin{enumerate}
    \item Run BFS on $\mathcal{G}$ from a root $r \in V$ to obtain spanning tree $T$.
    \item Let $\tilde{E}^* := E^* \setminus \{e^* : e \in T\}$ be the dual
          edges not crossed by any primal tree edge. Compute any spanning
          tree $C^*$ of $\tilde{E}^*$.
    \item Set $L := E \setminus \bigl(T \cup \{e \in E : e^* \in
          C^*\}\bigr)$. Assert $|L| = 2g$.
    \item Return $T$, $C^*$, $L$.
\end{enumerate}}
\end{algorithm}

\textbf{Homology Label Assignment (\algoref{alg:signatures})}. Given the generator edges $L$ and cotree $C^*$ from \algoref{alg:tree-cotree}, we assign a $2g$-bit label $\phi(e) \in \{0,1\}^{2g}$ to every edge $e \in E$, where bit $i$ records whether $e$ crosses the $i$-th generator cocycle (the closed loop of dual edges crossing generator $\ell_i$). In \algoref{alg:signatures} each generator edge $\ell_i$ is assigned a unique bitmask with only the $i$-th bit set to $1$. This bitmask is then propagated to all other edges by tracing the two faces (dual vertices) on either side of $\ell_i$ back to the cotree root $r^*$, and bit $i$ is flipped for every other primal edge crossed by this cocycle.
\begin{algorithm}[h]
\caption{Homology Label Assignment~\cite{Erickson_Whittlesey}}\label{alg:signatures}
\KwIn{}
\begin{enumerate}
    \item A dual spanning cotree $C^*$ with root $r^* \in V^*$, as produced
          by \algoref{alg:tree-cotree};
    \item Generator edges $L = \{\ell_1, \ldots, \ell_{2g}\}$, as produced
          by \algoref{alg:tree-cotree}.
\end{enumerate}
\KwOut{
A homology label map $\phi : E \to \{0,1\}^{2g}$, where bit $i$ of
$\phi(e)$ is $1$ if and only if primal edge $e$ crosses the
$i$-th generator cocycle, $P_{\ell_i}$.
}
\SetKwProg{Method}{Method}{:}{}
\Method{}{
\begin{enumerate}
    
    \item \textbf{Initialize the homology label map.}
    Set $\phi(e) := \mathbf{0} \in \{0,1\}^{2g}$ for all $e \in E$.

    \item \textbf{Assign a unique bitmask to each generator edge.}
    For each $i = 1, \ldots, 2g$, set $\phi(\ell_i) := b_i$, where
    $b_i \in \{0,1\}^{2g}$ is the unique $2g$-bit bitmask with only the
    $i$-th bit set:
    \[
        b_i[j] := \begin{cases} 1 & \text{if } j = i, \\ 0 &
        \text{otherwise.}\end{cases}
    \]
    These $2g$ bitmasks form a standard basis for $\{0,1\}^{2g}$ and will
    be propagated across the dual cotree $C^*$ in the following steps.

    \item \textbf{Propagate each generator's bit across the dual cotree.}
    For each $i = 1, \ldots, 2g$:
    \begin{enumerate}
        \item Let $F_A, F_B \in V^*$ be the two dual vertices (faces of the
              primal tiling) on either side of generator edge $\ell_i$.
        \item Trace both $F_A$ and $F_B$ back to the cotree root $r^*$ via
              the unique tree paths
              \[
                  P_A := \mathrm{TreePath}(C^*,\, F_A,\, r^*), \qquad
                  P_B := \mathrm{TreePath}(C^*,\, F_B,\, r^*).
              \]
        \item Form the XOR $P_A \oplus
              P_B$, consisting of dual edges touched an odd number of times.
              $P_{\ell_i} := (P_A \oplus P_B) \cup \{\ell_i^*\}$ is the
              generator cocycle crossing $\ell_i$.
        \item For each dual edge $e^* \in P_A \oplus P_B$, flip bit $i$ of the
              label of the corresponding primal edge $e$:
              \[
                  \phi(e)[i] \;:=\; \phi(e)[i] \oplus 1.
              \]
    \end{enumerate}

    \item \textbf{Result.} Every primal edge $e \in E$ carries a label
      $\phi(e) \in \{0,1\}^{2g}$, where $\phi(e)[i] = 1$ indicates that
      $e$ crosses the $i$-th generator cocycle. An edge $e$ with
      $\phi(e) = \mathbf{0}$ does not cross any generator cocycle.

    \item Return $\phi$.
\end{enumerate}}
\end{algorithm}

\subsection{Homologically Non-Trivial Cycle Enumeration}\label{sec:homology_cycle_enumeration}
With the homology label map $\phi : E \to \{0,1\}^{2g}$ established, we enumerate homologically non-trivial cycles on $\mathcal{G}$ and classify each by its homology class. 

We remark that Erickson and Whittlesey~\cite{Erickson_Whittlesey} find the shortest homology basis by re-running the tree-cotree decomposition from every vertex as root, each time generating only the $2g$ generator cycles associated with that choice of spanning tree, for a total of $O(gn)$ candidate cycles, where $n=|V|$. Their choice of $C^*$ as the maximum spanning tree weighted by BFS distance is essential for guaranteeing that the generator cycles produced at each root are as short as possible. In our algorithm, we perform the tree-cotree decomposition only once, from a single fixed root, solely to establish the homology label map $\phi$. The homology labels $\phi$ computed from this fixed decomposition are then used throughout Phase~2, which enumerates a much richer candidate set by running BFS from every vertex and collecting all cycles closed by non-tree edges, producing $O(n^2)$ candidate cycles in total. The choice of $C^*$ therefore has no effect on the lengths of the cycles we find in Phase~2.

\textbf{Cycle Enumeration (\algoref{alg:cycle-enum}).} We enumerate short 
homologically non-trivial cycles by running BFS from every vertex $v \in V$. For each root $v$, a BFS spanning tree $T_v$ is grown, and each non-tree edge $e_c = (u, w) \in E \setminus T_v$ closes a cycle
\[
    \gamma \;=\; \mathrm{TreePath}(T_v, v, u) \oplus \{e_c\} \oplus
    \mathrm{TreePath}(T_v, w, v),
\]
whose homology label is computed as $h(\gamma) = \bigoplus_{e \in \gamma} \phi(e)$. Cycles with $h(\gamma) = \mathbf{0}$ are trivial and discarded. Since BFS produces shortest-path trees, $\gamma$ is no longer than any closed walk through $v$ and $e_c$. Running BFS from every vertex produces $O(n^2)$ candidate cycles in total. Only the shortest representative per homology class is retained and the candidate set $\mathcal{L}$ is sorted by cycle length. Each cycle is stored compactly as a tuple $(|\gamma|,\, h(\gamma),\, v,\,e_c)$, where $|\gamma|$ is the number of edges in $\gamma$, $h(\gamma)\in \{0,1\}^{2g} \setminus \{\mathbf{0}\}$ is its homology label, $v \in V$ is the BFS root, and $e_c \in E$ is the non-tree edge that closes the cycle. The pair $(v, e_c)$ is sufficient to reconstruct $\gamma$ uniquely: given root $v$ and non-tree edge $e_c = (u, w)$, run BFS from $v$ to obtain $T_v$, then recover $\gamma = \mathrm{TreePath}(T_v, v, u) \oplus \{e_c\} \oplus \mathrm{TreePath}(T_v, w, v)$.

\begin{algorithm}[h]
\caption{Non-Trivial Cycle Enumeration}\label{alg:cycle-enum}
\KwIn{}
\begin{enumerate}
    \item A primal graph $\mathcal{G} = (V, E)$;
    \item A homology label map $\phi : E \to \{0,1\}^{2g}$, as produced by
          \algoref{alg:signatures}.
\end{enumerate}
\KwOut{
A set $\mathcal{L}$ of homologically non-trivial cycles, containing the
shortest enumerated representative of each homology class encountered, sorted in increasing
order of cycle length $|\gamma|$. Each cycle is stored as a tuple
$(|\gamma|,\, h(\gamma),\, v,\, e_c)$: $|\gamma|$ is the number of edges
in $\gamma$, $h(\gamma) \in \{0,1\}^{2g} \setminus \{\mathbf{0}\}$ is its
homology label, $v \in V$ is the BFS root, and $e_c \in E$ is the
closing edge.
}
\SetKwProg{Method}{Method}{:}{}
\Method{}{
\begin{enumerate}
    \item Set $\mathcal{L} := \emptyset$.
    \item For each vertex $v \in V$, run BFS on $\mathcal{G}$ from $v$ to obtain a
           spanning tree $T_v$.
    \item For each non-tree edge $e_c = (u, w) \in E \setminus T_v$, form
          the cycle
          \[
              \gamma \;:=\; \mathrm{TreePath}(T_v, v, u) \;\oplus\; \{e_c\}
                            \;\oplus\; \mathrm{TreePath}(T_v, w, v)
          \]
          and compute its homology label:
          \[
              h(\gamma) \;:=\; \bigoplus_{e \in \gamma} \phi(e).
          \]
    \item If $h(\gamma) \neq \mathbf{0}$, the cycle is non-trivial. If
          $\mathcal{L}$ contains no cycle with label $h(\gamma)$, add
          $(|\gamma|,\, h(\gamma),\, v,\, e_c)$ to $\mathcal{L}$. If
          $\mathcal{L}$ already contains a cycle $\gamma'$ with
          $h(\gamma') = h(\gamma)$ and $|\gamma'| > |\gamma|$, replace it
          with $\gamma$. Otherwise discard $\gamma$.
    \item Sort $\mathcal{L}$ by cycle length $|\gamma|$ in increasing
          order.
    \item Return $\mathcal{L}$.
\end{enumerate}}
\SetKwProg{Note}{Note}{:}{}
\Note{}{
The tuple $(v, e_c)$ is sufficient to reconstruct $\gamma$ uniquely: given
root $v$ and closing edge $e_c = (u, w)$, run BFS from $v$ to obtain
$T_v$; then $\gamma = \mathrm{TreePath}(T_v, v, u) \oplus \{e_c\} \oplus
\mathrm{TreePath}(T_v, w, v)$.
}
\end{algorithm}

\subsection{Automorphism Orbit Generation}\label{sec:automorphism_basis}
The candidate set $\mathcal{L}$ from \algoref{alg:cycle-enum} contains, for each homology class it encounters, the shortest enumerated representative. However, for the hyperbolic tilings we consider, we do not search for a canonical basis directly from $\mathcal{L}$. Instead, we exploit the orientation-preserving automorphism group $\mathrm{Aut}^+(\mathcal{T}) = \langle \rho, \sigma \rangle$ of the tiling to generate a structured candidate set $\mathcal{L}_{\mathrm{aut}}$ whose cycles are related by the tiling symmetries.

The motivation is architectural: the automorphisms $\rho$ and $\sigma$ are implemented as walking circuits on the semi-hyperbolic tiling, and logical operators related by these automorphisms can be mapped to each other by the corresponding circuits. A canonical basis drawn from the orbit of a seed cycle under $\mathrm{Aut}^+(\mathcal{T})$ therefore yields logical operators with this desirable symmetry structure.

\textbf{Seed Cycle Orbit Generation (\algoref{alg:aut_basis}).} We select one or more short cycles from $\mathcal{L}$ as \emph{seed cycles} and apply every element $\tau \in \mathrm{Aut}^+(\mathcal{T})$ to each seed, generating a family of image cycles $\{\tau(\gamma_s)\}$. Recall that $\mathrm{Aut}^+(\mathcal{T}) = G^+ = \langle \rho, \sigma \rangle$ is the orientation-preserving subgroup of the quotient group $G$, so that each $\tau$ acts by left multiplication, mapping the tile labeled by the group element $g$ to the tile labeled $\tau g$ (\Cref{sec:left_right_mult}). This permutation of the tiles is straightforward to compute by a traversal of the Cayley graph of $\mathrm{Aut}^+(\mathcal{T})$. Since automorphisms are isometries of the tiling, each image cycle inherits the length of the seed. We then compute the homology labels of the image cycles with $\phi$. The shortest representative per homology label is retained as before, producing $\mathcal{L}_{\mathrm{aut}}$. In practice we find that a small number of seed cycles can already generate a large fraction of the logical operators. For several of the examples in \Cref{tab:homology_automorphism_logicals}, a single seed cycle generates almost all of them.

\begin{algorithm}[h]
\caption{Seed Cycle Orbit Generation via Tiling Automorphisms}\label{alg:aut_basis}
\KwIn{}
\begin{enumerate}
    \item A set $\mathcal{L}$ of non-trivial cycles sorted by length,
          as produced by \algoref{alg:cycle-enum};
    \item The orientation-preserving automorphism group
          $\mathrm{Aut}^+(\mathcal{T}) = \langle \rho, \sigma \rangle$
          of the tiling $\mathcal{T}$.
\end{enumerate}
\KwOut{
A candidate set $\mathcal{L}_{\mathrm{aut}}$ containing, for each homology label occurring in the orbits of the seed cycles under $\mathrm{Aut}^+(\mathcal{T})$, the shortest such image cycle.
}
\SetKwProg{Method}{Method}{:}{}
\Method{}{
\begin{enumerate}
    \item Set $\mathcal{L}_{\mathrm{aut}} := \emptyset$.
    \item Select one or more short cycles from $\mathcal{L}$ as
          \emph{seed cycles}.
    \item For each seed cycle $\gamma_s$ and each group element
          $\tau \in \mathrm{Aut}^+(\mathcal{T})$, apply $\tau$ to
          $\gamma_s$ to obtain the image cycle $\tau(\gamma_s)$ and
          compute its homology label $h(\tau(\gamma_s))$.
    \item For each image cycle $\tau(\gamma_s)$: if
          $\mathcal{L}_{\mathrm{aut}}$ contains no cycle with label
          $h(\tau(\gamma_s))$, add $\tau(\gamma_s)$ to
          $\mathcal{L}_{\mathrm{aut}}$; if $\mathcal{L}_{\mathrm{aut}}$
          already contains a longer cycle with the same label, replace it.
    \item Return $\mathcal{L}_{\mathrm{aut}}$.
\end{enumerate}}
\end{algorithm}

\subsection{Canonical Homology Basis} \label{sec:canonical_homology_basis}

Given the candidate set $\mathcal{L}_{\mathrm{aut}}$ produced by \algoref{alg:aut_basis} (or alternatively the candidate set $\mathcal{L}$ produced by \algoref{alg:cycle-enum}), we now describe algorithms to find a canonical homology basis. 

\textbf{Cycle Intersection Matrix (\algoref{alg:crossing_matrix}).}
We compute the intersection numbers of the cycles in $\mathcal{L}_{\mathrm{aut}}$ (or $\mathcal{L}$). For each pair of cycles $\gamma_i, \gamma_j$ in the candidate set, the intersection number $\langle \gamma_i, \gamma_j \rangle \in \mathbb{F}_2$ records whether the two cycles cross an odd number of times on $\Sigma_g$, as defined in \Cref{sec:logical_basis}. This is a topological invariant of their homology classes. We therefore only need to count crossings geometrically for the $2g$ generator cycles of \algoref{alg:tree-cotree}: every other cycle is homologous to a symmetric difference of these, its homology label $h(\gamma)$ records which ones, and its intersection numbers can then be inferred algebraically from those of the generator cycles. Collecting them gives the cycle intersection matrix $M \in \{0,1\}^{m \times m}$ introduced above, where $m$ is the size of the candidate set.

\textbf{Symplectic pair search (\algoref{alg:max-clique}, \algoref{alg:backtrack}).} Given the intersection matrix $M \in \{0,1\}^{m \times m}$, we seek a maximum collection of mutually compatible symplectic pairs. A pair $(\gamma_i, \gamma_j)$ is a \emph{candidate symplectic pair} if $M_{ij} = 1$, i.e.\ the two cycles cross an odd number of times. Two candidate pairs $(\gamma_i, \gamma_j)$ and $(\gamma_k, \gamma_l)$ are compatible if none of the four cross terms between them is $1$: $M_{ik} = M_{il} = M_{jk} = M_{jl} = 0$. A canonical basis then corresponds to a choice of $2g$ cycles for which the $2g \times 2g$ submatrix of $M$, obtained by selecting those rows and the corresponding columns, equals the symplectic matrix $\Omega$ of \eqnref{eqn:symplectic_form_intro}.

While a canonical basis can always be constructed from any homology basis by Gaussian elimination over $\mathbb{F}_2$, this corresponds to taking symmetric differences of cycles, which can increase cycle lengths and yields cycles that are no longer images of a seed cycle under $\mathrm{Aut}^+(\mathcal{T})$, so they lose the symmetry structure and cannot be mapped onto one another by the automorphism circuits. The key constraint here is therefore that we must not alter the cycles themselves. We must instead find the symplectic structure by selecting $2g$ cycles from the candidate set and ordering them, so that the corresponding $2g \times 2g$ submatrix of $M$ equals $\Omega$ while the cycle lengths are preserved. This is equivalent to the induced matching problem on the graph defined by $M$, which is NP-complete in general \cite{induced_matchings}. 

In practice, our solvers find complete bases for the smaller tilings we consider, while for larger tilings we often obtain only $k_{\mathrm{found}} < g$ pairs within our computational budget. We use two exact search strategies: a reduction to a maximum clique search (\algoref{alg:max-clique}) and a backtracking search (\algoref{alg:backtrack}).

\algoref{alg:max-clique} reduces the search to a \emph{maximum clique problem} on a \emph{compatibility graph} $\mathcal{K}$ that we construct. The vertices of $\mathcal{K}$ are all candidate symplectic pairs $\mathcal{P} = \{(\gamma_i, \gamma_j) : M_{ij} = 1,\; i < j\}$, and an edge connects two vertices if and only if the corresponding pairs are compatible. A clique of size $g$ in $\mathcal{K}$ corresponds to a complete canonical basis of $g$ symplectic pairs, whereas a maximum clique of size $k_{\mathrm{found}} < g$ yields a partial basis. A maximum clique in $\mathcal{K}$ can then be obtained using standard clique-finding solvers~\cite{maximal_clique}.

\algoref{alg:backtrack} instead searches for the symplectic pairs directly. It builds up a canonical basis one pair at a time, adding a pair of crossing cycles at each step and backtracking whenever the remaining cycles can no longer be extended to $g$ mutually compatible pairs. If no complete basis exists it returns \textsc{Failure}; partial bases are obtained by searching for fewer pairs. This provides an alternative to the clique reduction of \algoref{alg:max-clique}, and in practice either algorithm suffices for the tilings we consider.

To summarize, we introduce computational methods to search for a symmetrical canonical homology basis. Enumerating homologically non-trivial cycles, classifying them by homology class, and generating the automorphism orbits all
run in polynomial time. The computational bottleneck is the symplectic pair search: finding cycles whose intersection submatrix matches the canonical
symplectic form is equivalent to finding a maximum induced matching, which is NP-complete in general~\cite{induced_matchings}. \algoref{alg:max-clique} and \algoref{alg:backtrack} solve it exactly, but their cost grows with the genus. For larger tilings, the search slows down and covering all $g$ pairs becomes increasingly difficult, so we do not always find a complete basis and instead obtain $k_{\mathrm{found}} \leq g$ pairs. Some examples of the bases we find are given in \Cref{tab:homology_automorphism_logicals}, and an extensive list of codes is provided in the extended tables section (\Cref{sec:promising_candidate_tables}). Even for the $8.8.8$-$400$ code $[[400,104,8]]$, a single weight-$12$ seed cycle covers $100$ of the $104$ logical qubits.

\begin{algorithm}[h]
\caption{Cycle Intersection Matrix}\label{alg:crossing_matrix}
\KwIn{}
\begin{enumerate}
    \item A set $\mathcal{L}$ or $\mathcal{L}_{\mathrm{aut}}$ of $m$
          non-trivial cycles $\{\gamma_1, \ldots, \gamma_m\}$, as
          produced by \algoref{alg:cycle-enum} and/or \algoref{alg:aut_basis}.
\end{enumerate}
\KwOut{
The cycle intersection matrix $M \in \{0,1\}^{m \times m}$, where
$M_{ij} = \langle \gamma_i, \gamma_j \rangle \in \mathbb{F}_2$ is $1$ if
$\gamma_i$ and $\gamma_j$ cross an odd number of times and $0$ otherwise,
with $M_{ii} = 0$.
}
\SetKwProg{Method}{Method}{:}{}
\Method{}{
\begin{enumerate}
    \item Set $M \in \{0,1\}^{m \times m}$ to the zero matrix.
    \item For each pair $i < j$, set
          \[
              M_{ij} \;=\; M_{ji} \;:=\; \langle \gamma_i, \gamma_j \rangle
              \;\in\; \mathbb{F}_2,
          \]
          the $\mathbb{F}_2$-valued intersection number of $\gamma_i$ and
          $\gamma_j$ as defined in \Cref{sec:logical_basis}.
    \item Return $M$.
\end{enumerate}}
\end{algorithm}

\begin{algorithm}[h]
\caption{Symplectic Cycle Pair Search via Maximum Clique}\label{alg:max-clique}
\KwIn{}
\begin{enumerate}
    \item A set $\mathcal{L}$ or $\mathcal{L}_{\mathrm{aut}}$ of $m$
          non-trivial cycles $\{\gamma_1, \ldots, \gamma_m\}$, as
          produced by \algoref{alg:cycle-enum} and/or \algoref{alg:aut_basis};
    \item The cycle intersection matrix $M \in \{0,1\}^{m \times m}$, as
          produced by \algoref{alg:crossing_matrix};
    \item The genus $g$ of the surface $\Sigma_g$.
\end{enumerate}
\KwOut{
A canonical homology basis $\mathcal{B} = \{(\alpha_1, \beta_1), \dots, (\alpha_{k_{\mathrm{found}}}, \beta_{k_{\mathrm{found}}})\}$ of $k_{\mathrm{found}} \leq g$ symplectic pairs,
where each pair $(\alpha_l, \beta_l) = (\gamma_{i_l}, \gamma_{j_l})$ satisfies
$\langle \alpha_l, \beta_l \rangle = 1$ and
$\langle \gamma, \gamma' \rangle = 0$ for any $\gamma \in \{\alpha_l, \beta_l\}$
and $\gamma' \in \{\alpha_{l'}, \beta_{l'}\}$ with $l \neq l'$. A
complete basis has $k_{\mathrm{found}} = g$; a partial basis $k_{\mathrm{found}} < g$ identifies $k_{\mathrm{found}}$ valid
symplectic pairs.
}
\SetKwProg{Method}{Method}{:}{}
\Method{}{
\begin{enumerate}
    \item \textbf{Identify symplectic pairs.} Let
          $\mathcal{P} := \{(\gamma_i, \gamma_j) : M_{ij} = 1,\; i < j\}$
          be the set of all candidate symplectic pairs.
          If $\mathcal{P} = \emptyset$, return \textsc{Failure}.
    \item \textbf{Build the compatibility graph.} Construct a graph $\mathcal{K}$
          with vertex set $\mathcal{P}$, where an edge connects
          $(\gamma_i, \gamma_j)$ and $(\gamma_k, \gamma_l)$ if and only if
          $M_{ik} = M_{il} = M_{jk} = M_{jl} = 0$, i.e.\ the two pairs are
          mutually non-crossing.
    \item \textbf{Find a maximum clique.} Apply the clique-finding algorithm
          of~\cite{maximal_clique} to $\mathcal{K}$. Return the largest maximal
          clique found as $\mathcal{B}$.
\end{enumerate}}
\end{algorithm}

\begin{algorithm}[H]
\caption{Symplectic Cycle Pair Search via Backtracking}\label{alg:backtrack}
\KwIn{}
\begin{enumerate}
    \item A set $\mathcal{L}$ or $\mathcal{L}_{\mathrm{aut}}$ of $m$
          non-trivial cycles $\{\gamma_1, \ldots, \gamma_m\}$, as
          produced by \algoref{alg:cycle-enum} and/or \algoref{alg:aut_basis};
    \item The cycle intersection matrix $M \in \{0,1\}^{m \times m}$, as
          produced by \algoref{alg:crossing_matrix};
    \item The genus $g$ of the surface $\Sigma_g$.
\end{enumerate}
\KwOut{
A canonical homology basis $\mathcal{B} = \{(\alpha_1, \beta_1), \dots, (\alpha_{k_{\mathrm{found}}}, \beta_{k_{\mathrm{found}}})\}$ of $k_{\mathrm{found}} \leq g$ symplectic pairs,
where each pair $(\alpha_l, \beta_l) = (\gamma_{i_l}, \gamma_{j_l})$ is selected from the candidate cycles,
or \textsc{Failure} if no solution is found within the time limit.
}
\SetKwProg{Method}{Method}{:}{}
\Method{}{
\begin{enumerate}
    \item \textbf{Preprocessing.} Iteratively remove any cycle $\gamma_i$ that
          does not cross any other remaining cycle, i.e.\ satisfying
          $M_{ij} = 0$ for all active $j$. Repeat until no further removals
          occur. Let $\mathcal{A} \subseteq \{1, \ldots, m\}$ denote the
          surviving indices. If $|\mathcal{A}| < 2g$, return
          \textsc{Failure}.
    \item \textbf{Row ordering heuristic.} Sort the active cycles in
          $\mathcal{A}$ in non-decreasing order of their crossing degree
          $\deg(i) := |\{j \in \mathcal{A} : M_{ij} = 1\}|$, so that
          cycles participating in fewer candidate pairs are tried first.
    \item \textbf{Backtracking search.} Call $\textsc{Solve}(1,\,
          \emptyset,\, \emptyset,\, \emptyset,\, g)$, where
          $\textsc{Solve}(t,\, R,\, C,\, F,\, g')$ is the recursive
          procedure:
    \begin{enumerate}
        \item If $g' = 0$, return the accumulated solution.
        \item If the number of remaining active cycles after position $t$ is
              less than $g'$, return \textsc{Failure} (pigeonhole pruning).
        \item \textbf{Degree pruning.} Among active cycles from
              position $t$ onward that are not in $R \cup C \cup F$,
              count how many participate in at least one candidate pair
              with another available cycle. If this count is less than
              $g'$, return \textsc{Failure}.
        \item For each cycle $\gamma_i$ from position $t$ onward (skipping
              if $i \in R \cup C \cup F$):
        \begin{enumerate}
            \item For each $\gamma_j$ with $j > i$, $j \notin R \cup C
                  \cup F$, and $M_{ij} = 1$:
            \begin{enumerate}
                \item Set $R' \leftarrow R \cup \{i\}$, $C' \leftarrow C
                      \cup \{j\}$.
                \item Compute the new conflict set
                      $F' \leftarrow F \cup \{l \neq j : M_{il} = 1\}
                      \cup \{l \neq i : M_{jl} = 1\}$.
                \item Recurse: call $\textsc{Solve}(i{+}1,\, R',\, C',\,
                      F',\, g'{-}1)$.
                \item If the recursive call succeeds, return its solution.
                      Otherwise, backtrack.
            \end{enumerate}
        \end{enumerate}
        \item Return \textsc{Failure}.
    \end{enumerate}
\end{enumerate}}
\end{algorithm}

\subsection{Lifting Homology Labels to Semi-Hyperbolic Tilings}
Let $\mathcal{T}$ be a hyperbolic tiling for which we find a canonical homology basis, and let $\mathcal{T}^*$ be its dual. We want to lift this basis to the fine-grained tiling $\mathcal{T}_{(L)} = (\mathcal{T}^*_{(L)})^*$, obtained by fine-graining $\mathcal{T}^*$ with parameter $L \in \mathbb{Z}_{>0}$ to get $\mathcal{T}^*_{(L)}$ and then taking its dual. Let $\mathcal{G} = (V, E)$ be the primal graph of $\mathcal{T}$ and let $\phi : E \to \{0,1\}^{2g}$ be the homology label map constructed from $\mathcal{T}$ by \algoref{alg:signatures}. Write $\mathcal{G}_{(L)} = (V_{(L)}, E_{(L)})$ for the primal graph of $\mathcal{T}_{(L)}$. Since fine-graining does not alter the topology of $\Sigma_g$, the homology group $H_1(\Sigma_g;\,\mathbb{F}_2)$ is unchanged and $\phi$ can be lifted to $\mathcal{G}_{(L)}$. Every edge $e \in E_{(L)}$ either crosses a segment of exactly one subdivided dual edge $\hat{e}^* \in E^*$ of $\mathcal{T}^*$, in which case it inherits the label $\phi(\hat{e})$, or lies entirely within a Euclidean patch and carries the zero label $\mathbf{0}$. This defines the \emph{lifted homology label map} $\phi_{(L)} : E_{(L)} \to \{0,1\}^{2g}$, constructed by \algoref{alg:lifting}.
\begin{algorithm}[H]
\caption{Homology Label Lifting}\label{alg:lifting}
\KwIn{}
\begin{enumerate}
    \item The hyperbolic tiling $\mathcal{T}$ with primal graph 
          $\mathcal{G} = (V, E)$;
    \item The homology label map $\phi : E \to \{0,1\}^{2g}$, as produced 
          by \algoref{alg:signatures};
    \item The fine-grained tiling $\mathcal{T}_{(L)} = (\mathcal{T}^*_{(L)})^*$
          with edge set $E_{(L)}$, obtained by fine-graining $\mathcal{T}^*$
          with parameter $L$ and taking the dual;
    \item For each edge $e \in E_{(L)}$, the identity of the dual 
          edge $\hat{e}^* \in E^*$ of $\mathcal{T}^*$ it crosses, if any.
\end{enumerate}
\KwOut{
A lifted homology label map $\phi_{(L)} : E_{(L)} \to \{0,1\}^{2g}$.
}
\SetKwProg{Method}{Method}{:}{}
\Method{}{
\begin{enumerate}
    \item Initialize $\phi_{(L)}(e) := \mathbf{0}$ for all 
          $e \in E_{(L)}$.
    \item For each edge $e \in E_{(L)}$ that crosses a dual 
          edge $\hat{e}^* \in E^*$ of $\mathcal{T}^*$, set
          \[
              \phi_{(L)}(e) \;:=\; \phi(\hat{e}).
          \]
    \item Return $\phi_{(L)}$.
\end{enumerate}}
\SetKwProg{Note}{Note}{:}{}
\Note{}{
The homology class of any cycle $\gamma \subseteq E_{(L)}$ on the 
fine-grained tiling is computed as
\[
    h(\gamma) \;=\; \bigoplus_{e \in \gamma} \phi_{(L)}(e) \;\in\; \{0,1\}^{2g},
\]
in exact analogy with the hyperbolic case. Cycles with $h(\gamma) = \mathbf{0}$ 
are trivial and discarded; non-zero labels identify the homology class.
}
\end{algorithm}

\section{Computing distances of semi-hyperbolic codes}\label{sec:distance_finding}

\begin{figure}
    \centering
    \includegraphics[width=0.7\linewidth]{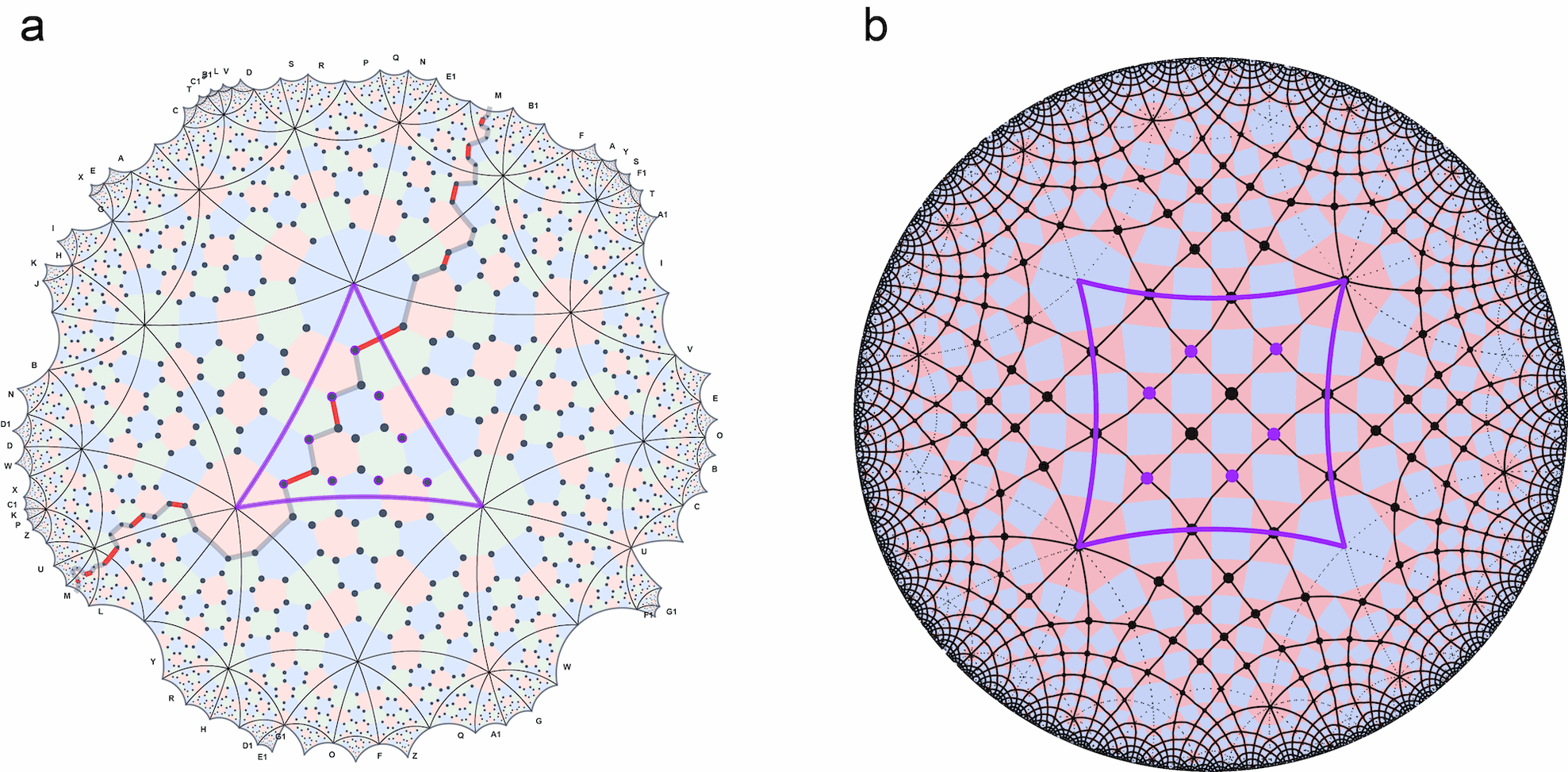}
    \caption{We find the distances of semi-hyperbolic color and surface codes by adapting the methods of \Cref{sec:homology_basis} and exploiting the symmetries of the tilings to improve efficiency. After establishing a homology coordinate system via the tree-cotree decomposition (\Cref{sec:homology_coordinate_system}), the shortest homologically non-trivial cycles are found via graph search initiated from a reduced source set $\mathcal{B}$. Global tiling automorphisms guarantee that at least one representative of a minimum-weight homologically non-trivial cycle passes through any chosen fundamental domain, allowing $\mathcal{B}$ to be restricted to a single tile rather than all vertices in the code. \textbf{a}, For semi-hyperbolic color codes, $\mathcal{B}$ is restricted to the vertices of the dual tiling $(\mathcal{T}^\triangle_L)^*$ within the fundamental triangle of the base tiling $\mathcal{T}^\triangle$ (purple boundary), or further restricted to those incident to inter-triangle crossing edges (purple vertices) without loss of accuracy. A red shortest stringlike error passing through this fundamental domain is highlighted. \textbf{b}, For semi-hyperbolic surface codes, $\mathcal{B}$ is restricted to the cell complex vertices associated with a single fundamental quadrilateral of the symmetry group (purple boundary), where candidate source vertices adjacent to the boundary of the quadrilateral (highlighted in purple) suffice to find shortest non-trivial cycles.}
    \label{fig:distance_finding_source_nodes}
\end{figure}

While the logical error rate under circuit-level noise with practical decoding is the most robust metric with which we assess the codes considered in this work, another important metric is the code distance.
Determining the distance of the semi-hyperbolic color and surface codes is important both for filtering viable code candidates, as well as for estimating the performance of codes (via extrapolation) at system sizes and error rates that are currently beyond the reach of our decoders or Monte Carlo sampling budget.
In our search for promising codes, we constructed over $10^4$ tilings, and for each tiling we constructed up to 7 semi-hyperbolic codes using $L\in \{1,2,\dots,7\}$.
Furthermore, many codes of interest have very large block lengths, some with as many as $n\approx10^5$ data qubits.
We emphasize that while these codes can be large, they can still be dense and therefore useful if they comprise a large fraction of the storage in a utility-scale quantum computer (and do not encode many more logical qubits than are needed).

Computing the distances of these large codes, not least doing so fast enough to filter through such a large database of tilings, was beyond the capabilities of the distance finding methods in the literature~\cite{webster2026distance}, including even the graphlike distance finding methods such as the stim method \texttt{stim.Circuit.shortest\_graphlike\_error}~\cite{Gidney2021stimfaststabilizer}.
Fortunately, we were able to develop fast and accurate distance finding methods by adapting the methods for finding short homology bases (see \Cref{sec:homology_basis}) while exploiting the symmetries of the tilings to dramatically reduce the search space without loss of accuracy.

\subsection{Monochromatic Cycle Distance Upper Bound for Semi-Hyperbolic Color Codes}\label{sec:color_distance_upper_bound}

For semi-hyperbolic color codes, the shortest logical error is not guaranteed to correspond to a monochromatic substring $\alpha_C$ of a homologically nontrivial cycle in the three-colorable tiling as defined in \Cref{sec:color_code_logical_basis} (i.e.~a cycle in a shrunk lattice of \cite{bombin_color}).
We will refer to the shortest error corresponding to a monochromatic substring $\alpha_C$ (for $C\in\{R,G,B\}$) as the shortest stringlike error.
The shortest stringlike color code logical, while not always minimum weight, is a very useful upper bound for the distance in practice and is usually tight (as we will show).
We now define the shortest stringlike error more explicitly.
As explained in \Cref{sec:color_code_logical_basis}, associated with a homologically non-trivial cycle $\alpha$ in the three-colorable tiling, there is both an $X$-type and a $Z$-type logical operator associated with each monochromatic substring $\alpha_C$ (for $C\in \{R,G,B\}$).
For example, the logical $\overline{X}_R^\alpha$ is the Pauli operator with $X$ support on the two qubits at the endpoints of each red edge in $\alpha$.
Concretely,
\begin{equation}
    \overline{X}_C^\alpha \coloneqq \bigotimes_{v \in V(\alpha_C)} X_v, \qquad
    \overline{Z}_C^\alpha \coloneqq \bigotimes_{v \in V(\alpha_C)} Z_v,
\end{equation}
where $\alpha_C \coloneqq \{e \in \alpha \mid c(e) = C\}$ denotes the set of edges of color $C$ contained in the cycle $\alpha$, and $V(\alpha_C) \subseteq V$ is the set of vertices incident to at least one edge in $\alpha_C$.

Because the edges in $\alpha_C$ are disjoint along any cycle $\alpha$ on a trivalent tiling, every edge $e = (u, w) \in \alpha_C$ contributes exactly two distinct vertices. 
The weight (number of physical data qubits in the support) of a monochromatic stringlike operator associated with color $C$ is therefore
\begin{equation}
    |\overline{X}_C^\alpha| \;=\; |\overline{Z}_C^\alpha| \;=\; |V(\alpha_C)| \;=\; 2\,|\alpha_C| \;.
\end{equation}
The weight of the shortest stringlike error associated with the cycle $\alpha$ is given by
\begin{equation}
    w_{\mathrm{string}}(\alpha) \;\coloneqq\; \min_{C \in \{R, G, B\}} 2\,|\alpha_C|,
\end{equation}
and the shortest stringlike error across the entire code is obtained by minimizing $w_{\mathrm{string}}(\alpha)$ over all homologically non-trivial cycles $\alpha$.

To compute the shortest stringlike error of a semi-hyperbolic color code efficiently, we adapt the cycle enumeration procedure of \Cref{sec:homology_basis}. We first establish a homology coordinate system on the three-colorable tiling via the tree-cotree decomposition (\Cref{sec:homology_coordinate_system}), which assigns a $2g$-bit homology label $\phi(e)$ to each edge. 

Instead of an unweighted BFS, which minimizes the total edge length $|\alpha|$, we determine the minimum-weight representative for each color $C \in \{R, G, B\}$ by running a shortest-path search on a weighted graph. Specifically, for a target color $C$, we define the edge weight function
\begin{equation}
    w_C(e) \;\coloneqq\; 2\,\delta_{c(e), C} \;=\; \begin{cases} 
    2 & \text{if } c(e) = C, \\ 
    0 & \text{if } c(e) \neq C.
    \end{cases}
\end{equation}
Under this weighting, the length of any cycle $\alpha$ under $w_C$ corresponds exactly to $|\overline{X}_C^\alpha|$. We then find the shortest homologically non-trivial cycle using Dijkstra's algorithm. Because edges with $c(e) \neq C$ carry zero weight, paths traversing connected components of non-$C$ edges do not accrue cost, which mirrors traversing the shrunk lattice of color $C$~\cite{bombin_color}.

Running this search from every vertex across large semi-hyperbolic lattices ($n \approx 10^5$) remains computationally expensive. To accelerate the search, we exploit the global symmetries of the tiling described in \Cref{sec:left_right_mult}. 
Recall that each base triangle in $\mathcal{T}^\triangle$ corresponds uniquely to an element $g \in G = \triangle(r,s,t)/\Gamma$. Because the data qubits of the semi-hyperbolic color code sit on the vertices of the dual tiling $(\mathcal{T}^\triangle_L)^*$, each physical vertex $v \in V$ lies within the interior of a base triangle. We can therefore define $\pi: V \to G$, which maps each vertex $v \in V$ to the group element labeling the base triangle that contains it. Left multiplication by an element $h \in G$ induces a global isometry $\mathcal{A}_h$ of the tiled surface, which acts as a code automorphism that preserves cycle weights, edge colors, and homological non-triviality.

Suppose $\alpha_C^{\min}$ is a minimum-weight stringlike error of color $C$. Because any cycle confined entirely within a single Euclidean base triangle is contractible (homologically trivial), $\alpha_C^{\min}$ must cross between distinct base triangles. Therefore, $\alpha_C^{\min}$ must contain at least one edge $(u, v)$ that crosses a base triangle edge, i.e.~such that $\pi(u) \neq \pi(v)$.

Applying the automorphism $\mathcal{A}_{\pi(u)^{-1}}$ associated with the inverse element $\pi(u)^{-1} \in G$ maps the base triangle $\pi(u)$ directly onto the identity fundamental domain $e$. The resulting image cycle
\begin{equation}
    \tilde{\alpha}_C \;\coloneqq\; \mathcal{A}_{\pi(u)^{-1}}(\alpha_C^{\min})
\end{equation}
has identical weight to $\alpha_C^{\min}$, while still belonging to a non-trivial homology class. 
The image vertex $\tilde{u} \coloneqq \mathcal{A}_{\pi(u)^{-1}}(u)$ satisfies $\pi(\tilde{u}) = e$ and is incident to an edge connecting the fundamental triangle $e$ to an adjacent base triangle.

As a result, we do not need to initiate the Dijkstra search from all $|V| = L^2|G|$ vertices in the fine-grained lattice. It is sufficient to restrict the candidate source vertices to
\begin{equation}
    \mathcal{B} \;\coloneqq\; \bigl\{ u \in V \;\big|\; \pi(u) = e \text{ and } \exists\, (u, v) \in E \text{ such that } \pi(v) \neq e \bigr\},
\end{equation}
which is the 1D ring of vertices situated along the perimeter of the fundamental triangle $e$ (\Cref{fig:distance_finding_source_nodes}a). Minimizing the resulting cycle weights over all source vertices $u \in \mathcal{B}$ and all three colors $C \in \{R, G, B\}$ is guaranteed to recover a globally shortest stringlike error. Because $|\mathcal{B}| = O(L)$, this restricts the search from $O(L^2|G|)$ sources to just $O(L)$ sources, reducing the total distance-finding runtime by a factor of $L|G|$ without compromising accuracy.

\subsection{Restricted Lattice Distance Lower Bound for Semi-Hyperbolic Color Codes}\label{sec:color_code_distance_lower_bound}

\begin{figure}
    \centering
    \includegraphics[width=0.8\linewidth]{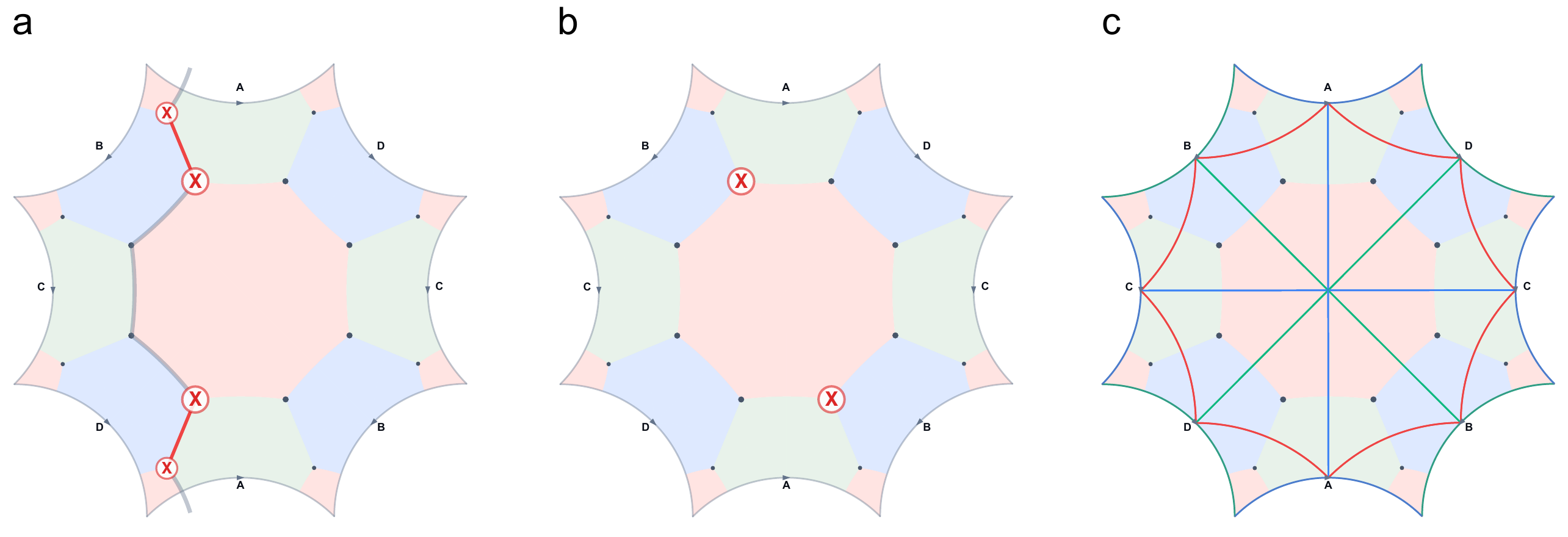}
    \caption{The minimum-weight stringlike error is an upper bound on the distance, but the true minimum-weight error may have lower weight, as shown here for the Bolza color code (with $L=1$). \textbf{a}, A minimum-weight stringlike error (weight 4), consisting of $X$ errors at the endpoints of the red edges $\alpha_R$ in a minimum-length homologically nontrivial cycle $\alpha$. \textbf{b}, The true distance of the Bolza color code is 2. A minimum-weight (non-stringlike) error is shown. Each $X$ error anti-commutes with the same triplet of stabilizers. \textbf{c}, The red, green and blue restricted lattices are shown on top of the tiling (combined, they form the dual tiling $\mathcal{T}^\triangle$). We will show how the shortest homologically nontrivial cycles in the restricted lattices can be used to provide a lower bound on the distance.}
    \label{fig:bolza_distance_example}
\end{figure}

While the weight of the shortest stringlike error $d_\mathrm{string}$ provides an efficient upper bound on the color code distance, $d\leq d_{\mathrm{string}}$, a general color code logical operator can branch (since each single-qubit error anti-commutes with three stabilizers), potentially resulting in a lower physical weight.
While we do find that $d=d_{\mathrm{string}}$ holds for the vast majority of codes we consider, we do find some codes (mostly distance 2) where it does not.
A simple example is the Bolza code, which has distance 2 despite its shortest stringlike error having weight 4, as shown in \Cref{fig:bolza_distance_example}.
In this section we show that we can also efficiently compute a useful \textit{lower bound} on the distance of the color code.
When this lower bound matches the stringlike upper bound $d_\mathrm{string}$, as is often the case, we can efficiently certify the true distance.

We will demonstrate the lower bound for the $X$ distance, namely for $X$-type errors $x\in\mathbb{F}_2^n$, and will denote the code's $Z$ parity check matrix by $H\in\mathbb{F}_2^{m\times n}$ and the $Z$ logical matrix by $L\in\mathbb{F}_2^{k\times n}$.
The $X$ distance $d$ is then
\begin{equation}
    d \coloneqq \min_{\substack{x \in \mathbb{F}_2^n \setminus \{0\} \\ Hx = 0 \\ L x \neq 0}} |x|.
\end{equation}
It is sufficient to consider the $X$ distance only, since color codes are self-dual, and hence have identical $X$ and $Z$ distances.

Our lower bound makes use of the red, green and blue \textit{restricted lattices} often used by color code decoders~\cite{delfosse2014decoding,Kubica2023efficientcolorcode}.
Before proving the distance lower bound more formally, we first sketch the argument, which goes as follows:
\begin{enumerate}
    \item We define the map (used in color code decoders) from an $X$-type error $x\in\mathbb{F}_2^n$ to a set of edges in each of the restricted lattices, its $C$-restriction $\mathrm{Res}_C(x)$.
    \item We show that the weight of the $C$-restriction cannot exceed the weight of the error itself, $|\operatorname{Res}_C(x)| \le |x|$.
    \item A logical error $x$, with trivial syndrome $Hx=0$ and non-trivial logical action $Lx\neq 0$, has a $C$-restriction that is homologically non-trivial for at least two of the three colors $C\in\{R,G,B\}$.
    \item This implies that the median of the systoles $s_C^{\mathrm{res}}$ for each of the restricted lattices is a lower bound on the distance $d_{\mathrm{res}} \coloneqq \operatorname{median}\bigl(s_R^{\mathrm{res}},\, s_G^{\mathrm{res}},\, s_B^{\mathrm{res}}\bigr)\leq d$.
\end{enumerate}

\begin{figure}
    \centering
    \includegraphics[width=0.95\linewidth]{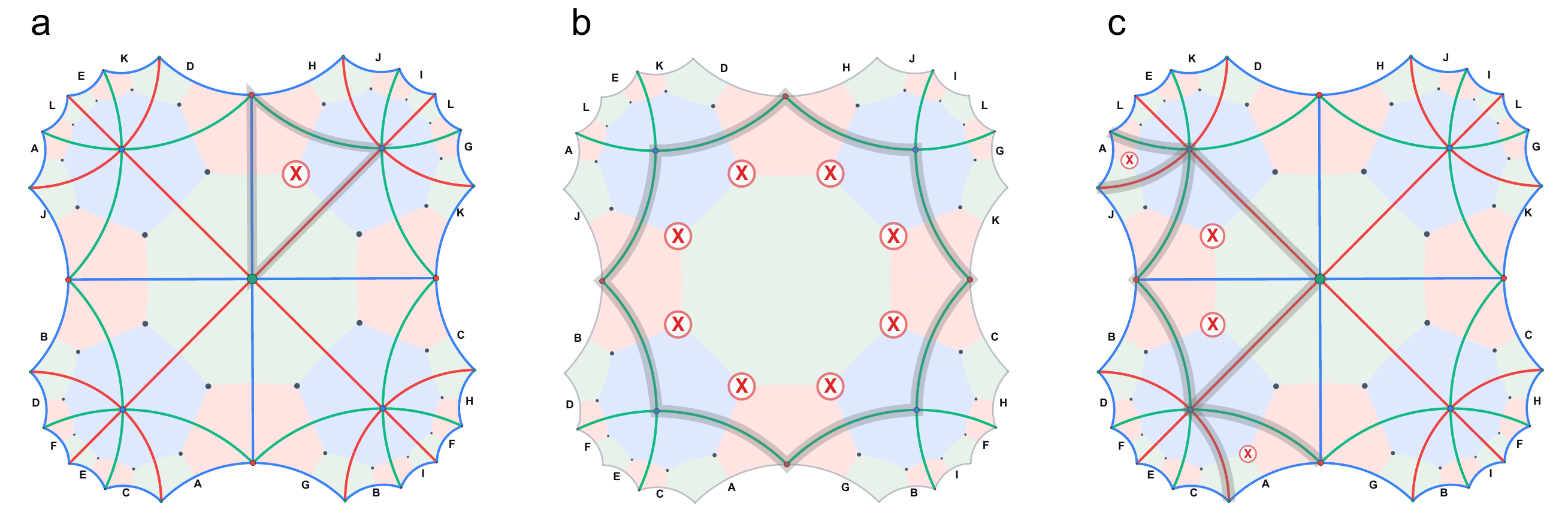}
    \caption{The restricted lattices of the 8.8.8 hyperbolic color code with parameters [[32,12,4]] and symmetry group $\triangle(4,4,4)_{40}$. \textbf{a}, The degree-3 tiling of octagons is the color code tiling $\mathcal{T}$, with a data qubit associated with each vertex. The dual triangular tiling $\mathcal{T}^\triangle$ is also shown, with each edge colored with the same color $C\in\{R,G,B\}$ as the primal edge in $\mathcal{T}$ that it crosses. Each vertex of $\mathcal{T}^\triangle$ inherits the color of the primal face it is dual to. An edge $e^* = (u, w) \in \mathcal{T}^\triangle$ connects vertices of two distinct colors, and its color $c(e^*)$ is assigned the third color, satisfying $\{c(e^*), c(u), c(w)\} = \{R, G, B\}$. An $X$ error on a single data qubit corresponding to a vertex $x\in\mathcal{T}$ is shown. The $C$-restriction $\mathrm{Res}_C(x)$ of an error on such a vertex $x\in\mathcal{T}$ is the unique adjacent edge $e^*$ of color $c(e^*)=C$ in $\mathcal{T}^\triangle$. We have highlighted in grey the edges $\mathrm{Res}_R(x)$, $\mathrm{Res}_G(x)$ and $\mathrm{Res}_B(x)$ of the $X$ error on $x$. \textbf{b}, The edges of the green restricted lattice $\mathcal{R}_G$ are shown. $\mathcal{R}_G$ is formed by removing each green vertex from $\mathcal{T}^\triangle$ and its incident red and blue edges to form a face. The edges of $\mathcal{R}_G$ are the green edges of $\mathcal{T}^\triangle$. The Pauli operator $y\in \mathbb{F}_2^n$ corresponding to a green $X$ stabilizer is shown, as well as its restriction $\mathrm{Res}_G(y)$, a face of the green restricted lattice. Any stabilizer has $C$-restrictions that are homologically trivial. \textbf{c}, A weight-four logical $X$-type error $z\in\mathbb{F}_2^n$ is shown (a blue monochromatic stringlike error $\overline{X}_B^\alpha$) as well as its $C$-restriction $\mathrm{Res}_C(z)$ on the three restricted lattices (all highlighted). A logical error has a homologically non-trivial $C$-restriction for at least two of the three colors. Here $\mathrm{Res}_R(z)$ and $\mathrm{Res}_G(z)$ are homologically non-trivial and $\mathrm{Res}_B(z)$ is trivial (an empty cycle). Since each single-qubit error contributes at most one edge in its $C$-restriction (and two errors with the same restriction cancel modulo 2), the weight of a $C$-restriction is a lower bound on the weight of the error, $|\mathrm{Res}_C(z)|\leq |z|$. Here $|\mathrm{Res}_R(z)|=|\mathrm{Res}_G(z)|=4$ and $|\mathrm{Res}_B(z)|=0$. The second smallest of the three systoles therefore lower bounds the distance. Concretely $d_{\mathrm{res}} \coloneqq \operatorname{median}\bigl(s_R^{\mathrm{res}},\, s_G^{\mathrm{res}},\, s_B^{\mathrm{res}}\bigr)\leq d$ where $s_C^{\mathrm{res}}$ is the systole of the restricted lattice $\mathcal{R}_C$. In this example $d_{\mathrm{res}}=4\leq d \leq d_{\mathrm{string}}=4$ certifies that the distance is 4.}
    \label{fig:restricted_lattcices_iter40}
\end{figure}

Let us first define the restricted lattices~\cite{delfosse2014decoding,Kubica2023efficientcolorcode}.
Recall that the base triangular tiling $\mathcal{T}^\triangle$, the dual of the color code tiling $\mathcal{T}$, has vertices colored $\{R,G,B\}$ corresponding to the faces of $\mathcal{T}$.
Each edge $e\coloneqq(u,v)\in\mathcal{T}^\triangle$ is assigned a color $c(e)\in \{R,G,B\}\setminus \{c(u),c(v)\}$ that complements the colors of the two vertices at its endpoints.
Equivalently, each edge in $\mathcal{T}^\triangle$ is assigned the same color as the edge it crosses in $\mathcal{T}$.
For each color $C\in\{R,G,B\}$ we form the \textit{$C$-restricted lattice} $\mathcal{R}_C$, whose edge set is all the edges of color $C$ in $\mathcal{T}^\triangle$.
Its vertices are the dual vertices of the other two colors $\{R,G,B\}\setminus \{C\}$. The vertices of color $C$ (and their incident edges) are deleted, each replaced with a face bounded by the remaining surrounding edges.
See \Cref{fig:restricted_lattcices_iter40} for an illustration of the restricted lattices.

Each data qubit corresponds to a vertex $v$ in $\mathcal{T}$ and the corresponding triangular face $\tau_v$ in the dual triangular lattice $\mathcal{T}^\triangle$. Each data qubit is therefore incident to exactly one edge in each restricted lattice $\mathcal{R}_C$.
We denote by $e_C(v)$ the $C$-colored edge belonging to $\tau_v$, i.e.~the edge in $\mathcal{R}_C$ adjacent to $v$ (see \Cref{fig:restricted_lattcices_iter40}).
Since color codes are self-dual, it is sufficient to consider only $X$-type errors, and then $Z$-type errors can be handled similarly.
We represent a physical Pauli $X$ error using a binary vector $x\in\mathbb{F}_2^n$, corresponding to the Pauli operator $\bigotimes_i X_i^{x_i}$.
The $C$-\textit{restriction} $\mathrm{Res}_C(x)$ of an error $x$ is a set of edges (1-chain) in the restricted lattice obtained by mapping each error on a qubit $v$ to its edge $e_C(v)$ in $\mathcal{R}_C$,
\begin{equation}
    \mathrm{Res}_C(x)\coloneqq \sum_{v\in V} x_v e_C(v)\in C_1(\mathcal{R}_C;\mathbb{F}_2).
\end{equation}
While each qubit error maps to a single edge in the restricted lattice $\mathcal{R}_C$, each edge in $\mathcal{R}_C$ is the image of two qubit errors, which can cancel modulo 2.
Concretely, $\left[\mathrm{Res}_C(x)\right]_{e^*}=x_u+x_v \mod 2$, where $e^*$ is the edge in $\mathcal{R}_C$ that is the dual of edge $e=(u,v)$ in $\mathcal{T}$.
See the blue restriction in \Cref{fig:restricted_lattcices_iter40}\textbf{c} for an example.
As a result, the weight of an error $|x|$ is lower bounded by the weight of its $C$-restriction,
\begin{equation}
    \left | \mathrm{Res}_C(x) \right | \leq |x|.
\end{equation}

We now show that any logical error $x\in\mathbb{F}_2^n$, by definition satisfying $Hx=0$ and $Lx\neq 0$, must have a $C$-restriction $\mathrm{Res}_C(x)$ that is homologically non-trivial in at least two of the three colors $C\in\{R,G,B\}$.
This follows directly from Theorem 6.4 of~\cite{delfosse2014decoding} and Theorem 2 of~\cite{Kubica2023efficientcolorcode}, however we present a self-contained argument here for completeness.
First we show that any error $x$ with trivial syndrome ($Hx=0$) has a $C$-restriction that is a cycle for every color $C\in\{R,G,B\}$.
First, notice that each vertex in $\mathcal{R}_R$ is the dual of a green or blue face in $\mathcal{T}$.
If an $X$ error $x$ commutes with a green or blue $Z$ stabilizer, it implies that there must be an even number of edges in its restriction $\mathrm{Res}_R(x)$ incident to the corresponding vertex in $\mathcal{R}_R$.
Therefore if $x$ commutes with all green and blue $Z$ stabilizers this implies that $\mathrm{Res}_R(x)$ must be a cycle in $\mathcal{R}_R$, and the same argument applied to the other restricted lattices suffices to show that all $C$-restrictions are cycles if $Hx=0$.
To see why the cycle must further be homologically non-trivial, we first observe that every face of $\mathcal{R}_C$ is the $C$-restriction of an $X$ stabilizer associated with a $C$-colored plaquette in $\mathcal{T}$ (see \Cref{fig:restricted_lattcices_iter40}\textbf{b}).
Therefore, if an error $x$ has homologically trivial restrictions in all three lattices, each $\operatorname{Res}_C(x)$ is a sum of faces of $\mathcal{R}_C$, and multiplying $x$ by a product of physical face stabilizers of each color yields an equivalent operator $x'$ whose restricted chains are all empty: $\operatorname{Res}_R(x') = \operatorname{Res}_G(x') = \operatorname{Res}_B(x') = 0$.
Because each dual edge of $\mathcal{R}_C$ carries the parity $x'_u + x'_v \pmod 2$ of the errors at the vertices of the primal edge $(u, v)$ of color $C$, an empty restriction forces $x'_u = x'_v$ across every primal edge of that color.
Having all three restrictions empty therefore propagates equality across every red, green, and blue edge of the tiling.
Since $\mathcal{T}$ is connected, $x'$ must be either $0$ (the identity) or $\mathbf{1}$ (the product of all red face stabilizers), both of which have trivial logical action $Lx' = 0$.
Therefore, by contrapositive, any non-trivial logical error ($Hx = 0$ and $Lx \neq 0$) must have a homologically non-trivial restriction $\operatorname{Res}_C(x)$ in at least one color $C$.
Now let $h_C \coloneqq [\operatorname{Res}_C(x)] \in H_1(\Sigma_g;\mathbb{F}_2)$ denote the homology class of $\mathrm{Res}_C(x)$.
Notice that the sum of the three $C$-restrictions of a single qubit error on qubit $v$ is the boundary of the corresponding triangle $\tau_v\in\mathcal{T}^\triangle$.
Therefore for any error $x$, the sum of the three homology classes is trivial:
\begin{equation}
    h_R+h_G+h_B=0\quad\mathrm{in}\quad H_1(\Sigma_g;\mathbb{F}_2).
\end{equation}
Consequently, a non-trivial logical operator cannot have exactly \textit{one} homologically non-trivial $C$-restriction, and therefore must be homologically non-trivial for at least two of the colors.

For each color $C$ with $h_C\neq 0$, we have $|x|\geq |\mathrm{Res}_C(x)|\geq s_C^{\mathrm{res}}$.
Because this holds for at least two colors, $|x|$ is bounded below by the second-smallest of the three systoles, which is their median:
\begin{equation}
    d_{\mathrm{res}}\coloneqq \mathrm{median}\left (s_R^{\mathrm{res}},s_G^{\mathrm{res}},s_B^{\mathrm{res}} \right)\leq d.
\end{equation}
Combining this with the stringlike upper bound we have
\begin{equation}
    d_{\mathrm{res}}\leq d \leq d_{\mathrm{string}}.
\end{equation}

The restricted lattice systoles can be computed efficiently using essentially the same methods as described for the upper bound computation in \Cref{sec:color_distance_upper_bound}, using unweighted BFS searches on the restricted lattices after establishing their homology coordinate system using a tree-cotree decomposition, and again restricting the BFS to source vertices contained within a fundamental triangle to speed up the search without loss of accuracy.
Whenever $d_{\mathrm{res}}=d_{\mathrm{string}}$, the code distance is certified exactly, using only efficient graph algorithms.

\subsection{Distance Computation for Semi-Hyperbolic Surface Codes}\label{sec:surface_code_distance_systole_search}

For semi-hyperbolic surface codes, the minimum-weight logical $Z$ ($X$) errors correspond to the shortest homologically non-trivial cycles in the primal (dual) cell complex $\mathcal{T}_{\textit{cell}}$ (\Cref{fig:surface_code_cell_complex_tiling}). We compute the code distance $d = \min(d_X, d_Z)$ by exploiting the symmetries of the tiling in an analogous way. After establishing a homology coordinate system on $\mathcal{T}_{\textit{cell}}$ and its dual via tree-cotree decomposition, we find the shortest non-trivial cycles using a BFS. 

As in the color code case, any homologically non-trivial cycle must cross between neighboring fundamental domains of the underlying tiling's symmetry group (whether a quadrilateral reflection group $\square(r,s,t,u)$ or a triangle group $\triangle(2,4,q)$). Global automorphisms therefore allow us to restrict the BFS source vertices to a small subset of boundary vertices $\mathcal{B}$ contained within a single fundamental domain (\Cref{fig:distance_finding_source_nodes}b). This preserves exactness while reducing the number of search sources from $O(L^2|G|)$ to $O(L)$.

\subsection{Performance Benchmarks for Distance Finding Methods}\label{sec:color_code_accuracy}

\begin{figure*}[t]
    \centering
    \includegraphics[width=\linewidth]{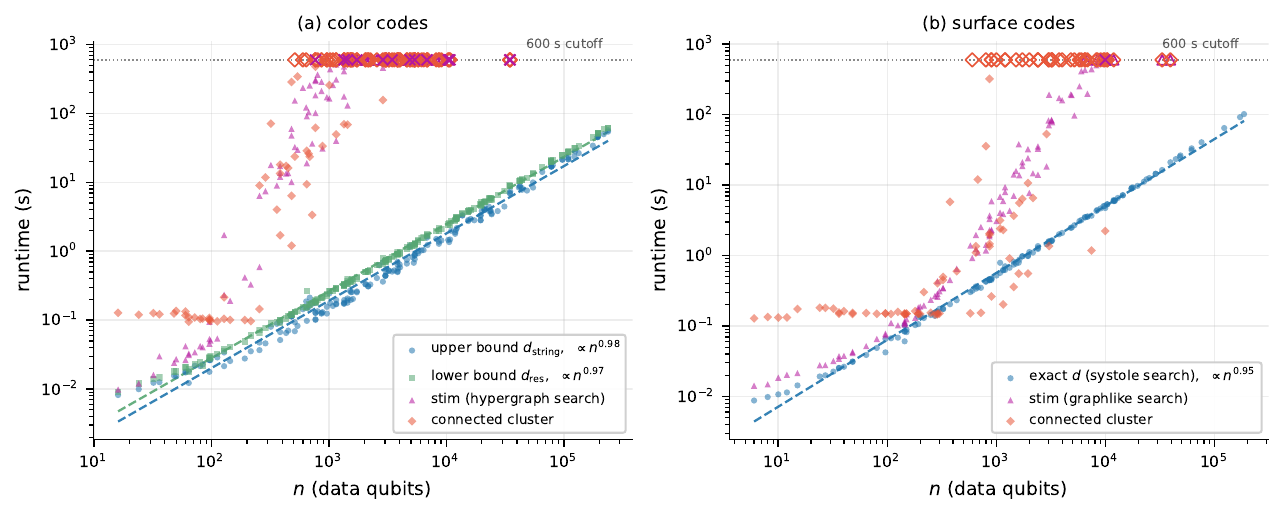}
    \caption{A comparison of the runtime of different distance finding methods for the color codes and surface codes in the tables in \Cref{sec:promising_candidate_tables}. For both color and surface codes we plot the runtime (measured using AMD EPYC 7B13 processors) against the number of data qubits in the code. In both plots, dashed lines show least-squares fits of $\log t$ against $\log n$ over all codes shown, corresponding to a power law $t\propto n^\alpha$ with the exponents given in the legend. We set a timeout of 600 seconds per code for all methods and open markers are runs that did not complete within this budget. \textbf{a}, Runtimes of distance-finding methods for semi-hyperbolic color codes. We compare the stringlike color code distance upper bound of \Cref{sec:color_distance_upper_bound} ($d_{\mathrm{string}}$), the restricted lattice lower bound of \Cref{sec:color_code_distance_lower_bound} ($d_{\mathrm{res}}$), the \texttt{stim.Circuit.search\_for\_undetectable\_logical\_errors} method from Stim~\cite{Gidney2021stimfaststabilizer}, and the connected cluster method of Ref.~\cite{dumer2017distance} as implemented in~\cite{dist_m4ri}. For the stim method we set \texttt{dont\_explore\_detection\_event\_sets\_with\_size\_above=5}, \texttt{dont\_explore\_edges\_with\_degree\_above=3} and \texttt{dont\_explore\_edges\_increasing\_symptom\_degree=False}. Note that for the stim and connected cluster methods there is a setup cost associated with how the methods are called which hides the true algorithmic scaling for small ($n<500$) codes.
    \textbf{b}, Runtimes of distance finding methods for semi-hyperbolic surface codes using the systole search method of \Cref{sec:surface_code_distance_systole_search}, compared to the Stim \texttt{stim.Circuit.shortest\_graphlike\_error} method and the connected cluster algorithm~\cite{dumer2017distance,dist_m4ri}.
    }
    \label{fig:distance_runtime_frontier}
\end{figure*}

We benchmarked the runtime and accuracy of our distance finding methods that we have described in this section, and compared them against the existing methods.
The methods we compare against for the semi-hyperbolic color codes are the connected cluster algorithm~\cite{dumer2017distance,dist_m4ri}, and the stim hypergraph search method \texttt{stim.Circuit.search\_for\_undetectable\_logical\_errors}~\cite{Gidney2021stimfaststabilizer}.
The connected cluster algorithm is exact but has complexity $n(\Delta-1)^{d-1}$ for a code with check weight $\Delta$ and distance $d$.
The stim hypergraph search method is a heuristic with tunable parameters to adjust the size of the search, and only produces an upper bound on the distance.
For the semi-hyperbolic surface codes, we compare our systole search method (\Cref{sec:surface_code_distance_systole_search}) against the stim method \texttt{stim.Circuit.shortest\_graphlike\_error} and connected cluster method.
We refer the reader to~\cite{webster2026distance} for a comparison of a broader range of existing exact and heuristic distance finding methods for hyperbolic surface and color codes with $n<1200$.

Our runtime comparisons are shown in \Cref{fig:distance_runtime_frontier}.
Empirically, across the system sizes we benchmarked, our systole search methods for both color and surface codes are an excellent fit to a runtime scaling of $t\propto n$.
In the limit of large system sizes we expect the complexity to become $t\propto |\mathcal{B}|n\lceil 2g/w\rceil$, where $\mathcal{B}$ is the set of source nodes (restricted to a fundamental domain).
By contrast, the stim and connected cluster methods have a much steeper scaling with $n$ (and are exponential in $d$), and are therefore too slow to be practical for system sizes above around $10^3$--$10^4$ qubits.
This demonstrates the importance of our efficient distance finding methods for evaluating codes with large system sizes relevant to utility-scale quantum computation.

The connected cluster method, being exact, is very useful for evaluating the accuracy (tightness) of our color code distance bounds for small system sizes.
In \Cref{tab:distance_bound_accuracy} we compare our efficient distance bounds with each other for 32,689 tilings we constructed, as well as comparing them to the connected cluster (exact) distance for the subset of codes we could run it on within a time budget of 300 seconds.
We find that the distance is efficiently certified ($d_{\mathrm{res}}=d_{\mathrm{string}}$) for a large fraction of the codes (88\% to 100\% of codes depending on $L$).
When comparing to the exact distance computed with connected cluster, we find that the exact distance matches the \textit{upper bound} $d_{\mathrm{string}}$ for all codes for which connected cluster finished at $L=3$ and $L=4$, and with a $>98\%$ agreement for $L=1$ and $L=2$.
This suggests that the stringlike distance $d_{\mathrm{string}}$ is an excellent estimate of the true distance for most codes, despite being only an upper bound.
Furthermore, in \Cref{tab:distance_bound_inheritance} we show that we only observe a loose bound $d_{\mathrm{string}}>d$ at $L=2$ for tilings where the $L=1$ code has true distance $d=2$ and the loose upper bound $d_{\mathrm{string}}=4$.

\begin{table*}[t]
\centering
\footnotesize
\setlength{\tabcolsep}{6pt}
\begin{tabular}{rrrrrr}
\toprule
 & \multicolumn{2}{c}{All codes} & \multicolumn{3}{c}{Connected cluster completed} \\
\cmidrule(lr){2-3} \cmidrule(lr){4-6}
$L$ & Codes & $d_{\mathrm{res}}=d_{\mathrm{string}}$ & Codes & $d_{\mathrm{string}}=d_{\mathrm{CC}}$ & $d_{\mathrm{res}}=d_{\mathrm{CC}}$ \\
 & & (\%) & & (\%) & (\%) \\
\midrule
1 & 32,689 & 88.73 & 24,413 & 98.46 & 92.27 \\
2 & 32,689 & 98.99 & 3,832 & 98.70 & 99.97 \\
3 & 32,689 & 100.00 & 1,127 & 100.00 & 100.00 \\
4 & 32,689 & 90.65 & 224 & 100.00 & 81.25 \\
5 & 32,689 & 98.96 & 0 & -- & -- \\
6 & 32,689 & 100.00 & 0 & -- & -- \\
7 & 32,582 & 90.98 & 0 & -- & -- \\
\midrule
All & 228,716 & 95.47 & 29,596 & 98.56 & 93.48 \\
\bottomrule
\end{tabular}
\caption{Accuracy of the distance bounds for the $228{,}716$ semi-hyperbolic color codes we constructed, split by side length $L$. We write $d_{\mathrm{CC}}$ for the exact distance returned by the connected cluster search~\cite{dumer2017distance}, which is exhaustive and therefore exact whenever it terminates within its budget. ``All codes'' covers every code at that $L$ and the condition $d_{\mathrm{res}}=d_{\mathrm{string}}$ certifies the distance exactly. ``Connected cluster completed'' is restricted to the codes for which the connected cluster algorithm~\cite{dumer2017distance,dist_m4ri} was run (we restricted the search to $n\leq 2000$) and finished (we set a timeout of 300 seconds).}
\label{tab:distance_bound_accuracy}
\end{table*}

\begin{table*}[t]
\centering
\footnotesize
\setlength{\tabcolsep}{6pt}
\begin{tabular}{lrrrr}
\toprule
 & & \multicolumn{2}{c}{$L=2$ determined} & \\
\cmidrule(lr){3-4}
$L=1$ bound & Tilings & Tight & Loose & $L=2$ undet. \\
\midrule
Tight & 24,037 & 3,698 & 0 & 20,339 \\
\midrule
Loose, 4 $\to$ 2 & 350 & 83 & 50 & 217 \\
Loose, 8 $\to$ 6 & 24 & 1 & 0 & 23 \\
Loose, 8 $\to$ 4 & 1 & 0 & 0 & 1 \\
Loose, 6 $\to$ 4 & 1 & 0 & 0 & 1 \\
\bottomrule
\end{tabular}
\caption{Tightness of the stringlike upper bound for a base ($L=1$) color code and for the fine-grained ($L=2$) code built from the same tiling, over the $24{,}413$ base tilings whose distance the connected cluster search was run on and for which it finished. Every fine-grained $L=2$ code with a tight upper bound also had a tight upper bound at $L=1$. Base codes with a loose bound are itemized by the gap $d_{\mathrm{string}}\to d_{\mathrm{CC}}$. ``$L=2$ undet.'' counts tilings for which the connected cluster algorithm did not complete (or was not attempted). Every fine-grained code we find with a loose bound is a fine-grained version of an $L=1$ code whose bound is already loose with a true distance $d_{\mathrm{CC}}=2$ and upper bound $d_{\mathrm{string}}=4$.}
\label{tab:distance_bound_inheritance}
\end{table*}

\section{Additional Numerical Simulations}

\subsection{Asymmetry of $X$ and $Z$ Bases for Modular Hyperbolic Surface Codes with Even Side Length $L$}\label{sec:asymmetry_X_Z}

\begin{figure}[h]
    \centering
    \includegraphics[width=0.8\linewidth]{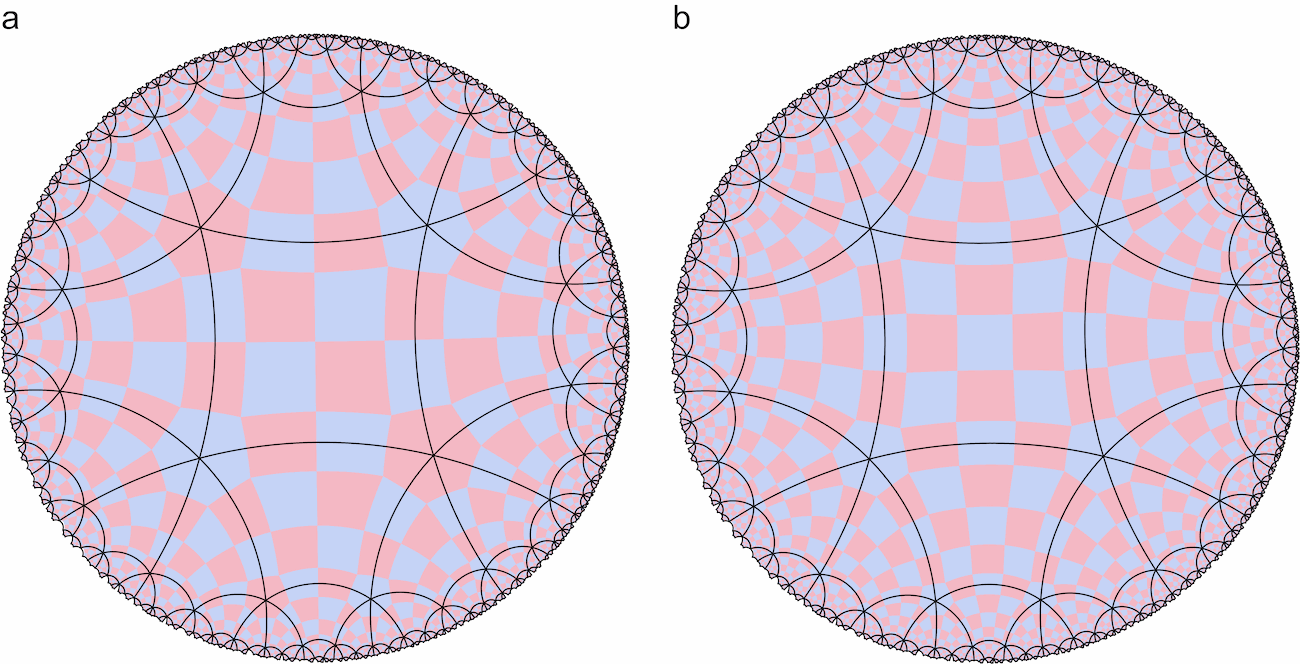}
    \caption{Stabilizer structure of modular hyperbolic surface codes derived from $\{4,6\}$ regular tilings. Each blue plaquette denotes a $Z$ stabilizer with support on its vertices, and similarly each red plaquette denotes an $X$ stabilizer. \textbf{a}, For odd side length (here $L=3$) the weight-6 stabilizers at the four corners of each base face alternate between $X$ and $Z$ type. \textbf{b}, For even side length (here $L=4$) the weight-6 stabilizers are all of the same Pauli type (here all weight-six stabilizers are $Z$-type).}
    \label{fig:regular_4_6_surface_L3_vs_L4_stabilizer_structure}
\end{figure}

\begin{figure}[H]
    \centering
    \includegraphics[width=0.7\linewidth]{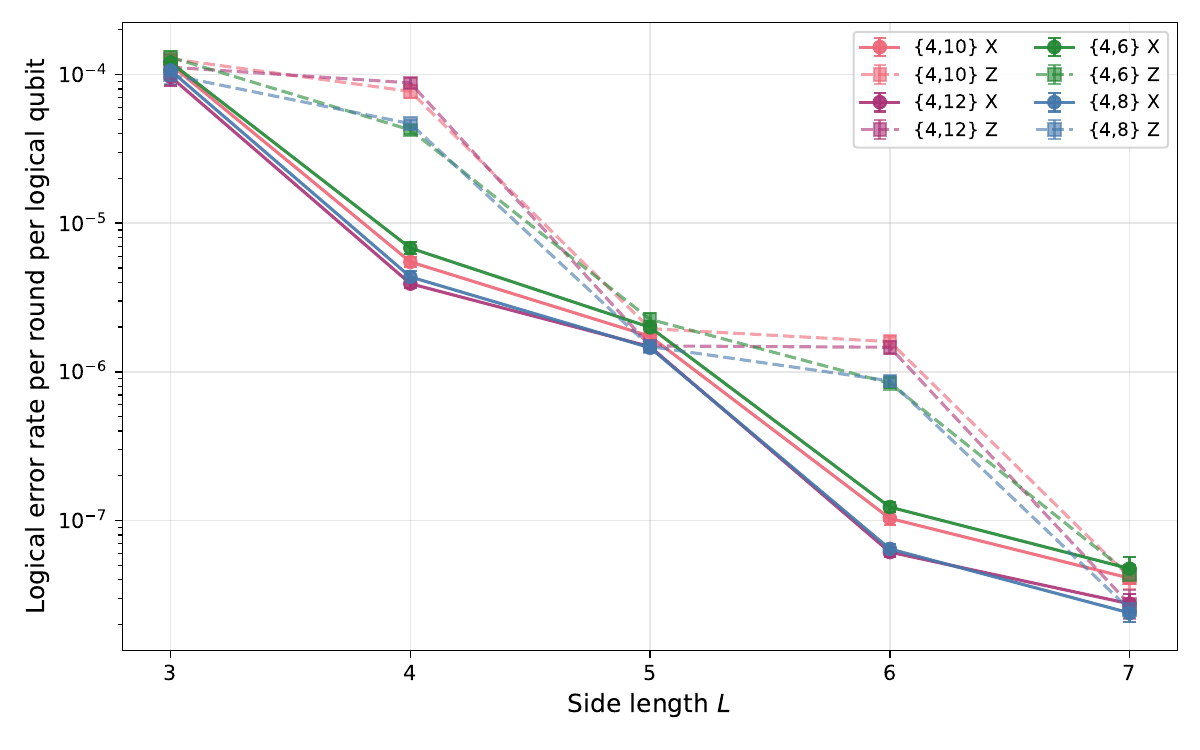}
    \caption{Logical error rate per round per logical vs.~side length $L$ for small modular hyperbolic surface codes derived from $\{4,s\}$ regular tilings. We use an SI1000 noise model with $p=0.001$ and $p_{\mathrm{seam}}=0.01$.}
    \label{fig:ler_vs_side_length}
\end{figure}

For quantum hardware with approximately isotropic noise (no $X/Z$ bias), it is generally desirable for a code to perform equally well in both the $X$ basis and $Z$ basis.
For modular hyperbolic \textit{color} codes this is guaranteed (at least in a code capacity setting), since the codes are all self-dual.
In this section we study $X/Z$ asymmetries of modular hyperbolic \textit{surface} codes, including how this differs for odd and even side length $L$.
As shown in \Cref{fig:regular_4_6_surface_L3_vs_L4_stabilizer_structure}, for modular hyperbolic surface codes derived from regular tilings, the structure of the $X$ and $Z$ stabilizers is identical for odd $L$.
In particular, the number of $X$ and $Z$ stabilizers of each size is identical, and the $X$ and $Z$ stabilizers can be swapped by applying a symmetry of the tiling.
In contrast, for even $L$, the structure of $X$ and $Z$ stabilizers is no longer identical.
This results in asymmetric $X$ and $Z$ logical error rates for even $L$, as shown in \Cref{fig:ler_vs_side_length}.
In fact, the overall logical error rate of an even-$L$ modular hyperbolic code (for these regular tilings) barely improves on the odd $(L-1)$ code derived from the same tiling.

We now study this asymmetry in more detail more generally for quadrilateral tilings.
Consider a quadrilateral tiling of a closed hyperbolic surface with face configuration V$a.b.c.d$, for example constructed from a $\square(r,s,t,u)$ quadrilateral group where $(a,b,c,d)=(2r,2s,2t,2u)$.
The vertices of each square have degrees $a$, $b$, $c$ and $d$ (going clockwise around the square).
We label each vertex based on which corner of the quadrilateral fundamental domain it is in the image of.
Specifically, we label a vertex as $a$-type if it is in the image of the degree-$a$ vertex of the fundamental domain, and similarly for $b$-type, $c$-type and $d$-type.
Therefore, in a slight abuse of notation, an $a$-type vertex has degree $a$, and similarly for $b$-type, $c$-type and $d$-type vertices.
Let $N_a$ be the number of $a$-type vertices (with degree $a$), and similarly for $N_b$, $N_c$ and $N_d$.
We now count the number of faces $|F|$, edges $|E|$ and vertices $|V|$ in the tiling.
First, since all faces are quadrilaterals we know that $4|F|=2|E|$.
Since each face is incident to one vertex of each type we also know that
\begin{equation}
    aN_a=bN_b=cN_c=dN_d=|F|.
\end{equation}
Therefore, written in terms of the number of faces $|F|$, there are $\frac{|F|}{x}$ $x$-type vertices and therefore the number of vertices $|V|$ in total is
\begin{equation}\label{eq:num_vertices_in_quad_tiling}
    |V|=|F|\left(\frac{1}{a}+\frac{1}{b}+\frac{1}{c}+\frac{1}{d}\right)
\end{equation}
and the number of edges $|E|=2|F|$.

We now count the number of $X$ and $Z$ stabilizers for a V$a.b.c.d$ tiling of side length $L$.
We start with the case of even $L$, for which all the ``corner'' stabilizers (each centered on a vertex of the base tiling) are of the same Pauli type.
Consistent with \Cref{fig:regular_4_6_surface_L3_vs_L4_stabilizer_structure}, we will assume that these $|V|$ corner stabilizers are all $Z$-type.
We next count the stabilizers that are fully internal to a face of the base tiling.
For an even side length $L$, these form an $(L-1)\times (L-1)$ square grid with $\frac{(L-1)^2+1}{2}$ $Z$ stabilizers and $\frac{(L-1)^2-1}{2}$ $X$ stabilizers.
Finally, along each edge of the base tiling we have $\frac{L}{2}$ $X$ stabilizers and $\left(\frac{L}{2}-1\right)$ $Z$ stabilizers.
Putting this together, the number of $X$ stabilizers $N_X^{\mathrm{even}}$ for even $L$ is
\begin{equation}
    N_X^{\mathrm{even}}=\frac{L^2}{2}|F|
\end{equation}
while the number of $Z$ stabilizers is
\begin{equation}
    N_Z^{\mathrm{even}}=\left(\frac{L^2}{2}+\frac{1}{a}+\frac{1}{b}+\frac{1}{c}+\frac{1}{d}-1\right)|F|.
\end{equation}

For odd $L$, the square grid of stabilizers fully contained within each base face contains $\frac{(L-1)^2}{2}$ $X$ stabilizers and the same number of $Z$ stabilizers.
There are also $\frac{L-1}{2}$ $X$ stabilizers (and the same number of $Z$ stabilizers) supported on each base edge of the tiling.
The Pauli basis of the corner stabilizers alternates around the four corners of each base face.
Let us assume that the $a$-type and $c$-type vertices each support $X$ stabilizers and the $b$-type and $d$-type vertices each support $Z$ stabilizers.
This leads to a number of $X$ stabilizers $N_X^\mathrm{odd}$ for odd $L$
\begin{equation}
    N_X^\mathrm{odd}=\left(\frac{L^2}{2}+\frac{1}{a}+\frac{1}{c}-\frac{1}{2}\right)|F|
\end{equation}
and a number of $Z$ stabilizers
\begin{equation}
    N_Z^\mathrm{odd}=\left(\frac{L^2}{2}+\frac{1}{b}+\frac{1}{d}-\frac{1}{2}\right)|F|.
\end{equation}

For even $L$, there are fewer $Z$ stabilizers than $X$ stabilizers, $N_Z^{\mathrm{even}}<N_X^{\mathrm{even}}$, provided
\begin{equation}
    \frac{1}{a}+\frac{1}{b}+\frac{1}{c}+\frac{1}{d}<1,
\end{equation}
which is exactly the condition under which the tiling is hyperbolic.
For odd $L$, we have that $N_Z^{\mathrm{odd}}<N_X^{\mathrm{odd}}$ provided that
\begin{equation}
    \frac{1}{b}+\frac{1}{d}<\frac{1}{a}+\frac{1}{c}.
\end{equation}
We can apply this to the tilings used in the simulations in \Cref{fig:ler_vs_side_length}.
The tilings are regular $\{4,s\}$ tilings, and so we can set $a=b=c=d=s$ in the formulas above.
For odd $L$, we have exactly the same number of $X$ and $Z$ stabilizers (of each weight), whereas for even $L$ we always have fewer $Z$ stabilizers (and they are higher weight), which we intuitively would expect to negatively impact the code's ability to detect and correct $X$ errors.
This is consistent with the numerical simulations, which have a higher logical error rate for the $Z$ memory experiments (which are sensitive to $X$-type logical errors).

\section{Tables of Modular Hyperbolic Codes}\label{sec:promising_candidate_tables}

In \Cref{tab:candidates_L1_color,tab:candidates_L1_surface}, we list hyperbolic color codes and surface codes derived from closed base tilings ($L=1$).
In \Cref{tab:candidates_L2_color_rgb8,tab:candidates_L3_color_ms2,tab:candidates_L4_color_ms2}, we list modular hyperbolic color codes across side lengths $L=2$ (with face valency 8 and octagon modules, $K=8$), $L=3$ (with quadrilateral modules, $K=2$), and $L=4$ ($K=2$).
In \Cref{tab:candidates_L5_surface_ms2,tab:candidates_L3_surface_ms4_quad4}, we list modular hyperbolic surface codes for side length $L=5$ (with rectangle modules, $K=2$) and $L=3$ (with big square modules, $K=4$ on quadrilateral tilings with vertex valency 4).
Each table lists codes on the Pareto frontier over $n$, $d$, $kd^2/n$, and maximum stabilizer weight.
The Pareto frontiers restricted to $8.8.8$ tilings are given for each side length in \Cref{tab:candidates_L2_color_888,tab:candidates_L3_color_ms2_888,tab:candidates_L4_color_ms2_888}, and are combined in \Cref{tab:candidates_888}.
Note that the Symmetry Group column labels the parent group used in the subgroup search, whereas the Tiling column gives the actual face or vertex sizes of the resulting quotient tiling (which can be smaller if a rotation generator has reduced order in the quotient).

\begin{table}[H]
\centering
\footnotesize
\setlength{\tabcolsep}{6pt}
\begin{adjustbox}{max width=\linewidth}
\begin{tabular}{cccrrrcr}
\toprule
Tiling & Symmetry Group & $[[n,k,d]]$ & $M_{8}$ & $N_{M8}$ & $N_{M8}/k$ & Orientable & $kd^2/n$ \\
\midrule
$8.8.8$ & $\triangle(4,4,4)_{32}$ & $[[64, 8, 6]]$ & $2$ & 144 & 18.0 & Yes & 4.50 \\
$8.8.8$ & $\triangle(4,4,4)_{40}$ & $[[128, 12, 8]]$ & $4$ & 288 & 24.0 & Yes & 6.00 \\
$8.8.8$ & $\triangle(4,4,4)_{58}$ & $[[512, 36, 12]]$ & $16$ & 1152 & 32.0 & Yes & 10.12 \\
$8.8.8$ & $\triangle(4,4,4)_{96}$ & $[[1344, 88, 14]]$ & $42$ & 3024 & 34.4 & Yes & 12.83 \\
$8.8.8$ & $\triangle(4,4,4)_{109}$ & $[[1600, 104, 14\text{--}16]]$ & $50$ & 3600 & 34.6 & Yes & 12.74--16.64 \\
$8.8.8$ & $\triangle(4,4,4)_{271}$ & $[[5376, 340, 18\text{--}20]]$ & $168$ & 12096 & 35.6 & Yes & 20.49--25.30 \\
$8.8.8$ & $\triangle(4,4,4)|\mathrm{PSL}(2,23)$ & $[[24288, 1522, 22]]$ & $759$ & 54648 & 35.9 & No & 30.33 \\
$8.8.8$ & $\triangle(4,4,4)|\mathrm{PSL}(2,25)$ & $[[31200, 1954, 22\text{--}24]]$ & $975$ & 70200 & 35.9 & No & 30.31--36.07 \\
\bottomrule
\end{tabular}%
\end{adjustbox}
\caption{Hyperbolic color codes on the Pareto frontier for 8.8.8 tilings and $L=2$. Here $M_8$ is the number of octagon modules and $N_{M8}$ is the number of physical qubits (data and ancilla) in the code using the octagon module partition. Orientable: whether the manifold is orientable. $kd^2/n$: code figure of merit. Distances quoted as a range are bounded below and above by the methods of \Cref{sec:distance_finding}.}
\label{tab:candidates_L2_color_888}
\end{table}

\begin{table}[H]
\centering
\footnotesize
\setlength{\tabcolsep}{6pt}
\begin{adjustbox}{max width=\linewidth}
\begin{tabular}{cccrrrlrccr}
\toprule
Tiling & Symmetry Group & $[[n,k,d]]$ & $M_{2}$ & $N_{M2}$ & $N_{M2}/k$ & Automorphism Bases & Cov. & Pair & Cayley & $kd^2/n$ \\
\midrule
$8.8.8$ & $\triangle(4,4,4)_{32}$ & $[[144, 8, 8]]$ & $8$ & 336 & 42.0 & $2/2\,(w{=}2{+}2;\,w_1{=}2{+}4)$ & 100\% & 0.5 & 1.2/2 & 3.56 \\
$8.8.8$ & $\triangle(4,4,4)_{40}$ & $[[288, 12, 12]]$ & $16$ & 672 & 56.0 & $3/3\,(w{=}3;\,w_1{=}4)$ & 100\% & 1.3 & 2.4/4 & 6.00 \\
$8.8.8$ & $\triangle(4,4,4)_{58}$ & $[[1152, 36, 16]]$ & $64$ & 2688 & 74.7 & $8/9\,(w{=}5;\,w_1{=}6) \mid 9/9\,(w{=}4{+}4{+}5;\,w_1{=}4{+}8{+}6)$ & 100\% & 1 & 3.3/6 & 8.00 \\
$8.8.8$ & $\triangle(4,4,4)_{96}$ & $[[3024, 88, 20]]$ & $168$ & 7056 & 80.2 & $17/22\,(w{=}6{+}6{+}6;\,w_1{=}6{+}6{+}8)$ & 77\% & 2.3 & 4.1/6 & 11.64 \\
$8.8.8$ & $\triangle(4,4,4)_{109}$ & $[[3600, 104, 22]]$ & $200$ & 8400 & 80.8 & $25/26\,(w{=}8;\,w_1{=}10)$ & 96\% & 1 & 4.4/6 & 13.98 \\
$8.8.8$ & $\triangle(4,4,4)_{120}$ & $[[4608, 132, 24]]$ & $256$ & 10752 & 81.5 & $21/33\,(w{=}6{+}7;\,w_1{=}8{+}8)$ & 64\% & 2.4 & 4.9/8 & 16.50 \\
$8.8.8$ & $\triangle(4,4,4)_{271}$ & $[[12096, 340, 28]]$ & $672$ & 28224 & 83.0 & $35/85\,(w{=}8{+}8;\,w_1{=}10{+}10)$ & 41\% & 4.1 & 6.0/10 & 22.04 \\
$8.8.8$ & $\triangle(4,4,4)|\mathrm{PSL}(2,23)$ & $[[54648, 1522, 32]]$ & $3036$ & 127512 & 83.8 & \text{Non-orientable} & --- & --- & --- & 28.52 \\
$8.8.8$ & $\triangle(4,4,4)|\mathrm{PGL}(2,23)$ & $[[109296, 3040, 36]]$ & $6072$ & 255024 & 83.9 & --- & --- & --- & --- & 36.05 \\
\bottomrule
\end{tabular}%
\end{adjustbox}
\caption{Hyperbolic color codes on the Pareto frontier for 8.8.8 tilings and $L=3$. Here $M_2$ is the number of quadrilateral modules and $N_{M2}$ is the number of physical qubits (data and ancilla) in the code using the quadrilateral module partition. The seed weights $w$ and $w_1$ and the Pair and Cayley columns are defined as in \Cref{tab:candidates_L3_color_ms2}, except that Pair and Cayley are computed for the listed basis with the lowest total seed weight $w$.}
\label{tab:candidates_L3_color_ms2_888}
\end{table}

\begin{table}[H]
\centering
\footnotesize
\setlength{\tabcolsep}{6pt}
\begin{adjustbox}{max width=\linewidth}
\begin{tabular}{cccrrrlrccr}
\toprule
Tiling & Symmetry Group & $[[n,k,d]]$ & $M_{2}$ & $N_{M2}$ & $N_{M2}/k$ & Automorphism Bases & Cov. & Pair & Cayley & $kd^2/n$ \\
\midrule
$8.8.8$ & $\triangle(4,4,4)_{32}$ & $[[256, 8, 12]]$ & $8$ & 576 & 72.0 & $2/2\,(w{=}2{+}2;\,w_1{=}2{+}4)$ & 100\% & 0.5 & 1.2/2 & 4.50 \\
$8.8.8$ & $\triangle(4,4,4)_{40}$ & $[[512, 12, 16]]$ & $16$ & 1152 & 96.0 & $3/3\,(w{=}3;\,w_1{=}4)$ & 100\% & 1.3 & 2.4/4 & 6.00 \\
$8.8.8$ & $\triangle(4,4,4)_{58}$ & $[[2048, 36, 20\text{--}24]]$ & $64$ & 4608 & 128.0 & $8/9\,(w{=}5;\,w_1{=}6) \mid 9/9\,(w{=}4{+}4{+}5;\,w_1{=}4{+}8{+}6)$ & 100\% & 1 & 3.3/6 & 7.03--10.12 \\
$8.8.8$ & $\triangle(4,4,4)_{96}$ & $[[5376, 88, 26\text{--}28]]$ & $168$ & 12096 & 137.5 & $17/22\,(w{=}6{+}6{+}6;\,w_1{=}6{+}6{+}8)$ & 77\% & 2.3 & 4.1/6 & 11.07--12.83 \\
$8.8.8$ & $\triangle(4,4,4)_{109}$ & $[[6400, 104, 30]]$ & $200$ & 14400 & 138.5 & $25/26\,(w{=}8;\,w_1{=}10)$ & 96\% & 1 & 4.4/6 & 14.62 \\
$8.8.8$ & $\triangle(4,4,4)_{120}$ & $[[8192, 132, 32]]$ & $256$ & 18432 & 139.6 & $21/33\,(w{=}6{+}7;\,w_1{=}8{+}8)$ & 64\% & 2.4 & 4.9/8 & 16.50 \\
$8.8.8$ & $\triangle(4,4,4)_{271}$ & $[[21504, 340, 36\text{--}38]]$ & $672$ & 48384 & 142.3 & $35/85\,(w{=}8{+}8;\,w_1{=}10{+}10)$ & 41\% & 4.1 & 6.0/10 & 20.49--22.83 \\
$8.8.8$ & $\triangle(4,4,4)_{786}$ & $[[65536, 1028, 40]]$ & $2048$ & 147456 & 143.4 & --- & --- & --- & --- & 25.10 \\
$8.8.8$ & $\triangle(4,4,4)|\mathrm{PSL}(2,23)$ & $[[97152, 1522, 42\text{--}44]]$ & $3036$ & 218592 & 143.6 & \text{Non-orientable} & --- & --- & --- & 27.64--30.33 \\
$8.8.8$ & $\triangle(4,4,4)|\mathrm{PSL}(2,25)$ & $[[124800, 1954, 42\text{--}46]]$ & $3900$ & 280800 & 143.7 & \text{Non-orientable} & --- & --- & --- & 27.62--33.13 \\
$8.8.8$ & $\triangle(4,4,4)|\mathrm{PGL}(2,23)$ & $[[194304, 3040, 48]]$ & $6072$ & 437184 & 143.8 & --- & --- & --- & --- & 36.05 \\
\bottomrule
\end{tabular}%
\end{adjustbox}
\caption{Hyperbolic color codes on the Pareto frontier for 8.8.8 tilings and $L=4$. Here $M_2$ is the number of quadrilateral modules and $N_{M2}$ is the number of physical qubits (data and ancilla) in the code using the quadrilateral module partition. The seed weights $w$ and $w_1$ and the Pair and Cayley columns are defined as in \Cref{tab:candidates_L3_color_ms2}, except that Pair and Cayley are computed for the listed basis with the lowest total seed weight $w$.}
\label{tab:candidates_L4_color_ms2_888}
\end{table}

\begin{table}[H]
\centering
\footnotesize
\setlength{\tabcolsep}{6pt}
\begin{tabular}{ccccr}
\toprule
Tiling & Symmetry Group & $[[n,k,d]]$ & Orientable & $kd^2/n$ \\
\midrule
$8.8.8$ & $\triangle(4,4,4)_{32}$ & $[[16, 8, 2]]$ & Yes & 2.00 \\
$6.8.8$ & $\triangle(3,4,4)_{8}$ & $[[24, 8, 4]]$ & Yes & 5.33 \\
$8.8.8$ & $\triangle(4,4,4)_{40}$ & $[[32, 12, 4]]$ & Yes & 6.00 \\
$6.12.12$ & $\triangle(3,6,6)_{20}$ & $[[36, 16, 4]]$ & Yes & 7.11 \\
$8.8.10$ & $\triangle(4,4,5)_{8}$ & $[[40, 16, 4]]$ & Yes & 6.40 \\
$8.12.12$ & $\triangle(4,6,6)_{48}$ & $[[48, 24, 4]]$ & Yes & 8.00 \\
$10.10.10$ & $\triangle(5,5,5)_{8}$ & $[[50, 24, 4]]$ & Yes & 7.68 \\
$10.12.12$ & $\triangle(5,6,6)_{12}$ & $[[60, 32, 4]]$ & Yes & 8.53 \\
$4.10.10$ & $\triangle(2,5,5)_{3}$ & $[[60, 10, 6]]$ & No & 6.00 \\
$8.16.16$ & $\triangle(4,8,8)_{93}$ & $[[64, 36, 4]]$ & Yes & 9.00 \\
$12.12.12$ & $\triangle(6,6,6)_{93}$ & $[[72, 40, 4]]$ & Yes & 8.89 \\
$10.16.16$ & $\triangle(5,8,8)_{10}$ & $[[80, 48, 4]]$ & Yes & 9.60 \\
$12.12.14$ & $\triangle(6,6,7)_{9}$ & $[[84, 48, 4]]$ & Yes & 9.14 \\
$12.16.16$ & $\triangle(6,8,8)_{78}$ & $[[96, 60, 4]]$ & Yes & 10.00 \\
$6.8.8$ & $\triangle(3,4,4)_{18}$ & $[[96, 20, 6]]$ & Yes & 7.50 \\
$14.14.14$ & $\triangle(7,7,7)_{10}$ & $[[98, 60, 4]]$ & Yes & 9.80 \\
$14.16.16$ & $\triangle(7,8,8)_{10}$ & $[[112, 72, 4]]$ & Yes & 10.29 \\
$8.8.8$ & $\triangle(4,4,4)_{58}$ & $[[128, 36, 8]]$ & Yes & 18.00 \\
$4.10.10$ & $\triangle(2,5,5)_{6}$ & $[[320, 36, 10]]$ & Yes & 11.25 \\
$6.6.8$ & $\triangle(3,3,4)_{10}$ & $[[360, 34, 10]]$ & No & 9.44 \\
$8.8.12$ & $\triangle(4,4,6)_{133}$ & $[[384, 132, 8]]$ & Yes & 22.00 \\
$8.12.12$ & $\triangle(4,6,6)_{260}$ & $[[480, 204, 8]]$ & No & 27.20 \\
$8.8.10$ & $\triangle(4,4,5)_{29}$ & $[[480, 148, 8]]$ & Yes & 19.73 \\
$4.10.12$ & $\triangle(2,5,6)_{14}$ & $[[480, 68, 10]]$ & Yes & 14.17 \\
$4.10.10$ & $\triangle(2,5,5)_{7}$ & $[[600, 64, 12]]$ & Yes & 15.36 \\
$12.12.12$ & $\triangle(6,6,6)_{707}$ & $[[648, 328, 8]]$ & Yes & 32.40 \\
$8.8.8$ & $\triangle(4,4,4)_{160}$ & $[[672, 172, 10]]$ & No & 25.60 \\
$14.14.14$ & $\triangle(7,7,7)_{16}$ & $[[686, 396, 8]]$ & Yes & 36.94 \\
$10.10.10$ & $\triangle(5,5,5)_{28}$ & $[[720, 292, 8]]$ & Yes & 25.96 \\
$12.16.16$ & $\triangle(6,8,8)_{864}$ & $[[768, 452, 8]]$ & Yes & 37.67 \\
$6.6.8$ & $\triangle(3,3,4)_{16}$ & $[[1008, 88, 12]]$ & Yes & 12.57 \\
$8.8.8$ & $\triangle(4,4,4)_{271}$ & $[[1344, 340, 10\text{--}12]]$ & Yes & 25.30--36.43 \\
$4.8.10$ & $\triangle(2,4,5)_{16}$ & $[[1440, 76, 16]]$ & Yes & 13.51 \\
$14.16.16$ & $\triangle(7,8,8)_{189}$ & $[[2688, 1636, 6\text{--}8]]$ & Yes & 21.91--38.95 \\
$4.8.12$ & $\triangle(2,4,6)_{79}$ & $[[2880, 244, 16]]$ & Yes & 21.69 \\
$8.12.12$ & $\triangle(4,6,6)|\mathrm{PGL}(2,17)|1$ & $[[4896, 2044, 8\text{--}10]]$ & No & 26.72--41.75 \\
$6.6.8$ & $\triangle(3,3,4)|\mathrm{PGL}(2,17)|1$ & $[[4896, 412, 16]]$ & Yes & 21.54 \\
$4.10.10$ & $\triangle(2,5,5)|\mathrm{PGL}(2,19)$ & $[[6840, 688, 16]]$ & Yes & 25.75 \\
$4.8.10$ & $\triangle(2,4,5)|\mathrm{PGL}(2,19)|1$ & $[[6840, 346, 18]]$ & No & 16.39 \\
$4.12.12$ & $\triangle(2,6,6)|\mathrm{PGL}(2,23)$ & $[[12144, 2028, 16]]$ & Yes & 42.75 \\
$6.8.8$ & $\triangle(3,4,4)|\mathrm{PGL}(2,23)$ & $[[12144, 2028, 14]]$ & Yes & 32.73 \\
$10.10.10$ & $\triangle(5,5,5)|\mathrm{PSL}(2,29)|1$ & $[[12180, 4876, 8\text{--}10]]$ & No & 25.62--40.03 \\
$6.10.10$ & $\triangle(3,5,5)|\mathrm{PSL}(2,29)|1$ & $[[12180, 3252, 12]]$ & No & 38.45 \\
$8.8.10$ & $\triangle(4,4,5)|\mathrm{PGL}(2,19) \times \mathrm{Z}_{2}|1$ & $[[13680, 4108, 8\text{--}12]]$ & Yes & 19.22--43.24 \\
$4.8.10$ & $\triangle(2,4,5)|\mathrm{PGL}(2,19) \times \mathrm{Z}_{2}|1$ & $[[13680, 688, 20]]$ & Yes & 20.12 \\
$8.8.8$ & $\triangle(4,4,4)|\mathrm{PSL}(2,25) \times \mathrm{Z}_{2}$ & $[[15600, 3904, 10\text{--}14]]$ & Yes & 25.03--49.05 \\
\bottomrule
\end{tabular}%
\caption{Hyperbolic color codes derived from base tilings ($L=1$). Orientable: whether the manifold is orientable. $kd^2/n$: code figure of merit.}
\label{tab:candidates_L1_color}
\end{table}

\begin{table}[H]
\centering
\footnotesize
\setlength{\tabcolsep}{6pt}
\begin{tabular}{ccccr}
\toprule
Tiling & Symmetry Group & $[[n,k,d]]$ & Orientable & $kd^2/n$ \\
\midrule
\{4,6\} & $\triangle(2,4,6)_{25}$ & $[[6, 4, 2]]$ & Yes & 2.67 \\
\{4,8\} & $\triangle(2,4,8)_{30}$ & $[[8, 6, 2]]$ & Yes & 3.00 \\
\{4,10\} & $\triangle(2,4,10)_{24}$ & $[[10, 8, 2]]$ & Yes & 3.20 \\
\{4,12\} & $\triangle(2,4,12)_{43}$ & $[[12, 10, 2]]$ & Yes & 3.33 \\
\{4,5\} & $\triangle(2,4,5)_{7}$ & $[[15, 5, 3]]$ & No & 3.00 \\
$V4.4.4.6$ & $\square(2,2,2,3)_{32}$ & $[[24, 4, 4]]$ & Yes & 2.67 \\
\{4,6\} & $\triangle(2,4,6)_{35}$ & $[[27, 11, 3]]$ & No & 3.67 \\
$V4.4.8.8$ & $\square(2,2,4,4)_{303}$ & $[[32, 10, 4]]$ & Yes & 5.00 \\
$V4.4.6.6$ & $\square(2,2,3,3)_{23}$ & $[[36, 8, 4]]$ & Yes & 3.56 \\
$V4.6.4.6$ & $\square(2,3,2,3)_{14}$ & $[[36, 8, 4]]$ & Yes & 3.56 \\
\{4,5\} & $\triangle(2,4,5)_{10}$ & $[[40, 10, 4]]$ & Yes & 4.00 \\
$V6.6.6.6$ & $\square(3,3,3,3)_{43}$ & $[[48, 18, 4]]$ & Yes & 6.00 \\
\{4,6\} & $\triangle(2,4,6)_{40}$ & $[[48, 18, 4]]$ & Yes & 6.00 \\
$V8.8.8.8$ & $\square(4,4,4,4)_{2558}$ & $[[64, 34, 4]]$ & Yes & 8.50 \\
\{4,5\} & $\triangle(2,4,5)_{11}$ & $[[80, 18, 5]]$ & Yes & 5.62 \\
$V8.8.12.12$ & $\square(4,4,6,6)_{2888}$ & $[[96, 58, 4]]$ & Yes & 9.67 \\
$V4.4.4.6$ & $\square(2,2,2,3)_{66}$ & $[[96, 10, 6]]$ & No & 3.75 \\
\{4,10\} & $\triangle(2,4,10)_{44}$ & $[[125, 77, 4]]$ & No & 9.86 \\
$V12.12.12.12$ & $\square(6,6,6,6)_{10445}$ & $[[144, 98, 4]]$ & Yes & 10.89 \\
\{4,5\} & $\triangle(2,4,5)_{13}$ & $[[150, 32, 6]]$ & Yes & 7.68 \\
$V4.4.6.8$ & $\square(2,2,3,4)_{237}$ & $[[192, 42, 6]]$ & No & 7.88 \\
$V4.4.4.6$ & $\square(2,2,2,3)_{103}$ & $[[192, 18, 8]]$ & Yes & 6.00 \\
\{4,7\} & $\triangle(2,4,7)_{14}$ & $[[273, 119, 5]]$ & No & 10.90 \\
$V6.6.6.6$ & $\square(3,3,3,3)_{369}$ & $[[288, 98, 6]]$ & Yes & 12.25 \\
$V4.4.6.6$ & $\square(2,2,3,3)_{327}$ & $[[576, 98, 8]]$ & No & 10.89 \\
$V6.8.6.8$ & $\square(3,4,3,4)_{1252}$ & $[[720, 302, 6]]$ & No & 15.10 \\
\{4,5\} & $\triangle(2,4,10)_{120}$ & $[[900, 182, 8]]$ & Yes & 12.94 \\
$V4.4.6.8$ & $\square(2,2,3,4)_{1726}$ & $[[1152, 242, 8]]$ & Yes & 13.44 \\
$V4.4.4.6$ & $\square(2,2,2,3)_{602}$ & $[[1320, 112, 10]]$ & No & 8.48 \\
$V4.4.8.8$ & $\square(2,2,4,4)|\mathrm{PGL}(2,9) \times \mathrm{Z}_{2}|3$ & $[[1440, 362, 8]]$ & Yes & 16.09 \\
$V4.4.10.10$ & $\square(2,2,5,5)_{774}$ & $[[1600, 482, 8]]$ & Yes & 19.28 \\
$V4.6.6.6$ & $\square(2,3,3,3)_{446}$ & $[[1944, 488, 8]]$ & Yes & 16.07 \\
$V4.4.4.6$ & $\square(2,2,2,3)_{843}$ & $[[1944, 164, 12]]$ & Yes & 12.15 \\
\{4,6\} & $\triangle(2,4,6)|\mathrm{PGL}(2,23) \times \mathrm{Z}_{2}$ & $[[3036, 1014, 8]]$ & Yes & 21.38 \\
$V4.4.4.8$ & $\square(2,2,2,4)|\mathrm{PGL}(2,17)$ & $[[4896, 614, d_X{=}10, d_Z{=}12]]$ & No & 12.54 \\
$V4.6.4.6$ & $\square(2,3,2,3)|\mathrm{PSL}(2,19) \times \mathrm{Z}_{2}|1$ & $[[6840, 1142, d_X{=}10, d_Z{=}12]]$ & Yes & 16.70 \\
\{4,10\} & $\triangle(2,4,10)|\mathrm{PGL}(2,31) \times \mathrm{Z}_{2}$ & $[[7440, 4466, 6]]$ & No & 21.61 \\
\{4,11\} & $\triangle(2,4,11)|\mathrm{PGL}(2,43)|2$ & $[[9933, 6323, 6]]$ & No & 22.92 \\
\bottomrule
\end{tabular}%
\caption{Hyperbolic surface codes derived from base tilings ($L=1$). Orientable: whether the manifold is orientable. $kd^2/n$: code figure of merit.}
\label{tab:candidates_L1_surface}
\end{table}

\begin{table}[H]
\centering
\footnotesize
\setlength{\tabcolsep}{3.5pt}
\resizebox{\textwidth}{!}{%
\begin{tabular}{cccrrrcr}
\toprule
Tiling & Symmetry Group & $[[n,k,d]]$ & $M_{8}$ & $N_{M8}$ & $N_{M8}/k$ & Orientable & $kd^2/n$ \\
\midrule
$8.8.8$ & $\triangle(4,4,4)_{32}$ & $[[64, 8, 6]]$ & $2$ & 144 & 18.0 & Yes & 4.50 \\
$8.8.8$ & $\triangle(4,4,4)_{40}$ & $[[128, 12, 8]]$ & $4$ & 288 & 24.0 & Yes & 6.00 \\
$8.8.12$ & $\triangle(4,4,6)_{47}$ & $[[192, 20, 8]]$ & $6$ & 432 & 21.6 & No & 6.67 \\
$8.16.16$ & $\triangle(4,8,8)_{93}$ & $[[256, 36, 8]]$ & $8$ & 576 & 16.0 & Yes & 9.00 \\
$6.8.8$ & $\triangle(3,4,4)_{18}$ & $[[384, 20, 12]]$ & $12$ & 864 & 43.2 & Yes & 7.50 \\
$8.8.8$ & $\triangle(4,4,4)_{58}$ & $[[512, 36, 12]]$ & $16$ & 1152 & 32.0 & Yes & 10.12 \\
$4.8.10$ & $\triangle(2,4,5)_{8}$ & $[[640, 12, 16]]$ & $20$ & 1440 & 120.0 & No & 4.80 \\
$6.8.8$ & $\triangle(3,4,4)_{22}$ & $[[768, 36, 14]]$ & $24$ & 1728 & 48.0 & Yes & 9.19 \\
$6.6.8$ & $\triangle(3,3,4)_{7}$ & $[[768, 20, 16]]$ & $24$ & 1728 & 86.4 & Yes & 6.67 \\
$8.12.12$ & $\triangle(4,6,6)_{163}$ & $[[1344, 144, 12]]$ & $42$ & 3024 & 21.0 & No & 15.43 \\
$8.8.8$ & $\triangle(4,4,4)_{96}$ & $[[1344, 88, 14]]$ & $42$ & 3024 & 34.4 & Yes & 12.83 \\
$8.10.10$ & $\triangle(4,5,5)_{10}$ & $[[1440, 130, 12]]$ & $45$ & 3240 & 24.9 & No & 13.00 \\
$6.6.8$ & $\triangle(3,3,4)_{10}$ & $[[1440, 34, 20]]$ & $45$ & 3240 & 95.3 & No & 9.44 \\
$6.8.8$ & $\triangle(3,4,4)_{32}$ & $[[1536, 68, 16]]$ & $48$ & 3456 & 50.8 & Yes & 11.33 \\
$8.8.8$ & $\triangle(4,4,4)_{109}$ & $[[1600, 104, 14\text{--}16]]$ & $50$ & 3600 & 34.6 & Yes & 12.74--16.64 \\
$8.16.16$ & $\triangle(4,8,8)_{954}$ & $[[2880, 364, 12]]$ & $90$ & 6480 & 17.8 & No & 18.20 \\
$8.8.10$ & $\triangle(4,4,5)_{47}$ & $[[3840, 292, 16]]$ & $120$ & 8640 & 29.6 & Yes & 19.47 \\
$6.8.8$ & $\triangle(3,4,4)_{45}$ & $[[3840, 164, 20]]$ & $120$ & 8640 & 52.7 & Yes & 17.08 \\
$6.6.8$ & $\triangle(3,3,4)_{16}$ & $[[4032, 88, 24]]$ & $126$ & 9072 & 103.1 & Yes & 12.57 \\
$8.8.12$ & $\triangle(4,4,6)_{340}$ & $[[5280, 444, 16]]$ & $165$ & 11880 & 26.8 & No & 21.53 \\
$8.8.8$ & $\triangle(4,4,4)_{271}$ & $[[5376, 340, 18\text{--}20]]$ & $168$ & 12096 & 35.6 & Yes & 20.49--25.30 \\
$6.8.8$ & $\triangle(3,4,4)_{79}$ & $[[10368, 436, 22]]$ & $324$ & 23328 & 53.5 & Yes & 20.35 \\
$4.8.10$ & $\triangle(2,4,5)|\mathrm{PGL}(2,11) \times \mathrm{Z}_{2}$ & $[[10560, 136, 32]]$ & $330$ & 23760 & 174.7 & Yes & 13.19 \\
$8.12.12$ & $\triangle(4,6,6)_{1384}$ & $[[10752, 1124, 16]]$ & $336$ & 24192 & 21.5 & No & 26.76 \\
$6.8.8$ & $\triangle(3,4,4)|\mathrm{PGL}(2,17)$ & $[[19584, 820, 24]]$ & $612$ & 44064 & 53.7 & Yes & 24.12 \\
$6.6.8$ & $\triangle(3,3,4)|\mathrm{PGL}(2,17)|1$ & $[[19584, 412, 32]]$ & $612$ & 44064 & 107.0 & Yes & 21.54 \\
$8.12.12$ & $\triangle(4,6,6)|\mathrm{PSL}(2,23)$ & $[[24288, 2534, 18]]$ & $759$ & 54648 & 21.6 & No & 33.80 \\
$8.8.8$ & $\triangle(4,4,4)|\mathrm{PSL}(2,23)$ & $[[24288, 1522, 22]]$ & $759$ & 54648 & 35.9 & No & 30.33 \\
$4.8.10$ & $\triangle(2,4,5)|\mathrm{PGL}(2,19)|1$ & $[[27360, 346, 36]]$ & $855$ & 61560 & 177.9 & No & 16.39 \\
$8.8.8$ & $\triangle(4,4,4)|\mathrm{PSL}(2,25)$ & $[[31200, 1954, 22\text{--}24]]$ & $975$ & 70200 & 35.9 & No & 30.31--36.07 \\
$6.8.8$ & $\triangle(3,4,4)|\mathrm{PGL}(2,23)$ & $[[48576, 2028, 28]]$ & $1518$ & 109296 & 53.9 & Yes & 32.73 \\
$4.8.10$ & $\triangle(2,4,5)|\mathrm{PGL}(2,19) \times \mathrm{Z}_{2}|1$ & $[[54720, 688, 40]]$ & $1710$ & 123120 & 179.0 & Yes & 20.12 \\
\bottomrule
\end{tabular}%
}
\caption{Modular hyperbolic color codes for side length $L=2$ with octagon modules ($K=8$ base triangles per module) on three-colorable tilings (color code tilings) where at least one of the polygons meeting each vertex in the base tiling is an octagon. $M_8$: number of octagon modules. $N_{M8}$: total physical qubits in the modular layout including both data qubits and ancilla qubits. $N_{M8}/k$: physical-to-logical qubit overhead. Orientable: whether the manifold is orientable. $kd^2/n$: code figure of merit.}
\label{tab:candidates_L2_color_rgb8}
\end{table}

\begin{table}[H]
\centering
\footnotesize
\setlength{\tabcolsep}{3.5pt}
\resizebox{\textwidth}{!}{%
\begin{tabular}{cccrrrlrccr}
\toprule
Tiling & Symmetry Group & $[[n,k,d]]$ & $M_{2}$ & $N_{M2}$ & $N_{M2}/k$ & Automorphism Bases & Cov. & Pair & Cayley & $kd^2/n$ \\
\midrule
$8.8.8$ & $\triangle(4,4,4)_{32}$ & $[[144, 8, 8]]$ & $8$ & 336 & 42.0 & $2/2\,(w{=}2{+}2;\,w_1{=}2{+}4)$ & 100\% & 0.5 & 1.2/2 & 3.56 \\
$12.12.12$ & $\triangle(6,6,6)_{54}$ & $[[216, 16, 8]]$ & $12$ & 504 & 31.5 & $4/4\,(w{=}2{+}3;\,w_1{=}2{+}4)$ & 100\% & 0 & 2.3/3 & 4.74 \\
$8.8.8$ & $\triangle(4,4,4)_{40}$ & $[[288, 12, 12]]$ & $16$ & 672 & 56.0 & $3/3\,(w{=}3;\,w_1{=}4)$ & 100\% & 1.3 & 2.4/4 & 6.00 \\
$8.8.12$ & $\triangle(4,4,6)_{47}$ & $[[432, 20, 12]]$ & $24$ & 1008 & 50.4 & \text{Non-orientable} & --- & --- & --- & 6.67 \\
$10.10.10$ & $\triangle(5,5,5)_{8}$ & $[[450, 24, 12]]$ & $25$ & 1050 & 43.8 & $6/6\,(w{=}3{+}3;\,w_1{=}4{+}6)$ & 100\% & 1.3 & 2.5/4 & 7.68 \\
$4.10.10$ & $\triangle(2,5,5)_{3}$ & $[[540, 10, 14]]$ & $30$ & 1260 & 126.0 & \text{Non-orientable} & --- & --- & --- & 3.63 \\
$12.12.12$ & $\triangle(6,6,6)_{93}$ & $[[648, 40, 12]]$ & $36$ & 1512 & 37.8 & $9/10\,(w{=}3{+}4{+}4;\,w_1{=}4{+}4{+}8) \mid 10/10\,(w{=}3{+}3{+}3{+}4;\,w_1{=}4{+}6{+}6{+}4)$ & 100\% & 1.2 & 3.1/6 & 8.89 \\
$6.8.8$ & $\triangle(3,4,4)_{18}$ & $[[864, 20, 16]]$ & $48$ & 2016 & 100.8 & --- & --- & --- & --- & 5.93 \\
$6.6.8$ & $\triangle(3,3,4)_{5}$ & $[[864, 12, 18]]$ & $48$ & 2016 & 168.0 & --- & --- & --- & --- & 4.50 \\
$14.14.14$ & $\triangle(7,7,7)_{10}$ & $[[882, 60, 12]]$ & $49$ & 2058 & 34.3 & $14/15\,(w{=}5{+}5{+}6;\,w_1{=}6{+}6{+}4)$ & 93\% & 1.5 & 3.6/6 & 9.80 \\
$16.16.16$ & $\triangle(8,8,8)_{326}$ & $[[1152, 84, 12]]$ & $64$ & 2688 & 32.0 & $16/21\,(w{=}3{+}3;\,w_1{=}4{+}6)$ & 76\% & 1.2 & 3.4/6 & 10.50 \\
$8.8.8$ & $\triangle(4,4,4)_{58}$ & $[[1152, 36, 16]]$ & $64$ & 2688 & 74.7 & $8/9\,(w{=}5;\,w_1{=}6) \mid 9/9\,(w{=}4{+}4{+}5;\,w_1{=}4{+}8{+}6)$ & 100\% & 2 & 3.2/6 & 8.00 \\
$6.6.10$ & $\triangle(3,3,5)_{5}$ & $[[1350, 24, 18]]$ & $75$ & 3150 & 131.2 & $6/6\,(w{=}6{+}6;\,w_1{=}6{+}6)$ & 100\% & 1.8 & 4.3/6 & 5.76 \\
$4.8.10$ & $\triangle(2,4,5)_{8}$ & $[[1440, 12, 22]]$ & $80$ & 3360 & 280.0 & \text{Non-orientable} & --- & --- & --- & 4.03 \\
$6.8.8$ & $\triangle(3,4,4)_{22}$ & $[[1728, 36, 20]]$ & $96$ & 4032 & 112.0 & $8/9\,(w{=}8;\,w_1{=}10) \mid 9/9\,(w{=}6{+}6{+}8;\,w_1{=}6{+}8{+}8)$ & 100\% & 2.1 & 3.8/7 & 8.33 \\
$6.6.8$ & $\triangle(3,3,4)_{7}$ & $[[1728, 20, 24]]$ & $96$ & 4032 & 201.6 & --- & --- & --- & --- & 6.67 \\
$6.10.10$ & $\triangle(3,5,5)_{7}$ & $[[2160, 68, 18]]$ & $120$ & 5040 & 74.1 & $15/17\,(w{=}5{+}6{+}6{+}6;\,w_1{=}6{+}6{+}6{+}6)$ & 88\% & 2.4 & 3.8/6 & 10.20 \\
$10.10.10$ & $\triangle(5,5,5)_{14}$ & $[[2250, 104, 16]]$ & $125$ & 5250 & 50.5 & $17/26\,(w{=}4{+}4;\,w_1{=}6{+}6)$ & 65\% & 2.1 & 3.6/6 & 11.83 \\
$8.12.12$ & $\triangle(4,6,6)_{163}$ & $[[3024, 144, 16]]$ & $168$ & 7056 & 49.0 & \text{Non-orientable} & --- & --- & --- & 12.19 \\
$8.8.8$ & $\triangle(4,4,4)_{96}$ & $[[3024, 88, 20]]$ & $168$ & 7056 & 80.2 & $17/22\,(w{=}6{+}6{+}6;\,w_1{=}6{+}6{+}8)$ & 77\% & 2.3 & 4.1/6 & 11.64 \\
$8.10.10$ & $\triangle(4,5,5)_{10}$ & $[[3240, 130, 18]]$ & $180$ & 7560 & 58.2 & \text{Non-orientable} & --- & --- & --- & 13.00 \\
$6.6.8$ & $\triangle(3,3,4)_{10}$ & $[[3240, 34, 30]]$ & $180$ & 7560 & 222.4 & \text{Non-orientable} & --- & --- & --- & 9.44 \\
$6.8.8$ & $\triangle(3,4,4)_{32}$ & $[[3456, 68, 24]]$ & $192$ & 8064 & 118.6 & $14/17\,(w{=}6{+}8;\,w_1{=}8{+}8)$ & 82\% & 3 & 4.9/7 & 11.33 \\
$8.8.8$ & $\triangle(4,4,4)_{109}$ & $[[3600, 104, 22]]$ & $200$ & 8400 & 80.8 & $25/26\,(w{=}8;\,w_1{=}10)$ & 96\% & 1 & 4.4/6 & 13.98 \\
$12.12.12$ & $\triangle(6,6,6)_{445}$ & $[[3888, 220, 16]]$ & $216$ & 9072 & 41.2 & $24/55\,(w{=}5{+}6;\,w_1{=}6{+}8)$ & 44\% & 2.5 & 4.1/6 & 14.49 \\
$8.8.8$ & $\triangle(4,4,4)_{120}$ & $[[4608, 132, 24]]$ & $256$ & 10752 & 81.5 & $21/33\,(w{=}6{+}7;\,w_1{=}8{+}8)$ & 64\% & 2.4 & 4.9/8 & 16.50 \\
$8.8.10$ & $\triangle(4,4,5)_{47}$ & $[[8640, 292, 24]]$ & $480$ & 20160 & 69.0 & $32/73\,(w{=}8{+}8{+}9;\,w_1{=}8{+}10{+}12)$ & 44\% & 3.8 & 5.4/8 & 19.47 \\
$6.8.8$ & $\triangle(3,4,4)_{45}$ & $[[8640, 164, 28]]$ & $480$ & 20160 & 122.9 & $25/41\,(w{=}8{+}10;\,w_1{=}10{+}12)$ & 61\% & 3.8 & 5.9/10 & 14.88 \\
$6.6.8$ & $\triangle(3,3,4)_{16}$ & $[[9072, 88, 36]]$ & $504$ & 21168 & 240.5 & $18/22\,(w{=}10{+}12{+}14;\,w_1{=}12{+}12{+}16)$ & 82\% & 3.6 & 6.9/12 & 12.57 \\
$8.8.12$ & $\triangle(4,4,6)_{340}$ & $[[11880, 444, 24]]$ & $660$ & 27720 & 62.4 & \text{Non-orientable} & --- & --- & --- & 21.53 \\
$6.8.8$ & $\triangle(3,4,4)_{52}$ & $[[11880, 224, 30]]$ & $660$ & 27720 & 123.8 & \text{Non-orientable} & --- & --- & --- & 16.97 \\
$8.8.8$ & $\triangle(4,4,4)_{271}$ & $[[12096, 340, 28]]$ & $672$ & 28224 & 83.0 & $35/85\,(w{=}8{+}8;\,w_1{=}10{+}10)$ & 41\% & 4.1 & 6.0/10 & 22.04 \\
$8.10.10$ & $\triangle(4,5,5)_{31}$ & $[[17280, 676, 24]]$ & $960$ & 40320 & 59.6 & \text{Non-orientable} & --- & --- & --- & 22.53 \\
$10.10.10$ & $\triangle(5,5,5)_{75}$ & $[[19440, 868, 24]]$ & $1080$ & 45360 & 52.3 & --- & --- & --- & --- & 25.72 \\
$4.6.14$ & $\triangle(2,6,7)_{21}$ & $[[19656, 56, 42]]$ & $1092$ & 45864 & 819.0 & $13/14\,(w{=}16;\,w_1{=}16)$ & 93\% & 8.2 & 14.7/24 & 5.03 \\
$6.8.8$ & $\triangle(3,4,4)_{79}$ & $[[23328, 436, 32]]$ & $1296$ & 54432 & 124.8 & $50/109\,(w{=}8{+}12;\,w_1{=}10{+}12)$ & 46\% & 4.7 & 7.3/11 & 19.14 \\
$6.6.10$ & $\triangle(3,3,5)_{20}$ & $[[23760, 356, 36]]$ & $1320$ & 55440 & 155.7 & $44/89\,(w{=}10{+}10;\,w_1{=}12{+}12)$ & 49\% & 5.2 & 7.9/12 & 19.42 \\
$4.8.10$ & $\triangle(2,4,5)|\mathrm{PGL}(2,11) \times \mathrm{Z}_{2}$ & $[[23760, 136, 44]]$ & $1320$ & 55440 & 407.6 & $24/34\,(w{=}14{+}16;\,w_1{=}14{+}16)$ & 71\% & 8.1 & 12.1/18 & 11.08 \\
$6.8.8$ & $\triangle(3,4,4)|\mathrm{PGL}(2,17)$ & $[[44064, 820, 36]]$ & $2448$ & 102816 & 125.4 & --- & --- & --- & --- & 24.12 \\
$6.6.8$ & $\triangle(3,3,4)|\mathrm{PGL}(2,17)|1$ & $[[44064, 412, 48]]$ & $2448$ & 102816 & 249.6 & $44/103\,(w{=}14;\,w_1{=}16)$ & 43\% & 7.2 & 9.5/15 & 21.54 \\
$8.8.8$ & $\triangle(4,4,4)|\mathrm{PSL}(2,23)$ & $[[54648, 1522, 32]]$ & $3036$ & 127512 & 83.8 & \text{Non-orientable} & --- & --- & --- & 28.52 \\
$8.8.8$ & $\triangle(4,4,4)|\mathrm{PGL}(2,23)$ & $[[109296, 3040, 36]]$ & $6072$ & 255024 & 83.9 & --- & --- & --- & --- & 36.05 \\
$6.8.8$ & $\triangle(3,4,4)|\mathrm{PGL}(2,23)$ & $[[109296, 2028, 40]]$ & $6072$ & 255024 & 125.8 & --- & --- & --- & --- & 29.69 \\
$4.8.10$ & $\triangle(2,4,5)|\mathrm{PGL}(2,19) \times \mathrm{Z}_{2}|1$ & $[[123120, 688, 56]]$ & $6840$ & 287280 & 417.6 & --- & --- & --- & --- & 17.52 \\
$6.6.10$ & $\triangle(3,3,5)|\mathrm{PSL}(2,31)$ & $[[133920, 1988, 42]]$ & $7440$ & 312480 & 157.2 & \text{Non-orientable} & --- & --- & --- & 26.19 \\
$4.6.14$ & $\triangle(2,3,7)|\mathrm{PGL}(2,27)$ & $[[176904, 472, 60]]$ & $9828$ & 412776 & 874.5 & --- & --- & --- & --- & 9.61 \\
\bottomrule
\end{tabular}%
}
\caption{Modular hyperbolic color codes for side length $L=3$ with quadrilateral modules ($K=2$ base triangles per module). $M_2$: number of quadrilateral modules. $N_{M2}$: total physical qubits in the modular layout including both data qubits and ancilla qubits. $N_{M2}/k$: physical-to-logical qubit overhead. Automorphism Bases: logical basis operator coverage $k_{\mathrm{found}}/g$ and generator seed weights $w$ (length of the shortest homologous dual cycle found for each seed) and $w_1$ (the same, restricted to dual edges of a single colour, with one colour per basis chosen to minimise the total), showing Pareto-optimal bases with $\ge 85\%$ coverage or the single highest-coverage basis if $<85\%$. Cov.: maximum basis coverage percentage. Pair: mean number of automorphism moves between measuring the two cycles of a symplectic pair with a fixed extractor. Cayley: mean and maximum of the same over all pairs of basis loops. Both are right Cayley graph distances among basis automorphisms, with the extractor placed to minimise Pair (then the maximum, then the mean), for the listed basis with the highest coverage. $kd^2/n$: code figure of merit.}
\label{tab:candidates_L3_color_ms2}
\end{table}

\begin{table}[H]
\centering
\footnotesize
\setlength{\tabcolsep}{3.5pt}
\resizebox{\textwidth}{!}{%
\begin{tabular}{cccrrrlrccr}
\toprule
Tiling & Symmetry Group & $[[n,k,d]]$ & $M_{2}$ & $N_{M2}$ & $N_{M2}/k$ & Automorphism Bases & Cov. & Pair & Cayley & $kd^2/n$ \\
\midrule
$8.8.8$ & $\triangle(4,4,4)_{32}$ & $[[256, 8, 12]]$ & $8$ & 576 & 72.0 & $2/2\,(w{=}2{+}2;\,w_1{=}2{+}4)$ & 100\% & 0.5 & 1.2/2 & 4.50 \\
$12.12.12$ & $\triangle(6,6,6)_{54}$ & $[[384, 16, 12]]$ & $12$ & 864 & 54.0 & $4/4\,(w{=}2{+}3;\,w_1{=}2{+}4)$ & 100\% & 0 & 2.3/3 & 6.00 \\
$8.8.8$ & $\triangle(4,4,4)_{40}$ & $[[512, 12, 16]]$ & $16$ & 1152 & 96.0 & $3/3\,(w{=}3;\,w_1{=}4)$ & 100\% & 1.3 & 2.4/4 & 6.00 \\
$12.12.12$ & $\triangle(6,6,6)_{88}$ & $[[768, 28, 14]]$ & $24$ & 1728 & 61.7 & $6/7\,(w{=}2{+}4{+}4;\,w_1{=}2{+}4{+}6) \mid 7/7\,(w{=}2{+}4{+}4{+}4;\,w_1{=}2{+}4{+}6{+}6)$ & 100\% & 0.7 & 1.7/3 & 7.15 \\
$8.8.12$ & $\triangle(4,4,6)_{47}$ & $[[768, 20, 16]]$ & $24$ & 1728 & 86.4 & \text{Non-orientable} & --- & --- & --- & 6.67 \\
$10.10.10$ & $\triangle(5,5,5)_{8}$ & $[[800, 24, 16]]$ & $25$ & 1800 & 75.0 & $6/6\,(w{=}3{+}3;\,w_1{=}4{+}6)$ & 100\% & 1.3 & 2.5/4 & 7.68 \\
$4.10.10$ & $\triangle(2,5,5)_{3}$ & $[[960, 10, 20]]$ & $30$ & 2160 & 216.0 & \text{Non-orientable} & --- & --- & --- & 4.17 \\
$8.16.16$ & $\triangle(4,8,8)_{93}$ & $[[1024, 36, 16]]$ & $32$ & 2304 & 64.0 & $8/9\,(w{=}2{+}3;\,w_1{=}2{+}4) \mid 9/9\,(w{=}2{+}3{+}4;\,w_1{=}2{+}4{+}4)$ & 100\% & 1.2 & 3.0/6 & 9.00 \\
$12.12.12$ & $\triangle(6,6,6)_{93}$ & $[[1152, 40, 16]]$ & $36$ & 2592 & 64.8 & $9/10\,(w{=}3{+}4{+}4;\,w_1{=}4{+}4{+}8) \mid 10/10\,(w{=}3{+}3{+}3{+}4;\,w_1{=}4{+}6{+}6{+}4)$ & 100\% & 1.2 & 3.1/6 & 8.89 \\
$6.8.8$ & $\triangle(3,4,4)_{19}$ & $[[1536, 20, 24]]$ & $48$ & 3456 & 172.8 & \text{Non-orientable} & --- & --- & --- & 7.50 \\
$14.14.14$ & $\triangle(7,7,7)_{10}$ & $[[1568, 60, 16]]$ & $49$ & 3528 & 58.8 & $14/15\,(w{=}5{+}5{+}6;\,w_1{=}6{+}6{+}4)$ & 93\% & 1.5 & 3.6/6 & 9.80 \\
$16.16.16$ & $\triangle(8,8,8)_{326}$ & $[[2048, 84, 16]]$ & $64$ & 4608 & 54.9 & $16/21\,(w{=}3{+}3;\,w_1{=}4{+}6)$ & 76\% & 1.2 & 3.4/6 & 10.50 \\
$8.8.8$ & $\triangle(4,4,4)_{58}$ & $[[2048, 36, 20\text{--}24]]$ & $64$ & 4608 & 128.0 & $8/9\,(w{=}5;\,w_1{=}6) \mid 9/9\,(w{=}4{+}4{+}5;\,w_1{=}4{+}8{+}6)$ & 100\% & 2 & 3.2/6 & 7.03--10.12 \\
$4.8.10$ & $\triangle(2,4,5)_{8}$ & $[[2560, 12, 32]]$ & $80$ & 5760 & 480.0 & \text{Non-orientable} & --- & --- & --- & 4.80 \\
$6.8.8$ & $\triangle(3,4,4)_{22}$ & $[[3072, 36, 28]]$ & $96$ & 6912 & 192.0 & $8/9\,(w{=}8;\,w_1{=}10) \mid 9/9\,(w{=}6{+}6{+}8;\,w_1{=}6{+}8{+}8)$ & 100\% & 2.1 & 3.8/7 & 9.19 \\
$6.6.8$ & $\triangle(3,3,4)_{7}$ & $[[3072, 20, 32]]$ & $96$ & 6912 & 345.6 & --- & --- & --- & --- & 6.67 \\
$6.10.10$ & $\triangle(3,5,5)_{7}$ & $[[3840, 68, 24]]$ & $120$ & 8640 & 127.1 & $15/17\,(w{=}5{+}6{+}6{+}6;\,w_1{=}6{+}6{+}6{+}6)$ & 88\% & 2.4 & 3.8/6 & 10.20 \\
$10.10.10$ & $\triangle(5,5,5)_{14}$ & $[[4000, 104, 22]]$ & $125$ & 9000 & 86.5 & $17/26\,(w{=}4{+}4;\,w_1{=}6{+}6)$ & 65\% & 2.1 & 3.6/6 & 12.58 \\
$6.12.12$ & $\triangle(3,6,6)_{64}$ & $[[4608, 100, 24]]$ & $144$ & 10368 & 103.7 & $17/25\,(w{=}5{+}6{+}7;\,w_1{=}6{+}6{+}8)$ & 68\% & 2.2 & 4.2/6 & 12.50 \\
$8.12.12$ & $\triangle(4,6,6)_{163}$ & $[[5376, 144, 22]]$ & $168$ & 12096 & 84.0 & \text{Non-orientable} & --- & --- & --- & 12.96 \\
$8.8.8$ & $\triangle(4,4,4)_{96}$ & $[[5376, 88, 26\text{--}28]]$ & $168$ & 12096 & 137.5 & $17/22\,(w{=}6{+}6{+}6;\,w_1{=}6{+}6{+}8)$ & 77\% & 2.3 & 4.1/6 & 11.07--12.83 \\
$8.10.10$ & $\triangle(4,5,5)_{10}$ & $[[5760, 130, 24]]$ & $180$ & 12960 & 99.7 & \text{Non-orientable} & --- & --- & --- & 13.00 \\
$6.6.8$ & $\triangle(3,3,4)_{10}$ & $[[5760, 34, 40]]$ & $180$ & 12960 & 381.2 & \text{Non-orientable} & --- & --- & --- & 9.44 \\
$6.8.8$ & $\triangle(3,4,4)_{32}$ & $[[6144, 68, 32]]$ & $192$ & 13824 & 203.3 & $14/17\,(w{=}6{+}8;\,w_1{=}8{+}8)$ & 82\% & 3 & 4.9/7 & 11.33 \\
$8.8.8$ & $\triangle(4,4,4)_{109}$ & $[[6400, 104, 30]]$ & $200$ & 14400 & 138.5 & $25/26\,(w{=}8;\,w_1{=}10)$ & 96\% & 1 & 4.4/6 & 14.62 \\
$12.12.12$ & $\triangle(6,6,6)_{445}$ & $[[6912, 220, 20\text{--}24]]$ & $216$ & 15552 & 70.7 & $24/55\,(w{=}5{+}6;\,w_1{=}6{+}8)$ & 44\% & 2.5 & 4.1/6 & 12.73--18.33 \\
$8.8.8$ & $\triangle(4,4,4)_{120}$ & $[[8192, 132, 32]]$ & $256$ & 18432 & 139.6 & $21/33\,(w{=}6{+}7;\,w_1{=}8{+}8)$ & 64\% & 2.4 & 4.9/8 & 16.50 \\
$14.14.14$ & $\triangle(7,7,7)_{16}$ & $[[10976, 396, 22\text{--}24]]$ & $343$ & 24696 & 62.4 & $36/99\,(w{=}4{+}5;\,w_1{=}6{+}8)$ & 36\% & 2.4 & 4.7/8 & 17.46--20.78 \\
$8.8.12$ & $\triangle(4,4,6)_{211}$ & $[[12288, 260, 26\text{--}28]]$ & $384$ & 27648 & 106.3 & $32/65\,(w{=}6{+}6;\,w_1{=}6{+}8)$ & 49\% & 3.2 & 5.3/8 & 14.30--16.59 \\
$8.8.10$ & $\triangle(4,4,5)_{47}$ & $[[15360, 292, 32]]$ & $480$ & 34560 & 118.4 & $32/73\,(w{=}8{+}8{+}9;\,w_1{=}8{+}10{+}12)$ & 44\% & 3.8 & 5.4/8 & 19.47 \\
$6.8.8$ & $\triangle(3,4,4)_{45}$ & $[[15360, 164, 38]]$ & $480$ & 34560 & 210.7 & $25/41\,(w{=}8{+}10;\,w_1{=}10{+}12)$ & 61\% & 3.8 & 5.9/10 & 15.42 \\
$6.6.8$ & $\triangle(3,3,4)_{16}$ & $[[16128, 88, 48]]$ & $504$ & 36288 & 412.4 & $18/22\,(w{=}10{+}12{+}14;\,w_1{=}12{+}12{+}16)$ & 82\% & 3.6 & 6.9/12 & 12.57 \\
$10.10.10$ & $\triangle(5,5,5)_{50}$ & $[[20000, 504, 28]]$ & $625$ & 45000 & 89.3 & $46/126\,(w{=}6{+}6;\,w_1{=}8{+}8)$ & 37\% & 3.1 & 5.4/9 & 19.76 \\
$8.8.12$ & $\triangle(4,4,6)_{340}$ & $[[21120, 444, 30\text{--}32]]$ & $660$ & 47520 & 107.0 & \text{Non-orientable} & --- & --- & --- & 18.92--21.53 \\
$6.8.8$ & $\triangle(3,4,4)_{52}$ & $[[21120, 224, 40]]$ & $660$ & 47520 & 212.1 & \text{Non-orientable} & --- & --- & --- & 16.97 \\
$4.10.10$ & $\triangle(4,5,5)_{25}$ & $[[21120, 136, 46]]$ & $660$ & 47520 & 349.4 & $24/34\,(w{=}10{+}10{+}14;\,w_1{=}8{+}10{+}14)$ & 71\% & 5 & 8.4/12 & 13.63 \\
$8.8.8$ & $\triangle(4,4,4)_{271}$ & $[[21504, 340, 36\text{--}38]]$ & $672$ & 48384 & 142.3 & $35/85\,(w{=}8{+}8;\,w_1{=}10{+}10)$ & 41\% & 4.1 & 6.0/10 & 20.49--22.83 \\
$10.12.12$ & $\triangle(5,6,6)_{130}$ & $[[23040, 676, 24\text{--}28]]$ & $720$ & 51840 & 76.7 & $49/169\,(w{=}6{+}6;\,w_1{=}8{+}8)$ & 29\% & 3.5 & 5.3/8 & 16.90--23.00 \\
$10.10.10$ & $\triangle(5,5,5)_{75}$ & $[[34560, 868, 32]]$ & $1080$ & 77760 & 89.6 & --- & --- & --- & --- & 25.72 \\
$12.12.12$ & $\triangle(6,6,6)_{2819}$ & $[[34944, 1096, 26\text{--}30]]$ & $1092$ & 78624 & 71.7 & --- & --- & --- & --- & 21.20--28.23 \\
$4.6.14$ & $\triangle(2,6,7)_{21}$ & $[[34944, 56, 56]]$ & $1092$ & 78624 & 1404.0 & $13/14\,(w{=}16;\,w_1{=}16)$ & 93\% & 8.2 & 14.7/24 & 5.03 \\
$6.8.8$ & $\triangle(3,4,4)_{79}$ & $[[41472, 436, 44]]$ & $1296$ & 93312 & 214.0 & $50/109\,(w{=}8{+}12;\,w_1{=}10{+}12)$ & 46\% & 4.7 & 7.3/11 & 20.35 \\
$6.6.10$ & $\triangle(3,3,5)_{20}$ & $[[42240, 356, 48]]$ & $1320$ & 95040 & 267.0 & $44/89\,(w{=}10{+}10;\,w_1{=}12{+}12)$ & 49\% & 5.2 & 7.9/12 & 19.42 \\
$4.8.10$ & $\triangle(2,4,5)|\mathrm{PGL}(2,11) \times \mathrm{Z}_{2}$ & $[[42240, 136, 64]]$ & $1320$ & 95040 & 698.8 & $24/34\,(w{=}14{+}16;\,w_1{=}14{+}16)$ & 71\% & 8.1 & 12.1/18 & 13.19 \\
$8.8.8$ & $\triangle(4,4,4)_{786}$ & $[[65536, 1028, 40]]$ & $2048$ & 147456 & 143.4 & --- & --- & --- & --- & 25.10 \\
$16.16.16$ & $\triangle(8,8,8)|\mathrm{PSL}(2,17) \times \mathrm{Z}_{2}|1$ & $[[78336, 3064, 24\text{--}28]]$ & $2448$ & 176256 & 57.5 & --- & --- & --- & --- & 22.53--30.67 \\
$8.12.12$ & $\triangle(4,6,6)|\mathrm{PGL}(2,17)|1$ & $[[78336, 2044, 32]]$ & $2448$ & 176256 & 86.2 & \text{Non-orientable} & --- & --- & --- & 26.72 \\
$6.8.8$ & $\triangle(3,4,4)|\mathrm{PGL}(2,17)$ & $[[78336, 820, 48]]$ & $2448$ & 176256 & 214.9 & --- & --- & --- & --- & 24.12 \\
$6.6.8$ & $\triangle(3,3,4)|\mathrm{PGL}(2,17)|1$ & $[[78336, 412, 64]]$ & $2448$ & 176256 & 427.8 & $44/103\,(w{=}14;\,w_1{=}16)$ & 43\% & 7.2 & 9.5/15 & 21.54 \\
$10.12.12$ & $\triangle(5,6,6)|\mathrm{S}_{7}|1$ & $[[80640, 2356, 30\text{--}32]]$ & $2520$ & 181440 & 77.0 & \text{Non-orientable} & --- & --- & --- & 26.29--29.92 \\
$12.12.12$ & $\triangle(6,6,6)|\mathrm{PSL}(2,23)$ & $[[97152, 3040, 26\text{--}32]]$ & $3036$ & 218592 & 71.9 & \text{Non-orientable} & --- & --- & --- & 21.15--32.04 \\
$8.8.8$ & $\triangle(4,4,4)|\mathrm{PSL}(2,23)$ & $[[97152, 1522, 42\text{--}44]]$ & $3036$ & 218592 & 143.6 & \text{Non-orientable} & --- & --- & --- & 27.64--30.33 \\
$4.8.10$ & $\triangle(2,4,5)|\mathrm{PGL}(2,19)|1$ & $[[109440, 346, 72]]$ & $3420$ & 246240 & 711.7 & \text{Non-orientable} & --- & --- & --- & 16.39 \\
$8.8.8$ & $\triangle(4,4,4)|\mathrm{PSL}(2,25)$ & $[[124800, 1954, 42\text{--}46]]$ & $3900$ & 280800 & 143.7 & \text{Non-orientable} & --- & --- & --- & 27.62--33.13 \\
$12.16.16$ & $\triangle(6,8,8)|\mathrm{PGL}(2,23)|3$ & $[[194304, 7088, 30\text{--}32]]$ & $6072$ & 437184 & 61.7 & \text{Non-orientable} & --- & --- & --- & 32.83--37.35 \\
$8.8.8$ & $\triangle(4,4,4)|\mathrm{PGL}(2,23)$ & $[[194304, 3040, 48]]$ & $6072$ & 437184 & 143.8 & --- & --- & --- & --- & 36.05 \\
$6.8.8$ & $\triangle(3,4,4)|\mathrm{PGL}(2,23)$ & $[[194304, 2028, 56]]$ & $6072$ & 437184 & 215.6 & --- & --- & --- & --- & 32.73 \\
$14.14.14$ & $\triangle(7,7,7)|\mathrm{PSL}(2,29)|11$ & $[[194880, 6964, 32]]$ & $6090$ & 438480 & 63.0 & \text{Non-orientable} & --- & --- & --- & 36.59 \\
$4.8.10$ & $\triangle(2,4,5)|\mathrm{PGL}(2,19) \times \mathrm{Z}_{2}|1$ & $[[218880, 688, 80]]$ & $6840$ & 492480 & 715.8 & --- & --- & --- & --- & 20.12 \\
$16.16.16$ & $\triangle(8,8,8)|\mathrm{PSL}(2,31)|1$ & $[[238080, 9304, 20\text{--}32]]$ & $7440$ & 535680 & 57.6 & \text{Non-orientable} & --- & --- & --- & 15.63--40.02 \\
$4.10.10$ & $\triangle(2,5,5)|\mathrm{PSL}(2,31)$ & $[[238080, 1492, 60]]$ & $7440$ & 535680 & 359.0 & \text{Non-orientable} & --- & --- & --- & 22.56 \\
\bottomrule
\end{tabular}%
}
\caption{Modular hyperbolic color codes for side length $L=4$ with quadrilateral modules ($K=2$ base triangles per module). $M_2$: number of quadrilateral modules. $N_{M2}$: total physical qubits in the modular layout including both data qubits and ancilla qubits. $N_{M2}/k$: physical-to-logical qubit overhead. Automorphism Bases: logical basis operator coverage $k_{\mathrm{found}}/g$ and generator seed weights $w$ (length of the shortest homologous dual cycle found for each seed) and $w_1$ (the same, restricted to dual edges of a single colour, with one colour per basis chosen to minimise the total), showing Pareto-optimal bases with $\ge 85\%$ coverage or the single highest-coverage basis if $<85\%$. Cov.: maximum basis coverage percentage. Pair: mean number of automorphism moves between measuring the two cycles of a symplectic pair with a fixed extractor. Cayley: mean and maximum of the same over all pairs of basis loops. Both are right Cayley graph distances among basis automorphisms, with the extractor placed to minimise Pair (then the maximum, then the mean), for the listed basis with the highest coverage. $kd^2/n$: code figure of merit.}
\label{tab:candidates_L4_color_ms2}
\end{table}

\begin{table}[H]
\centering
\footnotesize
\setlength{\tabcolsep}{3.5pt}
\resizebox{\textwidth}{!}{%
\begin{tabular}{cccrrrclrccr}
\toprule
Tiling & Symmetry Group & $[[n,k,d]]$ & $M_{2}$ & $N_{M2}$ & $N_{M2}/k$ & $G_{\mathrm{rot}}$ & Automorphism Bases & Cov. & Pair & Cayley & $kd^2/n$ \\
\midrule
\{4,6\} & $\triangle(2,4,6)_{25}$ & $[[150, 4, 10]]$ & $3$ & 348 & 87.0 & 3/3 & $2/2\,(w{=}8) \mid 2/2\,(w{=}2{+}2)$ & 100\% & 0.5 & 0.8/1 & 2.67 \\
\{4,8\} & $\triangle(2,4,8)_{30}$ & $[[200, 6, 10]]$ & $4$ & 464 & 77.3 & 4/4 & $3/3\,(w{=}2{+}2)$ & 100\% & 0.3 & 1.1/2 & 3.00 \\
\{4,10\} & $\triangle(2,4,10)_{24}$ & $[[250, 8, 10]]$ & $5$ & 580 & 72.5 & 5/5 & $4/4\,(w{=}2{+}2)$ & 100\% & --- & 0.7/1 & 3.20 \\
\{4,12\} & $\triangle(2,4,12)_{43}$ & $[[300, 10, 10]]$ & $6$ & 696 & 69.6 & 6/6 & $5/5\,(w{=}2{+}2)$ & 100\% & 0.2 & 1.6/3 & 3.33 \\
\{4,5\} & $\triangle(2,4,5)_{7}$ & $[[375, 5, 15]]$ & $7\,({+}1)$ & 873 & 174.6 & --- & \text{Non-orientable} & --- & --- & --- & 3.00 \\
$V4.4.4.6$ & $\square(2,2,2,3)_{32}$ & $[[600, 4, 20]]$ & $12$ & 1392 & 348.0 & 12/12 & $2/2\,(w{=}4)$ & 100\% & 2 & 2.0/3 & 2.67 \\
\{4,6\} & $\triangle(2,4,6)_{35}$ & $[[675, 11, 15]]$ & $13\,({+}1)$ & 1569 & 142.6 & --- & \text{Non-orientable} & --- & --- & --- & 3.67 \\
$V4.4.8.8$ & $\square(2,2,4,4)_{303}$ & $[[800, 10, 16]]$ & $16$ & 1856 & 185.6 & 16/16 & $5/5\,(w{=}4{+}6)$ & 100\% & 1.6 & 2.0/4 & 3.20 \\
$V4.4.4.8$ & $\square(2,2,2,4)_{212}$ & $[[800, 6, 20]]$ & $16$ & 1856 & 309.3 & 8/16 & $3/3\,(w{=}4{+}4)$ & 100\% & 1 & 1.7/3 & 3.00 \\
$V4.4.6.6$ & $\square(2,2,3,3)_{23}$ & $[[900, 8, 20]]$ & $18$ & 2088 & 261.0 & 9/18 & $4/4\,(w{=}4{+}4)$ & 100\% & 0.5 & 1.3/2 & 3.56 \\
$V4.6.4.6$ & $\square(2,3,2,3)_{14}$ & $[[900, 8, 20]]$ & $18$ & 2088 & 261.0 & 18/18 & $4/4\,(w{=}4{+}8)$ & 100\% & 0.5 & 1.6/3 & 3.56 \\
\{4,5\} & $\triangle(2,4,5)_{10}$ & $[[1000, 10, 20]]$ & $20$ & 2320 & 232.0 & --- & $5/5\,(w{=}16{+}30)$ & 100\% & --- & --- & 4.00 \\
$V6.6.6.6$ & $\square(3,3,3,3)_{43}$ & $[[1200, 18, 20]]$ & $24$ & 2784 & 154.7 & 24/24 & $8/9\,(w{=}6{+}8)$ & 89\% & 1.5 & 2.3/4 & 6.00 \\
\{4,6\} & $\triangle(2,4,6)_{40}$ & $[[1200, 18, 20]]$ & $24$ & 2784 & 154.7 & 24/24 & $8/9\,(w{=}30) \mid 9/9\,(w{=}6{+}8)$ & 100\% & 0.8 & 2.3/4 & 6.00 \\
\{4,5\} & $\triangle(2,4,5)_{11}$ & $[[2000, 18, 21]]$ & $40$ & 4640 & 257.8 & --- & --- & --- & --- & --- & 3.97 \\
$V6.8.6.8$ & $\square(3,4,3,4)_{73}$ & $[[2400, 42, 20]]$ & $48$ & 5568 & 132.6 & 48/48 & $16/21\,(w{=}4{+}4{+}4{+}8)$ & 76\% & 1.3 & 3.0/5 & 7.00 \\
$V6.6.8.8$ & $\square(3,3,4,4)_{152}$ & $[[2400, 42, 20]]$ & $48$ & 5568 & 132.6 & 48/48 & $16/21\,(w{=}4{+}8)$ & 76\% & 2 & 3.1/4 & 7.00 \\
$V4.4.4.6$ & $\square(2,2,2,3)_{66}$ & $[[2400, 10, 30]]$ & $48$ & 5568 & 556.8 & --- & \text{Non-orientable} & --- & --- & --- & 3.75 \\
$V6.8.10.8$ & $\square(3,4,5,4)_{32}$ & $[[3000, 60, 20]]$ & $60$ & 6960 & 116.0 & --- & \text{Non-orientable} & --- & --- & --- & 8.00 \\
\{4,8\} & $\triangle(2,4,8)_{71}$ & $[[3200, 66, 20]]$ & $64$ & 7424 & 112.5 & 32/64 & $23/33\,(w{=}22{+}26)$ & 70\% & 2.1 & 2.5/4 & 8.25 \\
$V8.8.8.8$ & $\square(4,4,4,4)_{10023}$ & $[[3200, 66, 20]]$ & $64$ & 7424 & 112.5 & 64/64 & $24/33\,(w{=}4{+}4{+}6)$ & 73\% & 1.2 & 3.1/6 & 8.25 \\
\{4,5\} & $\triangle(2,4,5)_{13}$ & $[[3750, 32, 30]]$ & $75$ & 8700 & 271.9 & --- & $14/16\,(w{=}24{+}38)$ & 88\% & --- & --- & 7.68 \\
$V8.8.10.10$ & $\square(4,4,5,5)_{196}$ & $[[4000, 90, 20]]$ & $80$ & 9280 & 103.1 & --- & \text{Non-orientable} & --- & --- & --- & 9.00 \\
$V4.4.4.6$ & $\square(2,2,2,3)_{103}$ & $[[4800, 18, 40]]$ & $96$ & 11136 & 618.7 & 96/96 & $8/9\,(w{=}16) \mid 9/9\,(w{=}8{+}16{+}16)$ & 100\% & 1.4 & 3.7/6 & 6.00 \\
\{4,10\} & $\triangle(2,4,10)_{60}$ & $[[6000, 146, 20]]$ & $120$ & 13920 & 95.3 & 120/120 & $36/73\,(w{=}28{+}28)$ & 49\% & 3.3 & 3.7/6 & 9.73 \\
\{4,7\} & $\triangle(2,4,7)_{14}$ & $[[6825, 119, 21]]$ & $136\,({+}1)$ & 15837 & 133.1 & --- & \text{Non-orientable} & --- & --- & --- & 7.69 \\
\{4,6\} & $\triangle(2,4,6)_{73}$ & $[[8250, 112, 25]]$ & $165$ & 19140 & 170.9 & --- & \text{Non-orientable} & --- & --- & --- & 8.48 \\
$V4.4.8.8$ & $\square(2,2,4,4)|\mathrm{PSL}(2,7) \times \mathrm{Z}_{2}|2$ & $[[8400, 86, 30]]$ & $168$ & 19488 & 226.6 & 168/168 & $21/43\,(w{=}8{+}8)$ & 49\% & 5.2 & 6.5/10 & 9.21 \\
$V4.6.6.6$ & $\square(2,3,3,3)_{74}$ & $[[9000, 92, 30]]$ & $180$ & 20880 & 227.0 & 180/180 & $25/46\,(w{=}8)$ & 54\% & 1.9 & 4.5/7 & 9.20 \\
\{4,12\} & $\triangle(2,4,12)_{211}$ & $[[9600, 258, 20]]$ & $192$ & 22272 & 86.3 & 96/192 & $53/129\,(w{=}28)$ & 41\% & 3.8 & 4.6/8 & 10.75 \\
\{4,6\} & $\triangle(2,4,6)_{94}$ & $[[12150, 164, 30]]$ & $243$ & 28188 & 171.9 & 243/243 & $45/82\,(w{=}8{+}12)$ & 55\% & --- & --- & 12.15 \\
$V6.6.6.6$ & $\square(3,3,3,3)_{950}$ & $[[12150, 164, 30]]$ & $243$ & 28188 & 171.9 & 243/243 & $28/82\,(w{=}8{+}10)$ & 34\% & 3.7 & 4.9/8 & 12.15 \\
$V4.4.6.6$ & $\square(2,2,3,3)_{316}$ & $[[14400, 98, 32]]$ & $288$ & 33408 & 340.9 & 144/288 & $22/49\,(w{=}10{+}12)$ & 45\% & 2.8 & 4.2/6 & 6.97 \\
$V4.6.4.6$ & $\square(2,3,2,3)_{173}$ & $[[18000, 122, 40]]$ & $360$ & 41760 & 342.3 & 360/360 & $31/61\,(w{=}10{+}16)$ & 51\% & 4.7 & 6.2/10 & 10.84 \\
\{4,5\} & $\triangle(2,4,10)_{120}$ & $[[22500, 182, 40]]$ & $450$ & 52200 & 286.8 & --- & --- & --- & --- & --- & 12.94 \\
$V6.6.6.8$ & $\square(3,3,3,4)_{533}$ & $[[25200, 380, 30]]$ & $504$ & 58464 & 153.9 & 504/504 & $59/190\,(w{=}8{+}10)$ & 31\% & 4.6 & 6.6/11 & 13.57 \\
$V4.4.6.8$ & $\square(2,2,3,4)_{1726}$ & $[[28800, 242, 36]]$ & $576$ & 66816 & 276.1 & 576/576 & $45/121\,(w{=}8{+}16)$ & 37\% & 3.9 & 5.7/9 & 10.89 \\
$V4.4.6.8$ & $\square(2,2,3,4)_{1730}$ & $[[28800, 242, d_X{=}38, d_Z{=}36]]$ & $576$ & 66816 & 276.1 & 576/576 & $41/121\,(w{=}12{+}16)$ & 34\% & 5.0 & 5.4/8 & 10.89 \\
$V4.4.4.6$ & $\square(2,2,2,3)_{602}$ & $[[33000, 112, 50]]$ & $660$ & 76560 & 683.6 & --- & \text{Non-orientable} & --- & --- & --- & 8.48 \\
$V4.10.6.10$ & $\square(2,5,3,5)_{236}$ & $[[48000, 738, 30]]$ & $960$ & 111360 & 150.9 & --- & \text{Non-orientable} & --- & --- & --- & 13.84 \\
$V4.4.4.6$ & $\square(2,2,2,3)_{843}$ & $[[48600, 164, 60]]$ & $972$ & 112752 & 687.5 & 972/972 & $44/82\,(w{=}16{+}20)$ & 54\% & 4.0 & 5.7/9 & 12.15 \\
\{4,7\} & $\triangle(2,4,7)|\mathrm{PGL}(2,27)$ & $[[61425, 1055, 30]]$ & $1228\,({+}1)$ & 142509 & 135.1 & --- & \text{Non-orientable} & --- & --- & --- & 15.46 \\
\{4,6\} & $\triangle(2,4,6)|\mathrm{PGL}(2,23) \times \mathrm{Z}_{2}$ & $[[75900, 1014, 40]]$ & $1518$ & 176088 & 173.7 & --- & --- & --- & --- & --- & 21.38 \\
$V4.4.4.8$ & $\square(2,2,2,4)|\mathrm{PGL}(2,17)$ & $[[122400, 614, d_X{=}52, d_Z{=}50]]$ & $2448$ & 283968 & 462.5 & --- & \text{Non-orientable} & --- & --- & --- & 12.54 \\
$V4.6.4.6$ & $\square(2,3,2,3)|\mathrm{PSL}(2,19) \times \mathrm{Z}_{2}|1$ & $[[171000, 1142, d_X{=}52, d_Z{=}50]]$ & $3420$ & 396720 & 347.4 & 3420/3420 & --- & --- & --- & --- & 16.70 \\
\{4,10\} & $\triangle(2,4,10)|\mathrm{PGL}(2,31) \times \mathrm{Z}_{2}$ & $[[186000, 4466, 30]]$ & $3720$ & 431520 & 96.6 & --- & \text{Non-orientable} & --- & --- & --- & 21.61 \\
\bottomrule
\end{tabular}%
}
\caption{Modular hyperbolic surface codes for side length $L=5$ with rectangle modules ($K=2$ base quadrilaterals per module). $M_2$: number of rectangle modules (residual single square modules in parentheses if unmatched). $N_{M2}$: total physical qubits in the modular layout including both data qubits and ancilla qubits. $N_{M2}/k$: physical-to-logical qubit overhead. $G_{\mathrm{rot}}$: order of the orientation-preserving rotational symmetry subgroup relative to target. Automorphism Bases: logical basis operator coverage $k_{\mathrm{found}}/g$ and generator seed weights $w$ (pruned to $\ge 85\%$ coverage or highest-coverage basis). Cov.: maximum basis coverage percentage. Pair: mean number of automorphism moves between measuring the two cycles of a symplectic pair with a fixed extractor. Cayley: mean and maximum of the same over all pairs of basis loops. Both are right Cayley graph distances among basis automorphisms, with the extractor placed to minimise Pair (then the maximum, then the mean), for the listed basis with the highest coverage. $kd^2/n$: code figure of merit.}
\label{tab:candidates_L5_surface_ms2}
\end{table}

\begin{table}[H]
\centering
\footnotesize
\setlength{\tabcolsep}{3.5pt}
\resizebox{\textwidth}{!}{%
\begin{tabular}{cccrrrccr}
\toprule
Tiling & Symmetry Group & $[[n,k,d]]$ & $M_{4}$ & $N_{M4}$ & $N_{M4}/k$ & $G_{\mathrm{rot}}$ & Orientable & $kd^2/n$ \\
\midrule
$V4.6.4.6$ & $\square(2,3,2,3)_{8}$ & $[[108, 4, 6]]$ & $3$ & 255 & 63.8 & 6/6 & Yes & 1.33 \\
$V4.8.4.8$ & $\square(2,4,2,4)_{184}$ & $[[144, 6, d_X{=}8, d_Z{=}6]]$ & $4$ & 340 & 56.7 & 8/8 & Yes & 1.50 \\
$V4.8.4.8$ & $\square(2,4,2,4)_{183}$ & $[[144, 6, 6]]$ & $4$ & 340 & 56.7 & 8/8 & Yes & 1.50 \\
$V4.4.8.8$ & $\square(2,2,4,4)_{179}$ & $[[144, 6, 6]]$ & $4$ & 340 & 56.7 & --- & No & 1.50 \\
$V4.10.4.10$ & $\square(2,5,2,5)_{8}$ & $[[180, 8, 6]]$ & $5$ & 425 & 53.1 & 10/10 & Yes & 1.60 \\
$V4.4.4.6$ & $\square(2,2,2,3)_{32}$ & $[[216, 4, 12]]$ & $6$ & 510 & 127.5 & 12/12 & Yes & 2.67 \\
$V4.4.8.8$ & $\square(2,2,4,4)_{303}$ & $[[288, 10, 10]]$ & $8$ & 680 & 68.0 & 16/16 & Yes & 3.47 \\
$V4.4.4.8$ & $\square(2,2,2,4)_{212}$ & $[[288, 6, 12]]$ & $8$ & 680 & 113.3 & 8/16 & Yes & 3.00 \\
$V4.4.6.6$ & $\square(2,2,3,3)_{23}$ & $[[324, 8, 12]]$ & $9$ & 765 & 95.6 & 9/18 & Yes & 3.56 \\
$V4.6.4.6$ & $\square(2,3,2,3)_{14}$ & $[[324, 8, 12]]$ & $9$ & 765 & 95.6 & 18/18 & Yes & 3.56 \\
$V4.6.6.6$ & $\square(2,3,3,3)|\mathrm{GL}(2,3)|2$ & $[[432, 14, 12]]$ & $12$ & 1020 & 72.9 & 24/24 & Yes & 4.67 \\
$V4.6.12.8$ & $\square(2,3,6,4)_{195}$ & $[[648, 29, 12]]$ & $18$ & 1530 & 52.8 & --- & No & 6.44 \\
$V4.6.6.8$ & $\square(2,3,3,4)_{52}$ & $[[648, 23, 12]]$ & $18$ & 1530 & 66.5 & --- & No & 5.11 \\
$V4.8.12.8$ & $\square(2,4,6,4)_{1399}$ & $[[864, 42, 12]]$ & $24$ & 2040 & 48.6 & 48/48 & Yes & 7.00 \\
$V4.6.12.12$ & $\square(2,3,6,6)_{330}$ & $[[864, 42, 12]]$ & $24$ & 2040 & 48.6 & 48/48 & Yes & 7.00 \\
$V4.6.8.8$ & $\square(2,3,4,4)_{150}$ & $[[864, 34, 12]]$ & $24$ & 2040 & 60.0 & 48/48 & Yes & 5.67 \\
$V4.8.6.8$ & $\square(2,4,3,4)_{174}$ & $[[864, 34, 12]]$ & $24$ & 2040 & 60.0 & 48/48 & Yes & 5.67 \\
$V4.4.4.6$ & $\square(2,2,2,3)_{66}$ & $[[864, 10, 18]]$ & $24$ & 2040 & 204.0 & --- & No & 3.75 \\
$V4.10.12.10$ & $\square(2,5,6,5)_{51}$ & $[[1080, 58, 12]]$ & $30$ & 2550 & 44.0 & --- & No & 7.73 \\
$V4.8.10.8$ & $\square(2,4,5,4)_{95}$ & $[[1080, 50, 12]]$ & $30$ & 2550 & 51.0 & --- & No & 6.67 \\
$V4.8.8.8$ & $\square(2,4,4,4)_{2771}$ & $[[1152, 50, 12]]$ & $32$ & 2720 & 54.4 & 64/64 & Yes & 6.25 \\
$V4.8.10.10$ & $\square(2,4,5,5)_{86}$ & $[[1440, 70, 12]]$ & $40$ & 3400 & 48.6 & --- & No & 7.00 \\
$V4.6.4.6$ & $\square(2,3,2,3)_{44}$ & $[[1512, 30, 18]]$ & $42$ & 3570 & 119.0 & --- & No & 6.43 \\
$V4.10.12.12$ & $\square(2,5,6,6)_{273}$ & $[[1620, 89, 12]]$ & $45$ & 3825 & 43.0 & --- & No & 7.91 \\
$V4.12.12.12$ & $\square(2,6,6,6)_{6604}$ & $[[1728, 98, 12]]$ & $48$ & 4080 & 41.6 & 96/96 & Yes & 8.17 \\
$V4.4.4.6$ & $\square(2,2,2,3)_{103}$ & $[[1728, 18, 24]]$ & $48$ & 4080 & 226.7 & 96/96 & Yes & 6.00 \\
$V4.10.10.10$ & $\square(2,5,5,5)_{51}$ & $[[2160, 110, 12]]$ & $60$ & 5100 & 46.4 & 120/120 & Yes & 7.33 \\
$V4.4.10.10$ & $\square(2,2,5,5)_{115}$ & $[[2880, 98, 14]]$ & $80$ & 6800 & 69.4 & --- & No & 6.67 \\
$V4.4.8.8$ & $\square(2,2,4,4)|\mathrm{PSL}(2,7) \times \mathrm{Z}_{2}|2$ & $[[3024, 86, 18]]$ & $84$ & 7140 & 83.0 & 168/168 & Yes & 9.21 \\
$V4.6.6.6$ & $\square(2,3,3,3)_{74}$ & $[[3240, 92, 18]]$ & $90$ & 7650 & 83.2 & 180/180 & Yes & 9.20 \\
$V4.4.6.6$ & $\square(2,2,3,3)_{327}$ & $[[5184, 98, 20]]$ & $144$ & 12240 & 124.9 & --- & No & 7.56 \\
$V4.4.8.10$ & $\square(2,2,4,5)_{517}$ & $[[5760, 178, 18]]$ & $160$ & 13600 & 76.4 & --- & No & 10.01 \\
$V4.4.10.10$ & $\square(2,2,5,5)_{352}$ & $[[6480, 218, 18]]$ & $180$ & 15300 & 70.2 & 360/360 & Yes & 10.90 \\
$V4.6.4.6$ & $\square(2,3,2,3)_{173}$ & $[[6480, 122, 24]]$ & $180$ & 15300 & 125.4 & 360/360 & Yes & 10.84 \\
$V4.4.12.12$ & $\square(2,2,6,6)|\mathrm{PSL}(2,13)|2$ & $[[9828, 366, 18]]$ & $273$ & 23205 & 63.4 & --- & No & 12.07 \\
$V4.4.6.8$ & $\square(2,2,3,4)_{1726}$ & $[[10368, 242, 22]]$ & $288$ & 24480 & 101.2 & 576/576 & Yes & 11.30 \\
$V4.4.6.8$ & $\square(2,2,3,4)_{1730}$ & $[[10368, 242, d_X{=}24, d_Z{=}22]]$ & $288$ & 24480 & 101.2 & 576/576 & Yes & 11.30 \\
$V4.6.10.8$ & $\square(2,3,5,4)_{316}$ & $[[11880, 475, 18]]$ & $330$ & 28050 & 59.1 & --- & No & 12.95 \\
$V4.8.6.8$ & $\square(2,4,3,4)|\mathrm{PGL}(2,11)|7$ & $[[11880, 442, d_X{=}20, d_Z{=}18]]$ & $330$ & 28050 & 63.5 & --- & No & 12.05 \\
$V4.4.4.6$ & $\square(2,2,2,3)_{602}$ & $[[11880, 112, 30]]$ & $330$ & 28050 & 250.4 & --- & No & 8.48 \\
$V4.8.8.8$ & $\square(2,4,4,4)|\mathrm{S}_{6} \times \mathrm{Z}_{2}|20$ & $[[12960, 542, 18]]$ & $360$ & 30600 & 56.5 & 720/720 & Yes & 13.55 \\
$V4.4.10.10$ & $\square(2,2,5,5)_{774}$ & $[[14400, 482, 20]]$ & $400$ & 34000 & 70.5 & 400/800 & Yes & 13.39 \\
$V4.6.10.10$ & $\square(2,3,5,5)_{248}$ & $[[17280, 738, 18]]$ & $480$ & 40800 & 55.3 & --- & No & 13.84 \\
$V4.10.6.10$ & $\square(2,5,3,5)_{236}$ & $[[17280, 738, 18]]$ & $480$ & 40800 & 55.3 & --- & No & 13.84 \\
$V4.4.4.6$ & $\square(2,2,2,3)_{843}$ & $[[17496, 164, 36]]$ & $486$ & 41310 & 251.9 & 972/972 & Yes & 12.15 \\
$V4.4.8.8$ & $\square(2,2,4,4)|\mathrm{PGL}(2,13)|15$ & $[[19656, 548, 22]]$ & $546$ & 46410 & 84.7 & --- & No & 13.49 \\
$V4.4.8.8$ & $\square(2,2,4,4)|\mathrm{PGL}(2,11) \times \mathrm{Z}_{2}|7$ & $[[23760, 662, 22]]$ & $660$ & 56100 & 84.7 & 1320/1320 & Yes & 13.49 \\
$V4.4.8.8$ & $\square(2,2,4,4)|\mathrm{PGL}(2,13) \times \mathrm{Z}_{2}|3$ & $[[39312, 1094, 24]]$ & $1092$ & 92820 & 84.8 & 2184/2184 & Yes & 16.03 \\
$V4.4.4.8$ & $\square(2,2,2,4)|\mathrm{PGL}(2,17)|1$ & $[[44064, 614, d_X{=}32, d_Z{=}30]]$ & $1224$ & 104040 & 169.4 & --- & No & 12.54 \\
$V4.6.4.6$ & $\square(2,3,2,3)|\mathrm{PSL}(2,19) \times \mathrm{Z}_{2}|1$ & $[[61560, 1142, d_X{=}32, d_Z{=}30]]$ & $1710$ & 145350 & 127.3 & 3420/3420 & Yes & 16.70 \\
\bottomrule
\end{tabular}%
}
\caption{Modular hyperbolic surface codes for side length $L=3$ with big square modules ($K=4$ base quadrilaterals per module) on quad tilings where at least one vertex in each base face has valency 4 (i.e. four quadrilaterals meet at the vertex). $M_4$: number of big square modules. $N_{M4}$: total physical qubits in the modular layout including both data qubits and ancilla qubits. $N_{M4}/k$: physical-to-logical qubit overhead. $G_{\mathrm{rot}}$: order of the orientation-preserving rotational symmetry subgroup relative to target. Orientable: whether the manifold is orientable. $kd^2/n$: code figure of merit.}
\label{tab:candidates_L3_surface_ms4_quad4}
\end{table}

\pagebreak
\makeatletter\edef\pre@bibdata{\pre@bibdata}\makeatother
\makeatletter\immediate\write\@auxoutSI{\string\citation{REVTEX42Control}}\makeatother
\makeatletter\immediate\write\@auxoutSI{\string\citation{apsrev42Control}}\makeatother
\makeatletter
\def\NAT@bibsetnum#1{\setlength{\topsep}{\z@}\NATx@bibsetnum{\ref{LastBibItemSI}}}
\def\endthebibliography{\auto@bib@innerbib\edef\@currentlabel{\arabic{NAT@ctr}}\label{LastBibItemSI}\endNAT@thebibliography\aftergroup\auto@bib@empty}
\makeatother
\bibliographystyleSI{apsrev4-2}
\bibliographySI{bibliography}
\end{document}